\documentclass[11pt,a4paper]{report}
\usepackage{graphicx,float}
\usepackage[hidelinks]{hyperref}
\usepackage[margin=1in]{geometry}
\usepackage[toc,page]{appendix}
\usepackage[utf8x]{inputenc}
\usepackage{subcaption}
\usepackage{amsthm}
\usepackage{amsmath}
\usepackage{nameref}
\usepackage{color}
\usepackage{cite}
\usepackage{multido}
\usepackage{comment}
\usepackage{mcite}
\usepackage{amssymb}
\usepackage{amstext}
\usepackage{leftidx}
\usepackage{esint}
\usepackage{footmisc}
\usepackage{datetime}
\usepackage[labelfont=bf,textfont=it]{caption}
\makeatletter

\ams@newcommand{\vardot}[2]{%
 {\mathop{#2\kern0pt}\limits^{\vbox to-1.4\ex@{\kern-\tw@\ex@
  \hbox{\normalfont\multido{}{#1}{.}}\vss}}}}
  
\theoremstyle{definition}
\newtheorem*{definition*}{Definition}

\newcommand{\Thesis}{Dark energy in quantum field theory: Implications on modern cosmology}
\newcommand{\Name}{Cristian Moreno Pulido}
\newcommand{\Supervisor}{Prof. Dr. Joan Sol\`a Peracaula} 
\newcommand{\Department}{Departament de F\'isica Qu\`antica i Astrof\'isica}
\newcommand{\University}{Universitat de Barcelona}

\newcommand{\Mpl}{M_{\rm Pl}}
\newcommand{\mpl}{m_{\rm Pl}}

\newcommand{\CC}{\Lambda}
\newcommand{\bv}{b_{\rm vac}}

\newcommand{\rDE}{\rho_{\rm DE}}

\newcommand{\nueff}{\nu_{\rm eff}}

\newcommand{\oD}{\omega_{\rm BD}}
\newcommand{\dpsi}{\dot{\psi}}
\newcommand{\ddpsi}{\ddot{\psi}}
\newcommand{\wBD}{\omega_{\rm BD}}

\newcommand{\fracdpsipsi}{\frac{\dpsi}{\psi}}

\newcommand{\xr}{x_{\rm r}}
\newcommand{\xm}{x_{\rm m}}
\newcommand{\xl}{x_\Lambda}
\newcommand{\xp}{x_\psi}
\newcommand{\w}{\omega_{\rm BD}}\newcommand{\p}{\prime}
\newcommand{\pp}{{\prime\prime}}

\newcommand{\eBD}{\epsilon_{\rm BD}}
\newcommand{\mPl}{m_{\rm Pl}}
\newcommand{\Geff}{G_{\rm eff}}
\newcommand{\dvphi}{\Delta\varphi}
\newcommand{\weff}{w_{\rm eff}}
\newcommand{\rvphi}{\rho_{\varphi}}
\newcommand{\pvphi}{p_{\varphi}}
\newcommand{\rv}{\rho_{\rm vac}}
\newcommand{\wv}{w_{\rm vac}}
\newcommand{\Pv}{P_{\rm vac}}
\newcommand{\rveff}{\rv^{\rm eff}}
\newcommand{\txi}{\tilde{\xi}}

\newcommand{\rI}{\rho_I}

\newcommand{\tHI}{\tilde{H}_I}
\newcommand{\tal}{\tilde{\alpha}}
\newcommand{\rvo}{\rho^0_{\rm vac}}

\newcommand{\Om}{\Omega_{\rm m}}
\newcommand{\zstar}{z_{*}}
\newcommand{\astar}{a_{*}}
\newcommand{\Omo}{\Omega^0_{\rm m}}

\newcommand{\OLo}{\Omega^0_{\Lambda}}

\newcommand{\rco}{\rho^0_{c}}
\newcommand{\rmo}{\rho_{m 0}}

\newcommand{\wm}{\omega_{\rm m}}

\newcommand{\rL}{\rho_{\CC}}

\newcommand{\cH}{\mathcal{H}}
\newcommand{\cpH}{\mathcal{H}^\prime}

\newcommand{\GB}{\mathfrak{G}}

\newcommand{\bk}{{\bf k}}
\DeclareUnicodeCharacter{2212}{-}
\def\blankpage{%
      \clearpage%
      \null%
      \clearpage}
\def\blankpageinitial{%
      \clearpage%
      \thispagestyle{empty}%
      \addtocounter{page}{-1}%
      \null%
      \clearpage}

\begin{document}

\renewcommand{\thechapter}{\arabic{chapter}}
\setcounter{section}{0}
\begin{titlepage}
\centering

\vspace{5pt}
\begin{LARGE}
\textbf{Doctoral Thesis}
\end{LARGE}
\vspace{12pt}
\vspace*{\fill}

\hrule
\vspace{5pt}
{\Huge \textbf{\Thesis}}
\vspace{12pt}
\hrule
\vspace*{\fill}

\begin{LARGE}

\textbf{\Name}\\
\Department\\
\University\\
\vspace{12pt}
\monthname[3], 2023
\end{LARGE}
\vspace*{\fill}

\begin{figure}[H]
\centering
\includegraphics[width=0.6\textwidth]{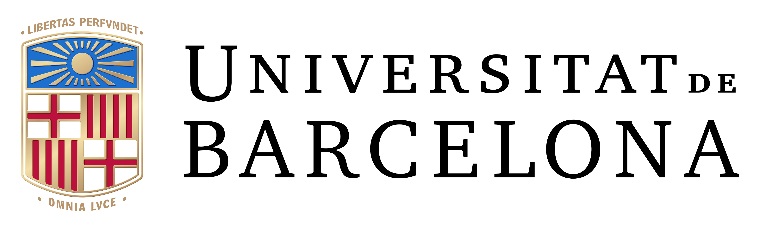}
\end{figure}
\vspace{\fill}

\end{titlepage}
\blankpageinitial

\begin{titlepage}
\centering

{\Large \textbf{\Thesis}}

\vspace{12pt}
{per}
\vspace{12pt}

\begin{LARGE}
\textbf{\Name}
\end{LARGE}

\vspace{12pt}

{Memòria presentada per optar al grau de doctor per la\\}
{Universitat de Barcelona.\\}

\vspace{12pt}

{Programa de doctorat en Física. Linia de recerca: Partícules i gravitació\\}

\vspace{12pt}

{Aquesta tesi ha estat realitzada sota la supervisió del\\}

\vspace{12pt}

\begin{LARGE}
\textbf{\Supervisor}
\end{LARGE}
\vspace{12pt}

catedràtic del departament de Física Quàntica i Astrofísica de la Universitat de Barcelona, i sota la tutela del {\bf Prof. Dr. Joan Soto Riera}, catedràtic del departament de Física Quàntica i Astrofísica de la Universitat de Barcelona.

\vspace{\fill}
{Aquesta tesi s'ha dut a terme al Departament de Física Quàntica i Astrofísica de la Universitat de Barcelona i a l'Institut de Ciències del Cosmos de la Universitat de Barcelona.}

\vspace{\fill}
\begin{LARGE}
\vspace{12pt}
Barcelona, Març de 2023
\end{LARGE}

\vspace{\fill}
\begin{figure}[H]
\centering
\includegraphics[width=0.6\textwidth]{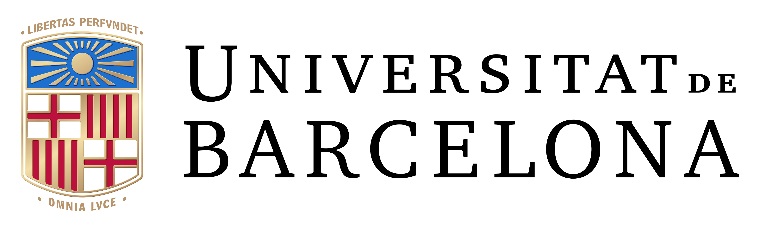}
\end{figure}
\vspace{\fill}

\end{titlepage}

\blankpageinitial
\thispagestyle{empty}
\clearpage
\vspace*{\fill}
\begin{center}
\begin{minipage}{.4\textwidth}
{\it A mi padre, por haberme descubierto el amor por la ciencia y las grandes preguntas.}
\end{minipage}
\end{center}
\vfill 
\clearpage

\blankpageinitial


\pagenumbering{arabic}


\chapter*{\centering Acknowledgements}

In 2018, I embarked on my PhD journey with great joy and passion for Physics and Cosmology, and these past few years have only strengthened my love for science. It has been an intense time, with better and worse moments, but overall, it has been a gratifying and enriching experience that I will treasure for the rest of my life. Without any doubt, I would not have come this far if it were not for the support of all the people around me.

First of all, let me start by thanking my PhD supervisor, Prof. Joan Solà Peracaula. I simply cannot express enough gratitude for his infinite patience, dedication, and guidance throughout these years. He has been an excellent example of how to be a great scientist in all aspects, and has taught me how to appreciate the beauty of physics beyond the mere instrument of formulas. Thank you for accepting to be my PhD supervisor, I will always remember that you gave me the opportunity to enter the world of research. I will always be grateful for teaching me how to do physics rigorously and with commitment, and for  showing me the personal and professional values that a physicist must possess. It has been a great honour to be your student, Joan.

Thanks to my great friend and collaborator, Javier de Cruz Pérez, for all of his help throughout these years. He is an excellent example of a hard worker and talented physicist, as well as an outstanding colleague. Thank you for your tips and detailed explanations whenever I required them. I have learned a lot from you, and you have been an inspiration to me.

Adrià Gómez-Valent has been an outstanding collaborator and colleague. It has been a pleasure to work with you throughout these years, thanks for all your help. You are another great example of the kind of physicist I aspire to be.

Samira, another excellent collaborator, thank you for everything. I admire your commitment and your ability to pursue your goals with determination. It has been a pleasure to collaborate with you over these years.

Thanks to Prof. Joan Soto for accepting to be the tutor of my thesis.

I am grateful to Prof. Nikolaos Mavromatos, Prof. Jaume Garriga, and Dr. Eleonora Di Valentino for agreeing to be part of my PhD committee. I also want to thank Prof. Federico Mescia, Dr. Jordi Salvadó, Dr. Jorge Casalderrey and Dr. Emmanuel Saridakis for agreeing to be substitute members of the committee and for their help along the years.

I would like to express my gratitude to Prof. Concha González and Prof. Josep Maria Paredes for their help as the principal investigator of the group and head of department, respectively.

Thanks to my beloved University of Barcelona, where I began my studies as an undergraduate many years ago. I have truly enjoyed my time here, and I am grateful to all the people who keep this place alive. From the professors I have had the pleasure of learning from, to the secretarial staff and cleaning personnel, everyone plays an essential role in making our research possible.

I would also like to express my sincere gratitude to the University of Sussex for hosting my three-month PhD stay and for the warm welcome I received there. It was a pleasure to spend time in Brighton (UK) and to meet so many wonderful people. Many thanks to Prof. Xavier Calmet for making this opportunity possible and for accepting me as a visitor student, and for his valuable contribution as an international referee of this thesis.

I am grateful to Prof. Leandros Perivolaropoulos from University of Ioannina  (Greece) for accepting the role of international referee for my thesis. 

My gratitude to the Catalan Government for funding my PhD research period.

Thank you to all the colleagues that I have met during my time as a graduate student: Carlos, Iñigo, Iván, Diego, Mikel, Paula, Marija, Marc, Itzi, Miguel, Aniol, Dani. Your friendship has been essential in helping me to fulfill my commitment.

A Rosi y a Domingo, mis padres, que nunca me han fallado, y que siempre han dado todo por sus hijos. Gracias por darme la vida y llenarla de alegría, por vuestro ejemplo, por todas vuestras enseñanzas que han permitido que llegue hasta donde estoy hoy y por estar a mi lado en cada momento importante de mi vida.

Sin ti, Lucía, esta tesis no habría sido posible. Has sido el faro que ilumina mi vida durante todos estos años, apoyándome en los malos momentos y siempre motivándome a seguir adelante. No hay palabras que puedan agradecerte suficientemente. La vida sin ti y sin Blava no tendría sentido.

Gracias por la familia que me ha tocado, por su apoyo incondicional a lo largo de todos estos años y por la unión que mantenemos bajo cualquier circunstancia.

Gracias a mi hermano Javi, por sus consejos, por su ayuda incondicional, por ser mi referente en la vida y ser el tipo de persona que me gustaría ser. Gracias a mi querida cuñada, por todo su apoyo durante todos estos años. A mi amado sobrino Oliver, y la maravillosa alegría que ha dado a nuestras vidas.

A mis queridos suegros, Jose y Emilio, por brindarnos todo su apoyo y amor y por ser un pilar muy importante en mi vida.

Muchas gracias a mi estimado amigo Rodrigo. If anything goes wrong, you will be my constant. Gracias también a tu maravillosa familia: Laura, mi amiga y vecina, y a mi querida ahijada Aina.

Dar las gracias a todos mis amigos de la infancia: David, Carreras, Dani, Pedro, Ariadna, Rubén, Javi, Luismi, Aitor, Jordi, Joel, Jose, Yael. Por todo vuestro amor y apoyo. Espero que nunca me faltéis.

Moltes gràcies als meus estimats amics Germán i Bernat per les nostres confidències i per ser elements indispensables de la meva vida, sé que sempre puc comptar amb vosaltres.

Gràcies a tots els meus companys del doble grau per la seva amistat i per tota la nostra trajectòria junts, no hauria arribat tan lluny si no fos per vosaltres.

To all the people who have made a positive impact on my life, whether big or small, thank you for being a part of my journey!

\pagebreak



\chapter*{\centering Preface}

Most of the energy density content of the Universe (around $70\%$) is not baryonic matter, radiation, or Dark Matter, but an entity or substance whose true nature is unclear and called {\it Dark Energy} (DE). The name reflects its lack of interaction with light and our ignorance regarding its provenance. To clarify the terminology, DE is responsible for the accelerated expansion of the Universe, which was confirmed 25 years ago, and its canonical formulation is a Cosmological Constant in Einstein's field equations. According to the standard vision of DE, it is a form of energy that possesses negative pressure, and its gravitational interaction with matter produces a repulsive effect. Since DE is currently the prominent component, it dominates the dynamical behavior of spacetime at large scales, resulting in the accelerated expansion mentioned earlier.

The most common belief is that  DE  may correspond to vacuum energy. This is a subtle concept in cosmology that has challenged theoretical physicists and cosmologists for many decades, especially with the advent of Quantum Theory. The problem stems from the interpretation of the {\it Cosmological Constant} (CC) or Cosmological Term, $\CC$, in Einstein's equations as a term connected with the notion of vacuum energy density (VED), $\rv$, a fundamental concept in quantum field theory (QFT). In particular, some contributions to the vacuum energy may come from the {\it zero-point energy} (ZPE) of quantum matter fields of the Standard Model of particle physics, as well as the vacuum expected value of the Higgs potential, among other contributions. However, the usual identification of DE with vacuum energy poses a troublesome enigma known as the CC Problem. This problem is traditionally formulated as the huge difference between its observed energy density associated value, $\rho_{\rm DE} \sim 10^{-47}$ ${\rm GeV}^4$, and the naive predictions of its value within the QFT framework, which may be proportional to the fourth power of Planck's mass, $\Mpl^4\sim 10^{76}$ ${\rm GeV}^4$. This accounts for a difference of 123 orders of magnitude! However, we do not have to reach such a high energy scale to face a gargantuan problem. Even a typical contribution from the Standard Model to the vacuum energy budget can pose a problem when compared to $\rho_{\rm DE}$. For instance, the zero-point energy of the electron field contributes around $\rho_{\rm ZPE}\sim m_{\rm e}^4 \sim 10^{-13}$ ${\rm GeV}^4$, while the ground state energy of the effective Higgs potential, $\left\langle V_{\rm eff} \right\rangle$, is around $-10^8$ ${\rm GeV}^4$. Even with this reduction from 123 to ``only'' 34 or 56 orders of magnitude, there is still a big difference that leads to the famous {\it fine-tuning} of the formulas to match our observations. In fact, the fine-tuning problem seems to persist in all forms of dark energy and is not limited only to the CC. This means that the essence of the problem does not lie in the simplicity of the CC model, but in the very conception of dark energy and its relation to fundamental physical theories. It is of paramount importance to point out this fact and to not only focus on the mathematical modeling of dark energy, but to go beyond and understand that the true problem resides in the correct interpretation of dark energy in any of its forms within the context of QFT or, eventually, in quantum extensions of gravitational theory. The approach presented in this thesis follows this philosophy: rather than try to solve the CC Problem  in its fullness, by means of intrincated mechanisms or with the help of ad hoc fields, we try to interpret the calculations in order to obtain a satisfactory physical result that can connect with reality and the cosmological observations.

The precise connection between the CC and the current value of dark energy is given by the equation $\rho_{\rm DE}^0=\CC/(8\pi G_N)$, where $G_N$ is the locally measured Newton's constant and the superindex 0 indicates that the quantity is taken at the present. It is important to note that in cosmological modeling, the CC is a mathematical term that does not have a clear physical interpretation. Rather, it is a parameter that is used to produce the desired effect in  Einstein's field equations. Despite initially being disregarded by evidence, the CC is now considered a fundamental component of the $\Lambda$-Cold-Dark-Matter model ($\Lambda$CDM), which is the Standard Model of Cosmology. The $\Lambda$CDM model assumes the existence of Dark Energy, Dark Matter, and an almost scale-invariant Primordial Power Spectrum. While the model is not without its flaws, it provides the most accurate phenomenological description of overall observations, including cosmic acceleration, chemical element abundances, structure formation, and the cosmic microwave background.

The CC problem is one of the biggest dilemmas that theoretical physics faces. There is a lot of uncertainty regarding the nature of $\Lambda$ and its origin at a fundamental level. However, as scientists, it is natural for us to seek to reconcile equations with reality, and it is inevitable to try to determine what kind of entity constitutes $\Lambda$ or, more generically, Dark Energy and how it behaves. Is it an exotic fluid that permeates the entire cosmos, or is it just a pure geometrical effect? Although the literature on this issue is vast, we may still be far from being able to answer the question.

Another intriguing aspect is the fact that DE and matter have comparable energy densities at the present time, namely $\rho_{\rm DE}^0/\rho_{\rm m}^0\sim \mathcal{O}(1)$. This ratio is strange because $\rho_{\rm m}$ and $\rho_{\rm DE}$ are expected to have very different evolutions with the expansion. While $\rho_{\rm DE}$ is usually assumed to be constant with time, $\rho_{\rm m}$ is inversely proportional to the physical volume, which increases with the expansion, and hence there is in principle no a priori reason why at this time it is of order of $\rho_{\rm DE}$. The unexpected naturalness of this ratio at our times is also is the so-called {\it coincidence problem} and is also a source of mystery.

Nowadays, we are in a golden era of abundant and accurate observations in cosmology. However, this seemingly ideal situation may also bring us nightmares, as there are observational problems that pose challenges to cosmologists, in addition to theoretical ones.  The precision data surveys have revealed conflicts between the results of different experiments, which could shake the foundations of the $\Lambda$CDM model. The most significant challenge is the value of the current Hubble parameter, $H_0$. The local observations from {\it SH0ES} collaboration, which stands for {\it Supernova, $H_0$, for the Equation of State of Dark Energy}, show a discrepancy of $\sim 5\sigma$ compared to the results of the {\it Planck Collaboration} studying the Cosmic Microwave Background, originated in the early Universe. Another crucial parameter that displays a significant discrepancy is $\sigma_8$, which is the root mean square of fluctuations in density perturbations at the 8 $h^{-1}$ Mpc scale and is closely related to the growth of structure. In this case, the discrepancy is less intense but still severe, at around $2-3\sigma$, and it arises from weak lensing surveys that conflict with the results of the {\it Planck Collaboration}. Other tensions, such as the lensing amplitude $A_{\rm L}$ and other minor anomalies in the Cosmic Microwave Background anisotropies, are also under debate. It is still unknown whether these {\it cosmological tensions} result from systematic errors in observations or are signals of the existence of new physics. If these issues were not enough, the community has also cast doubts on the Cosmological Principle itself due to the presence of unexpected large structures, the appearance of cosmic dipoles that cannot be entirely explained by kinematics, and other observations. Although we maintain a skeptical state and all the calculations of this work account for the Cosmological Principle, in some studies, the violation of this paradigm reaches the $\sim 4\sigma$ level, meaning that it is a vivid possibility that we have to keep in mind and could potentially relate to the former tensions.

This dissertation is a summary of our investigations on the nature of DE both from theoretical and phenomenological perspective under the direction of Prof. Joan Solà Peracaula. This means to go beyond the Standard Cosmological Model by exploring the possibility that DE is in fact a dynamical quantity in QFT in curved spacetime, namely a DE which is the result of the fluctuations of the quantum vacuum in the Universe, and hence evolving with the background expansion rather than being a true cosmological constant. Our aim is to go beyond the usual approach based on just ad hoc scalar fields, such as quintessence, phantom fields and the like, and rather treat the DE as quantum vacuum under appropriate QFT renormalization. This possibility is consistent with the cosmological principle, which requires homogeneity and isotropy at large scales, or at least to a great extent. But even if such a principle has been partially disputed recently, this is not in conflict with the possible dynamical evolution of the DE, quite the opposite, as in fact it could even acquire an enhanced protagonism in such an extended context, although a possible violation of the cosmological principle is not considered in this dissertation, as said. Our interest is in examining the implications of these ideas for the fundamental questions surrounding the cosmological constant and its connection with QFT, and hence with the fundamental principles of theoretical Physics. Despite the obvious relevance of these fundamental questions, our interest extends beyond technical and theoretical details, as we keep in mind that in Physics experimental evidence is the ultimate arbiter of reality.  Therefore, testing our models against available evidence is a must in our work, with particular attention to the aforementioned cosmological tensions.

The structure of the work is explained in the following paragraphs. The \hyperref[Chap:Intro]{first chapter} provides an introduction to Vacuum Energy in Cosmology, which includes the history of the cosmological constant and Vacuum Energy in QFT, and an outline of the main features of today's cosmology. In \hyperref[Chap:QuantumVacuum]{Chapter\,\ref{Chap:QuantumVacuum}}, the first part of our research involving the study of Vacuum Energy Density (VED) in QFT is shown. What is presented here is a derivation from first principles of the VED evolution with the expansion. It is based on an extension of the traditional adiabatic regularization technique present in many classical textbooks of QFT in curved spacetime. Taking that method as a starting point, we extend and systematize the procedure in a way that allows us to renormalize the vacuum energy-momentum tensor off-shell and hence the vacuum energy density itself in a manner which is unprecedented in the literature. We perform these computations for a free scalar field non-minimally coupled to curvature in full detail and by different paths, including the renormalization of the effective action. Despite the fact that our calculations are done in a specially simple context with a free field, we are convinced that our procedure captures the most essential points which, among other results, state that vacuum energy density in the expanding Universe has a dynamical character and evolves smoothly in terms of powers of the Hubble function, $H$, and its derivatives. This evolution corresponds to the family of Running Vacuum Models (RVM), which have been around for some years and were originally justified by general renormalization group arguments. Models exploring the possibility of DE evolving with a dynamical quantity are not new in the literature, but their rigorous derivation from QFT principles is not so common. Much less common is to revive the quantum vacuum and present it as the ultimate cause of the (dynamical) DE despite the aforementioned CC problem.  This is possible thanks to this new approach to the adiabatic renormalization procedure developed here.

Related with the previous results, in \hyperref[Chap:EoSVacuum]{Chapter\,\ref{Chap:EoSVacuum}} we investigate the structure of vacuum's pressure in an analogous procedure. This allows us to find an important result, the equation of state of the quantum vacuum (the relation between VED and its pressure), which is computed here from first principles, namely from QFT in curved spacetime in our case, and it is seen to depart from its traditional constant value: $w_{\rm vac}=-1$. Remarkably,  our investigations lead us to a new mechanism of inflation, which is completely unrelated to the traditional ones based on scalar fields (e.g. using inflatons or scalarons, etc.). The mechanism is again based on the quantum effects from the quantized matter fields in Friedmann-Lemaître-Robertson-Walker (FLRW) spacetime, which we have put forward here for the first time in our QFT context. In \hyperref[Chap:Fermions]{Chapter\,\ref{Chap:Fermions}} we generalize our previous results for non-minimally coupled scalar fields in the presence of an arbitrary number of species of them  and including also an arbitrary number of spin 1/2-fields using the same techniques described in previous chapters. Altogether it  implies a rather nontrivial extension of the calculation previously performed for the scalar fields non-minimally coupled to gravity since the calculations with fermions (spinor fields) in curved spacetime involve considerable technical difficulties.

The last two chapters, \hyperref[Chap:PhenomenologyofBD]{Chapter\,\ref{Chap:PhenomenologyofBD}} and \hyperref[Chap:PhenomenologyofRVM]{Chapter\,\ref{Chap:PhenomenologyofRVM}}, describe our phenomenological investigations. In \hyperref[Chap:PhenomenologyofBD]{Chap.\,\ref{Chap:PhenomenologyofBD}}, we present a detailed exploration of the {\it Brans-Dicke} (BD) model, which is the first and simplest example of a modified gravity theory in the context of scalar field theories. This modification of General Relativity includes a new scalar degree of freedom that mediates gravitational strength, with the scalar field providing a time-varying Newton's gravitational constant. We test the model using different datasets and scenarios and present the outcomes of the fits, discussing them in relation to cosmological tensions. In \hyperref[Chap:PhenomenologyofRVM]{Chap.\,\ref{Chap:PhenomenologyofRVM}}, we study a family of models called Ricci Running Vacuum Models (RRVM), a variation of the traditional RVM, where the vacuum energy density evolves with the Ricci scalar, $R$. We properly fit the models studied there and report promising results.

The final reflections of this dissertation are presented in the \hyperref[Conclusions]{Conclusions}, where we summarize the contents and the results obtained in the different chapters of the thesis and do some final comments regarding the CC Problem.

At the end, some appendices have been added for completeness and to avoid a flood of large formulas and some technical explanations that can be omitted in the main chapters.  \hyperref[Appendix:Conventions]{Appendix\,\ref{Appendix:Conventions}} contains useful formulas, sign conventions, and expressions that are used throughout the dissertation. \hyperref[Appendix:Dimensional]{Appendix\,\ref{Appendix:Dimensional}} explains how dimensional regularization may be used in conjunction with adiabatic regularization in calculations related to the renormalization of the VED in \hyperref[Chap:QuantumVacuum]{Chapter\,\ref{Chap:QuantumVacuum}}. \hyperref[Appendix:Abis]{Appendix\,\ref{Appendix:Abis}} explores the running of the VED and the gravitational constant in more detail than in the main text.

\hyperref[Appendix:AdiabExpFermModes]{Appendix\,\ref{Appendix:AdiabExpFermModes}} and \hyperref[Appendix:AdiabExpFermEMT]{Appendix\,\ref{Appendix:AdiabExpFermEMT}}  contain lengthy formulas related to the adiabatic expansion of the Fourier modes and the Energy-Momentum tensor of fermion fields of \hyperref[Chap:Fermions]{Chapter\,\ref{Chap:Fermions}}, respectively. We chose not to include these formulas in the main text for clarity. \hyperref[Appendix:Semi-Analytical]{Appendix\,\ref{Appendix:Semi-Analytical}} and \hyperref[Appendix:FixedPoints]{\ref{Appendix:FixedPoints}} provide two alternative ways to inspect the BD Model through semi-analytical and fixed points techniques of the equations of motion that govern this cosmological model. These techniques provide insight into the behavior of the model.

In \hyperref[Appendix:CosmoPerturbationsSynchronous]{Appendix\,\ref{Appendix:CosmoPerturbationsSynchronous}} and \hyperref[Appendix:PerturbationTheoryNewtonian]{Appendix\,\ref{Appendix:PerturbationTheoryNewtonian}} we summarize the perturbation equations of the Brans-Dicke model in Synchronous and Newtonian gauges, respectively, at linear order. Finally, \hyperref[Appendix:Bayesian]{Appendix\,\ref{Appendix:Bayesian}} is a small overview of Bayesian Statistics and Model Selection, while \hyperref[Appendix:Description]{Appendix\,\ref{Appendix:Description}} provides a brief description of the different data sources used in the analysis of the models.

The CC has been a double-edged sword in the cosmological puzzle, presenting an immeasurable challenge while also raising questions that have baffled the physics community for decades. Throughout this entire period of research, these disquisitions we explored were fascinating to examine. However, the results have also led to new questions and uncertainties, emphasizing the endless path we have yet to follow. As the author of this dissertation, I would like to express my gratitude to the reader for their interest in this work and hope that they will find it enjoyable to read.

\newpage

\chapter*{\centering Resum de la tesi}

La {\it Constant Cosmològica}, $\Lambda$, ha sigut un element controvertit des que Einstein les va introduir en les seves pròpies equacions de camp al 1917. Tot i mantenir una reputació irregular al llarg dels anys, el descobriment de l'expansió accelerada de l'Univers a finals del segle XX va confirmar que $\Lambda$ és un ingredient vital del trencaclosques cosmològic, formant part del model estàndard de cosmologia, el $\Lambda$CDM. Encara que el $\Lambda$CDM és capaç d'acomodar la major part de les observacions, presenta diversos problemes de caràcter teòric i fenomenològic que necessiten una atenció urgent.
\\
\\
\noindent
 La causa de l'expansió accelerada s'anomena de manera general {\it Energia Fosca}, la qual es modelitza matemàticament amb $\Lambda$, i que té un origen incert, tot i que s'acostuma a assumir que és energia de buit. Estimacions teòriques genèriques de la densitat d'energia de buit en el context de la teoria quàntica de camps es diferencien de les observacions fins a 123 ordres de magnitud en el pitjor dels casos. Addicionalment, els intents d'ajustar el seu valor col$\cdot$leccionant diverses contribucions de buit han sigut un fracàs, produint el denominat problema de {\it fine-tuning}. Aquesta incapacitat per derivar el resultat observacional de l'energia de buit constitueix el {\it Problema de la Constant Cosmològica}, un dels majors misteris al qual s'enfronta la física teòrica. El problema encara és més gran si un considera el fet que l'energia fosca i la matèria tenen una densitat del mateix ordre de magnitud en el present, tot i que la densitat d'energia de buit és constant, mentre que la matèria es dilueix amb l'expansió. Aquest problema és conegut com el problema de {\it coincidència}.
\\
\\
\noindent
Per si això no fos suficient, també hi ha problemes des de l'àmbit fenomenològic. Específicament, hi ha tensions cosmològiques entre les observacions de l'univers primigeni i observacions locals, afectant dos paràmetres importants en el model cosmològic. El primer és $H_0$ (la funció de Hubble o rati d'expansió en el present) i el segon és $\sigma_8$ (relacionat amb la formació d'estructura en l'Univers). Les discrepàncies poden arribar a $4-5\sigma$ i $2-3\sigma$, respectivament.
\\
\\
\noindent
Inspirats per aquests reptes, el treball realitzat sota la direcció del Prof. Joan Solà Peracaula ha seguit dos camins diferents, però íntimament relacionats que queden reflectits en dues parts diferenciades de la tesi. Després de la introducció al primer capítol, els Capítols 2, 3, 4 formen un primer bloc on mostrem les nostres investigacions respecte a la renormalització i regularització de la densitat d'energia de buit en el context de la teoria quàntica de camps a través d'un nou formalisme que estén la tradicional regularització adiabàtica. Hem obtingut resultats significatius i valuosos en relació amb el comportament dinàmic de l'energia de buit, la qual sembla evolucionar suaument amb l'expansió en termes de la funció  de Hubble, $\rho_{\rm vac}(H)$. Aquests resultats coincideixen amb els coneguts Running Vacuum Models (RVM), els quals han estat presents a la literatura des de fa uns anys, però que ara queden justificats amb arguments més rigorosos.
\\
\\
\noindent
En el segon bloc (capítols 5 i 6), amb les tensions cosmològiques en ment, vam confrontar dos models relacionats amb les investigacions teòriques prèvies contra un gran conjunt de dades cosmològiques per tal de restringir els paràmetres cosmològics:

\begin{itemize}

\item[1)] El model de Brans-Dicke, al capítol 5, consisteix en una modificació  de Relativitat General promocionant la constant gravitacional a un grau de llibertat escalar. Això pot ser reformulat com un model efectiu de Relativitat General amb una component de buit dinàmica, similar al RVM.

\item[2)] El  Ricci-RVM, al capítol 6, és una variació del tradicional RVM en el qual reemplacem la dependència en $H$ per $R$, l'escalar de Ricci. Això té alguns avantatges com ara no pertorbar les prediccions usuals de la teoria de Big Bang Nucleosynthesis.

\end{itemize}
\noindent

Les conclusions que segueixen a aquests capítols resumeixen els resultats principals i algunes reflexions finals. La tesi està complementada per una sèrie d'Apèndixs que estenen la informació dels capítols principals.

Resumint, aquesta tesi presenta una investigació rigorosa de la possible desviació respecte del paradigma del model $\Lambda$CDM, considerant la possibilitat que la densitat d'energia de buit sigui una quantitat dinàmica i amb una evolució ben determinada per la teoria quàntica de camps. Aquesta predicció es mostra de manera detallada i acabem obtenint resultats sorprenents i sense precedents en la literatura. Addicionalment, explorem dos models basats en aquesta dinàmica de buit contra diferents conjunts de dades i escenaris per tal d'aconseguir una perspectiva més àmplia. Els nostres fits mostren resultats prometedors envers una possible desviació respecte del tradicional model $\Lambda$CDM.

\newpage
\chapter*{\centering List of Publications}

List of publications during the PhD period:

\begin{itemize}
 \item[1)]{\it ``Brans–Dicke Gravity with a Cosmological Constant Smoothes Out $\Lambda$CDM Tensions.''}\\
 			J. Solà Peracaula, A. Gómez-Valent, J. de Cruz Pérez and C. Moreno-Pulido\\
 		    Astrophys. J. Lett. 886, L6 (2019); [arXiv:1909.02554]
 	     	    
 \item[2)]{\it ``Running vacuum in quantum field theory in curved spacetime: renormalizing $\rho_{\rm vac}$ without $\sim m^4$ terms.''}\\
 		   C. Moreno-Pulido and J. Solà Peracaula\\
 		   Eur. Phys. J.C 80 (2020); [arXiv:2005.03164]      
 
 \item[3)] {\it ``Brans-Dicke cosmology with a $ \Lambda$-term: a possible solution to $\Lambda$CDM tensions.''}\\
			J. Solà Peracaula, A. Gómez-Valent, J. de Cruz Pérez and C. Moreno-Pulido\\
			Class. Quant. Grav. 37, 245003 (2020); [arXiv:2006.04273]
			
  \item[4)] {\it ``Running vacuum against the $H_0$ and $\sigma_8$ tensions.''}\\
     		J. Solà Peracaula, A. Gómez-Valent, J. de Cruz Pérez and C. Moreno-Pulido\\
     		EPL 134 (2021) 1, 1900; [arXiv:2102.12758]
    
   \item[5)]  {\it `` Renormalizing the vacuum energy in cosmological spacetime: implications for the cosmological constant problem.''}\\
   			C. Moreno-Pulido and J. Solà Peracaula\\
   			 Eur.Phys.J.C 82 (2022) 6, 551; [arXiv:2201.05827]
   			 
   \item[6)] {\it ``Equation of state of the running vacuum.''}\\
   			C. Moreno-Pulido and J. Solà Peracaula\\
   		    Eur.Phys.J.C 82 (2022) 12, 1137; [arXiv:2207.07111]
   		    
   \item[7)] {\it ``Running Vacuum in QFT in FLRW spacetime: The dynamics of $\rho_{\rm vac}(H)$ from the quantized matter fields.''}\\
			C. Moreno-Pulido, J. Solà Peracaula and S.Cheraghchi\\
			    Eur.Phys.J.C 83 (2023) 7, 637; [arXiv:2301.05205]
			
	\item[8)] {\it ``Running vacuum in the Universe: phenomenological status in light of the latest observations, and its impact on the $\sigma_8$ and $H_0$ 					tensions.''}\\
     			J. Solà Peracaula, A. Gómez-Valent, J. de Cruz Pérez and C. Moreno-Pulido\\
     			    Universe 9 (2023) 6, 262; [arXiv:2304.11157]
			
\end{itemize}

This is the list of papers where I have participated within a large collaboration:

\begin{itemize}

 \item[9)] {\it ``Snowmass2021 - Letter of interest cosmology intertwined I: Perspectives for the next decade.''}\\
 		   E. Di Valentino {\it et al.}\\
 		   Astropart. Phys. 131 (2021), 102606; [arXiv:2008.11283] 
 	     	    
 \item[10)] {\it ``Snowmass2021 - Letter of interest cosmology intertwined II: The hubble constant tension.''}\\
 		   E. Di Valentino {\it et al.}\\
 		   Astropart. Phys. 131 (2021), 102605; [arXiv:2008.11284]      
 
 \item[11)] {\it ``Cosmology Intertwined III: $f \sigma_8$ and $S_8$''}\\
 		   E. Di Valentino {\it et al.}\\
 		   Astropart. Phys. 131 (2021), 102604; [arXiv:2008.11285]     
			
 \item[12)] {\it ``Snowmass2021 - Letter of interest cosmology intertwined IV: The age of the universe and its curvature.''}\\
 		   E. Di Valentino {\it et al.}\\
 		   Astropart. Phys. 131 (2021), 102607; [arXiv:2008.11286]    
 		    
  \item[13)] {\it ``Cosmology intertwined: A review of the particle physics, astrophysics, and cosmology associated with the cosmological tensions and 		anomalies.''}\\
 		   E. Abdalla {\it et al.}\\
 		   JHEAp 34 (2022), 49; [arXiv:2203.06142]     
   			 
\end{itemize}

This is the list of proceedings from conferences in which I have participated:
\begin{itemize}

 \item[14)] {\it ``Renormalized $\rho_{\rm vac}$ without $m^4$ terms''}\\
 		  C. Moreno-Pulido and J. Solà Peracaula\\
 		   16th Marcel Grossmann Meeting (2021); [arXiv:2110.08070] 
 		   
 \item[15)] {\it ``BD-$\Lambda$CDM and running vacuum models: Theoretical background and current observational status''}\\
 		  J. de Cruz Pérez,  J. Solà Peracaula, A. Gómez-Valent and C. Moreno-Pulido\\
 		   16th Marcel Grossmann Meeting (2021); [arXiv:2110.07569] 
 		   
  \item[16)]{\it ``Quantum vacuum, a cosmic chameleon''}\\
  		C. Moreno-Pulido, J. Solà Peracaula\\
  		 	Proc. of the Corfu Summer Institute 2022: Workshop on Tensions in Cosmology  
\end{itemize}

\tableofcontents

\blankpage
\blankpage

\chapter{Introduction}\label{Chap:Intro}

In this first introductory chapter, the aim is to contextualize what will be presented in subsequent chapters. The literature on the Cosmological Constant (CC) and Vacuum Energy is vast, so it is logical to review some historical and theoretical insights. On the other hand, there is currently strong research in Cosmology aimed at including new elements or modifying the traditional Standard Model of Cosmology\footnote{Throughout this work, we will refer to $\Lambda$CDM simply as the {\it Concordance Model} or {\it Standard Model}. Referring to it as the Standard Model may cause confusion, as the same short name is given to the Standard Model of Particle Physics. We will make it clear in the text which model we are referring to, if necessary.}, or $\Lambda$CDM, in a specific direction. This is also the case of our research. For this reason, the basics of the $\Lambda$CDM will be briefly reviewed, and we will argue the reasons that motivate us to go beyond it.

More specifically, the chapter starts with \hyperref[Sect:HistoryCosmConst]{Sect.\,\ref{Sect:HistoryCosmConst}}, where the history of the CC is summarized, starting with Sir Isaac Newton and leading up to its modern role. It is followed by \hyperref[Sect:VacuumEnergyReview]{Sect.\,\ref{Sect:VacuumEnergyReview}}, where we summarize the physical interpretation of vacuum energy in Quantum Field Theory (QFT) and Cosmology. In \hyperref[Sect:BasicsCosmology]{Sect.\,\ref{Sect:BasicsCosmology}}, the basics of Modern Cosmology, such as the $\Lambda$CDM and Inflation, are reviewed for completeness. In \hyperref[Sect:BeyondSM]{Sect.\,\ref{Sect:BeyondSM}}, we present the main reasons that cosmologists have to be suspicious of the validity of $\Lambda$CDM for explaining the totality of observations. Finally, the last section of this introduction is \hyperref[Sect:AlternativesSM]{Sect.\,\ref{Sect:AlternativesSM}}, where comments on some proposals as alternatives to the Standard Model, specifically those related to Dark Energy (DE), can be found.

\section{History of the Cosmological Constant}\label{Sect:HistoryCosmConst}

The history of the CC, or Cosmological Term, chronicles its periods of rise and fall in popularity as a major ingredient for describing the Universe. However, it is true that it has never been ruled out as a mathematical term in Einstein's field equations since it is allowed by general covariance, despite not always being welcomed from the phenomenological point of view throughout modern history.

The concept of the CC was proposed by Einstein in the early part of the last century. However, attempts to modify the laws of gravity to accommodate reality were made even earlier. The introduction of Newton's law of universal gravitation in the 17th century marked the unification of Heaven and Earth, making physical cosmology mathematically possible for the first time. Interestingly, there is a parallelism between Newtonian theory and General Relativity, as both were used by their authors for the exact same purpose of obtaining a model of a static Universe. Thus, our historical view of the CC starts in the 17th century, during the Age of Enlightenment, and ends in the current era where the Cosmological Constant is a fundamental building block of the $\Lambda$CDM model.

\subsection{Newtonian Cosmology}\label{SubSect:NewtonianCosmology}

In the times of Sir Isaac Newton (1642-1727), the question regarding the origin and physical structure of the Universe\,\cite{CosmoQuestionsNewtonSci,harrison1986newton} was unknown.  Newton did not seem to be concerned about the topic, at least not publicly. However, we are able to recover some of his contemplations on the matter through his private and unpublished documents.

In {\it De Gravitatione}\,\cite{newton1962gravitatione} (probably written around 1684-1685, although this date is a matter of debate\,\cite{ruffner2012newton}), Newton responds to the notion of an infinite material system embedded in an infinite space, which was defended by René Descartes in his 1644 publication  {\it Principles of Philosophy}\,\cite{descartes2017principles}. Sir Isaac claimed that space extended infinitely in all directions and was eternal in time. But, in his view, matter was clearly differentiated from space and was thought to be distributed along a finite volume, enveloped in an infinitely extended empty space. 

Later on, Newton further pondered and expanded upon his previous ideas in a series of letters\,\cite{newtonIoriginal,newtonIIoriginal,newtonIIIoriginal,newtonIVoriginal} exchanged with the cleric Robert Bentley at the beginning of the 1690s. Of course, in the 17th century, theology and physics were not separate from each other. In fact, Bentley used these letters to prepare a series of lectures against prevailing atheism, arguing that physical laws alone were insufficient to explain the system of the world, and that divine intervention was necessary. Nevertheless, these attempts to explore the cosmological consequences of the recently discovered laws of gravitation may be considered one of the earliest examples of physical cosmology, and these documents are of great value. In the first letter of this series to Robert Bentley, Newton wrote the following reflection:

{\it ``It seems to me that if the Matter of our Sun and Planets, and all the Matter of the Universe, were evenly scattered throughout all the Heavens, and every Particle had an innate Gravity towards all the rest, and the whole Space, throughout which this Matter was scattered, was but finite; the Matter on the outside of this Space would by its Gravity tend towards all the Matter on the inside, and by consequence fall down into the middle of the whole Space, and there compose one great spherical Mass. But if the Matter was evenly disposed throughout an infinite Space, it could never convene into one mass, but some of it would convene into one Mass and some into another, so as to make an infinite number of great masses, scattered at great distances from one to another throughout all that infinite Space.''}

In this text, Newton describes a finite Universe with a uniform distribution of masses that are gravitationally bound and surrounded by an immeasurable void, which would ultimately lead to collapse into a bigger mass. Although Newton do not consider the possibility of any random or systematic motion of the constituents that could lead to a potential dynamical equilibrium, the idea of collapse was a reasonable inference based on the law of gravitation. In the following sentences, he discusses the possibility of an infinite space filled with uniformly distributed stars. However, he assumes that the same scenario of local collapses would occur throughout the entire Universe. It is important to note that, in his view, space was absolute and at rest, and it is only when motion of matter takes place that it becomes relevant.

In a later letter, Richard Bentley proposed the thought-provoking idea of a perfect, immutable, and infinite Universe that was initially balanced by divine forces as an attempt to address the previous problem and provide a mechanical explanation for the immobility of distant stars:

{\it ``Every particle of matter in an infinite space has an infinite quantity of matter on all sides $\&$ by consequence an infinite attraction every way and therefore must rest in equilibrio because all infinites are equal. ''}

This idea was challenged by Newton, who argued that a naive interpretation of infinity could be problematic. He used arithmetic arguments to demonstrate that such a system is unbalanced and that any finite addition would produce an instability. However, despite these disquisitions, Newton ultimately concluded that the masses remain in equilibrium, although a physical explanation was lacking. At this point, Newton had changed his mind regarding the size of the Universe, becoming more inclined to consider an infinite Universe with uniformly distributed matter. In the second edition of the {\it Principia}\,\cite{newton1723philosophiae}, Newton introduced new ideas, particularly emphasizing the role of the Providence in the complexity of the world. God was responsible for the stability of the solar system, the possibility of distant stars forming their own systems similar to the Sun, and the perfect balance of stars at long distances to avoid mutual attraction and prevent a collapse.

Although Newton was satisfied with this qualitative picture, it was far from complete and had several gaps if we avoid the Providence argument. For instance, the regularization of such an unstable situation raised conceptual issues. At the end of the 19th century, both Hugo von Seeliger\cite{seeliger1896newton} (1849–1924) and Carl Neumann (1832-1925) (although Neumann was more interested in Coulomb forces\,\cite{neumann1896allgemeine}) objected that different methods of calculating the net force that a test mass would experience in a boundless Universe due to a uniformly distributed energy density could lead to a divergent integral. Therefore, within Newtonian physics, different approaches to the integral could produce any possible value of the force, finite or infinite. This problem can also be formulated in other terms, such as computing the force exerted by concentric spheres growing up to infinity and seeing how the result depends on the chosen coordinate origin\,\cite{horedt1989seeliger}.

Seeliger considered this a true paradox and thus a severe blow to Newtonian Gravity. In an attempt to show at least one solution to what is now modernly called the ``Seeliger's paradox'', he proposed (without any fundamental theory behind) that the gravitational attraction between two bodies was somehow diluted by the presence of matter between them, suggesting a new form for the Law of Universal Gravitation with an exponential decay\footnote{A similar kind of law was suggested for the first time by Laplace decades earlier\,\cite{laplace1799traite}.}:
\begin{equation}\label{Eq:Introduction.ModifiedGravitationalLaw}
|\vec{F}_{12}|=\frac{G_N M_1 M_2}{r_{12}^2}e^{-\alpha r_{12}}\,,
\end{equation}
where $r_{12}$ is the distance between two masses $M_1$ and $M_2$ and $\alpha$ is a positive constant with units of inverse of length and $G_N$ is Newton's Gravitational Constant\footnote{Its recomended value, at the time of writing, is reported to be $G_N=6.674 30(15)\times 10^{-11}$ m$^3$ kg$^{-1} $ s$^{-2}$ by CODATA committee\,\cite{Tiesinga:2021myr}.}. It is possible to show that with this modification, \eqref{Eq:Introduction.ModifiedGravitationalLaw} avoids the computational issues commented on previously, and the exponential term has a regularization effect. One may think that \eqref{Eq:Introduction.ModifiedGravitationalLaw} reflects, at some point, the ambiguous suggestion of Newton that the attraction between stars decays at sufficiently long distances, since a sufficiently small $\alpha$ assures that its effects are only of great impact at large cosmological distances and, as a consequence, do not disturb astronomical observations in an extended way.
 
We can analyze the situation from another point of view, more in the line of the calculations of Neumann. Poisson's equation for a full-filled space with constant mass density $\rho$ (modelling the sky plenty of uniformly distributed stars) is
\begin{equation} \label{eq:Introduction.PoissonEq}
\nabla^2 \varphi =4\pi G_N \rho \,,
\end{equation}
where $\nabla^2$ is the Laplacian operator and $\varphi$ is the Gravitational Potential. The acceleration condition of a test particle in that field configuration is
\begin{equation}\label{eq:Introduction.AccelerationEq}
\vec{a}(t)=-\vec{\nabla}\varphi=0 \,,
\end{equation}
so that to remain in a static situation. But it will suppose that eq.\,\eqref{eq:Introduction.PoissonEq} is also 0 and then the energy density should vanish. As a consequence both equations are inconsistent. Again, one may extend the Gravitational Laws to overcome the problem by modifying Poisson's equations\,\eqref{eq:Introduction.PoissonEq}. For instance, Neumann proposed the following,
\begin{equation} \label{eq:Introduction.PoissonEqModifiedI}
\nabla^2 \varphi -\Lambda_N \varphi= 4\pi G_N \rho \,.
\end{equation}
Now, by solving this equation inside a spherical distribution of radius $R$ the following result is yielded:
\begin{equation}\label{eq:Introduction.SolutionPoissonEqModifiedII}
\varphi (r)=-4\pi G_N \rho R^2\left(\frac{1}{R^2 \Lambda_N}+e^{-\sqrt{\Lambda_N} R}\frac{1+R^2\Lambda_N}{R^2\Lambda_N}\frac{e^{\sqrt{\Lambda_N} r}-e^{-\sqrt{\Lambda_N} r}}{2\sqrt{\Lambda_N} r}\right)\,,
\end{equation}
for $r<R$. If we now proceed to do the limit $R\rightarrow \infty$, the solution tends to a finite term,
\begin{equation}\label{eq:IntroductionSolutionPoissonEqModifiedInfinity}
\varphi_{\infty}\equiv\lim\limits_{R\rightarrow \infty} \varphi (r)=-\frac{4\pi G_N \rho}{\Lambda_N}\,.
\end{equation}
A constant potential implies that its gradient is zero, resulting in a net force of zero at any point in space. This is essential for a static Universe. By substituting $\varphi (r)=\varphi_\infty$ into \eqref{eq:Introduction.PoissonEqModifiedI}, a solution is obtained for an infinite extension of matter ($R=\infty$). The whole situation is equivalent to introducing a new force given by:
\begin{equation}\label{Eq:Introduction.ForceNew}
\vec{F}_\Lambda=-\frac{\Lambda_N \varphi_\infty}{3}r \hat{r}=\frac{4\pi G_N \rho}{3}r \hat{r}\,,
\end{equation}
where $r$ is the radial coordinate with respect to a test mass or an arbitrary origin point in the coordinate system. This force grows linearly with distance and is repulsive, counterbalancing the attractive force at large distances that has the opposite sign but the same magnitude. A problem of ``evaporation'' can emerge when a low potential at infinity results in stars being eventually ejected from the more dense region as they gain enough kinetic energy to escape the bounded system. A constant potential also solves  this problem since it avoids any preferred direction for the gradient inwards/outwards the system.

Although we have seen some modifications (see,\cite{norton1992paradox,norton1999cosmological} for a more complete set of solutions to Seeliger's paradox), there are no fundamental reasons to include them, and they do not provide a completely satisfactory answer. Overall, they are just ad-hoc mathematical terms added to the theory in an attempt to fit the pieces of the puzzle. In fact, some authors have followed Newton's position by denying that there is any problem at all: the problem that Seeliger tried to address may just be an artifact related more to the realm of mathematics and the Riemann rearrangement theorem than to any fundamental problem in the laws of gravitation\,\cite{sarasua2018seeliger,arrhenius1909unendlichkeit}. It was argued that the final physical solution should come from symmetry arguments. Moreover, some of these Modified Newtonian Gravity theories were tested against existing data and, in general, performed poorly. For instance, they were not able to solve the famous problem of the perihelion of Mercury\footnote{In 1859, U. Le Verrier, a French mathematician, reported that the observed precession of Mercury's orbit had a discrepancy of around $\sim 43$ arcseconds per year compared to the predictions of Newtonian theory\,\cite{le1859lettre}. Solving this discrepancy was one of the first successes of General Relativity.}, giving them scarce phenomenological support.

Advancing several years to the beginning of the 20th century, it was widely believed that the Universe was made up of the Milky Way, which was thought to be bounded, with the density of stars decreasing near the boundary\cite{young1889text,newcomb1906side}. The presence of nebulae among the stars had been observed since the 17$th$ century, but without further clues, they were commonly assumed to be part of the Milky Way. On the other hand, the {\it Island Universe theory} proposed the idea that some of those nebular objects in the sky were not part of our galaxy but were instead independent objects, requiring a revision of the Universe's structure\cite{gordon1969history}. Immanuel Kant (1724-1804) is usually credited with conceiving the theory\cite{kant1755allgemeine}, although the idea was already present in the works of Thomas Wright (1711-1786) and other researchers\cite{wright1750original}. The lack of powerful telescopes and more sophisticated techniques made it difficult to definitively discern whether other Island Universes, such as the Milky Way, existed. Closure on this matter did not arrive until much later. Firstly, Slipher (1875-1969) calculated the recession velocity of several of those nebulae\,\cite{slipher1912spectrum,slipher1913radial,slipher1915spectrographic,slipher1917nebulae} and observed that most of them were redshifted, moving away at great speeds compared to the typical star's velocity. This suggested that they were not gravitationally bound with the Milky Way. The final proof arrived in the 1920s, a decade after the emergence of General Relativity. This was thanks to the astronomer Edwin Hubble (1889-1953), who successfully calibrated the distances of Cepheid variables in spiral nebulae such as M33. The measurements were consistent with M33 being an independent object from the Milky Way, which confirmed the existence of other Island Universes, or {\it galaxies}\,\cite{hubble1925cepheids,hubble1926spiral,hubble1926extragalactic}. This discovery was of great importance in Astronomy and Cosmology. However, with respect to Seeliger's paradox, replacing an infinite web of stars with galaxies (if they are motionless) does not change the essential point. All in all, the stability of a Newtonian Universe seems to be compromised in the classical framework of Newtonian physics.

\subsection{The beginning of Relativistic Cosmology}\label{SubSect:RelativisticCosmology}

Let us advance to the beginning of the history of General Relativity (GR) at the end of 1915, when Einstein's field equations,
\begin{equation}\label{Eq:Introduction.OriEinsteinEq}
G_{\mu\nu}=\frac{8\pi G_N}{c^4} T_{\mu\nu}\,,
\end{equation}
were presented\,\cite{einstein1915feldgleichungen}. The preceding equations are the pivotal result of the theory, describing the relationship between the geometrical structure of the Universe and its material content. They consist of a set of non-linear, second-order differential equations on $g_{\mu\nu}$, the {\it metric tensor}. The left-hand side (LHS) of this equation is the geometrical part, where $G_{\mu\nu}\equiv R_{\mu\nu}-\frac{1}{2}g_{\mu\nu}R$ represents the so-called {\it Einstein tensor}. The {\it Ricci tensor} is denoted by $R_{\mu\nu}$ and $R\equiv R_{\mu\nu}g^{\mu\nu}$ is its trace, the {\it Ricci Scalar}. On the right-hand side (RHS), there is the matter part of Einstein's equations. $T_{\mu\nu}$ is the Energy-Momentum tensor (EMT) of matter.

Einstein's field equations constituted a generalization of Newtonian laws of gravitation, as the classical limit can be derived in the weak-field approximation and low velocities. This particular form\,\eqref{Eq:Introduction.OriEinsteinEq} of Einstein's field equations was not written in the original work, including the sign convention and notation. $G_{\mu\nu}$ was not present (Einstein reserved that notation for the Ricci tensor), and they were written equivalently in terms of the EMT trace\footnote{We do not pretend to present an exhaustive nor chronological overview of the history of GR and field equations here. For a historical approach, see\,\cite{janssen2007untying,janssen2015arch}.}. Throughout this dissertation, we will not use this original formulation. Besides, we will make use of natural units in subsequent equations, setting $c=1$ and $\hbar=1$ unless said otherwise. More details about the conventions and notations used in this work can be found in \hyperref[Appendix:Conventions]{Appendix A}.
 
A remarkable fact is that a static infinite Universe, similar the one imagined by Newton, is also not possible in the framework of the original Einstein's equations. It is important to clarify something: a static Universe was still very reasonable as in Newton's times. The existence of extra-galactic objects was still matter of debate, the size of the Universe was still an open question and the observed peculiar velocities of stars were not great. Einstein, at those times, was probably not aware of the aforementioned works of Slipher\,\cite{slipher1912spectrum,slipher1913radial,slipher1915spectrographic,slipher1917nebulae} about the redshifting of spiral Nebulae, which constituted the first evidence pointing out a possible dynamical behaviour rather than a static regime. So that, part of Newton's ideas were still very alive and it was natural to deal with an approximate motionless Universe. However, conceptually speaking, there is a huge difference regarding Newton's absolute space consideration: In GR the structure of spacetime is a much more complex than absolute space, it is not just the fixed scenario where matter's motion happens. Indeed, in GR, dynamical cosmological models emerge naturally, although it took several years to ponder these ideas seriously.

Einstein had a genuine interest in applying GR to the largest scales. He shown curiosity in the boundary conditions at infinity, the extension of the Universe and the origin of inertia as one can see from his letters with the mathematician William de Sitter and other colleagues\,\cite{ORaifeartaigh:2017uct,einstein1916letterdeSitter, kahn1975letters}. In fact, as affirmed by Einstein in\,\cite{einstein1918prinzipielles}, one of his foundational principles for GR was Mach’s Principle:

{\it ``The G-field is completely determined by the masses of the bodies [...] Mach’s principle (c) is a different story. The necessity to uphold it is by no means shared by all colleagues; but I myself feel it is absolutely necessary to satisfy it. With (c), according to the field equations of gravitation, there can be no G-field without matter. Obviously, postulate (c) is closely connected to the spacetime structure of the world as a whole, because all masses in the Universe will partake in the generation of the G-field.''}

With {\it G-field}, he means the metric. But the greatest signal pointing out he had in mind to do Cosmology is that {\it Kosmologische Betrachtungen zur allgemeinen Relativitaetstheorie}\,\cite{Einstein:1917ce} appeared shortly after the field equations of GR were released. We can consider this paper the kickoff of Relativistic Cosmology \cite{Einstein:1917ce}. He starts by recalling the intricacies of Poisson's equation \eqref{eq:Introduction.PoissonEq} to find a solution with boundary conditions such that the potential tends towards a constant value at spatial infinity, or the density decays vastly with distance. He announces that a similar problem arises in GR and peculiarly talks about the same solution that Neumann and Seeliger presented some years earlier\footnote{Without citing them, so it is assumed that he was not aware of their work at this point.}, i.e., equation \eqref{eq:Introduction.PoissonEqModifiedI} and its constant solution \eqref{eq:IntroductionSolutionPoissonEqModifiedInfinity}. He also reasoned that the extra term in the equation should be negligible in the presence of larger masses, like stars, which would lead to the recovery of the original form of Poisson's equation, and would not cause any significant deviation at local scales. This seemed to address all the issues with the Static Universe model within (an extended) Newtonian gravity. However, Einstein did not give much attention to this idea as it was only an analogy of what he aimed to achieve with GR.

Actually, his model of a static Universe is spatially finite, with an infinite temporal extension and whose spatial curvature is positive and constant: a closed Universe of radius $R_{\rm E}$, with no need for boundary conditions at far infinity. Matter is uniformly distributed at large scales, and the staticity condition is the existence of a reference frame in which matter is at rest. Although small peculiar motions of stars were not denied, assuming large volume scales, these effects become negligible. Sometimes this model is just called  {\it Einstein's Universe} or {\it Einstein's World}.

The heart of the paper is the proposal of extending Einstein's equations to 
\begin{equation}\label{Eq:Introduction.LambdaEinsteinEq}
G_{\mu\nu}+\Lambda g_{\mu\nu}=8\pi G_N T_{\mu\nu}\,.
\end{equation}
An additional new term containing the fundamental tensor $g_{\mu\nu}$ multiplied by a constant, $\Lambda>0$, is added. Naively speaking, the effects of attraction of ordinary matter may be counterbalanced by the repulsive effect produced by the term proportional to $\Lambda$, which posteriorly was known as {\it Cosmological Constant} (CC) or {\it Cosmological Term}. The computations then easily carried Einstein to calculate the radius of the Universe $R_{\rm E}$, the matter content and its relation with $\Lambda$,
\begin{equation}\label{Eq:Introduction.RadiusCurvature}
\frac{1}{R_{\rm E}^2}=\Lambda=4\pi G_N \rho\,.
\end{equation}
Einstein estimated the value of $R_{\rm E}$ based on rough estimations of the mean energy density in the Universe. In his paper, he mentioned that astronomers had determined the spatial density of matter, $\rho$, through star counting, and he claimed its value to be $\rho \sim 10^{-22} g/cm^3$. This value corresponds to $R_{\rm E}$ being $10^7$ ly. However, this value was not published but only appeared in private correspondence\cite{einstein1917lettertoEhrenfest,einstein1917letterFreundlich,einstein1917albertBesso,einstein1917letterDeSitterMarch12}. The reason why Einstein did not pay much attention to comparing his model with existing empirical data may have been due to the absence of a reasonable explanation for the fact that $R_{\rm E}$ was much larger than the farthest observed stars at around $10^4$ ly. It seems that Einstein lacked confidence in astronomical observations, which may have contributed to his little interest in comparing his model with the data available at the time.

At the end of the paper, Einstein admits that this extra term $\Lambda$ has the only purpose of describing a static Universe and does not have strong theoretical grounds, as he concludes:

{\it ``In order to arrive at this consistent view, we admittedly had to introduce an extension of the field equations of gravitation which is not justified by our actual knowledge of gravitation. It is to be emphasized, however, that a positive curvature of space is given by the presence of matter, even if the supplementary term is not introduced. That term is necessary only for the purpose of making possible a quasi-static distribution of matter, as required by the fact of the small velocities of the stars.''}

Some points must be remarked regarding the introduction of $\Lambda$ by Einstein:

\begin{itemize}
  \item $\Lambda$ was originally not admitted to be part of the energy/matter content of the Universe, but is a geometrical term that can be added to the field equations.
  
  \item[$\bullet$] As Einstein admits in the paper, their current knowledge of gravitation on those times was not able to give a justification for its inclusion. However, the addition of the new term $\Lambda g_{\mu\nu}$ does not remove the general covariance of the equations. What is more, now we know that, from the mathematical point of view, the addition of $\Lambda g_{\mu\nu}$ to the field equations is totally allowed by {\it Lovelock's theorem}\,\cite{lovelock1971einstein,lovelock1972four}.
  
  \item Bianchi identities ensure local covariant conservation of energy. This follows from the fact the Einstein tensor $G_{\mu\nu}$ is divergentless,
  \begin{equation}\label{Eq:Introduction.BianchiIdentity}
  \nabla^\mu G_{\mu\nu}=0 \Rightarrow \nabla^\mu T_{\mu\nu}=0.
  \end{equation}
  The inclusion of $\Lambda g_{\mu\nu}$ does not change anything since,
 \begin{equation}\label{Eq:Introduction.DerivativeOfLambdaTerm}
  \nabla^\mu \left(\Lambda g_{\mu\nu}\right) = 0 \Rightarrow \partial^\mu \Lambda= 0,
  \end{equation}
	if $\Lambda$ is constant. Here we are supposing a Levi-Civita connection, thus we demand metric compatibility of the covariant derivative, $\nabla_\gamma g_{\mu\nu}=0$. Thus, covariant conservation of the EMT still holds.
  
  \item[$\bullet$] If the value of $\Lambda$ is sufficiently small, it can be introduced without altering the former applications of GR to the Solar System scale that already existed in that epoch, such as the precise calculation of the perihelion of Mercury.
  
  \item The analogy with \eqref{eq:Introduction.PoissonEqModifiedI} that he presents in the paper is quite clear, although not exactly equivalent. A correct analogy would take this form:
  \begin{equation}\label{eq:Introduction.PoissonEqModifiedII}
  \nabla^2 \varphi+\Lambda=4\pi G_N\rho,
  \end{equation}
  which admits a constant solution for the potential when $\Lambda=4\pi G_N \rho$, as obtained previously in \eqref{Eq:Introduction.RadiusCurvature}. In fact, \eqref{eq:Introduction.PoissonEqModifiedII} is not only an analogy, but the Newtonian limit of the field equations for the Newtonian potential\,\cite{straumann2002history}.
\end{itemize}

One can recognize some parallelisms between the aims of Newtonian Gravity and Einsteinian gravity in finding a suitable system of the world adapted to the perceived reality of their times. Both started with the discovery of the laws of gravitation and applied them to cosmology. However, the inability to reconcile these laws with the paradigm of a static Universe led to efforts to modify the original framework. In Newton's case, theological arguments were used, but later authors such as Neumann introduced modifications to gravity by adding extra terms to the Newtonian force or Poisson equation. In Einstein's case, he introduced an ``ad-hoc'' cosmological term to modify the laws of gravitation.

Just a few months after the publication of Einstein's paper on the cosmological term, William de Sitter presented his own model of the Universe, the {\it de Sitter Universe}. At first, it was misunderstood as a static cosmological model, but it was reinterpreted as a non-static one some years later\footnote{For a detailed discussion on the change of interpretation of de Sitter's Universe, see \cite{north1965measure}, chapters 5 and 6.}. This model is characterized by a Universe completely dominated by a CC and devoid of matter, proposed in 1917, making it a vacuum solution. The idea of a Universe without any matter component was controversial for Einstein, as it contradicted Mach's principles and his notion of inertia being solely determined by matter. Nevertheless, Einstein admitted it as a correct mathematical solution. In its early days, the de Sitter Universe was of special interest because it was thought to provide an explanation for the observed redshift of the distant spiral nebulae found by Slipher around the time GR appeared\,\cite{straumann2002history}. Although it enjoyed some attention from the scientific community, it was later realized that it describes a Universe that expands forever and was gradually abandoned in favor of other frameworks of expanding Universes with material content. Nowadays, de Sitter Universes are of special interest because they may represent an inflationary period similar to the one that the Universe underwent during the so-called {\it inflation} (see \hyperref[SubSect:Inflationary]{Sect.\,\ref{SubSect:Inflationary}}).
 
After his famous paper on the Cosmological Constant, Einstein had identified a mathematical term that allowed him to model his cosmological ideas, but still had reservations about the physical interpretation and implications of adding $\Lambda$ to the equations. This is evident from his correspondence with colleagues\,\cite{einstein1917letterKlein, einstein1922spielen}. 

It was E. Schrödinger\,\cite{schrodinger1918losungssystem,harvey2012einstein} who first recognized, or at least communicated, that the cosmological constant in the field equations could be interpreted as a negative pressure fluid when transferred to the right-hand side of equation\,\eqref{Eq:Introduction.LambdaEinsteinEq}, 
 \begin{equation}\label{Eq:Introduction.TensorLambda}
 T^\mu_{\phantom{\mu}\nu}(\Lambda)=\left(\begin{matrix}
P_\Lambda & 0 & 0 & 0\\
0 & P_\Lambda & 0 & 0 \\
0 & 0 & P_\Lambda & 0\\
0 & 0 & 0 & P_\Lambda
\end{matrix}\right)\,,
 \end{equation}
where $P_\Lambda \equiv -\Lambda/(8\pi G_N)$. Einstein was not impresed by this fact as he had already noticed this when dealing with the modified field equations. Although the mathematical equivalence was clear, Einstein was aware that a dynamical quantity such as pressure requires an underlying theory. In his response to Schrödinger's work in 1918\,\cite{einstein1918bemerkung}, Einstein gave a mysterious physical interpretation of the CC as a property of space:
 
{\it ``Empty space takes the role of gravitating negative masses which are distributed all over the interstellar space.''}
 
Unfortunately, he did not elaborate on this topic or make an attempt to provide a proper mathematical description of this sentence in subsequent texts. Many years later, in 1945, he introduced an appendix in his book {\it The Meaning of Relativity}\,\cite{swann1945relativity} where he stated that for a static Universe to exist, 

{\it `` One has to introduce a negative pressure, for which there exists no physical justification. In order to make that solution possible, I originally introduced a new member into the equation.''} 

This is an advance of the fundamental problems of modern {\it Dark Energy} and Vacuum energy, although Einstein did not explicitly relate the Cosmological Term with a vacuum energy density with negative pressure in any of his letters\,\cite{Kragh:2014jaa}. It seems that Einstein was clueless with respect to a possible physical origin of the CC. Later on, he even tried to reformulate it by treating $\Lambda$ as a kind of integration constant\,\cite{einstein1922spielen}.

Returning to the historical picture, during the 1920s, observational and theoretical evidence started to accumulate, suggesting that the Universe was not static\,\cite{nussbaumer2011discovered,steer2012discovered}. Many authors began to consider this possibility as realistic. Friedmann (1888-1925) was the first to conceive non-static solutions to Einstein's equations\,\,\cite{friedman1922krummung}. However, his results were unnoticed by a great part of the community, and he did not attempt to connect his findings with astronomical observations. Friedmann realized that, depending on the magnitude of the CC, a Universe with plenty of ordinary matter could either expand or contract. In 1924, Knut Lundmark (1889 – 1958), a Swedish astronomer, published a distance-velocity diagram\,\cite{lundmark1924determination}. Georges Lemaître (1894-1966) independently discovered a set of dynamical Friedmann-like solutions in 1927. He applied these solutions to explain the redshifting of galaxies, suggesting that their receding implied the existence of an initial static state of the cosmos with a radius of $R=\Lambda^{-1/2}$, which eventually started to expand. He also theoretically demonstrated that the relation between the recession velocity $v$ and proper distance $D$ was given by:
\begin{equation}\label{Eq:Introduction.Hubble-Lemaitre}
  v=H_0 D
\end{equation}
where $H_0$ is a proportionality constant. Lemaître even provided an estimate of this value, which would be around $H_0\sim 600$ km/s/Mpc\,\cite{lemaitre1927univers}, based on Hubble's distances and Slipher's measures of galaxy redshifts. Unfortunately, Lemaître's work went largely unnoticed when it was released\,\cite{o2019eddington}. Finally, Hubble's work\,\cite{hubble1929relation,hubble1931velocity} provided definitive evidence of the dynamical nature of the Universe, based on observations of extragalactic nebulae (galaxies) receding according to the now called {\it Hubble-Lemaître Law}\,\eqref{Eq:Introduction.Hubble-Lemaitre}. In his honor, $H_0$ is called the {\it Hubble Constant}. The value of $H_0$ reported in\,\cite{hubble1929relation} was around $500$ km/s/Mpc for the 1929 work and around 560 km/s/Mpc for the 1931 work. Curiously, Hubble did not mention the ``Expanding Universe'' even once in his paper, and was actually doubtful regarding this fact\,\cite{way2011lema}. On the contrary, Lemaître was truly convinced of the expansion of the cosmos, as he demonstrated with his models and arguments.

It is useful to introduce now the Friedmann-Lemaître-Robertson-Walker (FLRW) metric\,\cite{friedman1922krummung,friedmann1924moglichkeit,lemaitre1927univers,lemaitre1937univers,robertson1935kinematics,robertson1936kinematics,walker1937milne}, which describes a homogeneous, isotropic, and dynamical background (expanding or contracting) spacetime. It was discovered independently by the four authors in the decades of 1920-1930 when they tried to develop dynamical cosmological models, although the idea of a non-static Universes was independently conceived by the first two authors, as said. However,  Howard P. Robertson (1903-1961) and Arthur G. Walker (1909-1931) were responsible for generalizing a mathematical expression for the metric tensor used in dynamical models. FLRW cosmology relies on one of the most fundamental assumptions about our Universe: the Cosmological Principle. This principle states that the Universe is isotropic and homogeneous at large scales. Within the FLRW framework, calculations over models that respect these symmetries are easier to perform.  The FLRW interval can be expressed in spherical coordinates as:
\begin{equation}\label{Eq:Introduction.FLRWMetric}
 ds^2=g_{\mu\nu} dx^\mu dx^\nu =-dt^2+a^2 (t) \left( \frac{dr^2}{1-kr^2}+r^2 d \theta^2 + r^2 \sin^2 \theta d\phi^2 \right)\,.
\end{equation}
Here $a(t)$ is the {\it scale factor}, a dimensionless parameter representing the relative expansion of the Universe between two different values of the cosmic time. Normally, it is normalized in such a way that at the present time $a=1$, although it is just a convention. The parameter $k$ is related to spatial curvature, with natural units $[k]=E^2$, $E$ being energy units. Spatial curvature is determined by the matter/energy content of the Universe. Sometimes $k$ is normalized as a dimensionless value depending on its sign: $+1,0,-1$. In this convention, the scale factor has units of length. 

The cosmological fluids are usually treated as perfect fluids,
\begin{equation}\label{Eq:Introduction.PerfectFluid}
T_{\mu\nu}=P g_{\mu\nu}+\left( \rho + P \right) u_\mu u_\nu\,,
\end{equation}
where $\rho$ and $P$ are the energy density and the pressure of the fluid and $u^\mu$ is the 4-velocity of the fluid. Matter at cosmological scales is modelled in terms of energy densities as if matter were distributed in a continuum fluid.

Having a mathematical object that encapsulates the cosmological principle is a powerful tool that makes it possible to formulate dynamical models in a more systematic way. 

Generalizing the sky filled with discrete stars that Bentley and Newton discussed can be achieved by assuming a constant energy density $\rho$ for matter, as shown in \eqref{eq:Introduction.PoissonEq} and subsequent equations. Notice that, by assuming a uniform distribution of matter, Einstein implicitly accepted the Universe to be isotropic and homogeneous at large scales, even without explicitly introducing the FLRW metric.

When introducing \eqref{Eq:Introduction.FLRWMetric} and \eqref{Eq:Introduction.PerfectFluid} in \eqref{Eq:Introduction.LambdaEinsteinEq} we are lead to the well-known {\it Friedmann Equations}:
\begin{equation}\label{Eq:Introduction.FriedmannI}
\left(\frac{\dot{a}}{a}\right)^2=\frac{8\pi G_N}{3}\rho+\frac{\Lambda}{3}-\frac{k}{a^2}
\end{equation}
and
\begin{equation}\label{Eq:Introduction.FriedmannII}
\frac{\ddot{a}}{a}=-\frac{4\pi G_N}{3}\left(\rho+3 P\right)+\frac{\Lambda}{3}\,.
\end{equation}
The above equations are the form that Einstein's equations take in cosmology when the Cosmological Principle is assumed. The dots represent derivatives with respect to cosmic time, where $\dot{\left(\right)}\equiv d\left(\right)/dt$. 

The former equations can describe, for instance, Einstein's world. In that way, we can understand the impossibility of a static Universe without the introduction of a Cosmological Term.  The staticity condition implies $\dot{a}=0$ and $\ddot{a}=0$. If $\Lambda=0$, then $k/a^2=8\pi G_N\rho /3$ and $\rho+3P=0$ must be satisfied\cite{Weinberg:2008zzc}. However, this last equality cannot hold for pressureless dustlike matter or radiation, as both have non-negative pressure. For $\Lambda > 0$ and assuming dustlike matter with $P=0$, we have $k/a^2=\Lambda=4\pi G_N\rho$. It is sometimes useful to define $\rho_\Lambda \equiv \Lambda/(8\pi G_N)$, as though $\Lambda$ were being interpreted as a fluid with negative pressure, $P_\Lambda = -\rho_\Lambda$. This was the point defended by Schrödinger, as explained earlier. We can also use the equations to study of de Sitter Universes, which is much more clear from this perspective. By using \eqref{Eq:Introduction.FriedmannI} and \eqref{Eq:Introduction.FriedmannII} for $k=0$ and $\rho=0$, the relative variation of the scale factor with cosmic time is exactly constant, $\dot{a}/a \equiv H_{\rm DS}\equiv \sqrt{\Lambda/3}$, and the scale factor admits exponential solutions as $a(t)\propto e^{\pm H_{\rm DS} t}$. For $k>0$, it admits a solution $a(t)\propto \cosh(H_{\rm DS}t)$, which expands from $R_0=1/H^2_{\rm DS}$ at $t=0$. Therefore, de Sitter Universes are actually not static.

Using this machinery, it is easy to realize that there is an additional problem that undermines the foundations of Einstein's Universe, a problem that Einstein was not aware of, which concerns the mathematical consistency of the model. This is a stability problem that was discovered by Arthur Eddington\,\cite{eddington1930instability}. In essence, he wrote the Einstein field equations for a FLRW metric, which depend on a scale factor $a(t)$ as the dynamic radius of a Universe that changes with time:
\begin{equation}\label{Eq:Introduction.Instability}
\frac{6}{a}\frac{d^2 a}{dt^2}=2\Lambda-8\pi G_N \left(\rho+3P\right)\,.
\end{equation}
For a static Universe with pressureless dust-matter (a  similar argument applies for radiation), $d^2 a/dt^2=0$ and thus $\Lambda=4\pi G_N \rho$. A uniform distribution of matter may not only be useful for computations but also very reasonable to conceive. However, it is also natural to expect small variations from uniformity, as $\rho$ represents only an average density. If there is a small fluctuation $\delta \rho$ in the energy density, such that $\delta \rho>0$, the RHS of \eqref{Eq:Introduction.Instability} becomes unbalanced, and $d^2 a/dt^2$ becomes negative. If the Universe contracts, the matter and energy density increase, assuming that the mass is conserved. Then $d^2 a/dt^2$ becomes even more negative, and the contracting effect is reinforced. Conversely, if $\delta \rho<0$, then an enhanced expansion begins, and equilibrium is not restored at all.

Einstein eventually abandoned the idea of an immutable Universe in 1931,\cite{ORaifeartaigh:2013khl,o2019eddington,nussbaumer2014einstein}. Despite the abundance of observational clues, Eddington's arguments about the instability of Einstein's world was the {\it coup de grace} to the model. Einstein also abandoned $\Lambda$ as the ``Biggest blunder'' of his life\footnote{This expression is extracted from an article by the Russian physicist George Gamow in 1956\,\cite{gamow1956evolutionary}, but it is unclear if Einstein literally said this to Gamow or if it is just embellishment.}. Einstein finally commented that his reason for refusing the use of a CC was that it could no longer provide a static stable model, and static models had been ruled out by observations\,\cite{einstein1931}.

Finally, let us mention that several other interesting models of the Universe have appeared since the publication of {\it Kosmologische Betrachtungen zur allgemeinen Relativitaetstheorie}\,\cite{Einstein:1917ce}. In particular, we will briefly summarize the {\it Einstein-de Sitter} cosmological models. After abandoning the idea of an immutable Universe, Einstein embraced the new paradigm of evolutionary Universes and exiled $\Lambda$ from his new models. Together with de Sitter\,\cite{einstein1932relation,einstein1933structure,o2015einstein}, they proposed this especially simple model in which, as a difference from other works, spatial curvature is set to zero and only matter is considered. In this sense, the destiny of the Cosmos has only to do with the density of matter. They did not claim any physical argument against curvature or the CC; they just set both to 0 following Occam's razor: without them, the model was still compatible with observations, but much simpler. In modern formulations, the relation between the energy density and the ``coefficient of expansion'', $H\equiv \dot{a}/a$, can be written as
\begin{equation}\label{Eq:Introduction.HubbleFunctionEinsteinDeSitter}
  H^2\equiv\left(\frac{\dot{a}}{a}\right)^2=\frac{8\pi G_N}{3}\rho\,.
\end{equation}
As a consequence, we have a clear description of the Universe in terms of two measurable parameters: $H$ and $\rho$. Surprisingly, their first paper had sparse content, and they did not study the evolution of the model or address questions about the beginning of spacetime or the boundary conditions at infinity. Because it was a simple model of a dynamical Universe and there was no evidence of non-null spatial curvature at the time, it gained popularity as a baseline model in the 20th century. Currently, an Einstein-de Sitter Universe is considered a good approximation for the period in which our Universe was dominated by cold dark matter prior to the current era.

\subsection{From the 1930's up to the end of the Century}\label{SubSect:30's}

Despite Einstein's abandonment of $\Lambda$, other authors did not give up easily and continued to develop their models of non-static Universes with $\Lambda$ still in the equations. Several motivations existed for this in the 1930s. First, although observations suggested that $\Lambda$ should be extremely small so as not to affect astrophysical considerations, the cosmological term was not required to be zero by any fundamental argument. Neglecting $\Lambda$ so easily may have been a mistake, as seen by some researchers such as R.C. Tolman\,\cite{Tolman1931} and de Sitter\,\cite{de1932kosmos}. Second, a non-zero $\Lambda$ was proposed in cosmological models by Eddington\,\cite{eddington1933expanding} and Lemaître\,\cite{lemaitre1933univers,lemaitre1934evolution,lemaitre1934evolutionII} in order to reconcile the short timespan of expansion predicted by many models without CC, which was much shorter than the estimated age of the oldest stars. An additional utility of Lemaître's models was that they naturally incorporated differentiated stages: cosmic acceleration, a stagnant phase, and deceleration. During the momentary balance between CC and gravitational attraction, a possible mechanism for structure formation could arise by the growth of matter perturbations into large structures such as galaxies or clusters. Finally, even though Friedmann's original solutions set a framework for an expanding Universe, there were no clues as to the origin and cause of this expansion. Eddington and de Sitter argued that the origin of the expansion should be the $\Lambda g_{\mu\nu}$ term. Lemaître went further and related the CC to a negative density $\rho_\Lambda$ corresponding to $-10^{-27} {\rm g}/{\rm cm}^3$ and a positive pressure $P_\Lambda=-\rho_\Lambda$\,\cite{lemaitre1934evolution,lemaitre1934evolutionII}.
  
After the 1930s, the cosmological term was invoked by various researchers on several occasions to address the complexities encountered by more established models of GR with $\Lambda=0$. Furthermore, some intriguing and conceptual ideas in cosmology emerged that were directly or indirectly related to the cosmological term. To provide clarity, let us list some events chronologically\footnote{For a detailed explanation of the history of the CC around these years, see\,\cite{o2018one} and references therein.}:

\begin{itemize}

\item[1948] The appearance of the {\it steady-state} models, which generalized the cosmological principle by assuming a Universe that, at large scales, not only has the same properties at different points and directions but also remains immutable in time\,\cite{bondi1948steady,hoyle1948new}. Thus, for mantaining the energy densities, the expansion of the Universe gets compensated by an uninterrupted creation of matter.  This model required the addition of a new tensor in Einstein's equations that had some similarities to a Cosmological Term.  Meanwhile, the Big Bang theory encapsulated the idea of Lemaître's primeval atom that evolves dynamically and was able to predict the genesis and abundances of light elements such as Hydrogen and Helium in a process called Big Bang Nucleosynthesis (BBN)\,\cite{alpher1948origin} and expected the observation of an imprint of an evolving Universe possessing an early hot stage\,\cite{gamow1948evolution,alpher1948evolution,alpher1949remarks,durrer2015cosmic}.
  
\item[1955] In the 1950s, the deceleration parameter $q(t)\equiv -\ddot{a}a / \dot{a}^2$ began to play a crucial role in observational cosmology. In 1955, Robertson calculated corrections to the relation between redshift and the apparent magnitude of galaxies\,\cite{robertson1955theoretical}, and $q_0\equiv q(t_0)$ (where $t_0$ is the present cosmic time) appeared in the expression. At that time, it was widely believed that the Universe would eventually stop expanding due to the gravitational effect of matter. Therefore, the parameter $q_0$ was introduced to measure the degree of deceleration. In 1956, F. Hoyle and A. Sandage\,\cite{hoyle1956second} used $q_0$ to mathematically classify the curvature of Einstein-de Sitter Universes and to determine whether the Universe was in a steady-state or in a Lemaître-Eddington ``exploding state'' with a positive cosmological term (corresponding to negative values of $q_0$). Precise evaluations of this term could have helped distinguish between models. However, contemporary measurements suffered from significant uncertainty, and there were observational challenges in reaching higher redshifts in the magnitude-redshift relation for galaxy clusters.
  
\item[1961] In his paper, A. Sandage studied models with $\Lambda>0$ in order to raise the predicted age of the Universe. The measured values of Hubble's Constant at those times were reduced to $H_0\sim 75$ km/s/Mpc, so the timespan of expansion raised considerably. Even in this case, they still encountered a problem of mismatch when comparing with older star's age\,\cite{sandage1961ability} by a factor 2, approximately.
  
\item[1965] The steady-state models were ultimately overturned by the Big Bang Theory after the discovery of the Cosmic Microwave Background (CMB) by Penzias and Wilson in 1965\cite{penzias1965measurement,dicke1965cosmic}. The CMB is an isotropic source of radiation, with fluctuations of only one part in $10^{5}$, and consists of a photon gas that behaves as a perfect black body with a temperature of $T_0=2.72548\pm 0.00057 {\rm K}$\,\cite{Fixsen:2009ug}. The origin of the CMB is in the recombination epoch of the early Universe. The steady-state hypothesis, which proposed an almost immutable Universe, was ultimately discarded in favor of the evidence of major events in the cosmological past revealed by the CMB. As a result, the Big Bang theory emerged as the dominant model. The CMB not only helped to differentiate between these two competing conceptions but also provided valuable and accurate information regarding the early Universe. It played a crucial role in triggering the development of precision cosmology and nowadays has become a fundamental tool for studying the physics of the early Universe and for the search of new physics.
  
\item[1967] Another study by V. Petrosian {\it et al}\,\cite{petrosian1967quasi} contemplated the possibility of a non-zero cosmological constant as a way to explain the anomalous high quasar count at redshifts around $z\approx 2$. Shortly after and with the same purpose in mind, N. Kardashev\,\cite{kardashev1967lemaitre} considered an expanding Lemaître Universe with positive curvature and a cosmological constant slightly above its value in an Einstein's world. 
  
\item[1968] Theoretical work by the Russian physicist Yakov Zel’dovich was important for understanding the cosmological constant and its possible value in the context of QFT, particularly in his papers from 1967 and 1968\,\cite{zel1967cosmological,zel1968cosmological}. Zel’dovich computed the Zero-Point Energy and found that it leads to an immense value of $\Lambda$, more than 44 orders of magnitude greater than the estimated values at the time. This was concerning, given that the effects of Zero-Point Energy had already been demonstrated through the discovery of the Casimir effect in the 1950s \cite{casimir1948attraction,sparnaay1957attractive}. Zel’dovich’s findings contributed to what is now known as the {\it cosmological constant problem}, which we will discuss in more detail below.
 
\item[1970] H. R. Dicke\,\cite{robertdicke1969,dicke1979big} formulated the {\it flatness problem}, which arises from the fact that even a small deviation from flatness in the early Universe would be amplified during cosmological evolution. Conversely, a curvature near 0 in the present time would imply an even closer to 0 value in the past with an unprecedented precision. 
 
\item[1970-74] An ongoing challenge for Einstein-de Sitter models of the Universe was the mismatch between the expected and observed fraction of energy associated with matter. Specifically, if we use the Hubble constant at present time, $H_0$, to define a critical energy density, then
\begin{equation}\label{Eq:Introduction.CriticalDensityFractional}
\rho_{\rm c}^0\equiv\frac{3H_0^2}{8\pi G_N}, \qquad \Omega_{\rm m}^0\equiv \frac{\rho_{\rm m}^0}{\rho_{\rm c}^0},\qquad q_0=\frac{\Omega_{\rm m}^0}{2}\,.
\end{equation}
The critical energy density in an Einstein-de Sitter Universe (with $\Lambda=0$) represents the exact value of the total energy density the Universe needs to be spatially flat. $\Omega_{\rm m}^0$ represents the fraction of energy associated with matter with respect to the critical density, and $q_0$ is the deceleration parameter at present. A flat Einstein-de Sitter Cosmos has $\Omega_{\rm m}^0=1$, an open Universe has $\Omega_{\rm m}^0<1$, while a closed one has $\Omega_{\rm m}^0>1$.

The first part of the decade saw several groups attempting to estimate the energy fraction contributed by galaxies to the cosmological budget. J. R. Gott {\it et al.} presented a comprehensive review\,\cite{gott1974unbound}, summarizing constraints on cosmological parameters inferred from contemporary studies, including local data (age of elements in meteorites, dynamics of nearby galaxies, etc.), and galaxies at large redshifts. Some of the physical assumptions used to constrain the parameters were inaccurate or model-dependent. However, the combination of all constraints tightly restrict the parameter space to low values of $\Omega_{\rm m}^0< 0.1$ within an Einstein-de Sitter Universe. This points to an open Universe, contrary to common beliefs at the time. Gott and collaborators concluded that the data were highly compatible with an Open Universe, although it could be a problem of {\it missing mass}: most of the mass in the Universe cannot form part of galaxy clusters and has yet to be detected.

\item[1975] In the paper by A.H. Gunn {\it et al.},\cite{gunn1975spectrophotometry,gunn1975accelerating}, they carried out a spectrophotometry study of galaxies in clusters. When introducing evolutionary effects in the modeling of elliptical galaxies, a negative value of the deceleration parameter arises that is not compatible with 0 at the present time. Interestingly, Gunn and collaborators did not discard so easily this result, opening the possibility of an accelerated cosmic expansion. The only known possibility within the context of GR that could lead to this accelerated expansion was a closed Lemaître-like Universe with a non-zero value of $\Lambda$.
 
\item[1981] The theory of {\it inflation} is considered as a possible explanation for the {\it horizon problem} (the fact that apparently causally disconnected regions in space can share almost identical properties, as observed in the small anisotropies of the CMB) and the aforementioned flatness problem. It was formulated by Guth in 1981,\cite{guth1981inflationary} and developed in subsequent years by many authors\,\cite{linde1982new,starobinsky1983physics}. Inflation describes an early stage in the cosmic history characterized by an abrupt process of expansion due to a tremendous vacuum energy density. The mechanism for inflation possesses some characteristics of a CC. Inflation can also explain the larger growth of primordial density fluctuations in linear perturbation theory and provides a mechanism to solve the flatness problem. For more detailed explanations on these topics, see \hyperref[SubSect:Inflationary]{Sect.\,\ref{SubSect:Inflationary}} and references therein. For now, let us mention that one of the results of inflation is to account for a Universe that is flat on hypersurfaces of constant time. The total energy density then adds up to the critical energy density, which may indicate a missing piece in the cosmological puzzle, such as the Cosmological Term or other possibilities, such as relativistic decays of massive relics\,\cite{peebles1984tests,turner1984flatness}.

\end{itemize}
 
Whilst we have enumerated some examples of uses and mentions of $\Lambda$ in the literature through those years, it is true that most part of the conundrums they faced disappeared in the light of new data, better understanding of phenomena or more accurate modelization. Over time, it can not be denied that the CC suffered from a loss of reputation after Einstein's rejection, nonetheless it was incorporated to the models with some regularity. The incorporation of a non-null CC to cosmological or astrophysical models was just seen as an alternative or an extra ingredient of a baseline model in contexts where the simpler Einstein-De Sitter framework did disappointing predictions. This started to change after the conception of inflation and its possible phenomenological consequences, although no definite clues were present yet. The transcendental turning point was about to appear, at the end of the 20$th$ century.

\subsection{Modern Cosmology}\label{SubSect:ModernCosmology}

At the beginning of the 1990s, there was a paradigm shift in cosmology due to the addition of some new elements. In particular, Cold Dark Matter (CDM)\footnote{Although Dark Matter is another milestone of modern cosmology, we did not mention it during this chapter. The reader may find a historical review in\,\cite{bertone2018history}.} was introduced to solve or alleviate the problem of Missing Mass, and it was widely believed that the Universe was flat (thus, the total energy density equaled the critical one). These features led to the birth of the Cold Dark Matter model (CDM model). However, researchers were also considering additional ideas. The theoretical convenience of inflation led some researchers\,\cite{peebles1984tests,efstathiou1990cosmological} to expand the CDM model by reintroducing $\Lambda\neq 0$. This allowed them to accommodate a low matter density Universe that is spatially flat, as favored by observations.

The improvement in instruments during the 1990s had a significant impact on the quality and quantity of cosmological data. One example of this is the Cosmic Background Explorer (COBE), a satellite launched in 1989 to investigate the anisotropies and spectrum of the CMB\,\cite{smoot1992structure, boggess1992cobe}. COBE's Far-Infrared Absolute Spectrophotometer (FIRAS) experiment confirmed the predictions of Big Bang models that the CMB has a thermal spectrum with no major deviations from a blackbody spectrum over the spectral range of 500 $\mu$m to 1 cm. FIRAS also detected the dipole anisotropy. Another experiment, the Differential Microwave Radiometers (DMR), found upper limits for the variation of different multipole temperature amplitudes, all of which were around $\Delta T/T \lesssim 10^{-5}$, and a scale-invariant fluctuations spectrum. In short, the basic assumptions of inflation appeared to be in agreement with observations.

Later on, the launch of the Hubble Telescope in 1990 enabled the measurement of the Hubble rate, $H_0$, to be $H_0 = 80\pm 17$ km/s/Mpc, by measuring distances to Cepheids in the Virgo Galaxy Cluster M100\,\cite{freedman1994distance}. This value, within the framework of Einstein-de Sitter Universes with energy density equal to the critical one, contradicts the inferred low age of the Universe, a problem that had been observed repeatedly in the past. Therefore, once again, the data prefered a model with low energy density and a non-negligible Cosmological Term.

The accumulated evidence started to point towards an extension of the CDM model to a new model called $\Lambda$CDM, which incorporates a non-zero cosmological constant\,\cite{carroll1992cosmological,krauss1995cosmological}. This model of low energy density seemed to be favored by other probes, such as gravitational lensing\,\cite{krauss1992gravitational,krauss1993angular}, large-structure observations\,\cite{loveday1992large}, and galaxy clustering\,\cite{bahcall1992galaxy}. However, the constraints were not yet sufficiently strict to claim a discovery. 

The final boost for $\Lambda$ cosmology arrived at the end of the 20th century when collaborations began to focus on measuring luminosity distances to type Ia supernovae (SNIa) in order to investigate the deceleration parameter and the possible presence of a non-zero cosmological constant. A brief description of these objects can be found in \hyperref[Appendix:Description]{Appendix\,\ref{Appendix:Description}}.

A precise measurement of their apparent brightness and redshift could lead to constraints on the parameters of the cosmological models of interest. Two different collaborations were successful in this quest and were able to determine the cosmic late-time acceleration and the positivity of the $\Lambda$ term at a great confidence level. In 1998, the High-Z Supernova Search Team (HSST) found 34 low-redshift SNIa and 16 high-redshift ones, independently and shortly after, the Supernova Cosmology Project (SCP) collaboration found 42 SNIa in the range $z=0.18-0.83$\,\cite{perlmutter1999measurements, riess1998observational}. The clarity of the results and the agreement between both parties were enough to quickly convince the members of the scientific community that such a surprising claim was reliable. The newly confirmed component responsible for the accelerated expansion of the Universe was called Dark Energy (DE)\,\cite{huterer1999prospects,turner1999cosmology}. The name does not only reflect its lack of interaction with electromagnetic radiation but also our deep ignorance about its nature.

Since then, richer and richer cosmological data have emerged from different experiments, leading us into an era of precision cosmology. To mention a couple of significant space satellites, WMAP (launched in 2001)\,\cite{WMAPWeb} and Planck (launched in 2009)\,\cite{PlanckCollabWebESA}, have improved our measurements regarding the CMB and its anisotropies, shaping what we know as the currently accepted standard model, widely known as the $\Lambda$CDM or concordance model. The fractional energy density of matter and Dark Energy in the Universe with $68\%$ errors from Planck 2018 baseline data (TT,TE,EE+lowE+lensing) analysis are (see Table 2 in\,\cite{aghanim2020planck}) $ \Omega_{\rm m}^0=0.3153\pm 0.0073$ and $\Omega_\Lambda^0=0.6847 \pm 0.0073$. Therefore, the fraction of energy density corresponding to DE (represented by a CC) is now around $70\%$, which is larger, but at the same order of magnitude, than the fraction of matter (Cold Dark Matter+Non Relativistic Matter), near $30\%$. The solution to the problem of Missing Mass in spatially flat Universes is to assume that our Universe is filled with Dark Matter and Dark Energy. The latter is assumed to be a sort of unidentified substance with some unusual properties, such as a repulsive gravitational effect. It is incredible how quickly this huge surprise has been widely accepted and incorporated into our prior knowledge. However, after all, this possibility has been chasing us for many decades, so we have had enough time to get used to it.

So far, we have summarized the history of the CC: its reminiscences to modifications of Newtonian Gravity, its appearance as the savior of static Cosmology, his fall and his lately victorious come back into fashion in light of new evidence as a model for DE. Now, it has become the dominant component and dictates the dynamics of the Cosmos at the largest scales.

\section{Aspects of Vacuum Energy Density}\label{Sect:VacuumEnergyReview}

In \hyperref[Sect:HistoryCosmConst]{Sect.\,\ref{Sect:HistoryCosmConst}}, we presented a summary of the history of the CC, which is a mathematical term in Einstein's field equations\,\eqref{Eq:Introduction.LambdaEinsteinEq}. Einstein's gravity with a Cosmological Term was also formulated in terms of a variational principle, so it admits a Lagrangian description as an extension of the Einstein-Hilbert Action (EH Action):
\begin{equation}\label{Eq:Introduction.EHAction1}
 S=\int d^4 x \sqrt{-g}\left[\frac{1}{16\pi G_N}R-\rho_\Lambda\right]+\int d^4 x \sqrt{-g} \mathcal{L}_{\rm mat.}\,,
\end{equation}
where $g$ is the determinant of the metric tensor, $R$ is the Ricci Scalar, $\rho_\Lambda \equiv \Lambda/(8\pi G_N)$, and $\mathcal{L}_{\rm mat.}$ is the matter Lagrangian that needs to be specified. In this interpretation, $\Lambda$ is a parameter in the gravitational action not being part of the matter described by $\mathcal{L}_{\rm mat.}$. It would  be written in the left-hand side (LHS) of equation \eqref{Eq:Introduction.LambdaEinsteinEq}. Thus, gravity is described by two different constants: $G_N$ and $\Lambda$. Alternatively, the former action can be written as follows,
\begin{equation}\label{Eq:Introduction.EHAction2}
 S=\int d^4 x \sqrt{-g}\frac{1}{16\pi G_N}R+\int d^4 x \sqrt{-g} \left[\mathcal{L}_{\rm mat.}-\rho_\Lambda\right]\,.
\end{equation}
In this manner, the CC is understood to have something to do with matter fields, as a shift to the matter lagrangian, rather than being a pure geometrical effect. This fact does not change any inch of other matter fields' physics, since $\Lambda=constant$ means that their equations of motion are unaltered. So $\Lambda$ should take part of the RHS of \eqref{Eq:Introduction.LambdaEinsteinEq}. The EMT can be computed to be
\begin{equation}\label{Eq:Introduction.EMTforLambda}
 T_{\mu\nu}^\Lambda=-\frac{2}{\sqrt{-g}}\frac{\delta S_\Lambda}{\delta g^{\mu\nu}}=-\rho_\Lambda g_{\mu\nu},
\end{equation}
where $S_\Lambda \equiv -\int d^4x \sqrt{-g}\rho_\Lambda$. Comparing it with the EMT for a perfect fluid \eqref{Eq:Introduction.PerfectFluid}, we see that $P_\Lambda=-\rho_\Lambda$. As a consequence, the equation of state of this $\Lambda$-fluid, i.e the parameter relating its energy density and pressure, is -1. A negative pressure has little to do with daily experience, but it is easily shown to have this repulsive effect at large scales, as we discussed after \eqref{Eq:Introduction.FriedmannII}. One more bizarre consequence is that a CC that is interpreted as a fluid is expected to have an energy density $\rho_\Lambda$ that is constant with the expansion. Other components, such as Non-Relativistic Matter or Radiation, decrease their energy densities when the scale factor $a$ gets bigger. Eventually, we can expect the Universe to be dominated entirely by a CC in this simple picture.

The discussion about placing $\rho_\Lambda$ in one side or another in the action may seem superfluous. Einstein had a similar opinion with respect where to place $\Lambda$ in the field equations and he expressed into his note to Schrödinger, as said in the last section. A geometrical or a material explanation are, of course, mathematically equivalent. But, maybe this question may have a deeper purpose since it evokes the true nature of the CC or, more in general, of DE. The description of DE as the responsible of late cosmic acceleration does not constitute in any manner a satisfactory description: we know its effects (in an axiomatical way) but a connection of $\Lambda$ or any other form of DE with reality is still missing. Is it just a pure geometrical effect or may be better described through an Energy Momentum Tensor (EMT) like \eqref{Eq:Introduction.TensorLambda}? More in general, there are several different Dark Energy conceptions and most of them can be summarized in one of the following options:

\begin{enumerate}

 \item DE corresponds to a fluid associated to Quantum or Classical field and can be described through an Energy-Momentum tensor. This may be similar to what Schrödinger had in mind. 
 
 \item DE is just a geometrical effect connected to a particular extension of Einstein's gravity.
 
 \item What we currently attribute to DE may simply be an ``illusion''; the apparent acceleration that we observe in our cosmos could be a consequence of local peculiarities, such as overdensities or voids. In other words, the cause of the observed cosmic acceleration may not be a separate entity or form of energy, but rather a result of the way matter is distributed in our local region of the Universe.
 
 \item DE is the result of different contributions from fields and/or geometry.
 
\end{enumerate}

So far, we have refrained from delving into the physical interpretations of CC in\,\hyperref[Sect:HistoryCosmConst]{Sect.\,\ref{Sect:HistoryCosmConst}}, and instead, preferred to focus on their use in Cosmology throughout history. We briefly mentioned the ideas of Lemaître, who considered $\Lambda$ as a property of empty space or connected to the vacuum, and Zel'dovich's computations relating it to Zero-Point Energy (ZPE), which we will discuss later. In this section, we will talk about vacuum energy, which is the candidate par excellence as the physical interpretation of Dark Energy. We will start with a summary of the first ideas about the nature of the vacuum in antiquity and leading up to its first mathematical formulation in the context of quantum mechanics. Afterward, we will explore the developments in the 20th century regarding the vacuum, Zero-Point energy, and related notions such as renormalization.

\subsection{From ether to Quantum theory}\label{SubSect:Ether}

Vacuum, as the total absence of matter and energy, has always been a fascinating source of mystery. Its existence and naturalness have been a matter of philosophical discussions and experimental tests for several centuries. Its conception has changed a lot since the very early philosophical discussions in antiquity to the modern idea of vacuum as a state of quantum field theories and UV extensions of gravity.

An early starting point may be found in pre-Socratic philosophy, particularly in the philosophy of atomism. The two major exponents of atomism were the philosophers Leucippus (born in the early 5th century BCE) and Democritus (c. 460 BCE - c. 370 BCE). They defended the idea that the Universe is composed of two components: the {\it void} (empty space of non-finite extent where motion takes place, ``what is not'') and {\it atoms} (constituents of matter that are indestructible and indivisible, ``what it is'')\,\cite{warren2014presocratics}. Void cannot be understood without atoms and vice versa: Atoms are plural and divided entities and are separated by the void, while void needs the atoms to be distinguished and plural to separate them. They gave the same degree of existence to void and atoms.

Aristotle (384 BCE - 322 BCE) is one of the most famous philosophers of antiquity. In his work {\it Physics}, he denies the existence of void by analyzing hypothetical situations and using reductio ad absurdum arguments\,\cite{kragh2014weight}. Void space is defined as not containing bodies, but being able to receive them. For instance, he reasoned that bodies falling in empty space should eventually acquire infinite velocity independently of their size and form, which is not observed at all. Aristotelian physics is interesting on its own from the perspective of modern physicists\,\cite{rovelli2015aristotle}. He differentiates between the matter of the heavens (beyond the Moon) and that of Earth. The former is made of a kind of exotic substance called the {\it fifth element}, which fills the space around celestial bodies and is immutable. The Stoic school of philosophy (founded around the beginning of the $3rd$ century BCE\,\cite{StandfordStoics}) has some similar ideas to Aristotle. They assume a finite world enveloped in an infinite void that cannot interact with the material world. Interestingly, this is quite similar to the very first ideas of Newton about the shape of the Universe that we summarized in the previous section.

The ideas of Aristotle were translated into Latin during the medieval era, and although they gained popularity, they also concerned some theologians who feared their conflict with the omnipotence of God. The existence of vacuum was commonly denied until the 17{\it th} century. Galileo Galilei (1564 – 1642) questioned the belief that nature abhorred vacuum and was particularly interested in the motion of bodies in vacuum rather than in its specific nature. He avoided theological or cosmological discussions. Vacuum was simply unable to oppose the motion of bodies traversing it\,\cite{grant1981much}. Evangelista Torricelli (1608-1647), who was Galileo's pupil, and Blaise Pascal (1623-1662) conducted experiments with the goal of creating artificial vacuums by evacuating the atmosphere above a mercury column. The success of these experiments convinced Pascal that, after all, nature does not abhor vacuum. In approximately the same period, Robert Boyle (1627-1691) and Otto von Guerike (1602-1686), inventors and researchers, were also interested in the properties of artificially created vacuums. Guerike distinguished between the three-dimensional vacuum artificially evacuated by machines and the nothingness, the origin of the world created by God. The image that Boyle had in mind was simpler: vacuum was merely defined as space emptied of air, avoiding sophisticated explanations, despite having genuine questions regarding the nature of the created vacuum.

The birth of physics as an experimental and theoretical science gave rise to the concept of vacuum being related to many other phenomena. In particular, the concept of {\it ether} emerged, with some similarities to Aristotle's fifth element or {\it quintessence}. It was thought to be a subtle substance rather than pure emptiness and was used to explain various physical processes, such as electricity and gravity. James Clerk Maxwell (1831-1879) defended the existence of an ether as a complement to his theory of electromagnetism,\cite{maxwell1873treatise}, since electromagnetic phenomena were observed to occur even in empty space. He even later wondered if this ether could also be the origin of gravitation, assuming the ether to possess a positive energy whose disturbance by dense bodies produces gravitational attraction. Many other scientists made efforts to construct more sophisticated theories of ether or provide mathematical grounds for its existence, and its presence was well-established in the heart of electromagnetic theory for a great part of the 19{\it th} century.

The famous experiment performed by A. Michelson (1852-1931) and Morley (1838-1923)\,\cite{michelson1887relative} was an experimental test to measure the relative motion of Earth with respect the luminiferous ether, responsible for the propagation of electromagnetic fields and light, through an interferometer invented by Michelson. Incidentally, their results were compatible with the absence of relative motion, so a stationary luminiferous ether was harshly discredited and the experiment constitute one of the clues pointing out towards Special Relativity. 

To be fair, the concept of ether was not entirely ruled out. In the first part of the 20{\it th} century, many scientist identified ether with {\it vacuum}, however this vacuum was not expected to be empty space. For instance, H.Lorentz ( 1853 — 1928) still had in mind ideas similar to Maxwell's ether, defining the vacuum as the seat of electromagnetic fields, its energy and vibrations. Another example is O.Lodge (1851 - 1940), who did not identify empty space with a space without any content but plenty of ether with an extremely highly energetic medium and where matter is embedded. Although his description was vague and did not go much further, assigning intrinsic energy to the medium is very similar to the idea of Zero-Point Energy (ZPE) or to the modern concept of {\it vacuum energy}. But these ideas had their origin in quantum theory and its nature and formulation are quite different from previous considerations. 

The concept of Quantum Vacuum has its foundational origin in the concept of {\it Nullpunktsenergie} or {\it Zero-Point Energy} (ZPE) back in the 1900's. Let us start with the {\it Raylegh-Jeans} Formula which is a classical prediction for the spectral energy density distribution of a Black Body\footnote{We will not use natural units for this section, as we would like to remain closer to the original formulas.},
\begin{equation}\label{Eq:Introduction.RayleghJeans}
 \rho (\nu,T)=\frac{8\pi \nu^2}{c^3}k_{\rm B} T \,,
\end{equation}
where $\nu$ is the frequency of the radiation, $c$ is the speed of light in vacuum, $k_{\rm B}$ is the Boltzmann constant and $T$ is Black Body's temperature. This result matches observations for lower frequencies but not for larger frequencies, as the quadratic increase with $\nu$ provides a huge value that departs from the experimental results. This problem was known as the Ultraviolet (UV) Catastrophe and was an open question at the time. The German physicist Max Planck (1858-1947) proposed a revolutionary theory in 1900 that could solve the UV catastrophe. He imagined the boundaries of the Black Body to be constituted by harmonic oscillators and realized that if $\nu$ is the frequency of the absorbed or emitted radiation, the energy of the system changed discretely as a multiple of the energy element or {\it quanta} $h\nu$:
\begin{equation}\label{Eq:Introduction.PlanckBlackBodyQuantum}
 E_n = n h\nu\,,
\end{equation}
where $n=0,1,2,\dots$ and $h$ is a new constant of nature currently know as Planck's Constant\footnote{The current recommended value from CODATA committee\,\cite{Tiesinga:2021myr} is $h=6.62607015\times 10^{-34}$ Js} which Planck estimated to be $h=6.55\times10^{-34}$ Js. Additionally,  he derived a formula for the spectral distribution different from $Raylegh-Jeans$ formula at high frequencies
\begin{equation}\label{Eq:Introduction.PlanckBlackBodyQuantumDistribution}
\rho (\nu,T)=\frac{8\pi \nu^2}{c^3}\frac{h\nu}{\exp\left(\frac{h\nu}{k_{\rm B} T}\right)-1}\,.
\end{equation}
Nevertheless, Eq.\,\eqref{Eq:Introduction.PlanckBlackBodyQuantumDistribution} not only reproduces eq. \eqref{Eq:Introduction.RayleghJeans} at low frequencies $h\nu\ll k_{\rm B} T$ but also reproduces the Stefan-Boltzmann Law and Wien's Displacement Law. Such an amazing result, however, did not appease Planck's distrust of the idea of the {\it quanta} of energy as physically realistic rather than a useful mathematical formulation. Some years later, he reformulated his theory in such a way that it resulted in a more recognizable form from a classical perspective\,\cite{planck1914theory,Planck1958}. In this new view, the emission of radiation is a probabilistic quantized phenomenon, although the absorption process has no restrictions and can occur in a continuous manner, we know this theory as his ``second theory''. He then derived an expression for the average energy of an oscillator with temperature $T$ vibrating at a frequency $\nu$,
\begin{equation}\label{Eq:Introduction.ZeroPointEnergy}
\bar{E}=\frac{h \nu}{2}+\frac{h \nu}{\exp \left(\frac{h\nu}{k_{\rm B} T}\right)-1}\,,
\end{equation}
with the appearance of the extra term $h \nu /2$ which remains in the limit $T\rightarrow 0$. This is the birth of the concept {\it zero-point energy} (ZPE). Planck was conscious that ZPE was something totally unfamiliar to classical physics and a non-measurable quantity. Trying to remove it from the equations made sense, but there was a good reason for not doing so. Thanks to the ZPE term, one can recover the average energy of an oscillator at a fixed temperature $T$ in thermal equilibrium, $\bar{E}=k_{\rm B} T$, from \eqref{Eq:Introduction.ZeroPointEnergy} in the classical limit $k_{\rm B } T\gg h\nu$.

Overall, the second theory of Planck was controversial and was disproved by many physicists such as Niels Bohr in light of his atomic theory. However, the concept of ZPE and whether this quantity was real or a spurious character in quantum theory attracted attention in the physics community from both theoretical and experimental sides\,\cite{einstein1913einige,debije1913einfluss,stern1913kinetischen,mulliken1925isotope}. Progressively, evidence for the existence of Zero-Point energy started to accumulate, but it was not until the appearance of quantum mechanics that the concept started to be backed up by a fundamental theory. Heisenberg\,\cite{heisenberg1925quantum} and Schrödinger\,\cite{schrodinger1926quantisierung} independently derived the energy levels of the quantum harmonic oscillator as
\begin{equation}\label{Eq:Introduction.EnergyLevels}
E_{\rm n}= \left(n+\frac{1}{2}\right) h \nu\,,
\end{equation}
recovering the ZPE result obtained by Planck's second theory back in 1911. Additionally, this result gained much more intuition thanks to Heisenberg's uncertainty principle, which means that even the ground state of the system should not possess null energy.

It is necessary to mention at this point that there were cosmological applications of the concept of ZPE prior to quantum mechanics. In 1916, Walther Nernst,\cite{nernst1916versuch} proposed a revised concept of the ether, which was thought to be filled with electromagnetic zero-point radiation. This etherial light medium stored a vast amount of energy in the form of ZPE, which had the capability to produce new atoms or absorb their decays. The ether was conceived to be in equilibrium with radiation inside the medium, in a continuous exchange of energy with the material content of the Universe, and the conservation of energy could only be understood statistically. The energy density of ZPE was obtained by applying the classical statistical mechanics Rayleigh-Jeans formula\,\eqref{Eq:Introduction.RayleghJeans}, but replacing $k_{\rm B}T$ with $h\nu$,
\begin{equation}\label{Eq:Introduction.RayleighII}
\rho_{\rm ZPE}(\nu)=\frac{8\pi h}{c^3}\nu^3\,.
\end{equation}
The previous formula, when integrated over all the possible frequencies, is quadratically divergent,
\begin{equation}\label{Eq:Introduction.cutoff}
\rho_{\rm ZPE}=\int_0^{\nu_{\rm max}} \rho_{\rm vac}(\nu)d\nu=\frac{2\pi h}{c^3}\nu_{\rm max}^4\,.
\end{equation}
In modern language, the introduction of $\nu_{\rm max}$ is called an {\it UV cutoff}, a regularization technique used in many works in posterior decades.

\subsection{Vacuum energy in Quantum Field theory and Cosmology}\label{SubSect:DEinQFT}

The ideas of Nernst were considered fringe at the time and somewhat vague, probably due to their lack of connection with General Relativity. Nevertheless, many people were aware of his work. For example, Wolfgang Pauli repeated Nernst's calculations but with a cutoff equal to the classical electron radius of about $10^{-13}$ cm. By relating the cosmological constant of Einstein's equation with the value of the calculated ZPE, it yielded a value of about 30 km for the radius of Einstein's world, which, as he claimed, ``would not even reach the moon''. It is also important to note a paper from Max Born, Werner Heisenberg, and Pascual Jordan\,\cite{born1925quantenmechanik,born1926quantenmechanik} in which they attempted to directly quantize the electromagnetic field. In addition to the thermal energy of the oscillators, a ZPE term appeared in their computations, $\sum h\nu_{k}/2$, summed over the $k$ degrees of freedom of the system, and resulting in the re-derivation of Einstein's formula for the statistical fluctuations of Planckian blackbody radiation. Jordan was still not convinced of the real nature of ZPE, first because of the lack of intuition regarding its physical origin, and second, because it still yielded a divergent contribution for an infinite number of degrees of freedom. To deal with this unnatural prediction, they tried a subtraction procedure\,\cite{born1926quantenmechanik} together with Born and Heisenberg. This involved removing the ZPE contribution of the oscillators to the Hamiltonian operator, thereby removing the infinite contribution in the example of a one-dimensional string. This constitutes the first example of a manifest divergence in quantum field theory (QFT) and also a primeval example of renormalization\,\cite{Schweber:1994qa}.

It became apparent from the works of Pauli and Jordan that they did not rely on the concept of ZPE in non-material systems such as the electromagnetic field. Rather, they viewed it more as a mathematical formality. In his work in 1933, Pauli derived a finite energy density associated with the electromagnetic field, dropping out the ZPE contribution in a procedure similar to the normal ordering in QFT which would be developed in the following decades. He stated that ZPE gave rise to an infinitely large energy per unit volume which was apparently unobservable, as it could neither be emitted nor absorbed, interact with other media, nor produce new particles out of the vacuum. As QFT developed, the concept of vacuum became more intricate, no longer as simple as previously imagined. One well-known example is the concept of Dirac's sea and its interpretation of the vacuum as a region full of negative energy states. However, we will not delve further into the history of QFT and the mathematical modeling of the vacuum state. 

The cosmological constant present in Einsten's equation and the Vacuum energy arising from QFT are two concepts with different origins. While the cosmological constant (CC) is a mathematical term introduced by Einstein in his field equations to describe his own model, vacuum energy is a rather obscure ingredient of QFT that may or may not emerge from the ZPE. The latter has been the subject of some naive calculations in the cosmological context.  But, as we mentioned in the previous section, one of the first clear interpretations of Vacuum Energy identified as the origin of the CC was given by Lemaître\,\cite{lemaitre1934evolution}:

{\it ``We must associate the pressure $p=-\rho c^2$ to the density of energy $\rho c^2$ of vacuum. This is essentially the meaning of the cosmical constant $\lambda$ which corresponds to a negative density of vacuum according to $\rho_0 = \lambda c^2 /4\pi G\approx 10^{-27} g/cm^3$.''}

Notice Lemaître's identificacion of Vacuum Energy with a positive cosmological constant corresponding to a negative energy density and positive pressure. He did not mention any ZPE nor QFT connection, though.  

An important result on the matter appeared in 1948, by the hand of Hendrik Casimir, discoverer of the {\it Casimir effect} as an attractive force acting on macroscopic uncharged boundaries due to the fluctuations of the ZPE associated to the electromagnetic field\,\cite{casimir1948attraction,lamoreaux2004casimir}. For instance, for two parallel plates, the force per unit area is proportional to $d^{-4}$, where $d$ is the distance between the plates.  The calculation of the Casimir force arises from the the computation of the total zero-point energy between the plates on a particular geometry compared with free space and assumes the assignation of ZPE of $h\nu/2$ to each mode of the electromagnetic fields. At the time, it was one of the first results associated to a direct effect of zero-point fluctuations whose experimental evidence arrived 10 years later\,\cite{sparnaay1958measurements}. Modern experiment with atomic force microscopy in\,\cite{Mohideen:1998iz}) confirmed this effect with higher precision. Whether if this effect is really a consequence of the ZPE emergence of the quantum fields or not\,\cite{schwinger1978casimir,Rugh:1999dmo} is out of the scope of this introduction. What we can say from the historical point of view is that the discovery of Casimir effect was another milestone in the history of vacuum energy and ZPE.

Yakov Zel'dovich, a russian cosmologist, was a capital figure in the theoretical understanding of the cosmological constant and its connection to the vacuum energy. He wrote a complete review on the matter\,\cite{zel1967cosmological,zel1968cosmological}, revisiting some old ideas regarding the Cosmological Term as well as incorporating new novelty ideas to the topic. He shown that a small, but non-zero, value of the CC did not contradict any cosmological observation but can accommodate the paper of a zero-point vacuum fluctuations in the cosmological realm. He starts his review mentioning the contemporany works of Petrosian, Salpeter, Szekeres, Shklovskii and Kardashev in dynamical models of Universe incorporating a cosmological constant, $\Lambda$, motivated by the discoveries of several remote quasars at a narrow redshift range around $z=1.95$. In the first part of the review he used the sentence 

{\it ``The genie has been let out of the bottle, and it is no longer easy to force it back in.''}

referring to the lack of fundamental arguments or observations that would set the value of $\Lambda$ to zero. It was noted that a small value of $|\Lambda| \lesssim 10^{-54}$ cm$^2$ was naturally expected, and its influence could only be relevant at the largest (cosmological) scales. In his paper, he defined an Energy-Momentum tensor equivalent of the cosmological constant, in which $\rho_\Lambda \equiv c^2 \Lambda /(8\pi G_N)$ and $P_\Lambda=-\rho_\Lambda$. This tensor preserves relativistic invariance since $\rho_\Lambda$ and $P_\Lambda$ have the same constant value at each coordinate point. Later, in an appendix of the same paper, he presented the regularization of the vacuum energy density associated with a scalar field from Zero Point fluctuations,
\begin{equation}\label{Eq:Introduction.Zel'dovichZPE}
\rho_\Lambda =K\int_0^\infty \sqrt{p^2+m^2c^2} p^2 dp \equiv K I(m) \,,
\end{equation}
where $K\equiv c/(4\pi^2 \hbar^3)$ and $m$ is the rest mass of the scalar field and, similarly, the pressure can be obtained from the diagonal $T_{11}$ component of the Vacuum Expected Value of the Energy-Momentum tensor,
\begin{equation}\label{Eq:Introduction.Zel'dovichPressure}
P_\Lambda =K\frac{1}{3}\int_0^\infty \frac{p^4}{\sqrt{p^2+m^2c^2}}dp\equiv K F(m)\,.
\end{equation}
Both integrals in \eqref{Eq:Introduction.Zel'dovichZPE} and \eqref{Eq:Introduction.Zel'dovichPressure} are manifestly divergent. A similar calculation follows for spin-1/2 fermions. After adding up the contributions from several fields, the regularization process of these quantities is presented by a promotion of the discrete sum to an integral and the constant $K$ to be an arbitrary function of the rest mass, $f(m)$. 
\begin{equation}\label{Eq:Introduction.Zel'dovichRegularize}
\rho_\Lambda =\int_0^\infty f(m)I(m)dm, \qquad P_\Lambda=\int_0^\infty f(m)F(m)dm\,.
\end{equation}
After introducing and UV cutoff much greater than the masses of the fields on $I(m)$ and $F(m)$ he realized that the former function $f(m)$ need to satisfy the following regularization conditions 
\begin{equation}\label{Eq:Introduction.Zel'dovichconditions}
\int_0^\infty f(m)dm =\int_0^\infty f(m)m^2 dm= \int_0^\infty f(m) m^4 dm=0\,,
\end{equation}
and, after performing the limit to infinity of the cutoff the remaining vacuum energy and pressure are
\begin{equation}\label{Eq:Introduction.Zel'dovichRegularized}
\rho_\Lambda=-P_\Lambda\equiv \frac{1}{8}\int_0^\infty f(m)m^4 \ln (m) dm\,.
\end{equation}
For Zel'dovich, this calculation showed that the ZPE was not equal to 0 and led to the natural equation of state, $w_\Lambda\equiv P_\Lambda/\rho_\Lambda=-1$, for the cosmological constant. At the end of the review, he presented his famous formulas for estimating the value of the cosmological constant and vacuum energy density. The first estimation is related with the lead contribution of the cutoff procedure,
\begin{equation}\label{Eq:Introduction.Zel'dovichEstimationI}
\rho_\Lambda=m_{\rm p}^4\frac{c^5}{\hbar^3}\sim 10^{50} \textrm{eV/cm}^3, \qquad \Lambda = 10^{-10} \textrm{cm}^{-2}\,,
\end{equation}
for $m_{\rm p}$, the mass of the proton.This guess is dimensionally correct but far away from the order of magnitude estimated by cosmological observations at the time, $|\Lambda|\lesssim 10^{-54}$ cm$^2$. In order to refine the previous estimate, Zel'dovich used the numerology of dimensionless ratios, as previously done by Dirac,\cite{dirac1937cosmological}, to obtain the quantity $Gm_{\rm p}^2/(\hbar c)\sim 10^{-38}$ (as a way to characterize the smallness of gravitational interaction). So that, 
\begin{equation}\label{Eq:Introduction.Zel'dovichEStimationII}
\rho_\Lambda=\frac{Gm_{\rm p}^2}{\hbar c}m_{\rm p}^4\frac{c^5}{\hbar^3}=\frac{Gm_{\rm p}^6 c^4}{\hbar^4}\sim 10^{12} \textrm{eV/cm}^3, \qquad \Lambda = 10^{-48} \textrm{cm}^{-2}\,,
\end{equation}
which is closer, but still some orders of magnitude above the cosmological measures. He ends the review by admitting the crucial connection between the question of the magnitude of the cosmological constant and elementary particle physics, as any use of typical particle physics scales seems to fail in order to match a prediction for the estimated value of $\Lambda$ from observations. We can say that this review is the first manifestation of the so-called {\it Cosmological Constant Problem} (CCP). The CCP is a gigantic problem that is still active today, probably more than ever, associated with the estimations of the value of the Cosmological Constant through calculations of the ZPE and other vacuum contributions. In a simplified scenario in QFT, such as the one studied by Zel'dovich, the leading contribution of the ZPE yields a gigantic contribution of order $m^4$, where $m$ is the mass of an elementary particle. When compared to the reported values of the critical energy density of the Universe, $\rho_{\rm c}=3 H_0^2 /(8\pi G_N)$ (or to the cosmological vacuum energy density, which we now know is of the same order of magnitude as $\rho_{\rm c}$), we are left with an huge difference of several orders of magnitude.

The fact that naive computations of the ZPE yield enormous values is just the beginning of the problem. It is natural to consider other contributions to the vacuum energy that could act like a cosmological constant. An additional vacuum contribution can shift the value of the cosmological constant, of course\,\cite{Weinberg:1989mp}. This can be done by introducing a VEV Energy-Momentum tensor in the form:
\begin{equation}\label{Eq:Introduction.AdditionalContribution}
\langle T_{\mu \nu} \rangle = -\langle \rho \rangle g_{\mu\nu}\,,
\end{equation}
which describes a contribution the vacuum energy in a Lorentz invariant form. Introducing again natural units, the effective vacuum energy density presents a shift from its original value
\begin{equation}\label{Eq:Introduction.CosmologicalConstantEffShift}
\rho_{\rm vac}^{\rm eff}\equiv \Lambda/(8\pi G_N)+\langle \rho \rangle\,.
\end{equation}
or 
\begin{equation}\label{Eq:Introduction.CosmologicalConstantEffShiftII}
\Lambda_{\rm eff}\equiv \Lambda+8\pi G_N\langle \rho \rangle \,,
\end{equation}
in terms of the effective cosmological constant, with $\rho_{\rm vac}^{\rm eff}=\Lambda_{\rm eff}/(8\pi G_N)$. For instance, in 1974, Joseph Dreitlein\,\cite{dreitlein1974broken} speculated that a broken symmetry may produce a non-zero Vacuum Expected Value (VEV) of the physical Energy-Momentum tensor, $\langle T_{\mu\nu}\rangle \neq 0$, appearing as a sort of effective cosmological term in Einstein's equations. As an example, we can illustrate this fact with the model of electroweak interaction (EW)\,\cite{weinberg1967model,salam1968elementary,glashow1961partial} and the mechanism of Spontaneous Symmetry breaking for generating the mass of elementary particles. In fact, the role of vacuum energy in cosmology acquired a new dimension after the success of the EW theory, and particle physicists started to ask about its applications to cosmology. In the early 80s, the proposal of the theory of {\it inflation}\,\cite{guth1981inflationary,linde1982new,albrecht1982cosmology,Starobinsky:1980te} was a great revolution for our view of the history of the early Universe. Basically, it is characterized by a short period of accelerated expansion in the early Universe carried by a high density of vacuum energy as a possible solution to several problems in cosmology, such as the {\it horizon} and {\it flatness} problems. We will say more about inflation and these problems in \hyperref[SubSect:Inflationary]{Sect.\,\ref{SubSect:Inflationary}}.

To study a simple model of Spontaneous Symmetry Breaking, let us consider a simplified scenario with a classical potential for a single scalar field $\varphi$,
\begin{equation}\label{Eq:Introduction.HiggsClassPotential}
V(\phi)=\frac{1}{2}m_{\phi}^2+\frac{\lambda}{4!}\phi^4\,.
\end{equation}
with $\lambda >0$. The classical action including the matter part is
\begin{equation}\label{Eq:Introduction.ActionTotal}
S=\int dx^4 \sqrt{-g} \frac{1}{16\pi G_N}R+S_{\rm matt}[\phi,\Lambda]\,,
\end{equation}
where 
\begin{equation}\label{Eq:Introduction.ActionMatterHiggs}%
S_{\rm matt}[\phi,\Lambda] \equiv -\int dx^4\sqrt{-g}\left[\frac{1}{2}g^{\mu\nu}\partial_\mu \phi \partial_\nu \phi + \left(V(\phi)+\rho_{\Lambda}\right)\right]\,.
\end{equation}
Here we have included the term $\rho_\Lambda\equiv \Lambda/(8\pi G_N)$ in the matter action. From the matter action, one can compute the matter Energy-Momentum tensor (EMT) to be $T_{\mu \nu}^{\rm matt}\equiv T_{\mu\nu}^\Lambda+T_{\mu\nu}^\phi$, being
\begin{equation}\label{Eq:Introduction.EMTphiandlambda}%
T_{\mu\nu}^\Lambda\equiv -\rho_{\Lambda} g_{\mu\nu}, \qquad T_{\mu\nu}^\phi \equiv \partial_\mu \phi \partial_\nu \phi-\frac{1}{2}g_{\mu \nu} \partial_\alpha \phi \partial^\alpha \phi  -V(\phi) g_{\mu\nu}\,.
\end{equation}
In the ground state of the field, no kinetic energy is remaining and we are left with
\begin{equation}\label{Eq:Introduction.EMTphiandlambdaVEV}%
\left\langle T_{\mu\nu}^\Lambda \right\rangle = -\left(\rho_\Lambda +\langle V(\phi ) \rangle\right) g_{\mu\nu}\equiv -\rho_{\rm vac}^{\rm eff}g_{\mu\nu}\,,
\end{equation}
as explained before \eqref{Eq:Introduction.CosmologicalConstantEffShift}. However, this is just the classical vacuum energy, since we still have to implement the symmetry breaking. For $m^2<0$, the ground state of the field is not trivial,
\begin{equation}\label{Eq:Introduction.GroundState}%
\left\langle \phi \right\rangle = \sqrt{-\frac{6m^2}{\lambda}}\,.
\end{equation}
The existence of a non-zero ground state generates an electrowak phase transition by the Higgs potential, inducing a contribution
\begin{equation}\label{Eq:Introduction.InducedEnergy}%
\rho_{\rm EW}=\left\langle V (\phi ) \right\rangle = -\frac{1}{8}M_{\mathcal{H}}^2 v^2\,.
\end{equation}
The physical mass of the field is can be recognized from the oscillations around the minimum of the potential,
\begin{equation}\label{Eq:Introduction.MassHiggs}%
M_{\cal H}\equiv \frac{\partial^2 V( \phi)}{\partial \phi^2}\Bigg|_{\phi = v}=-2m^2\,.
\end{equation}
The induced energy density can also be written in terms of the Fermi constant, $\rho_{\rm EW}\sim \langle V \rangle\sim -M_{\mathcal{H}}^2 /G_{\rm F}$, where $M_{\mathcal{H}}$ is the mass of the Higgs boson and $G_{\rm F}\sim 1.17\times 10^{-5}$ GeV$^{-2}$. The discovery in 2012 of the Higgs Boson\,\cite{ATLAS:2012yve,CMS:2012qbp}, of mass $M_{\mathcal{H}}\approx 125$ GeV, hence would suppose an additional contribution of $-10^8$ GeV$^4$ (now in natural units) to the vacuum energy density budget. What are the consequences of this fact? A current estimate of the critical density of the Universe is $\rho_{\rm c}^0 \sim 10^{-47}$ GeV$^4$. If we believe that the effective vacuum energy density, composed by different contributions, is to be identified by the cosmological observations then,
\begin{equation}\label{Eq:Introduction.CosmologicalSum}
\rho_{\rm c}^0\sim\rho_{\rm vac}^{\rm eff} = \rho_{\Lambda}+\rho_{\rm EW}\,.
\end{equation}
In this case, $\rho_{\Lambda}$ is related with the parameter of Einstein-Hilbert action \eqref{Eq:Introduction.EHAction1} as $\rho_{\Lambda}=\Lambda /(8\pi G_N)$. If we identify this quantity with the usual ZPE interpretation of Zel'dovich, the leading contribution to this term would be of order $m^4$ (coming from the leading term of the cutoff regularization). For instance, for the electron, this would suppose $\rho_{\Lambda}\sim m_{\rm e}^4\sim 6.8 \times 10^{-14}$ GeV$^{4}$, almost 34 order of magnitude bigger than $\rho_{\rm c}^0$. If we believe QFT up to the Planck scale, we have then $\rho_{\Lambda}\sim m_{\rm Pl}^4=2.2\times 10^{76}$ GeV$^4$, more than 120 orders of magnitude of difference! Of course, one can decide to left free the value of $\rho_{\Lambda}$ by the moment, but we should not forget the contribution of the SSB, which at the same time satisfies $|\rho_{\rm EW}/\rho_{\rm c}^0|=\mathcal{O}(10^{55})$.  
The decomposition of the vacuum energy in several terms of different orders of magnitude tells us  that they have to conspire to add up the remaining tiny term $\rho_{\rm c}^0$. Thus, we have to choose the original term $\rho_{\rm \Lambda}$ with a precision of at least $55$ decimal places in order to satisfy the equation\,\eqref{Eq:Introduction.CosmologicalSum}:
\begin{equation}\label{Eq:Introduction.CosmologicalSumII}
10^{-47} \textrm{GeV}^4=\rho_{\Lambda}-10^8 \textrm{GeV}^4\,.
\end{equation}
We call {\it fine tuning} to this unnatural adjust we perform in order to solve the CCP. This kind of formulation is common to each Dark Energy model, that is, it is not only proper of a model of Dark Energy modelized as a Cosmological constant, similar patologies can affect other models such as quintessence scalar fields\,\cite{Sola:2013gha}.

Of course, the previous situation is a quite simplified scenario that only serves to illustrate the problem. We should recall that the ZPE of quantum fields consists of all the vacuum-to-vacuum diagrams, those which do not possess external tails. This is a pure quantum effect whose loop order corrections can be labeled by powers of $\hbar$:
\begin{equation}\label{Eq:Introduction.ZeroPointCorrections}
\rho_{\rm ZPE}=\hbar \rho_{\rm ZPE}^{(1)}+\hbar^2\rho_{\rm ZPE}^{(2)}+\dots
\end{equation}
The first term is the one-loop approximation, that we already reviewed, see \eqref{Eq:Introduction.Zel'dovichZPE}. However, let us take a closer look to it:
\begin{equation}\label{Eq:Introduction.ZPEOneLoop}
\rho_{\rm ZPE}^{(1)}=\frac{1}{4\pi^2}\int_0^{M_{\rm UV}}dp p^2 \sqrt{p^2+m^2}=\frac{M_{\rm UV}^4}{16\pi^2}\left(1+\frac{m^2}{M_{\rm UV}^2}-\frac{1}{4}\frac{m^4}{M_{\rm UV}^4}\ln \frac{M^2_{\rm UV}}{m^2}+\dots\right)\,.
\end{equation}
Here $M_{UV}$ is an Ultraviolet cutoff mass scale. Eq.\,\eqref{Eq:Introduction.ZPEOneLoop} constitutes just a bare quantity, prior renormalization. The term proportional to $m^4 \ln m^2$ is independent of $M_{\rm UV}$ and is of capital importance since it will appear in the final renormalized result, which is expected to be cutoff independent and similar to
\begin{equation}\label{Eq:Introduction.ZPEOneLoopCutoffInd}
\rho_{\rm ZPE}^{(1)}=-\frac{m^4}{64\pi^2}\ln \frac{\mu^2}{m^2}+\dots
\end{equation}
In the previous formula, $\mu$ represents a mass or energy renormalization scale that persists after renormalization. However, the quartic term $m^4$ continues to provide a large contribution. For any standard choice of the mass $m$ within the Standard Model of particle physics, the resulting contribution to the ZPE would be enormous, exceeding the quantity we wish to estimate, $\rho_{\rm vac}^{\rm obs}$, which is comparable to the critical density $\rho_{\rm c}^0$.

There are renormalization schemes which can motivate the result\,\eqref{Eq:Introduction.ZPEOneLoopCutoffInd}. One of the most common ones is the Minimal Subtraction Scheme (MS-Scheme) under the {\it dimensional regularization} formalism. We replace the dimensionality of the integral in order to make explicit the divergences of the integral in the form of a simple pole,
\begin{equation}\label{Eq:Introduction.ZPEDR}
\rho_{\rm ZPE}^{(1)}=\frac{1}{2}\mu^{2\epsilon}\int \frac{d^{3-2\epsilon}k}{(2\pi)^{3-2\epsilon}}\sqrt{k^2+m^2}=\frac{1}{2}\frac{m^4}{2(4\pi)^2}\left(-\frac{1}{\epsilon}-\ln \frac{4\pi \mu^2}{m^2}+\gamma_{\rm E}-\frac{3}{2}\right)\,.
\end{equation}
The scale $\mu$  is the 't Hooft mass unit, necessary to preserve the correct dimensionality of $\rho_{\rm ZPE}^{(1)}$ and acts as a renormalization scale. The constant $\gamma_{\rm E}$ is the Euler-Mascheroni constant. We redefine the dimensionality of the integral through $\epsilon\equiv (4-n)/2$, and we understand the limit $\epsilon\rightarrow 0$. The reader may find more information about this computation in the appendices \hyperref[Sect:MasterInt]{\ref{Sect:MasterInt}} and \hyperref[Appendix:Dimensional]{\ref{Appendix:Dimensional}} of this work with useful formulas and details which explicitly implement the aforementioned calculation. In these appendices, we also present our perspective on this regularization procedure and its relation to our work. To avoid overwhelming the reader with explanations here, let us skip the details. In equation \eqref{Eq:Introduction.ZPEDR}, we see another example of regularization, which must be followed by a renormalization procedure. This process removes the former divergences and produces a final finite result that is closer to the physical one. An example of this procedure is the introduction of counterterms. The key idea is to consider the original Einstein-Hilbert action as a {\it bare} action\footnote{Strictly speaking, this action is not renormalizable in the context of QFT in curved spacetime. Short distance effects are important when studying the quantum fluctuations, and for this reason one has to incorporate higher-order tensors (such as $R^2, \Box R,\dots$) in the action of General relativity to make the theory renormalizable in semiclassical QFT in curved spacetime, where gravity is not quantized, only matter fields. Again, this is well understood in the main text of this dissertation (see \hyperref[Chap:QuantumVacuum]{Chap.\,\ref{Chap:QuantumVacuum}}) and we will not provide more details here, as we only need to focus in the low-energy regime for the sake of the explanations we plan to give in this introduction.}.
\begin{equation}\label{Eq:Introduction.EHBareAction}
S_{\rm EH}=-\int d^4 x \sqrt{-g}\left(\frac{1}{16\pi G^{({\rm b})}}R+\rho_\Lambda^{({\rm b})}\right)\,.
\end{equation}
The superindex ${({\rm b})}$ means that the couplings of the action are just bare quantities that will play a role in the renormalization process. When considering the full action the combined effect of $ \rho_\Lambda^{({\rm b})}+\rho_{\rm ZPE}^{{\rm (b)}}$ should be taken into account. The final result does not depend on the renormalization scale, and we  expect $\rho_\Lambda^{({\rm b})}+\rho_{\rm ZPE}^{{\rm (b)}}=\rho_\Lambda (\mu )+\rho_{\rm ZPE}(\mu )$. At this point, we should split $\rho_\Lambda^{({\rm b})}$ in the renormalized term plus a counterterm, $\rho_\Lambda^{({\rm b})}=\rho_\Lambda (\mu)+\delta \rho_\Lambda$. So that,
\begin{equation}\label{Eq:Introduction.SumBareQuan}
\rho_\Lambda^{({\rm b})}+\rho_{\rm ZPE}^{{\rm (b)}}=\rho_\Lambda (\mu)+\delta \rho_\Lambda+\rho_{\rm ZPE}^{({\rm b})}=\rho_\Lambda (\mu)+\delta \rho_\Lambda + \rho_{\rm ZPE}^{(1)}+\dots
\end{equation}
where the dots represent contribution beyond one-loop. In the (modified) Minimal Subtraction Scheme, denoted by $\overline{\rm MS}$, we define the counterterm as
\begin{equation}\label{Eq:Introduction.MSModCounterterm}
\delta\rho_\Lambda^{\overline{\rm MS}}\equiv -\frac{1}{2}\frac{m^4}{2(4\pi)^2}\left(\frac{1}{\epsilon}+\ln 4\pi-\gamma_{\rm E}\right)\,.
\end{equation}
By introducing this at \eqref{Eq:Introduction.SumBareQuan} we obtain the following result at one-loop:
\begin{equation}\label{Eq:Introduction.SumBareQuanII}
\rho_{\rm ZPE}^{(1)} (\mu )=-\frac{1}{2}\frac{m^4}{2(4\pi)^2}\left(\ln \frac{\mu^2}{m^2}+\frac{3}{2}\right)\,.
\end{equation}
The above expression is the renormalized result for the one-loop ZPE contribution to the vacuum energy, yielding a term of the style of \eqref{Eq:Introduction.ZPEOneLoopCutoffInd}. The combined renormalized effect of $\rho_\Lambda$ and ZPE is then,
\begin{equation}\label{Eq:Introduction.OneLoopVac}
\rho_{\rm vac}^{(1)}=\rho_{\Lambda}(\mu)+\frac{m^4 \hbar}{4(4\pi)^2}\left(\ln \frac{m^2}{\mu^2}-\frac{3}{2}\right)\,,
\end{equation}
the one-loop approximation of the vacuum energy density, which does not depend on the renormalization scale $\mu$, as mentioned before\,\eqref{Eq:Introduction.SumBareQuan}. The overall expression is $\mu$-independent, despite some of the terms are explicitely depending on this scale. The dependence on $\mu$ should cancel in the full renormalized effective action, and we obtain the beta function for the parameter $\rho_\Lambda$, after imposing the Renormalization group (RG) equation, $d\rho_{\rm vac}^{(1)}/d\ln \mu=0$,
\begin{equation}\label{Eq:Introduction.OneLoopBeta}
\beta_\Lambda^{(1)}\equiv \mu\frac{d\rho_\Lambda (\mu )}{d\mu}=\frac{\hbar m^4}{2 (4\pi)^2}\,.
\end{equation}
The result is quite standard in flat spacetime QFT, however it may have its own problems as we will discuss later on (cf. \hyperref[Sect:VEDMSS]{Sect.\,\ref{Sect:VEDMSS}}). More in general, there are more quantum effects in the effective potential beyond the ZPE (those taking into account not only the ZPE vacuum-to-vacuum diagrams, but also diagrams with external tails) that we have neglected previously. The calculational procedure is more involved (see\,\cite{Sola:2013gha} for a complete review on the matter), and the structure of the vacuum energy is then
\begin{equation}\label{Eq:Introduction.EntireRhoVacuum}
\rho_{\rm vac}=\rho_\Lambda^{\rm ren}+\left\langle V_{\rm eff}^{\rm ren} (\phi) \right\rangle\,,
\end{equation}
which is valid to all orders of perturbation theory and, again, the total vacuum energy is expected to satisfy the RG equation $d\rho_{\rm vac}/d\ln M=0$, so that the total dependence on the generical renormalization scale $M$ (which is not necessarily related to $\mu$ used in dimensional regularization  as it can come from alternative renormalization procedures) should not appear in the final result.

But, after all these historical and conceptual details, the final (and most important message) that we have to transmit in this section is that the Cosmological Constant Problem is not only the difficult problem of matching the naive prediction provided by the ZPE naive formulas like\,\eqref{Eq:Introduction.Zel'dovichEstimationI}. Once we start to take into account more vacuum related effects such as the mentioned Spontaneous Symmetry breaking of Electroweak interactions (EW) and Effective potential contributions at different loops in perturbation theory the problem starts to grows immesurably,
\begin{equation}\label{Eq:Introduction.CCPProblemFull}
\rho_{\rm c}^0 \sim  \rho_{\rm vac}^0=\rho_\Lambda+\rho_{\rm EW}+\left\langle V_{\rm eff}^{(1)} (\phi ) \right\rangle+\left\langle V_{\rm eff}^{(2)} (\phi )\right\rangle\dots\,
\end{equation}
the term $\rho_\Lambda$ has to be readapted order by order in perturbation theory against the other several contributions, at least up to 55 decimal places (because the EW vacuum contribution). 

And we are not even talking about of the curvature dependent effects of QFT in curved spacetime. In path integral formulation, the generating functional with a source $J$ is
\begin{equation}\label{Eq:Introduction.Functional}
Z[J]=\int D\phi \exp\left(\frac{i}{\hbar}\left[S[\phi,g_{\mu\nu}]+\int d^4x \sqrt{-g(x)}J(x)\phi (x)\right]\right)\,,
\end{equation}
from which we can obtain Green's functions by taking functional derivatives over $J$. The action $S[\phi,g_{\mu\nu}]$ here, for the example of a scalar field, is
\begin{equation}\label{Eq:Introduction.MatterAction}
S[\phi,g_{\mu\nu}]=-\int d^4 x \sqrt{-g(x)}\left(\frac{1}{2}g^{\mu\nu} \partial_\mu \phi \partial_\nu \phi+\frac{1}{2}\xi \phi^2 R+V(\phi)\right)\,,
\end{equation}
can be obtained by promoting the flat spacetime action for the matter field changing the measure to $\sqrt{-g}dx^4$ and incorporating a non-minimal coupling with gravity $\xi\phi^2 R$. We will not provide further elaboration on this matter in this section, as we intend to address it comprehensively in the following chapter and corresponding appendices. Nonetheless, it is worth noting that this consideration adds an additional facet to the cosmological problem.

Therefore, the problem is not only at the calculational level. We need a dynamical mechanism capable of adjusting the value of the vacuum energy in the present time from its enormous value that presumably had in the early Universe. For instance, a possibility suggested by the tininess of the measured value of $\Lambda$ was to find a framework in which its value is actually 0; however, the search of symmetries or other arguments justifying the vanishing of $\Lambda$ seems not to reach successful results. Some early hopes with Supersymmetry \cite{iliopoulos1974broken,zumino1975supersymmetry}, the maximum exponent of this philosophy, followed that direction. In that case, a global symmetry may lead to a null vacuum energy associated with the supersymmetric states, although the absence of evidence of supersymmetric partners in particle accelerators may suggest that the symmetry is extremely broken in nature or simply does not reflect reality. Such a mechanism is, then, still unveiled and, since Quantum theories of Gravity are still not a consistent theory yet, we should just accept to work within the context of QFT in curved spacetime. 

In this thesis, we show our works on this matter. Not with the goal of solving the cosmological constant problem in its ultimate form, but with the more humble objective of shedding light on new ideas related to the vacuum energy in QFT and its cosmological evolution in the expanding Universe as the Dark Energy candidate par excellence.

\section{Basics of Cosmology}\label{Sect:BasicsCosmology}

The $\Lambda$-Cold Dark Matter model ($\Lambda$CDM) has been the ``concordance model'' of modern cosmology for more than 20 years. Although not perfect, it is currently the best model we have to explain most of the observations and basic features of the Universe and its constituents\,\cite{Tsujikawa:2008uc, Dodelson:2003ft}. As the current paradigm in cosmology, despite its simplicity, the $\Lambda$CDM has passed many tests when confronted with available data from the Cosmic Microwave Background and structure formation in the Universe, among others. It describes an accelerated expanding Universe that evolved from a primeval hot and dense state in its early stages to its current condition. As two complementary master pillars for understanding such evolution, the $\Lambda$CDM can accommodate the ``Big Bang Nucleosynthesis'', a process for the genesis of light elements in the early Universe, and ``inflation'', a process of rapid exponential expansion at the beginning of its history (see \hyperref[SubSect:Inflationary]{Sect.\,\ref{SubSect:Inflationary}}). The model can be well described by six parameters, which can be selected in various ways. One possible selection is: the physical baryon density fraction ($\Omega_{\rm b}^0 h^2$), physical dark matter density fraction ($\Omega_{\rm dm}^0 h^2$), the age of the Universe ($t_0$), the spectral index ($n_{\rm s}$), the optical depth to the epoch of reionization ($\tau_{\rm reio}$), and the amplitude of primordial scalar curvature perturbations ($\Delta^2_{\cal R}$). The parameter $h$ is defined as $H_0\equiv 100h$ km/s/Mpc, where $H_0$ is the value of the Hubble parameter at the present time.

The mathematical framework on which the $\Lambda$CDM relies is General Relativity\footnote{The reader may find our conventions and useful geometric formulas in\,\hyperref[Appendix:Conventions]{Appendix\,\ref{Appendix:Conventions}}.}, characterized by the metric $g_{\mu\nu}$. This geometrical object is used to calculate the generalized distance between events in spacetime and describe the causal structure.:
\begin{equation}\label{Eq:Introduction.Interval}
ds^2=g_{\mu\nu}dx^\mu dx^\nu\,.
\end{equation}
In flat spacetime or Minkowski spacetime, as in special relativity, the metric is $g_{\mu\nu}=\eta_{\mu\nu}\equiv diag (-1,1,1,1)$. However, spacetime is naturally curved in the presence of matter and energy, and it is not expected to be static. Our Universe is in a state of accelerated expansion as discovered at the end of the 20th century.

The $\textit{Cosmological Principle}$ is implemented at the heart of the model, defining that at sufficiently large scales, the Universe is thought of as a smooth object, and the distribution of matter-energy is homogeneous and isotropic, with anisotropies restricted to the perturbation level. This principle is encoded in the structure of the so-called Friedmann-Lemaître-Robertson-Walker (FLRW) metric, which we introduced in \hyperref[SubSect:RelativisticCosmology]{Sect.\,\ref{SubSect:RelativisticCosmology}}, but let us elaborate a bit more. In Cartesian coordinates with cosmic time as the time coordinate, the metric can be written as:
\begin{equation}\label{Eq:Introduction.FLRWCartesian}
\begin{pmatrix}
-1 & 0 & 0 & 0\\
0 & a^2(t) & 0 & 0\\
0 & 0 & a^2 (t) & 0\\
0 & 0 & 0 & a^2 (t)\\
\end{pmatrix}
\end{equation}
Here, $a(t)$ is the \textit{scale factor}, a dimensionless quantity that parametrizes the relative growth of physical distance with respect to a reference distance, and $t$ is the cosmic time coordinate of expansion. The scale factor can be related to redshift $z$, which measures the relative change in the wavelength of light emitted from a receding object due to the expansion,
\begin{equation}\label{Eq:Introduction.RedshiftDefinition}
1+z \equiv \frac{\lambda_{\rm obs}}{\lambda_{\rm emit}}=\frac{a_0}{a}\,,
\end{equation}
where $a_0\equiv a(t_0)$ is the scale factor at the present time $t_0$. The scale factor is usually normalized to be $a_0=1$, as we will do. 

Alternative, we can use spherical coordinates to describe the interval:
\begin{equation}\label{Eq:Introduction.FLRWMetricSpherical}
 ds^2=g_{\mu\nu} dx^\mu dx^\nu =-dt^2+a^2 (t) \left( \frac{dr^2}{1-kr^2}+r^2 \delta \theta^2 + r^2 \sin^2 \theta d\phi^2 \right)\,.
\end{equation}
The parameter $k$ is related to spatial curvature, with natural units $[k]=E^2$, where $E$ is in energy units. Spatial curvature is determined by the matter/energy content of the Universe. Sometimes $k$ is normalized as a dimensionless value depending on its sign: $+1$, $0$, or $-1$. In this convention, the scale factor has units of length. Throughout this work, we restrict ourselves to the particular value of $k=0$ for a Universe that is flat on constant time hypersurfaces, as it is the common assumption. Although we will adopt $k=0$ for the rest of the dissertation, it is worth mentioning that the possibility of an open ($k<0$) or closed ($k>0$) Universe cannot be completely ruled out, and the spatial curvature of the Universe is still subject to study \cite{Ooba:2018dzf, DiValentino:2019qzk, Handley:2019tkm, Khadka:2020vlh, Cao:2020evz, Cao:2021ldv, Ratra:2022ksb, deCruzPerez:2022hfr}.

The coordinates $(r,\theta, \phi)$ of an object appearing in \eqref{Eq:Introduction.FLRWMetricSpherical} are the comoving coordinates, which do not change with cosmological expansion (only due to possible peculiar motions of the object independent of the Hubble flow). In subsequent chapters, we will use the so-called {\it conformal time}, which corresponds to the comoving distance that light can travel since $t=0$ (assuming it does not interact at any point),
\begin{equation}\label{Eq:Introduction.ComovingDistanceConformal}
\eta (t) \equiv \int_0^t \frac{dt^\prime}{a(t^\prime)}\,.
\end{equation}
Conformal time can serve as a parametrization of the time variable, just like $t$ or the scale factor $a$. However, each variable has distinct interpretations and may be more appropriate in different contexts. Regions separated by a comoving distance bigger than $\eta$ are causally disconnected, as $\eta$ is defined as the maximum comoving distance that light could have traveled since the beginning of the expansion. The comoving distance between a distant emitter (emitting at cosmic time $t$) and us (at present time) is
\begin{equation}\label{Eq:Introduction.ComovingDistanceBetween}
\chi (t)=\eta (t_0)-\eta (t)=\int_{t(a)}^{t_0} \frac{dt^\prime}{a (t^\prime)}=\int_{a(t)}^1 \frac{da^\prime}{\left(a^\prime \right)^2 H \left(a^\prime \right)}\,.
\end{equation}
On the other hand, {\it proper} or {\it physical} distance, takes into account the evolution of the distance and is evaluated at a specific cosmological times. Proper distance is related to comoving distance as $d (t) =a(t) \chi (t)$.

In a non-expanding Universe, there are several equivalent ways to calculate distances. However, in an expanding Universe, this is no longer the case. Therefore, alternative distance measures need to be defined, and we need to understand how they are related to one another if we wish to connect experimental observations with model parameters. The first example of such an alternative distance measure is the {\it angular diameter distance}, which is the ratio of the known physical size of an object, $l$, to the angle it subtends in the sky, $\theta$. This can be expressed as:
\begin{equation}\label{Eq:Introduction.AngularDiameterDistance}
d_{\rm A} \equiv \frac{l}{\theta}\,.
\end{equation}
In the spatially flat $\Lambda$CDM, the angular diameter distance turns out to be
\begin{equation}\label{Eq:Introduction.AngularDiameterDistanceFlat}
d_{\rm A}^{\rm flat}(z) \equiv \frac{\chi (z)}{1+z}\,.
\end{equation}
A second example is the luminosity distance, $d_{\rm L}$, which relates the observed light flux from an astronomical source to the distance of the object. This relation can be obtained by tracking the evolution of a spherical shell in an expanding background, along with the redshift of the emitted photons. Accounting for these effects, the flux is
\begin{equation}
F=\frac{L}{4\pi d_{\rm L}^2(a)}\,.
\end{equation}
In the flat space scenario, $d_{\rm L}$ can be expressed as $\chi (a)/a$. For expressions in the case of arbitrary spatial curvature and more details, standard textbooks such as \cite{Tsujikawa:2008uc, Dodelson:2003ft} can be consulted.

\subsection{Cosmological Background Equations}\label{Sect:CosmologicalBackEq}

The expanding Universe's background evolution is described using the Friedmann equations, which are a form of Einstein's equations specifically for cosmology. These equations are derived by assuming that the FLRW metric with a spherical symmetry (as given in Equation \eqref{Eq:Introduction.FLRWMetricSpherical}) is the solution. In addition to the Friedmann equations, the Cosmological principle plays an important role in modeling matter at cosmological scales. To describe the distribution of matter, we often use energy densities as if matter were a continuum fluid. This approach is typically based on treating matter as perfect fluids, which are described using an Energy-Momentum Tensor of the form
\begin{equation}\label{Eq:Introduction.PerfectFluidModel}
T_{\mu\nu}=P g_{\mu\nu}+\left( \rho + P \right) u_\mu u_\nu\,,
\end{equation}
where $\rho$ and $P$ are the energy density and the pressure of the fluid and $u^\mu$ is the 4-velocity of the fluid. If we use cosmic time as time coordinate then $u^\mu = (1,0,0,0)$ and $u_\mu = (-1,0,0,0)$. 

The first Friedmann equation is
\begin{equation}\label{Eq:Introduction.FirstFriedmannEq}
H^2=-\frac{k}{a^2}+\frac{8\pi G_N}{3}\rho (t) +\frac{\Lambda}{3}\,,
\end{equation}
where $G_N$ is the Gravitational Constant, $\rho (t) \equiv \sum_{\rm N} \rho_{\rm N} (t)$ is the total energy density of matter in the Universe for the different components labeled by N, $\Lambda$ is the Cosmological Constant and $H$ is the Hubble function, which accounts for the expansion rate of the Universe. Measures of the Hubble rate at the present time, $H_0\equiv H(t_0)$, are oftenly parametrized as
\begin{equation}
H_0 =100 h  \textrm{ km/s/Mpc }\,.
\end{equation}
The parameter $h$ is dimensionless and is typically expected to have a value of around $0.7$. In natural units, the Hubble rate at the present time $H_0$ is approximately $2\times 10^{-33}$ eV. This is an extremely small energy scale when compared to any energy scale in the Standard Model of particle physics.
The second Friedmann equation is
\begin{equation}\label{Eq:Introduction.SecondFriedmannEq}
2 \dot{H}+ 3 H^2=-\frac{k}{a^2}-8\pi G_N P(t) + \Lambda \,,
\end{equation}
which introduces the total pressure, $P(t) \equiv \sum_{\rm N} P_{\rm N} (t)$, defined as the sum of the pressures of the different components labeled by N. We use the dot for indicating the derivative with respect cosmic time, $\dot{()}\equiv d/dt()$. Both, the pressure and the density are expected to evolve with matter, as dictated by the Cosmological principle. One of the former equations\,\eqref{Eq:Introduction.FirstFriedmannEq} or\,\eqref{Eq:Introduction.SecondFriedmannEq} can be exchanged by the Bianchi identity,$\nabla^\mu G_{\mu \nu}=0$. That identity enforces the matter part to satisfy the covariant conservation equation,$\nabla^\mu T_{\mu\nu}=0$, or, in terms of the densities and pressures,
\begin{equation}\label{Eq:Introduction.BianchiIdentityII}
\sum_{\rm N}\left[\dot{\rho}_{\rm N}+3H\left(\rho_{\rm N}+P_{\rm N}\right)\right]=0\,.
\end{equation}
In addition to the Friedmann equations, one can impose the condition that all species are self-conserved and that there are no interactions between different components at the background level. This is given by the equation $\dot{\rho}_{\rm N} + 3H(\rho_{\rm N} + P_{\rm N}) = 0$ for each species labeled by N. This equation expresses the conservation of the energy-momentum tensor for each component in the absence of external sources or interactions with other components. The combination of the Friedmann equations and the conservation equations provides a complete set of equations that describe the evolution of the Universe at the background level. It is also useful to define the equation of state of a fluid, $w_{\rm N}$ as the ratio  between its pressure and its density,
\begin{equation}\label{Eq:Introduction.EqOfState}
w_{\rm N} \equiv P_{\rm N} / \rho_{\rm N}\,.
\end{equation}
For a constant equation of state and assuming covariant self-conservation, each species have a well defined background evolution with the expansion in terms of the scale factor,
\begin{equation}\label{Eq:Introduction.DensityEvolution}
\rho_{\rm N}=\rho_{\rm N}^0 a^{-3(1+w_{\rm N})}\,,
\end{equation}
being $\rho_{\rm N}^0 \equiv \rho_{\rm N} (t_0)$.

\subsection{Cosmic inventory}

Now we have given some context to the concordance model and written the basic equations, let us talk about its material content.  In the early Universe, reactions between different species occurred rapidly, resulting in particles being in thermal equilibrium and sharing a common temperature. To describe the energy densities of different species, it is useful to introduce distribution functions, denoted by $f_{\rm N}(\vec{p})$, which account for the statistical distribution of particles in a volume of phase space around a particular physical momentum $\vec{p}$. The total energy density of a species labeled by N can be obtained by calculating the number of particles per phase space volume with a particular energy $E(p) = \sqrt{p^2 + m^2}$ (where $p$ is the modulus of momentum), integrated over all momenta and weighted by $E(p)$, i.e.,
\begin{equation}\label{Eq:Introduction.EnergyDensityOccupationStates}
\rho_{\rm N}=\frac{g_{\rm N}}{(2\pi)^3}\int d^3 p f_{\rm N}(\vec{p}) E(p)\,,
\end{equation}
where $g_{\rm N}$ accounts for the number of internal degrees of freedom of the species labeled by N. The distribution functions allow us to describe the occupation of different energy states by particles and are used to derive the energy densities and pressures of the different components of the Universe. The equilibrium distribution function takes the form
\begin{equation}\label{Eq:Introduction.DistributionFunctions}
f_{\rm N}=\frac{1}{\exp \left(\frac{E(p)-\mu_{\rm N}}{T}\right)\pm 1}\,,
\end{equation}
where we choose $-1$ in the denominator for Bosons such as photons and $+1$ for spin-1/2 fermions such as electrons. The quantity $\mu_{\rm N}$ is the chemical potential, which can be neglected if the temperature is high enough, a condition fullfilled in the early Universe. Similarly, the pressure is
\begin{equation}
P_{\rm N}=\frac{g_{\rm N}}{(2\pi)^3}\int d^3 p f_{\rm N}(\vec{p}) \frac{p^2}{3 E(p)}\,.
\end{equation}
Using the previous expressions, the total entropy density takes the following form:
\begin{equation}
s\approx\frac{1}{T}\sum_{\rm N} \left(\rho_{\rm N}+P_{\rm N}\right)\,.
\end{equation}
Equipped with the previous formulas, we can now proceed describe the energy budget of the background Universe.  

\vspace{5mm}
{\large \textbf{Photons}}

The almost isotropic radiation that arrives from the Cosmic Microwave Background (CMB) is composed of photons with a temperature of $T_0=2.72548\pm 0.00057$ K \cite{Fixsen:2009ug}. The energy of the photons decreases with the inverse of the scale factor, and therefore, the temperature is inversely proportional to the scale factor,
\begin{equation}\label{Eq:Introduction.TemperaturePh}
T(a)=T_0/a(t)\,.
\end{equation}
The energy density associated to the CMB photons can be obtained from\,\eqref{Eq:Introduction.EnergyDensityOccupationStates} for a relativistic boson species,
\begin{equation}\label{Eq:Introduction.EnergyDensityPhotons}
\rho_\gamma=2 \int \frac{d^3p}{\left(2\pi\right)^3}\frac{p}{e^{p/T}-1}\,,
\end{equation}
with no spatial dependence at the background level.
At a good approximation, the Cosmic Microwave Background (CMB) can be considered as having a perfect black body spectrum with zero chemical potential due to the thermalization process resulting from various interactions, including Compton scattering and bremsstrahlung. Deviations from the black body spectrum have been well constrained to $|\mu/T|\lesssim 10^{-4}$ based on pioneering measurements by FIRAS \cite{Fixsen:1996nj,Bianchini:2022dqh}. After some manipulations, the former integral\,\eqref{Eq:Introduction.EnergyDensityPhotons} can be written as
\begin{equation}\label{Eq:Introduction.EnergyDensityPhotonsII}
\rho_\gamma=2 \frac{8\pi T^4}{\left(2\pi \right)^3} \int_0^\infty dx\frac{x^3}{e^x-1}=\frac{\pi^2}{15}T^4\approx 2\times 10^{-51} \left(\frac{\textrm{GeV}}{a}\right)^4\,,
\end{equation}
in natural units. Therefore, it follows from equation\,\eqref{Eq:Introduction.DensityEvolution} that $w_\gamma=1/3$. The energy fraction at  is
\begin{equation}
\Omega_\gamma = \frac{\rho_\gamma^0}{\rho_{\rm c}^0}a^{-4}=\frac{8\pi G_N\rho_\gamma^0}{3 H_0^2} a^{-4}\approx 5\times 10^{-5} a^{-4}\,,
\end{equation}
for $h=0.7$. Therefore, the contribution of photons to the current energy budget is quite limited.

\vspace{5mm}
{\large \textbf{Baryons}}

Baryons are part of the non-relativistic contribution to the total energy density. They cannot be traced through their temperature since they do not behave as a gas. Instead, we know that their energy density should scale as $a^{-3}$, as their energy correspond to their rest mass, $E\approx m$:
\begin{equation}\label{Eq:Introduction.DensityofBaryons}
\rho_{\rm b}=\rho_{\rm b}^0 a^{-3}\,.
\end{equation}
Here $\rho_{\rm b}^0$ is the energy density of baryons at the present time. The corresponding equation of state is $w_{\rm m}=0$. We need to rely on cosmological measurements of baryon density, such as those obtained from CMB observations\,\cite{aghanim2020planck} or Big Bang Nucleosynthesis constraints\,\cite{Cooke:2016rky}. 

\vspace{5mm}
{\large \textbf{Dark Matter}}

The largest fraction of non-relativistic matter in the Universe does not correspond to baryons. In fact, the total energy density associated with non-relativistic matter is about 5-6 times larger than that of baryonic matter. The remaining energy may come from a substance that we do not have direct evidence for, as it appears to not interact electromagnetically. We conclude that most of the matter in the Universe is not in the form of baryons but is a different entity that we call Dark Matter (DM), usually together with the adjective cold because we expect a large fraction of it to behave non-relativistically. Its nature and origin are matters of vivid debate not only in cosmology but also in particle physics since several candidates for DM can represent extensions of the Standard Model of Particle Physics or be connected to the existence of primordial black holes,\cite{Garrett:2010hd,Feng:2010gw,Carr:2016drx}. Let us just say that some indirect ways to measure the fractional energy of matter (incorporating both baryonic matter and DM) exist, for instance, generalizing the fraction $\Omega_{\rm b}^0/\Omega_{\rm m}^0$ of galaxy clusters to be representative of the Universe as a whole\,\cite{White:1993wm}, or studying the anisotropies of the CMB\,\cite{aghanim2020planck}, which are sensitive to the combined quantity $\Omega_m^0 h^2$.

Same arguments as for baryons apply here, as we cannot infer the energy density of DM from its temperature, and we also need to find the energy density (or the fraction of energy) of DM through parameter inference in the light of cosmological data. Since direct observations are not possible, the evidence of its existence relies on its gravitational effects on the growth of large-scale structure. Let us simply remark that the evolution of DM follows the same expression as that of baryons:
\begin{equation}
\rho_{\rm DM}=\rho_{\rm DM}^0 a^{-3}\,.
\end{equation}

\vspace{5mm}
{\large \textbf{Neutrinos}}

Fermions are light particles that rarely interact with matter, however they were expected to be in equilibrium with the rest of the components of the primordial plasma in the early Universe. Neutrinos are, at first approximation, expected to be relativistic since their masses are expected to be extremely tiny and are often totally neglected. They are classified in three different generations: electron neutrino ($\nu_{\rm e}$), muon neutrino ($\nu_{ \mu}$) and tau neutrino ($\nu_{\tau}$). There is only one spin degree of freedom for each neutrino.

As fermions, neutrinos follow a Fermi-Dirac distribution with null chemical potential in good approximation and with a temperature decaying with the scale factor equally as for the photons, behaving as radiation for most part of the cosmological history. The decoupling of neutrinos from the cosmic plasma happened before electron-positron annihilation, earlier than Big Bang Nucleosynthesis (BBN) happened. Following a similar computations to the one performed for photons we arrive to the expression of the energy density in terms of the temperature of neutrinos,
\begin{equation}
\rho_\nu = N_{\rm eff} \frac{7 \pi^2}{120}T_{\rm \nu}^4\,.
\end{equation}
In principle, $N_{\rm eff}\equiv g_{\nu}/2$ accounts for the number of internal degrees of freedom. We should expect $N_{\rm eff}=6/2=3$ for the standard model with 3 generation of relativistic neutrinos and antineutrinos. But it is usual to leave $N_{\rm eff}$ as a free parameter in order to allow for extra degrees of freedom that this simple description cannot capture. It is possible to constrain to the effective parameter $N_{\rm eff}$ from the abundances of light elements predicted by BBN\,\cite{Mangano:2005cc}. Current estimations set its value  to be slightly greater than 3, $N_{\rm eff}=3.04$.

We can relate, the neutrino and photon temperature today. It can be done, using the conservation of entropy before and after electron and positron annihilation. It is also important the fact that entropy density $s$ scales as $s\propto a^{-3}$.  This yields a relation between photon and neutrino's temperature:
\begin{equation}\label{Eq:Introduction.RelTnuTg}
\frac{T_\nu}{T_\gamma}=\left(\frac{4}{11}\right)^{1/3}\,.
\end{equation}
With this information one may construct the energy density of neutrinos as compared to photons by taking also into account the degeneracy number for neutrinos and the difference in the thermal distribution function between photons and neutrinos\,\cite{Dodelson:2003ft}:
\begin{equation}\label{Eq:Introduction.rhonuneff}
\rho_\nu=N_{\rm eff}\frac{7}{8} \left(\frac{4}{11}\right)^{4/3} \rho_\gamma \approx 0.68\rho_\gamma \,.
\end{equation}
The enegy fraction from radiation (relativistic neutrinos plus CMB photons) is
\begin{equation}\label{Eq:Introduction.Omegar}
\Omega_{\rm r} =\frac{\rho_\gamma+\rho_\nu}{\rho_{\rm c}^0}\approx 1.68 \Omega_\gamma^0 a^{-4}\approx 8\times 10^{-5}a^{-4}\,.
\end{equation}

Nevertheless, the former is just an approximation of the energy density for when neutrinos behaved relativistically. Neutrinos at some point started to behave non-relativistically because of the drop in temperature with the cosmological expansion. We will not elaborate more on this topic, since it is out of the scope of this simple introduction. It is worth mentioning, however, that modeling the change of regime from relativistic to non-relativistic neutrinos produces the absence of an analytical form for $\rho_\gamma$ with respect to the scale factor for the entire history. As a result, numerical methods are usually necessary to deal with this issue.

\vspace{5mm}
{\large \textbf{Dark Energy}}

Beyond radiation and matter, the most predominant component in the Universe is something else. The combination of radiation and matter only accounts for approximately 0.3 of the total energy density, as determined by observations. Additionally, observations suggest that the spatial curvature of the Universe is close to zero, implying that the total energy density is expected to be equal, or at least extremely close, to the critical energy density. Therefore, we expect the total fraction of matter and energy to add up to 1. The remaining energy fraction is attributed to the presence of a mysterious and smooth substance known as Dark Energy (DE), which is the responsible for the accelerated expansion of the Universe. Although its nature and origin are still unknown, DE has been confirmed by various independent probes such as the luminosity distance dependence on redshift from Supernovae SnIa, Baryon Acoustic Oscillations (BAO), and Cosmic Microwave Background (CMB). As a result, DE is considered to be a key piece of the cosmological puzzle and an essential part of the standard model of cosmology.

In the concordance model, DE is modeled through a cosmological constant $\Lambda$, with an associated energy density $\rho_\Lambda =\Lambda/(8\pi G_N)$. This implies that the equation of state for DE is $w_{\rm DE}=P_{\rm DE}/\rho_{\rm DE}=-1$, with negative pressure. Dark energy is responsible for the accelerated expansion of the Universe because the source of acceleration for a particular component is given by $\rho+3P$,
\begin{equation}
\frac{\ddot{a}}{a}=- \frac{4 \pi G_N }{3} \left(\rho + 3 P\right)\,.
\end{equation}
A fluid with $w<-1/3$ would produce $\ddot{a}>0$ if it were the unique component. More generally, if DE's equation of state were a function of the scale factor, DE energy density would evolve with time as
\begin{equation}\label{Eq:Introduction.DEEvolution}
\rho_{\rm DE} (a)=\rho_{\rm DE}^0 \exp \left\{ -3\int_1^a \frac{1+w_{\rm DE} (\tilde{a} )}{\tilde{a}} d\tilde{a} \right\}\,.
\end{equation}

\subsection{Inflationary Universe}\label{SubSect:Inflationary}

The $\Lambda$CDM, even if able to explain a large set of observations, is an incomplete image. There are some problems that the vanilla Standard Model is unable to deal with\,\cite{Dodelson:2003ft,Tsujikawa:2008uc, Liddle:2000cg}
 
 \begin{itemize}
 \item Setting initial conditions to primordial fluctuations. To accurately describe the evolution of the large-scale structure in our Universe, we must solve the perturbation equations that govern its growth around a smooth background. However, to do this, we first need to establish initial conditions for primordial fluctuations that occur in the very early Universe. The primordial fluctuations, which are thought to have originated from quantum processes, provide the seeds for the formation of cosmic structures like galaxies and clusters. Thus, we need a mechanism to explain how tiny quantum fluctuations were stretched to cosmic scales, generating the primordial density fluctuations that are observed today.
 
 \item The {\it horizon problem}. This disquisition has many facets, but it can be formulated in a simple manner. On the one hand, the temperature map of the cosmic microwave background (CMB) is nearly isotropic, exhibiting a relative difference of $\Delta T/T\sim 10^{-5}$ at all scales. On the other hand, the photons that we receive today from the last scattering surface were initially separated by distances much larger than the cosmological horizon. As a result, these large scales were unable to reach thermal equilibrium since they were not causally connected until later in cosmic history. Nevertheless, we observe equal temperatures for photons reaching us from different points in the sky.
 
The comoving horizon, $\eta$, provides a reference distance for determining the causal connection between objects separated by a certain distance. If their comoving distance is greater than $\eta$, then the objects were not causally connected in the past, since information could not travel the entire path from one point to another by definition. From Eq.\,\eqref{Eq:Introduction.ComovingDistanceConformal}, we see that $\eta$ depends on the integration of the comoving Hubble radius, $r_H = 1/(a H)$. An estimation of the comoving Hubble radius during the radiation-dominated epoch (RDE) yields $r_H \propto a^2$, so the comoving horizon received only very small contributions in the early Universe, of order $\eta \propto a^2$. In other words, if we naively extrapolate this dependence to the beginning of the cosmos (extending the RDE up to $a=0$), photons could not have traveled large distances in the past, and the sky would consist of non-causally connected patches. This means that solving the horizon problem within the traditional paradigm of an RDE followed by a matter-dominated epoch (MDE) is not possible.
 
 \item The {\it flatness problem}. The Universe is commonly assumed to have an almost negligible spatial curvature. For instance, the Baseline dataset from Planck 2018\,\cite{aghanim2020planck} in combination with BAO data constrains the curvature parameter to be $\Omega_k^0=0.0007\pm 0.0019$. If $\Omega_k (a)=-k/(a H(a))^2$ was not forced to be exactly 0 since the very beginning, this would be difficult to explain, since $|\Omega_k (a)|$ is expected to grow during the matter and radiation dominated epochs. One option is to fine-tune the initial value of $\Omega_k$ to an unreasonably small value around 0 in such a way that it can provide the still tiny value it has today, but this would seem unnatural. Another option is to consider an early phase of expansion ($\ddot{a}$) in the remote past that reduced the Hubble radius and, as a consequence, the curvature parameter by several orders of magnitude. In order to achieve the present level of flatness, we require a cosmic inflationary phase prior to the RDE, where the scale factor grows by a factor of more than 60 e-folds, or $e^{60}$ times.
 
 \item The {\it monopole problem}. The Standard Model of Particle Physics has some extensions that predict hypothetical particles produced in the very early Universe, such as magnetic monopoles, which behave as isolated magnetic charges\,\cite{tHooft:1974kcl}. The difficulty in detecting magnetic monopoles (and other hypothetical extensions of the Standard Model) can be explained if their formation occurred prior to the inflationary period. In this scenario, after the exponential expansion, the density of these particles became so diluted that it is now impossible to detect them in experiments.
 
\end{itemize}

It is clear that the common factor in the solution to these problems is the existence of a short period of exponential expansion in the early Universe, which effectively behaves as a De Sitter Universe. This is known as the theory of {\it inflation}\,\cite{Dodelson:2003ft,guth1981inflationary,linde1982new,Starobinsky:1980te}. It is interesting to note that in order for the Universe to undergo an accelerating expansion stage, the energy budget must be dominated by a negative pressure fluid (or a mechanism that behaves similarly to it), similar to Dark Energy. However, this process cannot be explained by a simple cosmological constant, as it only had a very short duration.

An active field of research is to describe the physics of the underlying mechanism capable of producing the inflationary process. The very early epoch is not currently accessible by experiments, since it involves very high energy scales and temperatures. This means that particle physics cannot make robust affirmations, despite the fact that there are several well-studied insights, such as {\it Grand Unified Theories} (GUT)\,\cite{linde1982new,guth1981inflationary}, characterized by energy scales beyond $10^{15-16}$ ${\rm GeV}$. However, the most common mechanism for inflation is carried by a scalar field, since it consists of a simple model that may describe the essential features of inflation. This scalar field $\phi$ of this model is usually called the {\it inflaton}, but it is important to remark that it has not been identified with any field of nature yet. Schematically, in the model of the inflaton, the scale factor is expected to evolve exponentially as $a\propto \exp \left( H_I t\right)$, where $H_I$ is the Hubble function during inflation. The scalar field is thought to possess a self-interacting potential $V(\phi)$, and the pressure and energy density associated to the scalar field are
\begin{equation}\label{Eq:Introduction.RhoandPressureOfPhi}
P_\phi \equiv \frac{1}{2}\dot{\phi}^2-V(\phi)\,, \qquad \rho_\phi \equiv \frac{1}{2}\dot{\phi}^2+V(\phi )\,.
\end{equation}
Then, the equation of state of the scalar field is
\begin{equation}\label{Eq:Introduction.RhoandPressureOfPhiII}
w_\phi \equiv \frac{\frac{1}{2}\dot{\phi}^2-V(\phi)}{\frac{1}{2}\dot{\phi}^2+V(\phi )}\,.
\end{equation}
Here $\dot{\phi}\equiv d\phi /dt$. In the {\it slow-roll} regime, where the field is expected to carry a negligible kinetic energy, and the equation of state is dominated by the huge value of the potential and is $w_\phi \approx -1$, similar to a cosmological constant. When inflation stops, $V(\phi) \sim \dot{\phi}^2$ and the scalar field oscillates around the minimum of the potential, releasing radiation and different particles. This process is known as {\it reheating} and leads to the traditional Big Bang picture and subsequent RDE.

For completeness it is also necessary to mention {\it Starobinsky model}\,\cite{Starobinsky:1980te,Starobinsky:1981vz}. In this case, the inflationary process is carried out by the higher-order correction $R^2$ into the Einstein-Hilbert Action, where $R$ is the Ricci scalar. This term and similar ones will appear in subsequent chapters when trying to renormalize our theory in the context of General Relativity, so that correcting Einstein's field equations with higher-derivative terms.

\section{Why going beyond the Standard Model?}\label{Sect:BeyondSM}

We have already stressed the fact that the concordance $\Lambda$CDM model is afflicted by several theoretical conundrums. The most prominent one is the CC problem, which actually affects all forms of dark energy (DE)\,\cite{weinberg1989cosmological, witten2001cosmological, sahni2000case, peebles2003cosmological, padmanabhan2003cosmological, copeland2006dynamics, Amendola:2015ksp, Sola:2013gha, SolaPeracaula:2022hpd}. Jointly with the aforementioned coincidence problem, we urgently require a non-trivial explanation that can answer these questions in their entirety, possibly through physics beyond the Standard Model of Particle Physics or with the development of a successful quantum description of gravity. At present, we are restricted to the more pedestrian Quantum Field Theory, which, as we will see in the following chapters, may have something to say regarding these intriguing problems after all.

On the other hand, one may feel relieved because, despite the fact that we do not control all the fundamental pillars of the model, from an effective phenomenological perspective, the $\Lambda$CDM seems to accommodate the overall cosmological observations with success. Nonetheless, the observational situation in recent years does not present a fully positive picture of the phenomenological status of the $\Lambda$CDM anymore. Namely, several discrepancies in crucial parameters plague the landscape. These discrepancies emerge in various forms, ranging from mere curiosities in some parameters related to the cosmological probes to the level of true cosmological tensions in crucial parameters that threaten the viability and trust in the concordance model\,\cite{Perivolaropoulos:2021jda}.

This is particularly true concerning two main observables\,\cite{verde2019tensions, Macaulay:2013swa, Nesseris:2017vor, DiValentino:2020zio, DiValentino:2020vvd, Abdalla:2022yfr}. The most noticiable and perturbing one is the so-called {\it Hubble tension} and is related to the dissimilar indepedently obtained measurements of the Hubble parameter $H_0$ from local and early Universe data coming from the Cosmic Microwave Background (CMB). The constrains imposed by the CMB predict a low value of the parameter, $H_0=67.36\pm 0.54$ km/s/Mpc in the baseline scenario presented in Planck collaboration's results\,\cite{aghanim2020planck}. On the other hand, local geometrical estimations relying in the distance ladder method\,\cite{Riess:2016jrr, Riess:2018uxu, Riess:2019cxk, Reid:2019tiq, Riess:2021jrx} infer a greater value of the current value of the parameter. We can take as a reference the value $H_0=73.04\pm 1.04$ km/s/Mpc in the baseline scenario from\,\cite{Riess:2021jrx}, the lattest result from SH0ES Team. It is also remarkable  that the missmatch arisen by the two results is not a recent trend, the discrepancy has escalated with time since the original publications of SH0ES were released\,\cite{Riess:2018uxu, Riess:2016jrr} and even enhanced when combined with other cosmological probes. For instance, when the value $H_0=73.5\pm 1.4$ km/s/Mpc from\,\cite{Reid:2019tiq} is combined with other independent observation, such as the Strong-Lensing results from the H0LICOW collaboration\,\cite{Wong:2019kwg}, which leads to $H_0=73.3^{+1.7}_{-1.8}$ km/s/Mpc, the combination produces $H_0= 73.42\pm 1.09 $ km/s/Mpc, and this result amounts to an astounding $\sim 5\sigma$ tension with respect to the Planck 2018 result. This tension is in the line of the $\sim 5\sigma$ level in light of the aforementioned result from\,\cite{Riess:2021jrx}.

The second one is also worrysome and related to the structure formation data. The concordance model predicts a value of $\sigma_8$ (the current matter density rms fluctuations within spheres of radius $8h^{-1}$,Mpc, where $h\simeq 0.7$ is the reduced Hubble constant) in excess by $2-3\sigma$ over the direct data values at low redshifts\,\cite{Macaulay:2013swa,Nesseris:2017vor,Kazantzidis:2018rnb,DiValentino:2018gcu,SolaPeracaula:2018xsi,Skara:2019usd,DiValentino:2020vvd}. Specifically, the tension is between measurements of the amplitude of the power spectrum of density perturbations inferred using CMB data and those directly measured by large-scale structure (LSS) formation on smaller scales, from redshift space distortions (RSD) {(see e.g.\,\cite{Gil-Marin:2016wya}) and weak-lensing (WL) data\,\cite{Hildebrandt:2016iqg,Joudaki:2017zdt,Kohlinger:2018sxx,Wright:2020ppw}.}

Whether all these tensions are the result of yet unknown systematic errors or really hint at some underlying new physics is still unclear and constitutes a hot topic in modern-day cosmology. There is still a strong possibility that these discrepancies may just be a signal of a deviation from the $\CC$CDM model. This would mean exploring alternatives to the $\Lambda$CDM that arise after slight modifications of the former, i.e., by the introduction of new degrees of freedom with respect to the vanilla concordance model. These extensions cannot be arbitrarily far away from the $\Lambda$CDM, as it has the ability to explain and accommodate several phenomena over the entire expansion history and explain the main features of cosmological observations. In general, the best practice for judging satisfactory approaches is to consider models that can ameliorate the $H_0$ tension without enhancing the $\sigma_8$ one and vice versa. This ``golden rule'' will be our guiding principle for finding promising models in chapters\,\hyperref[Chap:PhenomenologyofBD]{\ref{Chap:PhenomenologyofBD}} and\,\hyperref[Chap:PhenomenologyofRVM]{\ref{Chap:PhenomenologyofRVM}}.

Of course, there are many attempts to solve these tension in the literature. Condensing all of them here is simply impossible, and any list would be surely incomplete and unavoidably biased. Nevertheless, let us mention some of them, e.g. see\,\cite{Zhao:2017cud , DiValentino:2019dzu , DiValentino:2017rcr , DiValentino:2017iww , DiValentino:2017zyq , DiValentino:2016hlg , Martinelli:2019dau , Salvatelli:2014zta,Costa:2016tpb, Anand:2017wsj , An:2017crg , Li:2015vla , Li:2014cee , Li:2014eha , Hazra:2018opk , Yan:2019gbw,Liao:2020zko,WangChen2020,Jedamzik:2020krr,Vagnozzi:2019ezj,Calderon:2020hoc}. In general these attempts are purely phenomenological,  e.g. those using phantom equation of state for the DE since they have no consistent theoretical support and usually they further spoil the $\sigma_8$-tension\,\cite{DiValentino:2019jae,Alestas:2020mvb}. This drawback was repeatedly emphasized in different existing analyses in the literature\,\cite{SolaPeracaula:2018wwm, Sola:2016jky, SolaPeracaula:2016qlq, SolaPeracaula:2017esw , Gomez-Valent:2018nib  , Sola:2017znb ,  Sola:2016hnq}. There are also attempts to introduce early dark energy (EDE), but different works seem to reach different conclusions on the effectiveness of these models to improve the tensions, see e.g. \cite{Poulin:2018cxd,Hill:2020osr,Chudaykin:2020acu,Braglia:2020bym}. For reviews on the tensions or their solutions, please see\,\cite{DiValentino:2020zio,DiValentino:2020vvd,Abdalla:2022yfr,Perivolaropoulos:2021jda,Schoneberg:2021qvd} and references therein.

\section{Some alternatives to the \texorpdfstring{$\CC$CDM}{LCDM}}\label{Sect:AlternativesSM}

The quest for extensions of the $\Lambda$CDM is a well-motivated task, as discussed in the previous section. However, there are many valid approaches to this problem.  Together with my PhD supervisor, Prof. Joan Solà Peracaula, we have chosen a particular direction in this plethora of possibilities, focusing our studies on finding theoretical results on dynamical DE (or vacuum energy density) and related cosmological models that emerge from this framework. 

This line of research was initiated by my PhD supervisor's devoted research on the cosmological constant problem some decades ago. Our motivation regarding dynamical DE is twofold. First, the Cosmological Problem has an unstimable interest on its own. Second, there is a potential solution to the cosmological tensions in the form of dynamical DE models in cosmology. Notice that exploring this particular direction is quite natural, since if DE can evolve with time, it is easier to understand how it can adapt its value at different epochs in the cosmological expansion than if it were stuck at a constant value, i.e., a cosmological constant.

In our works (see the subsequent chapters of this thesis) we do not consider general models where the cosmological constant is simply promoted to a generical expression of cosmic time, $\Lambda (t)$. Instead, we seek fundamental motivations for the running of the vacuum coming from Quantum Field Theory, which provides strong theoretical grounds for our models, particularly for the so-called family of ``Running Vacuum Models'' (RVM). 

Probably, the very first suggestion that the cosmological constant can be, in fact, be  a function of time was given by the Russian physicist Matvei Bronstei\,\cite{bronstein1933expanding,  o2018one}. In his paper of 1933, he treated the cosmological constant as a new kind of matter-energy fluid, capable of interact with ordinary matter. This proposal envisioned a time-varying cosmological constant, a prediction that can be seen as a precursor of modern ideas such as {\it quintessence}. More in general, the idea that the DE could be not just the CC of Einstein's equations but a dynamical variable, or some appropriate function of the cosmic time, has been explored since long ago and sometimes on purely phenomenological grounds\,\cite{Overduin:1998zv}, see particularly\,\cite{Ozer:1985ws, Ozer:1985wr, Bertolami:1986bg, Freese:1986dd, Carvalho:1991ut, sahni2000case,peebles2003cosmological,padmanabhan2003cosmological,copeland2006dynamics,Amendola:2015ksp}. In this last section of our introductory chapter, we will summarize some of these possibilities.

\subsection{Generical Parametrizations of the equation of state}

In the introduction of the basic background equations in\,\hyperref[Sect:CosmologicalBackEq]{Sect.\,\ref{Sect:CosmologicalBackEq}} we shown that the covariance conservation for each species individually implies a differential equation for the energy density in terms of the equation of state (EoS) of the fluid. If the EoS is constant, then the energy density of the fluid evolves as a power of the scale factor, see eq.\,\eqref{Eq:Introduction.DensityEvolution}. For Dark Energy, the traditional scenario is a Cosmological Constant term, i.e. $w_{\rm DE}=-1$ all along the cosmological history, and thefore the DE density remains constant as one may see from\,\eqref{Eq:Introduction.DensityEvolution}. But this condition is not in any way mandatory, and cosmologies with $\dot{w_{\rm DE}}\neq 0$ are a possible extension of the vanilla $\Lambda$CDM. The simplest way to implement this approach consists in a Taylor expansion  of $w_{\rm DE}$, as a function of the scale factor around the present time, $a=1$,
\begin{equation}\label{Eq:Introduction.EoSExpansion}
w_{\rm DE} (a)=w_{\rm DE} (a=1)+\frac{d w_{\rm DE}}{da}\Bigg|_{a=1}(a-1)+\dots
\end{equation}
The dots indicate higher order beyond the linear one in the Taylor expansion. If we neglect linear orders and keep the first constant term, $w_X\equiv w_{\rm DE} (a=1)$, we are left with the traditional picture of a constant EoS, however there is no need to adjust this constant to its former value $-1$, typical for a cosmological constant and $w_X$ may be treated as a free parameter of the model. Therefore, the DE energy density evolves with the scale factor as
\begin{equation}\label{Eq:Introduction.rhoX}
\rho_{\rm X}=\rho_{\rm X}^0 a^{-3(1+w_{\rm X})}\,,
\end{equation}
where $\rho_{\rm X}^0$ is the energy density of DE at the present time. This parametrization is called XCDM (also known as $w$CDM) cosmology\,\cite{Turner:1998ex}. 

If we go to linear order in\,\eqref{Eq:Introduction.EoSExpansion} we encounter the very well-known Chevallier-Polarski-Linder (CPL) parametrization\,\cite{Linder:2002et,Chevallier:2000qy}. If we truncate the series at linear order and define $w_0\equiv w_{\rm DE} (a=1)$, $w_a \equiv -\frac{d w_{\rm DE}}{da}|_{a=1}$, then the EoS is parametrized as the following simple expression:
\begin{equation}\label{Eq:Introduction.CPLParametrization}
w_{\rm DE} (a)=w_0+w_a (1-a)\,.
\end{equation}
which has the nice property of not growing uncontrolabily for values of the scale factor close to 0. Since the covariance conservation of DE is
\begin{equation}\label{Eq:Introduction.CovConsDE}
\dot{\rho}_{\rm DE}+3H(1+w_{\rm DE})\rho_{\rm DE}=0\,,
\end{equation}
plugging eq.\,\eqref{Eq:Introduction.CPLParametrization} yields the following expression for the energy density
\begin{equation}
\rho_{\rm DE}(a)=\rho_{\rm DE}^0 a^{-3(1+w_0+w_a )}\exp\{3w_a (a-1)\}
\end{equation}
and again $\rho_{\rm DE}^0$ is the DE energy density at the present time.

Both the XCDM and CPL do not rely in fundamental theory nor have a microphysical explanation supporting them, but represent the most generical evolution one may expect from $w_{\rm DE}$, and they are sometimes used as a reference model, meaning that when testing more sophisticated models they have to prove to perform better than the most generical and simple parametrizations.

\subsection{Scalar Field cosmology}

Scalar fields have been widely used in cosmology for different purposes because of their simplicity and versatility. Some of these examples include attempts to cancel out or relax the cosmological constant contribution through a dynamical mechanism. Some examples are\,\cite{dolgov1983very,Barr:1986ya,Wetterich:1987fm,Peccei:1987mm}. For what we are
 interested in, let us focus in scalar fields in the postinflationary regime including a potential\,\cite{peebles1988cosmology,ratra1988cosmological}. In there, the energy density of the homogeneous scalar field $\phi (t)$, considered to be a remnant of the inflatonary field, resembles a non-null cosmological constant density with a slow evolution, though. Such models were conceived to reconcile with observational data, rather than as mechanisms for the cancellation of the cosmological constant. On the contrary, the mimicking of scalar fields to the cosmological constant is crucial.

The action of the model is
\begin{equation}\label{Eq:Introduction.TotalActionScalarField}
S_{\rm Tot} = \int d^4x \sqrt{-g}\frac{1}{16\pi G_N}\left(R-\frac{1}{2} g^{\mu\nu} \partial_\mu \phi \partial_\nu \phi-V(\phi) \right)+S_{\rm m}\,.
\end{equation}
The matter part is represented by $S_{\rm m}$. The equation of state (EoS) for the ideal fluid approximation for the scalar field is different from $-1$:
\begin{equation}\label{Eq:Introduction.EqStateScalarField}
w_\phi= \frac{P_\phi}{\rho_\phi}=\frac{\frac{\dot{\phi}^2}{2}-V(\phi )}{\frac{\dot{\phi}^2}{2}+V(\phi )}\,.
\end{equation}
The kinetic term is $\dot{\phi}/2$, where the time derivative is indicated by the dot. The potential should be chosen to ensure that it is sufficiently small in the past and that the energy density only dominates at recent times, so as not to disturb Big Bang Nucleosynthesis nor the growth of matter structure on small scales in earlier stages. If we want the EoS to behave as a cosmological constant, the potential term should dominate the kinetic one, i.e. $\dot{\phi}^2 \ll V(\phi)$. A typical example is $V(\phi)\sim \phi^{-\alpha}$, with $\alpha >0$, although there are other possibilities. On the other hand, since the field is not expected to interact with ordinary matter (or at least in a noticiable manner) the Klein-Gordon equation for the scalar field is just a rewritting of the covariant conservation of the energy density of the field
\begin{equation}\label{Eq:Introduction.}
\ddot{\phi}+3H\dot{\phi}+\frac{dV}{d\phi}=0\,.
\end{equation}
A final comment is necessary. Even if we could think that the dynamical nature of the scalar field may prevent some of the pathologies of the traditional cosmological constant approach, namely fine-tuning, this is not true. In fact, the greatest difficulty of the cosmological constant problem is that it seems to afflict any know form of DE. As a simple argument, in order to identify the rolling scalar field with DE, we should have $\rho_\phi\sim \rho_\Lambda^0\sim 10^{-47}$ GeV$^4$. Since the initial value of the field is fixed shortly after inflation, and the kinetic term is very small, it approximately mantains its value during the expanding history. A typical energy scale in inflation is given by $M_X\sim 10^{16}$ GeV, from Gran Unified Theories. That means one should adjust carefully  the parameters of the potential $V(\phi)$ to compare such different energy scales, $M_X$ and $\rho_\Lambda^{1/4}$, which leads to unnatural values for the dimensionless parameters of the potential or the mass of the scalar field. 

\subsection{Running Vacuum Models}

Running Vacuum Models (RVM) conform a family of models in which DE (identified as vacuum energy) runs with the cosmic expansion as a series of powers of the Hubble function $H$ and its derivatives, $\rho_{\rm vac}(H)$. They represent an essential part of our investigations along the years, and of course, of this thesis. Its original conception was formulated in terms of generic arguments of Renormalization Group formalism, as we are about to explain, but one of the main outcomes presented in this thesis is how this structure $\rho_{\rm vac}(H)$ naturally emerges from the rigorous computations in Quantum Field Theory presented in detail in chapters\,\hyperref[Chap:QuantumVacuum]{\ref{Chap:QuantumVacuum}},\,\hyperref[Chap:EoSVacuum]{\ref{Chap:EoSVacuum}} and\,\hyperref[Chap:Fermions]{\ref{Chap:Fermions}}. However, let us stick here to the more intuitive and general arguments and reserve the details for later.

The key idea of Running Vacuum Models came from the very fundamental ideas of General Relativity and Standard Cosmology: general covariance of the theory and the cosmological principle. When using the FLRW metric, the curvature invariants $R$, $R^2 $, $R_{\mu\nu}R^{\mu\nu}$ are limited to be composed from $H$ and its derivatives, the available energy scale characteristic of the expanding Universe. Take for instance the Ricci scalar
\begin{equation}\label{Eq:Introduction.RicciScalar}
R=-12H^2+6\dot{H}\,.
\end{equation}
It depends on $H^2 = \dot{a}^2/a^2$ and $\ddot{a} / a$, both containing two derivatives of the cosmic time, which we will simply denote as $\mathcal{O} (H^2)$. On the other hand $R^2, R_{\mu\nu} R^{\mu\nu},\dots $ depend on $H^4, \dot{H}^2, \dot{H} H^2$ (similary, collectively denoted by $\mathcal{O} (H^4)$). Each of the former invariants contain terms with an even total number of time derivatives of the scale factor $\mathcal{O} (H^{2n})$. Equivalently, they all scale as even powers of energy in natural units\footnote{One cannot discard, a priori, the appearence of other terms such as $R^{1/2} \sim \mathcal{O} (H)$. However, we will not consider such possibility here. First, because it would complicate the general ideas and second, General Relativity and its usual UV renormalizable extensions in cosmology contain only terms of the desired form $\mathcal{O} ( H^{2n} )$.}. No odd terms can be present in covariant theory can be present, then. These simple arguments suggest that the form of the cosmological effective action may should be written in the following general way\,\cite{Sola:2013gha},
\begin{equation}\label{Eq:Introduction.EffectiveActPart}
\mathcal{L}_{\rm eff}= c_0 +c_1 H^2+c_2 H^4 + \mathcal{O} (H^6)+ \dots 
\end{equation}
where $c_0$, $c_1$, $c_2$, etc. are  constants. If\,\eqref{Eq:Introduction.EffectiveActPart} is interpreted as the part of the effective action describing the corrections to the vacuum energy density (VED), it means that we can expect the VED to inhenerit the same analytical description, leading to a Renormalization Group (RG) Equation of the form
\begin{equation}\label{Eq:Introduction.RGEVacuum}
\frac{d \rho_{\rm vac}(H)}{d\ln H^2}=\frac{1}{(4\pi)^2}\sum\limits_{\rm N}\left[A_{\rm N} M_{\rm N}^2 H^2+B_{\rm N} H^4+\mathcal{O}\left(\frac{H^6}{M_{\rm N}^2}\right)\right]\,.
\end{equation}
Here, $A_{\rm N}$ and $B_{\rm N}$ are dimensionless constants, $N$ is supposed to run among the different massive fields, and $M_{\rm N}$ is the mass of these fields. Higher-order terms, represented by $\mathcal{O}(H^6/M_{\rm N}^2)$, decouple for $M_{\rm N}\rightarrow \infty$, as it should happen in a physically renormalizable theory. The role of $H$ in the previous expansion is to act as a running scale in the cosmological scenario. Therefore, the former equation describes the running of vacuum energy density, triggered by quantum fluctuations of the matter fields, with the background cosmological expansion.

Rather than predicting the exact value of the Cosmological Constant or Vacuum Energy at a particular time,  equation\,\eqref{Eq:Introduction.RGEVacuum} describes a more modest but still powerful result: if we are provided with an observational value of the vacuum energy density at a particular renormalization point, i.e. at a particular time, then \eqref{Eq:Introduction.RGEVacuum} would leave us to predict its value at another renormalization point. This is not at all an unprecedented result: as we know, the RG formalism in flat spacetime already does afirmations of similar nature in the context of Particle Physics. For instance, is able to do predictions of the electromagnetic structure fine at a particular scale $\alpha_{\rm EM}(\mu_1)$, once we know its value at another renormalization scale $\alpha_{\rm EM}(\mu_2)$. However, it is totally unable to do any prediction of $\alpha_{\rm EM}(\mu_1)$ nor $\alpha_{\rm EM}(\mu_2)$ from first principles if one of the two is not measured experimentally. This is well-stablished in particle physics in flat spacetime and \eqref{Eq:Introduction.RGEVacuum} may be interpreted in a similar philosophical current, although with some extra intricacies, such as working in a curved background\,\cite{Shapiro:2009dh}.

Phenomenologically speaking, we usually use a low-energy regime version derived from  equation\,\eqref{Eq:Introduction.RGEVacuum},
\begin{equation}\label{Eq:Introduction.RVM(H)}
\rho_{\rm vac} (H)=c_0+\frac{3 \nu}{8\pi}M_{\rm Pl}^2 H^2\,.
\end{equation}
where $M_{\rm Pl}\sim 10^{19}$ GeV is Planck's mass and $\nu\equiv 1/(6\pi)\sum_{\rm N}A_{\rm N} M_{\rm N}^2 / M_{\rm Pl}^2$ is a small parameter (since $M_{\rm N}^2 \ll M_{\rm Pl}^2$ for each mass of the Standard Model of particle Physics) that emerges from the solution of the RGE. The term $c_0$ is of extreme importance, if $\nu$ were exactly 0 we would recover the standard $\rho_{\rm vac}$=constant of the standard model. Additionally, if we want to retrieve standard results, a transition from a decelerated state to an accelerated one is only possible for $c_0 \neq 0$. Another way to write eq.\,\eqref{Eq:Introduction.RVM(H)} is subtracting the same equation at the present time, $H=H_0$. The result explicitates what was mentioned before, the prediction $\rho_{\rm vac} (H)$ can be done if we have total knowledge of $\rho_{\rm vac}^0\equiv \rho_{\rm vac}(H_0)$:
\begin{equation}\label{Eq:Introduction.RVM(H)Sub}
\rho_{\rm vac} (H)=\rho_{\rm vac}^0+\frac{3 \nu}{8\pi}M_{\rm Pl}^2 (H^2-H_0^2)\,.
\end{equation}
Up to this point, there is nothing that prevents us to change $\nu H^2$ in the previous formulas with a linear combination $\nu H^2+\tilde{\nu} \dot{H}$. After all, $\dot{H}$ is also ``allowed'' by the aforementioned general arguments of general covariance and RG. From the point of view of the phenomenology this could be an interesting possibility too, however let us just consider the simpler version, i.e. $\tilde{\nu}=0$. The energy conservation equation dictated by the Bianchi identity is
\begin{equation}\label{Eq:Introduction.BianchiNonCons}
\frac{d}{dt}\left[G ( \rho_{\rm m}+\rho_\Lambda ) \right] +3GH( \rho_{\rm m}+P_{\rm m} )=0\,.
\end{equation}
Notice that we have not imposed self-conservation of matter and vacuum separatedly, besides we admit the possibility of $G$, the gravitational coupling, being a function of the renormalization scale too. Starting from here, seveal possibilities arise\,\cite{Sola:2013gha}. Some examples are:
\begin{enumerate}
\item[$\bullet$] {\bf Model I.} $G$=const.\,, $\rho_{\rm vac}$=const.\\
This is the traditional $\Lambda$CDM where Dark Energy is represented by a cosmological constant.
\item[$\bullet$]{\bf Model II.} $G$=const.\,, $\dot{\rho}_{\rm vac} \neq 0$.\\
Matter can interact with vacuum and exchange energy:
\begin{equation}\label{Eq:Introduction.BianchiNonConsMII}
\dot{\rho}_{\rm m}+\dot{\rho}_\Lambda +3H(\rho_{\rm m}+P_{\rm m} )=0\,.
\end{equation}
We can solve $\rho_{\rm m}(z)$, because of the existence of the running vacuum law\,\eqref{Eq:Introduction.RVM(H)Sub}
\item[$\bullet$]{\bf Model III.} $\dot{G}\neq 0$\,, $\dot{\rho}_{\rm vac} \neq 0$ and matter conserved.\\
Vacuum can exchange energy with the background gravitational field:
\begin{equation}\label{Eq:Introduction.BianchiNonConsMIII}
\dot{G}\rho_{\rm vac}+G\dot{\rho}_{\rm vac}=0\,.
\end{equation}
We can solve $G(H)$, because of the existence of the running vacuum law\,\eqref{Eq:Introduction.RVM(H)Sub}.
\end{enumerate}
This just shows a small part of the rich and non-trivial phenomenology that these models have. In \hyperref[Chap:PhenomenologyofRVM]{chapter\,\ref{Chap:PhenomenologyofRVM}}, we fully explore a particular subfamily of models where one replaces the Hubble function by the Ricci scalar, $\rho_{\rm vac}(R)$, which has the virtue of not perturbing the well-known physics of Big Bang Nucleosynthesis. To sum up, running vacuum models can potentially alleviate cosmological tensions and clarify some fundamental questions regarding the cosmological constant.
%



\chapter{Renormalizing the vacum energy in cosmological spacetime: implications for the cosmological constant problem}\label{Chap:QuantumVacuum}

The cosmological constant (CC) term, $\Lambda$, in Einstein's equations has been for some three decades a fundamental building block of the `concordance' or standard $\Lambda$CDM model of cosmology\,\cite{peebles1993principles}. The model, however, was phenomenologically favored only as of the time $\CC$ became a physically measured quantity some twenty years ago \,\cite{riess1998observational,perlmutter1999measurements}. Nowadays $\Lambda$, or more precisely the associated current cosmological parameter $\OLo=\rL/\rco$ became a precision quantity\,\cite{aghanim2020planck}. Here $\rL \equiv \CC/(8\pi G_N)$ is the (vacuum) energy density induced by $\CC$, $G_N$ is Newton's constant and $\rco=3H_0^2/(8\pi G_N)$ is the current critical density. The accurate knowledge of $\OLo$ around $0.7$ is an important observational achievement, but it does not mean that we fully understand its nature and origin at a fundamental level. In the introduction of this dissertation we have already talked about the cosmological constant problem. It is a preeminent example of a fundamental theoretical conundrum. But, the abstruse theoretical problems, though, are not the only nagging ones afflicting the concordance model (see \hyperref[Sect:BeyondSM]{Sect.\,\ref{Sect:BeyondSM}}). In practice, the $\Lambda$CDM appears to be currently in tension with some important measurements, most significantly the discordant values of the current Hubble parameter $H_0$ obtained independently from measurements of the local and the early Universe. Whether these tensions are the result of as yet unknown systematic errors is not known, but there remains perfectly upright the possibility that a deviation from the $\CC$CDM model could provide an explanation for such discrepancies\cite{Riess:2019cxk}. As it has been shown in the literature, models mimicking a time-evolving $\CC$ (and hence a dynamical vacuum energy density $\rL$) could help in alleviating these problems, see e.g.\,\cite{Sola:2016jky,SolaPeracaula:2018wwm, Basilakos:2012ra,Gomez-Valent:2014rxa,Gomez-Valent:2014fda,Gomez-Valent:2015pia,Sola:2017znb,SolaPeracaula:2016qlq,SolaPeracaula:2017esw,SolaPeracaula:2020vpg,SolaPeracaula:2019zsl,Rezaei:2019xwo} and \,\cite{DiValentino:2017iww,DiValentino:2016hlg,DiValentino:2019jae,DiValentino:2017iww,DiValentino:2016hlg,Ooba:2018dzf,Park:2017xbl,Park:2018bwy,Park:2018fxx,Martinelli:2019dau,Salvatelli:2014zta,Costa:2016tpb,Li:2015vla,Li:2014cee,Li:2014eha}. Although the potential phenomenological implications of dynamical VED or, more in general, a dynamical DE may have, it is true that such an evolution is not often well-motivated from fundamental reasons and is rather an ad-hoc hypothesis added to the cosmological models.

In this chapter we present the research performed on theoretical aspects of dynamical {\it vacuum energy density} (VED) in the context of Quantum Field Theory (QFT) in curved spacetime\,\cite{birrell1984quantum,parker2009quantum,fulling1989aspects,mukhanov2007introduction}. In the works in which this chapter is based\,\cite{Moreno-Pulido:2020anb,Moreno-Pulido:2022phq,Moreno-Pulido:2022upl, SolaPeracaula:2022hpd} we tried to explain from first principles the dynamical behaviour of VED as a consequence of the running of ZPE of quantum fields in an expanding background. The renormalization of the vacuum energy in QFT is usually plagued with theoretical conundrums related not only with the renormalization procedure itself, but also with the fact that the final result leads usually to very large (finite) contributions incompatible with the measured value of $\Lambda$ in cosmology that we mentioned previously. As a consequence, one is bound to extreme fine-tuning of the parameters and so to sheer unnaturalness of the result and of the entire approach. 

Up to the moment there is not a completely successful quantum theory of gravity that combines QFT and General Relativity (GR) that can dispense the problem straightforwardly. However, there are many simplifications and techniques in the literature to study the quantum fields in the context of a curved spacetime\,\cite{birrell1984quantum, parker2009quantum}, which can help us to get over this adversity. Here, we encompass our works related with one of these techniques, a renormalization method based on the WKB approximation of the field modes in the expanding Universe called {\it adiabatic regularization}. This method has been applied successfully since its original description many decades ago and it is present in all the classic textbooks on QFT in Curved spacetime \,\cite{birrell1984quantum,parker2009quantum}. However, in this chapter we present a novelty application of how to use this machinery in our particular area of interest, VED in a dynamical expanding background. The prescription dictated by adiabatic regularization, together with our definition of EMT renomalization is called {\it Adiabatic Regularization Procedure} (ARP). Above all we wish to focus on the dynamics associated to the running vacuum model (RVM)\,\cite{Shapiro:2000dz,Shapiro:2003kv,Sola:2007sv,Shapiro:2009dh}; for a review, see\,\cite{Sola:2013gha, Sola:2014tta, Sola:2011qr, Sola:2015rra, GomezValent:2017kzh} and references therein. For related studies, see e.g.\,\cite{Babic:2004ev,Maggiore:2010wr,Maggiore:2011hw,Hollenstein:2011cz} and \,\cite{Kohri:2016lsj,Kohri:2017iyl,Antipin:2017pbt,Ferreiro:2018oxx,Ferreiro:2020zyl}, some of them extending the subject to the context of supersymmetric theories\,\cite{Bilic:2011zm,Bilic:2010xd,Bilic:2011rj} and also to supergravity\,\cite{Basilakos:2015yoa}. More recently the matter has also been addressed successfully in the framework of the effective action of string theories\,\cite{Basilakos:2019acj,Basilakos:2019mpe,Basilakos:2020qmu}. Here, however, we aim at the computation of the VED in QFT in a curved background, specifically in the spatially flat Friedmann-Lema\^\i tre-Robertson-Walker (FLRW) metric. Recall that in the momentum subtraction scheme, the renormalized Green's functions and running couplings are obtained by subtracting their values at a renormalization point $p^2=M^2$ (space-like in our metric, which becomes an Euclidean point after Wick rotation) or at the time-like one $p^2=-M^2$ (depending on the kinematical region involved). Here the situation is similar, but since for vacuum diagrams we do not have external momenta, we renormalize the ZPE by subtracting (to its on-shell value) the corresponding value computed to $4th$ adiabatic order, which we will introduce later on, but at an arbitrary mass scale $M$. This suffices both to eliminate the divergent terms in the first four adiabatic orders, which are the only ones that can be divergent in the renormalization of the EMT\,\cite{Moreno-Pulido:2020anb} and to relate the ZPE at different scales. However, if we wish to compute the zero-point energy at the value of the particle mass $m$, then the renormalization is to be performed on-shell, {\it i.e.} at $M=m$. In such a case it is evident that a subtraction at $4th$-order would give a vanishing result. Therefore, since general covariance of the effective action requires that the vacuum energy can only be expanded in even adiabatic orders, the leading contribution to the on-shell renormalized ZPE must appear at the sixth order of adiabaticity.

Hence, our first goal here is to compute the Zero-Point Energy (ZPE) of quantum fields and then, our second goal, is to determine the running of VED as well as the running of the other couplings in the theory. In particular, we perform the computation for a free real scalar field, whose inclusion is just for practical reasons, as it simplifies the calculations. However, we replicate the computation for a free Dirac fermion in \hyperref[Chap:Fermions]{chapter\,\ref{Chap:Fermions}}.
The corresponding VED involves the cosmological term $\CC$ (or, more precisely, the parameter $\rL$) in the Einstein-Hilbert (EH) action and the zero-point energy (ZPE) of the quantum matter fluctuations. We should emphasize from the beginning that we do not address here the issues of quantum gravity (QG) and the functional integration over metrics, with or without the $\CC$ term, see e.g.\,\cite{Christensen:1979iy,Christensen:1980ee,Christensen:1978gi,Christensen:1978md}. While these are potentially important matters, in the current context gravity is treated as a classical background field and hence the QG considerations in connection to quantizing the metric lie out of the scope of our semiclassical approach. Needless to say, the QFT calculation of vacuum-to-vacuum diagrams implies to perform renormalization since we meet UV-divergent integrals. The usual procedures to account for the regularization and renormalization of divergent quantities in QFT, such as e.g. the minimal subtraction (MS) scheme\,\cite{Bollini:1972ui,tHooft:1973mfk}, do not yield a sensible answer for the VED. Despite producing finite results, these are useless and incongruent with the physical facts. They lead to a quartic dependence on the mass of the fields ($\sim m^4$) and this enforces a very serious (in fact incommensurable) fine-tuning among the parameters, this being so both in Minkowski and in curved spacetime\,\footnote{See e.g.\,\cite{Sola:2013gha,SolaPeracaula:2022hpd} for a pedagogical introduction to the ZPE calculation in flat and curved spacetime.}. In this chapter, we forgo making use of such an unsuccessful method and rather adhere to the adiabatic renormalization procedure (ARP) mentioned above\,\cite{birrell1984quantum,parker2009quantum,fulling1989aspects}, in order to renormalize the ZPE in a more meaningful way\footnote{The method was introduced for the study of the running couplings in curved spacetime in\,\cite{Ferreiro:2018oxx}, although it was not applied to the VED nor to the study of the running of this quantity throughout the cosmic evolution. This was done for the first time in\,\cite{Moreno-Pulido:2020anb}.} The connection with the general framework of the running vacuum model (RVM) mentioned earlier it is also studied in detail in the main text and in an appendix.

Noticiably, we perform the calculation in three different ways: one through a modified form of the ARP\,\cite{Ferreiro:2018oxx}, the second (presented in one appendix) involving dimensional regularization (DR) and the third through the heat-kernel technique which lead us to the computation of the effective action. The common result is that the properly renormalized running VED, obtained upon inclusion of the renormalized value of $\rL$ at a given scale, does not contain the unwanted contributions proportional to the fourth power of the particle masses ($\sim m^4$) and hence it is free from large induced corrections to the VED. This is tantamount to subtracting the Minkowskian contribution from the curved spacetime result, as we show. In addition, we find that the final expression for the VED adopts the RVM form for the current Universe, namely it contains not only the usual constant term but also one that evolves with the square of the Hubble rate ($\sim \nu H^2$, with $|\nu|\ll 1$). The latter represents only a mild (dynamical) correction to the constant contribution and it can mimic quintessence or phantom DE depending on the sign of $\nu$.

Finally, let us specify the structure of this chapter. First, we will try to motivate and contextualize the framework of these calculations in \hyperref[Sect:EMTScalarField]{Sect.\,\ref{Sect:EMTScalarField}}, which consists of a neutral scalar field non-minimally coupled to gravity and with no self-interactions. We also compute the classical energy-momentum tensor (EMT). In \hyperref[Sect:AdiabaticVacuum]{Sect.\,\ref{Sect:AdiabaticVacuum}}, we define our quantum field theoretical model under study, which consists in the quantum fluctuations of the mentioned scalar field. We assume a spatially flat Friedmann-Lema\^\i tre-Robertson-Walker (FLRW) background and solve for the mode functions of the scalar field using the WKB approximation up to $4th$ and $6th$ adiabatic orders. This enables us to compute, in \hyperref[Sect:AdRegEMT]{Sect.\,\ref{Sect:AdRegEMT}}, the vacuum expectation value of the energy-momentum tensor (EMT) within QFT in curved spacetime up to the same orders. For renormalization purposes, it is only necessary the calculation up to $4th$ order suffices. In \hyperref[Sect:RenormZPE]{Sect.\,\ref{Sect:RenormZPE}}, we employ an off-shell generalization of the usual adiabatic renormalization procedure to compute the zero-point energy (ZPE) of the quantum fluctuations and show that the scaling evolution of the vacuum energy density (VED), $\rv$, is free from quartic powers of the masses. We discuss the absence of fine-tuning in \hyperref[Sect:RenormalizedVED]{Sect.\,\ref{Sect:RenormalizedVED}}. Remarkably, we find that $\rv$ carries a dynamical component $\sim H^2$ which is characteristic of the running vacuum model (RVM) at low energies as stated in \hyperref[Sect:RunningConnection]{Sect.\,\ref{Sect:RunningConnection}}. In \hyperref[Sect:Trace]{Sect.\,\ref{Sect:Trace}}, we compute the trace of the EMT up to $6th$ adiabatic order, which will be used to extract the quantum vacuum pressure at the same adiabaticity order. In passing we verify, as a useful cross-check, that our results correctly reproduce the trace anomaly. Next, with the purpose of bolt securing the renormalization results for the EMT that have been obtained from direct calculation of the expansion modes in the previous sections, we recompute them anew in \hyperref[Sect:EffectiveActionQFT]{Sect.\,\ref{Sect:EffectiveActionQFT}} within the effective action formalism using the heat-kernel expansion of the propagator with the DeWitt-Schwinger technique. We use the effective action to compute the running couplings, in particular the $\beta$-function and renormalization group equation (RGE) for the vacuum energy density (VED) itself, $\rv(H)$, showing it to be consistent with the absence of quartic mass scales in the running. To our knowledge, this is the first time that the dynamical VED is derived from first principles. The final discussion and a summary of the conclusions is presented in \hyperref[Sect:ConclusionsChapterRVQFT]{Sect.\,\ref{Sect:ConclusionsChapterRVQFT}}. Three appendices at the end furnish complementary material. Specifically, \hyperref[Appendix:Conventions]{Appendix\,\ref{Appendix:Conventions}} defines our conventions and collects some useful formulas. \hyperref[Appendix:Dimensional]{Appendix\,\ref{Appendix:Dimensional}} reconsiders the main parts of the renormalization of the EMT using Dimensional Regularization and the standard counterterm procedure, starting of course from the same WKB expansion of the field modes. Finally, \hyperref[Appendix:Abis]{Appendix\,\ref{Appendix:Abis}} presents more details on our computations on the running graviational constant and the low energy regime of the VED near the presents time, besides more details on the dynamical evolution of the latter. 

\section{Classical Energy-Momentum tensor for a non-minimally coupled scalar field}\label{Sect:EMTScalarField}

We start from the EH action for gravity plus matter:
\begin{equation}\label{Eq:QuantumVacuum.EH}
S_{\rm EH+m}= S_{\rm EH}+S_{\rm m}=\frac{1}{16\pi G}\int d^4 x \sqrt{-g}\, R - \int d^4 x \sqrt{-g}\, \rL + S_{\rm m}\,.
\end{equation}
The (constant) term $\rL$ has dimension of energy density and is it usually identified with the vacuum energy density. However, we will not call it that way here since it is not yet the physical vacuum energy density (VED), $\rv$, as we shall see. The term $\rL$ is at this point just a bare parameter of the EH action, as the gravitational constant $G$ itself. We prefer not to introduce special notations by now. The physical values will be identified only after renormalizing the bare theory\,\footnote{If we would write $\rL=\CC/(8\pi G_N)$, where $G_N$ is the value of $G$ as measured in Cavendish-like experiments at the surface of the Earth, the parameter $\CC$ could not be interpreted as a physical cosmological constant, but just as the bare cosmological term. As mentioned in the introduction, the quantity that can be associated with the physically measured cosmological constant, $\CC$ , is defined through $\rv=\Lambda/(8\pi G_N)$, where $\rv$ and $G_N$ are the physical quantities. The latter, however, can only be identified after properly renormalizing the QFT calculation.}. 

Varying the part involving the $\rL$ term yields
\begin{equation}\label{Eq:QuantumVacuum.SrL}
\delta S_{\CC}= -\int d^4 x \,\delta \sqrt{-g}\, \rL = -\frac{1}{2}\int d^4 x\, \sqrt{-g}\, \left(- \rL\, g_{\mu\nu}\right)\,\delta g^{\mu\nu}\,.
\end{equation}
Together with the variation of the EH and matter terms of \eqref{Eq:QuantumVacuum.EH}, the gravitational field equations read
\begin{equation}\label{Eq:QuantumVacuum.EinsteinEq}
\frac{1}{8\pi G}G_{\mu \nu}=-\rho_\Lambda g_{\mu \nu}+T_{\mu \nu}^{\rm m}\,,
\end{equation}
where $G_{\mu\nu}=R_{\mu\nu}-(1/2) g_{\mu\nu} R$ is the usual Einstein tensor and $ T_{\mu \nu}^{\rm m}$ is the stress-energy-momentum tensor, or just energy-momentum tensor (EMT for short) of matter\,\footnote{A list of geometric quantities of interest here are shown in \hyperref[Appendix:Conventions]{Appendix\,\ref{Appendix:Conventions}}, where we also define our conventions.}
\begin{equation}\label{Eq:QuantumVacuum.deltaTmunu2}
 T_{\mu \nu}^{\rm m}=-\frac{2}{\sqrt{-g}}\frac{\delta S_{\rm m}}{\delta g^{\mu\nu}}\,.
\end{equation}
In eq. \eqref{Eq:QuantumVacuum.EinsteinEq}, $\rho_\Lambda \equiv \Lambda/(8\pi G_N)$ can be thought as the VED associated to $\CC$. Alternatively, the latter can be cast as a contribution to the total EMT as a term $T_{\mu \nu}^\Lambda \equiv -\rho_\Lambda g_{\mu \nu}$. However, in general, there will be also other contributions to the total VED, in particular those associated to the quantum fluctuations of the fields, and also to their classical ground state energy (if it is non-vanishing). For simplicity we will suppose that there is only one (matter) field contribution to the EMT on the right hand side of \eqref{Eq:QuantumVacuum.EinsteinEq} in the form of a real scalar field, $\phi$, with mass $m$. Such contribution will be denoted $T_{\mu \nu}^{\phi}$. Hence the total EMT reads $T^{\rm tot}_{\mu\nu}=T_{\mu \nu}^\Lambda+T_{\mu \nu}^{\phi}$. We neglect for the moment the incoherent matter contributions (e.g. from dust and radiation). They can be added a posteriori without altering the pure QFT aspects on which we wish to focus right now.
Suppose that the scalar field is non-minimally coupled to gravity and that it does not couple to itself. The part of the action involving $\phi$, then, reads
\begin{equation}\label{Eq:QuantumVacuum.Sphi}
 S[\phi]=-\int d^4x \sqrt{-g}\left(\frac{1}{2}g^{\mu \nu}\partial_{\nu} \phi \partial_{\mu} \phi+\frac{1}{2}(m^2+\xi R)\phi^2 \right)\,,
\end{equation}
where $\xi$ is the non-minimal coupling of $\phi$ to gravity. In the special case $\xi=1/6$, the massless ($m=0$) action is conformally invariant in $4$ spacetime dimensions, {\it i.e.} symmetric under simultaneous local Weyl rescalings of the metric and the scalar field: $g_{\mu\nu}\to e^{2\alpha(x)}g_{\mu\nu}$ and $\phi\to e^{-\alpha(x)}\phi$, for any local spacetime function $\alpha(x)$.
However, we will keep $\xi$ general since our scalar field will be massive.

In our study, no classical potential for $\phi$ is present in our analysis so we do not need not consider the quantum corrections and corresponding renormalization of the effective potential. Here we wish to concentrate mainly on the {\it zero-point energy} (ZPE) of the quantum fields, which in itself is already rather cumbersome. In general, the non-minimal coupling $\xi$ is needed for renormalization since it is generated by loop effects even if it is absent in the classical action\,\cite{birrell1984quantum}. However, $\xi$ is not needed for the renormalization of the action in the present case since the scalar field is free as a quantum field, its interaction being only with the classical geometric/gravitational background. Under these conditions $\xi$ is not necessary for renormalizing the theory. Even so, by keeping $\xi\neq 0$ we can obtain more general results, which will be particularly useful for the connection with the Running Vacuum Model (RVM) framework in \hyperref[Sect:RunningConnection]{Sect.\,\ref{Sect:RunningConnection}}. Furthermore, it allows us to perform a non-trivial test of our calculations by reproducing the conformal anomaly for the quantum corrected action that we present in \hyperref[Sect:TraceAnomaly]{Sect.\,\ref{Sect:TraceAnomaly}}. In addition, the presence of a non-minimal coupling is expected in a variety of contexts of extended gravity theories\,\cite{Sotiriou:2008rp,Capozziello:2007ec,Capozziello:2011et,faraoni2011beyond}. For instance, $f(R)$ gravity is equivalent to scalar-tensor theory, and also to Einstein theory coupled to an ideal fluid\,\cite{Capozziello:2005mj}. The non-minimal coupling is crucially involved in models of Higgs-induced inflation\,\cite{Barvinsky:2008ia}. Furthermore, higher order and non-minimally coupled terms can be transformed, by means of a conformal transformation, into Einstein gravity plus one or more scalar fields minimally coupled to curvature. These are only a few examples in QFT, see e.g.\,\cite{faraoni2011beyond} and references therein. Let us also mention that non-minimal coupling of dilaton fields to curvature are also common in the context of the effective action of string theory at low energies (we do some comments on an interesting connection of the RVM with strings in \hyperref[SubSect:RVMInflation]{Sect.\,\ref{SubSect:RVMInflation}}). Nevertheless, as previously indicated, even in the absence of $V(\phi)$ the presence of $\xi$ can be very useful.

From hereon in, we will exclusively target the adiabatic renormalization of the ZPE of $\phi$, which in itself is already quite involved in curved spacetime. In the case of general non-minimal coupling $\xi$, the classical EMT can be computed through the functional derivative with respect to the metric of \eqref{Eq:QuantumVacuum.Sphi}:
\begin{equation}\label{Eq:QuantumVacuum.EMTScalarField}
\begin{split}
T_{\mu \nu}^{\phi}=&-\frac{2}{\sqrt{-g}}\frac{\delta S[\phi]}{\delta g^{\mu\nu}}= (1-2\xi) \partial_\mu \phi \partial_\nu\phi+\left(2\xi-\frac{1}{2} \right)g_{\mu \nu}\partial^\sigma \phi \partial_\sigma\phi\\
& -2\xi \phi \nabla_\mu \nabla_\nu \phi+2\xi g_{\mu \nu }\phi \Box \phi +\xi G_{\mu \nu}\phi^2-\frac{1}{2}m^2 g_{\mu \nu} \phi^2.
\end{split}
\end{equation}
For $\xi=0$ we recover the trivial result for the free and minimally coupled scalar field. The field $\phi$ obeys the Klein-Gordon (KG) equation in curved spacetime, which follows from varying the action \eqref{Eq:QuantumVacuum.Sphi} with respect to $\phi$:
\begin{equation}\label{Eq:QuantumVacuum.KG}
(\Box-m^2-\xi R)\phi=0\,,
\end{equation}
where $\Box\phi=g^{\mu\nu}\nabla_\mu\nabla_\nu\phi=(-g)^{-1/2}\partial_\mu\left(\sqrt{-g}\, g^{\mu\nu}\partial_\nu\phi\right)$ is the standard box operator in curved spacetime. As it is well-known, the time and space variables in the KG equation can be separated, {\it i.e.} placed in the form $\phi(t,x)\sim \sum_k\psi_k(x)\phi_k(t)$ (the sum usually being in the continuum limit) provided the metric is conformally static\,\cite{fulling1989aspects}, which means $ds^2=-dt^2+a^2(t)\gamma_{ij}(x)dx^idx^j$ in cartesian coordinates, where $a(t)$ is the scale factor and $\gamma_{ij}$ is the metric of any three-dimensional Riemannian manifold as its basic spatial section. The spacetime metric can then be put in the form $ds^2= C(\tau)(-d\tau^2+\gamma_{ij}(x)dx^idx^j)$, where the conformal scale factor $C(\tau)=a^2(\tau)$ is a function of the conformal time $\tau$. The latter is connected to the cosmic time through $\tau=\int dt/a$. The separability condition certainly holds for any FLRW metric. In the following, however, we will focus on the spatially flat three-dimensional case, $\gamma_{ij}=\delta_{ij}$. The FLRW line element is then conformally static and even conformally flat, and can be written $ds^2=a^2(\tau)\eta_{\mu\nu}dx^\mu dx^\nu$, where $\eta_{\mu\nu}={\rm diag} (-1, +1, +1, +1)$ is the Minkowski metric in our conventions. The derivative with respect to the conformal time will be denoted $^\prime\equiv d/d\tau$ and thus the Hubble rate in conformal time reads $\mathcal{H}(\tau)\equiv a^\prime /a$. Since $dt=a d\tau$, the relation between the Hubble rate in cosmic and conformal times is $\mathcal{H}(\tau)=a H(t)$, where $H(t)=\dot{a}/a$ (with $\dot{}\equiv d/dt$) is the usual Hubble rate. We will present most of our calculations in terms of the conformal time, but at the end it will be useful to express the VED in terms of the usual Hubble rate $H(t)$, as this will ease the comparison with the RVM results in the literature.

Because our metric is conformally flat, $g_{\mu\nu}=a^2(\tau)\eta_{\mu\nu}$, we have the inverse $g^{\mu\nu}=a^{-2}(\tau)\eta^{\mu\nu}$ and $\sqrt{-g}=a^4(\tau)$, and as a result the action \eqref{Eq:QuantumVacuum.Sphi} can be rewritten as
\begin{equation}\label{Eq:QuantumVacuum.Sphi2}
 S[\phi]=\frac12\int d\eta\,d^3x\, a^2 \left(\phi^{\prime 2} -(\nabla\phi)^2 - a^2(m^2+\xi R)\phi^2 \right)\,.
\end{equation}
If we perform the field redefinition $\phi=\varphi/a$ and disregard total derivatives, the previous action becomes the following functional of $\varphi$:
\begin{equation}\label{Eq:QuantumVacuum.Svarphi}
 S[\varphi] =\frac12\int d\eta d^3x \left\{\varphi^{\prime 2} -(\nabla\varphi)^2 - a^2 \left[m^2 + \left(\xi-\frac16\right) R \right] \varphi^2\right\}\,,
\end{equation}
where we have used (cf. \hyperref[Appendix:Conventions]{Appendix A}) $R=6a^{\prime\prime}/a^3$. The above field redefinition enables us to have a simpler field equation for $\varphi$ as if we were in Minkowski space (with conformal time) and an effective time-dependent mass different from that in \eqref{Eq:QuantumVacuum.KG}. Computing $\delta S[\varphi]/\delta\varphi=0$ from \eqref{Eq:QuantumVacuum.Svarphi} we find:
\begin{equation}
(\tilde{\Box}-m^2_{\rm eff}(\eta))\varphi=0\,,\ \ \ \ \ \ \ m^2_{\rm eff}(\eta)\equiv a^2(\eta) \left[m^2 + \left(\xi-\frac16\right) R(\eta) \right]\,,\label{Eq:QuantumVacuum.KGvarphi}
\end{equation}
where $\tilde{\Box}\varphi \equiv \eta^{\mu\nu}\partial_\mu\partial_\nu\varphi=-\varphi''+\nabla^2\varphi$ is the Minkowskian box operator acting on $\varphi$ in conformal coordinates $x^\mu=(\eta, {\bf x})$. The above equation for $\varphi$ is, of course, equivalent to \eqref{Eq:QuantumVacuum.KG} for the original field $\phi$, as one can check by computing the curved spacetime box operator in the conformal metric.
 
Altenatively, another way to present the KG equation \eqref{Eq:QuantumVacuum.KG} in conformally flat coordinates is as follows:
\begin{equation}\label{Eq:QuantumVacuum.KGexplicit}
 \phi^{\prime\prime}+2\cH\phi^\prime-\nabla^2\phi+a^2(m^2+\xi R)\phi=0\,,
\end{equation}
where $\Box\phi=-a^{-2}\left(\phi''+2\cH\phi-\nabla^2\phi\right)$.
The separability condition in these coordinates, namely the factorization $\phi(\tau,x)\sim \int d^3k \ A_{\bf k}\psi_k({\bf x})\phi_k(\tau)+cc$, can be implemented with $\psi_k(x)=e^{i{\bf k\cdot x}}$, but in contradistinction to the Minkowskian case we cannot take $\phi_k(\tau)=e^{\pm i\omega_k \tau}$ since the frequencies of the modes are no longer constant. The precise form of the modes $\phi_k(\tau)$ in the curved spacetime case are determined by the KG equation. In fact, starting from the Fourier expansion with separated space and time variables
\begin{equation}\label{Eq:QuantumVacuum.FourierModes}
\phi(\tau,{\bf x})=\int d^3{k} \left[ A_\bk u_k(\tau,{\bf x})+A_\bk^\ast u_k^*(\tau,{\bf x}) \right]=\int\frac{d^3{k}}{(2\pi)^{3/2}} \left[ A_\bk e^{i{\bf k\cdot x}} \phi_k(\tau)+A_\bk^\ast e^{-i{\bf k\cdot x}} \phi_k^*(\tau) \right]
\end{equation}
($A_\bk $ and $A_\bk^\ast$ being the Fourier coefficients, treated still classically at this point) and substituting it into \eqref{Eq:QuantumVacuum.KGexplicit} we find that the mode functions $\phi_k(\tau)$ are determined by the non-trivial differential equation
\begin{equation}\label{Eq:QuantumVacuum.KGFourier}
 \phi_k^{\prime\prime}+2\cH\phi^{\prime}_k+\left(\omega_k^2(m)+a^2\xi R\right)\phi_k=0\,.
\end{equation}
Because $\omega_k^2(m)\equiv k^2+a^2m^2 $, the mode functions depend only on the modulus $k\equiv|\bk|$ of the momenta, where $k$ is the comoving momentum and $\tilde{k}=k/a$ the physical one. The frequencies are seen to be functions of the time-evolving scale factor $a=a(\tau)$. This is the first unmistakable sign that a particle interpretation will become hard in this context, or in other words, it amounts to the phenomenon of particle creation in a time-dependent gravitational field\,\cite{parker1967creation,Parker:1968mv,Parker:1969au} -- for a review see e.g. \,\cite{DeWitt:1975ys,parker1979aspects,Ford:1997hb,Ford:2021syk}. If we perform the change of field mode variable $\phi_k=\varphi_k/a$ the above equation simplifies to a more amenable one in which the damping term is absent:
\begin{equation}\label{Eq:QuantumVacuum.KGFourierVarphi}
\varphi_k^{\prime \prime}+\left(\omega_k^2(m)+a^2\,(\xi-1/6)R)\right)\varphi_k=0\,.
\end{equation}
Despite it being the equation of an harmonic oscillator, it has a time-dependent frequency and cannot be solved analytically except in a few cases. For example, for conformally invariant matter, {\it i.e.} for massless scalar field ($m=0$) and conformal coupling ($\xi=1/6$), the above equation boils down to the form $\varphi_k^{\prime \prime}+k^2\varphi_k=0$, whose positive- and negative-energy solutions are just $e^{- ik\tau}$ and $e^{+ ik\tau}$, respectively. These are the very same solutions as in the massless Minkowskian case (for which $R=0$), which is ultimately the reason why no particles are created in the quantized version of the theory (in which $A_\bk $ and $A_\bk^\dagger$ -- the latter replacing $A_\bk^\ast$ -- become the annihilation and creation operators) in the conformally invariant case. In the massless case with minimal coupling ($\xi=0$) Eq.\,\eqref{Eq:QuantumVacuum.KGFourier} takes on the form
\begin{equation}\label{Eq:QuantumVacuum.KGFourierConformal}
\varphi_k^{\prime \prime}+(k^2-a^2R/6)\varphi_k=0\,.
\end{equation}
In the radiation epoch ($a\propto\tau$, thus $R=6a^{\prime\prime}/a^3=0$) we find once more the trivial modes $\varphi_k(\tau)=e^{\pm ik\tau}$. On the other hand, both in the de Sitter ($a=-1/(H\tau)$, $H=$const.) and matter-dominated ($a\propto\tau^2$) epochs we have $a^2R=12/\tau^2$, which leads to $\varphi_k''+(k^2-2/\tau^2)\varphi_k=0\,.$ This equation is non-trivial but admits an exact (positive-energy) solution in terms of Bessel functions. In the de Sitter case ($\tau<0$) one may impose the Bunch-Davies vacuum limit $\sim e^{-ik|\tau|}$ in the far remote past ($\tau\to-\infty$)\footnote{For a given mode $k$ this condition insures $k|\tau|\gg1$ and hence the modes can be thought of as being essentially insensitive to curvature effects, since $a^2R=12/\tau^2\to 0$ in this limit. In this way we are free to fix the convenient initial condition $\phi_k(\tau)\sim e^{ik\tau}=e^{-ik|\tau|}$ in the remote past. } and one finds the solution in terms of Bessel/Hankel functions: $\varphi(\tau)\propto\sqrt{k|\tau|}\left(J_{3/2}(k|\tau|)-iJ_{-3/2}(k|\tau|)\right) =\sqrt{k|\tau|}H_{3/2}^{(2)}(k|\tau|)$. Because of the half-integer order of these functions in this case, it leads to a close analytic form: $\varphi_k(\tau)\propto (1-i/(k|\tau|))e^{-ik|\tau|}$. The same solution is valid for the matter-dominated era (for which $\tau>0$). The corresponding solutions for $\phi_k$ are of course $\phi_k(\tau)=\varphi_k(\tau)/a(\tau)$ for each relevant epoch. For $m\neq0$ and/or $\xi\neq 1/6$ a solution in terms of modified Bessel functions is also possible in the de Sitter epoch. 

In general, however, there is no analytic solution of \eqref{Eq:QuantumVacuum.KGFourier} for the whole cosmic expansion history of the Universe up to the current DE epoch. Therefore, we are generally led to search for a WKB (Wentzel-Kramers-Brillouin) expansion of the solution. But before doing that let us take up the quantization of the scalar field $\phi$, since we are mainly interested in computing the vacuum fluctuations.

\section{Quantum fluctuations and adiabatic vacuum}\label{Sect:AdiabaticVacuum}

Let us now move from classical to quantum field theory. We can take into account the quantum fluctuations of the field $\phi$ by considering the expansion of the field around its background (or classical mean field) value $\phi_{\rm b}$:
\begin{equation}\label{Eq:QuantumVacuum.ExpansionField}
\phi(\eta,x)=\phi_{\rm b}(\eta)+\delta \phi (\eta,x). 
\end{equation}
We wish to compute the vacuum expectation value (VEV) of the EMT of $\phi$, {\it i.e.} $\langle T_{\mu \nu}^\phi \rangle\equiv \langle 0 |T_{\mu \nu}^\phi |0 \rangle$. The VEV of the field is identified with the background value, $\langle 0 | \phi (\tau, x) | 0\rangle=\phi_{\rm b} (\tau)$, whereas we assume zero VEV for the fluctuation: $\langle 0 | \delta\phi | 0\rangle =0$. Not so, of course, the VEV of the bilinear products of fluctuations, e.g. $\langle \delta\phi^2 \rangle\neq0$. These and other bilinear VEV's will be responsible for the ZPE of the field. We will define vacuum state for the QFT in a curved background, called the adiabatic vacuum\,\cite{Bunch:1980vc}, we are referring to with more precision below.

For an appropriate definition of the ZPE, given the above field decomposition into a classical plus fluctuating part the corresponding EMT decomposes itself as $\langle T_{\mu \nu}^\phi \rangle=\langle T_{\mu \nu}^{\phi_{\rm b}} \rangle+\langle T_{\mu \nu}^{\delta_\phi}\rangle$, where $\langle T_{\mu \nu}^{\phi_{b}} \rangle\equiv \langle 0 | T_{\mu \nu}^{\delta\phi}| 0\rangle=T_{\mu \nu}^{\phi_{b}}$ is the contribution from the classical or background part, whilst $\langle T_{\mu \nu}^{\delta\phi}\rangle\equiv \langle 0 | T_{\mu \nu}^{\delta\phi}| 0\rangle$ is the genuine vacuum contribution from the field fluctuations $\delta\phi$. The $00$-component of the latter is connected with the zero-point energy (ZPE) density of the scalar field in the FLRW background. Because $\rho_\Lambda$ is also part of the vacuum action \eqref{Eq:QuantumVacuum.EH}, the total vacuum contribution to the EMT reads
\begin{equation}\label{Eq:QuantumVacuum.EMTvacuum}
\langle T_{\mu \nu}^{\rm vac} \rangle= T_{\mu \nu}^\Lambda+\langle T_{\mu \nu}^{\delta \phi}\rangle=-\rho_\Lambda g_{\mu \nu}+\langle T_{\mu \nu}^{\delta \phi}\rangle\,.
\end{equation}
The above equation says that the total vacuum EMT is made out of the contributions from the cosmological term and of the quantum fluctuations of the field. However, since these quantities are formally UV-divergent, the physical vacuum contribution can only be identified upon suitable regularization and renormalization of our calculation. We will use later on a renormalized version of this equation and extract a relation satisfied by the renormalized VED. Rather than using minimal subtraction, as it has been customary in addressing the vacuum problem in QFT, we will use the adiabatic method. We remind the reader that $\rho_\Lambda=\Lambda/(8\pi G_N)$ denote a parameter in the Einstein-Hilbert action. This is not yet the physical vacuum energy density, $\rho_{\rm vac}$, which we are aiming at. The latter is obtained from the $00$-component of the LHS of \eqref{Eq:QuantumVacuum.EMTvacuum}, see \hyperref[Sect:RenormalizedVED]{Sect.\,\ref{Sect:RenormalizedVED}} for its precise definition. In this respect, let us note that it is common in the literature to denote the physical quantity in the conventional form $\rho_\Lambda$, but this should not be confused with the more precise notations used hereafter.

The field \eqref{Eq:QuantumVacuum.ExpansionField} obeys the curved spacetime KG equation \eqref{Eq:QuantumVacuum.KG} independently by the classical and quantum parts. Similarly, $\varphi$ and $\delta\varphi$ obey separately the Minkowskian KG equation \eqref{Eq:QuantumVacuum.KGvarphi}. Let us concentrate on the fluctuation $\delta\varphi$. Denoting the frequency modes of the fluctuating part $\delta\varphi$ by $h_k(\tau)$, we can write
\begin{equation} \label{Eq:QuantumVacuum.QuantumFourierModes}
\delta \varphi(\tau,{\bf x})=\int \frac{d^3{k}}{(2\pi)^{3/2}} \left[ A_\bk e^{i{\bf k\cdot x}} h_k(\tau)+A_\bk^\dagger e^{-i{\bf k\cdot x}} h_k^*(\tau) \right]\,.
\end{equation}
Since $\phi=\varphi/a$, the expansion of $\delta\phi$ is, of course, the same as that of \eqref{Eq:QuantumVacuum.QuantumFourierModes} but divided by the scale factor $a$. Here $A_\bk$ and $A_\bk^\dagger $ are no longer the classical Fourier coefficients but are now promoted to be (time-independent) annihilation and creation quantum operators, which satisfy the usual commutation relations
\begin{equation}\label{Eq:QuantumVacuum.CommutationRelation}
[A_\bk,{A_{\bf q}}^\dagger]=\delta({\bf k}-{\bf q}), \qquad [A_\bk,A_{\bf q}]=0. 
\end{equation}
Notice that $A_\bk$ and $h_k$ have mass dimensions $-3/2$ and $-1/2$ in natural units, respectively.
Upon substituting the Fourier expansion \eqref{Eq:QuantumVacuum.QuantumFourierModes} in $(\tilde{\Box}-m^2_{\rm eff}(\eta))\delta\varphi=0$ we find that the frequency modes of the fluctuations satisfy the (linear) differential equation
\begin{equation}\label{Eq:QuantumVacuum.KGModes}
h_k^{\prime \prime}+ \Omega_k^2 h_k=0\,, \ \ \ \ \ \ \ \ \ \ \Omega_k^2(\tau) \equiv k^2+m^2_{\rm eff}(\tau)=\omega_k^2(m)+a^2\, (\xi-1/6)R\,, 
\end{equation}
with $\omega_k^2(m)\equiv k^2+a^2 m^2$. As we can see, $h_k$ depends only on the modulus $k\equiv|\bk|$ of the momentum. Notice that $\Omega_k(\tau)$ is a non-trivial function of the conformal time, so the modes cannot be found in a simple form except in the simple cases mentioned at the end of \hyperref[Sect:EMTScalarField]{Sect.\,\ref{Sect:EMTScalarField}}. However, one can generate an approximate solution from a recursive self-consistent iteration based on the phase integral ansatz
\begin{equation}\label{Eq:QuantumVacuum.WKBSolution}
h_k(\tau)=\frac{1}{\sqrt{2W_k(\tau)}}\exp\left(-i\int^\tau W_k(\tilde{\tau})d\tilde{\tau} \right)\,.
\end{equation}
The latter is normalized through the Wronskian condition
\begin{equation}\label{Eq:QuantumVacuum.WrosnkianCondition}
h_k^\prime h_k^* - h_k^{} h_k^{*\prime}=i, 
\end{equation}
which insures that the standard equal-time commutation relations between the field operator $\varphi$ and its canonical momentum, $\pi_\varphi=\varphi^\prime$, are preserved.
The effective frequency function $W_k$ in the above ansatz follows from the differential equation obtained from inserting \eqref{Eq:QuantumVacuum.WKBSolution} into \eqref{Eq:QuantumVacuum.KGModes}:
\begin{equation}\label{Eq:QuantumVacuum.WKBIteration}
W_k^2=\Omega_k^2 -\frac{1}{2}\frac{W_k^{\prime \prime}}{W_k}+\frac{3}{4}\left( \frac{W_k^\prime}{W_k}\right)^2\,. 
\end{equation}
Although this equation is non-linear, it can be solved using the WKB approximation or Carlini-Liouville–Green approximation\,\cite{fulling1989aspects}. The leading term holds when the time variation of the frequency $W_k(\tau)$ is supposed to be very small as compared to $k$. In that case, the derivative terms on the RHS of \eqref{Eq:QuantumVacuum.WKBIteration} can be neglected and the phase integral in \eqref{Eq:QuantumVacuum.WKBSolution} with $W_k(\tau)\simeq\Omega_k(\tau)$ furnishes a sufficient approximation. The remaining terms of \eqref{Eq:QuantumVacuum.WKBIteration} improve the accuracy and can be computed by iterating the procedure in what is known as the adiabatic expansion. The implementation in the gravitational context is well-known since long\,\,\cite{parker2009quantum,fulling1989aspects}. Taking into account that the WKB solution is valid for large $k$ ({\it i.e.} for short wave lengths, as e.g. in geometrical Optics) the function $\Omega_k$ is slowly varying for weak fields. In our case such a regime is appropriate to study the short-distance behavior of the theory, {\it i.e.} the UV-divergences and the renormalization procedure. Because the general mode functions $h_k(\tau)$ are not the canonical $\varphi_k(\tau)=e^{\pm i\omega_k\tau}$ anymore, particles with definite frequencies cannot be strictly defined in a curved background. This motivates a notion of vacuum\footnote{A simple physical example in hydrodynamics is that of small amplitude adiabatic acoustic waves in an otherwise homobaric fluid ({\it i.e.} whose unperturbed pressure is constant). If the equilibrium/background state (the counterpart of the adiabatic ``vacuum'' in QFT) varies only little over the characteristic lengthscale $\lambda=1/k$ of variation of the wave, then a ``wave-like solution'' can be found through the WKB method for the pressure perturbation $\delta p$ \,\cite{Gough:2007bt}.} called the {\it adiabatic vacuum}\,\cite{Bunch:1980vc}, see also\,\cite{birrell1984quantum,parker2009quantum,fulling1989aspects,mukhanov2007introduction}. Our VEV's actually refer to that adiabatic vacuum. Rather than formulating it as the state without particles, we can at least say it is a state essentially empty of high frequency modes. Indeed, particles with definite frequencies cannot be strictly defined in a curved background, since $\Omega_k(\eta)$ is a function of time. Nonetheless an approximate Fock space interpretation is still possible, and the adiabatic vacuum can be formally defined as the quantum state which is annihilated by all the operators $A_k$ of the above Fourier expansion, see\,\cite{birrell1984quantum,parker2009quantum,fulling1989aspects,mukhanov2007introduction} for details. Our VEV's actually refer to that adiabatic vacuum. In such conditions, the minimal excited state is $h_k\simeq e^{ik\eta}/\sqrt{2k}$, with $k\simeq\Omega_k$, and hence one can maintain an approximate particle interpretation of the quantized fields in a curved background provided the geometry is slowly varying. However, in general, the physical interpretation of the modes (\ref{Eq:QuantumVacuum.KGModes}) with time varying frequencies must be sought in terms of field observables rather than in particle language. In practice, the adiabatic vacuum approximation assumes both short wavelengths and weak (or at least non strong) gravitational fields, such that the effective frequencies $\Omega_k$ are slowly varying functions of time around the Minkowskian values defined through the masses and momenta. Therefore both $m_{\rm eff}^2\equiv a^2\left(m^2+(\xi-1/6)R\right)$ and $\Omega_k^2$ remain safely positive in our domain of study. Simple estimates show that this is so for the most accessible part of the cosmic history, starting from the radiation-dominated epoch (where $R=0$) until the present time and into the future, in which $R\sim H^2$ is very small as compared to the usual particle masses (squared). The Bunch-Davies vacuum mentioned above was a particular form of adiabatic vacuum for the case of the de Sitter space. We emphasize that in all cases, including the situation with the stronger gravitational fields in the inflationary epoch (see \hyperref[SubSect:RVMInflation]{Sect.\,\ref{SubSect:RVMInflation}} for further discussion), in the absence of a clear-cut particle interpretation, a more physical approach to the vacuum effects of the expanding Universe can be obtained by computing the vacuum part of the EMT of the scalar field in the cosmologically expanding background. To accomplish this task, we need to insert the above Fourier expansions in \eqref{Eq:QuantumVacuum.EMTScalarField} and compute the VEV in Fourier space, hence integrating over all modes, $\int\frac{d^3k}{(2\pi)^3}(...) $. In the process we must use the expansion of \eqref{Eq:QuantumVacuum.WKBIteration} in order to compute the explicit form of the modes, and this yields UV-divergent integrals. 

\section{Adiabatic Regularization of the Energy-Momentum tensor}\label{Sect:AdRegEMT}

Sometimes the following notation is used in the literature. Let $T$ be a dimensionless parameter and let us replace $\Omega_k\to T\Omega_k$. The power of $T^{-1}$ defines the {\it adiabaticity order}. Upon rescaling, this is equivalent to replace $a(\tau)\to a(\tau/T)$. Then the derivatives of $\Omega_k(a(\tau))$ with respect to $\tau$ all go to $0$ for $T\rightarrow\infty$. The number of derivatives coincides with the power of $T^{-1}$, {\it i.e.} the adiabaticity order. This shows that the condition of validity of the expansion is that $\Omega_k(a)$ varies slowly in time, whence the name adiabatic expansion. In practice we will not keep $T$ explicitly, it suffices to count the number of time derivatives of the various terms of the expansion and we will indicate adiabaticity order $N$ of a quantity by a superindex $(N)$. But the power $T^{-N}$ of a given term serves as a practical book-keeping device to identify the $Nth$ adiabaticity order of such a term.

As mentioned in the last section, we need to renormalize the VEV of the EMT. We will see that, to accomplish this task, it would be necessary to appropriately subtracting the first four adiabatic orders, which are UV-divergent. Adiabatic orders higher than $4$ decay sufficiently quick at large momentum $k$ (short-distances) so as to make the corresponding integrals convergent. This is a reflex of the Appelquist-Carazzone decoupling theorem\,\cite{Appelquist:1974tg}. For an adiabatic (slowly varying) $\Omega_k$, we can use \eqref{Eq:QuantumVacuum.WKBIteration} as a recurrence relation to generate an (asymptotic) series solution. In the gravitational context, such WKB approximation is organized through adiabatic orders and constitutes the basis for the adiabatic regularization procedure (ARP)\footnote{The ARP was first introduced for minimally coupled (massive) scalar fields in\,\cite{Parker:1974qw,Fulling:1974zr,fulling1974conformal,bunch1979feynman} and subsequently generalized for arbitrary couplings\,\cite{Bunch:1980vc}. For a review, see e.g. the classic books\,\cite{birrell1984quantum,parker2009quantum}. The method has been applied to related studies of QFT in curved backgrounds\,\cite{Ferreiro:2018oxx,Kohri:2016lsj,Kohri:2017iyl} and has also been extended for spin one-half fields in\,\cite{Landete:2013axa, Landete:2013lpa, delRio:2014cha, BarberoG:2018oqi}.}. Notice that the adiabatic expansion is an asymptotic expansion, and therefore going to higher and higher orders (which become extremely cumbersome in practice) does not necessarily imply a degree of better convergence of the series. We will reach 6{\it th} adiabatic order in the computation of the EMT, which is already messy, but is feasible and necessary for the study of the on-shell renormalized theory and other properties of the quantum vacuum.

Before going to compute the EMT in this perturbative way, we need to seek the expansion of the modes $h_k$. In order to do that, let us specify a little bit more about the adiabatic orders in the FLRW context. The quantities that are taken to be of adiabatic order 0 are: $k^2$ and $a$. Of adiabatic order 1 are: $a^\prime$ and $\mathcal{H}$. Of adiabatic order 2: $a^{\prime \prime},a^{\prime 2},\mathcal{H}^\prime$ and $\mathcal{H}^2$. Each additional derivative increases the adiabatic order by one unit.
Therefore, the solution of the ``effective frequency'' $W_k$ is found from a WKB-type asymptotic expansion in powers of the adiabatic order:
\begin{equation}\label{Eq:QuantumVacuum.WkExp}
W_k=\omega_k+\omega_k^{(2)}+\omega_k^{(4)}+\omega_k^{(6)}+\dots,
\end{equation}
where each $\omega_k^{(j)}$ is an adiabatic correction of order $j$ (and $\omega_k \equiv \omega_k^{(0)}$). Then the above series \eqref{Eq:QuantumVacuum.WkExp} can be regarded as an expansion in $T^{-2}$ for $T\rightarrow\infty$. In this way we obtain an adiabatic expansion of the mode functions $h_k$ in powers of even order adiabatic terms ($0, 2, 4,..$.), such as $a$, $a^{\prime\prime} \propto R$, $\left({\omega_k}^\prime\right)^2$, ${\omega_k}^{\prime\prime}$, $\left({\omega_k}^{\prime\prime}\right)^2$, $R^2$, etc. The non-appearance of odd adiabatic orders is justified by arguments of general covariance, which forbid tensors of odd adiabatic order in the effective action and in gravitational field equations. The $\omega_k^{(j)}$ can be expressed in terms of $\Omega_k(\tau)$ and its time derivatives. However, since $\Omega_k(\tau)$ in our case adopts the explicit form indicated in \eqref{Eq:QuantumVacuum.KGModes}, with $R$ being of adiabatic order $2$, to insure that the adiabaticity order is preserved it suffices that the derivatives in the terms on the RHS of \eqref{Eq:QuantumVacuum.WKBIteration} are performed on $\omega_k(\tau)$ only. We will see this feature in the formulas given below.
\subsection{Relating different renormalization scales through the ARP}\label{Sect:RelDiffRenScales}

We start by defining the first term of the above WKB expansion and compute its first two derivatives:
\begin{equation}\label{Eq:QuantumVacuum.omegak0}
\begin{split}
\omega_k^{(0)} \equiv \omega_k&= \sqrt{k^2+a^2 M^2},\\ \omega_k^\prime=a^2\mathcal{H}\frac{M^2}{\omega_k}, \qquad \omega_k^{\prime \prime}&=2a^2\mathcal{H}^2\frac{M^2}{\omega_k}+a^2\mathcal{H}^\prime \frac{M^2}{\omega_k}-a^4\mathcal{H}^2\frac{M^4}{\omega_k^3}\,.
\end{split}
\end{equation}
Notice that in this approach the WKB expansion is performed off-shell, {\it i.e.} we use the arbitrary mass scale $M$ instead of the original mass $m$. In this fashion the ARP can be formulated in such a way that we can relate the adiabatically renormalized theory at two scales\,\cite{Ferreiro:2018oxx}. The mass scale $M$ is an 0{\it th} order adiabatic quantity which plays a role similar to the scale $\mu$ in Dimensional Regularization (DR), but it can be given a more physical meaning\footnote{We distinguish $M$ from 't Hooft's mass unit $\mu$ in DR, which can be used together with $M$ in \hyperref[Sect:EffectiveActionQFT]{Sec.\,\ref{Sect:EffectiveActionQFT}} to regularize the UV divergences of the effective action. The parameter $\mu$ is unphysical and is used in the MS scheme with DR to define the renormalization point. We should stress, however, that we do not use such a scheme at all in our calculation, even if we make some (optional) use of DR in certain parts. In our physical results, $\mu$ always cancels out and the final renormalized quantities depend on $M$ only.}. When $M$ is fixed at the physical mass of the quantized field ($M = m$) we expect to obtain the renormalized theory on-shell. By keeping the $M$-dependence we can subtract the EMT at such value, thus obtaining the renormalized theory at $M$.
Physically, it means that we can explore other scales away from the one associated to the mass of the particle.
In the subtraction procedure, the divergences will be cancelled and the quadratic mass differences $\Delta^2\equiv m^2-M^2$ will appear in the correction terms relating the theory at the two renormalization scales. These differences must be reckoned as being of adiabatic order 2 since it is mandatory for the renormalization procedure off-shell\,\cite{Ferreiro:2018oxx}. For $\Delta = 0$ we recover $M = m$ and corresponds to the usual ARP (where one renormalizes the theory only at the scale of the particle mass)\,\cite{birrell1984quantum,parker2009quantum}.
We will use this procedure to explore the behavior of the VED throughout the cosmological evolution. In principle, the masses $m$ could be associated to fields of the Standard Model of particle physics, but it is also possible (and maybe convenient) to consider also heavier fields, such as the ones present in some Grand Unified Theories (GUT) and explore the behavior in the low energy domain $M^2\ll m^2$. Needless to say, for the sake of simplicity, we model here all particles in terms of (real) scalar fields.

From these elementary differentiations in \eqref{Eq:QuantumVacuum.omegak0} one can then compute the more laborious derivatives appearing in the above expressions, such as $ (\omega_k^{(2)})^\prime, (\omega_k^{(2)})^{\prime\prime}, (\omega_k^{(4)})^\prime, (\omega_k^{(4)})^{\prime\prime}$, etc.
The explicit form with all of the terms after computing the various derivatives and expanding the products and powers of the different terms leads to a rather formidable output. We refrain from quoting it here, but of course it will be used for the computation of the EMT up to ${\cal O}(T^{-6})$. One can see immediately that the adiabatic expansion becomes an expansion in powers of $\mathcal{H}$ and its time derivatives. For instance, an even number of derivatives of $\omega_k$ (hence of even adiabatic order) is equivalent to an expansion in even powers of $\cH$ and odd powers of $\cpH$ (notice e.g. that $\cH^2$ and $\cpH$ are homogeneous), as in both cases it involves an even number of derivatives of the scale factor. In this way, the expansion is compatible with general covariance, as indicated above. This is a noticeable property which will be of paramount importance for our considerations. Notice that if the final formulas for the physical quantity (in our case the ZPE) are written in terms of the ordinary Hubble function, $H(t)$, no factor of $a$ can remain. All the terms with $n$ cosmic time derivatives of the scale factor in different ways are of adiabatic order $N$. For example, for $N=4$ one can have, in principle, $5$ possible combinations: $H^4, \dot{H}^2,\, H^2 \dot{H},\vardot{3}{H}$ and $ H\ddot{H}$, all of them being ${\cal O}(T^{-4})$; and for $N=6$ we can have $11$ structures of order ${\cal O}(T^{-6})$, to wit: $H^6, H^4\dot{H}, \dot{H}^3, H^3\ddot{H}, H^2\vardot{3}{H}, \dot{H}\vardot{3}{H},\ddot{H}^2, H\vardot{4}{H}, H^2\dot{H}^2, H\dot{H}\ddot{H}$ and $\vardot{5}{H}$. We shall find explicitly all the actual terms. Somewhat unexpectedly, though, we will find that terms of a given order in the list do not show up in the final result. So the correct adiabaticity order is a necessary but not a sufficient condition to appear in the final result. Notice also that, for the current Universe, the powers $\cH^2$ and $\cpH$ are sufficient for the phenomenological description, as it is obvious from the fact that $R= (6/a^2)(\cH^2+\cpH)$, whereas the higher powers bring corrections which can be important in the early Universe. 

\subsection{Computing the adiabatic orders and the regularized ZPE}\label{Sect:CompAdOrdersZPE}

The ZPE is related to the 00 component of the EMT associated to the quantum vacuum fluctuations in curved spacetime with FLRW metric. We have now all the necessary ingredients to calculate it. Our interest is to compute the field modes up to $6th$ order, for reasons that will be clarified later on. To obtain the different orders, we start with the initial solution $W_k\approx \omega_k^{(0)}$ indicated in Eq.\eqref{Eq:QuantumVacuum.omegak0}. For $a=1$ this would yield the standard Minkowski space modes. But since $a=a(\eta)$ we have to find a better approximation. Introducing that initial solution on the RHS of \eqref{Eq:QuantumVacuum.WKBIteration} and expanding it in powers of $\omega_k^{-1}$ we may collect the new terms up to adiabatic order 2 to find $\omega_k^{(2)}$. Next we iterate the procedure by introducing $W_k\approx \omega_k^{(0)}+\omega_k^{(2)}$ on the RHS of the same equation, expand again in $\omega_k^{-1}$ and collect the terms of adiabatic order 4, etc. Since this mathematical procedure implies an expansion in powers of $\omega_k^{-1}\sim 1/k\sim\lambda$ ({\it i.e.} a short wavelength expansion) it is obvious that the UV divergent terms of the ARP are the ones containing the first lowest powers of $1/\omega_k$, and hence are concentrated in the first adiabatic orders, whilst the higher adiabatic orders represent finite contributions\,\cite{Bunch:1980vc,Parker:1974qw,Fulling:1974zr,fulling1974conformal,bunch1979feynman}. The result is intuitive: for any given physical quantity, the UV divergences are concentrated in the first adiabatic orders whereas the higher orders must decay sufficiently quick at high momentum so as to make the corresponding integrals convergent and yielding a suppressed contribution. For the main quantity at stake in our case, the EMT, its regularization implies to work, at least, up to $4th$ adiabatic order, as we shall show in detail below in \hyperref[Sect:RenormZPE]{Sect.\,\ref{Sect:RenormZPE}}. Upon renormalization, we will obtain a finite expression for the EMT.

The starting procedure is to insert the decomposition \eqref{Eq:QuantumVacuum.ExpansionField} of the quantum field $\phi$ in the EMT as given in Eq.\,\eqref{Eq:QuantumVacuum.EMTScalarField} and select only the fluctuating parts $\delta\phi$ decomposed as in \eqref{Eq:QuantumVacuum.ExpansionField}. However, we are interested just on the contribution from the fluctuations, so we pick out the quadratic fluctuations in $\delta\phi$ only since, as previously indicated, we have zero VEV for the fluctuation itself. By the same token, the crossed terms with the background part $\phi_{\rm b}$ and the fluctuation $\delta\phi$ vanish, since they are independent. 

Using \eqref{Eq:QuantumVacuum.omegak0} and working out up to 6 {\it th} adiabatic order\footnote{We are not aware that his result has been previously reported in the literature} in \eqref{Eq:QuantumVacuum.WKBIteration}, one finds
\begin{equation}\label{Eq:QuantumVacuum.WKBexpansions}
\begin{split}
\omega_k^{(0)}&=\omega_k,\\
\omega_k^{(2)}&= \frac{a^2 \Delta^2}{2\omega_k}+\frac{a^2 R}{2\omega_k}(\xi-1/6)-\frac{\omega_k^{\prime \prime}}{4\omega_k^2}+\frac{3\omega_k^{\prime 2}}{8\omega_k^3}\,,\\
\omega_k^{(4)}&=-\frac{1}{2\omega_k}\left(\omega_k^{(2)}\right)^2+\frac{\omega_k^{(2)}\omega_k^{\prime \prime}}{4\omega_k^3}-\frac{\omega_k^{(2)\prime\prime}}{4\omega_k^2}-\frac{3\omega_k^{(2)}\omega_k^{\prime 2}}{4\omega_k^4}+\frac{3\omega_k^\prime \omega_k^{(2)\prime}}{4\omega_k^3}\,\\
\omega_k^{(6)}&=\frac{\omega_k^{\prime\prime}\omega_k^{(4)}}{4\omega_k^3}-\frac{\omega_k^{\prime\prime}\left(\omega_k^{(2)}\right)^2}{4\omega_k^4}+\frac{\left(\omega_k^{(2)}\right)^{\prime\prime}\omega_k^{(2)}}{4\omega_k^3}-\frac{\left(\omega_k^{(4)}\right)^{\prime\prime}}{4\omega_k^2} -\frac{3 \left(\omega_k^\prime\right)^2\omega_k^{(4)}}{4\omega_k^4}+\frac{9\left(\omega_k^\prime\right)^2\left(\omega_k^{(2)}\right)^2}{8\omega_k^5},\\
&+\frac{3\left(\left(\omega_k^{(2)}\right)^\prime\right)^2}{8\omega_k^3}
 +\frac{3\omega_k^\prime \left(\omega_k^{(4)}\right)^\prime}{4\omega_k^3}-\frac{3\omega_k^\prime \left(\omega_k^{(2)}\right)^\prime\omega_k^{(2)}}{2\omega_k^4}-\frac{\omega_k^{(2)}\omega_k^{(4)}}{\omega_k}\,.
\end{split}
\end{equation}
We are now ready to compute the energy density associated to the quantum vacuum fluctuations in curved spacetime with FLRW metric, {\it i.e.} the ZPE. We start from the EMT given in Eq.\,\eqref{Eq:QuantumVacuum.EMTScalarField} with $\phi$ decomposed as in \eqref{Eq:QuantumVacuum.ExpansionField}. However, we are interested just on the fluctuating part, and select the quadratic fluctuations in $\delta\phi$ only since, as previously indicated, we have zero VEV for the fluctuation itself. For the $00$-component, related to the energy density of the vacuum fluctuations, we find
\begin{equation}\label{Eq:QuantumVacuum.EMTInTermsOfDeltaPhi}
\begin{split}
\left\langle T_{00}^{\delta \phi} \right\rangle = & \left\langle \frac{1}{2} \left( \delta\phi^{\prime}\right)^2+\left(\frac{1}{2}-2\xi\right)\sum_i\partial_i \delta \phi\partial_i \delta \phi+6\xi\mathcal{H}\delta \phi \delta \phi^\prime \right.\\
&\phantom{xx}-\left. 2\xi\delta\phi\sum_i \partial_{ii}\delta\phi+3\xi\mathcal{H}^2\delta\phi^2+\frac{a^2m^2}{2}(\delta\phi)^2 \right \rangle\,.
\end{split}
\end{equation}
To clarify the notation, notice that $\left(\delta\phi^{\prime}\right)^2\equiv\left(\delta\partial_0\phi\right)^2= \left(\partial_0\delta\phi\right)^2$. We may now substitute the Fourier expansion of $\delta\phi=\delta\varphi/a$, as given in \eqref{Eq:QuantumVacuum.QuantumFourierModes}, into Eq.\,\eqref{Eq:QuantumVacuum.EMTInTermsOfDeltaPhi} and apply the commutation relations \eqref{Eq:QuantumVacuum.CommutationRelation}.
After symmetrizing the operator field product $\delta\phi \delta\phi^\prime$ with respect to the creation and annihilation operators,
\begin{equation}\label{Eq:QuantumVacuum.Symmetrization}
\delta\phi \delta\phi^\prime \rightarrow \frac{1}{2}\left(\delta\phi \delta\phi^\prime+\delta\phi^\prime \delta\phi \right)\,,
\end{equation}
we end up with the following expression in terms of the amplitudes of the Fourier modes of the scalar field:
\begin{equation}\label{Eq:QuantumVacuum.T00}
\begin{split}
\left\langle T_{00}^{\delta \phi}\right\rangle &=\frac{1}{4\pi^2 a^2}\int dk k^2 \left[ \left|h_k^\prime\right|^2+(\omega_k^2+a^2\Delta^2)\left|h_k\right|^2 \right.\\
&\left. +\left(\xi-\frac{1}{6}\right)\left(-6\mathcal{H}^2\left|h_k\right|^2+6\mathcal{H}\left(h_k^\prime h_k^{*}+h_k^{*\prime}h_k\right)\right)\right]\,,
\end{split}
\end{equation}
where we have integrated $ \int\frac{d^3k}{(2\pi)^3}(...)$ over solid angles and expressed the final integration in terms of $k=|\bk|$.

We expand the various terms of the above integral consistently up to $6th$ order using the WKB approximations \eqref{Eq:QuantumVacuum.WKBexpansions}. After some tedious calculations, we find
\begin{equation}\label{Eq:QuantumVacuum.exphk2}
\begin{split}
&|h_k|^2=\frac{1}{2W_k}=\frac{1}{2\omega_k}-\frac{\omega_k^{(2)}}{2\omega_k^2}-\frac{\omega_k^{(4)}}{2\omega_k^2}-\frac{\omega_k^{(6)}}{2\omega_k^2}+\frac{\left(\omega_k^{(2)}\right)^2}{2\omega_k^3}
+\frac{\omega_k^{(2)}\omega_k^{(4)}}{\omega_k^3}-\frac{\left(\omega_k^{(2)}\right)^3}{2\omega_k^4}\,,\\
\end{split}
\end{equation}
\begin{equation}\label{Eq:QuantumVacuum.exphkp2}
\begin{split}
\left| h_k^\prime \right|^2=\frac{\left(W_k^\prime\right)^2}{8W_k^3}+\frac{W_k}{2}&=\frac{\omega_k}{2}\left(1 +\frac{\omega_k^{(2)}}{\omega_k} + \frac{\omega_k^{(4)}}{\omega_k} +\frac{\omega_k^{(6)}}{\omega_k}\right)\\
& +\frac{\left(\omega_k^\prime\right)^2}{8\omega_k^3}\left(1 -\frac{3\omega_k^{(2)}}{\omega_k} - \frac{3\omega_k^{(4)}}{\omega_k} + 6\frac{\left(\omega_k^{(2)}\right)^2}{\omega_k^2}\right)\\
&+\frac{\left(\left(\omega_k^{(2)}\right)^\prime\right)^2}{8\omega_k^3}
+ \frac{\left(\omega_k^{(2)}\right)^\prime\omega_k^\prime}{4\omega_k^3}\left(1 -\frac{3\omega_k^{(2)}}{\omega_k}\right) +\frac{\left(\omega_k^{(4)}\right)^\prime\omega_k^\prime}{4\omega_k^3}\,,
\end{split}
\end{equation}
\begin{equation}\label{Eq:QuantumVacuum.exphkphk}
\begin{split}
\begin{split}
 h_k^\prime h_k^*+\left(h_k^*\right)^\prime h_k= -\frac{W_k^\prime}{2W_k^2}&=-\frac{\omega_k^\prime}{2\omega_k^2}\left(1 - \frac{2\omega_k^{(2)}}{\omega_k} -\frac{2\omega_k^{(4)}}{\omega_k} + \frac{3\left(\omega_k^{(2)}\right)^2}{\omega_k^2}\right)\\
&-\frac{\left(\omega_k^{(2)}\right)^\prime}{2\omega_k^2}\left(1 - \frac{2\omega_k^{(2)}}{\omega_k} \right) -\frac{\left(\omega_k^{(4)}\right)^\prime}{2\omega_k^2}\,.
\end{split}
\end{split}
\end{equation}
Upon substituting the above WKB expansions in \eqref{Eq:QuantumVacuum.T00} and using the relations \eqref{Eq:QuantumVacuum.WKBexpansions} and \eqref{Eq:QuantumVacuum.omegak0}, the result can be phrased as follows after a significant amount of algebra:
\begin{equation}\label{Eq:QuantumVacuum.EMTFluctuations}
\begin{split}
\left\langle T_{00}^{\delta \phi} \right\rangle^{(0-4)} & =\frac{1}{8\pi^2 a^2}\int dk k^2 \Bigg[ 2\omega_k+\frac{a^4M^4 \mathcal{H}^2}{4\omega_k^5}-\frac{a^4 M^4}{16 \omega_k^7}(2\mathcal{H}^{\prime\prime}\mathcal{H}-\mathcal{H}^{\prime 2}+8 \mathcal{H}^\prime \mathcal{H}^2+4\mathcal{H}^4)\\
&\phantom{xxxxxxxxxxxxx}+\frac{7a^6 M^6}{8 \omega_k^9}(\mathcal{H}^\prime \mathcal{H}^2+2\mathcal{H}^4) -\frac{105 a^8 M^8 \mathcal{H}^4}{64 \omega_k^{11}}\\
&\phantom{xxxxxxxxxxxxx}+\left(\xi-\frac{1}{6}\right)\left(-\frac{6\mathcal{H}^2}{\omega_k}-\frac{6 a^2 M^2\mathcal{H}^2}{\omega_k^3}+\frac{a^2 M^2}{2\omega_k^5}(6\mathcal{H}^{\prime \prime}\mathcal{H}-3\mathcal{H}^{\prime 2}+12\mathcal{H}^\prime \mathcal{H}^2)\right. \\
&\phantom{xxxxxxxxxxxxxxxxxxxxxxx}\left. -\frac{a^4 M^4}{8\omega_k^7}(120 \mathcal{H}^\prime \mathcal{H}^2 +210 \mathcal{H}^4)+\frac{105a^6 M^6 \mathcal{H}^4}{4\omega_k^9}\right)\\
&\phantom{xxxxxxxxxxxxx}+ \left(\xi-\frac{1}{6}\right)^2\bigg(-\frac{1}{4\omega_k^3}(72\mathcal{H}^{\prime\prime}\mathcal{H}-36\mathcal{H}^{\prime 2}-108\mathcal{H}^4)\\
&\phantom{xxxxxxxxxxxxxxxxxxxxxxxxx}+\frac{54a^2M^2}{\omega_k^5}(\mathcal{H}^\prime \mathcal{H}^2+\mathcal{H}^4) \bigg)\Bigg]\\
&+\frac{1}{8\pi^2 a^2} \int dk k^2 \left[ \frac{a^2\Delta^2}{\omega_k} -\frac{a^4 \Delta^4}{4\omega_k^3}+\frac{a^4 \mathcal{H}^2 M^2 \Delta^2}{2\omega_k^5}-\frac{5}{8}\frac{a^6\mathcal{H}^2 M^4\Delta^2}{\omega_k^7} \right.\\
& \phantom{xxxxxxxxxxxxxx}\left. +\left( \xi-\frac{1}{6} \right) \left(-\frac{3a^2\Delta^2 \mathcal{H}^2}{\omega_k^3}+\frac{9a^4 M^2 \Delta^2 \mathcal{H}^2}{\omega_k^5}\right)\right]\,.
\end{split}
\end{equation}
Here $(0-4)$ means that the EMT has been computed up to the 4{\it th} order. Let us note the presence of the $\Delta$-dependent terms in the last two rows, which contribute at second ($\Delta^2$) and fourth ($\Delta^4$) adiabatic order. As expected, only even powers of $\cal H$ remain in the final result. Mind that $k$ in the above formulas is the comoving momentum, whereas the physical momentum is $\tilde{k}=k/a$.

In view of these explicit results it is obvious that the VEV \eqref{Eq:QuantumVacuum.T00} is UV-divergent, specifically the integrals $\int dk k^2 \left|h_k^\prime\right|^2$ and $\int dk k^2 \omega_k^2\left|h_k\right|^2$ in it are both quartically divergent, $\int dk k^2 \left|h_k\right|^2$ is quadratically divergent and $\int dk k^2 \left(h_k^\prime h_k^{*}+h_k^{*\prime}h_k\right)$ is logarithmically divergent. No terms can be left in the EMT being linear in $\cal H$, nor any odd power of it, as they would violate the covariance of the result. Only even powers of $\cH$ can remain in the final result (strictly speaking, terms with an even number of derivatives of the scale factor), as we shall further reconfirm below.

The result \eqref{Eq:QuantumVacuum.EMTFluctuations} constitutes the WKB approximation up to $4th$ adiabatic order. It is enough to encompass all the UV-divergences that appear in the WKB expansion of the ZPE. However, we need to continue such an expansion one more step since we want to compute the on-shell value of the ZPE and, as it will be clear in the next section, the effort is necessary. Thus, we can reach higher orders and the result can be conveniently split into the various contribution up to $6th$ adiabatic order (plus higher orders, if necessary, but not in our case):
 \begin{equation}\label{eq:T00expansion}
 \left\langle T_{00}^{\delta \phi} \right\rangle=\left\langle T_{00}^{\delta \phi }\right\rangle^{(0)}+\left\langle T_{00}^{\delta \phi } \right\rangle^{(2)}+\left\langle T_{00}^{\delta \phi }\right\rangle^{(4)} +\left\langle T_{00}^{\delta \phi }\right\rangle^{(6)} +... =\left\langle T_{00}^{\delta \phi } \right\rangle^{(0-4)} +\left\langle T_{00}^{\delta \phi}\right\rangle^{ (6)} +...
 \end{equation}
 where for convenience we have collected the contribution from the terms up to $4th$ adiabatic order in the expression $T_{00}^{\delta \phi (0-4)}\equiv T_{00}^{\delta \phi (0)}+T_{00}^{\delta \phi (2)} +T_{00}^{\delta \phi (4)}$. 
We now move on to the calculation of the $6th$-order contribution, $\langle T_{00}^{\delta \phi }\rangle^{(6)}$, which is more cumbersome than the contributions up to $4th$-order, Eq.\,\eqref{Eq:QuantumVacuum.EMTFluctuations}. We will quote the expression only at the on-shell point $M=m$ (so all of the terms proportional to $\Delta$ vanish in this case). Our renormalization procedure is based in a subtraction between different scales. There is no need to compute $\langle T_{00}^{\delta \phi }\rangle^{(6)}$ at an arbitrary scale $M$ since no subtraction is needed for a contribution which is fully convergent, piece by piece. We will see all the details in \hyperref[Sect:RenormZPE]{Sect.\,\ref{Sect:RenormZPE}}. In Fourier space, it reads as follows:
\pagebreak
\begin{equation}\label{Eq:QuantumVacuum.Tsixthhorder}
\begin{split}
 \left\langle T_{00}^{\delta \phi} \right\rangle^{\rm (6)} (m)=\frac{1}{4\pi^2 a^2}\int dk k^2 &\Bigg[ \frac{a^4 m^4}{128\omega_k^9}\Bigg(16\mathcal{H}^6 +96 \mathcal{H}^4 \mathcal{H}^\prime+84 \mathcal{H}^2\left(\mathcal{H}^\prime\right)^2-12 \left(\mathcal{H}^\prime\right)^3\\
&\phantom{aaaaaaaa}+56\mathcal{H}^3\mathcal{H}^{\prime\prime}+36\mathcal{H}\mathcal{H}^\prime\mathcal{H}^{\prime\prime}+\left(\mathcal{H}^{\prime\prime}\right)^2+16\mathcal{H}^2 \mathcal{H}^{\prime\prime\prime}\\
&\phantom{aaaaaaaa}-2 \mathcal{H}^\prime \mathcal{H}^{\prime\prime\prime}+2\mathcal{H}\mathcal{H}^{\prime\prime\prime\prime}\Bigg)\\
&\phantom{}+\frac{a^6 m^6}{32\omega_k^{11}}\bigg(-152\mathcal{H}^6-396\mathcal{H}^4\mathcal{H}^\prime-114\mathcal{H}^2\left(\mathcal{H}^\prime \right)^2+5\left(\mathcal{H}^\prime\right)^3\\
&\phantom{aaaaaaaaa}-102\mathcal{H}^3\mathcal{H}^{\prime\prime}-15\mathcal{H}\mathcal{H}^\prime\mathcal{H}^{\prime\prime}-9\mathcal{H}^2\mathcal{H}^{\prime\prime\prime}\bigg)\\
&+\frac{33a^8m^8}{256\omega_k^{13}}\left(212\mathcal{H}^6+264\mathcal{H}^4\mathcal{H}^\prime+27\mathcal{H}^2\left(\mathcal{H}^\prime\right)^2+26\mathcal{H}^3\mathcal{H}^{\prime\prime}\right)\\
&-\frac{3003 a^{10}M^{10}}{128\omega_k^{15}}\left(2\mathcal{H}^6+\mathcal{H}^4\mathcal{H}^\prime\right)+\frac{25025 a^{12}\mathcal{H}^6 m^{12}}{1024\omega_k^{17}}\Bigg]\\
+\frac{\left(\xi-\frac{1}{6}\right)}{4\pi^2 a^2}\int dk k^2 &\Bigg[\frac{3a^2 m^2}{16\omega_k^7}\Bigg(-32\mathcal{H}^4\mathcal{H}^\prime-32\mathcal{H}^2\left(\mathcal{H}^\prime\right)^2+8\left(\mathcal{H}^\prime\right)^3-24\mathcal{H}^3\mathcal{H}^{\prime\prime}\\
&\phantom{aaaaaaaa}-\left(H^{\prime\prime}\right)^2-8\mathcal{H}^2\mathcal{H}^{\prime\prime\prime}+2\mathcal{H}^\prime\mathcal{H}^{\prime\prime\prime}-2\mathcal{H}\mathcal{H}^{\prime\prime\prime\prime}\Bigg)\\
&+\frac{21a^4 m^4}{32\omega_k^9}\bigg(76\mathcal{H}^6+232\mathcal{H}^4\mathcal{H}^\prime+73\mathcal{H}^2\left(\mathcal{H}^\prime\right)^2-4\left(\mathcal{H}^\prime\right)^3+74\mathcal{H}^3\mathcal{H}^{\prime\prime}\\
&\phantom{aaaaaaaaa}+12\mathcal{H}\mathcal{H}^\prime\mathcal{H}^{\prime\prime}+8\mathcal{H}^2\mathcal{H}^{\prime\prime\prime}\bigg)\\
\phantom{\left\langle T_{00}^{\delta \phi} \right\rangle^{\rm (6)} (m)=\frac{1}{4\pi^2 a^2}\int dk k^2 }&-\frac{63 a^6 m^6}{16\omega_k^{11}}\left(92\mathcal{H}^6+123\mathcal{H}^4\mathcal{H}^\prime+13\mathcal{H}^2\left(\mathcal{H}^\prime\right)^2+14\mathcal{H}^3\mathcal{H}^{\prime\prime}\right)\\
&+\frac{693a^8m^8}{128\omega_k^{13}}\left(123\mathcal{H}^6+64\mathcal{H}^4\mathcal{H}^\prime\right)-\frac{45045 a^{10}m^{10}\mathcal{H}^6}{128\omega_k^{15}}\Bigg]\\
+\,\frac{\left(\xi-\frac{1}{6}\right)^2}{4\pi^2 a^2}\int dk k^2 &\Bigg[ \frac{9}{8\omega_k^5}\bigg(-4\mathcal{H}^2\left(\mathcal{H}^\prime\right)^2-4\left(\mathcal{H}^\prime\right)^3-8\mathcal{H}^3\mathcal{H}^{\prime\prime}+12\mathcal{H}\mathcal{H}^\prime\mathcal{H}^{\prime\prime}\\
&\phantom{aaaaaa}+\left(\mathcal{H}^{\prime\prime}\right)^2-2\mathcal{H}^\prime\mathcal{H}^{\prime\prime\prime}+2\mathcal{H}\mathcal{H}^{\prime\prime\prime\prime}\bigg)\\
&+\frac{45a^2 m^2}{4\omega_k^7}\bigg(-8\mathcal{H}^4\mathcal{H}^\prime-9\mathcal{H}^2\left(\mathcal{H}^\prime\right)^2+\left(\mathcal{H}^\prime\right)^3-8\mathcal{H}^3\mathcal{H}^{\prime\prime}\\
&\phantom{aaaaaaaaaa}-3\mathcal{H}\mathcal{H}^\prime\mathcal{H}^{\prime\prime}-2\mathcal{H}^2\mathcal{H}^{\prime\prime\prime}\bigg)\\
&+\frac{315 a^4 m^4}{16 \omega_k^9}\bigg( 13\mathcal{H}^6+32\mathcal{H}^4\mathcal{H}^\prime+7\mathcal{H}^2\left(\mathcal{H}^\prime \right)^2+6\mathcal{H}^3\mathcal{H}^{\prime\prime}\bigg)\\
&-\frac{2835 a^6 m^6}{8\omega_k^{11}}\mathcal{H}^4\left(\mathcal{H}^2+\mathcal{H}^\prime\right)\Bigg]\\
+\frac{\left(\xi-\frac{1}{6}\right)^3}{4 \pi^2 a^2}\int dk k^2 &\Bigg[ \frac{9}{2\omega_k^5}\left(-15\mathcal{H}^6+9\mathcal{H}^2\left(\mathcal{H}^\prime\right)^2-6\left(\mathcal{H}\right)^3+18\mathcal{H}^3\mathcal{H}^{\prime\prime}+18\mathcal{H}\mathcal{H}^\prime\mathcal{H}^{\prime\prime}\right)\\
&-\frac{405 a^2 m^2}{2\omega_k^7}\left(\mathcal{H}^6+2\mathcal{H}^4\mathcal{H}^\prime+\mathcal{H}^2\left(\mathcal{H}^\prime\right)^2\right)\Bigg].
\end{split}
\end{equation}
Some of the integrals in \eqref{Eq:QuantumVacuum.EMTFluctuations} are UV-divergent, as we have seen, and others are convergent. The integrals in \eqref{Eq:QuantumVacuum.Tsixthhorder}, instead, are all convergent. One possibility is to compute/regularize every single integral in these formulas (convergent or divergent) using the master formula for DR in \hyperref[Sect:MasterInt]{Sect.\,\ref{Sect:MasterInt}}, but this is not mandatory since an alternative regularization procedure is going to be presented.

Let us consider the ZPE part of the EMT, as given by Eq.\,\eqref{Eq:QuantumVacuum.EMTFluctuations}. We can split it into two parts as follows:
\begin{equation}\label{Eq:QuantumVacuum.DecompositionEMT}
\left\langle T_{00}^{\delta \phi}\right\rangle (M)= \left\langle T_{00}^{\delta \phi}\right\rangle_{\rm Div}(M)+\left\langle T_{00}^{\delta \phi}\right\rangle_{\rm Non-Div}(M), 
\end{equation}
where
\begin{equation}\label{Eq:QuantumVacuum.DivergentPart}
\begin{split}
\left\langle T_{00}^{\delta \phi}\right\rangle_{\rm Div}(M) &\equiv \frac{1}{8\pi^2 a^2}\int dk k^2 \Bigg[ 2\omega_k +\frac{a^2 \Delta^2}{\omega_k}-\frac{a^4\Delta^4}{4\omega_k^3} -\left(\xi-\frac{1}{6}\right)6\mathcal{H}^2\left(\frac{1}{\omega_k}+\frac{a^2 M^2}{\omega_k^3}+\frac{a^2 \Delta^2}{2\omega_k^3}\right) \\
& -\left(\xi-\frac{1}{6}\right)^2\frac{9}{\omega_k^3}(2\mathcal{H}^{\prime\prime}\mathcal{H}-\mathcal{H}^{\prime 2}-3\mathcal{H}^4) \Bigg]
\end{split}
\end{equation}
is the UV-divergent contribution, which involves $\omega_k =\sqrt{k^2+a^2M^2}$ and the powers $1/\omega_k^n$ up to $n=3$.

The terms in \eqref{Eq:QuantumVacuum.EMTFluctuations} which are not in \eqref{Eq:QuantumVacuum.DivergentPart} are the ones which are finite (as they involve powers of $1/\omega_k$ higher than $3$), and constitute the $\langle T_{00}^{\delta \phi}\rangle_{\rm Non-Div}(M)$ part of \eqref{Eq:QuantumVacuum.DecompositionEMT}. Computing the (manifestly convergent) integrals with the help of Eq.\,\eqref{Eq:Conventions.DRFormula} (for $\epsilon=0$) in \hyperref[Sect:MasterInt]{Sect.\,\ref{Sect:MasterInt}}, the final result reads
\begin{equation}\label{Eq:QuantumVacuum.Non-DivergentPart}
\begin{split}
\left\langle T_{00}^{\delta \phi}\right\rangle_{\rm Non-Div}(M) &=\frac{m^2 \mathcal{H}^2}{96\pi^2}-\frac{1}{960\pi^2 a^2}\left( 2\mathcal{H}^{\prime \prime}\mathcal{H}-\mathcal{H}^{\prime 2}-2\mathcal{H}^4\right)\\
&+\frac{1}{16\pi^2 a^2}\left(\xi-\frac{1}{6}\right) \left(2\mathcal{H}^{\prime \prime}\mathcal{H}-\mathcal{H}^{\prime 2}-3\mathcal{H}^4\right)+\frac{9}{4\pi^2 a^2}\left(\xi-\frac{1}{6}\right)^2 (\mathcal{H}^\prime \mathcal{H}^2 +\mathcal{H}^4)\\
&+ \left(\xi-\frac{1}{6}\right)\frac{3\Delta^2 \mathcal{H}^2}{8\pi^2}+\dots
\end{split}
\end{equation}
where the dots in the last expression correspond to higher adiabatic orders. Since the complete adiabatic series is an asymptotic series representation of Eq.\,\eqref{Eq:QuantumVacuum.EMTInTermsOfDeltaPhi}, there is some arbitrariness in the way of choosing the leading adiabatic order because, independently of our choice, such series does not really converge and only serves as an approximation, which is obtained after one cuts the series at some particular order. There is, however, a minimum number of steps to do in order to obtain a meaningful result. To start with, let us set the arbitrary scale $M$ to the on-shell mass value of the quantized scalar field, $M=m$, hence $\Delta=0$. In such a case, the divergent part \eqref{Eq:QuantumVacuum.DivergentPart} reduces to
\begin{equation}\label{Eq:QuantumVacuum.DivergentPartClassic}
\begin{split}
\left\langle T_{00}^{\delta \phi}\right\rangle_{\rm Div}(m)&=\frac{1}{8\pi^2 a^2}\int dk k^2 \Bigg[ 2\omega_k(m) -\left(\xi-\frac{1}{6}\right)6\mathcal{H}^2\left(\frac{1}{\omega_k(m)}+\frac{a^2m^2}{{\omega_k^3(m)}}\right) \\
& -\left(\xi-\frac{1}{6}\right)^2\frac{9}{{\omega_k^3(m)}}(2\mathcal{H}^{\prime\prime}\mathcal{H}-\mathcal{H}^{\prime 2}-3\mathcal{H}^4) \Bigg]\,.
\end{split} 
\end{equation}
Again, \eqref{Eq:QuantumVacuum.DivergentPartClassic} is a bare integral, formally divergent and does not depend on any renormalization scale. It possesses divergences which are similar to the ones of \eqref{Eq:QuantumVacuum.DivergentPart}, but not exactly the same. This is related with our normalization prescription. In fact, what we are going to follow in order to renormalize the ZPE (and, in general, the EMT) is somehow reminiscent of the momentum subtraction scheme, although is certainly different in many respects. In the latter the renormalized Green's functions and running couplings are obtained by subtracting their values at a renormalization point $p^2=M^2$ (space-like in our metric, which becomes an Euclidean point after Wick rotation) or at the time-like one $p^2=-M^2$ (depending on the kinematical region involved)\,\cite{donoghue2014dynamics,Manohar:1996cq}. Since for vacuum diagrams we do not have external momenta, here, instead, we renormalize the ZPE by subtracting the terms that appear up to $4th$ adiabatic order at the arbitrary mass scale $M$. This suffices to eliminate the divergent terms through the ARP, as it is amply discussed in the literature\,\cite{birrell1984quantum,parker2009quantum,fulling1989aspects}.

\subsection{Particular case: ZPE with minimal coupling and in Minkowski spacetime} \label{Sect:ParticularCases}

Before analyzing the renormalization of the ZPE, let us analyze what is the situation in Minkowski space. Suppose we fix the scale $M$ at the physical mass of the particle ($M=m$), so that the $\Delta$-terms vanish. Let us project the UV-divergent terms of order $\cH^2$ and neglect those of higher adiabatic order. The first two adiabatic orders $T_{00}^{\delta \phi (0-2)}\equiv T_{00}^{\delta \phi (0)}+T_{00}^{\delta \phi (2)}$ can be easily identified:
\begin{equation}\label{Eq:QuantumVacuum.T002}
 \left\langle T_{00}^{\delta \phi}\right\rangle^{ (0-2)} (m) =\frac{1}{8\pi^2 a^2}\int dk k^2 \left[ 2\omega_k(m)+\frac{a^4m^4 \mathcal{H}^2}{4\omega_k^5(m)}
-\left(\xi-\frac{1}{6}\right)\left(\frac{6\mathcal{H}^2}{\omega_k(m)}+\frac{6 a^2 m^2\mathcal{H}^2}{\omega_k^3(m)}\right)\right]\,,
\end{equation}
where $\omega_k(m)\equiv \sqrt{k^2+a^2 m^2}$. Let us next project the UV-divergent terms of this formula only and assume that the non-minimal coupling to gravity is absent ($\xi=0$). We are then left with
\begin{equation}\label{Eq:QuantumVacuum.LowOrderZPE1}
\left.\left\langle T_{00}^{\delta \phi} \right\rangle^{ (0-2)}_{\rm Div}(m) \right|_{\xi=0}=\frac{1}{8\pi^2 a^2}\int dk k^2\left( 2\omega_k(m)+\frac{\cH^2}{\omega_k(m)}+\frac{a^2m^2 \cH^2}{\omega_k^3(m)}\right)\,.
\end{equation}
Formula \eqref{Eq:QuantumVacuum.LowOrderZPE1} is in agreement with previous results found in the literature for $\xi=0$, in the ${\cal O}(H^2)$ approximation\,\cite{Parker:1974qw,Fulling:1974zr,fulling1974conformal} -- see also\,\cite{Maggiore:2010wr,Maggiore:2011hw,Hollenstein:2011cz,Bilic:2011zm,Bilic:2010xd}. 
Notice that $k$ is the comoving momentum, whereas the physical momentum is $\tilde{k}=k/a$. Defining the physical energy mode $\tilde\omega_k(m)=\sqrt{\tilde{k}^2+m^2}$, and keeping in mind that $\cH=a H$, we can re-express the above result as
\begin{equation}\label{Eq:QuantumVacuum.LowOrderZPE2}
 \left.\left\langle T_{00}^{\delta \phi}\right\rangle^{ (0-2)}_{\rm Div}(m) \right|_{\xi=0}=\frac{a^2}{4\pi^2 }\int d\tilde{k}\tilde{k}^2\left[\tilde\omega_k(m)+\frac{H^2}{2\tilde\omega_k(m)}\left(1+\frac{m^2}{\tilde\omega_k^2(m)}\right)\right]\,.
\end{equation}
Finally, the Minkowskian spacetime result is obtained for $a=1$ ($H=0)$:
\begin{equation}\label{Eq:QuantumVacuum.Minkowski}
 \left\langle T_{00}^{\delta \phi}\right\rangle^{\rm Mink} (m) =\frac{1}{4\pi^2}\int dk k^2 \omega_k = \int\frac{d^3k}{(2\pi)^3}\,\left(\frac12\,\hbar\,\omega_k\right)\,,
\end{equation}
where $\hbar$ has been restored only in the trailing term for a better identification of the result. The last quantity is the vacuum energy density of the quantum fluctuations in flat spacetime, {\it i.e.} the ZPE in Minkowski spacetime\,\cite{Sola:2013gha, Sola:2014tta, Sola:2011qr, Akhmedov:2002ts}. It is of course the traditional contribution found in usual calculations. It is quartically UV-divergent. Usual attempts to regularize and renormalize this result by e.g. cancelling the corresponding UV-divergence against the bare $\rL$ term in the action \eqref{Eq:QuantumVacuum.EH} within the context of a simple cutoff method or appealing to the Pauli-Villars regularization \cite{pauli1971}, more formal, procedure; or even using the Minimal subtraction (MS) scheme and related ones, leads in all these cases to the well-known ugly fine-tuning problem inherent to the CCP, see \hyperref[Sect:VEDMSS]{Sec.\,\ref{Sect:VEDMSS}} for a summarized discussion. We will certainly not proceed in this way here. We seek out (and will find) an alternative way for renormalizing the above result \eqref{Eq:QuantumVacuum.EMTFluctuations} in its full general form\,\footnote{Let us note that Supersymmetry is not sufficiently helpful for solving the CCP since the cancellation of quartic divergences (warranted e.g. in the Wess-Zumino model\,\cite{Zumino:1974bg}, cf. also\,\cite{Barvinsky:2018lyi}) does not guarantee the cancellation of the subleading ones, e.g. the quadratic divergences\,\cite{Bilic:2011zm}. The quadratic parts are of the form $\Lambda_c^2 H^2$ (where $\Lambda_c$ can serve as a UV cutoff). See\,\cite{Maggiore:2010wr,Maggiore:2011hw,Hollenstein:2011cz,Visser:2016mtr,Donoghue:2020hoh} for a discussion in non-supersymmetric contexts. For $\Lambda_c$ around the Planck mass, it can be phenomenological acceptable only if $\Lambda_c^2 H^2$ carries a small coefficient, as it was noticed much earlier in\,\cite{Shapiro:2000dz,Sola:2007sv}. The current calculation substantiate these results for the first time in a rigorous QFT context, see \hyperref[Sect:RenormalizedVED]{Sec.\,\ref{Sect:RenormalizedVED}}.}.

The previous formulas \eqref{Eq:QuantumVacuum.Minkowski} require appropriate regularization and renormalization. But, the ZPE in FLRW spacetime is even more involved, containing different kinds of divergences as we have seen in Eq.\,\eqref{Eq:QuantumVacuum.EMTFluctuations}, which constitutes a WKB approximation up to $4th$ adiabatic order.

\section{Renormalization of the ZPE in the FLRW background}\label{Sect:RenormZPE}

We may compare the evolving vacuum energy density (VED) of cosmological spacetime with a Casimir device wherein the parallel plates slowly move apart (“expand”)\,\cite{Sola:2014tta, Sola:2011qr}.
While the total vacuum energy density cannot be measured, the `differential' effect associated to the presence of the plates, and then also to their increasing separation with time, it can. Similarly, in the expanding FLRW spacetime there is a genuine non-vanishing spacetime curvature, $R$, as compared to Minkowskian spacetime and such a curvature is changing with the expansion. The VED must vary accordingly and we naturally expect that there is a contribution proportional to $R$, hence to $H^2$ and $\dot{H}$ (plus higher derivative (HD) effects $R^2$, $R^{\mu\nu}R_{\mu\nu}$, etc. in the early Universe). Both spacetimes, Minkowski and FLRW, are obviously similar at short distances, in the sense that the curved background is locally flat. However, the short distance singularities are not really identical since the curvature carries additional ones related to the non-trivial geometric structures.

More formally, the energy-momentum tensor (EMT), Eq.\,\eqref{Eq:QuantumVacuum.EMTScalarField}, is a quadratic functional of the field $\phi(x)$. However, in the context of QFT, $\phi(x)$ is an operator-valued distribution and hence terms like $g_{\mu\nu} m^2\phi^2(x)$, $ \partial_\mu \phi(x) \partial_\nu\phi(x)$, etc. in EMT are not well defined at a given point $x$ since a square of a distribution is not generally defined. This is ultimately the source of the UV-divergences of QFT in configuration space. For this reason it is advisable to consider a bilinear functional replacing the original EMT, which we may denote as $T_{\mu\nu}(x,x')=T_{\mu\nu}(\phi(x)\phi(x'))$, where a point-splitting has been operated in order to avoid the UV-divergence\,\cite{birrell1984quantum,parker2009quantum,fulling1989aspects}. The coincidence point limit in configuration space amounts, of course, to the UV-limit in momentum space. In practice, we need the VEV of that bilinear functional and the point splitting regularization of $\langle T_{\mu\nu}(x,x')\rangle$ is carried out through a (differential) operator ${D}_{\mu\nu}$ acting on an appropriate two-point (Green's, Hadamard's, etc.) function $G(x,x')$ as follows: $\langle T_{\mu\nu}(x,y)\rangle={D}_{\mu\nu} G(x,y)$. The operator ${ D}_{\mu\nu}$ can be easily identified from the terms involved in \eqref{Eq:QuantumVacuum.EMTScalarField} and is usually expressed in a symmetrized form. For instance, the VEV of the first term on the RHS of \eqref{Eq:QuantumVacuum.EMTScalarField} is treated as
\begin{equation}\label{Eq:QuantumVacuum.pointsplit}
 (1-2\xi)\left\langle \nabla_\mu \phi \nabla_\nu\phi\right\rangle\longrightarrow (1-2\xi)\frac12\left(\nabla_\mu\nabla_{\nu^\prime}+\nabla_{\mu^\prime}\nabla_{\nu}\right) G(x,x^\prime)\,,
\end{equation}
where the derivatives with primed indices are assumed to act on $x^\prime$ and those without primes on $x$. In the jargon of QFT, this part would be regularization.
The renormalization of the EMT is then performed by subtracting the vacuum expectation value through the coincidence limit $x\prime\to x$. In the simplest case of Minkowski space, with Minkowskian vacuum $|0\rangle^{\rm Mink}$, it would be natural to define the renormalized EMT operator as
\begin{equation}\label{Eq:QuantumVacuum.RenormEMTpointsplit}
T_{\mu\nu}(x)= \lim\limits_{x^{\prime}\to x} \left[T_{\mu\nu} (x,x^\prime)-\left\langle 0|T_{\mu\nu}(x,x^\prime)|0\right\rangle^{\rm Mink}\right]\,,
\end{equation}
since in this case the VEV of the renormalized EMT is expected to be zero for sound physical reasons. In our Casimir example, the short-distance behavior in the region between the plates is the same as that outside the plates and the limit gives a finite result. However, as warned above, curved spacetime induces new types of infinities as compared to Minkowskian spacetime. The latter, however, are still there and may still carry the core of the quantum vacuum problem if the Minkowskian result is not renormalized to zero (cf. \hyperref[Sect:VEDMSS]{Sect.\,\ref{Sect:VEDMSS}}).

The generalization of \eqref{Eq:QuantumVacuum.RenormEMTpointsplit} in curved spacetime is more delicate, but under appropriate conditions it is natural to use a similar definition where we replace the Minkowskian vacuum $|0\rangle^{\rm Mink}$ with the adiabatic vacuum, simply denoted $|0\rangle$ as we have been doing in the previous sections. We may define the renormalized EMT operator performing a suitable subtraction, but in this case we should not presume a zero result for the VEV of the renormalized EMT. We would rather extract the non-vanishing renormalized vacuum energy density and pressure in curved spacetime as a function of the background itself, in such a way that when the background is Minkowskian we ought to recover the previous vanishing VEV. There are, however, some additional specifications to handle correctly the UV-divergences. Moreover, we wish to provide an off-shell definition enabling us to explore the VED at different scales. Thereby we define the renormalized EMT operator in $n$-dimensional curved spacetime (with $n-1$ spatial dimensions) up to adiabatic order $N\geq n$ through the following off-shell subtraction prescription (which we shall refer to also as the off-shell ARP):
\begin{equation}\label{Eq:QuantumVacuum.RenormEMToperator}
\begin{split}
T^{(0-N)}_{\mu\nu}(x)_{\rm ren}(M) = T^{(0-N)}_{\mu\nu}(x)(m) - \left\langle 0\left| T^{(0-n)}_{\mu\nu}(x) \right|0\right\rangle (M)\,.
\end{split}
\end{equation}
In this equation, $ T^{(0-N)}_{\mu\nu}(x)(M)$ refers to the computation of the renormalized ETM to adiabatic order $N\geq n$ at the scale $M$ (not necessarily equal to the on-shell mass value $m$), whereas $ \langle 0| T^{(0-n)}_{\mu\nu}(x)|0\rangle (M)$ is the VEV of the EMT computed up to adiabatic order $n$ (the dimension of spacetime, {\it i.e.} $n=4$ in our context). The on-shell value is just $T^{(0-N)}_{\mu\nu}(x)(m)$, of course. The subtraction is, therefore, performed upon that on-shell value. By virtue of general covariance, the adiabatic orders involved in the EMT must be even ($N=0,2,4,6,...$). However, irrespective of the adiabaticity order $N$ at which the on-shell value is computed, the subtracted quantity at the scale $M$, {\it i.e.} $T^{(0-n)}_{\mu\nu}(x)(M)$, must include just the first $\frac{n}{2}+1$ (non-vanishing) even orders $N=0,2,4\cdots n$, as these are the only ones which are UV-divergent (in $n$ spacetime dimensions). In $n=4$, this means that $T^{(0-4)}_{\mu\nu}(x)(M)$ must contain the first three even adiabatic orders $N=0,2,4$.

We have mentioned point-splitting regularization\,\cite{DeWitt:1975ys}, see also\,\cite{Christensen:1976vb,Christensen:1978yd,Bunch:1978yq}, because it illustrates very clearly the origin of the UV-divergences in QFT computations and because it is in general a consistent, and physically justified, covariant procedure to define the rnormalized EMT. Proceeding in this way, however, can be rather cumbersome. Indeed, in the general case one has to start from the adiabatic expansion of the Green function $G(x,x')$ and the structure of divergences is not apparent until the mode integral has been performed. Examples are well described in the literature\,\cite{birrell1984quantum,parker2009quantum,fulling1989aspects}. Fortunately, however, the ARP procedure defined above for renormalizing the EMT and other local quantities can be shown to be equivalent to the point-splitting procedure\,\cite{birrell1978application}. In particular, when the field equation can be solved by separation of variables, as in the case under study, one can resort to a simpler method of renormalization which is to perform a mode by mode subtraction process under the integral sign in Eq.\,\eqref{Eq:QuantumVacuum.QuantumFourierModes} using the adiabatic expansion of the modes\,\cite{fulling1989aspects}. In Sec.\,\ref{Sect:EMTScalarField} we have seen that the properly normalized form of these modes is
 \begin{equation}\label{Eq:QuantumVacuum.NormalizedModes}
 u_k(\tau,{\bf x})=(2\pi)^{-3/2}\, a^{-1}(\tau)\,e^{i{\bf k}\cdot{\bf x}}\,\varphi_k(\tau)\,,
 \end{equation}
in which the space and time variables are separated and the time-evolving part $\varphi_k(\tau)$ obeys the non-trivial Eq.\,\eqref{Eq:QuantumVacuum.KGFourier}. Similarly for the equations satisfied by the fluctuating parts, \eqref{Eq:QuantumVacuum.QuantumFourierModes} and \eqref{Eq:QuantumVacuum.KGModes}.
When the field modes can be expressed in separated form it is possible to arrange for the explicit cancellation of UV-divergences before the mode integral is computed. The advantage is clear since the arrangement of terms can be made inside the subtracted integrand such that no UV-divergence is present and the integral appears manifestly convergent \textit{ab initio}.

The VEV of the $00th$-component of $ T^{(0-N)}_{\mu\nu}(x)(m)$ in \eqref{Eq:QuantumVacuum.RenormEMToperator} is precisely given by Eq.\,\eqref{Eq:QuantumVacuum.EMTInTermsOfDeltaPhi} in our case.
Thus, the renormalized vacuum EMT up to ${\cal O}(T^{-N})$ in $n=4$ spacetime dimensions reads
 \begin{equation}\label{Eq:QuantumVacuum.RenormEMTAdiabatic}
\begin{split}
\left\langle 0 \left| T^{(0-N)}_{\mu\nu}(x)\right|0\right\rangle_{\rm ren}(M) &= \left\langle 0\left|T^{(0-N)}_{\mu\nu}(x)\right|0\right\rangle (m) - \left\langle 0\left| T^{(0-4)}_{\mu\nu}(x)\right|0\right\rangle (M) \,,
\end{split}
\end{equation}
where it is supposed, of course, that the mode expansion has been performed to adiabatic order $N$.
Since the EMT structure is made of quadratic expressions of the fields, they are expanded at that order in terms of the above mentioned modes \eqref{Eq:QuantumVacuum.NormalizedModes} and the creation and annihilation operators, and finally one can move to momentum space by integrating $\int d^3k (...) $ the result. The detailed computational results of this procedure have already been given in \hyperref[Sect:AdRegEMT]{Sec.\,\ref{Sect:AdRegEMT}}. Here we just discuss the formal procedure and furnish the practical recipe \eqref{Eq:QuantumVacuum.RenormEMTAdiabatic}, which is necessary to achieve a renormalized finite result using such a mode by mode subtraction at any order.

\subsection{Off-shell renormalization of the EMT}\label{Sect:RenEMToffshell}

It goes without saying that to call Eq.\,\eqref{Eq:QuantumVacuum.RenormEMToperator} `renormalized' EMT and \eqref{Eq:QuantumVacuum.RenormEMTAdiabatic} its VEV is almost unnecessary since the mode by mode subtraction in the integrand makes the integral manifestly finite. The ARP procedure (based on the adiabatic expansion) defines automatically the renormalized quantity. However, as mentioned above, while in the usual adiabatic regularization method\,\cite{birrell1984quantum,parker2009quantum,fulling1989aspects} the subtraction is always performed on-shell, here we shall instead perform the subtraction off-shell, {\it i.e.} at a scale $M$ which is generally different from the mass of the particle. This enables us to test the scale dependence of the renormalized result \eqref{Eq:QuantumVacuum.RenormEMTAdiabatic}. We will use these results to extract physical consequences as to the scale dependence of the ZPE.

If dimensional regularization is used, the needed counterterms to cancel the poles can be generated from the basic three parameters $G^{-1}$, $\rho_\Lambda$ and $\alpha$ appearing in the generalized form of Einstein's equations (compare with the original form \eqref{Eq:QuantumVacuum.EinsteinEq}):
\begin{equation}\label{Eq:QuantumVacuum.MEEs}
\frac{1}{8\pi G} G_{\mu \nu}+\rho_\Lambda g_{\mu \nu}+\alpha\ \leftidx{^{(1)}}{\!H}_{\mu \nu}= T_{\mu \nu}\,.
\end{equation}
Here $\leftidx{^{(1)}}{\!H}_{\mu \nu}$ is the HD tensor which appears from the metric variation of the $R^2$-term in the higher derivative vacuum action for FLRW spacetime. We remind the reader that the term emerging from the variation of the square of the Ricci tensor, called $\leftidx{^{(2)}}{\!H}_{\mu \nu}$, is not necessary in our case since it is not independent of $\leftidx{^{(1)}}{\!H}_{\mu \nu}$ for FLRW spacetimes (see \hyperref[Appendix:Conventions]{Appendix\,\ref{Appendix:Conventions}}).
All three couplings $G^{-1}$, $\rho_\Lambda$ and $\alpha$ are necessary to generate the counterterms that cancel all the divergences in the regularized EMT:
\begin{equation}\label{Eq:QuantumVacuum.splitcounters}
\begin{split}
G^{-1}&= G^{-1}(M)+\delta_\epsilon G^{-1},\\
\rho_\Lambda&=\rho_\Lambda(M)+\delta_\epsilon\rho_\Lambda,\\
\alpha&=\alpha(M)+\delta_\epsilon \alpha.
\end{split}
\end{equation}
The counterterms are denoted with the subscript $\epsilon$ to emphasize that they depend on the regulator $\epsilon$ and become infinite for $\epsilon\to 0$ (see below and \hyperref[Appendix:Dimensional]{Appendix\,\ref{Appendix:Dimensional}}). The subscript is also useful to distinguishing this notation from other quantities introduced in Sec.\,\ref{Sect:RenormalizedVED} which bare some notational resemblance.
The specific forms of the three counterterms mentioned above is:
\begin{equation}\label{Eq:QuantumVacuum.counters}
\begin{split}
 \delta_\epsilon G^{-1} &=-\frac{m^2}{2\pi}\,\left(\xi-\frac{1}{6}\right)\, D_\epsilon\,,\\
 \delta_\epsilon\rho_\Lambda &=+\frac{m^4}{64\pi^2}\, D_\epsilon\,,\\
 \delta_\epsilon \alpha &= -\frac{1}{32\pi^2}\,\left(\xi-\frac{1}{6}\right)^2\, D_\epsilon\,,
 \end{split}
\end{equation}
with
\begin{equation}\label{Eq:QuantumVacuum.Depsilon}
 D_\epsilon\equiv\frac{1}{\epsilon}-\gamma_E+\ln 4\pi=\frac{2}{4-n}-\gamma_E+\ln 4\pi\,.
\end{equation}
The pole is at $n-1=3$ space (resp. $n=4$ spacetime) dimensions, where $\epsilon=0$.
No more counterterms are needed in the present calculation. In particular, we do not need the non-minimal coupling $\xi$ to generate an additional counterterm for the free scalar field theory that we are addressing here. Even so it is useful to keep a non-vanishing value of $\xi$ in the action \eqref{Eq:QuantumVacuum.Sphi} for the general reasons explained in \hyperref[Sect:EMTScalarField]{Sect.\,\ref{Sect:EMTScalarField}} and for more specific ones that we will consider in the coming chapters. Overall the results obtained using the counterterm method and renormalization of constants in the generalized Einstein's equations is identical to that of performing the mode by mode subtraction directly in the integrand until evincing the convergent nature of the integrals. The reader may find the details on the counterterm method in full in \hyperref[Appendix:Dimensional]{Appendix\,\ref{Appendix:Dimensional}}, were a lengthy explanation of this alternative is presented. We have just reminded the reader that in these cases the two procedures are equivalent.

After computing the adiabatic WKB expansion of the integrand of the divergent integrals a subtraction is carried out at an arbitrary scale $M$, {\it i.e.} we apply the off-shell ARP\,\eqref{Eq:QuantumVacuum.RenormEMTAdiabatic}.
Taking into account that in four spacetime dimensions the only adiabatic orders that are divergent in the case of the EMT are the first four ones, the subtraction at the scale $M$ is performed only up to the fourth adiabatic order. The on-shell value of the EMT can be computed of course at any order, all terms beyond $4th$-order being finite.
Let us apply this procedure to the UV-divergent ZPE as given by Eq.\,\eqref{Eq:QuantumVacuum.EMTFluctuations}. 

In view of the previous considerations, we will define the renormalized ZPE in curved spacetime at the scale $M$ as follows:
\begin{equation}\label{Eq:QuantumVacuum.EMTRenormalizedDefinition}
\begin{split}
\left\langle T_{00}^{\delta \phi}\right\rangle_{\rm Ren}(M) & \equiv \left\langle T_{00}^{\delta \phi} \right\rangle(m)-\left\langle T_{00}^{\delta \phi}\right\rangle^{(0-4)}(M)\\
\end{split}
\end{equation}
where we have used the fact that $\langle T_{00}^{\delta \phi}\rangle_{\rm Non-Div}(m)-\langle T_{00}^{\delta \phi}\rangle_{\rm 
Non-Div}^{(0-4)}(M)$ yields precisely the last term of \eqref{Eq:QuantumVacuum.EMTRenormalizedDefinition} before the dots (which represent higher orders), as it follows immediately from \eqref{Eq:QuantumVacuum.Non-DivergentPart}. This subtraction prescription is, of course, equally valid for any component of the EMT, as it is obvious from Eq.\,\eqref{Eq:QuantumVacuum.RenormEMTAdiabatic}. In the above equation and hereafter we omit the adiabaticity order $N$ up to which the EMT is computed. In our context, the spacetime dimension is $n=4$ and hence it is understood that $N\geq 4$. The value $N=4$ is the minimum one which is necessary to perform the renormalization of the EMT, but for some applications we will consider also up to $N=6$. We already presented the 6{\it th} order of the EMT on-shell in eq.\,\eqref{Eq:QuantumVacuum.Tsixthhorder}

To ease the presentation of the result, it proves convenient to use a more explicit notation in order to distinguish explicitly between the off-shell energy mode $\omega_k(M)=\sqrt{k^2+a^2 M^2}$ (formerly denoted just as $\omega_k$) and the on-shell one $\omega_k(m)=\sqrt{k^2+a^2 m^2}$. With this notation, calculations lead to the following result up to fourth adiabatic order:
\begin{equation}\label{Eq:QuantumVacuum.RenormalizedEMT}
\begin{split}
\left\langle T_{00}^{\delta \phi}\right\rangle^{\rm (0-4)}_{\rm ren}(M)&=\frac{1}{8\pi^2 a^2}\int dk k^2 \left[ 2 \left(\omega_k (m)- \omega_k (M)\right)-\frac{a^2 \Delta^2}{\omega_k (M)}+\frac{a^4 \Delta^4}{{4\omega^3_k (M)}}\right]\\
&- \frac{3\left(\xi-\frac{1}{6}\right)\mathcal{H}^2}{4\pi^2 a^2}\int dk k^2 \left[\frac{1}{\omega_k(m)}-\frac{1}{\omega_k (M)}-\frac{a^2 M^2}{{\omega^3_k (M)}}-\frac{a^2 \Delta^2}{2{\omega^3_k (M)}}+\frac{a^2 m^2}{{\omega^3_k (m)}} \right]\\
&- \frac{9\left(\xi-\frac{1}{6}\right)^2\left(2 \mathcal{H}^{\prime \prime}\mathcal{H}-\mathcal{H}^{\prime 2}-3 \mathcal{H}^{4}\right)}{8\pi^2 a^2}\int dk k^2 \left[ \frac{1}{{\omega^3_k (m)}}-\frac{1}{{\omega^3_k (M)}}\right]\\
&-\left(\xi-\frac{1}{6}\right)\frac{3\Delta^2 \mathcal{H}^2}{8\pi^2}.
\end{split}
\end{equation}
Even though some of the individual terms in the integrand of \eqref{Eq:QuantumVacuum.RenormalizedEMT} look formally UV-divergent, one can check upon careful inspection that the overall integral is not, and this explains why the final result is perfectly finite. For instance, the expression under square brackets in the first line of \eqref{Eq:QuantumVacuum.RenormalizedEMT} can be written
\begin{equation}\label{Eq:QuantumVacuum.ExampleConvergence}
2 (\omega_k (m)- \omega_k (M))-\frac{a^2 \Delta^2}{\omega_k (M)}+\frac{a^4 \Delta^4}{{4\omega^3_k (M)}}= \Delta^6 a^6 \frac{\omega_k (m)+3\omega_k (M)}{4\omega^3_k (M)(\omega_k (m)+\omega_k (M))^3}\,,
\end{equation}
where the RHS of the equality goes as $\Delta^6 a^6 \frac{1}{k^5}$ as $k\rightarrow \infty$, so that it is now manifestly UV-convergent. Through these algebraic steps we can write all the integrals in a manifestly convergent way since the power counting for all of them leads to, at most, $\sim \int dk k^{-3}$ in the UV region. After computing these integrals, we arrive to this expression for \eqref{Eq:QuantumVacuum.RenormalizedEMT}
\begin{equation}
\begin{split}\label{Eq:QuantumVacuum.ExplicitRenormalized}
\left\langle T_{00}^{\delta \phi}\right\rangle_{\rm ren}(M)&=\frac{a^2}{128\pi^2 }\left(-M^4+4m^2M^2-3m^4+2m^4 \ln \frac{m^2}{M^2}\right)\\
&-\frac{3 \left(\xi-\frac{1}{6}\right) \mathcal{H}^2 }{16 \pi^2 }\left(m^2-M^2-m^2\ln \frac{m^2}{M^2} \right)\\
&+ \frac{9\left(\xi-\frac{1}{6}\right)^2\left(2 \mathcal{H}^{\prime \prime} \mathcal{H}- \mathcal{H}^{\prime 2}- 3 \mathcal{H}^{4}\right)}{16\pi^2 a^2}\ln \frac{m^2}{M^2}\,+\cdots
\end{split} 
\end{equation}
The dots in the former expression represent higher adiabatic orders. While we want to consider terms of 6{\it th} adiabatic order in our analysis, the divergences are located up to 4{\it th} adiabatic order.

As noted, the same result can be obtained from the counterterm procedure, see \hyperref[Appendix:Dimensional]{Appendix\,\ref{Appendix:Dimensional}}. The counterterms take the precise form \eqref{Eq:QuantumVacuum.splitcounters}, which only depends on the physical mass $m$ of the particle, not on the arbitrary scale $M$. Hence they cancel in the subtraction \eqref{Eq:QuantumVacuum.EMTRenormalizedDefinition}. However, the counterterms can also be used to cancel the poles and write down the generalized Einstein's equations \eqref{Eq:QuantumVacuum.MEEs} fully in terms of finite, renormalized, quantities at the scale $M$, as we shall do in the next section.
Let us emphasize that the last expression in \eqref{Eq:QuantumVacuum.ExplicitRenormalized} is not yet the renormalized vacuum energy density, it is only the renormalized ZPE.

\subsection{The full renormalized ZPE up to $6th$ adiabatic order}\label{Sect:ZPE6th}

The explicit form of the $6th$ adiabatic order is obtained by computing the integrals in the expression\,\eqref{Eq:QuantumVacuum.Tsixthhorder}, which is one of the main objectives of this chapter. In the absence of the $6th$-order terms, the $4th$-order result that we have obtained, Eq.\,\eqref{Eq:QuantumVacuum.RenormalizedEMT}, vanishes on-shell ({\it i.e.} for $M=m$), as it should be expected from the definition itself in \eqref{Eq:QuantumVacuum.RenormalizedEMT}. As a matter of fact, this is the reason why we need to include the next non-vanishing adiabatic order so as to get the first non-vanishing contribution to the on-shell value of the ZPE. The higher order finite effects must satisfy the Appelquist-Carazzone decoupling theorem\,\cite{Appelquist:1974tg} since they must be suppressed for large values of the physical mass $m$ of the quantum field. We may now compute these finite contributions. We may use the master integral formulas given in \hyperref[Sect:MasterInt]{Appendix\,\ref{Sect:MasterInt}} for this, and the final renormalized result computed up to $6th$-order is:
\begin{equation}
\begin{split}\label{Eq:QuantumVacuum.renormalized6th}
\left\langle T_{00}^{\delta \phi}\right\rangle^{\rm (0-6)}_{\rm ren}(M)&=\frac{a^2}{128\pi^2}\left(-M^4+4m^2M^2-3m^4+2m^4\ln \frac{m^2}{M^2}\right)\\
&-\left(\xi-\frac{1}{6}\right)\frac{3 \mathcal{H}^2 }{16 \pi^2 }\left(m^2-M^2-m^2\ln \frac{m^2}{M^2} \right)\\
&+\left(\xi-\frac{1}{6}\right)^2 \frac{9\left(2 \mathcal{H}^{\prime \prime} \mathcal{H}- \mathcal{H}^{\prime 2}- 3 \mathcal{H}^{4}\right)}{16\pi^2 a^2}\ln \frac{m^2}{M^2}\\
&+\frac{1}{20160 \pi^2 a^4m^2}\left(4\left(\mathcal{H}^\prime\right)^3-24\mathcal{H}^3\mathcal{H}^{\prime\prime}-6\mathcal{H}^\prime\mathcal{H}^{\prime\prime\prime}+96\mathcal{H}^4\mathcal{H}^\prime -12\mathcal{H}\mathcal{H}^\prime\mathcal{H}^{\prime\prime}\right.\\
&\left. \phantom{xxxxxxxxxxxxx} +6\mathcal{H}\mathcal{H}^{\prime\prime\prime\prime} +3\left(\mathcal{H}^{\prime\prime}\right)^2-12\mathcal{H}^2\left(\mathcal{H}^\prime\right)^2-24\mathcal{H}^2\mathcal{H}^{\prime\prime\prime}\right)\\
&+\left(\xi-\frac{1}{6}\right)\frac{1}{160\pi^2a^4m^2}\bigg(12\mathcal{H}^3\mathcal{H}^{\prime\prime}-\left(\mathcal{H}^{\prime\prime}\right)^2-40\mathcal{H}^4\mathcal{H}^\prime+2\mathcal{H}^\prime\mathcal{H}^{\prime\prime\prime}-2\mathcal{H}\mathcal{H}^{\prime\prime\prime\prime}\\
&\phantom{xxxxxxxxxxxxxxxxxxx}+10\mathcal{H}^2\left(\mathcal{H}^{\prime}\right)^2+8\mathcal{H}^2\mathcal{H}^{\prime\prime\prime}\bigg)\\
&+\left(\xi-\frac{1}{6}\right)^2\frac{3}{32\pi^2 a^4 m^2}\left(4\mathcal{H}^6+48\mathcal{H}^4\mathcal{H}^\prime-12\mathcal{H}^2\left(\mathcal{H}^\prime\right)^2-16\mathcal{H}^3\mathcal{H}^{\prime\prime}+\left(\mathcal{H}^{\prime\prime}\right)^2\right.\\
&\left.\phantom{xxxxxxxxxxxxxxxxxxx} -8\mathcal{H}^2\mathcal{H}^{\prime\prime\prime} -2\mathcal{H}^\prime\mathcal{H}^{\prime\prime\prime}+2\mathcal{H}\mathcal{H}^{\prime\prime\prime\prime}\right)\\
&-\left(\xi-\frac{1}{6}\right)^3\frac{9}{8\pi^2a^4m^2}\left(\mathcal{H}^2+\mathcal{H}^\prime\right)\left(11\mathcal{H}^4+\mathcal{H}^2\mathcal{H}^\prime+
2\left(\mathcal{H}^\prime\right)^2-6\mathcal{H}\mathcal{H}^{\prime\prime}\right)\,.
\end{split}
\end{equation}
The first three lines of this expression embody just the $4th$-order renormalized result. We can easily convince ourselves that the remaining terms of \eqref{Eq:QuantumVacuum.renormalized6th} are of $6th$ adiabatic order, {\it i.e.} ${\cal O}(T^{-6})$, where we are using the notation introduced at the beginning of \hyperref[Sect:AdRegEMT]{Sect.\,\ref{Sect:AdRegEMT}} for the adiabaticity order. Moreover, they are all suppressed by two powers of the particle's mass $m$, {\it i.e.} they fall off as $\sim 1/m^2$, or to be more precise as ${\cal O}(T^{-6})/m^2$. Thus, as formerly announced, the $6th$-order terms satisfy the Appelquist-Carazzone decoupling theorem\footnote{An alternative way to express this decoupling result for large $m$ is to say that in the opposite limit ($m\to 0$) the higher order adiabatic terms beyond $N>4$ (all of them of even order owing to covariance, $N=6,8,...$) are infrared divergent for $m\to 0$ in four dimensions. This is a well-known behavior expected from the effective action\,\cite{birrell1984quantum}, which e.g. can be immediately appraised in the explicit form of the $6th$ order adiabatic integral \eqref{Eq:QuantumVacuum.Tsixthhorder}. Although we are not affected by IR effects ($m$ is in our case very large), the IR limit of ARP must be treated with care\,\cite{Animali:2022lig}.} for large $m$\,\cite{Appelquist:1974tg}. The next adiabatic order would be the $8th$ one. These terms also fulfill the decoupling theorem and are further suppressed as ${\cal O}(T^{-8})/m^4$. We shall not be concerned with them.

We should also note that the $6th$-order terms do not depend on the arbitrary mass scale $M$, but only on the mass of the particle, $m$. The reason is that $M$ enters only the terms up to adiabatic order $4$, which are the only ones which are subtracted (because they are the only ones which are originally UV-divergent), as it is obvious from the definition \eqref{Eq:QuantumVacuum.EMTRenormalizedDefinition}. As a consequence, the on-shell value of \eqref{Eq:QuantumVacuum.renormalized6th} is now non-vanishing and it is exclusively determined by the higher order adiabatic terms beyond ${\cal O}(T^{-4})$. It is convenient to express the result in terms of the ordinary Hubble rate defined in terms of the cosmic time, $H(t)$, with $\mathcal{H}(\tau)=a H(t)$. We may now use the conversion relations between the derivatives of $\mathcal{H}$ with respect to the conformal time and the derivatives of $H$ with respect to the cosmic time (see \hyperref[Appendix:Conventions]{Appendix\,\ref{Appendix:Conventions}}). After some algebra we find the following expression for the renormalized on-shell value ($M=m$) at 6th adiabatic order, {\it i.e.} at ${\cal O}(T^{-6})$ (which, we stress again, is the first and hence the leading non-vanishing order in the on-shell case):
\begin{equation}\label{Eq:QuantumVacuum.OnshellHigher}
\begin{split}
\left\langle T_{00}^{\delta \phi}\right\rangle^{(6)}_{\rm ren}(m)&=
\frac{a^2}{20160 \pi^2 m^2}\left(-8H^6-36H^4\dot{H}-20\dot{H}^3+42H^3\ddot{H}+3\ddot{H}^2-6\dot{H}\vardot{3}{H}\right.\\
&\phantom{XXXXXXXX}\left.+84H^2\dot{H}^2+36H^2\vardot{3}{H}+60H\dot{H}\ddot{H}+6H\vardot{4}{H}\right)\\
&+\left(\xi-\frac{1}{6}\right)\frac{a^2}{160\pi^2m^2}\left(2H^6+12H^4\dot{H}+8\dot{H}^3-14H^3 \ddot{H}-\ddot{H}^2+2\dot{H}\ddot{H}-34H^2\dot{H}^2\right.\\
&\phantom{XXXXXXXXXXX}\left.-12H^2\vardot{3}{H}-24H\dot{H}\ddot{H}-2H\vardot{4}{H}\right)\\
&+\left(\xi-\frac{1}{6}\right)^2\frac{3 a^2}{32\pi^2 m^2}\left(-24H^4\dot{H}-8\dot{H}^3+10H^3\ddot{H}+\ddot{H}^2-2\dot{H}\vardot{3}{H}+32H^2\dot{H}^2\right.\\
&\phantom{XXXXXXXXXXX}\left.+12H^2\vardot{3}{H}+24H\dot{H}\ddot{H}+2H\vardot{4}{H}\right)\\
&-\left(\xi-\frac{1}{6}\right)^3\frac{9 a^2}{8\pi^2 m^2}\left(2H^2+\dot{H}\right)\left(2H^4-19H^2\dot{H}+2\dot{H}^2-6H\ddot{H}\right)\,.
\end{split}
\end{equation}

\section{Renormalized vacuum energy density}\label{Sect:RenormalizedVED}

Renormalization theory is concerned with the relations of renormalized couplings, operators and Green's functions at different renormalization points. It is not our intention to compute any of these quantities from first principles, in particular the VED. Ultimately this is an input from experiment at a given scale, and once it is given one can predict its value at another scale.

It is also important to remark that, in order to make possible the renormalization program in the context of QFT in curved spacetime, we need to count on the higher derivative (HD) terms in the classical effective action of vacuum\,\cite{birrell1984quantum}, in addition to the usual Einstein-Hilbert (EH) term with a cosmological constant, $\CC$. In four dimensions, the HD part is composed of the ${\cal O}(R^2)$ terms, {\it i.e.} the squares of the curvature and Ricci tensors: $R^2$ and $R_{\mu\nu}R^{\mu\nu}$. No more HD terms are needed in our case since the one associated to the square of the Riemann tensor, $R_{\mu\nu\rho\sigma}R^{\mu\nu\rho\sigma}$, is not independent owing to the topological nature of the Euler's density in $4$ dimensions, which involves all these HD terms together: $E=R_{\mu\nu\rho\sigma}R^{\mu\nu\rho\sigma}-4R_{\mu\nu}R^{\mu\nu}+R^2$. Moreover the square of the Weyl tensor, $C^2=R_{\mu\nu\rho\sigma}R^{\mu\nu\rho\sigma}-2R_{\mu\nu}R^{\mu\nu}+(1/3)\,R^2$, exactly vanishes for conformally flat spacetimes such as FLRW. The full action, therefore, boils down to the relevant EH+HD terms mentioned above plus the matter part (consisting here of the scalar field $\phi$ only) with a non-minimal coupling to gravity, Eq.\,\eqref{Eq:QuantumVacuum.Sphi}. Its variation of the action with respect to the metric provides the modified Einstein's equations, which become extended as compared to their original form \eqref{Eq:QuantumVacuum.EinsteinEq} as follows:
\begin{equation}\label{Eq:QuantumVacuum.MEEsHD}
\Mpl^2(M) G_{\mu \nu}+\rho_\Lambda (M) g_{\mu \nu}+\alpha(M)\ \leftidx{^{(1)}}{\!H}_{\mu\nu}= \left\langle T_{\mu\nu}^{\delta \phi}\right\rangle_{\rm ren}(M)\,.
\end{equation}
where we use renormalized quantities and hence we have indicated explicitly the dependence of the parameters and of the EMT on the subtraction point $M$. The background (classical) part of the EMT does not depend on it. Baring in mind that we wish to relate the theory at different renormalization points we can safely ignore any classical contribution present in \eqref{Eq:QuantumVacuum.MEEsHD} since it will cancel when doing the subtraction.

The higher order tensor $H_{\mu \nu}^{(1)}$ is obtained by functionally differentiating $R^2$ with respect to the metric (see \hyperref[Appendix:Conventions]{Appendix\,\ref{Appendix:Conventions}}). A further simplification is possible here since the corresponding term associated to the functional differentiation of the square of the Ricci tensor, called $H_{\mu \nu}^{(2)}$, is not necessary since it is not independent of $H_{\mu \nu}^{(1)}$ for FLRW spacetimes\,\cite{birrell1984quantum}. This follows from the aforementioned properties of the Euler density and the Weyl tensor for conformally flat spacetimes. The higher order tensor $H_{\mu \nu}^{(1)}$ is indeed to be included in the extended field equations since it is anyway generated by the quantum fluctuations and is therefore indispensable for the renormalizability of the theory.

The fact that Eq.\,\eqref{Eq:QuantumVacuum.MEEsHD} has been written with all couplings defined at some arbitrary mass scale $M$ is because we have shown that the EMT used in our calculation is the renormalized one at that scale following the ARP. However, in the \hyperref[Appendix:Dimensional]{Appendix\,\ref{Appendix:Dimensional}} we offer an alternative approach based on the more conventional counterterm procedure, starting from the bare parameters of the action.

\subsection{Vacuum energy density at different scales. Absence of $\sim m^4$ terms.}\label{Sect:TotalVED}

The renormalized expression for the vacuum fluctuations, $\langle T_{\mu\nu}^{\delta \phi}\rangle_{\rm ren}(M)$, is not yet the final one for the renormalized VED. As indicated in \eqref{Eq:QuantumVacuum.EMTvacuum}, the latter is obtained upon including the contribution from the $\rL$-term in the Einstein-Hilbert action \eqref{Eq:QuantumVacuum.EH}. This term is initially a bare quantity, as in the traditional counterterm method with Eq.\,\eqref{Eq:QuantumVacuum.splitcounters}, but we take its renormalized value at the same scale $M$. Therefore, the renormalized vacuum EMT at the scale $M$ is given by
\begin{equation}\label{Eq:QuantumVacuum.RenEMTvacuum}
\left\langle T_{\mu\nu}^{\rm vac} \right\rangle_{\rm ren}(M)=-\rho_\Lambda (M) g_{\mu \nu}+\left\langle T_{\mu \nu}^{\delta \phi}\right\rangle_{\rm ren}(M)\,.
\end{equation}
For the considerations in this section we will use only the renormalized expressions up to $4th$ adiabatic order, since these suffice to discuss the renormalization of the EMT. The renormalized VED is obtained from extracting the $00th$-component of the expression \eqref{Eq:QuantumVacuum.RenEMTvacuum} is
\begin{equation}\label{Eq:QuantumVacuum.RenVDE}
\rv(M)= \frac{\left\langle T_{00}^{\rm vac}\right\rangle_{\rm ren}(M)}{a^2}=\rho_\Lambda (M)+\frac{\left\langle T_{00}^{\delta \phi}\right\rangle_{\rm ren}(M)}{a^2}\,,
\end{equation}
where we have used the fact that $g_{00}=-a^2$ in the conformal metric that we are using. The above equation stems from treating the vacuum as a perfect fluid, namely with an EMT of the form
\begin{equation}\label{Eq:QuantumVacuum.VacuumIdealFluid}
\left\langle T_{\mu\nu}^{{\rm vac}}\right\rangle=\Pv g_{\mu \nu}+\left(\rv+\Pv\right)u_\mu u_\nu\,,
\end{equation}
where $u^\mu$ is the $4$-velocity. In conformal coordinates in the comoving cosmological (FLRW) frame, $u^\mu=(1/a,0,0,0)$ and hence $u_\mu=(-a,0,0,0)$. By the moment, there are no clues pointing out any relation between vacuum's pressure and vacuum's energy density. Nevertheless, taking the $00th$-component of \eqref{Eq:QuantumVacuum.VacuumIdealFluid}, the relation $\langle T_{00}^{{\rm vac}}\rangle=-a^2 \Pv+\left(\rv+\Pv\right) a^2=a^2\rho_{\rm vac}$ follows, irrespective of $\Pv$. Finally, inserting the $00th$-component of \eqref{Eq:QuantumVacuum.RenEMTvacuum} into \eqref{Eq:QuantumVacuum.VacuumIdealFluid} we obtain Eq.\,\eqref{Eq:QuantumVacuum.RenVDE}, as desired. Equating the last expression to Eq.\,\eqref{Eq:QuantumVacuum.EMTvacuum}, and taking the $00$-component of the equality (keeping also in mind that $g_{00}=-a^2(\eta)$ in the conformal frame), we obtain
\begin{equation}\label{Eq:QuantumVacuum.Totalrhovac}
\rho_{\rm vac}(M)=\rho_{\Lambda}(M)+\frac{\left\langle T_{00}^{\delta \phi}\right\rangle_{\rm ren}(M )}{a^2}.
\end{equation}
Notice that we distinguish between VED and ZPE. When they are both renormalized quantities, the sum \eqref{Eq:QuantumVacuum.RenVDE} provides the physically measurable quantity which includes the dependence on the renormalization point since we are using the renormalized theory at that scale. The total VED at an arbitrary scale $M$ is then the sum of the renormalized contributions from the cosmological term plus that of the quantum fluctuations of the scalar field at that scale ({\it i.e.} the renormalized ZPE). In a symbolic way, we can write
\begin{equation}\label{Eq:QuantumVacuum.Symbolic}
 {\rm VED}=\rL+{\rm ZPE}\,.
\end{equation}
More explicitly, we can write it out on taking cognizance of the important result presented in Eq.\,\eqref{Eq:QuantumVacuum.ExplicitRenormalized}:
\begin{equation}\label{Eq:QuantumVacuum.RenVDEexplicit}
\begin{split}
\rv(M)&= \rho_\Lambda (M)+\frac{1}{128\pi^2 }\left(-M^4+4m^2M^2-3m^4+2m^4 \ln \frac{m^2}{M^2}\right)\\
&+\left(\xi-\frac{1}{6}\right)\frac{3 \mathcal{H}^2 }{16 \pi^2 a^2}\left(M^2-m^2+m^2\ln \frac{m^2}{M^2} \right)\\
& +\left(\xi-\frac{1}{6}\right)^2 \frac{9\left(2 \mathcal{H}^{\prime \prime} \mathcal{H}- \mathcal{H}^{\prime 2}- 3 \mathcal{H}^{4}\right)}{16\pi^2 a^4}\ln \frac{m^2}{M^2}+\cdots
\end{split}
\end{equation}
Here the dots denote that we have not written higher order adiabatic orders beyond the 4th one. It is important to remark that these terms do not depend on the renormalization parameter $M$. The simplified notation $\rho_{\rm vac}(M)$ should not obscure the fact that the VED in FLRW spacetime is dynamical, as it rests on the expansion rate of the Universe and its derivatives, apart from the scale $M$,
{\it i.e.} $\rv(M)\equiv\rv(M,\cH, \cH',\cH'',...)$. We note that $M$ itself is dynamical in cosmology since we will associate $M$ with a cosmological variable. Thus, the dynamical character of the VED enters both through the explicit dependence in the Hubble rate and also implicitly through $M$ (cf. \hyperref[Sect:RunningConnection]{Sect.\,\ref{Sect:RunningConnection}}).

Now it is time to relate different renormalization points. Subtracting the renormalized result at two scales, $M$ and $M_0$, we find:
\begin{equation}\label{Eq:QuantumVacuum.VEDscalesMandM0}
\begin{split}
\rv(M)-\rv(M_0)&=\frac{\left\langle T_{00}^{\rm vac}\right\rangle_{\rm ren}(M)-\left\langle T_{00}^{\rm vac}\right\rangle_{\rm ren}(M_0)}{a^2}\\
&=\rL(M)-\rL(M_0)+\frac{\left\langle T_{00}^{\delta \phi}\right\rangle_{\rm ren}(M)-\left\langle T_{00}^{\delta \phi}\right\rangle_{\rm ren}(M_0)}{a^2}\,,
\end{split}
\end{equation}
where
\begin{equation}\label{Eq:QuantumVacuum.RenT00vacuumMM0}
\begin{split}
\left\langle T_{00}^{\delta \phi}\right\rangle_{\rm ren}(M)-\left\langle T_{00}^{\delta \phi}\right\rangle_{\rm ren}(M_0)&=-\frac{a^2}{128\pi^2}\left(M^4-M_0^{4}-4m^2(M^2-M_0^{2})+2m^4\ln \frac{M^{2}}{M_0^2}\right)\\
&+\frac{3\left(\xi-\frac16\right)\cH^2}{16\pi^2}\,\left(M^2 - M_0^{2} -m^2\ln \frac{M^{2}}{M_0^2}\right)\\
&+\frac{9\left(\xi-\frac16\right)^2}{16 \pi^2 a^2}\left(\mathcal{H}^{\prime 2}-2\mathcal{H}^{\prime \prime}\mathcal{H}+3 \mathcal{H}^4 \right)\ln \frac{M^2}{M_0^{2}}\,.\\
\end{split}
\end{equation}
To account for this difference we have just used the $4th$-order form \eqref{Eq:QuantumVacuum.ExplicitRenormalized} since, as noticed, the $6th$ adiabatic contribution does not carry along any new dependency on the scale $M$ (it only adds up new contributions which depend on the mass $m$); and hence all the higher order adiabatic effects beyond $4th$-order cancel out when one is just relating scales rather than computing the result at a given scale. The same subtraction can be performed using the generalized Einstein's equations \eqref{Eq:QuantumVacuum.MEEs}. We take now these equations in vacuo and in terms of the renormalized couplings at the scale $M$. 

Apart from $\rL(M)$ and $\alpha(M)$ we have defined another renormalized coupling at the scale $M$,
\begin{equation}\label{Eq:QuantumVacuum.RenCouplings}
\Mpl^2(M)=\frac{G^{-1}(M)}{8\pi}\,,
\end{equation}
which is nothing but the reduced Planck mass squared at that scale. Its relation with the ordinary Planck mass is $\Mpl (M)=\mpl (M) /\sqrt{8\pi}$. As we shall further discuss in what follows (see also \hyperref[Appendix:Abis]{Appendix\,\ref{Appendix:Abis}}), the setting $M=H$ is the most appropriate one to make contact between the renormalized value of a parameter and its physical value at the epoch $H$. In accordance with this prescription, the measured local value of gravity, $G_N$, is obtained when $M$ is set to the current value of the Hubble parameter, {\it i.e.} $G(H_0)\equiv G_N=1/\mpl^2$, where $\mpl\simeq 1.2\times10^{19}$ GeV. Performing the subtraction of the $00th$-component of \eqref{Eq:QuantumVacuum.MEEsHD} at the two scales $M$ and $M_0$, we find:
\begin{equation}\label{Eq:QuantumVacuum.RenT00vacuumMM0Eeqs}
\begin{split}
\left\langle T_{00}^{\delta \phi}\right\rangle_{\rm ren}(M)-\left\langle T_{00}^{\delta \phi}\right\rangle_{\rm ren}(M_0)&=- a^2\left(\rL(M)-\rL(M_0)\right)+\left(\Mpl^2(M)-\Mpl^2(M_0)\right)G_{00}\\
&+\left(\alpha(M)-\alpha(M_0)\right)\leftidx{^{(1)}}{\!H}_{00}\,,
\end{split}
\end{equation}
where in the first line we have used once more that $g_{00}=-a^2$. Comparison between \eqref{Eq:QuantumVacuum.RenT00vacuumMM0} and \eqref{Eq:QuantumVacuum.MEEsHD} yields the important relation
\begin{equation}\label{Eq:QuantumVacuum.SubtractionrL}
\delta\rL(m,M,M_0)\equiv\rL(M)-\rL(M_0)=\frac{1}{128\pi^2}\left(M^4-M_0^{4}-4m^2(M^2-M_0^{2})+2m^4\ln \frac{M^{2}}{M_0^2}\right).
\end{equation}
Upon using the known form of $G_{00}$ and $\leftidx{^{(1)}}{\!H}_{00}$ in the conformal metric (given in \hyperref[Appendix:Conventions]{Appendix\,\ref{Appendix:Conventions}}) we collect also the two relations:
\begin{equation} \label{Eq:QuantumVacuum.SubtractionMPl}
\delta\Mpl^2(m,M,M_0)\equiv\Mpl^2(M)-\Mpl^2(M_0)=\left(\xi-\frac{1}{6}\right)\frac{1}{16\pi^2}\left[M^2 - M_0^{2} -m^2\ln \frac{M^{2}}{M_0^2}\right]
\end{equation}
and\,\footnote{The scale shifts quoted in equations \eqref{Eq:QuantumVacuum.SubtractionrL}-\eqref{Eq:QuantumVacuum.Subtractionalpha} are finite quantities in our renormalization scheme and should not be confused with counterterms, such as those in \eqref{Eq:QuantumVacuum.splitcounters}-\eqref{Eq:QuantumVacuum.counters}. Strictly speaking, we do not need counterterms in the ARP since we perform a subtraction of UV-divergent quantities at two scales, and this renders a finite result.}
\begin{equation} \label{Eq:QuantumVacuum.Subtractionalpha}
\delta\alpha(M,M_0)\equiv\alpha(M)-\alpha(M_0)= -\frac{1}{32\pi^2}\left(\xi-\frac{1}{6}\right)^2 \ln \frac{M^2}{M_0^{2}}\,.
\end{equation}
These relations are important not only because they furnish the scaling laws of the couplings $\Mpl^2(M)$ and $\alpha(M)$ in the modified Einstein's equations, but also because they help to properly identify the various terms in Eq.\,\eqref{Eq:QuantumVacuum.RenT00vacuumMM0Eeqs}. In particular they contribute to isolate the shift of the renormalized vacuum parameter $\rL(M)$, Eq.\,\eqref{Eq:QuantumVacuum.SubtractionrL}. Using \eqref{Eq:QuantumVacuum.SubtractionrL} we can rewrite \eqref{Eq:QuantumVacuum.RenT00vacuumMM0} in the following form
\begin{equation}\label{Eq:QuantumVacuum.RenT00vacuumMM0rhos}
\begin{split}
\frac{\left\langle T_{00}^{\delta \phi}\right\rangle_{\rm ren}(M)-\left\langle T_{00}^{\delta \phi}\right\rangle_{\rm ren}(M_0)}{a^2}&=-\delta\rL(m,M,M_0)\\
&+\left(\xi-\frac16\right)\frac{3\cH^2}{16\pi^2 a^2}\,\left(M^2 - M_0^{2} -m^2\ln \frac{M^{2}}{M_0^2}\right)\\
&+\left(\xi-\frac16\right)^2\frac{9}{16 \pi^2 a^4}\left(\mathcal{H}^{\prime 2}-2\mathcal{H}^{\prime \prime}\mathcal{H}+3 \mathcal{H}^4 \right)\ln \frac{M^2}{M_0^{2}}\,.\\
\end{split}
\end{equation}
Finally, on combining equations \eqref{Eq:QuantumVacuum.VEDscalesMandM0} and \,\eqref{Eq:QuantumVacuum.RenT00vacuumMM0rhos} we see that the expression $\delta\rL(m,M,M_0)$ exactly cancels and we are left with
\begin{equation}\label{Eq:QuantumVacuum.VEDscalesMandM0Final}
\begin{split}
\rv(M)&=\rv(M_0)+\left(\xi-\frac16\right)\frac{3\cH^2}{16\pi^2 a^2}\,\left(M^2 - M_0^{2} -m^2\ln \frac{M^{2}}{M_0^2}\right)\\
&+\left(\xi-\frac16\right)^2\frac{9}{16 \pi^2 a^4}\left(\mathcal{H}^{\prime 2}-2\mathcal{H}^{\prime \prime}\mathcal{H}+3 \mathcal{H}^4 \right)\ln \frac{M^2}{M_0^{2}}\\\
&=\left(\xi-\frac{1}{6}\right) \frac{3H^2 }{16\pi^2}\,\left[M^2 - M_0^{2}-m^2\ln \frac{M^{2}}{M_0^2}\right]\\
&+\left( \xi-\frac{1}{6}\right)^2\frac{9}{16\pi^2}\, \left(\dot{H}^2 - 2 H\ddot{H} - 6 H^2 \dot{H} \right)\ln \frac{M^2}{M_0^2}\,,\\
\end{split}
\end{equation}
where in the second equality we have written the result in terms of the Hubble rate in the ordinary cosmic time, and for this reason it does not depend explicitly on the scale factor. The formal quantity $\rL(M)-\rL(M_0)$ indeed never shows up physically and hence only the difference of VED's at the two renormalization points $M$ and $M_0$ remains, Eq.\,\eqref{Eq:QuantumVacuum.VEDscalesMandM0Final}. The exact cancellation of the quantity \eqref{Eq:QuantumVacuum.SubtractionrL} in Eq.\,\eqref{Eq:QuantumVacuum.VEDscalesMandM0Final} is very important since such a term is precisely the potentially conflicting quantity carrying all of the awkward quartic powers of the mass scales. Its remarkable cancellation in our renormalization setup, however, shows that the values of $\rv(M)$ and $\rv(M_0)$ differ only by a small quantity proportional to $H^2$ and another which is of ${\cal O}(H^4)$, both small in the current Universe (the latter being utterly irrelevant for the entire FLRW regime). Such a physically measurable quantity is a smooth function $\sim m^2 H^2$ of the cosmic evolution. Thus, no fine-tuning is needed to relate $\rv(M)$ and $\rv(M_0)$ in the present renormalization framework. This is also true in the special case of Minkowskian spacetime, where neither $\rL(M)$ nor the ZPE can be measured in an isolated way, just the sum, which in this case is exactly zero (see next section). The foregoing considerations show that, in the context of the running vacuum model, the dark energy that we observe is just the (non-constant)vacuum energy density predicted within QFT in FLRW spacetime. At any cosmic time $t$ characterized by $H(t)$ there is a (different) `CC' term $\CC(H)=8\pi G_N\rv(H)$ acting (approximately) as a cosmological constant for a long period around that time, but there is no true CC valid at all times!

The result \eqref{Eq:QuantumVacuum.VEDscalesMandM0Final} is the value of the VED at the scale $M$ once we know its value at another scale $M_0$, {\it i.e.} it expresses the `running' of the VED. Only in the case of conformally invariant fields ($\xi=1/6$) the result would be the same at all scales, if the VED received only contributions from scalar fields. But in general, this is not the case since one has to add the contribution from fermions and vector boson fields, which we do not consider here, so in general the total VED appears a running quantity with the expansion. As mentioned, the running is slow for small $H$, as it depends on terms of the form ${\cal O}(H^2)$ times a mass scale squared, and on ${\cal O}(H^4)$ contributions, but not on quartic mass scales. We repeat once more that $\rho_{\rm vac}(M)$ cannot be computed from the above expression, even if the expansion rate $H$ is known at a given time since we do not know the value of the renormalized parameter $\rL(M)$ in the action. As usual, we need some experimental input, we will comment more on this fact in \hyperref[Sect:RunningConnection]{Sect.\,\ref{Sect:RunningConnection}}. We should not forget that the main aim of renormalization theory is not so much to predict the value of a quantity, for example the VED, at a certain scale (and time, in cosmology) but to relate it at different scales or renormalization points, and hence to account for its evolution with the scale $M$.

\subsection{Equivalent approach: subtracting the Minkowskian contribution}\label{Sect:SubtractMinkowski}

It cannot be overstated that the above result \eqref{Eq:QuantumVacuum.VEDscalesMandM0Final} is free from quartic powers of the masses. These would still be present if we had subtracted just the ZPE part at different scales without including the renormalized $\rL$. This is obvious from Eq.\,\eqref{Eq:QuantumVacuum.RenT00vacuumMM0}, where we can see that the problem actually stems from Minkowskian spacetime, see\,\cite{Sola:2013gha, Sola:2014tta, Sola:2011qr} for a discussion. The renormalized ZPE in flat spacetime is obtained from Eq.\,\eqref{Eq:QuantumVacuum.RenT00vacuumMM0} in the limit $a=1$ (which implies that $\mathcal{H}$ and all its derivatives are zero). Only the first term of it remains, although it is the one carrying the mentioned quartic powers. This term vanishes for $M=m$ since the renormalized on-shell value was computed only up to fourth adiabatic order. As previously emphasized (cf. \hyperref[Sect:RenEMToffshell]{Sect.\,\ref{Sect:RenEMToffshell}}), this does not mean that the exact renormalized ZPE in curved spacetime vanishes on-shell, of course. One still has to add the higher order adiabatic terms, but they are finite and subleading since they decouple for larger and larger values of the physical mass $m$ ({\it i.e.} they satisfy the decoupling theorem\,\cite{Appelquist:1974tg}), and we have not tracked them explicitly. Our main aim here is to pick out just the leading contributions to the renormalized ZPE up to $4th$ adiabatic order. Since we compute the total VED, defined as the sum of the renormalized value of $\rL$ and the renormalized ZPE, the difference of VED values at two scales is free from the quartic powers of mass scales. Of course this is possible owing to the renormalized form for the ZPE that we have used, Eq.\,\eqref{Eq:QuantumVacuum.EMTRenormalizedDefinition}, which involves a subtraction of the on-shell value at another arbitrary mass scale. In the \hyperref[Appendix:Dimensional]{Appendix\,\ref{Appendix:Dimensional}}, we provide an alternative calculation leading to the same result \eqref{Eq:QuantumVacuum.VEDscalesMandM0Final} and further comments on this fact.

In Minkowski space we should expect zero vacuum energy, as in such a case we can apply the normal ordering of the quantum operators in the canonical formalism. In our context we encounter the same result. To start with, the renormalized ZPE in Minkowski space is the value of $\langle T_{00}^{\delta \phi}\rangle_{\rm ren}(M)$, given by Eq.\,\eqref{Eq:QuantumVacuum.ExplicitRenormalized} for $a=1$ and $\cH=0$:
\begin{equation}\label{Eq:QuantumVacuum.ZPEMinkowski}
\begin{split}
\left\langle T_{00}^{\delta \phi}\right\rangle_{\rm ren}^{\rm Mink}(M)=\frac{1}{128\pi^2 }\left(-M^4+4m^2M^2-3m^4+2m^4 \ln \frac{m^2}{M^2}\right)\,.
\end{split}
\end{equation}
However, this quantity is purely formal and does not appear in physical results since it cancels exactly against $\rL(M)$. The correct renormalized VED result can be viewed as subtracting the Minkowskian contribution in analogy with the Casimir effect. Namely, one expects that if we compute the expression for $\langle T_{\mu\nu}^{\rm vac}\rangle$ in Minkowskian spacetime and subtract it from its equivalent in curved spacetime the result should depend only on the curvature of the latter and hence evolve only mildly with the cosmic evolution through a function of the Hubble rate (which is the key term providing the departure of the FLRW background from Minkowskian spacetime)\,\cite{Sola:2013gha,Sola:2014tta,Sola:2011qr}. In fact, the subtraction of the Minkowskian spacetime result has been argued from different perspectives\,\cite{Maggiore:2010wr,Maggiore:2011hw,Hollenstein:2011cz,Bilic:2011zm,Bilic:2010xd}. In the Minkowski limit, the subtraction of scales in\,\eqref{Eq:QuantumVacuum.RenT00vacuumMM0Eeqs} leaves only the term $\delta{\rho_\CC}(m,M,M_0) g_{\mu \nu}=\delta{\rho_\CC}(m,M,M_0)\eta_{\mu\nu}$ on its RHS. Taking the $00$-component (with $\eta_{00}=-1$ in our conventions), we find
\begin{equation}\label{Eq:QuantumVacuum.MinkowskiSubtracted}
\begin{split}
\left\langle T_{00}^{\delta \phi}\right\rangle_{\rm ren}^{\rm Mink}(M)-\left\langle T_{00}^{\delta \phi}\right\rangle_{\rm ren}^{\rm Mink}(M_0)& = -\delta{\rho_\CC}(m,M,M_0)\\
&=-\frac{1}{128\pi^2}\left(M^4-M_0^{4}-4m^2(M^2+M_0^{2})+2m^4\ln \frac{M^{2}}{M_0^2}\right)\,.
\end{split}
\end{equation}
Following the above proposal, we define now the physical VED in the expanding Universe as the outcome of subtracting the Minkowskian ZPE from its value in FLRW spacetime:
\begin{equation}\label{Eq:QuantumVacuum.VESubtractedMinkowski}
\begin{split}
&\rho_{\rm vac}(M)\equiv\frac{\left\langle T_{00}^{\delta \phi}\right\rangle_{\rm ren}( M )}{a^2}-\left[\frac{\left\langle T_{00}^{\delta \phi}\right\rangle_{\rm Ren}(M )}{a^2}\right]^{\rm Mink}
=\frac{\left\langle T_{00}^{\delta \phi}\right\rangle_{\rm Ren}(M )}{a^2}-\left\langle T_{00}^{\delta \phi}\right\rangle_{\rm Ren}^{\rm Mink}(M)\,.
\end{split}
\end{equation}
Thus, inserting equations \eqref{Eq:QuantumVacuum.RenT00vacuumMM0Eeqs} and \eqref{Eq:QuantumVacuum.MinkowskiSubtracted} in the above relation and recalling again that $g_{00}=-a^2$, we are led to
\begin{equation}\label{Eq:QuantumVacuum.AlternativeDef}
\begin{split}
\rho_{\rm vac}(M)&=\frac{\left\langle T_{00}^{\delta \phi}\right\rangle_{\rm ren}(M_0)}{a^2}-\left\langle T_{00}^{\delta \phi}\right\rangle_{\rm ren}^{\rm Mink}(M_0)\\
&+\frac{\delta \rho_\CC (m,M,M_0)}{a^2} g_{00}+\frac{\delta \Mpl^2 (m,M,M_0)}{a^2} G_{00}\\
&+\delta\alpha(m,M,M_0){a^2} H^{(1)}_{00} + \delta \rho_\CC (m,M,M_0)\\
&=\rho_{\rm vac}(M_0)+\frac{3\cH^2}{a^2}\,\delta \Mpl^2 (m,M,M_0)-\frac{18}{a^4}\left(\mathcal{H}^{\prime 2}-2\mathcal{H}^{\prime \prime}\mathcal{H}+3 \mathcal{H}^4 \right)\,\delta\alpha(m,M,M_0)\,.
\end{split}
\end{equation}
The result for the total VED is indeed the same as in Eq.\,\eqref{Eq:QuantumVacuum.VEDscalesMandM0Final} after we cast $\cH$ and its derivatives in terms of the ordinary Hubble rate, $H$. In other words, we can reach again the same relation between the values of VED at two different scales, which does not involve $\sim m^4$ contributions.

The two quantities $\rL(M)$ and $\langle T_{00}^{\delta \phi}\rangle_{\rm ren}^{\rm Mink}(M)$ both carry quartic dependencies on the mass scales, but they exactly conspire to sum up to zero in Minkowski spacetime since \eqref{Eq:QuantumVacuum.MEEsHD} implies that
\begin{equation}\label{Eq:QuantumVacuum.EinsteinFieldMinkowski}
 \left\langle T_{00}^{\delta \phi}\right\rangle_{\rm ren}^{\rm Mink}(M)=\eta_{00} \rL(M)=-\rL(M),
\end{equation}
so its sum became the VED in Minkowski spacetime as dictated by \eqref{Eq:QuantumVacuum.Totalrhovac},
\begin{equation}\label{Eq:QuantumVacuum.ZeroVEDMinkowski}
 \rho_{\rm vac}^{\rm Mink}(M)=\rho_\Lambda (M)+\left[\frac{\left\langle T_{00}^{\delta \phi}\right\rangle_{\rm ren}(M)}{a^2}\right]^{\rm Mink}=\rho_\Lambda (M)+\left\langle T_{00}^{\delta \phi}\right\rangle_{\rm ren}^{\rm Mink}(M)=0.
\end{equation}
Hence the detailed structure of these formal quantities plays no role in the physically measured quantities. In contrast to the Minkowski case, in curved spacetime such quantities cannot be isolated since the sum is not zero, see Eq.\,\eqref{Eq:QuantumVacuum.RenVDEexplicit}, but it yields a smooth quantity mildly evolving with the cosmological evolution. For $a\to 1$ ($\cH\to 0$) the LHS of \eqref{Eq:QuantumVacuum.RenVDEexplicit} goes to zero, as we have seen, and hence the RHS goes to zero too. At this point we retrieve the Minkowskian space result \eqref{Eq:QuantumVacuum.ZeroVEDMinkowski} from the curved spacetime case \eqref{Eq:QuantumVacuum.RenVDEexplicit}. But at any intermediate stage of this limit we cannot determine $\rL(M)$ separately from the ZPE, only the sum is physically relevant and it defines the dynamical vacuum energy density in curved spacetime. Recall that the renormalization point itself $M$ is dynamical in curved spacetime. As it was previously indicated, the scale setting prescription $M=H$ is an appropriate ansatz for testing the cosmological evolution of the VED at different stages of the expansion history (cf. \hyperref[Sect:RunningConnection]{Sect.\,\ref{Sect:RunningConnection}} and \hyperref[Sect:Abis1]{Appendix\,\ref{Sect:Abis1}}). It is remarkable that the VED of the expanding Universe despite it being currently very small (of order $\rvo$) is dynamical and such dynamics could be measured since it is of ${\cal O}(H^2)$. The VED is exactly zero only in Minkowski spacetime, where the vacuum energy plays no cosmological role. In actual fact, the scale $M$ becomes in this case a purely formal quantity devoid of any physical meaning, much the same as the artificial mass unit $\mu$ employed in DR (as discussed in the next section). There is no dynamics of gravity in Minkowski space and therefore nothing can physically run with $M$ (or $\mu$). In contrast, in FLRW spacetime the gravitational field is dynamical and hence the prescription $M=H$ is physically meaningful and enables exploring the running of the vacuum energy density with the cosmic expansion (cf. and \hyperref[Sect:Abis1]{Appendix\,\ref{Sect:Abis1}} for a thorough discussion). This is actually the original point of view of the RVM from the renormalization group approach\,\cite{Sola:2013gha,SolaPeracaula:2022hpd}.

On the other hand, remember that the divergences associated to our calculation are of course of UV type, hence short-distance effects. The leading effects of this kind are similar to the ones of QFT in Minkowski spacetime and therefore are independent from the possible boundary effects of the cosmological spacetime. We have just seen that an alternative way to renormalize the energy-momentum tensor is precisely to subtract the Minkowskian contribution following the adiabatic regularization procedure up to fourth order. Furthermore, if one takes into account only wavelengths under the horizon ({\it i.e.} for $\tilde{k}^2\gg H^2$, with $\tilde{k}=k/a$ the physical momentum defined in \hyperref[Sect:ParticularCases]{Sect.\,\ref{Sect:ParticularCases}}), the situation remains as in the Minkowskian spacetime, namely the integrals with low inverse powers of the function $\omega_k=\sqrt{k^2+a^2 M^2}$, corresponding to the lowest adiabatic orders, are still divergent in the UV. The short-distance region where the UV effects are encountered is of course contained within the horizon. The presence of a causal horizon can only produce long distance effects, and therefore they can be related with IR (infrared) divergences. The IR behavior of gravity theories can indeed be non-trivial in some cases but we do not address these aspects in our work as they are out of its scope. However, if we would consider effects of this kind in our momentum integrals they would rather be related with the lower limits of integration, which should be of order $H$, since the (apparent) horizon is of order $1/H$ (in fact, it is exactly so in the spatially flat case, which we are considering) and the effects that could produce are subleading. To see this, take for instance the simple cases analyzed in \hyperref[Sect:ParticularCases]{Sect.\,\ref{Sect:ParticularCases}}, say Eq.\,\eqref{Eq:QuantumVacuum.LowOrderZPE2}. Since the physical momenta satisfy $\tilde{k}^2\ll m^2$ in the IR, the contribution from these integrals in the IR region provides powers of $H$ higher than $H^2$ involving also masses, e.g. $m H^3$, $H^5/m$ etc. Similar terms carrying suppressed powers would appear if the more complete expression \eqref{Eq:QuantumVacuum.EMTFluctuations} would be used. The presence of odd powers of $H$ is not surprising since we have put boundaries to an otherwise covariant integration.

\subsection{Minimal Subtraction scheme and the fine-tuning problem}\label{Sect:VEDMSS}

The former are certainly properties we should expect from a correct, physically meaningful, renormalization of the vacuum energy density, in contrast to other formal treatments in the literature in the framework of different renormalization schemes. In particular, the absence of fine-tuning among the different terms is much welcome as well as the vanishing value of the VED in Minkowskian spacetime, which in the simplified notation \eqref{Eq:QuantumVacuum.Symbolic} introduced above reads $\rL+$ZPE$=0$. This condition can be thought of as a necessary condition for the physical renormalization prescription for the quantum vacuum energy, that is Eq.\,\eqref{Eq:QuantumVacuum.RenormEMTpointsplit}, and is encoded in the general form \eqref{Eq:QuantumVacuum.RenVDE}. Many other approaches and renormalization prescriptions have been advocated to deal with the vacuum energy, see\,\cite{Shapiro:2000dz, Sola:2007sv, Shapiro:2009dh,Babic:2001vv,Babic:2004ev,Maggiore:2010wr,Bilic:2011zm,Bilic:2011rj,Domazet:2012tw,Ward:2010qs,Antipin:2017pbt,Kohri:2016lsj,Kohri:2017iyl,Ferreiro:2018oxx,Ferreiro:2020zyl,Animali:2022lig,birrell1978application,Akhmedov:2002ts,Ossola:2003ku,Visser:2016mtr,Donoghue:2020hoh,Christensen:1976vb,Christensen:1978yd,Bunch:1978yq}, for instance. Arguably, the simplest treatments are those based on the Minimal Subtraction (MS) scheme\,\cite{Bollini:1972ui,tHooft:1973mfk} (cf.\,\cite{Manohar:1996cq,Sterman:1993hfp,Brown:1992db,collins1984renormalization} for further explanations and practical applications). Its use was soon extended to QFT in curved spacetime\,\cite{Bunch:1979mq}. But in this context, simplicity does not necessarily mean adequacy to the physical purposes, and in fact the MS scheme does not lead to a physically acceptable approach to the renormalization of the VED, cf.\,\cite{Shapiro:2009dh} and references therein. Let us briefly summarize the situation of the fine-tuning problem in the MS scheme (we refer the reader e.g. to \,\cite{Sola:2013gha,SolaPeracaula:2022hpd} for more details). It will suffice to focus on Minkowskian spacetime for this consideration.

In flat spacetime ($a=1$, $\cH=0$) the ZPE \eqref{Eq:QuantumVacuum.EMTFluctuations} shrinks just to the compact form \eqref{Eq:QuantumVacuum.Minkowski} (for $\xi=1/6$), where we shall continue with $\hbar=1$ in natural units. Using DR in Minkowskian $n$- spacetime (with $n-1$ spatial dimensions), a simple calculation with the notation and formulas of \hyperref[Sect:MasterInt]{Sect.\,\ref{Sect:MasterInt}} in \hyperref[Appendix:Conventions]{Appendix\,\ref{Appendix:Conventions}} leads to the result\,\cite{Sola:2013gha,SolaPeracaula:2022hpd}
\begin{equation}\label{Eq:QuantumVacuum.MinkoskiMSS}
\begin{split}
\left\langle T_{00}^{\delta \phi}\right\rangle^{\rm Mink}(m)&= \int\frac{\mu^{2\epsilon}d^{n-1}k}{(2\pi)^{n-1}}\,\left(\frac12\,\omega_k(m)\right)=\frac{1}{2}\, I_{n-1}(p=-1,Q=m)\\
&= \frac{m^4}{4(4\pi)^2}\,\left(-D_\epsilon+ \ln\frac{m^2}{\mu^2}-\frac32\right)\,,
\end{split}
\end{equation}
where $D_\epsilon$ contains the pole at $n-1=3$ ({\it i.e.} at $\epsilon=0$, or equivalently at $n=4$ spacetime dimensions) as given by Eq.\,\eqref{Eq:Dimensional.Depsilon}. It is natural to assume that the VED in Minkowskian space is given by a similar equation to \eqref{Eq:QuantumVacuum.RenVDE}, but with the bare quantities at this point since \eqref{Eq:QuantumVacuum.MinkoskiMSS} is divergent, {\it i.e.} $\rv^{\rm Mink}=\rL+\langle T_{00}^{\delta \phi}\rangle^{\rm Mink}$. We next split the bare term $\rL$ into the renormalized quantity $\rL(\mu)$ plus the counterterm, as shown in Eq.\,\eqref{Eq:QuantumVacuum.splitcounters}. As we know, in the MS scheme the running scale is usually represented by means of the arbitrary 't Hooft's mass unit $\mu$, which displays dimensions away from $n=4$ and keeps control of dimensional analysis.
Using the MS scheme (or its variant $\overline{\rm MS}$\,\cite{Bardeen:1978yd,Manohar:1996cq}) to deal with the UV-divergences is very tempting for we can choose the counterterm $\delta\rL$ in the form given in Eq.\,\eqref{Eq:QuantumVacuum.counters}, which precisely cancels the pole in \eqref{Eq:QuantumVacuum.MinkoskiMSS} and this allows to define the renormalized ZPE in Minkowski space, $\langle T_{00}^{\delta \phi}\rangle^{\rm Mink}_{\rm Ren}(\mu)$. One may then be tempted to interpret that the MS-renormalized VED in Minkowskian spacetime is the finite expression
\begin{equation}\label{Eq:QuantumVacuum.VEDMSS}
\rv^{\rm Mink}=\rL(\mu)+\left\langle T_{00}^{\delta \phi}\right\rangle^{\rm Mink}_{\rm Ren}(\mu)=\rL(\mu)+\frac{m^4}{4(4\pi)^2}\,\left( \ln\frac{m^2}{\mu^2}-\frac32\right)\,.
\end{equation}
However, in spite of its formal simplicity the above formula leads to the usual fine-tuning nightmare associated to the CCP, which is brought about by the fact that the renormalized ZPE is proportional to the quartic power of the mass of the particle $\sim m^4$. As a result, the MS-renormalized term $\rL(\mu)$ must be fine tuned in a preposterous way against the renormalized ZPE contribution so as to get a VED value in a reasonable phenomenological range, see\,\cite{Sola:2013gha,SolaPeracaula:2022hpd} for a detailed exposition of the fine-tuning problem. Since $\mu$ was introduced on mere dimensional grounds there is no special physical meaning to be ascribed to it. To set $\mu=H$ does not make much sense here since the above formula applies to Minkowskian spacetime, where $H=0$. Besides, any attempt to make sense of the above VED formula (using DR or Pauli-Villars regularization\,\cite{pauli1971}, for example) leads to nowhere. No matter what physical quantity is chosen for $\mu$ or the type and number of fields involved, the numerical results are completely astrayed\,\cite{Visser:2016mtr,Koksma:2011cq}. Let us stress once more that one should \textit{not} aim at a prediction of the value of the VED at present, as this is out of the scope of renormalization theory. Equation \eqref{Eq:QuantumVacuum.VEDMSS} is certainly not the VED neither in flat nor in curved spacetime. Such an expression is unphysical, it just describes the mathematical running of the parameter $\rL(\mu)$ and the renormalized ZPE with $\mu$ in such a way that their sum remains equal to the original bare (hence RG-invariant) parameter $\rL$ in the action. There is not an inch of physics in it since $\mu$ cannot be related to any quantity of cosmological interest; we reiterate that \eqref{Eq:QuantumVacuum.VEDMSS} was derived in Minkowki space, where we have seen that the VED is just zero. In \hyperref[Sect:RGE-VED]{Sect.\,\ref{Sect:RGE-VED}} we come back to this point, after discussing the running couplings in curved spacetime.

Nothing of this sort occurs in our renormalization scheme, where the VED is given by \eqref{Eq:QuantumVacuum.RenVDEexplicit}. To start with, the value of that expression in Minkowski space is exactly zero, as we have shown above, in stark contrast with the MS formula \eqref{Eq:QuantumVacuum.VEDMSS}. Furthermore, as long as we hold on to the aforementioned prescription for Minkowskian spacetime, the implication on the corresponding calculation for curved spacetime is that the VED is no longer zero but a mildly dynamical quantity, which evolves smoothly (without fine-tuning) from one scale to another throughout the cosmic evolution following the `running law' \eqref{Eq:QuantumVacuum.VEDscalesMandM0Final}\,\footnote{If one naively extends the MS renormalization to curved spacetime, the fine-tuning problem persists unmodified, see\,\cite{Sola:2013gha,SolaPeracaula:2022hpd} for a summarized account. The two fine-tuning-generating pieces on the RHS of \eqref{Eq:QuantumVacuum.VEDMSS} remain exactly as they are. The curved background only adds purely geometric terms ${\cal O}(R, R^2, R^{\mu\nu}R_{\mu\nu},...)$ and the essence of the fine-tuning problem embodied in the Minkowskian spacetime replicates identically in curved spacetime in such scheme, see\,\cite{Bunch:1979mq} for more technical details. Renormalization of the VED \textit{à la} MS seems to be completely hopeless in cosmology.\label{FN.QuantumVacuum.MS}}.
Thus, while we do not aim at a prediction of the value of the VED at present from pure renormalization theory, a prediction is made of its value at some scale, given its value e.g. at present. Such an evolution is governed by the Hubble flow and a quadratic (not quartic) dependence on the mass scale, which is made extremely smooth since it is accompanied by $H^2$ and thereby evolving as $\sim m^2 H^2$.
In what follows we take up what are the implications for the late time Universe and in particular for our present time.

\section{Running vacuum connection}\label{Sect:RunningConnection}

Eq.\,\eqref{Eq:QuantumVacuum.VEDscalesMandM0Final} constitutes our desired outcome for relating a dynamical VED between two different renormalization points, but does not provide the calculated value of the vacuum energy density at a given scale, e.g. it says nothing on the value of $\rho_{\rm vac}(M_0)$ and hence it has no implication on the cosmological constant problem mentioned in the Introduction. That is to say, it has no bearing on it if such problem is meant to be the computation of the value itself of the VED at some point in the history of the Universe. However, our result can be useful to explore the `running' of the VED when we move from one scale to another. In other words, if $\rho_{\rm vac}$ is known at some scale $M_0$, we can use the obtained relation to compute the value of $\rho_{\rm vac}$ at another scale $M$. Such connection of values from one renormalization point to another is what we have been calling ``running'' of the VED, and in fact it was suggested long ago from the point of view of the renormalization group in curved spacetime from different perspectives\,\cite{Shapiro:2000dz,Shapiro:2003kv,Sola:2007sv,Shapiro:2009dh} -- for a review of the running vacuum model (RVM), see\,\cite{Sola:2013gha,Sola:2014tta,Sola:2011qr,Sola:2015rra,GomezValent:2017kzh} and references therein. Interestingly enough, it can provide also a framework for the possible time variation of the so-called fundamental constants of nature\,\cite{Fritzsch:2012qc,Fritzsch:2015lua,Fritzsch:2016ewd}.

Before entering into the study of the VED in the current Universe, we recommend the reader to peruse \hyperref[Appendix:Abis]{Appendix\,\ref{Appendix:Abis}}, where the technical items are presented. Here we are focused in the physical details. While for the present ($H=H_0$) we may neglect all terms of order ${\cal O}(H^4)$ (which comprise also $\dot{H}^2, H\ddot{H}$ and $H^2 \dot{H}$) the piece proportional to $H^2$ in \eqref{Eq:QuantumVacuum.VEDscalesMandM0Final} may be significant in the present Universe. It entails that the vacuum energy density is dynamical and such a dynamics is amenable to being measured. This property leads to the notion of `running VED'. By running we mean that the VED is not static but changing with the cosmic evolution. A good tracking of that evolution in the FLRW context is provided by the Hubble rate $H$. In our aim to find a proper physical interpretation of Eq.\,\eqref{Eq:QuantumVacuum.VEDscalesMandM0Final} setting $M=H$ may be a reasonable election. At any given cosmic time, $H(t)$ (Hubble rate) constitutes a characteristic energy of the expanding Universe and may serve to parametrize the scaling evolution of the VED, see \,\cite{Sola:2013gha,Sola:2014tta,SolaPeracaula:2022hpd,Sola:2011qr,Sola:2015rra} and references therein for the old connection with the renormalization group arguments\,\cite{Shapiro:2000dz,Sola:2007sv}. 

Within this approach, if $\rv$ is known at some reference scale $M_0$, $M_0$ is taken to be the current Hubble parameter $H_0$, we can use the relation Eq.\,\eqref{Eq:QuantumVacuum.VEDscalesMandM0Final} to compute the value of $\rv$ at another scale $M$ associated to some other epoch $H$. If , then $\rv(H_0)=\rvo$ can be identified as being the presently observed value of the VED at $H=H_0$. We may relate $\rvo$ with the value $\rv(M=H)$ at another scale in the past corresponding to the cosmic epoch $H$, which we typically select within the accessible FLRW cosmic history. On \hyperref[Sect:Abis1]{Sect.\,\ref{Sect:Abis1}} in \hyperref[Appendix:Abis]{Appendix\,\ref{Appendix:Abis}} we present the notations and details of how to use Eq.\,\eqref{Eq:QuantumVacuum.VEDscalesMandM0Final} to connect two values of the VED which lead to the Late-Universe regime:
\begin{equation}\label{Eq:QuantumVacuum.RVM2}
\rv(H)\simeq \rvo+\frac{3\nueff}{8\pi}\,(H^2-H_0^2)\,\mpl^2=\rvo+\frac{3\nueff}{\kappa^2}\,(H^2-H_0^2)\,,
\end{equation}
where $\kappa^2=8\pi G_N$. Here we have defined the `running parameter' $\nueff$, which is approximately given by
\begin{equation}\label{Eq:QuantumVacuum.nueffAprox}
\nueff\simeq\frac{1}{2\pi}\,\left(\xi-\frac{1}{6}\right)\,\frac{m^2}{\mpl^2}\ln\frac{m^2}{H_0^2}\,.
\end{equation}
The more detailed treatment in \hyperref[Sect:Abis1]{Appendix\,\ref{Sect:Abis1}} shows that $\nueff$ is actually a slowly changing (logarithmic) function of the Hubble rate. But, to a fairly good approximation, $\nueff$ can be taken essentially as the constant value given above for values of $H$ corresponding to the relatively recent Universe.
As it was foreseen from the beginning, the structure of the RVM vacuum does not necessarily require the non-minimal coupling of matter to the external gravitational field.

Let us mention that different extensions of gravity can mimic the effective behavior of the running vacuum model (RVM). This is a fact confirmed in a variety of contexts. For instance, in the context of Brans-Dicke Theory with a cosmological term, it has been shown that a kind of RVM behavior emerges when one tries to rewrite the theory in a GR-like picture\,\cite{SolaPeracaula:2018dsw,deCruzPerez:2018cjx}. This turns out to be phenomenologically very favorable, as the reader can see in  \hyperref[Chap:PhenomenologyofBD]{chapter\,\ref{Chap:PhenomenologyofBD}}, based in our papers\,\cite{SolaPeracaula:2020vpg,SolaPeracaula:2019zsl}. Another potentially interesting example can be found in gravity theories with torsion, see e.g.\,\cite{Cai:2015emx} and references therein. Since the torsion scalar $T$ differs only by a total derivative with respect to the Ricci scalar, the EH action with $R$ replaced by $T$ is equivalent to GR. One may generalize the action structure through the replacement $T\rightarrow T+f(T)$, with $f(T)$ a function of the torsion scalar. This is characteristic of teleparallel gravity theories\,\cite{Cai:2015emx}. Since $T=-6H^2$ in the FLRW background, by an appropriate choice of $f(T)$ one may, in principle, mimic the RVM as well. In \hyperref[SubSect:RVMInflation]{Sect.\,\ref{SubSect:RVMInflation}} we include some comments on another example, in this case in the context of the low-energy effective action of string theory, which also behaves as the RVM.

Remarkably, for general $\xi$ the structure obtained for $\nueff$ is very close to that obtained within the RVM approach, see\,\cite{Sola:2013gha,Sola:2014tta,Sola:2011qr}. In such context, it defines the coefficient of the one-loop $\beta$-function for the renormalization group equation of $\rho_{\rm vac}$. The presence of the additional logarithmic piece $\ln {m^2}/{H_0^2}$ appears in the direct QFT calculation employed here, but it does not make any difference in practice since it is constant and $\nueff$ must be fitted directly to the observations as an effective coefficient. In our case we have simplified the theoretical calculation by considering just the contribution from one single scalar field to $\nueff$. We expect it to be small, {\it i.e.} $|\nueff|\ll1$, owing to the ratio $m^2/\mpl^2\ll 1$. We can see from \eqref{Eq:QuantumVacuum.nueffAprox} that $\xi=0$ does \textit{not} imply $\nueff=0$. The vanishing of $\nueff$ and hence of the dynamical $\sim H^2$ part of \eqref{Eq:QuantumVacuum.RVM2} is obtained only for $\xi=1/6$. In that specific case, there are no corrections to the vacuum energy density from scalars since we are then in the conformal limit of QFT. If the scalar field pertains to a typical GUT, {\it i.e.} $m\sim M_X\sim 10^{16}$ GeV, the ratio $m^2/\mpl^2\sim 10^{-6}$ remains sizeable. Taking into account that the parameter $\xi$ can be, in principle, arbitrary and that the multiplicity of states in a GUT is usually high, the value of $\nueff$ can actually be much larger. For $\xi\neq 0$, the sign of $\nueff$ depends entirely on the value of $\xi$ (if only a scalar field would contribute), so we can provide a discussion within a more general class of theories and also carrying potential phenomenological consequences. As indicated in \hyperref[Sect:TotalVED]{Sect.\,\ref{Sect:TotalVED}}, $\xi$ does not automatically determine the sign of $\nueff$. Other contributions (e.g. from fermion fields) should be added in our calculation. A detailed account of the fermionic quantum effects on the RVM vacuum structure will be provided in \hyperref[Chap:Fermions]{chapter\,\ref{Chap:Fermions}}. Ultimately, the final value could have either sign and be much larger since $\nueff$ depends also on the multiplicity and nature of the fields involved, so we cannot predict $\nueff$ with precision on mere theoretical grounds. It has to be determined by fitting the model to data. However, we understand that the basic facts derived from the renormalization procedure followed here should also hold in the general case.

As we can see from Eq.\eqref{Eq:QuantumVacuum.RVM2}, for $\nueff>0$ the vacuum can be conceived as decaying into matter since the vacuum energy density is larger in the past (where $H>H_0$), whereas if $\nueff<0$ the opposite occurs. The former situation, however, is more natural from a thermodynamical point of view, for if the vacuum decays into matter one can show that the Second Law of Thermodynamics is satisfied by the general RVM, see\,\cite{SolaPeracaula:2019kfm} for a detailed discussion. Moreover, for $\nueff>0$ the RVM effectively behaves as quintessence since the vacuum energy density decreases with time. For $\nueff<0$ the behavior is that of phantom DE. One may also interpret here that $G$ is changing with time owing to vacuum decay. Both possibilities have been discussed within the RVM in Ref.\,\cite{Fritzsch:2012qc,Fritzsch:2015lua,Fritzsch:2016ewd}, see also \hyperref[Sect:RenormalizedFriedmann]{Sect.\,\ref{Sect:RenormalizedFriedmann}}.

Recall that we expect $|\nueff|\ll1$ from the theoretical structure \eqref{Eq:QuantumVacuum.nueffAprox}; we cannot hope for observing dramatic deviations from the standard $\CC$CDM model. This would actually not be welcome, given the considerable success of the concordance cosmology. Remarkably enough, we confirm it from the phenomenological fits of RVM against the overall cosmological data\,\cite{Sola:2017znb,SolaPeracaula:2016qlq,SolaPeracaula:2017esw,SolaPeracaula:2021gxi}, whereby we do not observe dramatic deviations from the standard $\CC$CDM model. But the fact that the fitting results point to a positive $\nueff= {\cal O}(10^{-3})$ suggests that the effects are not necessarily negligible and in fact they can be helpful to cure or alleviate some of the existing tensions in the context of the $\CC$CDM model, as actually shown in the aforementioned references and also in the framework of alternative cosmological models which also mimic the RVM behavior\,\cite{Rezaei:2019xwo, Rezaei:2021qwd, SolaPeracaula:2020vpg,SolaPeracaula:2019zsl, SolaPeracaula:2021gxi}. The models presented in \hyperref[Chap:PhenomenologyofBD]{chapter\,\ref{Chap:PhenomenologyofBD}} and \hyperref[Chap:PhenomenologyofRVM]{chapter\,\ref{Chap:PhenomenologyofRVM}} are a good example of it.  Notice that the standard model particles is not expected to make significant contributions to $\nueff$, since for all of the known particle masses, $m^2/\mpl^2\ll M_X^2/\mpl^2\sim 10^{-6}$, where $M_X\sim 10^{16} \textrm{GeV}$ is the typical mass scale associated to GUT theories. In this sense, a moderate value of $\nueff$ may be an indication of the presence of extremely massive fields, for instance around the order of the Great Unified Theories (GUT) scale. Curiously, the accurate determination of $\nueff$ determines the low-energy regime of evolution of VED in the Late-Universe and, at the same time, can serve as a probe of physics beyond the Standard Model of Particle Physics. A recent analysis of Big Bang nucleosynthesis (BBN) constraints points to the same order of magnitude mentioned before for $\nueff$, although is not sensitive to the sign\,\cite{Asimakis:2021yct}. However an extended study on this topic it is going to be carried in the future, incorporating some of the new features regarding the VED that are presented in this chapter and in the coming ones.

\section{Trace of the vacuum EMT in curved spacetime} \label{Sect:Trace}

We may now explicitly compute the VEV of the trace of the vacuum part of the EMT. There are, at least two very good reasons to do that. First, it can be used to compute the pressure of the vacuum fluid, in analogy to the computation performed in former sections for the VED. This will, in fact, guide us to the computation of the equation of state of the quantum vacuum, which presumably can deviate from -1 due to the quantum effects. Secondly, we may try to recover the well-known result of the trace anomaly, {\it i.e.} the non-vanishing VEV trace associated to quantum fluctuations in the massless limit in curved spacetime\,\cite{birrell1984quantum}.

We start by computing the trace of the classical EMT, which we denote $T^{\rm cl.}\equiv T^{\mu}_{\ \mu}$. Using \eqref{Eq:QuantumVacuum.EMTScalarField}, it can be expressed as
\begin{equation}\label{Eq:QuantumVacuum.ClassicTrace}
\begin{split}
T^{\rm cl.}&=\left(6\xi-1\right)\ \nabla^\mu\phi \nabla_\mu\phi+6\xi\phi \Box \phi-\xi R\phi^2-2m^2\phi^2\\
&=\left(6\xi-1\right) \nabla^\mu\phi \nabla_\mu\phi+\left(6\xi-1\right)\phi\Box \phi-m^2\phi^2\,,
\end{split}
\end{equation}
where in the last step we have used the equation of motion \eqref{Eq:QuantumVacuum.KG}. This last form is useful since it makes transparent that the trace is null in the conformal limit ($m=0$ and $\xi=1/6$), as it should be (in four spacetime dimensions). An alternative form which will be more helpful for our purposes and still makes apparent the previous property, is obtained by trading $\phi\Box\phi$ for $R\phi^2$ as follows:
\begin{equation}\label{Eq:QuantumVacuum.ClassicTrace2}
\begin{split}
T^{\rm cl.}&=\left(6\xi-1 \right)\nabla^\mu \phi\nabla_\mu \phi +2(3\xi-1)m^2\phi^2+ \xi \left(6\xi-1\right) R \phi^2\\
&=\left(6\xi-1 \right) g^{\mu\nu}\nabla_\mu\phi\nabla_\mu \phi +2(3\xi-1)m^2\phi^2+6\left(\xi-\frac{1}{6} \right)^2R \phi^2+\left(\xi-\frac{1}{6} \right)R \phi^2\,,
\end{split}
\end{equation}
where the last rearrangement is just for convenience.
On examining the quantum fluctuations \eqref{Eq:QuantumVacuum.ExpansionField} about the background field, we note that the vacuum expectation value (VEV) of the trace $T_{\mu \nu}^{\phi}$ with the quantum field $\phi$, can only comprise terms quadratic (or, more rigorously, bilinear under the coincidence limit) on its fluctuations $\delta\phi$. Denoting by $\langle T^{\delta \phi} \rangle\equiv \langle 0| T^{\delta \phi}|0\rangle$ such a result, we find
\begin{equation}\label{Eq:QuantumVacuum.QuantumTrace}
\left\langle T^{\delta \phi} \right\rangle=\left\langle \left(6\xi-1 \right)g^{\mu\nu}\nabla_\mu\delta \phi\nabla_\mu \delta \phi +2(3\xi-1)m^2\delta \phi^2+6\left(\xi-\frac{1}{6} \right)^2R\delta \phi^2+\left(\xi-\frac{1}{6} \right)R\delta \phi^2\right\rangle\,.
\end{equation}
This result for the vacuum trace (i.e. the vacuum expectation value of the trace) adopts the same form as \eqref{Eq:QuantumVacuum.ClassicTrace2}, with $\phi$ replaced by its fluctuating part $\delta\phi$. By using the Fourier decompositions of the field fluctuation $\delta\phi$ in the mode functions $h_k(\tau)$ and also utilizing the commutation relations between the creation and annihilation operators, {\it i.e.} we proceed along the lines we already followed in \hyperref[Sect:AdiabaticVacuum]{Sect.\,\ref{Sect:AdiabaticVacuum}} with the components of the EMT, we can rewrite Eq.\,\eqref{Eq:QuantumVacuum.QuantumTrace} as follows:
\begin{equation}\label{Eq:QuantumVacuum.TraceVEV}
\begin{split}
\left\langle T^{\delta \phi} \right\rangle&=
-\frac{\left(6\xi-1\right)}{(2\pi)^3 a^4}\left(\mathcal{H}^2\int d^3 k |h_k|^2+\int d^3 k |{h}_k^\prime|^2-\mathcal{H}\int d^3k \left(h_k h_k^{\prime*}+h_k^{\prime} h_k^* \right) \right)\\
&+\frac{\left(6\xi-1\right)}{a^2}\frac{1}{(2\pi)^3 a^2}\int d^3k k^2 |h_k|^2+2(3\xi-1)m^2\frac{1}{(2\pi)^3 a^2}\int d^3k |h_k|^2\\
&+6\left(\xi-\frac{1}{6} \right)^2 R\frac{1}{(2\pi)^3 a^2}\int d^3k |h_k|^2+\left(\xi-\frac{1}{6} \right) R\frac{1}{(2\pi)^3 a^2}\int d^3k |h_k|^2\\
&=\frac{1}{(2\pi)^3 a^2}\int d^3 k\left((6\xi-1)\frac{k^2-\mathcal{H}^2}{a^2}+2(3\xi-1)m^2+6\left(\xi-\frac{1}{6} \right)^2 R+\left(\xi-\frac{1}{6} \right) R \right) |h_k|^2\\
&-\frac{(6\xi-1)}{a^2}\frac{1}{(2\pi)^3 a^2}\int d^3k |h_k^\prime|^2+\frac{\left(6\xi-1 \right)}{a^2}\frac{\mathcal{H}}{(2\pi)^3a^2}\int d^3k \left(h_k h_k^{\prime*}+h_k^{\prime} h_k^* \right)\,.
\end{split}
\end{equation}
The first equality makes it clearer the structure of the result. For instance, using the fact that $g^{\mu\nu}\nabla_\mu\delta \phi\nabla_\mu \delta \phi =-a^{-2}\left((\delta\phi')^2-\nabla^2\delta\phi\right)$ and taking into account that the expansion of $\delta\phi'=(\delta\phi)'$ involves the calculation of $(d/d\tau)(h_k(\tau)/a)=(h'_k-\cH h_k)/a)$, it is easy to understand the origin of the first line of Eq.\,\eqref{Eq:QuantumVacuum.TraceVEV}, and similarly with the other terms.
Up to this point this result is generic and no approximation has been performed (apart from using the adiabatic vacuum, on which the creation and annihilation operators act upon). We must now expand the above VEV with respect to such vacuum state up to the $6th$-order. To this aim we employ the $6th$-order adiabatic expansions of the mode functions given in equations \eqref{Eq:QuantumVacuum.exphk2}-\eqref{Eq:QuantumVacuum.exphkphk} in combination with the relations \eqref{Eq:QuantumVacuum.omegak0}. On substituting them in the above formula the result is a rather lengthy expression. Thanks to the use of Mathematica\,\cite{Mathematica} the computation can be performed exactly. It is better to show the complete formula up to 6{\it th} order decomposed in some pieces, organized in powers of $\left(\xi-1/6\right)$:
\begin{equation}\label{Eq:QuantumVacuum.TraceDecomposition}
\begin{split} 
 \left\langle T^{\delta \phi} \right\rangle^{(0-6)} &= \left\langle T^{\delta \phi} \right\rangle_A+\left(\xi-\frac{1}{6}\right) \left\langle T^{\delta \phi} \right\rangle_B+\left(\xi-\frac{1}{6}\right)^2 \left\langle T^{\delta \phi} \right\rangle_C+\left(\xi-\frac{1}{6}\right)^3 \left\langle T^{\delta \phi} \right\rangle_D
\end{split}
\end{equation}
Each one of these pieces contain terms of adiabatic order from 0 to 6. Their expressions are:
\begin{equation}\label{Eq:QuantumVacuum.TraceDecompositionA}
\begin{split}
\left\langle T^{\delta \phi}\right\rangle_A =\frac{1}{4\pi^2 a^4}\int dk k^2 & \Bigg\{ -\frac{a^2 M^2}{\omega_k}-\frac{a^4 M^4}{4 \omega_k^5}\left(2 \mathcal{H}^2+\mathcal{H}^\prime \right)\\
&+\frac{a^4 M^4}{16\omega_k^7}\left(8\mathcal{H}^4+24\mathcal{H}^2 \mathcal{H}^\prime+6(\mathcal{H}^\prime)^2+8\mathcal{H}\mathcal{H}^{\prime \prime}+\mathcal{H}^{\prime \prime \prime} \right)\\
&+\frac{5 a^6 M^6}{8\omega_k^7} \mathcal{H}^2-\frac{7a^6 M^6}{32\omega_k^9}\left(28 \mathcal{H}^4+36\mathcal{H}^2 \mathcal{H}^\prime+3 (\mathcal{H}^\prime)^2+4\mathcal{H} \mathcal{H}^{\prime \prime} \right)\\
&-\frac{a^4M^4}{64\omega_k^9}\bigg(32\mathcal{H}^6+240\mathcal{H}^4\mathcal{H}^\prime+60\left(\mathcal{H}^{\prime}\right)^3+160\mathcal{H}^3\mathcal{H}^{\prime \prime}+20\left(\mathcal{H}^{\prime\prime}\right)^2\\
&\phantom{xxxxxxxx}+30\mathcal{H}^\prime \mathcal{H}^{\prime \prime \prime}+360\mathcal{H}^2 \left(\mathcal{H}^\prime\right)^2+60\mathcal{H}^2\mathcal{H}^{\prime \prime \prime}+240\mathcal{H}\mathcal{H}^\prime\mathcal{H}^{\prime\prime}\\
&\phantom{xxxxxxxx}+12\mathcal{H}\mathcal{H}^{\prime \prime\prime\prime}+\mathcal{H}^{\prime\prime\prime\prime\prime}\bigg)\\
&+\frac{3a^6 M^6}{128\omega_k^{11}}\bigg(1264 \mathcal{H}^6+1512 \mathcal{H}^3\mathcal{H}^{\prime \prime}+23\left(\mathcal{H}^{\prime\prime}\right)^2+228\left(\mathcal{H}^\prime\right)^3+38\mathcal{H}^\prime\mathcal{H}^{\prime\prime\prime}\\
&\phantom{xxxxxxxxx}+940\mathcal{H}\mathcal{H}^{\prime}\mathcal{H}^{\prime\prime}+18\mathcal{H}\mathcal{H}^{\prime\prime\prime\prime}+4672\mathcal{H}^4\mathcal{H}^\prime+3276\mathcal{H}^2\left(\mathcal{H}^{\prime}\right)^2\\
&\phantom{xxxxxxxxx}+256\mathcal{H}^2\mathcal{H}^{\prime\prime\prime}\bigg)\\
&+\frac{231 a^8 M^8}{32 \omega_k^{11}}\left(2\mathcal{H}^4+\mathcal{H}^\prime \mathcal{H}^2 \right)-\frac{1155 a^{10} M^{10} \mathcal{H}^4}{128 \omega_k^{13}}\\
&-\frac{11a^8 M^8}{128\omega_k^{13}}\bigg(3152\mathcal{H}^6+61\left(\mathcal{H}^\prime\right)^3+1116\mathcal{H}^3\mathcal{H}^{\prime\prime}+258\mathcal{H}\mathcal{H}^\prime\mathcal{H}^{\prime\prime}\\
&\phantom{xxxxxxxxxx}+2364\mathcal{H}^2\left(\mathcal{H}^\prime\right)^2+75\mathcal{H}^2\mathcal{H}^{\prime\prime\prime}+6660\mathcal{H}^4\mathcal{H}^\prime\bigg)\\
&+\frac{429a^{10}M^{10}\mathcal{H}^2}{256\omega_k^{15}}\left(492\mathcal{H}^4+572\mathcal{H}^2\mathcal{H}^\prime+83\left(\mathcal{H}^\prime\right)^2+40\mathcal{H}\mathcal{H}^{\prime \prime}\right)\\
&-\frac{255255a^{12}M^{12} \mathcal{H}^4}{512 \omega_k^{17}}\left(2\mathcal{H}^2+\mathcal{H}^\prime\right)+\frac{425425a^{14}M^{14}\mathcal{H}^6}{1024\omega_k^{19}}\\
&+\Delta^2\bigg(-\frac{a^2}{\omega_k}+\frac{a^4 M^2}{2\omega_k^3}-\frac{a^4 M^2 \left(2\mathcal{H}^2+\mathcal{H}^\prime\right)}{2\omega_k^5}+\frac{5a^6 M^4}{8\omega_k^7}\left(5\mathcal{H}^2+\mathcal{H}^\prime \right)\\
&\phantom{xxxxxx}-\frac{35a^8M^6}{16\omega_k^9}\mathcal{H}^2\bigg)\\
&+\Delta^4\left(\frac{a^4}{2\omega_k^3}-\frac{3a^6M^2}{8\omega_k^5}\right)\Bigg\},\\
\end{split}
\end{equation}
\begin{equation}
\begin{split}\label{Eq:QuantumVacuum.TraceDecompositionB}
\left\langle T^{\delta \phi} \right\rangle_B &=\frac{1}{4\pi^2 a^4}\int dk k^2 \Bigg\{ \frac{6 \mathcal{H}^\prime}{\omega_k}+\frac{3 a^2 M^2}{\omega_k^3} \left(\mathcal{H}^2 +2\mathcal{H}^\prime\right)-\frac{9a^4 M^4 \mathcal{H}^2}{ \omega_k^5}\\
&\phantom{xxxxxxxxxxxxxx}-\frac{a^2 M^2}{2 \omega_k^5}\left(3\mathcal{H}^{\prime \prime \prime}+12\mathcal{H} \mathcal{H}^{\prime \prime}+9(\mathcal{H}^\prime )^2+12 \mathcal{H}^2 \mathcal{H}^\prime \right)\\
&\phantom{xxxxxxxxxxxxxx}+\frac{3a^2M^2}{8\omega_k^7}\bigg(32\mathcal{H}^4\mathcal{H}^\prime+96\mathcal{H}^2\left(\mathcal{H}^\prime\right)^2+24\left(\mathcal{H}^\prime\right)^3+40\mathcal{H}^3\mathcal{H}^{\prime\prime}+92\mathcal{H}\mathcal{H}^\prime\mathcal{H}^{\prime\prime}\\
&\phantom{xxxxxxxxxxxxxxxxxxxxxx}+13\left(\mathcal{H}^{\prime\prime}\right)^2+20\mathcal{H}^2\mathcal{H}^{\prime\prime\prime}+18\mathcal{H}^\prime \mathcal{H}^{\prime \prime\prime}+6\mathcal{H}\mathcal{H}^{\prime\prime\prime\prime}+\mathcal{H}^{\prime\prime\prime\prime\prime}\bigg)\\
&\phantom{xxxxxxxxxxxxxx}+\frac{a^4 M^4}{4\omega_k^7}\left( 210 \mathcal{H}^4+390 \mathcal{H}^2 \mathcal{H}^\prime+45(\mathcal{H}^\prime )^2+60 \mathcal{H} \mathcal{H}^{\prime \prime}\right)\\
&\phantom{xxxxxxxxxxxxxx}-\frac{a^6 M^6}{8 \omega_k^9}\left(1365 \mathcal{H}^4+840\mathcal{H}^2 \mathcal{H}^\prime \right)\\
&\phantom{xxxxxxxxxxxxxx}-\frac{21a^4M^4}{16\omega_k^9}\bigg(152\mathcal{H}^6+57\left(\mathcal{H}^\prime\right)^3+288\mathcal{H}^3\mathcal{H}^{\prime \prime}+7\left(\mathcal{H}^{\prime \prime}\right)^2\\
&\phantom{xxxxxxxxxxxxxxxxxxxxxxx}+12\mathcal{H}^\prime\mathcal{H}^{\prime\prime\prime}+232\mathcal{H}\mathcal{H}^{\prime}\mathcal{H}^{\prime\prime}+6\mathcal{H}\mathcal{H}^{\prime\prime\prime}+724\mathcal{H}^4\mathcal{H}^\prime\\
&\phantom{xxxxxxxxxxxxxxxxxxxxxxx}+642\mathcal{H}^2\left(\mathcal{H}^\prime\right)^2+61\mathcal{H}^2\mathcal{H}^{\prime\prime\prime}\bigg)+\frac{945a^8 M^8\mathcal{H}^4}{8\omega_k^{11}}\\
&\phantom{xxxxxxxxxxxxxx}+\frac{63a^6 M^6}{32\omega_k^{11}}\bigg(1332\mathcal{H}^6+40\left(\mathcal{H}^\prime\right)^3+636\mathcal{H}^3\mathcal{H}^{\prime\prime}+172\mathcal{H}\mathcal{H}^\prime\mathcal{H}^{\prime\prime}\\
&\phantom{xxxxxxxxxxxxxxxxxxxxxxx}+1359\mathcal{H}^2\left(\mathcal{H}^\prime\right)^2+52\mathcal{H}^2\mathcal{H}^{\prime\prime\prime}+3276\mathcal{H}^4\mathcal{H}^\prime\bigg)\\
&\phantom{xxxxxxxxxxxxxx}-\frac{693a^8 M^8\mathcal{H}^2}{64\omega_k^{13}}\left(860\mathcal{H}^4+1117\mathcal{H}^2 \mathcal{H}^\prime+180\left(\mathcal{H}^\prime\right)^2+88\mathcal{H}\mathcal{H}^{\prime\prime}\right)\\
&\phantom{xxxxxxxxxxxxxx}+\frac{9009 a^{10}M^{10}\mathcal{H}^4}{128\omega_k^{15}}\left(173\mathcal{H}^2+94\mathcal{H}^\prime\right)-\frac{675675a^{12}M^{12}\mathcal{H}^6}{128\omega_k^{17}}\\
&\phantom{xxxxxxxxxxxxxx}+\Delta^2\bigg(\frac{3a^2}{\omega_k^3}\left(\mathcal{H}^2+\mathcal{H}^\prime\right)-\frac{9a^4M^2}{2\omega_k^5}\left(5\mathcal{H}^2+2\mathcal{H}^\prime\right)\\
&\phantom{xxxxxxxxxxxxxxxxxxx}+\frac{45a^6 M^4}{2\omega_k^5}\mathcal{H}^2 +\frac{45a^6 M^4 \mathcal{H}^2}{2\omega_k^7}\bigg)\Bigg\},
\end{split}
\end{equation}

\begin{equation}\label{Eq:QuantumVacuum.TraceDecompositionD}
\begin{split}
\left\langle T^{\delta \phi} \right\rangle_D =\frac{1}{4\pi^2 a^4}\int dk k^2 & \Bigg\{ \frac{81}{\omega_k^5}\left(5\mathcal{H}^4\mathcal{H}^\prime-\left(\mathcal{H}^\prime\right)^3-4\mathcal{H}\mathcal{H}^\prime\mathcal{H}^{\prime\prime}-\left(\mathcal{H}^{\prime\prime}\right)^2-\mathcal{H}^{\prime\prime\prime}\left(\mathcal{H}^2+\mathcal{H}^\prime\right)\right)\\
&+\frac{135a^2M^2}{2\omega_k^7}\left(\mathcal{H}^2+\mathcal{H}^\prime \right) \left(\mathcal{H}^4+29\mathcal{H}^2\mathcal{H}^\prime+4 \left(\mathcal{H}^\prime\right)^2+12\mathcal{H}\mathcal{H}^{\prime\prime}\right)\\
&-\frac{2835a^4M^4\mathcal{H}^2}{2\omega_k^9}\left(\mathcal{H}^2+\mathcal{H}^\prime\right)^2 \Bigg\}.
\end{split}
\end{equation}
\newpage
In the previous equation some terms of adiabatic order $6$ have been omitted. Namely, those which are proportional to $\mathcal{O}\left(H^4\right)\Delta^2$, $\mathcal{O}\left(H^2\right)\Delta^4$ and $\Delta^6$, where $\mathcal{O}(H^n)$ encompass terms with $n$ derivatives of the scale factor. In other words, the collection of all the contributions of adiabatic order 6 or higher which are constructed from the product of $\Delta^2$ times other contributions of adiabatic order $4th$ such as $\Delta^2 H^4$, $\Delta^2 \left(\mathcal{H}^\prime\right)^2$, $\Delta^2 \mathcal{H}^{\prime\prime\prime}$, $\Delta^4 \mathcal{H}^2$, $\Delta^4 \mathcal{H}^\prime$ , $\Delta^6$, $\dots$ We do not include these terms in our analysis because they vanish on-shell, thereby playing no role in the renormalization of $\langle T^{\delta \phi} \rangle$ according to the subtraction prescription we will provide below.

\subsection{Trace renormalization}\label{Sect:TraceRenorm}

It is important to realize that the vacuum trace \eqref{Eq:QuantumVacuum.TraceVEV} is UV-divergent and therefore needs renormalization. For instance, similar to the situation with Eq.\,\eqref{Eq:QuantumVacuum.T00}, the integrals in the first line of \eqref{Eq:QuantumVacuum.TraceVEV} are quadratically, quartically and logarithmically UV-divergent, respectively, cf. Eqs.\,\eqref{Eq:QuantumVacuum.exphk2}-\eqref{Eq:QuantumVacuum.exphkphk}. We are going to proceed analogously to \hyperref[Sect:RenormZPE]{Sect.\,\ref{Sect:RenormZPE}} with $\langle T^{\delta \phi} \rangle$ instead of $\langle T_{\mu\nu}^{\delta \phi} \rangle$. We start by recognizing in the above expression Eq.\,\eqref{Eq:QuantumVacuum.TraceDecomposition} the mentioned UV-divergent terms. All of them are of 4th adiabatic order or lower, exactly as stated before Eq.\,\eqref{Eq:QuantumVacuum.EMTFluctuations}:
\begin{equation}\label{Eq:QuantumVacuum.UVdivergentTrace}
\begin{split}
\left\langle T^{\delta \phi} \right\rangle_{\rm Div} &\equiv \frac{1}{4\pi^2 a^4}\int dk k^2 \Bigg\{ -\frac{a^2 M^2}{\omega_k}+\left(\xi-\frac{1}{6}\right) \left(\frac{6}{\omega_k}\mathcal{H}^\prime +\frac{3 a^2 M^2}{\omega_k^3}\left(\mathcal{H}^2+2\mathcal{H}^\prime \right)\right)\\
& \phantom{xxxxxxxxxxxxx}+ \left( \xi-\frac{1}{6}\right)^2\left(\frac{1}{\omega_k^3}\left(-54 \mathcal{H}^2 \mathcal{H}^\prime+9\mathcal{H}^{\prime \prime \prime} \right) \right)-\frac{a^2 \Delta ^2}{\omega_k}\\
&\phantom{xxxxxxxxxxxxx}+\frac{a^4}{2\omega_k^3}\left(M^2 \Delta^2+\Delta^4 \right)+\left(\xi-\frac{1}{6}\right)\frac{3a^2 \Delta^2}{\omega_k^3}\left(\mathcal{H}^2+\mathcal{H}^\prime \right) \Bigg\}\,.
\end{split}
\end{equation}
The remaining terms, given of course by $\langle T^{\delta \phi} \rangle_{\rm Non-Div}\equiv\langle T^{\delta \phi} \rangle-\langle T^{\delta \phi} \rangle_{\rm Div}$, are finite. In order to meet a well defined renormalized expression within the ARP we must perform the subtraction of the trace of the EMT up to the $4th$ adiabatic order at an arbitrary scale $M$. Let us emphasize that we are following the same prescription as for the $00th$-component of the EMT (cf. \hyperref[Sect:AdRegEMT]{Sect.\,\ref{Sect:AdRegEMT}}), namely the subtraction is performed in this case over all of the terms of $\langle T^{\delta \phi} \rangle_{\rm ren}(M)$, whether UV-divergent or UV-convergent:
\begin{equation}\label{Eq:QuantumVacuum.TraceEMTsubtracted}
\left\langle T^{\delta \phi} \right\rangle_{\rm ren}(M)=\left\langle T^{\delta \phi} \right\rangle (m)-\left\langle T^{\delta \phi} \right\rangle^{(0-4)} (M)\,.
\end{equation}
This is of course the same subtraction prescription that we have already followed with the components of the EMT, see \eqref{Eq:QuantumVacuum.EMTRenormalizedDefinition}. 
This overall subtraction is crucial to insure the consistency of the procedure, namely to avoid that the net finite part that remains in the subtractions turns out to be ambiguous.
After the subtraction the integrations left in that expression are finite and can be performed with the help of the formulas of \hyperref[Appendix:Conventions]{Appendix\,\ref{Appendix:Conventions}}, in \hyperref[Sect:MasterInt]{Sect.\,\ref{Sect:MasterInt}}. Albeit the overall integration left is indeed convergent it is not evident if one considers the isolated pieces. Following the method of \hyperref[Appendix:Dimensional]{Appendix\,\ref{Appendix:Dimensional}} one can either proceed by explicitly exhibiting the divergent parts of these isolated pieces (for example through the poles in DR) and showing that they cancel out altogether, or one can rearrange that expression to show that the integrals can be put in a convergent form as done in previous sections, Eq.\,\eqref{Eq:QuantumVacuum.ExplicitRenormalized}. If one wishes to follow the last procedure, the following relations prove useful to show that upon performing the subtraction \eqref{Eq:QuantumVacuum.TraceEMTsubtracted} on the integral \eqref{Eq:QuantumVacuum.UVdivergentTrace} one finds manifestly convergent expressions:
\begin{equation}\label{Eq:QuantumVacuum.AlgebraicRelTraceI}
\begin{split}
&-\frac{a^2 m^2}{\omega_k (m)}+\frac{a^2 M^2}{\omega_k (M)}+\frac{a^2 \Delta^2}{\omega_k (M)}-\frac{a^4}{2\omega_k^3 (M)}\left(M^2 \Delta^2+\Delta^4 \right) \\
&=-\frac{a^6 m^2 \Delta^4 }{2\omega_k^3(M) \omega_k(m)\left(\omega_k (m)+\omega_k(M)\right)}\left(1+\frac{\omega_k (M)}{\omega_k (M)+\omega_k (m)}\right),
\end{split}
\end{equation}
and
\begin{equation}\label{Eq:QuantumVacuum.AlgebraicRelTraceII}
\begin{split}
&\frac{6 \mathcal{H}^\prime}{\omega_k (m)}+\frac{3a^2 m^2}{\omega_k^3 (m)}(\mathcal{H}^2+2\mathcal{H}^\prime)-\frac{6\mathcal{H}^\prime}{\omega_k (M)}-\frac{3a^2 M^2}{\omega_k^3 (M)}(\mathcal{H}^2+2\mathcal{H}^\prime)-\frac{3a^2 \Delta^2}{\omega_k^3 (M)}(\mathcal{H}^2+\mathcal{H}^\prime)\\
&=-\mathcal{H}^\prime\Bigg[\frac{3a^4\Delta^4}{\omega_k^2 (m)\omega_k (M)(\omega_k (m)+\omega_k (M))}\left(\frac{1}{\omega_k (m)}+\frac{1}{\omega_k (m)+\omega_k(M)}\right)\\
&\phantom{xxxxxx}-\frac{3a^4(m^4-M^4)}{\omega_k^2 (M)\omega_k (m)}\left(\frac{1}{\omega_k^2 (m)}+\frac{1}{\omega_k (M)(\omega_k (m)+\omega_k (M))}\right)\Bigg]\\
&-\mathcal{H}^2\frac{3a^4 m^2\Delta^2}{\omega_k^2 (M)\omega_k (m)}\left(\frac{1}{\omega_k^2 (m)}+\frac{1}{\omega_k (M)(\omega_k (m)+\omega_k (M))}\right).\\
\end{split}
\end{equation}
As we can see after these rearrangements, the integration $\int dk k^2$ of these expressions leads to convergent integrals. They all behave as $\sim \int dk k^2/k^5\sim \int dk /k^3$ in the UV region, similarly as in the situation indicated before Eq.\,\eqref{Eq:QuantumVacuum.ExplicitRenormalized}.
The remaining task is to compute these converging integrals, which is not completely trivial. With the help of Mathematica\,\cite{Mathematica} the final result can be expressed as in Eq.\,\eqref{Eq:QuantumVacuum.TraceIntegrated} of the main text. Alternatively, if one performs the calculation in DR one may account for all the integrals (whether UV-divergent or convergent) on using the master formula \eqref{Eq:Conventions.DRFormula} of \hyperref[Sect:MasterInt]{Sect.\,\ref{Sect:MasterInt}}. Anyway, computational details are lengthy and involved. Fortunately, the final form of the renormalized trace of the vacuum EMT, exact up to 6th adiabatic order, can be cast in a relatively compact form as follows:
\newpage
\begin{equation}\label{Eq:QuantumVacuum.TraceIntegrated}
\begin{split}
\left\langle T^{\delta \phi} \right\rangle^{{\rm (0-6)}}_{\rm ren}(M)&=\frac{1}{32\pi^2}\left(3m^4-4m^2M^2+M^4-2m^2\ln \frac{m^2}{M^2}\right)\\
&+\frac{3\left(\xi-\frac{1}{6}\right)}{8\pi^2}\left(m^2-M^2-m^2\ln \frac{m^2}{M^2}\right)\left(2H^2+\dot{H}\right)\\
&-\frac{9}{8\pi^2}\left(\xi-\frac{1}{6}\right)^2\left(12H^2 \dot{H}+4\dot{H}^2+7H\ddot{H}+\vardot{3}{H}\right)\ln \frac{m^2}{M^2}\\
&+\frac{1}{10080\pi^2 m^2}\left(16H^6+96H^4\dot{H}-44\dot{H}^3-66H^3\ddot{H}-54\dot{H}\vardot{3}{H}-96H^2\dot{H}^2\right.\\
&\phantom{xxxxxxxxxxxx}\left.-93H^2\vardot{3}{H}-267H\dot{H}\ddot{H}-30H\vardot{4}{H}-36\ddot{H}^2-3\vardot{5}{H}\right)\\
&-\frac{\left(\xi-\frac{1}{6}\right)}{80\pi^2 m^2}\left(4H^6+30H^4\dot{H}-44H^2\dot{H}^2-18\dot{H}^3-22H^3\ddot{H}-103H\dot{H}\ddot{H}\right.\\
&\left.\phantom{xxxxxxxxx}-14\ddot{H}^2-31H^2\vardot{3}{H}-20\dot{H}\vardot{3}{H}-10H \vardot{4}{H}-\vardot{5}{H}\right)\\
&+\frac{3\left(\xi-\frac{1}{6}\right)^2}{16\pi^2 m^2}\left( 48H^4\dot{H}-16\dot{H}^3-8H^3\ddot{H}-14\ddot{H}^2-20\dot{H}\vardot{3}{H}-16H^2\dot{H}^2\right.\\
&\left.\phantom{xxxxxxxxxxx}-29H^2\vardot{3}{H}-95H\dot{H}\ddot{H}-10H\vardot{4}{H}-\vardot{5}{H} \right)\\
&-\frac{9\left(\xi-\frac{1}{6}\right)^3}{4\pi^2 m^2}\left(-8H^6+60H^4\dot{H}+11\dot{H}^3+42H^3\ddot{H}+45H\dot{H}\ddot{H}+3\ddot{H}^2+3\dot{H}\vardot{3}{H}\right.\\
&\phantom{xxxxxxxxxxx}\left.+102H^2 \dot{H}^2+6H^2\vardot{3}{H}\right)\,.\\
\end{split}
\end{equation}
We have used once more the conversion relations between the derivatives of $\mathcal{H}$ with respect to the conformal time and the derivatives of $H$ with respect to the cosmic time (see \hyperref[Appendix:Conventions]{Appendix\,\ref{Appendix:Conventions}}) so as to express the final result in terms of $H=H(t)$. Notice that the first three lines of the above expression comprise the terms up to the 4th adiabatic order while the remaining lines stand for the complete $6th$-order contributions. The subsequent contribution would be of adiabatic order $8th$, which we are not interested in here.

\subsection{Trace anomaly}\label{Sect:TraceAnomaly}

Let us use our results to reproduce the famous trace or conformal anomaly of the energy-momentum tensor in curved spacetime. In our case it is obtained upon computing Eq.\,\eqref{Eq:QuantumVacuum.TraceVEVconformal1} in the limit $m \to 0$. While this is not the interesting limit for our purposes, it serves nevertheless as a non-trivial calculational check. We know, at the classical level the trace of the EMT vanishes in the massless ($m=0$) conformal limit ($\xi=1/6$). Indeed, using equation \eqref{Eq:QuantumVacuum.ClassicTrace2} it is obvious that for ($\xi=1/6$) we get the result
\begin{equation}\label{Eq:QuantumVacuum.ConformalClassicalLimit}
T^{\rm cl.}=-m^2\phi^2 \,.
\end{equation}
Moreover, it is evident that $\lim_{m\rightarrow 0} T^{\rm cl.}=0$, and hence the classical trace of the EMT vanishes in the massless conformal limit, a well-known result which corresponds to the Noether identity following from the conformal invariance of the theory in that limit. However, if we move to the part of the trace inherent to the quantum fluctuations, we find from \eqref{Eq:QuantumVacuum.QuantumTrace} that in the conformal limit the trace takes the form $\langle T^{\delta \phi} \rangle=-m^2\langle \delta\phi^2 \rangle$, still consistent with \eqref{Eq:QuantumVacuum.ConformalClassicalLimit} since equations \eqref{Eq:QuantumVacuum.QuantumTrace} and \eqref{Eq:QuantumVacuum.ClassicTrace2} are formally identical, as we noted. We may naturally wonder if $\lim_{m\rightarrow 0} \langle T^{\delta \phi} \rangle=0$ holds good too. The naive answer is yes, but the correct (and also well-known) answer is no. This is the origin of the famous trace anomaly (also called conformal anomaly), see e.g.\,\cite{birrell1984quantum} and references therein. Basically, what happens is that the quantum fluctuation $\langle \delta\phi^2 \rangle$ involves terms $\sim 1/m^2$, which make the limit non-vanishing and independent of $m$. Setting $\xi=1/6$ in Eq.\,\eqref{Eq:QuantumVacuum.TraceVEV} we find the reduced result:
\begin{equation}\label{Eq:QuantumVacuum.TraceVEVconformal1}
\begin{split}
&\left.\left\langle T^{\delta \phi} \right\rangle\right|_{\xi=1/6}=
-\frac{m^2}{(2\pi)^3 a^2}\int d^3k |h_k|^2=-\frac{m^2}{2\pi^2 a^2}\int dk k^2 |h_k|^2\,.
\end{split}
\end{equation}
The explicit form of the result remains non-vanishing for $m\to 0$ and can be easily extracted from Eq.\,\eqref{Eq:QuantumVacuum.TraceDecompositionA} (see also Eq.\,\eqref{Eq:QuantumVacuum.TraceDecomposition}) above by setting $\xi=1/6$ and $M=m$. Notice that only the $4th$ adiabatic order terms contribute to the limit for $m\to 0$ since the $6th$ adiabatic order must decouple for $m \to\infty$, so it cannot be independent of $m$. Then, we have to pick out just the terms of $4th$ adiabatic order which are independent of $m$ under an appropriate change of integration variable (see below). These are the following:
\begin{equation}\label{Eq:QuantumVacuum.TraceVEVconformal2}
\begin{split}
 \left.\lim\limits_{m\to 0} \left\langle T^{\delta \phi} \right\rangle\right|_{(\xi=1/6,M=m )} &=\lim\limits_{m\to 0}\int \frac{dk k^2}{4\pi^2 a^4} \Bigg\{\frac{a^4 m^4}{16\omega_k^7}\left(8\mathcal{H}^4+24\mathcal{H}^2 \mathcal{H}^\prime+6(\mathcal{H}^\prime)^2+8\mathcal{H}\mathcal{H}^{\prime \prime}+\mathcal{H}^{\prime \prime \prime} \right)\\
&\phantom{xxxxxxxxxxxxxx} -\frac{7a^6 m^6}{32\omega_k^9}\left(28 \mathcal{H}^4+36\mathcal{H}^2 \mathcal{H}^\prime+3 (\mathcal{H}^\prime)^2+4\mathcal{H} \mathcal{H}^{\prime \prime} \right)\\
&\phantom{xxxxxxxxxxxxxx}+\frac{231 a^8 m^8}{32 \omega_k^{11}}\left(2\mathcal{H}^4+\mathcal{H}^\prime \mathcal{H}^2 \right)-\frac{1155 a^{10} m^{10} \mathcal{H}^4}{128 \omega_k^{13}}\Bigg\}\\
\end{split}
\end{equation}
The involved integrals are actually independent of $m$. To see that, let us make the change of variable $k= a m x$ in the last equation. Then, $\omega_k=\sqrt{k^2+a^2m^2}=am\sqrt{1+x^2}$, and we realize that the powers of $m$ in the numerator of all the above terms (plus the three added ones from $dk k^2 $) exactly cancel against those in the denominator, so the integrals do not actually depend on $m$ and hence the expression \eqref{Eq:QuantumVacuum.TraceVEVconformal2} cannot vanish for $m\to 0$:
\begin{equation}\label{Eq:QuantumVacuum.TraceVEVconformal2Expl}
\begin{split}
 \left.\lim\limits_{m\to 0} \left\langle T^{\delta \phi} \right\rangle\right|_{(\xi=1/6,M=m )} &=\lim\limits_{m\to 0} \int_0^\infty \frac{dx x^2}{4\pi^2 a^4}\left\{\frac{8\mathcal{H}^4+24\mathcal{H}^2 \mathcal{H}^\prime +6\left(\mathcal{H}^\prime\right)^2 +8\mathcal{H} \mathcal{H}^{\prime\prime}+\mathcal{H}^{\prime\prime\prime}}{16\left(1+x^2\right)^{7/2}} \right.\\
& \phantom{xxxxxxxxxxxxxxx}-\frac{7(28\mathcal{H}^4+36\mathcal{H}^2\mathcal{H}^\prime+3\left(\mathcal{H}^\prime\right)^2+4\mathcal{H}\mathcal{H}^{\prime\prime})}{32\left(1+x^2\right)^{9/2}}\\
&\phantom{xxxxxxxxxxxxxxx}\left.+\frac{231\left(2\mathcal{H}^4+\mathcal{H}^\prime\mathcal{H}^2\right)}{32\left(1+x^2\right)^{11/2}}-\frac{1155\mathcal{H}^4}{128\left(1+x^2\right)^{13/2}}\right\}.\\
\end{split}
\end{equation}
The result after some computations (with the help of the master integral formula in \hyperref[Sect:MasterInt]{Sect.\,\ref{Sect:MasterInt}}) reads
\begin{equation}\label{Eq:QuantumVacuum.TraceVEVconformal2res}
\begin{split}
 \left.\lim\limits_{m\to 0} \left\langle T^{\delta \phi} \right\rangle\right|_{(\xi=1/6,M=m )} &=\frac{1}{480\pi^2 a^4}\left(-4\mathcal{H}^2\mathcal{H}^\prime+\mathcal{H}^{\prime\prime\prime}\right)=\frac{1}{480\pi^2}\left(\frac{\ddot{a}^2}{a^2}+\frac{\vardot{4}{a}}{a}+3\frac{\dot{a}\vardot{3}{a}}{a^2}-3\frac{\dot{a}^2\ddot{a}}{a^3}\right).
\end{split}
\end{equation}
The latter is the result associated to the finite part of the effective action. Since the above calculation comes up from the unrenormalized part of the EMT, the trace anomaly is just minus the above result since the vacuum trace of the total EMT derived from the full effective action must be zero in the massless conformally coupled limit\,\cite{birrell1984quantum}. The latter can be computed first in the conformal metric and subsequently expressed in a covariant form, with the final result (for more details see the next section, particularly \hyperref[Sect:TraceAnomalyComp]{Sect.\,\ref{Sect:TraceAnomalyComp}}):
\begin{equation}\label{Eq:QuantumVacuum.TraceAnomaly}
\left.\lim\limits_{m\to 0} \left\langle T^{\delta \phi} \right\rangle\right|_{(\xi=1/6,M=m )}^{\rm anomaly} =\frac{1}{480\pi^2 a^4}\left(4\mathcal{H}^2\mathcal{H}^\prime-\mathcal{H}^{\prime\prime\prime}\right)=\frac{1}{2880\pi^2}\left[R^{\mu\nu}R_{\mu\nu}-\frac{1}{3} R^2+\Box R\right]\,.
\end{equation}
The last equation coincides with the standard covariant formulation of the anomaly, except that the square of the Weyl tensor (see its definition in \hyperref[Appendix:Conventions]{Appendix\,\ref{Appendix:Conventions}}) does not show up here (while it appears for more general backgrounds\,\cite{birrell1984quantum}) since the FLRW spacetime is conformal to Minkowski spacetime (i.e. it is a conformally flat spacetime) and hence the Weyl tensor vanishes identically.

\section{Effective action of QFT in curved spacetime}\label{Sect:EffectiveActionQFT}
The effective action describing the quantum matter vacuum effects of QFT in curved spacetime, $W$, is defined through its relation to the VEV of the EMT\,\cite{birrell1984quantum,parker2009quantum,fulling1989aspects}. In our conventions,
\begin{equation}\label{Eq:QuantumVacuum.DefW}
\langle T^{\mu\nu}\rangle=\frac{2}{\sqrt{-g}} \,\frac{\delta W}{\delta g_{\mu\nu}}\ \ \ \ \ \Longleftrightarrow\ \ \ \ \ \ \langle T_{\mu\nu}\rangle=-\frac{2}{\sqrt{-g}} \,\frac{\delta W}{\delta g^{\mu\nu}}\,.
\end{equation}
Such an effective action provides the quantum matter vacuum effects on top of the classical action. These quantum effects can be computed through a loopwise expansion in powers of $\hbar$. Thus, if the expansion is truncated at the one loop level it contains all terms of the complete theory to order $\hbar$.
It is well-known that at leading (one loop) order the value of $W$ for the free theory is essentially given by the trace of the logarithm of the inverse of the Green's function. More specifically:
\begin{equation}\label{Eq:QuantumVacuum.EAW}
\begin{split}
W= &\frac{i\hbar}{2}Tr \ln (-G_F^{-1})=-\frac{i\hbar}{2}Tr \ln (-G_F)\\
=& -\frac{i\hbar}{2} \int d^4 x \sqrt{-g}\lim\limits_{x\to x'} \ln\left[ -G_F(x,x')\right]\equiv \int d^4 x \sqrt{-g}\, L_W\,,
\end{split}
\end{equation}
wherein in the second line we have indicated the precise computational meaning of the trace in the spacetime continuum. The last equality defines the Lagrangian density $\sqrt{-g}\, L_W$, and for simplicity we will call the piece $L_W$ the (effective) quantum vacuum Lagrangian, as it accounts for the quantum vacuum effects from the quantized matter fields (in our case just the scalar field $\phi$). We retain $\hbar$ in the above expression just to emphasize the aforementioned fact that the above formula describes a pure quantum effect at one loop. From now on, however, we continue with $\hbar=1$ (as it has been done in most of the dissertation). The above action gives the vacuum effects, {\it i.e.} the effects originating from the matter vacuum-to-vacuum (`bubble') diagrams. These are closed loop diagrams without external tails, thereby rendering the zero-point energy contributions (ZPE). Let us note that, in the case of the free field theory based on the action \eqref{Eq:QuantumVacuum.Sphi}, we have no self-interactions of $\phi$ since there is no effective potential. In this situation we have one single bubble diagram and the one-loop effective action is the exact result since in the absence of matter interactions we cannot have additional vertices to insert in the bubble diagram to produce higher order loops.

The ZPE is usually discarded in flat spacetime on account of normal ordering of the operators (in the operator formulation) or by normalizing to one the generating functional of the Green's functions at zero value of the source (in the functional approach to QFT). In the context of gravity, none of these arbitrary settings is permitted. Although we have already performed the computation of the ZPE by directly computing the VEV of the enery-momentum tensor, {\it i.e.} the LHS of Eq.\,\eqref{Eq:QuantumVacuum.DefW}, here we wish to dwell further on these considerations in the context of the effective action approach by computing $W$ and then re-deriving the vacuum EMT using Eq.\,\eqref{Eq:QuantumVacuum.DefW}. While the procedure is well-known\,\cite{Bunch:1979uk,birrell1984quantum,parker2009quantum,fulling1989aspects}, we wish to discuss the changes introduced in it when we compute the effective action off-shell, as this is convenient to better understand the connection with the previous sections, in which we subtracted the EMT off-shell.

In curved spacetime, the Feynman propagator, $G_F$, is the solution to the following distributional differential equation\,\cite{birrell1984quantum,parker2009quantum,fulling1989aspects}:
\begin{equation}\label{Eq:QuantumVacuum.KGPropagatorOnShell}
\left(\Box_x-m^2-\xi R(x)\right)G_F(x,x^\prime)=-\left(-g(x)\right)^{-1/2}\delta^{(n)}(x-x^\prime)\,,
\end{equation}
where $\delta^{(n)}$ is the Dirac $\delta$ distribution in $n$ spacetime dimensions. For all practical purposes in our work, $n=4$. Notwithstanding we can keep $n$ general at the moment since DR will be employed for regularizing the UV divergences in the calculations presented in this section. We now reformulate the above on-shell equation in an appropriate form as follows:
\begin{equation}\label{Eq:QuantumVacuum.KGPropagatorOffShell}
\left(\Box_x-M^2-\Delta^2-\xi R(x)\right)G_F(x,x^\prime)=-\left(-g(x)\right)^{-1/2}\delta^{(n)}(x-x^\prime)\,.
\end{equation}
Here we have introduced a new scale, $M$, and also the important quantity:
\begin{equation}\label{eq:Delta2}
\Delta^2\equiv m^2-M^2\,.
\end{equation}
Although we have already introduced $\Delta^2$ in the context of the WKB expansion (cf. \hyperref[Sect:RelDiffRenScales]{Sect.\,\ref{Sect:RelDiffRenScales}}), we now endow it with a different perspective that may help to better understand its meaning. The strategy behind Eq.\,\eqref{Eq:QuantumVacuum.KGPropagatorOffShell} is to delve into the solution to the propagator equation (and hence of the effective action) for an arbitrary mass scale $M$. We can recover the on-shell case $M=m$ by simply setting $\Delta=0$. But if the quantity $\Delta^2$ is to be used to explore the off-shell regime it must be dealt with as being of adiabatic order higher than $M$ (which is of order zero). Hence $\Delta^2$ must be conceived as being of adiabatic order $2$, which is the next-to-leading order compatible with general covariance. Taking into account that the term $\xi R$ in \eqref{Eq:QuantumVacuum.KGPropagatorOffShell} is also of adiabatic order $2$, the combination $\Delta^2+\xi R$ can be treated as a block of adiabatic order $2$. This adiabaticity assignment for $\Delta^2$ is consistent with our former considerations in \hyperref[Sect:RelDiffRenScales]{Sect.\,\ref{Sect:RelDiffRenScales}} and can be regarded as an alternative justification for it. As long as the adiabaticity order of the terms must be hierarchical respected, the fact that the mass scale $M$ is of adiabatic order zero whereas the special quantity $\Delta^2$ is of adiabatic order $2$ is precisely what makes the solution to the Green's function equation \eqref{Eq:QuantumVacuum.KGPropagatorOffShell} different from the solution to the original (on-shell) equation \eqref{Eq:QuantumVacuum.KGPropagatorOnShell}. The adiabatic expansion of the solution to Eq.\,\eqref{Eq:QuantumVacuum.KGPropagatorOffShell} will generate new ($\Delta^2$-dependent) terms which are genuinely distinct as compared to the adiabatic expansion of the solution to \eqref{Eq:QuantumVacuum.KGPropagatorOnShell}. In what follows we work out such an expansion of the Green's function for a scalar field in curved spacetime\,\cite{birrell1984quantum}, with the purpose of identifying the (extra) $\Delta$-dependent terms characteristic of our off-shell subtraction procedure. See also the approach of\,\cite{Ferreiro:2020zyl}, which is however slightly different as we will comment later on.

To solve the Green's function equation in curved spacetime, Eq.\,\eqref{Eq:QuantumVacuum.KGPropagatorOffShell}, is not such a simple task as to find the corresponding solution in the flat spacetime case. Here we summarize the well-known procedure\,\cite{Bunch:1979uk,birrell1984quantum,parker2009quantum,fulling1989aspects} putting special emphasis on highlighting the differences introduced by the parameter $\Delta^2\equiv m^2-M^2$, Eq.\,\eqref{eq:Delta2}, which is crucial in our off-shell approach, see also\,\cite{Ferreiro:2018oxx} for a related formulation.

 \subsection{Computing the effective Lagrangian from the heat-kernel}\label{SubSect:HeatKernel}
 
The solution to \eqref{Eq:QuantumVacuum.KGPropagatorOffShell} can be obtained from the adiabatic expansion of the Green's function. The method is well-known\,\,\cite{birrell1984quantum,parker2009quantum} but is outlined here. A traditional method to circumvent the difficulty of dealing with a curved spacetime manifold has been to expand the metric around Minkowski space. A suitable implementation of this idea is to make a local expansion of the metric in Riemann normal coordinates, up to four derivatives of the metric (hence up to fourth adiabatic order).
In these coordinates, the metric admits the following expansion up to $4th$ order\cite{Bunch:1979uk}:
\begin{equation}
\begin{split}\label{Eq:QuantumVacuum.NormalCoordExp}
g_{\mu\nu}(y)&=\eta_{\mu \nu}-\frac{1}{3}R_{\mu \alpha \nu \beta}y^\alpha y^\beta-\frac{1}{6}R_{\mu \alpha \nu \beta;\gamma}y^\alpha y^\beta y^\gamma\\
&+\left[-\frac{1}{20}R_{\mu\alpha\nu\beta ; \gamma \delta}+\frac{2}{45}R_{\alpha\mu\beta\lambda}R^\lambda_{\ \gamma\nu\delta}\right]y^\alpha y^\beta y^\gamma y^\delta+\dots
\end{split}
\end{equation}
Here $y$ stands for the difference between the spacetime coordinate $x$ and the source point $x^\prime$ taken as a reference point in normal coordinates, {\it i.e.} $y=x-x^\prime$. The different curvature tensors and its derivatives (for instance $R_{\mu\alpha\nu\beta}$) that appear in the expansion above are assumed to be computed at the source point $x^\prime$ (i.e. at $y=0$). The same is true for the expansion of the determinant and the inverse of the metric.
For simplicity it is easier to define
\begin{equation}\label{Eq:QuantumVacuum.Redefine}
\mathcal{G}_F(x,x^\prime)=\left(-g(x)\right)^{1/4}G_F(x,x^\prime)\,.
\end{equation}
We can operate using the standard definition of curved spacetime box operator:
\begin{equation}\label{Eq:QuantumVacuum.boxoperator}
\begin{split}
\Box_x G_F(x,x^\prime)&=\Box_x \left((-g(x))^{-1/4}\mathcal{G}_F(x,x^\prime)\right)\\
&=\frac{1}{(-g(x))^{1/2}}\partial_\mu\left((-g(x))^{1/2}\partial^\mu \left((-g(x))^{-1/4}\mathcal{G}_F(x,x^\prime)\right)\right)\\
&=(-g(x))^{1/4}\Bigg[\frac{3}{16}\mathcal{G}_F(x,x^\prime)\frac{\partial_\mu (-g(x))\partial^\mu (-g(x))}{(-g(x))^2}-\frac{\partial_\mu \left(\mathcal{G}_F(x,x^\prime)\right)\partial^\mu (-g(x))}{4(-g(x))}\\
&\phantom{xxxxxxxxxxx}-\frac{\mathcal{G}_F(x,x^\prime)\partial_\mu\partial^\mu (-g(x))}{4(-g(x))}+ \frac{\partial^\mu \left(\mathcal{G}_F(x,x^\prime)\right)\partial_\mu (-g(x))}{4(-g(x))}\\
&\phantom{xxxxxxxxxxx}+\partial_\mu \partial^\mu \left(\mathcal{G}_F(x,x^\prime)\right)\Bigg]\,.
\end{split}
\end{equation}
In order to continue, we need to know the expansion of the determinant of the metric as well its inverse. For convenience, we define
\begin{equation}\label{Eq:QuantumVacuum.MetricExpansion}
g_{\mu\nu}(y)=\eta_{\mu \nu}+h_{\mu \nu}\,,
\end{equation}
where the deviation $h_{\mu \nu}$ from flat spacetime is written in powers of the normal coordinate $y$ according to \eqref{Eq:QuantumVacuum.NormalCoordExp}. We denote it as follows:
\begin{equation}\label{Eq:QuantumVacuum.hmunu}
h_{\mu \nu}=h^{(1)}_{\mu \nu}+h^{(2)}_{\mu \nu}+h^{(3)}_{\mu \nu}+h^{(4)}_{\mu \nu}+\cdots= h^{(2)}_{\mu \nu}+h^{(3)}_{\mu \nu}+h^{(4)}_{\mu \nu}+\cdots
\end{equation}
where $h^{(i)}_{\mu\nu}$ stands for the $ith$-term in the indicated order of \eqref{Eq:QuantumVacuum.NormalCoordExp}.
The missing term in the second equality is because \eqref{Eq:QuantumVacuum.NormalCoordExp} tells us that $h^{(1)}_{\mu \nu}=0$. Such linear term is missing because the expansion refers to a local inertial (Lorentz) frame, which is the tangent Lorentz frame at the point $x^\prime$ of the curved spacetime manifold. This demands not only $g_{\mu\nu}(0)=\eta_{\nu\nu}$ but also $\partial_{\alpha}g_{\mu\nu}(0)\equiv\partial g_{\mu\nu}/\partial y^{\alpha}(0)=0$, both being satisfied at $x^\prime$ (i.e. at $y=0$).
From the above expansion of the metric we can Taylor expand the corresponding determinant of it, $g(y)$:
\begin{equation}\label{Eq:QuantumVacuum.DeterminantMetricExp}
g(y)=g(0)+\frac{\partial g}{\partial g_{\mu \nu}}\Bigg|_{y=0} \,h_{\mu \nu}
+\frac{1}{2!}\,\frac{\partial^2 g}{\partial g_{\gamma \lambda}\partial g_{\mu \nu}}\Bigg|_{y=0}\,h_{\mu \nu}\,h_{\gamma \lambda}+\cdots\,,
\end{equation}
with $g(0)=-1$ and $h_{\mu \nu}$ given by \eqref{Eq:QuantumVacuum.NormalCoordExp}.
The derivatives of the determinant can be computed as follows:
\begin{equation}\label{Eq:QuantumVacuum.DerivativeDeterminantMetric}
\frac{\partial g}{\partial g_{\mu \nu}}=g(y) g^{\mu \nu}(y), \qquad
\frac{\partial^2 g}{\partial g_{\gamma \lambda}\partial g_{\mu \nu}}=g(y) g^{\gamma\lambda}(y)g^{\mu \nu}(y)-g(y) g^{\mu \gamma}(y) g^{\lambda \nu}(y).
\end{equation}
Furthermore, the expansion of the inverse of the metric in powers of the normal coordinate reads
\begin{equation}\label{Eq:QuantumVacuum.InverseMetric}
g^{ab}=\eta^{ab}-\eta^{a\mu}\eta^{b\nu}\,h_{\mu \nu}+\frac{1}{2!}\left(\eta^{a\lambda}\eta^{\mu \rho}\eta^{b\nu}+\eta^{a\mu}\eta^{b\lambda}\eta^{\nu \rho}\right)\,h_{\mu \nu}\,h_{\rho \lambda}+\cdots
\end{equation}
Thus, the previous calculations can be expanded up to fourth order as follows:
\begin{equation}
\begin{split}\label{Eq:QuantumVacuum.NormalCoordExpDet}
g(y)=&g(0)+h_{\mu\nu}\frac{\partial g}{\partial g_{\mu \nu}}\Bigg|_{y=0}+\frac{1}{2!}h^{(2)}_{\mu \nu}h^{(2)}_{\rho \lambda}\frac{\partial^2 g}{\partial g_{\rho \lambda}\partial g_{\mu \nu}}\Bigg|_{y=0}+\cdots\\
=&g(0)-\left( h^{(2)}_{\mu \nu}+h^{(3)}_{\mu \nu}+h^{(4)}_{\mu \nu}+\cdots\right) \eta^{\mu\nu}-\frac{1}{2!}h^{(2)}_{\mu \nu}h^{(2)}_{\rho \lambda}\left(\eta^{\rho\lambda}\eta^{\mu \nu}-\eta^{\mu \rho}\eta^{\lambda \nu}\right)+\cdots
\end{split}
\end{equation}
and
\begin{equation}\label{Eq:QuantumVacuum.NormalCoordExpInv}
\begin{split}
&g^{ab}=\eta^{ab}-\eta^{a\mu}\eta^{b\nu}\left(h_{\mu \nu}^{(2)}+h_{\mu \nu}^{(3)}+h_{\mu \nu}^{(4)}\right)+\frac{1}{2!}\left(\eta^{a\lambda}\eta^{\mu \rho}\eta^{b\nu}+\eta^{a\mu}\eta^{b\lambda}\eta^{\nu \rho}\right) h_{\mu \nu}^{(2)}h_{\rho \lambda}^{(2)}+\cdots
\end{split}
\end{equation}
Using \eqref{Eq:QuantumVacuum.NormalCoordExp} and the previous results we find, after some calculations:
\begin{equation}\label{Eq:QuantumVacuum.NormalCoordExpDetExplicit}
\begin{split}
g(y)&=-1+\frac{1}{3}R_{\alpha\beta}y^\alpha y^\beta+\frac{1}{6}R_{\alpha\beta;\gamma}y^\alpha y^\beta y^\gamma \\
&+\left[\frac{1}{20}R_{\alpha\beta ; \gamma \delta}-\frac{1}{18}R_{\alpha \beta}R_{\gamma \delta}+\frac{1}{90}R^\mu_{\ \alpha\beta\lambda} R^{\lambda}_{\ \gamma\delta \mu}\right]y^\alpha y^\beta y^\gamma y^\delta+\cdots
\end{split}
\end{equation}
and
\begin{equation}\label{Eq:QuantumVacuum.NormalCoordExpInvExplicit}
\begin{split}
g^{ab}&=\eta^{ab}+\frac{1}{3}\eta^{a\mu}\eta^{b\nu}R_{\mu\alpha\nu\beta}y^\alpha y^\beta+\frac{1}{6}\eta^{a\mu}\eta^{b\nu}R_{\mu\alpha\nu\beta;\gamma}y^\alpha y^\beta y^\gamma\\
&+\left[\frac{1}{20}\eta^{a\mu}\eta^{b\nu}R_{\mu\alpha\nu\beta;\gamma\delta}-\frac{2}{45}\eta^{a\mu}\eta^{b\nu}R_{\alpha\mu\beta\lambda}R^\lambda_{\ \gamma\nu\delta}\right.\\
&\left.\phantom{xx}+\frac{1}{18}\left(\eta^{b\nu}\eta^{ak}\eta^{\lambda\mu}+\eta^{a\mu}\eta^{bk}\eta^{\lambda \nu}\right)R_{\mu\alpha\nu\beta}R_{k\gamma\lambda\delta}\right]y^\alpha y^\beta y^\gamma y^\delta+\cdots
\end{split}
\end{equation}
It will be useful to consider the Fourier integrals and transforms
\begin{equation}\label{Eq:QuantumVacuum.FourierTransform1}
\mathcal{G}_F(x,x^\prime)=\frac{1}{(2\pi)^n}\int d^n k e^{iky}\mathcal{G}_F(k),
\end{equation}
\begin{equation}\label{Eq:QuantumVacuum.FourierTransform2}
i\eta^{\alpha \beta}y_\beta\mathcal{G}_F(x,x^\prime)=\frac{1}{(2\pi)^n}\int d^n k e^{iky}\frac{\partial}{\partial k_\alpha}\mathcal{G}_F(k)\,,
\end{equation}
with $ky\equiv \eta^{\alpha\beta}y_\alpha k_\beta$. Note that in normal coordinates we can raise and lower indices with the Minkowskian metric, as it can be easily shown from \eqref{Eq:QuantumVacuum.NormalCoordExp}. We organize the solution in adiabatic orders, {\it i.e.} counting the number of time derivatives of the metric:
\begin{equation}\label{Eq:QuantumVacuum.ExpansionPropagator}
\mathcal{G}_F(k)=\mathcal{G}_F^{(0)}(k)+\mathcal{G}_F^{(1)}(k)+\mathcal{G}_F^{(2)}(k)+\mathcal{G}_F^{(3)}(k)+\mathcal{G}_F^{(4)}(k)+\cdots
\end{equation}
Introducing this expansion in the propagator equation \eqref{Eq:QuantumVacuum.KGPropagatorOffShell} one can generate a solution of it in terms of an adiabatic series. The results, up to $4th$-order, are
\begin{equation}\label{Eq:QuantumVacuum.ExpansionPropagatorTerms}
\begin{split}
\mathcal{G}_F^{(0)}(k)&=\frac{1}{k^2+M^2}\,,\\
\mathcal{G}_F^{(1)}(k)&=0\,, \\
\mathcal{G}_F^{(2)}(k)&=-\frac{1}{(k^2+M^2)^2}\left(\left(\xi-\frac{1}{6}\right)R+\Delta^2\right)\,,\\
\mathcal{G}_F^{(3)}(k)&=-\frac{i}{2}\left(\xi-\frac{1}{6}\right)R_{;\alpha}\frac{\partial}{\partial k_\alpha}\left(\frac{1}{(k^2+M^2)^2}\right),\\
\mathcal{G}_F^{(4)}(k)&=\frac{1}{3}Q_{\alpha \beta}\frac{\partial^2}{\partial k_\alpha \partial k_\beta}\left(\frac{1}{(k^2+M^2)^2}\right)\\
&+\left[\left(\xi-\frac{1}{6}\right)^2R^2+\Delta^4+2\Delta^2R \left(\xi-\frac{1}{6}\right) -\frac{2}{3}{Q^\lambda}_\lambda\right]\frac{1}{(k^2+M^2)^3}\,,
\end{split}
\end{equation}
where we have defined
\begin{equation}\label{Eq:QuantumVacuum.Qalphabeta}
Q_{\alpha\beta}\equiv\frac{1}{2}\left(\xi-\frac{1}{6}\right)R_{;\alpha\beta}+\frac{1}{120}R_{;\alpha\beta}-\frac{1}{40}{R_{\alpha\beta;\lambda}}^\lambda+\frac{1}{30}{R_{\alpha}}^\lambda R_{\lambda \beta}-\frac{1}{60}{{{R^\kappa}_\alpha}^\lambda}_\beta R_{\kappa \lambda}-\frac{1}{60}{R^{\lambda \mu \kappa}}_\alpha R_{\lambda \mu \kappa \beta}.
\end{equation}
As we know, of the two parameters $M^2$ and $\Delta^2=m^2-M^2$ entering the propagator equation \eqref{Eq:QuantumVacuum.KGPropagatorOffShell}, the former is of adiabatic order $0$ whereas the latter is of adiabatic order $2$. One can easily recognize that the terms $\mathcal{G}_F^{(i)}(k)$ are of adiabatic orders $i=0,1, 2,3,4$, respectively, and represent successive corrections to the propagator solution up to $4th$-order.

The obtained solution represents an adiabatic expansion of the propagator in momentum space. Using Fourier integral formulas such as \eqref{Eq:QuantumVacuum.FourierTransform1}-\eqref{Eq:QuantumVacuum.FourierTransform2} we can transfer the solution to position space. Integrating by parts and neglecting the boundary terms, we find:
\begin{equation}\label{Eq:QuantumVacuum.FourierGF}
\begin{split}
\mathcal{G}_F (x,x^\prime)=\frac{1}{(2\pi)^n}\int d^n k e^{iky}&\Bigg\{ \hat{a}_0 (x,x^\prime)+\hat{a}_1 (x,x^\prime)\left(-\frac{\partial}{\partial M^2}\right)\\
& +\hat{a}_2(x,x^\prime)\left(-\frac{\partial}{\partial M^2}\right)^2+\cdots\Bigg\}\left(\frac{1}{k^2+M^2}\right),
\end{split}
\end{equation}
with
\begin{equation}\label{Eq:QuantumVacuum.bilocalWScoeff}
\begin{split}
&\hat{a}_0 (x,x^\prime)=1,\\
&\hat{a}_1 (x,x^\prime)=-\left(\xi-\frac{1}{6}\right)R-\Delta^2-\frac{1}{2}\left(\xi-\frac{1}{6}\right)R_{;\alpha}y^\alpha-\frac{1}{3}Q_{\alpha\beta}y^\alpha y^\beta ,\\
&\hat{a}_2 (x,x^\prime)=\frac{1}{2}\left(\xi-\frac{1}{6}\right)^2 R^2+\frac{\Delta^4}{2}+\Delta^2 R \left(\xi-\frac{1}{6}\right)-\frac{1}{3}{Q^\lambda}_\lambda\,.
\end{split}
\end{equation}
As we can see, these bilocal coefficients receive $\Delta^2$-dependent corrections in our case.
The quantity ${Q^\lambda}_\lambda$ in the last expression can be found explicitly by taking the trace of \eqref{Eq:QuantumVacuum.Qalphabeta}:
\begin{equation}\label{Eq:QuantumVacuum.traceQ}
{Q^{\lambda}}_\lambda= -\frac{1}{60} R^{\alpha\beta\gamma\delta}R_{\alpha\beta\gamma\delta}+\frac{1}{60} R^{\alpha\beta}R_{\alpha\beta}+\frac12 \left(\xi-\frac15\right)\Box R\,.
\end{equation}
Using the Euler's density $E$ and the square of the Weyl tensor ($C^2$) -- see \hyperref[Appendix:Conventions]{Appendix\,\ref{Appendix:Conventions}} --
we can rewrite \eqref{Eq:QuantumVacuum.traceQ} as follows:
\begin{equation}\label{Eq:QuantumVacuum.traceQ2}
\frac{1}{3}{Q^\lambda}_\lambda =-\frac{1}{120}C^2+\frac{1}{360}E+\frac{1}{6}\left(\xi-\frac{1}{5}\right)\Box R\,.
\end{equation}
The pole in \eqref{Eq:QuantumVacuum.FourierGF} must be shifted $M^2\rightarrow M^2-i\epsilon$ in order to have a time ordered product. In addition, we employ Schwinger's proper time representation\,\cite{Schwinger:1951nm,Schwinger:1951xk} of the zeroth order propagator through the following identity and corresponding derivatives with respect to the scale $M$:
\begin{equation}\label{Eq:QuantumVacuum.IntRepresent}
\begin{split}
&(k^2+M^2-i\epsilon)^{-1}=i\int_0^\infty ds e^{-is(k^2+M^2-i\epsilon)},\\
&\left(-\frac{\partial}{\partial M^2}\right)^j (k^2+M^2-i\epsilon )^{-1}=i\int_0^\infty (is)^j e^{-is(k^2+M^2-i\epsilon )}ds.
\end{split}
\end{equation}
This is the basis for subsequently obtaining the DeWitt-Schwinger representation of the sought-for Green's function in curved spacetime\,\cite{DeWitt:1975ys}, originally derived by DeWitt\,\cite{dewitt1965dynamical} following the work of Schwinger\,\cite{Schwinger:1951nm,Schwinger:1951xk}.
Using the integral representations \eqref{Eq:QuantumVacuum.IntRepresent} in the expression \eqref{Eq:QuantumVacuum.FourierGF} we can interchange the order of integration and perform first the following Gaussian integral in momentum space
\begin{equation}\label{Eq:QuantumVacuum.GaussianRep}
\int d^nk e^{iky-isk^2}=i\left(\frac{\pi}{is}\right)^{n/2}e^{-\sigma(x,x^\prime)/(2is)}\,,
\end{equation}
where the characteristic function $\sigma(x,x^\prime)$ (sometimes called the world function\,\cite{fulling1989aspects}) is one-half of the square of the geodesic distance between $x$ and $x^\prime$: $\sigma(x,x^\prime )= \frac{1}{2}y_\alpha y^\alpha\equiv \frac{1}{2}\,(x-x^\prime)^2$.
In this way the desired final form for the proper time representation of the Green's function \eqref{Eq:QuantumVacuum.Redefine} ensues:
\begin{equation}\label{Eq:QuantumVacuum.HK}
G_F(x,x^\prime)=\frac{i{\cal D}^{1/2}(x,x^\prime)}{(4\pi)^{n/2}}\int_0^\infty i ds \frac{e^{-iM^2 s-\sigma /(2is)}}{(is)^{n/2}}\left[\hat{a}_0(x,x^\prime)+is \hat{a}_1 (x,x^\prime)+(is)^2 \hat{a}_2 (x,x^\prime)+\cdots\right],
\end{equation}
where ${\cal D} (x,x^\prime)\equiv -\det\left(-\partial_\mu \partial_{\nu^\prime} \sigma (x,x^\prime) \right)/\sqrt{g(x)g(x^\prime)}$ is the general expression for the Van Vleck-Morette determinant, which reduces to ${\cal D}(x,x^\prime)=(-g(x))^{-1/2}$ for the case of normal coordinates. This, of course, agrees with the redefinition we made in \eqref{Eq:QuantumVacuum.Redefine}. One can easily recognize in \eqref{Eq:QuantumVacuum.HK} a generalized form of the fundamental solution of the heat (or diffusion) equation, {\it i.e.} its integral kernel.
Once we have the proper time representation of the propagator we may compute the effective Lagrangian $L_W$ associated to the quantum vacuum effective action,
\begin{equation}\label{Eq:QuantumVacuum.Waction}
W=-\frac{i}{2}Tr \ln (-G_F)=\int d^4 x \sqrt{-g} L_W\,.
\end{equation}
The trace in this expression is to be computed as specified in Eq.\,\eqref{Eq:QuantumVacuum.EAW}. Proceeding now in the standard way\,\cite{birrell1984quantum} the effective Lagrangian in $n$ spacetime dimensions can finally be put in the form of a DeWitt-Schwinger expansion at the arbitrary scale $M$:
\begin{equation}\label{Eq:QuantumVacuum.LWDR}
L_W(M)=\frac{\mu^{4-n}}{2(4\pi)^{n/2}}\sum_{j=0}^\infty \hat{a}_j (x) \int_0^\infty (is)^{j-1-n/2}e^{-iM^2 s}ids,
\end{equation}
where $\mu$ is 't Hooft's mass unit introduced by dimensional purposes (viz. in this case to maintain $L_W$ with natural dimension $4$ in $n$ spacetime dimensions) and $\hat{a}_j(x)\equiv\hat{ a}_j(x,x)$ are the corresponding DeWitt-Schwinger coefficients, which appear after computing the coincidence limits $x\to x^\prime$ (i.e. $y \to 0$) of the bilocal coefficients \eqref{Eq:QuantumVacuum.bilocalWScoeff}. Upon implementing this limit, the final DeWitt-Schwinger coefficients carry $\Delta^2$-dependent correction terms, as follows:
\begin{equation}\label{Eq:QuantumVacuum.ModifDWScoeff}
\begin{split}
&\hat{a}_0 (x)=1=a_0 (x),\\
&\hat{a}_1 (x)=-\left(\xi-\frac{1}{6}\right)R-\Delta^2=a_1(x)-\Delta^2 ,\\
&\hat{a}_2 (x)=\frac{1}{2}\left(\xi-\frac{1}{6}\right)^2R^2+\frac{\Delta^4}{2}+\Delta^2 R \left(\xi-\frac{1}{6}\right)-\frac{1}{3}Q^\lambda_{\ \lambda}=a_2(x)+\frac{\Delta^4}{2}+\Delta^2 R \left(\xi-\frac{1}{6}\right)\,,
\end{split}
\end{equation}
where include the zero, second and fourth adiabatic orders, respectively. The hatless $a_j(x)$ represent the ordinary DeWitt-Schwinger coefficients when $\Delta=0$ (on-shell expansion),
\begin{equation}\label{Eq:QuantumVacuum.ClassicDWScoeff}
\begin{split}
&a_0 (x)=1,\\
&a_1(x)=-\left(\xi-\frac{1}{6}\right)R ,\\
&a_2(x)=\frac{1}{2}\left(\xi-\frac{1}{6}\right)^2R^2-\frac{1}{3}{Q^\lambda}_\lambda\,.
\end{split}
\end{equation}
Expressed in this way we can more clearly see what is the effect of performing the expansion off-shell.
Computing the integral involved in \eqref{Eq:QuantumVacuum.LWDR} with the help of the Euler $ \Gamma$ function, we find
\begin{equation}\label{Eq:QuantumVacuum.LWDR2}
\begin{split}
L_W=&\frac{\mu^{4-n}}{2(4\pi)^{n/2}} \sum_{j=0}^\infty \hat{a}_j (x) \int_0^\infty (is)^{j-1-n/2}e^{-iM^2 s}ids\\
=&\frac{1}{2(4\pi)^{2+\frac{\varepsilon}{2}}}\left(\frac{M}{\mu}\right)^{\varepsilon}\sum_{j=0}^\infty \hat{a}_j (x) M^{4-2j}\Gamma \left(j-2-\frac{\varepsilon}{2}\right)\,,
\end{split}
\end{equation}
where $\varepsilon\equiv n-4$ and the limit $\varepsilon \rightarrow 0$ is understood; $\mu$ is 't Hooft's mass unit to keep the effective Lagrangian with natural dimension $+4$ of energy in $n$ spacetime dimensions. The final results will not depend on it. The sum is over $j=0,1,2,...$ and includes the even adiabatic orders only.

\subsection{Renormalization of effective Lagrangian of QFT in curved spacetime}\label{SubSect:RenormEffLag}

We wish to pay special attention to the modification introduced by the presence of the extra terms $\Delta^2$ on top of the usual procedure. The final result \eqref{Eq:QuantumVacuum.LWDR2} allows us to determine the effective Lagrangian defined in Eq.\,\eqref{Eq:QuantumVacuum.EAW} and express it in the form of an asymptotic DeWitt-Schwinger expansion\,\cite{DeWitt:1975ys}. We have explicitly checked that one can obtain the same results with the subtraction procedure used to renormalize the EMT in the previous sections. The effective Lagrangian \eqref{Eq:QuantumVacuum.LWDR2} and corresponding effective action are UV-divergent quantities since the Euler's $\Gamma$-function is divergent for $j=0,2,4$ in $n=4$ spacetime dimensions.

Let us now consider some renormalization issues. Our starting point was the Einstein-Hilbert action \eqref{Eq:QuantumVacuum.LWDR2} together with the quantum matter action \eqref {Eq:QuantumVacuum.SrL}. The former is associated to the Lagrangian
\begin{equation}
L_{EH}=-\rho_\Lambda+\frac{1}{16\pi G} R=-\rho_\Lambda+\frac{1}{2}\Mpl^2 R\,,
\end{equation}
which represents the starting (classical) vacuum action. We have nevertheless observed in our discussion on the EMT renormalization in \hyperref[Sect:RenormZPE]{Sect.\,\ref{Sect:RenormZPE}} that, even though we did not start with higher derivative (HD) terms in the action, such as $R^2(x)$, $R^{\mu\nu}(x)R_{\mu\nu}(x),\dots$ these purely geometric structures are generated by the quantum fluctuations of the matter field, which probe the short distances around $x$. Therefore, renormalizability of QFT in the FLRW background requires that the more general classical action comprises also these HD geometric structures. Let us write the extended classical gravitational Lagrangian for the vacuum with all the necessary terms in two alternative ways as follows:
\begin{equation}\label{Eq:QuantumVacuum.LEHHD}
\begin{split}
L_G^{\rm cl.}=L_{EH}+L_{HD}=&-\rho_\Lambda +\frac{1}{2}\Mpl^2R+\alpha_Q \frac{{Q^\lambda}_\lambda}{3}+\alpha_2 R^2\\
=&-\rho_\Lambda+\frac{1}{2}\Mpl^2 R+\alpha_1 C^2+\alpha_2 R^2+\alpha_3 E+\alpha_4\Box R\,,
\end{split}
\end{equation}
in which the notation in $L_G^{\rm cl.}$ indicates that this is the classical Lagrangian part of the gravitational field, to which we still have to add the quantum vacuum effects. Coefficients $\alpha_i$ for $i=1,3,4$ in the second expression can be easily related with the coefficient $\alpha_Q$ if we take into account that the combined HD structure ${Q^\lambda}_\lambda$ can be phrased in terms of the square of the Weyl tensor ($C^2$), the Euler density ($E$) and a total derivative term as follows (cf. \hyperref[Appendix:Conventions]{Appendix\,\ref{Appendix:Conventions}}):
\begin{equation}\label{Eq:QuantumVacuum.traceQ1}
\frac{1}{3}{Q^\lambda}_\lambda =-\frac{1}{120}C^2+\frac{1}{360}E+\frac{1}{6}\left(\xi-\frac{1}{5}\right)\Box R\,.
\end{equation}
These HD terms did not appear when we renormalized the EMT in \hyperref[Sect:RenormZPE]{Sect.\ref{Sect:RenormZPE}} since we used a restricted generalization of Einstein's equations, viz. Eq.\,\eqref{Eq:QuantumVacuum.MEEs}, which is sufficiently general for the FLRW spacetime. The three terms \eqref{Eq:QuantumVacuum.traceQ1} appear in a natural way in the effective action approach since they are involved as part of the DeWitt-Schwinger coefficient $a_2$ of Eq.\eqref{Eq:QuantumVacuum.ClassicDWScoeff}, so we have just computed them within the natural flow of the effective action procedure, but none of these terms actually plays any role for FLRW spacetime since the latter in conformal to the Minkowski metric and hence the Weyl tensor vanishes identically. The other two are also irrelevant at the level of the action since $E$ leads to a topological invariant in $n=4$ dimensions, the Gauss-Bonnet term $\GB$ (cf. \hyperref[Appendix:Conventions]{Appendix\,\ref{Appendix:Conventions}}), and $\Box R$ is a total derivative. We have carried along these HD terms up to this point just for completeness, but in effect the only HD term which stays in the FLRW background is $R^2$, as we warned in \hyperref[Sect:RenormZPE]{Appendix\,\ref{Sect:RenormZPE}}. We will nonetheless still keep these terms in the next section so as to close our discussion on the effective action method in a more complete way.

Starting from $L_W$ and following a procedure similar to our definition of adiabatically renormalized EMT, see Eq.\,\eqref{Eq:QuantumVacuum.EMTRenormalizedDefinition}, we define now the renormalized quantum vacuum Lagrangian at the scale $M$. It is obtained by subtracting the divergent adiabatic orders at this scale from the on-shell value $L_W (m)$:
\begin{equation}\label{Eq:QuantumVacuum.LWrenormalized}
L_W^{\rm ren}(M )= L_W (m)-L_W^{(0-4)}(M)\equiv L_W (m)-L_{\rm div} (M)\,,
\end{equation}
where $L_{\rm div}(M)\equiv L_W^{(0-4)}(M)$ is the divergent part of Eq.\,\eqref{Eq:QuantumVacuum.LWDR2}; by this we mean that $L_{\rm div}$ is that part of $L_W$ involving only the terms $j=0,1,2$, {\it i.e.} up to fourth adiabatic order. Of course both $L_W (m)$ and $L_{\rm div} (M)$ are divergent, but the former is assumed to involve the full DeWitt-Schwinger expansion at the scale $m$, whilst the latter stops the expansion at $j=2$ and is evaluated at a different scale $M$. This subtraction prescription for the quantum vacuum Lagrangian is the exact analogue of the off-shell ARP that we used for the EMT and it is sufficient to make $L_W^{\rm ren}(M)$ a finite quantity. The above renormalized Lagrangian describes the vacuum effects from the quantum matter (in this case, the scalar field $\phi$) and it must be added up to the classical vacuum Lagrangian so as to form the total vacuum Lagrangian. We do this in the next section.

Upon expanding Euler's $\Gamma$-function in the limit $\varepsilon\to 0$ and using the explicit form of the modified DeWitt-Schwinger coefficients \eqref{Eq:QuantumVacuum.ModifDWScoeff}, we find after a relatively lengthy but straightforward calculation the following result:
\begin{equation}\label{Eq:QuantumVacuum.LwrenM}
\begin{split}
L_W^{\rm ren}(M)=\delta \rho_\Lambda(M)-\frac{1}{2}\delta\Mpl^2(M) R-\delta \alpha_Q(M) \frac{{Q^\lambda}_\lambda}{3}-\delta \alpha_2(M) R^2+\cdots\,,
\end{split}
\end{equation}
where the dots stand for subleading contributions which decouple at large $m$, and
\begin{equation}\label{Eq:QuantumVacuum.deltacouplings}
\begin{split}
&\delta\rL(M)=\frac{1}{8\left(4\pi\right)^2}\left(M^4-4m^2M^2+3m^4-2m^4 \ln \frac{m^2}{M^2}\right),\\
&\delta\Mpl^2(M) =\frac{\left(\xi-\frac{1}{6}\right)}{(4\pi)^2}\left(M^2-m^2+m^2\ln \frac{m^2}{M^2}\right),\\
&\delta \alpha_Q(M)=-\frac{1}{2(4\pi)^2}\ln\frac{m^2}{M^2},\\
&\delta{\alpha_2}(M)=\frac{\left(\xi-\frac{1}{6}\right)^2}{4(4\pi)^2}\ln\frac{m^2}{M^2}.
\end{split}
\end{equation}
As promised, the dependence on $\mu$ fully cancelled out along with the poles at $n=4$. We have used DR to verify the cancellation of the UV-divergences (similarly to the procedure used in the \hyperref[Appendix:Dimensional]{Appendix\,\ref{Appendix:Dimensional}}). We emphasize that the use of DR is auxiliary here, it can be done with other regulators, the final result has no memory of this intermediate step. The chief difference here is not so much about regularization but about renormalization\,\footnote{We emphasize that the subtracted term $L_{\rm div}(M)$ at the scale $M$ in \eqref{Eq:QuantumVacuum.LWrenormalized} involves not just the UV-divergences but the full expression obtained from the sum of the first three terms ($j=0,1,2$) in the DeWitt-Schwinger expansion\,\eqref{Eq:QuantumVacuum.LWDR2}, including their finite parts (cf. Eq.\,\eqref{Eq:QuantumVacuum.LdivMdef}), hence fully in consonance with the procedure Eq.\,\eqref{Eq:QuantumVacuum.EMTRenormalizedDefinition} utilized for the EMT. This renormalization prescription is, of course, entirely different from MS renormalization.}. The quantities \eqref{Eq:QuantumVacuum.deltacouplings} are finite renormalization effects associated to the quantum vacuum Lagrangian $L_W$.

\subsection{Running couplings}\label{SubSect:RunningCouplings}

We are now ready to modify the classical or background vacuum Lagrangian \eqref{Eq:QuantumVacuum.LEHHD} by including the quantum matter effects generated in our scalar field model and in this way to track the shift received by each parameter as a function of the renormalization point $M$. This will allow us to derive the running couplings. The full effective Lagrangian from which we can extract physical information up to one loop (actually the complete result at the quantum level, in the absence of scalar self-interactions) is obtained by adding the extended classical Lagrangian of gravity plus the (renormalized) quantum effects, {\it i.e.} the sum of equations \eqref{Eq:QuantumVacuum.LEHHD} and \eqref{Eq:QuantumVacuum.LwrenM}:
\begin{equation}\label{Eq:QuantumVacuum.Full-Leff1}
\begin{split}
L_{\rm eff}&=L_G^{\rm cl.}(M)+L_W^{\rm ren}(M)=-\rL(M)+\frac{1}{2}\Mpl^2(M) R+\alpha_1(M) C^2+\alpha_2(M) R^2+\alpha_3(M) E\\
&+\alpha_4(M)\Box R+\delta \rL(M)-\frac{1}{2}\delta\Mpl^2(M) R-\delta \alpha_Q(M) \frac{{Q^\lambda}_\lambda}{3}-\delta \alpha_2(M) R^2+\cdots\\
\end{split}
\end{equation}
where the dots represent the subleading finite pieces emerging from the DeWitt-Schwinger expansion \,\eqref{Eq:QuantumVacuum.LWDR2} presented earlier\footnote{We have not computed these terms in the effective action formalism (in contrast to the calculation that we have previously performed within the direct EMT approach, where we have reached up to the (finite) $6th$ adiabatic order. Here we just want to cross-check the core design of the renormalization procedure within the effective action method and confirm that we obtain the same results.}.
Notice that the couplings of the classical part are dependent on the renormalization scale $M$ since the above expression represents the full effective renormalized Lagrangian of the theory. Overall it is independent of $M$ (i.e. RG-invariant), but each coupling `runs' (scales) with $M$ even though there is a net internal compensation among all the scaling dependencies.
It is convenient to rearrange \eqref{Eq:QuantumVacuum.Full-Leff1} as follows:
\begin{equation}\label{Eq:QuantumVacuum.Full-Leff}
\begin{split}
L_{\rm eff}&=\left[-\rL(M)+\delta\rL(M) \right]+\frac12\left[\Mpl^2(M)-\delta\Mpl^2(M)\right] R\\
&+\left[\alpha_1 (M)+\frac{1}{120} \delta\alpha_Q(M)\right] C^2+\left[\alpha_3 (M)-\frac{1}{360}\delta\alpha_Q(M)\right] E\\
&+\left[\alpha_4 (M)-\frac{1}{6}\left(\xi-\frac15\right) \delta\alpha_Q(M)\right]\Box R+\left[\alpha_2 (M)-\delta\alpha_2(M)\right] R^2+\dots\\
\end{split}
\end{equation}
where we have used Eq.\,\eqref{Eq:QuantumVacuum.traceQ1}. As previously remarked, the full effective Lagrangian $L_{\rm eff}$ must be independent of the renormalization point $M$. It follows that each one of the quantities in the square brackets of \eqref{Eq:QuantumVacuum.Full-Leff} must be independent of the scale $M$, and this allows us to readily compute the $\beta$-functions for each of the couplings:
\begin{equation}\label{Eq:QuantumVacuum.BetaFunctionrL}
\beta_{\rL} (M)=\frac{1}{2(4\pi)^2}(M^2-m^2)^2
\end{equation}
\begin{equation}\label{Eq:QuantumVacuum.BetaFunctionMPl}
\beta_{\Mpl^2} (M)=\frac{\left(\xi-\frac{1}{6}\right)}{8\pi^2} (M^2-m^2)
\end{equation}
and
\begin{equation}\label{Eq:QuantumVacuum.BetaFunctions12}
\begin{split}
\beta_{\alpha_1}=-\frac{1}{120(4\pi)^2}\ \ \ \ \ \
 \beta_{\alpha_2}=-\frac{\left(\xi-\frac{1}{6}\right)^2}{2(4\pi)^2}
 \end{split}
\end{equation}
 \begin{equation}\label{Eq:QuantumVacuum.BetaFunctions34}
\begin{split}
 \beta_{\alpha_3}=\frac{1}{360 (4\pi)^2}\ \ \ \ \ \ \
 \beta_{\alpha_4}=\frac{\xi-\frac15}{6(4\pi)^2}\,.
\end{split}
\end{equation}
We have used the explicit expressions \eqref{Eq:QuantumVacuum.deltacouplings} for the calculation of the $\beta$-functions through
\begin{equation}\label{Eq:QuantumVacuum.BetaFunction}
 \beta_i=M \frac{\partial \lambda_i(M)}{\partial M}
\end{equation}
for each of the couplings $(\lambda_i=\rL,\Mpl^2,\alpha_1,...,\alpha_4)$ present in the effective lagrangian\,\footnote{Related formulas have been considered in\,\cite{Ferreiro:2020zyl}. Let us, however, note that they differ from ours in that we consider the scale $M$ as the primary off-shell quantity from which to parameterize the quantum effects, rather than the difference $\Delta^2$ (called $-\mu^2$ in their case).}.

Let us note that in our approach the decoupling effects of physical quantities, such as the vacuum energy density itself, satisfy the Appelquist-Carazzone theorem\,\cite{Appelquist:1974tg}. This is apparent in our $6th$-order formulas, in the limit of large $m$, see e.g. Eq.\,\eqref{Eq:QuantumVacuum.renormalized6th}. This is not to be expected for the couplings in general, as they do not have the same level of physical significance. For example, we know that $\rL(M)$, which satisfies the first renormalization group equation (RGE) above, is a formal quantity which does not appear in the physical results. Only the vacuum EMT has physical meaning, and in particular the VED, so there is no need in general for the couplings to satisfy manifest decoupling.

It is straightforward to integrate the corresponding RGE's and derive the explicit running of the couplings with the renormalization point $M$, assuming that they are defined at some initial value $M_0$:
\begin{equation}\label{Eq:QuantumVacuum.RGEscouplings}
\begin{split}
&\rho_\Lambda(M)=\rho_\Lambda(M_0 )+\frac{1}{8(4\pi)^2}\left(M^4-M_0^4-4m^2(M^2-M_0^2)+2m^4\ln \frac{M^2}{M_0^2}\right),
\\
&\Mpl^2(M)=\Mpl^2(M_0)+ \frac{\left(\xi-\frac{1}{6}\right)}{(4\pi)^2}\,\left(M^2-M_0^2-m^2\ln \frac{M^2}{M_0^2}\right),\\
&\alpha_1 (M)=\alpha_1 (M_0)-\frac{1}{240(4\pi)^2}\ln \frac{M^2}{M_0^2},
\\
&\alpha_2 (M)=\alpha_2 (M_0)-\frac{\left(\xi-\frac{1}{6}\right)^2}{4(4\pi)^2}\ln \frac{M^2}{M_0^2},\\
&\alpha_3 (M)=\alpha_3 (M_0)+\frac{1}{720(4\pi)^2 }\ln \frac{M^2}{M_0^2},\\
&\alpha_4 (M)=\alpha_4 (M_0)+\frac{\xi-\frac15}{12(4\pi)^2}\ln \frac{M^2}{M_0^2}.
\end{split}
\end{equation}
The equation for the running (reduced) Planck mass squared $\Mpl^2(M)= 1/\left(8\pi G(M)\right)$ given above can also be cast in terms of the running Newton's constant:
\begin{equation}\label{Eq:QuantumVacuum.RGENewton}
G(M)=\frac{G(M_0)}{1+\frac{\left(\xi-\frac{1}{6}\right)}{2\pi}G(M_0)\left(M^2-M_0^2-m^2\ln \frac{M^2}{M_0^2}\right)}\,.
\end{equation}
The previous equation can be related to the physical running of the gravitational coupling during the cosmological expansion. In \hyperref[Sect:RenormalizedFriedmann]{Sect.\,\ref{Sect:RenormalizedFriedmann}}, in the next chapter, we further dwell upon the running of the gravitational coupling in combination with that of the VED, and will come back to Eq.\,\eqref{Eq:QuantumVacuum.RGENewton}.

We can see that the first two Renormalization Group Equations (RGE) solutions and the fourth one in \eqref{Eq:QuantumVacuum.RGEscouplings} are nothing but equations \eqref{Eq:QuantumVacuum.SubtractionrL}, \eqref{Eq:QuantumVacuum.SubtractionMPl} and \eqref{Eq:QuantumVacuum.Subtractionalpha}, respectively (with $\alpha_2=\alpha/2$) which we found in the process of renormalization of the EMT. Overall we have met at this point a rather non-trivial consistency check between the renormalization procedure of the EMT from which we started our calculation in \hyperref[Sect:RenormZPE]{Sect.\,\ref{Sect:RenormZPE}}, and the alternative approach based on the renormalization of the effective action (and corresponding effective Lagrangian), which we have undertaken in this section\,\footnote{Although the RGEs for $\alpha_1,\alpha_3$ and $\alpha_4$ were not discussed in \hyperref[Sect:RenormZPE]{Sect.\,\ref{Sect:RenormZPE}}, we have derived them en route in the effective action approach only for completeness.}. In other words, it confirms that the renormalized couplings that we have now computed from the effective Lagrangian method are indeed the same parameters which appeared in the renormalized EMT following from the original prescription \eqref{Eq:QuantumVacuum.EMTRenormalizedDefinition} and performing the corresponding subtraction $\delta X(M,M_0)= X(M)-X(M_0)$ in the renormalized Einstein's equations \eqref{Eq:QuantumVacuum.MEEsHD}. In a similar way, we can easily check that the relations \eqref{Eq:QuantumVacuum.deltacouplings} can be recovered now as a particular case of the above running solutions for the case $M_0=m$ upon defining $\delta X(M)\equiv\delta X(M,m)= X(M)-X(m)$ for each of the parameters $X=\rL, \Mpl^2, \alpha_i$. For instance, using the first relation in \eqref{Eq:QuantumVacuum.RGEscouplings} we find
\begin{equation}\label{Eq:QuantumVacuum.RGErL}
\delta\rL(M)=\rL(M)-\rL(m)=\frac{1}{8\left(4\pi\right)^2}\left(M^4-4m^2M^2+3m^4-2m^4 \ln \frac{m^2}{M^2}\right)\,,
\end{equation}
which matches the first one of \eqref{Eq:QuantumVacuum.deltacouplings}. Similarly for the other parameters.
Let us finally pause at this point to observe that there is a long way mediating between these two approaches, namely, the one based on tackling a direct renormalization of the EMT by means of the adiabatic procedure and the other based on computing the effective action from the DeWitt-Schwinger expansion. However different they are, they appear to be fully consistent. This fact is, of course, very much welcome as it demonstrates the cogency and congruence of the results obtained in our calculation. The touchstone of such a consistency can be made even more transparent if we compute the functional derivative of the action associated to the renormalized vacuum effective Lagrangian with respect to the metric, {\it i.e.} if we show how to recover the renormalized EMT obtained in previous sections using the effective action itself, Eq.\,\eqref{Eq:QuantumVacuum.DefW}.

\subsection{Renormalizing the effective action}\label{sec:RenW}

The effective vacuum action $W$ and corresponding Lagrangian $L_W$ obtained in the previous section shows up in the form of a DeWitt-Schwinger expansion. However, it is divergent
since the first terms $j=0,1,2$ are UV-divergent and include the contributions up to $4th$ adiabatic order. This so-called divergent part of $L_W$ at the scale $M$ is defined through
\begin{equation}\label{Eq:QuantumVacuum.LdivMdef}
L_{\rm div}(M)\equiv L_W^{(0-4)}(M)=\lim\limits_{\varepsilon \rightarrow 0}\frac{1}{2(4\pi)^{2+\varepsilon}}\left(\frac{M}{\mu}\right)^{\varepsilon}\sum_{j=0}^2 \hat{a}_j (x) M^{4-2j}\Gamma \left(j-2-\frac{\varepsilon}{2}\right)\,.
\end{equation}
The divergent character is apparent since the $\Gamma$ function has poles for $j\leq 2 $ in the limit $\epsilon\to 0$. Therefore it requires renormalization. Since we are tracking the poles through DR, it is convenient to expand $L_{\rm div}$ for $\epsilon\to 0$. We find
\begin{equation}\label{Eq:QuantumVacuum.LdivM}
\begin{split}
L_{\rm div}(M)&=\frac{1}{2\left(4\pi\right)^2}\left(1+ \frac{\epsilon}{2}\ln\frac{M^2}{4\pi\mu^2}+\mathcal{O}\left(\epsilon^2\right)\right)\Bigg[\hat{a}_0 (x)M^4\left(-\frac{1}{\epsilon}-\frac{\gamma_E}{2}+\frac{3}{4}+\mathcal{O}\left(\epsilon\right)\right)\\
&+\hat{a}_1(x)M^2\left(\frac{2}{\epsilon}+\gamma_E-1+\mathcal{O}\left(\epsilon\right)\right)+\hat{a}_2(x)\left(-\frac{2}{\epsilon}-\gamma_E+\mathcal{O}\left(\epsilon\right)\right)\Bigg]\\
&=\frac{1}{2\left(4\pi\right)^2}\Bigg[\frac{1}{\epsilon}\left(-\hat{a}_0(x)M^4+2\hat{a}_1(x)M^2-2\hat{a}_2(x)\right)+\gamma_E\left(-\frac{1}{2}\hat{a}_0(x)M^4+\hat{a}_1(x)M^2-\hat{a}_2(x)\right)\\
&+\hat{a}_0(x)M^4\left(\frac{3}{4}-\frac{1}{2}\ln\frac{M^2}{4\pi\mu^2}\right)+\hat{a}_1(x)M^2\left(-1+\ln\frac{M^2}{4\pi\mu^2}\right)-\hat{a}_2(x)\ln\frac{M^2}{4\pi\mu^2}\Bigg]\,.
\end{split}
\end{equation}
To perform the renormalization, we could generate UV-divergent counterterms by splitting the parameters of the extended classical Lagrangian (including the HD terms) into a renormalized parameter plus an UV-divergent counterterm -- cf. Eq.\,\eqref{Eq:QuantumVacuum.splitcounters} -- and then cancel the divergences of $L_W$ leaving some arbitrary finite parts. However, we do not want to use this procedure (MS scheme or variations thereof) since it does not produce acceptable results in this context. Instead, we wish to renormalize the effective action $W$ and corresponding effective Lagrangian in the same way as we did with the EMT, namely by performing a subtraction at another scale. Thus, we define the renormalized vacuum effective Lagrangian $L_W$ at the scale $M$ through the subtraction prescription
\begin{equation}\label{Eq:QuantumVacuum.LWrenormalizedII}
L_W^{\rm ren}(M)= L_W (m)-L_{\rm div} (M),
\end{equation}
with $L_{\rm div} (M)$ the divergent part of $L_W$, as defined in \eqref{Eq:QuantumVacuum.LdivMdef}-\eqref{Eq:QuantumVacuum.LdivM}. The latter involves terms only up to adiabatic order $4$, precisely as in the case of the definition of the renormalized EMT -- see Eq.\,\eqref{Eq:QuantumVacuum.EMTRenormalizedDefinition}. Thus, $L_W^{\rm ren}(M)$ is a finite quantity. Notice that $L_W^{\rm ren}(m )= L_W (m)-L_{\rm div} (m)$ is also finite, of course: it is zero if $L_W(m)$ is evaluated up to $j=2$ but is non-vanishing if $L_W(m)$ is evaluated beyond $j=2$ (i.e. beyond $4th$ adiabatic order).

To exhibit the finiteness of the renormalized Lagrangian, let us compute $L_W^{\rm ren}(M)$ explicitly. Notice that
\begin{equation}\label{Eq:QuantumVacuum.LWRenaprox}
L_W^{\rm ren}(M)= L_W (m)-L_{\rm div} (M)= L_{\rm div} (m)-L_{\rm div} (M)+\cdots
\end{equation}
The dots in this expression represent finite subleading terms (viz. higher than $4th$ adiabatic order) emerging from the DeWitt-Schwinger expansion\,\eqref{Eq:QuantumVacuum.LWDR2} of $L_W (m)$. These subleading terms decouple for large values of the mass $m$ of the scalar field, as can be easily seen from Eq.\,\eqref{Eq:QuantumVacuum.LWDR2} for $j>2$ and $M=m$. Thus, if we are just interested in tracking the cancellation of divergences and the finite parts left in the process, it is enough to compute $L_{\rm div} (m)-L_{\rm div} (M)$. For the sake of convenience concerning other formulas used in the main text, it will be more useful to first perform the subtraction between two arbitrary scales $M$ and $M_0$:
\begin{equation}\label{Eq:QuantumVacuum.Lren}
L_W^{\rm ren}(M)-L_W^{\rm ren}(M_0)= L_{\rm div} (M_0)-L_{\rm div} (M)\,.
\end{equation}
Although the calculation of this quantity is straightforward it is a bit laborious, as there are many terms. In particular, one has to use the explicit form of the modified DeWitt-Schwinger coefficients \eqref{Eq:QuantumVacuum.ModifDWScoeff}. Notice that these coefficients depend on the quantity $\Delta^2(M)=m^2-M^2$ when we perform the calculation of $L_{\rm div}(M)$, whereas they depend on $\Delta^2(M_0)=m^2-M_0^2$ when we compute $L_{\rm div}(M_0)$. All these terms must be tracked carefully, as they are responsible for the precise expressions \eqref{Eq:QuantumVacuum.deltacouplingsB} quoted in the final result given below. Most important, one has to check that the poles cancel in the subtraction and that no trace is left of the arbitrary mass unit $\mu$ either.
Some terms can be shown immediately to vanish in this subtraction, e.g. it is easy to check that the overall coefficient of $\gamma_E$ in \eqref{Eq:QuantumVacuum.LdivM}, namely $-\frac{1}{2}\hat{a}_0(x)M^4+\hat{a}_1(x)M^2-\hat{a}_2(x)$, does not depend on $M$ and therefore this term will automatically cancel in the subtraction. Other terms require more work and one has to go through all the details. After some tedious algebra one finds that the poles which appear in the limit $\epsilon\to 0$ indeed cancel along with the dependence on the arbitrary mass unit $\mu$, and the final result can be cast in the compact form
\begin{equation}\label{Eq:QuantumVacuum.divMdivM0B}
\begin{split}
 L_{\rm div}(M_0)-L_{\rm div}(M)&=\delta\rL(m,M,M_0)-\frac{1}{2}\delta\Mpl^2(m,M,M_0) R\\
 &-\delta \alpha_Q(M,M_0) \frac{{Q^\lambda}_\lambda}{3}- \delta{\alpha_2}(M,M_0)R^2\,,
\end{split}
\end{equation}
where the various contributions read as follows:
\begin{equation}\label{Eq:QuantumVacuum.deltacouplingsB}
\begin{split}
&\delta\rL(m,M,M_0)=\frac{1}{8\left(4\pi\right)^2}\left(M^4-M_0^4-4m^2M^2+4m^2M_0^2+2m^4\ln\frac{M^2}{M_0^2}\right),\\
&\delta\Mpl^2(m,M,M_0) =\frac{\left(\xi-\frac{1}{6}\right)}{(4\pi)^2}\left(M^2-M_0^2-m^2\ln \frac{M^2}{M_0^2}\right),\\
&\delta \alpha_Q(M,M_0)=\frac{1}{2(4\pi)^2}\ln\frac{M^2}{M_0^2},\\
&\delta{\alpha_2}(M,M_0)=-\frac{\left(\xi-\frac{1}{6}\right)^2}{4(4\pi)^2}\ln\frac{M^2}{M_0^2}.
\end{split}
\end{equation}
In the light of these general subtraction formulas we may now evaluate the leading terms involved in our original expression \eqref{Eq:QuantumVacuum.LWRenaprox} by just setting $M_0=m$ in the above equations. We can easily check that the result is precisely given by the equations \eqref{Eq:QuantumVacuum.SubtractionrL}, \eqref{Eq:QuantumVacuum.SubtractionMPl}.

In the original renormalization approach to the EMT some of these quantum effects appeared as parameter differences computed at the two scales under consideration, i. e.
$\delta X\equiv X(M)-X(M_0)$, for the various couplings $X$ and using the renormalized form of Einstein's equations \eqref{Eq:QuantumVacuum.MEEsHD}, see \hyperref[Sect:RenormZPE]{Sect.\ref{Sect:RenormalizedVED}}. This is because we renormalized the EMT following the subtraction prescription defined in Eq.\,\eqref{Eq:QuantumVacuum.EMTRenormalizedDefinition}. We can indeed recognize the first two expressions in \eqref{Eq:QuantumVacuum.deltacouplingsB} as being identical to the parameter subtractions \eqref{Eq:QuantumVacuum.SubtractionrL}-\eqref{Eq:QuantumVacuum.SubtractionMPl}. The third and fourth expressions in \eqref{Eq:QuantumVacuum.deltacouplingsB} are related to the coefficients of the HD terms $Q^\lambda_{\ \lambda}$ and $R^2$. In particular, $\delta{\alpha_2}(M,M_0)$ is just one half of $\delta\alpha$ given in Eq.\,\eqref{Eq:QuantumVacuum.Subtractionalpha}. The factor of $1/2$ is because the parameter $\alpha_2$ in the Lagrangian \eqref{Eq:QuantumVacuum.LEHHD} is related to the parameter $\alpha$ in the generalized Einstein's equations \eqref{Eq:QuantumVacuum.MEEs} through $\alpha_2=\alpha/2$. Recall that $\alpha$ is the coefficient of the HD tensor $\leftidx{^{(1)}}{\!H}_{\mu \nu}$ in these equations, and that tensor is given by the functional derivative of $R^2$ with respect to the metric, see \hyperref[Appendix:Conventions]{Appendix\,\ref{Appendix:Conventions}}.

At this point we have fully justified the important Eq.\eqref{Eq:QuantumVacuum.LwrenM}, which gives the renormalized effective Lagrangian of vacuum. From here we may construct the full effective Lagrangian \eqref{Eq:QuantumVacuum.Full-Leff} and reproduce the remaining considerations. In particular, we can obtain the coefficients of the $\beta$-functions for the various couplings \eqref{Eq:QuantumVacuum.BetaFunctionrL}-\eqref{Eq:QuantumVacuum.BetaFunctions34} and solve the corresponding renormalization group equations, with the result \eqref{Eq:QuantumVacuum.RGEscouplings}.

\subsection{Full consistency between the EMT and effective action results}

We take up the renormalized quantum vacuum Lagrangian defined in the previous section, Eq.\,\eqref{Eq:QuantumVacuum.LWrenormalized}. The effective action associated to such Lagrangian is
\begin{equation}\label{Eq:QuantumVacuum.Wren}
W_{\rm ren}(M)\equiv\int d^4 x \sqrt{-g} \ L_{\rm W}^{\rm ren}(M)=\int d^4 x \sqrt{-g} \ \left(L_W (m)-L_{\rm div} (M)\right)\,.
\end{equation}
Let us now show that with this action we can recompute the renormalized vacuum EMT that we have previously found in \hyperref[Sect:RenEMToffshell]{Sect.\,\ref{Sect:RenEMToffshell}}.
In fact, on inserting Eq.\,\eqref{Eq:QuantumVacuum.LwrenM} in it we find
\begin{equation}\label{Eq:QuantumVacuum.effActionLren}
\begin{split}
W_{\rm ren}(M)=\int d^4 x\sqrt{-g} \left( \delta \rL(M)-\frac{1}{2}\delta\Mpl^2(M) R-\delta \alpha_Q(M) \frac{{Q^\lambda}_\lambda}{3}-\delta \alpha_2(M) R^2\right)\,.
\end{split}
\end{equation}
The renormalized vacuum EMT now follows from
\begin{equation}\label{Eq:QuantumVacuum.DefWMM0}
\left\langle T_{\mu\nu}^{\delta \phi}\right\rangle_{\rm ren}(M)=-\frac{2}{\sqrt{-g}} \,\frac{\delta W_{\rm ren}(M)}{\delta g^{\mu\nu}}\,.
\end{equation}
Using \eqref{Eq:QuantumVacuum.effActionLren} on the RHS of \eqref{Eq:QuantumVacuum.DefWMM0} we may compute the metric functional variation. In performing the variation of the HD term $\frac13 {Q^\lambda}_\lambda$ as given in Eq.\,\eqref{Eq:QuantumVacuum.traceQ1}, we can use some of the formulas quoted in \hyperref[Appendix:Conventions]{Appendix\,\ref{Appendix:Conventions}}. In particular, we drop the contribution from the Euler density $E$ (since the metric functional variation of the Gauss-Bonnet term $\GB$ is exactly zero in $n=4$ spacetime dimensions) and of course that of the total derivative term $\Box R$. Therefore, using the mentioned appendix,
\begin{equation}\label{Eq:QuantumVacuum.VariationQ}
\frac{1}{\sqrt{-g}}\frac{\delta\left( {Q^\lambda}_\lambda/3\right)}{\delta g^{\mu\nu}}=\frac{1}{\sqrt{-g}}\frac{\delta}{\delta g^{\mu\nu}}\left( -\frac{1}{120} C^2\right)=-\frac{1}{60}\left( \leftidx{^{(2)}}{\!H}_{\mu\nu}-\frac13 \leftidx{^{(1)}}{\!H}_{\mu\nu}\right)\,.
\end{equation}
With this proviso, the sought-for metric functional variation can be easily performed and \eqref{Eq:QuantumVacuum.DefWMM0} can be written in the compact form
\begin{equation}\label{Eq:QuantumVacuum.TrenMm}
\begin{split}
\left\langle T_{\mu\nu}^{\delta \phi}\right\rangle_{\rm ren}(M)&= \delta\Mpl^2(M) G_{\mu\nu}+ \delta \rL(M) g_{\mu\nu}+\delta\alpha(M)\leftidx{^{(1)}}{\!H}_{\mu\nu}\\
&-\frac{1}{30}\delta\alpha_Q(M)\left(\leftidx{^{(2)}}{\!H}_{\mu\nu}-\frac13 \leftidx{^{(1)}}{\!H}_{\mu\nu}\right)\,,
\end{split}
\end{equation}
where we have used the fact that $2\delta\alpha_2=\delta\alpha$ and we recall that the coefficients of the various tensor expressions on the RHS of the previous formula are given explicitly by Eqs.\,\eqref{Eq:QuantumVacuum.deltacouplings}. For conformally flat spacetimes (which comprise, in particular, all the FLRW backgrounds) the two HD tensors $\leftidx{^{(1)}}{\!H}_{\mu\nu}$ and $\leftidx{^{(2)}}{\!H}_{\mu\nu}$ are related in the form $\leftidx{^{(2)}}{\!H}_{\mu\nu}=\frac13 \leftidx{^{(1)}}{\!H}_{\mu\nu}$ -- cf. Eq.\,\eqref{Eq:Conventions.ConfFlat} of \hyperref[Appendix:Conventions]{Appendix\,\ref{Appendix:Conventions}} for more details\footnote{The notations `(1)' and `(2)' as upper indices on the left for these HD tensors is standard\cite{birrell1984quantum}, it has nothing to do with adiabatic orders.}. As a consequence, the previous equation simplifies into
\begin{equation}\label{Eq:QuantumVacuum.TrenMm0FLRWMm}
\begin{split}
\left\langle T_{\mu\nu}^{\delta \phi}\right\rangle_{\rm ren}(M)=
& \delta \rL(M) g_{\mu\nu}+\delta\Mpl^2(M) G_{\mu\nu}+\delta\alpha(M)\leftidx{^{(1)}}{\!H}_{\mu\nu}\,.
 \end{split}
\end{equation}
If we take the $00th$-component of this result and use the formulas given in \hyperref[Appendix:Conventions]{Appendix\,\ref{Appendix:Conventions}} we find
\begin{equation}\label{Eq:QuantumVacuum.RenormalizedfromAction}
\begin{split}
\left\langle T_{00}^{\delta \phi}\right\rangle_{\rm ren}(M)&=\delta \rL(M) g_{00}+\delta\Mpl^2(M) G_{00}+\delta\alpha(M)\leftidx{^{(1)}}{\!H}_{00}\\
&=\frac{a^2}{128\pi^2 }\left(-M^4+4m^2M^2-3m^4+2m^4 \ln \frac{m^2}{M^2}\right)\\
&-\left(\xi-\frac{1}{6}\right)\frac{3 \mathcal{H}^2 }{16 \pi^2 }\left(m^2-M^2-m^2\ln \frac{m^2}{M^2} \right)\\
&+\left(\xi-\frac{1}{6}\right)^2 \frac{9\left(2 \mathcal{H}^{\prime \prime} \mathcal{H}- \mathcal{H}^{\prime 2}- 3 \mathcal{H}^{4}\right)}{16\pi^2 a^2}\ln \frac{m^2}{M^2}+\dots
\end{split}
\end{equation}
The upshot is that we are led once more to Eq.\,\eqref{Eq:QuantumVacuum.ExplicitRenormalized}, as it should be. The corresponding result for the VED obtains now from \eqref{Eq:QuantumVacuum.RenormalizedfromAction} using the relation \eqref{Eq:QuantumVacuum.RenVDE}. In this way we reach again the final result \eqref{Eq:QuantumVacuum.RenVDEexplicit} and hence we have demonstrated the perfect consistency between the two renormalization procedures.

We can perform a similar computation to find the scaling evolution of the EMT between the renormalization point $M$ and $M_0$. One option is to use Eq.\,\eqref{Eq:QuantumVacuum.RenormalizedfromAction} to reproduce the results we have already found in \hyperref[Sect:RenormalizedVED]{Sect.\,\ref{Sect:RenormalizedVED}}. But we may also repeat the above procedure ab initio, now using the subtracted effective action at the two mentioned scales:
\begin{equation}\label{Eq:QuantumVacuum.effActionExplicit2}
\begin{split}
W_{\rm ren}(M)-W_{\rm ren}(M_0)=&\int d^4 x \sqrt{-g} \ \left(L_{\rm W}^{\rm ren}(M)-L_{\rm W}^{\rm ren}(M_0)\right)\\
=&\int d^4 x \sqrt{-g} \ \left(L_{\rm div} (M_0)-L_{\rm div} (M)\right)\\
=&\int d^4 x\sqrt{-g} \left( \delta \rL(m,M,M_0)-\frac{1}{2}\delta\Mpl^2(m,M,M_0) R\right.\\
&\left.\phantom{xxxxxxxxxx}-\delta \alpha_Q(M,M_0) \frac{{Q^\lambda}_\lambda}{3}-\delta \alpha_2(M,M_0) R^2\right)\,.
\end{split}
\end{equation}
Here we have used Eq.\,\eqref{Eq:QuantumVacuum.divMdivM0B} and we note that the coefficients of the various tensor expressions on the RHS of the previous formula are not the same as in \eqref{Eq:QuantumVacuum.effActionLren} but are given explicitly in Eq.\,\eqref{Eq:QuantumVacuum.deltacouplingsB}. The difference of vacuum EMT values at the two scales reads
\begin{equation}\label{eq:DefWMM0bis}
\delta\left\langle T_{\mu\nu}^{\delta \phi}\right\rangle\equiv \left\langle  T_{\mu\nu}^{\delta \phi}\right\rangle_{\rm ren}(M)-\left\langle  T_{\mu\nu}^{\delta \phi}\right\rangle_{\rm ren}(M_0)=-\frac{2}{\sqrt{-g}} \,\frac{\delta}{\delta g^{\mu\nu}}\left(W_{\rm ren}(M)-W_{\rm ren}(M_0)\right)\,.
\end{equation}
and upon computing the metric functional variation we find
\begin{equation}\label{Eq:QuantumVacuum.TrenMm0General}
\begin{split}
\delta\left\langle T_{\mu\nu}^{\delta \phi}\right\rangle &=
\,\delta\Mpl^2(m,M.M_0) G_{\mu\nu}+ \delta \rL(m,M,M_0) g_{\mu\nu}+\delta\alpha(M,M_0) \leftidx{^{(1)}}{\!H}_{\mu\nu}\\
&-\frac{1}{30}\,\delta\alpha_Q(M,M_0)\left(\leftidx{^{(2)}}{\!H}_{\mu\nu}-\frac13\,\leftidx{^{(1)}}{\!H}_{\mu\nu}\right).
 \end{split}
\end{equation}
For conformally flat spacetimes we can repeat the same argument as given above and the above result boils down to
\begin{equation}\label{Eq:QuantumVacuum.TrenMm0FLRWl}
\begin{split}
\delta\left\langle T_{\mu\nu}^{\delta \phi}\right\rangle=
& \,\delta\Mpl^2(m,M,M_0) G_{\mu\nu}+ \delta \rL(m,M,M_0) g_{\mu\nu}+\delta\alpha(M,M_0)\leftidx{^{(1)}}{\!H}_{\mu\nu}\,.
 \end{split}
\end{equation}
The obtained expression is just the subtracted form of Eq.\,\eqref{Eq:QuantumVacuum.MEEsHD} at the two scales $M$ and $M_0$. Thus, if we take the $00th$-component of this result and use the formulas given in \hyperref[Appendix:Conventions]{Appendix\,\ref{Appendix:Conventions}} to perform the identifications on both sides and the definition of VED we encounter once more the important Eq.\,\eqref{Eq:QuantumVacuum.VEDscalesMandM0Final} which gives the smooth evolution of the VED between the two scales with the total absence of quartic mass contributions. This corroborates the perfect consistency between the two approaches. Having found the very same renormalization results with the effective action formalism, all of the discussions made in \hyperref[Sect:RenormalizedVED]{Sec.\,\ref{Sect:RenormalizedVED}} can be iterated exactly as they are there.

\subsection{Renormalization group equation for the VED}\label{Sect:RGE-VED}

To compute the RGE for $\rv(M)$ we have to take into account that only up to the 4{\it th} adiabatic order carry $M$-dependence since the higher orders are finite and hence need not be subtracted. It follows that the exact $\beta$-function for the VED can be obtained from Eq.\,\eqref{Eq:QuantumVacuum.RenVDEexplicit} as follows:
\begin{equation}\label{Eq:QuantumVacuum.RGEVED1}
\begin{split}
\beta_{\rv}&=M\frac{\partial\rv(M)}{\partial M}=\beta_{\rL}+\frac{1}{128\pi^2}\left(-4M^4+8 m^2 M^2- 4 m^4\right)\\
&-\left(\xi-\frac{1}{6}\right)\frac{3 \mathcal{H}^2 }{16 \pi^2 a^2}\left(-2M^2+2m^2\right)+\left(\xi-\frac{1}{6}\right)^2 \frac{9\left(2 \mathcal{H}^{\prime \prime} \mathcal{H}- \mathcal{H}^{\prime 2}- 3 \mathcal{H}^{4}\right)}{16\pi^2 a^4}(-2)\\
&=\left(\xi-\frac{1}{6}\right)\frac{3 \mathcal{H}^2 }{8 \pi^2 a^2}\left(M^2-m^2\right)+\left(\xi-\frac{1}{6}\right)^2 \frac{9\left(\mathcal{H}^{\prime 2}-2\mathcal{H}^{\prime \prime}\mathcal{H}+3 \mathcal{H}^4\right)}{8\pi^2 a^4}\,.
\end{split}
\end{equation}
In the first line we have used the $\beta$-function for $\rL(M)$ that we have just obtained in \eqref{Eq:QuantumVacuum.BetaFunctionrL}. It is seen that $\beta_{\rL}$ exactly cancels against the contribution from the second term on the RHS of the above equation. This cancellation is most welcome, as it leaves the $\beta$-function of the VED completely free from quartic mass contributions. It follows that the running of $\rv(M)$ rests only on the presence of quadratic mass scales in the final result. Integrating the above RGE we find
\begin{equation}\label{eq:VEDintegrated}
\begin{split}
\rv(M)&=\rv(M_0)+\left(\xi-\frac16\right)\frac{3\cH^2}{16\pi^2 a^2}\,\left(M^2 - M_0^{2} -m^2\ln \frac{M^{2}}{M_0^2}\right)\\
&+\left(\xi-\frac16\right)^2\frac{9}{16 \pi^2 a^4}\left(\mathcal{H}^{\prime 2}-2\mathcal{H}^{\prime \prime}\mathcal{H}+3 \mathcal{H}^4 \right)\ln \frac{M^2}{M_0^{2}}\,.\\
\end{split}
\end{equation}
Thus we have recovered the expected result \eqref{Eq:QuantumVacuum.VEDscalesMandM0Final}, which gives the evolution of the VED with the scale $M$, starting from another scale $M_0$, and such relation involves only quadratic mass scales which in leading order are highly tempered by the presence of quadratic powers of the Hubble rate. In other words, rather than the hard $\sim m^4$ behavior we obtain the much softened one $\sim m^2H^2$.

The following observation is now in order. As we warned in \hyperref[Sect:RenormalizedVED]{Sect.\,\ref{Sect:RenormalizedVED}}, the renormalization of the EMT and in particular of the VED involves the renormalization of formal quantities which do not ever play a role in the physical interpretation of the VED. Such is the case of the quantity \eqref{Eq:QuantumVacuum.SubtractionrL}, which carries the quartic powers of the masses. This quantity cancels exactly in the important expression \eqref{Eq:QuantumVacuum.VEDscalesMandM0Final}, which physically relates the VED at the two scales $M$ and $M_0$ and hence no dependence is left of the unwanted terms $\sim m^4$. As we know, this is the clue to avoid the need for fine-tuning in our renormalization procedure. Now we can see that the primary reason for that stems from the soft behavior of the VED $\beta$-function \eqref{Eq:QuantumVacuum.RGEVED1}.

In contrast, the $\beta$-function for the parameter $\rL(M)$ is proportional to the quartic power of the particle mass, as indicated in Eq.\,\eqref{Eq:QuantumVacuum.BetaFunctionrL}, and for this reason the solution to the corresponding RGE, as represented by the first equation in \eqref{Eq:QuantumVacuum.RGEscouplings}, is also proportional to those quartic terms. However, the running of $\rL$ with $M$ has no physical implication since these terms exactly cancel out in the VED, as we have just seen. This situation can be compared to the running of $\rL$ with the unphysical mass unit $\mu$ in the MS approach to the VED, as discussed in \hyperref[Sect:VEDMSS]{Sect.\,\ref{Sect:VEDMSS}}. Yet there is an important difference: in the MS case\footnote{Recall the footnote on page \pageref{FN.QuantumVacuum.MS}} one usually interprets that the renormalized VED is given by Eq.\,\eqref{Eq:QuantumVacuum.VEDMSS}. If so, then, as a (purported) physical quantity one is enforced to fine tune $\rL(\mu)$ against the large $\sim m^4$ contribution (represented by the second term in that expression). The counterpart of these formulas in our calculation is just given by a single piece of our Lagrangian \eqref{Eq:QuantumVacuum.Full-Leff} (since all the others are geometric contributions from curved spacetime), to wit: it is just (minus) the first term, or $\rL(M)-\delta\rL(M)$. The last equation indeed contains, among others, the terms involved in Eq. \eqref{Eq:QuantumVacuum.VEDMSS}. This can be checked from Eq.\,\eqref{Eq:QuantumVacuum.RGErL}, conveniently rewritten as
\begin{equation}\label{Eq:QuantumVacuum.deltarhocommon}
-\delta\rL(M)=\frac{m^4}{64\pi^2}\left(\ln\frac{m^2}{M^2}-\frac32-\frac{M^4}{2m^4}+\frac{2M^2}{m^2}\right)\,,
\end{equation}
upon replacing $M$ with $\mu$ and neglecting the last two terms since we consider $M^2=H^2\ll m^2$.
The RG-invariant expression $\rL(M)-\delta\rL(M)=\rL(m)$ is not at all the VED. In Minkowski spacetime, we saw that the correctly renormalized VED is zero in our framework (cf. \hyperref[Sect:SubtractMinkowski]{Sect.\,\ref{Sect:SubtractMinkowski}}). What is more, on comparing the VED at two different scales the effect of $\rL(M)$ always cancels against the quartic terms emerging from the renormalized ZPE, and the net result is free from the influence of the quartic masses, cf. Eq.\,\eqref{eq:VEDintegrated}. Thanks to this crucial fact the observable running of the VED in curved spacetime depends only on the quadratic mass scales times the Hubble rate square, {\it i.e.} $\sim m^2H^2$, as shown in that expression. The presence of $H^2$ makes the running rate much more temperate: it just follows the evolution of the cosmic flow itself. In fact, this is nothing but the characteristic running law of the RVM\,\cite{Sola:2013gha,SolaPeracaula:2022hpd}.

\subsection{Trace anomaly and effective action}\label{Sect:TraceAnomalyComp}

The trace anomaly can also be elucidated in the framework of the effective action, $W$\,\cite{birrell1984quantum}. The latter is related to the VEV of the EMT, as noted in \eqref{Eq:QuantumVacuum.DefW}.
The effective action is purely geometric and involves the quantum effects of $\phi$ in our case. Since the UV-divergences are inherent to short-distance effects, they all involve the behavior of geometric tensors $R^2, R_{\mu\nu} R^{\mu\nu},\dots$ at short distances. The non-trivial local behavior of the curved spacetime at the level of the effective action is the counterpart to the UV behavior of the field modes in the EMT. The two languages lead to the same answer. Thus, although one can use $W$ and the UV-divergences associated to these geometric terms to derive the trace anomaly\,\cite{birrell1984quantum}, here we have used directly the VEV of the EMT corresponding to the quantum matter field $\phi$. The divergences of $W$ are of course the same as those of the vacuum EMT. So, if we write $W=W_{\rm div}+W_{\rm ren}$, the divergent and renormalized parts of the vacuum EMT must correspond respectively to $W_{\rm div}$ and $W_{\rm ren}$.
This means that the obtained expression \eqref{Eq:QuantumVacuum.TraceVEVconformal2Expl} can be identified with the vacuum trace emerging from the divergent part of the effective action, which in the massless conformal limit turns out to be finite. Thus,
\begin{equation}\label{eq:EMisner:1973prbdiv}
\left.\lim\limits_{m\to 0} \left\langle T^{\delta \phi} \right\rangle\right|_{(\xi=1/6,M=m )}=\frac{2}{\sqrt{-g}}\, g_{\mu\nu}\,\frac{\delta W_{\rm div}}{\delta g_{\mu\nu}}\,.
\end{equation}
Because the vacuum trace of the total EMT derived from the effective action must vanish in the massless conformally coupled limit\,\cite{birrell1984quantum}, the trace associated to $W_{\rm ren}$ (the so-called renormalized part of the effective action) is given by minus the previous result \eqref{eq:EMisner:1973prbdiv}, and this defines the trace anomaly:
\begin{equation}\label{Eq:QuantumVacuum.FormalTraceAnomaly}
\left.\lim\limits_{m\to 0} \left\langle T^{\delta \phi} \right\rangle\right|_{(\xi=1/6,M=m )}^{\rm anomaly} =\frac{2}{\sqrt{-g}}\, g_{\mu\nu}\,\frac{\delta W_{\rm ren}}{\delta g_{\mu\nu}}=-\frac{2}{\sqrt{-g}}\, g_{\mu\nu}\,\frac{\delta W_{\rm div}}{\delta g_{\mu\nu}}=- \left.\lim\limits_{m\to 0} \left\langle T^{\delta \phi} \right\rangle\right|_{(\xi=1/6,M=m )}\,.
\end{equation}
It can be expressed in an invariant form, showing that the anomaly is a general coordinate scalar. With the help of the geometric relations given in \hyperref[Appendix:Conventions]{Appendix\,\ref{Appendix:Conventions}} one can readily show that the obtained expression can be written in a covariant way as follows:
\begin{equation}\label{Eq:QuantumVacuum.TraceInvariant}
\begin{split}
\left.\lim\limits_{m\to 0} \left\langle T^{\delta \phi} \right\rangle\right|_{(\xi=1/6,M=m )}^{\rm anomaly} &=-\left.\lim\limits_{m\to 0} \left\langle T^{\delta \phi} \right\rangle\right|_{(\xi=1/6,M=m )}=-\frac{1}{480\pi^2 a^4}\left(-4\mathcal{H}^2\mathcal{H}^\prime+\mathcal{H}^{\prime\prime\prime}\right)\\
&=+\frac{1}{2880\pi^2}\left[R^{\mu\nu}R_{\mu\nu}-\frac{1}{3} R^2+\Box R\right]\,.
\end{split}
\end{equation}
It is well known that there is no contribution from the square of the Weyl tensor $C^2=C^{\alpha\beta\gamma\delta}C_{\alpha\beta\gamma\delta}$ for conformally flat spacetimes since that tensor vanishes identically for them.
The above expression is the form which we have quoted in the main text, see Eq.\,\eqref{Eq:QuantumVacuum.TraceAnomaly}. In general the conformal anomaly can also be written in a very succinct way in terms of the DeWitt-Schwinger coefficient of adiabatic order $4$ -- cf. \hyperref[SubSect:HeatKernel]{Sect.\,\ref{SubSect:HeatKernel}} and\,\cite{birrell1984quantum}. Borrowing equations \eqref{Eq:QuantumVacuum.traceQ2} and \eqref{Eq:QuantumVacuum.ModifDWScoeff}, one finds
\begin{equation}\label{Eq:QuantumVacuum.TraceInvariant2}
\begin{split}
\left.\lim\limits_{m\to 0} \left\langle T^{\delta \phi} \right\rangle\right|_{(\xi=1/6,M=m )}^{\rm anomaly} &=+\left.\frac{a_2}{16\pi^2}\right|_{\xi=1/6}=-\left.\frac{1}{48\pi^2} {Q^\lambda}_\lambda\right|_{\xi=1/6}= +\frac{1}{1920\pi^2}\left[C^{2}-\frac13 E+\frac{2}{3}\Box R\right]\,.
\end{split}
\end{equation}
Using \eqref{Eq:Conventions.R2R4}, with $C_{\alpha\beta\gamma\delta}=0$ for FLRW spacetime, the previous expression boils down to the particular form \eqref{Eq:QuantumVacuum.TraceInvariant}. Notice that $\hat{a}_2=a_2$ in this case, since $M=m$ and hence $\Delta=0$ in \eqref{Eq:QuantumVacuum.ModifDWScoeff}.

\section{Discussion of the chapter}\label{Sect:ConclusionsChapterRVQFT}

We have devoted this chapter to investigate the possible dynamics of vacuum in the context of quantum field theory in the expanding FLRW spacetime and the possible connection with the cosmological term in Einstein's equations. The quantum field theoretical context is well-known\,\cite{birrell1984quantum,parker2009quantum,fulling1989aspects,mukhanov2007introduction} but, the difficulties in reconciling QFT (and string theory) predictions with cosmological observations in connection to this subject are at the basis of the so-called Cosmological Constant Problem (CCP)\,\cite{weinberg1989cosmological,witten2001cosmological}. Such mystery is perhaps the greatest conceptual challenge faced by theoretical physics ever, owing to the mind boggling discrepancy existing between the measured value of the vacuum energy density (VED) and the typically predicted one by our most cherished QFT's, say quantum chromodynamics and specially the electroweak standard model, both being essential parts of what we call the standard model of particle physics, which in itself is a highly successful theory of the fundamental interactions. Time and again this is one of the main problems of theoretical physics and cosmology that demands an urgent explanation at a fundamental level. The methods to deal with QFT in curved spacetime are well-known since long, and yet some of the most pressing problems of modern cosmology still remain unaccounted. The CCP is certainly a focus issue for any formal theory of the cosmic evolution. It is a must to be addressed in this kind of field theoretical studies since the physical interpretation of the cosmological term $\CC$ has traditionally been linked to the current value of the VED, $\rvo$, through Lema\^\i tre's formula $\rvo=\CC/(8\pi G_N)$. As we have mentioned in the introduction, the discrepancy existing between the measured value of $\rvo$ and the generic prediction made in field theories of the fundamental interactions (e.g. the standard model of particle physics) is utterly disproportionate. Such an appalling clash of theoretical concurring ideas versus direct astrophysical observations is at the root of the CCP. Furthermore, irrespective of the fact that there are many sources of vacuum energy in QFT (which are, in principle, uncorrelated), each one of them is very large as compared to $\rvo\sim 10^{-47}$ GeV$^4$, and hence the possible compensation among these sources leads to hopeless fine-tuning among the parameters of the theory. If that is not enough, the adjustment must be redone order by order in perturbation theory. Such an unending process of tuning and retuning makes the CCP even harsher, in fact unacceptable as a natural solution\,\cite{Sola:2013gha,SolaPeracaula:2022hpd}. 

Even though tackling such problems may require the concepts and the sophisticated theoretical tools underlying quantum gravity (QG) and string theory\,\cite{witten2001cosmological}, QG does not exist as a consistent theory yet; and string theory somehow abhors de Sitter space, as `swampland' conjectures preclude the construction of metastable de Sitter vacua in the string framework\,\cite{SolaPeracaula:2019kfm,Vafa:2005ui,Agrawal:2018own,Obied:2018sgi}. We may say that difficulties appear in all fronts. While we remain agnostic about these problems a lot of exciting QG phenomenology is still possible with the advent of the multi-messenger era, characterized by a steep increase in the quantity and quality of experimental data that are being obtained from
the detection of the various cosmic messengers (photons, neutrinos, cosmic rays and gravitational waves) from numerous origins\,\cite{Addazi:2021xuf}. The bare truth, however, is that neither one of them has succeeded in improving significantly the situation for the time being. In the meantime, we expect that some sort of provisional result should perhaps be possible within the -- much more pedestrian -- semiclassical QFT approach, in which quantum matter fields interact with an external gravitational field may still shed some light on those pending issues in the cosmological arena, in particular on the vacuum energy and its renormalization. This has been the main aim guiding our task here.

Specifically, in this chapter we have reconsidered the calculation of the renormalized energy-momentum tensor (EMT) of a real quantum scalar field non-minimally coupled to the FLRW background. Even though a full-blown calculation of the VED in QFT cannot be faced at this point, here we have focused on the simplest, and yet non-trivial, QFT model interacting with the FLRW spacetime background that we can think of, namely a scalar field $\phi$ non-minimally coupled to gravity. For the sake of a more simple presentation and, therefore, to avert `not seeing the forest for the trees', we have assumed that $\phi$ has no self-interactions and hence no spontaneous symmetry breaking. This assumption allows us not only to avoid dealing with the renormalization of the corresponding effective potential but also to concentrate on the computation of the zero-point energy (ZPE) part, which is a pure quantum effect and thereby constitutes the most genuine quantum vacuum piece within the whole VED structure. With these basic assumptions, we have undertaken the renormalization of the corresponding energy-momentum tensor (EMT) using the adiabatic regularization and renormalization method. A key point in our approach has been to implement an appropriate renormalization of the EMT by performing a subtraction of its on-shell value at an arbitrary renormalization point $M$. The presence of this floating scale brings into play the renormalization group flow. Since the renormalized EMT becomes a function of $M$, we can compare the renormalized result at different epochs of the cosmic history characterized by different energy scales, which we have tested with the value of $H$ (the Hubble rate) at each epoch. This is certainly along the lines of the original RG approach\,\cite{Sola:2013gha,SolaPeracaula:2022hpd} but goes well beyond it since it provides a more formal and explicit QFT calculation. The renormalization program described here is based on adiabatic orders, and we presented the EMT up to the 6{\it th} order. This is a computationally demanding task, but it has allowed us to determine the on-shell renormalized form of the EMT and will help us in the derivation of the equation of state (EoS) of the quantum vacuum presented in \hyperref[Chap:EoSVacuum]{chapter\,\ref{Chap:EoSVacuum}}.

We have performed the calculation following three different lines of approach. Two of them are based on adiabatic regularization/renormalization of the EMT and another one based the on effective action through the Heat-Kernel. The perfect match between the three approaches strengthen our conclusion, so our analysis seems to be robust. Let us start talking about the first two cases. We started from the WKB expansion of the field modes in the FLRW spacetime. Then we defined an appropriately renormalized EMT by performing a subtraction of its on-shell value (i.e. the value defined at the mass $m$ of the quantized field) at an arbitrary renormalization point $M$. The resulting EMT becomes finite because we subtract the first four adiabatic orders (the only ones that can be divergent). Since the renormalized EMT becomes a function of the arbitrary scale $M$, we can compare the renormalized result at different epochs of the cosmic history characterized by different energy scales. In the main text we have shown by direct calculation that the renormalized EMT defined in that way is finite. In \hyperref[Appendix:Dimensional]{Appendix\,\ref{Appendix:Dimensional}} we presented and alternative path and used Dimensional Regularization (DR) to subtract the poles of the low adiabatic orders. In particular, we use the more conventional method based on cancelling the poles using the counterterms associated to the fundamental parameters $\rL,G_N^{-1}$ and $a_1$ (the coefficient of $R^2$). We have subsequently corroborated all of this with a third approach: from the perspective of the effective action formalism. It means that we have solved the curved spacetime Feynman propagator of the non-minimally coupled scalar field to gravity using the adiabatic method and computed the effective action using the heat-kernel expansion. Since that expansion has also been performed off-shell (i.e. at the arbitrary scale $M$ rather than at the physical mass $m$), it was necessary to compute the corresponding corrections induced on the DeWitt-Schwinger coefficients. With the help of the effective action we have rederived the renormalized EMT and all three approaches concur to the same renormalized result for the Zero-Point Energy of the Quantum field, {\it i.e.} the 00{\it th} component of the EMT associated to the quantum fluctuations of the fields. A renormalization program for the effective Lagrangian, totally analogous to the one used for the ZPE, has been used.

The next important point is the extraction of the VED from the renormalized EMT, which is composed not only of the ZPE part (involving the quantum fluctuations of the scalar field) but also of $\rL(M)$, the renormalized value of $\rL$ at the scale $M$. We find it remarkable that when we compute the evolution of the VED from one scale to another within our renormalization framework, the result is free from quartic contributions $\sim m^4$, which are usually responsible for the exceedingly large contributions to the VED. This is in stark contrast with other renormalization schemes in which the $\sim m^4$ effects are present and hence are badly in need of extreme fine-tuning arrangements. In our opinion, even if being well aware of the many other difficulties ahead of us, with the absence of these terms in our framework we might be inching into an eventual solution of the CCP. Additionally, we have shown that the renormalized VED obtained from this QFT calculation takes on approximately the usual form of the running vacuum models (RVM's)\,\cite{Sola:2013gha,Sola:2014tta, Sola:2011qr}, in which $\rv=\rv(H)$ appears in the manner of an additive constant plus a series of powers of $H$ (the Hubble rate) and its time derivatives. Originally, the RVM approach was motivated from general considerations involving the renormalization group in QFT in curved spacetime (cf.\,\cite{Sola:2013gha,Sola:2014tta, Sola:2011qr,SolaPeracaula:2022hpd} and references therein). With the present QFT calculations we have provided for the first time a solid foundation of the RVM, in which the dynamical structure of the VED is seen to ensue from first principles, namely from the quantum effects associated with the proper renormalization of the EMT.  In it, all the terms made out of powers of $H$ (and its time derivatives) are of even adiabatic order. This means that all these powers effectively carry an even number of time derivatives of the scale factor, which is essential to preserve the general covariance of the action. 

The lowest order dynamical component of the vacuum energy density (VED) consists of an additive constant together with a small dynamical component $\sim \nueff H^2$. The dimensionless parameter $\nueff$ is predicted to be small ($|\nueff|\ll1$), but it must ultimately be determined experimentally by confronting the model with cosmological data. This parameter is sufficient to describe the dynamics of the vacuum in the current Universe, while the higher-order terms can play a role in the early Universe, particularly in describing inflation. We will comment more on the latter at the end of the next chapter. On the other hand, $\nueff$ is seen to be proportional to the coefficient of the $\beta$-function of the running VED, although they have opposite signs. This is a reflection of the fact that the two leading effects on the late-time dynamics of the VED, namely one from the scaling evolution with $M$ (before we fix its value) and the other from the expansion rate $H$, are actually opposed. However, the second one is dominant, and this fact implies that the net sign of the slope of the VED is fixed by the sign of $\nueff$. For $\nueff>0$ (resp. $\nueff<0$), the evolving VED mimics quintessence (resp. phantom dark energy). In previous works, the model has been phenomenologically fitted to a large amount of cosmological data, and the running parameter $\nueff$ has been found to be positive and in the ballpark of $\sim 10^{-3}$\,\,\cite{Sola:2015wwa,Basilakos:2009wi,Grande:2010vg,Sola:2016jky,SolaPeracaula:2018wwm,Basilakos:2012ra,Gomez-Valent:2014rxa,Gomez-Valent:2014fda,Gomez-Valent:2015pia,Sola:2017znb,SolaPeracaula:2016qlq,SolaPeracaula:2017esw}.

In a completely analogous procedure, we have been able to compute the vacuum expectation value (VEV) of the trace of the EMT, as well as its renormalized value. This will be used to compute the pressure associated with the vacuum in the next chapter. Additionally, we have explicitly checked that we were able to recover the standard calculation of the trace anomaly for the scalar field. An extension of our discussions regarding the trace anomaly will be seen in\,\hyperref[Chap:Fermions]{chapter\,\ref{Chap:Fermions}}.

As a bonus we have extracted the renormalization group equations (RGE's) for the couplings and also for the VED itself. For this we have had to find out the explicit form of $\beta$-function for the VED running, Eq.\,\eqref{Eq:QuantumVacuum.RGEVED1}. The latter appears to be free from quartic mass scales, which otherwise would recreate the usual (unfathomable) fine-tuning problem which we wanted to eschew. The smoothly behaving RGE for the VED that we have found was long suspected from semi-qualitative RG arguments, see\,\cite{Sola:2013gha,SolaPeracaula:2022hpd} and references therein, but on \cite{Moreno-Pulido:2020anb,Moreno-Pulido:2022phq} (the works in which this chapter is based) we demonstrated for the first time in the literature in a full-fledged QFT context. Besides, we provide the RGE for the gravitational coupling, although we reserve the details for \hyperref[Chap:EoSVacuum]{chapter\,\ref{Chap:EoSVacuum}}.

Let us also mention that even though our QFT calculation has been simplified by the use of a single (real) quantum scalar field and just focused on the ZPE. But further investigations will be needed to generalize these results for multiple fields, interacting fields and Spontaneous Symmetry Breaking contributions. By the moment, in \hyperref[Chap:Fermions]{chapter\,\ref{Chap:Fermions}} we will extend this method also to fermions, encountering the same conclusions shown here. Up to computational details, however, we expect that a similar dynamical structure should emerge from the VED in the general case since the expansion of the full effective action in the context of FLRW spacetime should result in an even power series of the Hubble rate (owing once more to general covariance).

\chapter{Equation of state of the quantum vacuum and other features}\label{Chap:EoSVacuum}

In the last chapter, we presented a lengthy computation that resulted in a running law for the vacuum energy density (VED) in curved spacetime. Appropriate renormalization of the energy-momentum tensor shows that the VED is a smooth function of the Hubble rate and its derivatives: $\rv=\rv(H, \dot{H},\ddot{H},...)$. However, we did not exhaustively explore this idea. This chapter is based on our works from\,\cite{Moreno-Pulido:2022phq, Moreno-Pulido:2022upl}, and its aim is to analyze some of the consequences of the computational machinery described in preceding pages.

For example, we did not mention anything regarding the vacuum's pressure or the equation of state (EoS) of the quantum vacuum. We should not presume that the vacuum's EoS is exactly $\Pv=-\rv$, as we must first carefully evaluate the quantum effects. Obviously, the EoS cannot depart too much from the traditional one, at least around the present time, but we will see that it is not exactly $-1$. The vacuum pressure is defined in a way similar to the vacuum energy density \eqref{Eq:QuantumVacuum.RenVDE}. Assuming the vacuum to be a homogeneous and isotropic medium (it should preserve the cosmological principle), we may define the pressure using any diagonal $ii$-component of the renormalized vacuum stress tensor. This is not the only interesting continuation of the previous chapter. The Einstein field equations within our framework, i.e., the Friedmann equations incorporating not only a dynamical VED but also a running $G(H)$, are presented. The running of the VED means that the conservation equations should also be revisited.

The contents of this chapter can be summarized in the following manner. In the first section, \hyperref[Sect:RenormPressure]{Sect.\,\ref{Sect:RenormPressure}}, we present the pressure of the quantum vacuum and show an effective approach to generalized Running Vacuum Models through the Friedmann equation's. Such a picture may have interesting phenomenological applications, however a formal treatment is mandatory to show rigorously the mathematical behaviour of Vacuum's EoS. In the second section, \hyperref[Sect:RenormalizedFriedmann]{Sect.\,\ref{Sect:RenormalizedFriedmann}}, we study Friedmann's equation in the presence of the running $\rv(H)$ and observe that the gravitational coupling $G$ is also a running quantity, although evolving only logarithmically with the expansion rate: $G=G( \ln H)$. We analyze the Bianchi identity for this particular scenario and the local conservation law for $\rv$ and verify (as a robust check of our calculation) that it only depends on the $4th$ adiabatic terms (all of the $6th$ order effects cancel non-trivially in it). The vacuum pressure is used in the next section, \hyperref[Sect:EquationStateOfVacuum]{Sect.\,\ref{Sect:EquationStateOfVacuum}}, to find out the equation of state (EoS) of the quantum vacuum up to 6{\it th} adiabatic order. We also do a good estimation of the EoS as a function of redshift $z$ which encapsulates its behaviour up to deep Radiation Dominated Epoch (RDE). We observe that it is not stuck to $-1$, in contrast to the usual situation. Owing to the quantum effects, the EoS becomes dynamical. In particular we present a simpler version of this equation which applies for the late Universe, for very low redshifts, and mimics quintessence near the present. {In \hyperref[Sect:PhenoImplications]{Sect.\,\ref{Sect:PhenoImplications}}, we present the possible phenomenological implications of the RVM}. First, we relate its implications in the later Universe with previous studies\,\cite{SolaPeracaula:2021gxi} (this is connected to \hyperref[Chap:PhenomenologyofRVM]{chapter\,\ref{Chap:PhenomenologyofRVM}}). In the second part of the section our $6th$ order calculation is instrumental to unveil a generalized form of the RVM at high energies with potential implications for the physics of the very early Universe. Particularly, we explore the idea of Running Vacuum as framework for a mechanism of inflation. Such a mechanism is specially simple and it is based on the higher powers of $H$ in the adiabatic expansion of VED which may enhance the magnitude of VED in the Early Universe, enough to produce an exponential inflationary period of expansion.  Finally, our discussion of the chapter is delivered in \hyperref[Sect:eos.Discussion]{Sect.\,\ref{Sect:eos.Discussion}}.

\section{Renormalized Vacuum Pressure}\label{Sect:RenormPressure}

In order to find a suitable expression for Vacuum's Pressure, our starting point are the results from the last chapter. Adopting once more the perfect fluid form\,\eqref{Eq:QuantumVacuum.VacuumIdealFluid} for the vacuum EMT, we may infer the expression for the vacuum pressure by following the same logic as for the vacuum energy density \eqref{Eq:QuantumVacuum.RenVDE}. We start taking the $11th$-component, $T_{11}^{\rm vac}$, of the mentioned EMT. As we said, any $ii$th-component would do equally well owing to isotropy, and we find $T_{11}^{\rm vac}=a^2\Pv$ in the conformal metric. Mind that since we are using again the comoving cosmological frame there is no contribution from the $4$-velocity part. We subsequently equate this result to the $11th$-component of \eqref{Eq:QuantumVacuum.EMTvacuum}, $\langle T_{11}^{\rm vac} \rangle=-\rho_\Lambda g_{11}+\langle T_{11}^{\delta \phi}\rangle=-\rL a^2+\langle T_{11}^{\delta \phi}\rangle$. Thus, the renormalized vacuum pressure at the scale $M$ is given by
\begin{equation}\label{Eq:eos.VacuumPressureDef}
\Pv(M)\equiv \frac{\left\langle T_{11}^{\rm vac}\right\rangle_{\rm ren}(M)}{a^2}= -\rho_\Lambda (M)+ \frac{\left\langle T_{11}^{\delta \phi} \right\rangle_{\rm ren}(M)}{a^2}\,,
\end{equation}
which looks similar to the renormalized VED, Eq.\,\eqref{Eq:QuantumVacuum.RenVDE}, up to a sign in the $\rL$ term. This sign points to the expected EoS for the vacuum, but we need to proceed carefully before unveiling the final result. Having computed the $00th$-component of the EMT and its trace in the previous sections, the isotropy condition enables us to compute the $11th$-component of the EMT simply by means of the relation
\begin{equation}\label{Eq:eos.VacuumT11}
\frac{\left\langle T_{11}^{\delta \phi} \right\rangle_{\rm ren}(M)}{a^2}=\frac{1}{3}\left(\left\langle T^{\delta \phi} \right\rangle_{\rm ren}(M)+\frac{\left\langle T_{00}^{\delta \phi} \right\rangle_{\rm ren}(M)}{a^2}\right)\,.
\end{equation}
Using now our definition \eqref{Eq:QuantumVacuum.RenVDE} of VED, we can eliminate $\rL(M)$ in favor of $\rv(M)$ in the above equations, and we find
\begin{equation}\label{Eq:eos.VacuumPressureDef2}
\Pv(M)=-\rv(M)+\frac{1}{3}\left( \langle T^{\delta \phi} \rangle_{\rm ren}(M)+4\frac{\langle T_{00}^{\delta \phi} \rangle_{\rm ren}(M)}{a^2}\right)\,.
\end{equation}
This equation clearly shows that the EoS of the quantum vacuum is not exactly $-1$, and the departure from this value can be obtained from the previously computed expressions. We can provide a rather precise result by including terms up to $6th$ adiabatic order\,\footnote{The reader may carefully track the calculation and observe that there is once more an exact cancellation of the quartic mass scales in the sum of the two terms in parenthesis on the RHS of Eq.\,\eqref{Eq:eos.VacuumPressureDef2}. To check this one has to use equations \eqref{Eq:QuantumVacuum.ExplicitRenormalized} and \eqref{Eq:QuantumVacuum.TraceIntegrated}. It follows that the scaling evolution of the vacuum pressure is also free from quartic mass dependencies. This is of course reassuring and shows the consistency of our calculation.}:
\begin{equation}\label{Eq:eos.VacuumPressureFull}
\begin{split}
\Pv(M)=&-\rv(M)+\frac{\left(\xi-\frac{1}{6}\right)}{8\pi^2}\dot{H}\left(m^2-M^2-m^2\ln\frac{m^2}{M^2}\right)\\
&-\frac{3}{8\pi^2}\left(\xi-\frac{1}{6}\right)^2\left(6\dot{H}^2+3H\ddot{H}+\vardot{3}{H}\right)\ln \frac{m^2}{M^2}\\
&+\frac{1}{10080\pi^2 m^2}\left(8H^4\dot{H}-28\dot{H}^3+6H^3\ddot{H}-10\ddot{H}^2-22\dot{H}\vardot{3}{H}\right.\\
&\phantom{xxxxxxxxxxxx}\left.+24H^2 \dot{H}^2-7H^2 \vardot{3}{H}-49H \dot{H}\ddot{H}-6H \vardot{4}{H}-\vardot{5}{H}\right)\\
&+\frac{\left(\xi-\frac{1}{6}\right)}{240\pi^2 m^2}\left(-6H^4\dot{H}+34\dot{H}^3-6H^3\ddot{H}+12\ddot{H}^2+24\dot{H}\vardot{3}{H}-24H^2\dot{H}^2\right.\\
&\phantom{xxxxxxxxxx}\left.+7H^2\vardot{3}{H}+55H\dot{H}\ddot{H}+6H\vardot{4}{H}+\vardot{5}{H}\right)\\
&-\frac{\left(\xi-\frac{1}{6}\right)^2}{16\pi^2 m^2}\left(32\dot{H}^3-12H^3\ddot{H}+12\ddot{H}^2+24\dot{H}\vardot{3}{H}-48H^2\dot{H}^2\right.\\
&\phantom{xxxxxxxxxx}\left.+5H^2\vardot{3}{H}+47H\dot{H}\ddot{H}+6H\vardot{4}{H}+\vardot{5}{H}\right)\\
&+\frac{9\left(\xi-\frac{1}{6}\right)^3}{4\pi^2m^2}\left(4H^4\dot{H}-5\dot{H}^3-6H^3\ddot{H}-11H\dot{H}\ddot{H}-\ddot{H}^2\right.\\
&\phantom{xxxxxxxxxxx}\left.-\dot{H}\vardot{3}{H}-24H^2\dot{H}^2-2H^2\vardot{3}{H}\right)+\dots
\end{split}
\end{equation}
where $\dots$ represent the $8th$-order contributions and above, which we shall not consider at all. The obtained expression takes the generic form
\begin{equation}\label{Eq:QuantumVacuum.VacuumPressureFullsplit}
\Pv(M)=-\rv(M)+f_2(M,\dot{H})+ f_4(M,H,\dot{H},...,\vardot{3}{H})+f_6(\dot{H},...,\vardot{5}{H})+\cdots\,,
\end{equation}
in which $f_2$, $f_4$ and $f_6$ involve second, fourth and sixth adiabatic contributions, respectively. Near the present, they represent a small correction to the canonical relation $\Pv(M)=-\rv(M)$ for the vacuum EoS and therefore, strictly speaking, make the quantum vacuum a quasi-vacuum state.
Notice that the adiabatic contributions $f_i$ are specific effects on the pressure not present in the vacuum energy density, which in its own also contains contributions to all these orders.

\subsection{Generalized RVM at low energies}\label{SubSect:GeneralizedRVM}

The result \eqref{Eq:eos.VacuumPressureFull}, derived in previously, reveals an interesting new feature. Among the various terms that we have collected on its RHS (all of which are contributions to the vacuum pressure beyond those entering the VED), the ones of adiabatic order 2 are particularly worth noticing, namely the term
\begin{equation}\label{Eq:eos.f2}
\begin{split}
f_2(M,\dot{H})=\frac{\left(\xi-\frac{1}{6}\right)}{8\pi^2}\dot{H}\left(m^2-M^2-m^2\ln\frac{m^2}{M^2}\right)\,.
\end{split}
\end{equation}
This term can have implications on the vacuum dynamics at low energy since $\dot{H}$ is of the same order as $H^2$. To see this, let us write down the two ordinary Friedmann's equations for flat three-dimensional space and in the presence of a dominant matter component and vacuum energy (neglecting any dynamical effect on $G$ and assuming a constant value $G=G_N$):
\begin{equation}\label{Eq:eos.FriedmanEqs}
\begin{split}
&3H^2= 8\pi G_N (\rho_{\rm m}+\rv)\,, \\
&2\dot{H}+3H^2=- 8\pi G_N (p_{\rm m}+\Pv)\,,
\end{split}
\end{equation}
where $\rho_{\rm m}$ and $p_{\rm m}$ are the density and pressure of the dominant matter component (relativistic or non-relativistic). From these two equations one can derive the differential equation that is satisfied by the Hubble rate:
\begin{equation}\label{Eq:eos.DiffH1}
\dot{H}+\frac32\,(1+\wm)\,H^2=4\pi\,G_N \left(w_{\rm m}\rv-\Pv\right)\,,
\end{equation}
where $w_{\rm m}=p_{\rm m}/\rho_{\rm m}$ is the EoS of the dominant matter component. For the present Universe, we have $w_{\rm m}=0$ and the above equation reduces to
\begin{equation}\label{Eq:eos.DiffH}
\dot{H}+\frac32\,H^2=-4\pi\,G_N \Pv\equiv 4\pi\,G_N \rveff\,.
\end{equation}
Here we have defined an effective vacuum pressure $\rveff=-\Pv$ as if the EoS of the quantum vacuum would be exactly $-1$ . However, as we know, this does not imply $\rv=-\Pv$. We use $\rveff$ only to mimic the situation in the $\CC$CDM, but in reality the quantum vacuum contributes with a term that produces a departure of the EoS from the usual value. From Eq.\,\eqref{Eq:eos.VacuumPressureFull} we have the dominant contribution \eqref{Eq:eos.f2} at the scale $M$, which is of second adiabatic order, and can still be sizeable in the current Universe. In fact, we have two pieces of second adiabatic order on the RHS of \eqref{Eq:eos.VacuumPressureFull}, one contained in $\rv$ and the other given by $f_2$, which when combined lead to
\begin{equation}\label{Eq:eos.VacuumPressureCurrentU}
\begin{split}
\rveff(M)&=-\Pv(M)=\rv(M)-f_2(\dot{H})+ ...\\
&\thickapprox \left(\xi-\frac{1}{6}\right)\frac{3 {H}^2 }{16 \pi^2}\left(M^2-m^2 +m^2\ln \frac{m^2}{M^2} \right)\\
&+\left(\xi-\frac{1}{6}\right)\frac{\dot{H}}{8\pi^2}\left(M^2-m^2+m^2\ln\frac{m^2}{M^2}\right)+...
\end{split}
\end{equation}
Here we have used Eq.\,\eqref{Eq:QuantumVacuum.RenVDEexplicit} and neglected the higher order adiabatic terms. The meaning of $\thickapprox$ is that we have also omitted the first two terms of the mentioned equation, since we know that when we compare the VED at two scales in the same manner as we did in \eqref{Eq:QuantumVacuum.VEDscalesMandM0Final} these terms will exactly cancel each other and only the indicated terms of \eqref{Eq:eos.VacuumPressureCurrentU} will contribute. In fact, when we insert Eq.\,\eqref{Eq:eos.VacuumPressureCurrentU} on the RHS of \eqref{Eq:eos.DiffH} and solve for $H$, it will all occur as though the effective vacuum energy density contains not only the $\sim H^2$ dynamical component but also the new one proportional to $\dot{H}$. We can repeat a very similar argument to that in \,\hyperref[Sect:RunningConnection]{Sect.\,\ref{Sect:RunningConnection}}, with the two scales $M=H$ and $M_0=H_0$, and we find that the effective expression for the vacuum energy density in the present Universe can be expressed in a generic form as follows:
\begin{equation}\label{Eq:eos.RVMgeneralized2}
\rveff(H,\dot{H})=\rvo+\frac{3\nueff}{8\pi G_N}\,(H^2-H_0^2)+ \frac{3\tilde{\nu}_{\rm eff}}{8\pi G_N}\,(\dot{H}-\dot{H}_0)\,.
\end{equation}
We have normalized this relation such that $\rveff(H=H_0, \dot{H}=\dot{H}_0)=\rvo$ at the present time, where $H_0$ and $\dot{H}_0$ stand for the respective current values of $H$ and $\dot{H}$. Notice also that we have placed two generic coefficients $\nueff$ and $\tilde{\nu}_{\rm eff}$ for each of the two terms of adiabatic order $2$, $H^2$ and $\dot{H}$, rather than the specific ones in \eqref{Eq:eos.VacuumPressureCurrentU} (which would entail just $\tilde{\nu}_{\rm eff}=(2/3){\nu}_{\rm eff}$ for a single scalar field and no other matter field) because in general we expect that these coefficients will receive contributions from different sorts of fields, fermions and bosons. These coefficients are naturally small in magnitude, viz. of order $\sim m^2/\mpl^2\ll1$, cf. Eq.\,\eqref{Eq:QuantumVacuum.nueffAprox}, but not hopelessly small if $m$ is the mass of a GUT particle. In the limit $\nueff,\tilde{\nu}_{\rm eff}\to 0$ we just recover the $\CC$CDM with constant $\rvo=\CC/(8\pi G_N)$, but if they are small though non-vanishing they can impact non-trivially on the phenomenology of the dark energy. Here we have computed the effect from a single scalar field only, but in general we have to sum over the concomitant contributions from other bosons and fermions\,\cite{Samira2022}.

The above formula \eqref{Eq:eos.RVMgeneralized2} obviously extends the structure of Eq.\,\eqref{Eq:QuantumVacuum.RVM2}.
In this way we have found a justification for a generalized form of the RVM. Basically, one expects an extended form for the vacuum structure \eqref{Eq:QuantumVacuum.EMTvacuum} such that it comprises more geometric structures which are not possible in Minkowski spacetime but are certainly available in curved spacetime. Namely, one may naturally conceive a generalization of the form
 \begin{equation}\label{Eq:eos.GeneralVDE1}
 \left\langle T_{\mu\nu}^{\rm vac} \right\rangle= -\rL g_{\mu\nu} +\left\langle T_{\mu\nu}^{\delta\phi} \right\rangle + \alpha_1 R g_{\mu\nu} +\alpha_2 R_{\mu\nu}+\mathcal{O}(R^2)\,.
\end{equation}
In the above expression, $\mathcal{O}(R^2)$ represents possible contributions from geometric tensors of adiabatic order 4, that is $R^2$, $R_{\mu\nu}R^{{\mu\nu}},\dots$, and $\alpha_i$ are parameters of dimension $+2$ in natural units. In a more realistic picture, contributions from all fields (bosons and fermions) are expected, and general covariance leads to a generic form as represented by \,\eqref{Eq:eos.RVMgeneralized2}. In fact, the prospect for new terms in the effective vacuum action has been discussed in various ways in the literature\,\cite{Maggiore:2010wr,Bilic:2011zm,Bilic:2011rj}. These terms are also expected in the aforementioned stringy version of the RVM, see\,\cite{Mavromatos:2020kzj,Mavromatos:2021urx}.

Even though the vacuum dynamics from cosmological observations will receive contributions from all fields at a time, and in this sense the values of the coefficients $\nueff$ and $\tilde{\nu}_{\rm eff}$ can only be determined observationally, what matters here is that the theoretical framework leads to small values for them, as we have seen. After all the $\CC$CDM with a rigid cosmological constant works relatively well. Even so we know that the latter is afflicted with persisting tensions which call for an explanation. The RVM seems to encode the key theoretical features for such an explanation and appears phenomenologically preferred as well\,\footnote{The fitting results with different data sets confirm that the coefficients $\nueff$ and $\tilde{\nu}_{\rm eff}$ are of order $10^{-3}$ at most, see e.g.\,\cite{Sola:2017znb,SolaPeracaula:2016qlq,SolaPeracaula:2017esw,Perico:2016kbu,Geng:2017apd,SolaPeracaula:2020vpg,SolaPeracaula:2019zsl,Rezaei:2021qwd,Rezaei:2019xwo,Tsiapi:2018she,Singh:2021jrp,Asimakis:2021yct,Yu:2021djs,Grande:2011xf}. This suffices to have a non-trivial impact on the $\sigma_8$ and $H_0$ tensions\,\cite{SolaPeracaula:2021gxi}. }.

\section{Friedmann's equations and conservation laws with running vacuum}\label{Sect:RenormalizedFriedmann}

Friedmann's equations in the presence of running vacuum are a bit more complicated than usual. They have been dealt with previously in \eqref{Eq:eos.FriedmanEqs} assuming the traditional EoS for Vacuum and neglecting a possible evolution with the expansion of the gravitational constant, $G$. We see that VED which boils down to \eqref{Eq:eos.RVMgeneralized2} at low energy. This is sufficient for an effective treatment of the RVM since they offer the possibility to confront the predictions with the data and put bounds to the $\nueff$ parameter. This has led to a fruitful phenomenology, cf.\,\cite{Sola:2015wwa,Sola:2016jky,Gomez-Valent:2014rxa,Gomez-Valent:2014fda,Gomez-Valent:2015pia,Basilakos:2009wi,Grande:2010vg,Basilakos:2012ra,Sola:2017znb,SolaPeracaula:2016qlq,SolaPeracaula:2017esw,Perico:2016kbu,Geng:2017apd,SolaPeracaula:2020vpg,SolaPeracaula:2019zsl,Rezaei:2021qwd,Rezaei:2019xwo,Tsiapi:2018she,Singh:2021jrp,Asimakis:2021yct,Yu:2021djs,SolaPeracaula:2021gxi}, for instance. The phenomenological approach is very useful because there may be many QFT models (even string models\,\cite{Mavromatos:2020kzj,Mavromatos:2021urx}) whose effective behaviour leads to a vacuum energy density of the form \eqref{Eq:QuantumVacuum.RVM2}, with even higher corrections of order $\mathcal{O}(H^4)$. However, it is also interesting to study the exact form of Friedmann's equations of the RVM using directly the field variables involved in the QFT model under consideration, which in the present instance is based on a non-minimally coupled scalar field with action \eqref{Eq:QuantumVacuum.Sphi}. As can be expected, this part is more cumbersome but it reveals some new clues on the internal consistency of our calculation. Assuming an FLRW background, the starting point is Einstein's equations in renormalized form, see \eqref{Eq:QuantumVacuum.MEEsHD}, which we have to combine with the explicit formulas that we have derived in the previous sections for the vacuum energy density and pressure in our adiabatically renormalization approach. 

\subsection{Field equations and matter conservation law}

In the context of the model \eqref{Eq:QuantumVacuum.Sphi}, we have to distinguish between the background field density and pressure and their fluctuating or vacuum components (cf. \hyperref[Sect:AdiabaticVacuum]{Sect.\,\ref{Sect:AdiabaticVacuum}}). We denote by $(\rho_\phi, P_\phi)$ the background components. The fluctuating parts of these quantities have been object of devoted study in the previous sections and are represented by the quantities $(\rv, \Pv)$, which have been computed up to $6th$ adiabatic order. The generalized Friedmann's equation emerging from the $00th$-component of Eq.\,\eqref{Eq:QuantumVacuum.MEEsHD} can be written as follows,
\begin{equation}\label{Eq:eos.FriedmannDensity}
H^2=\frac{8\pi}{3}G(M)\left[\rv(M)+\rho_\phi+\rho_X(M)\right],
\end{equation}
where the running gravitational coupling $G(M)$ is related to the parameter $\Mpl^2(M)$ frequently used in the previous sections through Eq.\,\eqref{Eq:QuantumVacuum.RenCouplings}. Needless to say, if this equation were to apply to the current Universe we would need to add baryons and CDM, but here we just want to illustrate the interplay between the field $\phi$ and the vacuum without introducing more elements. In fact, the main actor here are the quantum vacuum effects produced by $\phi$. Its background part is not the main focus, but we include it for completeness and self-consistency.
In this context, we have got also the gravitational contribution from the HD tensor $\leftidx{^{(1)}}{\!H}_{\mu\nu}$ in the generalized Einstein's equations, which contributes the term $\rho_X$ in the above Friedmann's equation as follows:
\begin{equation}\label{Eq:eos.rhoX}
\rho_X \equiv -\alpha(M)\frac{\leftidx{^{(1)}}{\!H}_{00}}{a^2}= 18\alpha (M) (\dot{H}^2-2H\ddot{H}-6H^2\dot{H})\,.
\end{equation}
This effective energy density (acting as an effective fluid $X$) stems from the $00th$-component of the mentioned HD tensor (cf. \hyperref[Appendix:Conventions]{Appendix\,\ref{Appendix:Conventions}}).
Similarly, the generalized pressure equation within the Friedmann's pair can be written as
\begin{equation}\label{Eq:eos.FriedmannPressure}
3H^2+2\dot{H}=-8\pi G (M)\left[P_{\rm vac}(M)+P_\phi+P_X\right],
\end{equation}
where
\begin{equation}\label{Eq:eos.PX}
P_X \equiv -\alpha(M)\frac{\leftidx{^{(1)}}{\!H}_{11}}{a^2}= \alpha(M)\left(108H^2\dot{H}+54\dot{H}^2+72H\ddot{H}+12 \vardot{3}{H} \right)\,.
\end{equation}
Combining the two generalized Friedmann's equations given above we find
\begin{equation}\label{Eq:eos.dotHgen}
\dot{H}=-4\pi G(M) \left[P_{\rm vac}(M)+\rv(M)+P_\phi+\rho_\phi+P_X+\rho_X\right]\,.
\end{equation}
The conservation equation for the fluid X reads
\begin{equation}\label{Eq:eos.ConservationOfX}
\dot{\rho}_X+3H\left(\rho_X+P_X\right)=18\dot{\alpha}\left(\dot{H}^2-2H\ddot{H}-6H^2\dot{H}\right)=\frac{\dot{\alpha}}{\alpha}\rho_X.
\end{equation}
For completeness, we show the local conservation law for the background field $\phi_{\rm b}$ (denoted $\phi$ here for simplicity) which ensues from the fact that $\nabla^\mu T_{\mu\nu}^\phi=0 $. Indeed, using the explicit form of the classical EMT, Eq.\,\eqref{Eq:QuantumVacuum.EMTScalarField}, we find
\begin{equation}\label{Eq:eos.MatterCovConservation}
\begin{split}
\nabla^\mu T_{\mu\nu}^\phi &=\left(1-2\xi\right)\left(\Box \phi\right) \left( \nabla_\nu \phi\right)+\left(1-2\xi\right)\left(\nabla_\mu \phi \right)\left(\nabla^\mu \nabla_\nu \phi\right)+\left(2\xi-\frac{1}{2}\right)\nabla_\nu \left(\nabla^\alpha \phi \nabla_\alpha \phi\right)\\
&-2\xi\left(\nabla^\mu \phi\right)\left( \nabla_\mu \nabla_\nu \phi\right)-2\xi\phi \nabla^\mu \nabla_\mu \nabla_\nu \phi+2\xi \nabla_\nu \left(\phi \Box \phi\right)+\xi \left(\nabla^\mu G_{\mu\nu}\right)\phi^2\\
&+\xi G_{\mu\nu}\nabla^\mu \phi^2-\frac{1}{2}m^2\nabla_\nu \phi^2\\
&=\left(\Box-m^2\right)\phi \nabla_\nu \phi+2\xi \phi \left(G_{\mu\nu}\nabla^\mu \phi+\nabla_\nu \Box \phi-\Box\nabla_\nu \phi\right)\\
&=\xi R\phi \nabla_\nu \phi+2\xi \phi \left(G_{\mu\nu}\nabla^\mu \phi-R_{\mu\nu}\nabla^\mu \phi\right)=0\,.
\end{split}
\end{equation}
In the above derivation we have used the Klein Gordon equation \eqref{Eq:QuantumVacuum.KG} and the Bianchi identity $\nabla^\mu G_{\mu\nu}=0$.
At the same time we have made use of the formula \eqref{Eq:Conventions.CommutationNablaBox} in the \hyperref[Appendix:Conventions]{Appendix\,\ref{Appendix:Conventions}} in order to commute the covariant box operator $\Box$ and $\nabla_\nu$.
Equation \eqref{Eq:eos.MatterCovConservation} for $\nu=0$ can be rephrased in terms of the energy density and pressure:
\begin{equation}\label{Eq:eos.MatterConservation1}
\nabla^\mu T_{\mu\nu}^\phi=g^{\mu\alpha}\left(\partial_\alpha T_{\mu\nu}-\Gamma^\sigma_{\alpha\mu}T_{\sigma\nu}-\Gamma^\sigma_{\alpha\nu}T_{\mu \sigma}\right)=-\rho_\phi^\prime-3\mathcal{H}\left(\rho_\phi+P_\phi\right)=0\,,
\end{equation}
which, if rewritten in cosmic time differentiation (using $d/d\tau=a (d/dt)$), implies that
\begin{equation}\label{Eq:eos.MatterConservation}
\dot{\rho}_{\phi}+3H\left(\rho_\phi+P_\phi\right)=0\,,
\end{equation}
with
\begin{equation}\label{Eq:eos.rhophi}
\begin{split}
\rho_\phi\equiv\frac{T_{00}^{\phi_{\rm b}}}{a^2}=\frac{1}{2}\dot{\phi}^2+\frac{1}{2}m^2\phi^2+3\xi\left(2H\phi\dot{\phi}+H^2\phi^2\right)
\end{split}
\end{equation}
and
\begin{equation}\label{Eq:eos.Pphi}
\begin{split}
P_\phi\equiv\frac{T_{11}^{\phi_{\rm b}}}{a^2}=\frac{1}{2}\dot{\phi}^2-\frac{1}{2}m^2\phi^2-\xi\left(2\dot{\phi}^2+4H\phi\dot{\phi}+2\phi\ddot{\phi}+3H^2\phi^2+2\dot{H}\phi^2\right)\,.
\end{split}
\end{equation}
The ratio $w_\phi=P_\phi/\rho_\phi$ from the last two equations defines the EoS of the non-minimally coupled ($\xi\neq 0$) scalar field $\phi$, which is seen to be non-trivial. All in all, we have found that the background matter field $\phi$ does not interact with the vacuum, and hence its energy density is covariantly self-conserved during the expansion, cf. \eqref{Eq:eos.MatterConservation}.

It is easy to see that the local conservation law \eqref{Eq:eos.MatterConservation} is just another way to write the Klein-Gordon equation for the background field $\phi$:
\begin{equation}\label{Eq:eos.DifferentialEqPhi}
\ddot{\phi}+3H\dot{\phi}+(m^2+\xi R)\phi=\ddot{\phi}+3H\dot{\phi}+m^2\phi+\xi\left(12H^2+6\dot{H}\right)\phi=0\,.
\end{equation}
This equation is, of course, the same as Eq.\,\eqref{Eq:QuantumVacuum.KGexplicit}, but written in terms of the cosmic time and after having neglected the term $\nabla^2\phi$ owing to homogeneity and isotropy for the background field $\phi$ -- which, as advertised, corresponds to $\phi_{\rm b}(t)$ in Eq.\,\eqref{Eq:QuantumVacuum.ExpansionField}.

We may ask ourselves for the vacuum contribution of the background scalar field. For instance, a suitable constant field configuration $\phi_{\rm b}=\phi_{\rm b}^c$ in the minimal coupling case, reduces the background EMT \eqref{Eq:QuantumVacuum.EMTScalarField} to
\begin{equation}\label{Eq:eos.EMTScalarFieldxinull}
 T_{\mu\nu}^{\phi_{\rm b}}=-\frac{1}{2}m^2\left( \phi_{\rm b}^c\right)^2 g_{\mu\nu}.
\end{equation}
It has exactly the same EMT that it is associated to the CC defined in \hyperref[Sect:EMTScalarField]{Sect.\,\ref{Sect:EMTScalarField}}. Hence, it is tantalizing to include this contribution to our vacuum EMT \eqref{Eq:QuantumVacuum.EMTvacuum}. However, even if it is the case, notice that after doing the subtraction of scales \eqref{Eq:QuantumVacuum.RenT00vacuumMM0Eeqs} the independence of the former quantity on the renormalization point scale $M$ yields no explicit contribution to the running law. The contribution of $\phi_{\rm b}$ gets encapsulated in $\rho_{\rm vac}^0$ in \eqref{Eq:QuantumVacuum.RVM2}, {\it i.e.} the input value of the running formula. This points out once more the fact that QFT fails in the task of yielding the value of $\rho_{\rm vac}$ at any point in time from first principles. Only comparison between two different renormalization points is possible. Hence, all the background contributions get eventually camouflaged in the RVM law, so that we can just ignore them in our manipulations since the subtraction procedure, performed in the renormalization of VED, eventually get rid of them.

\subsection{Conservation equation for the quantum vacuum}

The vacuum, however, does not obey the same conservation equation as matter in general. In point of fact, it is not generally conserved. We find
\begin{equation}\label{Eq:eos.NonConserVED1}
\begin{split}
\dot{\rho}_{\rm vac}+3H\left(\rv+P_{\rm vac}\right)&=\frac{3\dot{M}}{8\pi^2 M} \left(\xi-\frac{1}{6}\right)H^2(M^2-m^2)\\
&+\frac{9\dot{M}}{8\pi^2 M}\left(\xi-\frac{1}{6}\right)^2\left(\dot{H}^2-2H\ddot{H}-6H^2\dot{H}\right) \,.
\end{split}
\end{equation}
It is remarkable that this equation can be written very succinctly in terms of the $\beta$-function of the running vacuum obtained in \eqref{Eq:QuantumVacuum.RGEVED1}:
\begin{equation}\label{Eq:eos.NonConserVED2}
\dot{\rho}_{\rm vac}+3H\left(\rv+P_{\rm vac}\right)=\frac{\dot{M}}{M}\,\beta_{\rv}\,.
\end{equation}
Here we have taken into account that the scale $M$ in cosmology is associated to a dynamical variable ($H$ in our case, although we do not implement any particular choice at this point), and hence it evolves with the cosmic time, $\dot{M}\neq0$. The compact form \eqref{Eq:eos.NonConserVED2} illustrates the fact that the non-conservation of the VED is due to both the running of $\rv$ with $M$ (i.e. the fact that $\beta_{\rv}\neq 0$) and to the cosmic time dependence of $M$. This feature is in contradistinction to ordinary gauge theories of strong and electroweak interactions\,\cite{Fritzsch:2012qc,Fritzsch:2015lua,Fritzsch:2016ewd}, and allows us to probe the effect of the time-dependence of $M$ in the running couplings and in particular in the VED. This is possible and even necessary in cosmology since the scale $M$ should be linked with cosmological variables changing with the cosmic time. When one studies situations where the ordinary gauge couplings participate in cosmological problems, it is perfectly possible to find out that they run both with the (time-independent) 't Hooft's mass unit $\mu$ and also with the cosmic (time-dependent) scale $M$, which is associated to $H$\,\cite{Fritzsch:2012qc,Fritzsch:2015lua,Fritzsch:2016ewd}. In \hyperref[Sect:appendixAbisbis1]{Appendix\,\ref{Sect:appendixAbisbis1}} we use Eq.\,\eqref{Eq:eos.NonConserVED2} to further investigate the time evolution of the VED. We show that the result is consistent with Eq.\,\eqref{Eq:QuantumVacuum.RVM2}, as it should.

The following comments are pertinent at this point. We remark that equation \,\eqref{Eq:eos.NonConserVED1}, or equivalently \eqref{Eq:eos.NonConserVED2}, is exact, that is to say, fulfilled to all adiabatic orders. This must be so since the scale dependence of the running quantities stops at order four. The higher order adiabatic terms do not bring additional $M$-dependent terms. However, let us not forget that the time-dependence is carried by all orders through the powers of $H(t)$ and its derivatives. We have used this fact to explicitly check that even in the presence of the complicated $6th$ order contributions in the structure of $\rv$ and $\Pv$ -- see Sections \hyperref[Sect:ZPE6th]{Sect.\,\ref{Sect:ZPE6th}} and \hyperref[Sect:RenormPressure]{Sect.\,\ref{Sect:RenormPressure}} -- Eq.\,\eqref{Eq:eos.NonConserVED1} is exactly satisfied and the $6th$ order effects in it just cancel out precisely. The calculation has been performed in the following way. The quantity $\rv(M)$ depends on time implicitly through $M$ but also through the many terms which depend on the Hubble rate and its time derivatives: $H,\dot{H}, \ddot{H}\dots \vardot{5}{H}$. Let's separate the first term on the LHS of \eqref{Eq:eos.NonConserVED1} as follows :
\begin{equation}\label{Eq:eos.chainruletM}
\dot{\rho}_{\rm vac}+3H\left(\rv+P_{\rm vac}\right)=\dot{M}\frac{\partial\rv}{\partial M}+ \frac{\partial\rv}{\partial t}\Bigg|_{M}+3H\left(\rv+P_{\rm vac}\right)\,,
\end{equation}
where $\dot{M}=dM/dt$ and $|_{\rm M}$ in the second term on the RHS is to emphasize that we keep $M$ constant in time when performing such a differentiation. The first term on the RHS, $\dot{M}\frac{\partial \rv}{\partial M}$, is entirely responsible for the result \eqref{Eq:eos.NonConserVED1}.
The last two terms on the RHS of \eqref{Eq:eos.chainruletM} can be shown to yield an identically vanishing result:
\begin{equation}\label{Eq:eos.TraditionalNonConserVED}
\frac{\partial \rho_{\rm vac}}{\partial t}\Bigg|_{M}+3H\left(\rho_{\rm vac}+P_{\rm vac}\right)=0\,.
\end{equation}
We have verified the exactness of this equation. Carrying out the check explicitly is a bit ponderous as it implies using the full structure (up to $6th$ adiabatic order) of the expressions for the vacuum density and pressure, {\it i.e.} equations \eqref{Eq:QuantumVacuum.renormalized6th} and \eqref{Eq:eos.VacuumPressureFull}. For this reason we have performed it with the help of Mathematica\,\cite{Mathematica}. We believe it constitutes a pretty robust consistency check of our formulas. The net outcome is just the expression on the RHS of \eqref{Eq:eos.NonConserVED1}, which involves effects up to $4th$ adiabatic order. All the remaining contributions from higher order cancel identically.

As emphasized above, the VED is not locally conserved since the scale $M$ evolves with the cosmic time and the VED runs with $M$. The integration of Eq.\,\eqref{Eq:eos.NonConserVED1} yields, of course, the characteristic RVM evolution law \eqref{Eq:QuantumVacuum.RenVDEexplicit}. The full local conservation equation containing all the ingredients is more complicated. Let us find it. We first extend the generalized Einstein's equations \eqref{Eq:QuantumVacuum.MEEsHD} by including also the background EMT contribution from the scalar field (i.e. by inserting the term $T_{\mu\nu}^{\phi}$ on its RHS), as in this way we take into account all of the components exchanging energy in the system:
\begin{equation}\label{Eq:eos.EqsVacExt}
\Mpl^2(M) G_{\mu \nu}+\rho_\Lambda (M) g_{\mu \nu}+\alpha(M) \leftidx{^{(1)}}{\!H}_{\mu\nu}= \langle T_{\mu\nu}^{\delta \phi}\rangle_{\rm ren}(M)+T_{\mu\nu}^{\phi}\,.
\end{equation}
We multiply next this equation by $8\pi G(M)=1/\Mpl^2(M)$ and take the covariant divergence on both sides (i.e. we apply the operator $\nabla^\mu$ on each term). Taking into account that $G_{\mu\nu}$ is a conserved tensor (i.e. we have the Bianchi identity $\nabla^\mu G_{\mu \nu}=0$) and that the HD tensor $ \leftidx{^{(1)}}{\!H}_{\mu\nu}$ is also conserved ($\nabla^\mu\, \leftidx{^{(1)}}{\!H}_{\mu\nu}=0$, see \hyperref[Appendix:Conventions]{Appendix\,\ref{Appendix:Conventions}}, we are left with a reduced expression where $\nabla^\mu$ acts now only on the running parameters and on the vacuum part of the EMT:
\begin{equation}\label{Eq:eos.EqsVacExt2}
\begin{split}
&\nabla^\mu\left(G(M)\rho_\Lambda (M)\right) g_{\mu \nu}+\nabla^\mu\left(G(M)\alpha(M)\right) \leftidx{^{(1)}}{\!H}_{\mu\nu} \\
&= \nabla^\mu\left(G(M)\langle T_{\mu\nu}^{\delta \phi}\rangle_{\rm ren}(M)\right)+\nabla^\mu\left(G(M)\right) T_{\mu\nu}^{\phi}\,.
\end{split}
\end{equation}
We have used $\nabla^\mu T_{\mu\nu}^\phi=0 $ as well ---cf. Eqs.\,\eqref{Eq:eos.MatterConservation1}-\eqref{Eq:eos.MatterConservation}. Performing the remaining derivatives and writing down the $\nu=0$ component of the final result, one finds after some calculations the following expression:
\begin{equation}
\begin{split}\label{Eq:eos.MixedConservation}
&\dot{G}(M)\left(\rho_\phi+\rv\right)+G(M)\dot{\rho}_{\rm vac}+3HG(M)\left(\rv(M)+P_{\rm vac}(M)\right)\\
&=\left( \alpha(M)\dot{G}(M)+G(M)\dot{\alpha}(M)\right) \frac{\leftidx{^{(1)}}{\!H}_{00}}{a^2}\,.
\end{split}
\end{equation}
With this result it is straightforward to show that the generalized Friedmann's equations and local conservation laws given above lead to the following overall conservation law involving all of the ingredients entering our quantum matter system non-minimally coupled to gravity:
\begin{equation}\label{Eq:eos.BianchiIdentity_all}
\begin{split}
&\frac{d}{dt}\left[G(M)\left(\rho_\phi+\rv(M)+\rho_X\right)\right]\\
&+3HG(M)\left(\rho_\phi+\rv(M)+\rho_X+P_\phi+P_{\rm vac}(M)+P_X\right)=0.
\end{split}
\end{equation}
\subsection{Running gravitational coupling}\label{SubSect:runningG}

If we neglect the effect of the HD term in the current Universe, the RHS of \eqref{Eq:eos.MixedConservation} can be set to zero and we are left with
\begin{equation}\label{Eq:eos.MixedConservationApprox1}
\dot{G}(M)\left(\rho_\phi+\rv(M)\right)+G(M)\dot{\rho}_{\rm vac}(M)+3HG(M)\left(\rv(M)+P_{\rm vac}(M)\right)=0\,,
\end{equation}
where we have used the conservation law for the background component of the scalar field, Eq.\,\eqref{Eq:eos.MatterConservation}.
A further simplification can be obtained if we assume that the EoS of the quantum vacuum is exactly $\Pv=-\rv$. We shall check right next what is the effect of the correction we have found in \,\hyperref[Sect:RenormPressure]{Sect.\,\ref{Sect:RenormPressure}}. In the meantime, if we just take the standard EoS of vacuum it allows us to dispense with the last term in the above equation. Because the two terms left involve derivatives with respect to the cosmic time we can write down \eqref{Eq:eos.MixedConservationApprox1} as a simple differential form:
\begin{equation}\label{Eq:eos.MixedConservationApprox2}
\left(\rho_\phi+\rv\right) dG+G d\rv=0\,.
\end{equation}
This equation can be used together with Friedmann's equation \eqref{Eq:eos.FriedmannDensity} in the same approximation (i.e. neglecting the HD terms and hence ignoring the $\rho_X$ component in it). The two equations can be easily combined in the nicely separable form,
\begin{equation}\label{Eq:eos.diffeqG}
\frac{3H^2}{8\pi G} dG+Gd\rv=\frac{3H^2}{8\pi G} dG+G\frac{3\nueff}{4\pi} \mpl^2{H} dH=0 \,,
\end{equation}
in which the sum $\rho_\phi+\rv$ has been replaced with $3H^2/(8\pi G)$ thanks to Friedmann equation, and $d\rv$ has been computed from \eqref{Eq:QuantumVacuum.RVM2} within the approximation of constant $\nueff$ -- cf.\,Eq.\,\eqref{Eq:QuantumVacuum.nueffAprox}. Notice that $G$ varies with $H$ and our aim is to find this function in this approximation. Dividing out the above equation by $G$ and upon identifying $G(H_0)=G_N=1/\mpl^2$ with the current local value measured in Cavendish-type experiments, we can solve for the function $G(H)$, with the result
\begin{equation}\label{Eq:eos.GH}
G(H)=\frac{G_N}{1+\nueff \ln\frac{H^2}{H_0^2}}\,,
\end{equation}
where $\nueff$ here is given by the constant coefficient \eqref{Eq:QuantumVacuum.nueffAprox}. This equation was previously met in\,\cite{Fritzsch:2012qc,Fritzsch:2015lua,Fritzsch:2016ewd} within a more simplified theoretical context and used to study the potential variation of the fundamental constants of Nature. However, equation \eqref{Eq:eos.GH} is only approximate in our context. Remember that we assumed that $\nueff$ is constant in its derivation, but we know from \hyperref[Sect:Abis1]{Appendix\,\ref{Sect:Abis1}} that it is not a strict constant.

It is natural to compare the above formula with Eq.\,\eqref{Eq:QuantumVacuum.RGENewton} at this point. The latter stems from the existence of running couplings, which is of course a direct reflex of the RG invariance of the effective action. In it, $M$ can be arbitrary since the effective action is independent of $M$.
Moreover, our assumption that $\Pv=-\rv$, which we also used in the above derivation, is not a sufficiently good approximation since we know from \hyperref[Sect:RenormPressure]{Sect.\,\ref{Sect:RenormPressure}} that there is indeed a departure of the quantum vacuum EoS from $-1$, see Eq.\,\eqref{Eq:eos.VacuumPressureFull}. Admittedly the last equation is a bit cumbersome, but if we consider only the contributions that can be relevant for the current Universe, the departure is given by the term $f_2(M,\dot{H})$ on the RHS of Eq.\,\eqref{Eq:QuantumVacuum.VacuumPressureFullsplit}. This is the same approximation used in our discussion of the EoS of the quantum vacuum for the current Universe, see \hyperref[Sect:EquationStateOfVacuum]{Sect.\,\ref{Sect:EquationStateOfVacuum}}. Thus, from Eq.\,\eqref{Eq:eos.f2} and setting $M=H$, according to our usual prescription (cf. \hyperref[Sect:Abis1]{Appendix\,\ref{Sect:Abis1}} for details), we find
\begin{equation}\label{Eq:eos.EoSaprox}
\Pv(H)+\rv(H)\simeq \frac{\left(\xi-\frac{1}{6}\right)}{8\pi^2}\dot{H}\left(m^2-H^2-m^2\ln\frac{m^2}{H^2}\right)\simeq \frac{\left(\xi-\frac{1}{6}\right)}{8\pi^2}\dot{H} m^2\left(1-\ln\frac{m^2}{H^2}\right)\,,
\end{equation}
where we have neglected a term of ${\cal O}(\dot{H}H^2)={\cal O}(H^4)$ but we have kept the terms proportional to $m^2\dot{H}$ as they are not necessarily negligible in the present Universe. The above expression gives the leading deviation of the quantum vacuum EoS from $-1$ at low energy (check next section \hyperref[Sect:EquationStateOfVacuum]{\ref{Sect:EquationStateOfVacuum}} for a complete discussion regarding Vacuum's EoS. 
), being such a deviation of the same order of magnitude as the $\sim H^2$ term involved in $\rv(H)$. The above correction to the quantum vacuum EoS genuinely originates from our calculation of the vacuum pressure in \hyperref[Sect:RenormPressure]{Sect.\,\ref{Sect:EquationStateOfVacuum}}. Therefore, it must be considered on equal footing with the dynamical term of $\rv(H)$ in the correct calculation. 
By duly taking into account Eq.\,\eqref{Eq:eos.EoSaprox} in the calculation of $G(H)$ and using the exact function $\nueff(H)$ given in the \hyperref[Sect:Abis1]{Appendix\,\ref{Sect:Abis1}} rather than just inserting the approximate constant result \eqref{Eq:QuantumVacuum.nueffAprox}, we find after some calculations the following expression for the running of the gravitational coupling (see the details in \hyperref[Sect:appendixAbis2]{Appendix\,\ref{Sect:appendixAbis2}}):
\begin{equation}\label{Eq:eos.GNHfinal}
G(H)=\frac{G_N}{1-\frac{\left(\xi-\frac{1}{6}\right)}{2\pi}\frac{m^2}{m^2_{\rm Pl}}\ln \frac{H^2}{H_0^2}}\,.
\end{equation}
This formula is not only more rigorous than Eq.\,\eqref{Eq:eos.GH} in our context, but in contrast to the latter it is entirely consistent with the running coupling formula \eqref{Eq:QuantumVacuum.RGENewton} when we set $M=H$ and $M_0=H_0$ in it and use the fact that $H^2-H_0^2$ is fully negligible versus $m^2\ln\left(H^2/H_0^2\right)$ for all $H$ (in post-inflationary times). In fact, for $H$ around the current value $H_0$, this follows from
\begin{equation}\label{Eq:eos.ratio}
 \frac{H^2-H_0^2}{m^2\ln\frac{H^2}{H_0^2}}=\frac{H_0^2}{m^2}\left(1+\frac12 x+{\cal O}(x^2)\right)\ll1 \ \ \ \ \ \ \ \ \left( 0\leq |x|<1\right)\,,
\end{equation}
with $x\equiv (H^2-H_0^2)/H_0^2$ and $H_0^2/m^2\ll1$ for any known particle. On the other hand, for large values of $H$ we also have $H^2/m^2\ll1$ since $m$ is assumed to be a mass of a typical GUT particle. Notice that the running of $G(H)$ from Eq.\eqref{Eq:eos.GNHfinal} is very mild, not only because it is a logarithmic running but also because the coefficient of the log is of order $m^2/\mpl^2\ll 1$, which holds good even for $m$ in the GUT range, {\it i.e.} $m\sim M_X\sim 10^{16}$ GeV, assuming that $\xi$ is not very big.

We should emphasize that the setting $M=H$ used above to study the running of $G(H)$ is the same one employed to infer the running vacuum formula (\hyperref[Sect:Abis1]{Appendix\,\ref{Sect:Abis1}}). The fully consistent derivation of \eqref{Eq:eos.GNHfinal} from two diverse roads; namely one (more physical) relying on the overall local conservation law \eqref{Eq:eos.MixedConservationApprox1}, and the other (more formal) based on the running coupling formula \eqref{Eq:QuantumVacuum.RGENewton} -- and ultimately on the RG-invariance of the effective action -- is a most remarkable feature. At the end of the day, the scale setting prescription $M=H$ proves to be the clue for exploring the physical consequences of our renormalization framework. Overall the obtained results speak up of the full mathematical and physical consistency of our approach.

Finally, we should note that although our discussion in this section has focused on the background and vacuum effects from the single quantum matter field $\phi$, this does not exclude the possibility that additional contributions from incoherent dust matter and radiation become involved. In the discussion of \hyperref[Sect:appendixAbis2]{Sect.\,\ref{Sect:appendixAbis2}} we show that the presence of ordinary matter does not alter at all the results presented above, provided one assumes that such an ordinary matter is conserved. Apart from that, there is also the possibility that matter components interact with the running vacuum. The safest possibility would be to assume potential interactions between Cold Dark Matter (CDM) and vacuum, as in this way the most sensitive and well known components of the Universe (baryons and photons) are unaffected. These new types of interaction between Dark Matter (DM) and vacuum can certainly be important but are model dependent, as they rely on introducing new parameters in the theory.

\section{EoS of the quantum vacuum and the CCP problem}\label{Sect:EquationStateOfVacuum}

Let us start with the traditional, on-shell approach to the study of the EoS of the Quantum Vacuum. The specific $f_6$ terms shown above in \hyperref[Sect:RenormPressure]{Sect.\,\ref{Sect:RenormPressure}} for the pressure are essential if we want to compute the EoS on-shell since $f_2=0$ and $f_4=0 $ for $M=m$, as it is obvious from the first two lines of \eqref{Eq:eos.VacuumPressureFull}. Therefore, at leading order, the on-shell value of the vacuum EoS is
\begin{equation}\label{Eq:eos.onshellEoS}
\Pv(m)=-\rv(m)+f_6=-\rho_\Lambda (m)-\frac{\langle T_{00}^{\delta \phi}\rangle^{(6)}_{\rm ren}(m)}{a^2}+f_6\,,
\end{equation}
where $\langle T_{00}^{\delta \phi}\rangle^{(6)}_{\rm ren}(m)$ is given by Eq.\,\eqref{Eq:QuantumVacuum.renormalized6th}. The EoS ``parameter'' therefore reads
\begin{equation}\label{Eq:eos.WVOnShell}
\begin{split}
\wv(m)&=\frac{\Pv(m)}{\rv(m)}=-1+\frac{f_6(m)}{\rv(m)}\\
&=-1+\frac{1}{10080\pi^2 m^2 \rv(m)}\left(8H^4\dot{H}-28\dot{H}^3+6H^3\ddot{H}-10\ddot{H}^2-22\dot{H}\vardot{3}{H}\right.\\
&\left.\phantom{xxxxxxxxxxxxxxxxxxxxxx}+24H^2 \dot{H}^2-7H^2 \vardot{3}{H}-49H \dot{H}\ddot{H}-6H \vardot{4}{H}-\vardot{5}{H}\right)\\
&+\frac{\left(\xi-\frac{1}{6}\right)}{240\pi^2 m^2 \rv(m) }\left(-6H^4\dot{H}+34\dot{H}^3-6H^3\ddot{H}+12\ddot{H}^2+24\dot{H}\vardot{3}{H}-24H^2\dot{H}^2\right.\\
&\left.\phantom{xxxxxxxxxxxxxxxx}+7H^2\vardot{3}{H}+55H\dot{H}\ddot{H}+6H\vardot{4}{H}+\vardot{5}{H}\right)\\
&-\frac{3\left(\xi-\frac{1}{6}\right)^2}{48\pi^2 m^2\rv(m) }\left(32\dot{H}^3-12H^3\ddot{H}+12\ddot{H}^2+24\dot{H}\vardot{3}{H}-48H^2\dot{H}^2\right.\\
&\phantom{xxxxxxxxxxxxxxx}\left.+5H^2\vardot{3}{H}+47H\dot{H}\ddot{H}+6H\vardot{4}{H}+\vardot{5}{H}\right)\\
&+\frac{9\left(\xi-\frac{1}{6}\right)^3}{4\pi^2 m^2 \rv(m)}\left(4H^4\dot{H}-5\dot{H}^3-6H^3\ddot{H}-11H\dot{H}\ddot{H}-\ddot{H}^2\right.\\
&\left.\phantom{xxxxxxxxxxxxxxx}-\dot{H}\vardot{3}{H}-24H^2\dot{H}^2-2H^2\vardot{3}{H}\right)\,.\\
\end{split}
\end{equation}
Quite obviously $w_{\rm vac}(m)=w_{\rm vac}(m,H, \dot{H}, \ddot{H},,...) $ is actually a function of $H$ and its derivatives. These terms can be relevant at the very early stages of the cosmological evolution. However, even during the short inflationary period deviations from the vacuum EoS $w_{\rm vac}=-1$ are tiny since all the terms that can trigger a departure depend on derivatives of $H$, but $H$ remains essentially constant during inflation. We discuss RVM-inflation in \,\hyperref[SubSect:RVMInflation]{Sect.\,\ref{SubSect:RVMInflation}}. At this point the important result that we have just obtained must be emphasized in a twofold manner, quantitatively and qualitatively. First, quantitatively, we have just proven that the on-shell EoS of the quantum vacuum is essentially the expected one, {\it i.e.} close to $w_{\rm vac}=-1$, and hence this is no longer an assumption or imposition; second, qualitatively, we have found that the quantum vacuum is not a static state but is dynamical: it changes very slowly at present but it could well have been a powerful driving force in the past. Both of these conclusions are perfectly reasonable for a primeval vacuum which might have been highly ``creative'' in the past but became much more tempered at present.

\subsection{Derivation of the quantum vacuum EoS for the FLRW regime}\label{SubSect:QuantumVacuumEoSFLRW}

Despite of the fact that the we have presented the complete expression for the equation of state (EoS) of the quantum vacuum on-shell, it seems natural to extend its computation off-shell as to use the renormalized expressions of both vacuum Presure and VED. In that sense, we can make use of Eq.\,\eqref{Eq:eos.VacuumPressureFull} to relate them. Quantum effects (starting at the end of the inflationary epoch) trigger a fully dynamical behavior of $\wv$ during the subsequent cosmic evolution within the conventional FLRW regime. As a result, the EoS of the quantum vacuum does not remain stuck to the classical value $\wv=-1$ and indeed changes throughout different epochs. Such an evolution can be explicitly derived from the QFT framework.

Our goal is to provide some details about the derivation of the important EoS formula within the renormalization approach presented above, which is valid for the post-inflationary epoch, {\it i.e.} for the whole FLRW regime. For all the considerations made during the FLRW regime we will neglect the quantum corrections of order ${\cal O}(H^4)$ or above, which can only be relevant for the inflationary epoch. Thus, for the EoS determination during the post-inflationary epoch, it suffices to keep the terms of adiabatic orders $2$ in Eq.\,\eqref{Eq:QuantumVacuum.VacuumPressureFullsplit} only. We find
\begin{equation}\label{Eq:eos.wvFLRW}
\begin{split}
\wv(H)=\frac{\Pv(H)}{\rv(H)}=-1+\frac{1}{\rho_{\rm vac}(H)} \frac{\left(\xi-\frac{1}{6}\right)}{8\pi^2}\dot{H}m^2\left(1-\ln\frac{m^2}{H^2}\right) + {\cal O}(H^4)
\end{split}
\end{equation}
where $\rv(H)$ in the denominator of the above formula is given by Eq.\,\eqref{Eq:QuantumVacuum.RVM2}. The $ {\cal O}(H^4)$ terms are to be neglected hereafter.
We can see from Eq.\,\eqref{Eq:eos.wvFLRW} that at leading order the vacuum EoS is  coincident with that of the $\CC$CDM  ($\wv=-1$),  as it could not be otherwise. Up to second adiabatic order, it reads
\begin{equation}\label{Eq:eos.wvFLRWapprox}
\wv(H) =-1+\frac{\epsilon \mpl^2}{4\pi\rho_{\rm vac}(H_0)}\dot{H}\left(1-\ln\frac{m^2}{H_0^2}\right)\simeq  -1-\nueff\, \mpl^2\,\frac{\dot{H}}{4\pi\rvo}\,,
\end{equation}
where the small parameter $\epsilon$ is defined by 
\begin{equation}\label{Eq:eos.nuandepsilon}
  \nueff \equiv \nueff (H_0)\simeq \epsilon \ln \frac{m^2}{H_0^2},\qquad \epsilon\equiv\frac{1}{2\pi}\left(\xi-\frac{1}{6}\right)\frac{m^2}{\mpl^2}.
\end{equation}
We have set $H=H_0$ in the log since the change is extremely slow within long cosmological periods, for example around our time, and used $\ln\frac{m^2}{H_0^2}\gg 1$ in the last step.  This expression is the result at ${\cal O}(\nueff)$ for very low redshift. By simple manipulations upon using \eqref{Eq:eos.nuandepsilon} and the $\CC$CDM form for $\dot{H} \approx-3\Omega_{\rm m}^0 H_0^2\left(1+z\right)^3/2$ (which is consistent at this order) we can reach now a beautiful and compact expression for the current EoS of the quantum vacuum, which just depends on $\nueff$ and on the current cosmological parameters $ \Omega_{\rm m}^0$ and $\Omega_\CC^0$, and can be written in terms of the cosmological redshift $z$:
\begin{equation}\label{Eq:eos.EoSDeviation}
\wv (z) \simeq  -1 + \nueff \frac{\Omega_{\rm m}^0}{ \Omega_{\rm vac}^0 }(1+z)^3\,, \qquad z < \mathcal{O}(1).
\end{equation}
Here, $\Omega_{\rm vac} \equiv \rho_{\rm vac}^0 / \rho_c =8\pi G_N \rho_{\rm vac} / (3H_0^2)$ and similarly for $ \Omega_{\rm m}^0$.
However, we would like to generalize that formula for arbitrarily large redshift within the FLRW epoch and for this we cannot approximate the denominator of  \eqref{Eq:eos.wvFLRW} through the constant $\rvo=\rv(H_0)$ as we did before.  We need to use now  the dynamical form of the VED during the FLRW epoch, {\it i.e.} Eq\,\eqref{Eq:RVM.RVMcanonical}, in which the  parameter $\nueff$  itself is running:
\begin{equation}\label{Eq:eos.nueff2}
\nueff(H)\equiv \epsilon\left(-1+\ln \frac{m^2}{H^{2}}-\frac{H_0^2}{H^2-H_0^2}\ln \frac{H^2}{H_0^2}\right)\,.
\end{equation}
Its approximately constant form for $H$ in the late time Universe is given by \eqref{Eq:eos.nuandepsilon}. To find out the vacuum EoS such that it be valid for any redshift from now up to the initial stages of the radiation-dominated epoch, we have to insert Eq.\,\eqref{Eq:eos.nueff2} into the canonical RVM form for the VED, that the reader may find in  Eq.\,\eqref{Eq:RVM.RVMcanonical} of\,\hyperref[Appendix:Abis]{appendix\,\ref{Appendix:Abis}}, and use the latter in the denominator of the EoS equation\,\eqref{Eq:eos.wvFLRW}.  To further proceed we need an explicit form for $H$. For $\nueff$ strictly constant, the RVM can be solved analytically\,\cite{Sola:2015wwa,Sola:2016jky,SolaPeracaula:2016qlq,SolaPeracaula:2017esw}. However, the QFT form of the RVM is more complicated since the effective parameter \eqref{Eq:eos.nueff2} is a function of $H$ and then an exact analytical solution is not feasible. Even so, taking into account that  $\nueff(H)$ is a slowly varying function of $H$ and that  $|\epsilon|\ll 1$,  the function $\nueff(H)$ remains always small,  and hence we can obtain a very good approximate solution for the full FLRW regime by expanding the solution in the small parameter $\epsilon$.  In this way we will be able to split the corrected $H^2$ (involving the QFT effects) into the leading  $\CC$CDM part plus ${\cal O}(\epsilon)$ corrections or higher.  The standard or concordance $\CC$CDM model part of $H^2$ is simply
\begin{equation}\label{Eq:eos.H2LCDM}
{H^2_{\Lambda\rm{CDM}}(z)} = H_0^2\left[ \Omega_{\rm m}^0 (1+z)^3+ \Omega_{\rm r}^0 (1+z)^4+\Omega_{\rm vac}^0\right]\,.
\end{equation}
Now upon inserting Eq.\,\eqref{Eq:RVM.RVMcanonical} into Friedmann's equation and separating the $\CC$CDM contribution, we find the following result:
\begin{equation}\label{Eq:eos.H2Oepsilon}
H^2=\frac{8\pi G(H)}{3}\rho \simeq H^2_{\Lambda \rm{CDM}}+\epsilon \left(H^2_{\Lambda \rm{CDM}}-H_0^2\right)\left(-1+\ln \frac{m^2}{H_0^2}\right)+{\cal O}(\epsilon^2)\,,
\end{equation}
where $\rho=\rho_{\rm m} (z)+\rho_{\rm r}(z)+\rho_{\rm vac}(H)+\cdots$ is the total density and the dots stand for the neglected ${\cal O}(H^4)$ corrections to Friedmann's equation in the present Universe \eqref{Eq:eos.rhoX} (which we neglect in all the considerations of this section). In the above expression, the term departing from the $\CC$CDM result has been calculated up to order ${\cal O}(\epsilon)$, but we should remark that $G(H)$ in \eqref{Eq:eos.H2Oepsilon} is given by by Eq.\,\eqref{Eq:eos.GNHfinal} and hence it had also to be expanded to ${\cal O}(\epsilon)$ so as to obtain the complete ${\cal O}(\epsilon)$ correction indicated in Eq.\,\eqref{Eq:eos.H2Oepsilon}.  In a similar way we find
\begin{equation} \label{Eq:eos.DerivativeOfHExpanded}
\dot{H}=\dot{H}_{\Lambda \rm{CDM}}+\epsilon\dot{H}_{\Lambda \rm{CDM}}\left(-1+\ln\frac{m^2}{H_0^2}\right)+{\cal O}(\epsilon^2)\,.
\end{equation}
Finally, introducing the above equations in Eq.\,\eqref{Eq:eos.wvFLRW},  we arrive after some calculations at the formula
\begin{equation}\label{Eq:eos.EoSepsilon1}
\begin{split}
&\wv(z)\simeq -1+\frac{\nu_{\rm eff}\,\left(1-\frac{\ln E_{\Lambda \rm{CDM}}^2}{\ln \frac{m^2}{H_0^2}}\right)\left( \Omega_{\rm m}^0 (1+z)^3+\frac{4}{3} \Omega_{\rm r}^0 (1+z)^4\right)}{\Omega_{\rm vac}^0+ \nu_{\rm eff}\left[-1+E_{\Lambda \rm{CDM}}^2(z)\left(1-\frac{\ln E_{\Lambda \rm{CDM}}^2(z)}{\ln \frac{m^2}{H_0^2}}\right)\right]}\,,
\end{split}
\end{equation}
in which $E_{\Lambda \rm{CDM}}^2(z)\equiv H^2_{\Lambda \rm{CDM}}(z)/H_0^2$,  with $\nueff$ given by \eqref{Eq:eos.nuandepsilon}. Once more we have used $\ln m^2 / H_0^2\gg 1$ to simplify the final result. In practice, it is sufficient to use the even more simplified form
\begin{equation}\label{Eq:eos.EoSepsilon2}
\wv(z)= -1+\frac{\nu_{\rm eff}\left( \Omega_{\rm m}^0 (1+z)^3+\frac{4}{3} \Omega_{\rm r}^0 (1+z)^4\right)}{\Omega_{\rm vac}^0+ \nu_{\rm eff}\left[-1+E_{\Lambda \rm{CDM}}^2(z)\right]}\,,
\end{equation}
since $\ln E_{\Lambda \rm{CDM}}^2/\ln \left(m^2 / H_0^2\right) \ll 1$ in the entire FLRW regime, as it can be easily checked. It is fully model-independent as the mass of the scalar particle has been absorbed by the generalized coefficient $\nueff$ (within the very good approximation used to derive it). Moreover, as indicated above, for small redshift values Eq.\,\eqref{Eq:eos.EoSepsilon2} trivially reduces to the much simpler form \eqref{Eq:eos.EoSDeviation}.

The above EoS formula for the quantum vacuum can still be further refined to include the next-to-leading ${\cal O}(\nueff^2)$ terms. This implies more work since we need to consistently collect all of $\epsilon^2$ terms and in particular also those from expanding up to that order the running gravitational coupling \eqref{Eq:eos.GNHfinal}. We shall omit the details of this lengthier calculation. The result stays, however, rather compact and we find that up to the next-to-leading order in $\epsilon$ we have
\begin{equation}\label{Eq:eos.H2Oepsilon2}
H^2(z)=H_0^2+ \left(H^2_{\Lambda \rm{CDM}}-H_0^2\right)\left(1+\epsilon\left(-1+\ln \frac{m^2}{H_0^2}\right)+\epsilon^2\left(-1+\ln \frac{m^2}{H_0^2}\right)^2\right)
\end{equation}
or
\begin{equation}\label{Eq:eos.H2Oepsilon2b}
E^2(z)\equiv \frac{H^2(z)}{H_0^2}\simeq \, E_{\Lambda \rm{CDM}}^2(z)+\nueff \left(E_{\Lambda \rm{CDM}}^2(z)-1\right) +\nueff^2 \left(E_{\Lambda \rm{CDM}}^2(z)-1\right)
\end{equation}
and
\begin{equation} \label{Eq:eos.DerivativeOfHOepsilon2}
\begin{split}
\dot{H}&=\dot{H}_{\Lambda \rm{CDM}}+\epsilon\dot{H}_{\Lambda \rm{CDM}}\left(-1+\ln\frac{m^2}{H_0^2}\right)+\epsilon^2\dot{H}_{\Lambda \rm{CDM}}\left(-1+\ln\frac{m^2}{H_0^2}\right)^2\\
&\simeq \dot{H}_{\Lambda \rm{CDM}}+\nueff\dot{H}_{\Lambda \rm{CDM}}+\nueff^2\dot{H}_{\Lambda \rm{CDM}}\,.
\end{split}
\end{equation}
These expressions obviously extend the previous ones up to ${\cal O}(\epsilon^2)$. We can use them to compute the EoS at this order. Once more we see that the expansion in $\epsilon$ is such that at leading order it can be expressed as an expansion in $\nueff$. The final result for the EoS to ${\cal O}(\nueff^2)$ takes on the form in  Eq.\,\eqref{Eq:eos.EoSepsilon1} with only the replacement $\nueff\to \nueff(1+\nueff)$ in the parameter $\nueff$ of its numerator. Thus, since  $0<\nueff\ll 1$, the next-to-leading ${\cal O}(\nueff^2)$ terms obviously imply a tiny correction to the ${\cal O}(\nueff)$ formula, which in practice can be neglected.

\begin{figure}
\begin{center}
\includegraphics[scale=0.55]{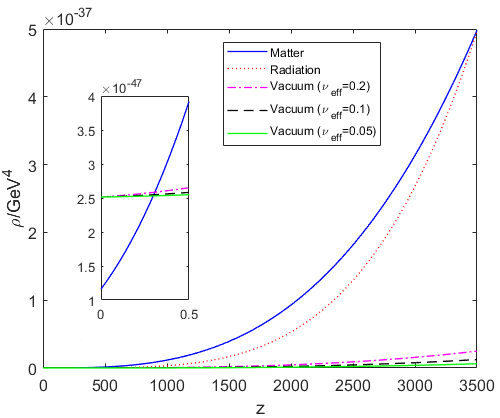}
\includegraphics[scale=0.55]{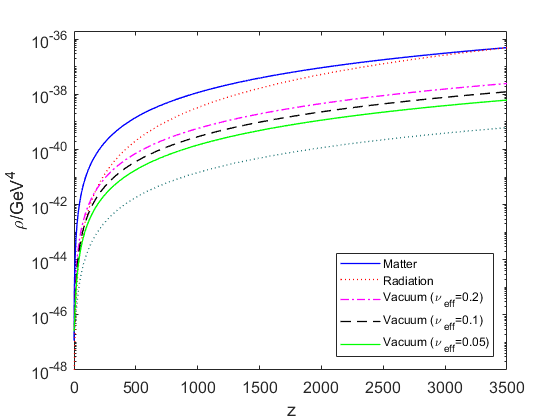}
\end{center}
\caption{Evolution of the different energy densities with the expansion in the RVM context. The plot on the right provides a complementary view in a (vertical) logarithmic scale. The VED exhibits a very mild dynamics up to the early radiation dominated epoch. Parameters $H_0$ and $\Omega_{\rm m}^0$ are taken from the best-fit values of\,\protect\cite{aghanim2020planck}. The vacuum evolution is very mild and to make it more apparent it is shown for different values of $\nu_{\rm eff}$. The additional (gren-dotted) line on the right plot corresponds to $\nu_{\rm eff}=0.0005$. Since it could not be appreciated on the left plot (which uses a linear scale), it has not been marked there.}\label{Fig:eos.DensitatLogDensitatsPaper}
\end{figure}

We remark that the model at this point is solved. Indeed, from Eq.\,\eqref{Eq:eos.H2Oepsilon2b} the quantum correction to the ordinary $\CC$CDM parameter $\Omega_{\rm vac}^0$ can be expressed directly in terms of the redshift as follows:
\begin{equation}\label{Eq:eos.rvz}
\Omega_{\rm vac}(z)\simeq \, \Omega_{\rm vac}^0+\nueff \left(E_{\Lambda \rm{CDM}}^2(z)-1\right) +\nueff^2 \left(E_{\Lambda \rm{CDM}}^2(z)-1\right)\,.
\end{equation}
Obviously $\Omega_{\rm vac}(z=0)= \Omega_{\rm vac}^0$ is satisfied, as it should be. Interestingly enough, to within ${\cal O}(\nueff)$ this expression is similar to the one found in previous calculations based on the phenomenological RVM, see e.g.\,\cite{Sola:2015wwa,Sola:2016jky,SolaPeracaula:2016qlq,SolaPeracaula:2017esw}, except that here we have derived the fundamental RVM formulas, including the quantum vacuum EoS, from QFT in curved spacetime within the framework put forward in earlier sections. The above equation can be written to ${\cal O}(\nueff)$ in terms of the vacuum energy density itself as follows:
\begin{equation}\label{Eq:eos.VEDz}
\rv(z)\simeq  \rvo+\nueff\,\rco \left(E_{\Lambda \rm{CDM}}^2(z)-1\right)\,,
\end{equation}
where $\rco=3H_0^2/(8\pi G_N)$ is the current critical density. This expression has been used for the VED plots in \hyperref[Fig:eos.DensitatLogDensitatsPaper]{Fig.\,\ref{Fig:eos.DensitatLogDensitatsPaper}}.
\subsection{Chameleonic Vacuum}

A sufficiently accurate approximation to the quantum vacuum EoS during the entire FLRW cosmic stretch can be derived directly from \eqref{Eq:eos.EoSepsilon2}:
\begin{equation}\label{Eq:eos.EoSChameleon}
\wv(z) = -1+\frac{\nu_{\rm eff}\left(\Omega_{\rm m}^0 (1+z)^3+\frac{4}{3}\Omega_{\rm r}^0 (1+z)^4\right)}{\Omega_{\rm vac}^0+ \nu_{\rm eff}\left[-1+\Omega_{\rm m}^0 (1+z)^3+\Omega_{\rm r}^0 (1+z)^4+\Omega_{\rm vac}^0\right]}\,,
\end{equation}
where $\Omega_{\rm vac}^0=\rvo/\rco\simeq 0.7$ is the current vacuum cosmological parameter,  whereas  $\Omega_{\rm m}^0=\rmo/\rco\simeq 0.3 $  and $ \Omega_{\rm r}^0=\rho_{\rm r}^0/\rco\sim 10^{-4}$  are the corresponding matter and radiation parts.
\begin{figure}
\begin{center}
\includegraphics[scale=0.75]{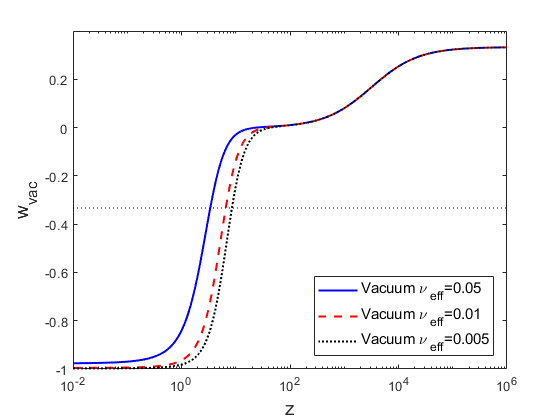}
\end{center}
\caption{Vacuum EoS for different (positive) values of $\nueff$. Some regimes to be noted: i) $\wv\simeq-1$ for very low redshift, ii)  $-1<\wv<-1/3$, vacuum mimics quintessence for low and intermediate redshift (the horizontal dotted line marks off the DE threshold $\wv=-1/3$), iii) $\wv=0$ plateau, vacuum imitates dust matter, and iv) $\wv=1/3$ plateau, vacuum mimics radiation. The quantum vacuum behaves as a cosmic chameleon.}\label{Fig:eos.EosEntire}
\end{figure}
The above EoS formulas depend on the crucial coefficient $\nueff$, which we have computed in QFT but it must ultimately be fitted to the cosmological data\,\cite{SolaPeracaula:2021gxi,Sola:2015wwa,Sola:2016jky,SolaPeracaula:2016qlq,SolaPeracaula:2017esw,Gomez-Valent:2018nib,Gomez-Valent:2017idt,CosmoTeam2023}. These analyses show that $\nueff\sim 10^{-2}-10^{-3}$  and that  $\nueff>0$ is the preferred sign. From the foregoing considerations, we find that the quantum vacuum never has the exact EoS $\wv=-1$ during the FLRW stage,  not even at $z=0$, where
\begin{equation}\label{Eq:eos.EoSnow}
\wv(0) \simeq -1+\nueff \frac{ \Omega_{\rm m}^0}{\Omega_{\rm vac}^0}\gtrsim -1 \ \ \ \ \ \ \ \ \ \ (\nueff>0)\,.
\end{equation}
Thus, amazingly,  the quantum vacuum currently behaves as quintessence or phantom\,\footnote{Equation \eqref{Eq:eos.EoSDeviation} resembles previous effective EoS forms for the dynamical VED derived phenomenologically in \cite{Sola:2005et,Sola:2005nh,Basilakos:2013vya}, although it is fundamentally different from them since it predicts a quintessence behavior of the quantum vacuum already at $z=0$,  in contrast to the aforementioned forms which predict the conventional behavior $\wv=-1$ at $z=0$. Not to mention, of course, that the EoS we have found here is derived from explicit QFT calculations.}. Such an effective behavior is triggered by the quantum effects and there is no need to introduce \textit{ad hoc} quintessence fields (nor \textit{ad hoc} inflatons, as we will see in the next section).
In \hyperref[Fig:eos.EosEntire]{Fig.\,\ref{Fig:eos.EosEntire}} we provide a detailed plot of the EoS \eqref{Eq:eos.EoSChameleon} for a large window of the FLRW regime well beyond the inflationary epoch.  The plot is performed for different values of $\nueff$ in a wide redshift range spanning from the present time up to the radiation epoch. The approximate EoS \eqref{Eq:eos.EoSDeviation} is only valid for the most recent Universe and deviates significantly from the more accurate one \eqref{Eq:eos.EoSChameleon}  for intermediate or large values of $z$. This can be clearly seen in \hyperref[Fig:eos.EoSApvsEx]{Fig.\,\ref{Fig:eos.EoSApvsEx}} where the two formulas are plotted to ease the comparison and to evince the large deviation at higher and higher redshifts. Notice that the detailed plot of the vacuum EoS in \hyperref[Fig:eos.EosEntire]{Fig.\,\ref{Fig:eos.EosEntire}} interpolates in a numerical way the results that can be directly inferred analytically from Eq.\,\eqref{Eq:eos.EoSChameleon} for the different redshift intervals all the way from the radiation epoch, down to the matter-dominated epoch until reaching the current epoch. Denoting by $z_{\rm eq}= \Omega_{\rm m}^0/ \Omega_{\rm r}^0-1\simeq 3300$ the equality point between matter and radiation, we find for $\nueff\neq 0$:
 \begin{equation}\label{eq:EoSregimes}
\wv(z)=\left\{
\begin{array}{ll}
 \frac13 \ \ \ \ \text{for}\ \ z\gg z_{\rm eq}\ \ \text{with}\ \ \  \Omega_{\rm r}^0 (1+z)\gg \Omega_{\rm m}^0, \quad \text{radiation behavior}, \\
 & \\
\hspace{0.0cm} 0 \ \ \ \ \text{for}\ \ {\cal O}(1)<z\ll z_{\rm eq}\ \ \text{with}\ \ \  \Omega_{\rm m}^0\gg  \Omega_{\rm r}^0 (1+z), \quad {\text{dust behavior}},\\
& \\
-1+\nueff \frac{ \Omega_{\rm m}^0}{\Omega_{\rm vac}^0}(1+z)^3\ \ \ \ \text{for}\ \ \ -1<z<{\cal O}(1)\,,
\quad \text{phantom/quint. behavior}. \\
\end{array}
\right.
\end{equation}
As we can see, the quantum vacuum EoS follows the EoS of relativistic matter in the radiation-dominated epoch, then the EoS of non-relativistic (dust) matter in the matter-dominated epoch, the EoS of quintessence at present (for $\nueff>0$) and asymptotes to de Sitter era in the future ($z\to-1$), where $\omega_{\rm vac}\to -1$.

In the presence of the quantum vacuum effects, the deceleration parameter $q = -1 -\dot{H}/H^2$ can be easily derived. Using the expression for the quantum corrected $H$ up to order ${\cal O}(\nueff)$, eq.\,\eqref{Eq:eos.H2Oepsilon}, and requiring that $q=0$ we find that the transition redshift from deceleration to acceleration becomes slightly shifted with respect to that of the concordance model ($\CC$CDM), as follows:
\begin{equation}\label{Eq:eos.TransitionPoint}
z_t=\left(\frac{2\left(\Omega_{\rm vac}^0-\nueff\right)}{ \Omega_{\rm m}^0 (1+\nueff)}\right)^{1/3} - 1\,.
\end{equation}
As expected, the $\CC$CDM result is recovered for $\nueff=0$. Since, however, $\nueff$ is small and  $z_t$ cannot be measured with precision yet, it is not the ideal signature. What it really acts as a useful signature of the quantum vacuum is its effective behavior as quintessence in the low redshift range, as we have seen above.
\begin{figure}
\begin{center}
\includegraphics[scale=0.55]{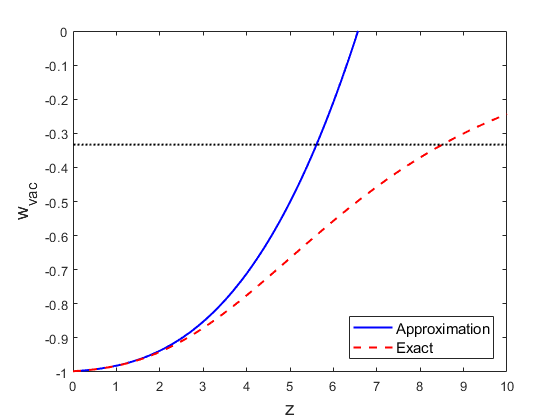}
\end{center}
\caption{Vacuum EoS at low redshifts including the quantum effects presented in this chapter. We show both the curve corresponding to the of the approximate vacuum EoS formula \protect\eqref{Eq:eos.EoSDeviation} with respect to the more precise one given by Eq.\,\protect\eqref{Eq:eos.EoSChameleon} for a typical value $\nueff=0.005$, observing a deviation at low redshifts. Because the fitting analyses\,\protect\cite{SolaPeracaula:2021gxi} generally favor the sign $\nu_{\rm eff}>0$, the EoS deviates $\omega_{\rm vac}=-1$ already at $z=0$ and mimics quintessence. }\label{Fig:eos.EoSApvsEx}
\end{figure}
Indeed, amazingly enough, the quantum vacuum is kind of `chameleonic'. It  behaves as `true' vacuum ($\wv=-1$) only in the very early times when it  triggers inflation. It then  remains silent for eons  (hidden as if being  relativistic or  pressureless matter). Today,  it appears as (dynamical)  dark energy (DE), specifically as  quintessence  ($-1<\wv\lesssim-1/3$), cf. \hyperref[Fig:eos.EosEntire]{Fig.\,\ref{Fig:eos.EosEntire}}.  As a result of this multifaceted behavior, it may crucially help in solving the $\sigma_8$ and $H_0$ tensions afflicting the $\CC$CDM model, see e.g.\,\cite{DiValentino:2020vvd,DiValentino:2020vvd,DiValentino:2021izs,Perivolaropoulos:2021jda,Vagnozzi:2019ezj,Abdalla:2022yfr,SolaPeracaula:2020vpg,SolaPeracaula:2019zsl} and the long list of references therein on different models and alternative points of view.  We will comment more on these phenomenological implications in the next section, but first some comments are required.

Through the former derivations we have set $\rho_{\rm m}=\rho_{\rm m}^0 a^{-3}=\rho_{\rm m}^0 (1+z)^{3}$, just as in the matter conservation law for the $\CC$CDM. This is justified to ${\cal O}(\nueff)$ as we now argue. Recall that in the presence of running vacuum the matter conservation law can be affected in some cases. For example, if there is an exchange between vacuum and cold dark matter (CDM), the matter conservation law takes corrections of the type
$\rho_{\rm cdm}^0a^{-3(1-\nueff)}$\,\cite{Sola:2017znb,SolaPeracaula:2016qlq,SolaPeracaula:2017esw}. However, even in these cases it does not modify the leading form of the EoS that we have found, as it only amounts to add a ${\cal O}(\nueff^2)$ correction to it. In other situations, such as e.g. the one that will be studied in \hyperref[Chap:PhenomenologyofRVM]{chapter\,\ref{Chap:PhenomenologyofRVM}} (so-called type-II) the gravitational coupling also runs with the cosmic expansion but the running is of the form $\sim \nueff \ln H$. This term would induce once more a negligible ${\cal O}(\nueff^2)$ correction to the equation of state \eqref{Eq:eos.EoSChameleon}. As it turns out, therefore, the EoS that ensues from our QFT approach is pretty universal for the RVM in its various implementations, at least to order ${\cal O}(\nueff)$ and for the indicated low redshift range.
A more detailed treatment of the EoS in the general regime will be presented elsewhere.

In summary, Eq.\,\eqref{Eq:eos.EoSDeviation} reconfirms what we had already advanced in Sec.\ref{Sect:RunningConnection}, namely that for $\nueff>0$ (resp. $\nueff<0$) the evolving VED mimics quintessence (resp. phantom DE). This result is a bit provocative and comes as something of a surprise. The usual picture of the vacuum is that it is a kind of medium with a strict EoS equal to $-1$. However, the QFT analysis of vacuum in curved spacetime shows that it is not so. The so-called quintessence or phantom fields (and in fact DE in general) could be nothing else but a manifestation of the (dynamical) quantum vacuum, and if so {there is no need of \textit{ad hoc} fields with particular potentials to explain the DE. Vacuum fluctuations of quantum matter fields could just make it since upon proper renormalization they lead to small contributions of order $m^2 H^2$.} The current fits to $\nueff$ suggest that it should be in the ballpark of $\sim 10^{-3}$ and positive\,\,\cite{Sola:2015wwa,Sola:2016jky,Gomez-Valent:2014rxa,Gomez-Valent:2014fda,Gomez-Valent:2015pia,Basilakos:2009wi,Grande:2010vg,Basilakos:2012ra,Sola:2017znb,SolaPeracaula:2016qlq,SolaPeracaula:2017esw,Perico:2016kbu,Geng:2017apd,SolaPeracaula:2020vpg,SolaPeracaula:2019zsl,Rezaei:2021qwd,Rezaei:2019xwo,Tsiapi:2018she,Singh:2021jrp,Asimakis:2021yct,Yu:2021djs,SolaPeracaula:2021gxi}, and hence the favored dynamical DE mimicked by the quantum vacuum seems to be that of quintessence. A related result was already hinted at long ago but on much more phenomenological grounds\,\cite{Sola:2005et}. Surprisingly, it also holds in Brans-Dicke theory with a cosmological constant (see \hyperref[Chap:PhenomenologyofBD]{chapter\,\ref{Chap:PhenomenologyofBD}}), as this context has recently been shown to mimic the RVM, see\,\cite{SolaPeracaula:2018dsw,deCruzPerez:2018cjx} and\,\cite{SolaPeracaula:2020vpg,SolaPeracaula:2019zsl}. After we have studied the impact of the new pressure terms of adiabatic order 2 at low energies, in the next section we also consider the impact of the higher adiabatic terms at high energies.

\section{Running vacuum: some phenomenological implications}\label{Sect:PhenoImplications}

We have devoted most of this chapter and the previous one to put the foundations of the running vacuum model (RVM) on a sound theoretical basis within QFT in curved spacetime. In this last section, which precedes the discussion of this chapter, we would like to put forward some phenomenological considerations in order to illustrate the possible physical impact of the quantum running vacuum in the light of the observational data and its implications in the early Universe, providing a simple mechanism for inflation.

\subsection{Post-Inflationary implications of RVM}\label{SubSect:PostInflationary}

Although a more detailed phenomenological analysis will be presented elsewhere\,\cite{CosmoTeam2023}, here we just highlight a few potentially significant consequences of our study. We have already mentioned that the RVM has been tested in the past in a variety of phenomenological contexts, where the basic parameter $\nueff$ has been fitted to different sets of cosmological data\,\cite{Sola:2015wwa,Sola:2016jky,Gomez-Valent:2014rxa,Gomez-Valent:2014fda,Gomez-Valent:2015pia,Basilakos:2009wi,Grande:2010vg,Basilakos:2012ra,Sola:2017znb,SolaPeracaula:2016qlq,SolaPeracaula:2017esw,Perico:2016kbu,Geng:2017apd,SolaPeracaula:2020vpg,SolaPeracaula:2019zsl,Rezaei:2021qwd,Rezaei:2019xwo,Tsiapi:2018she,Singh:2021jrp,Asimakis:2021yct,Yu:2021djs,SolaPeracaula:2021gxi}. The fact that we have now been able to account for the structure of the RVM on QFT grounds, it obviously strengthens its position. Let us focus in a particular variation of the RVM that we recently investigated in one of our works\,\cite{SolaPeracaula:2021gxi}, the RRVM ({\it Ricci RVM}). It is fully revisited \hyperref[Chap:PhenomenologyofRVM]{chapter\,\ref{Chap:PhenomenologyofRVM}}, however we would like to advance some aspects since some imprints of these theoretical derivations are reflected on the results of this model. The RRVM arises from the generalized expression for the VED which we have predicted in \hyperref[SubSect:GeneralizedRVM]{Sec.\,\ref{SubSect:GeneralizedRVM}}. It includes the two low energy terms $H^2$ and $\dot{H}$, each one with independent coefficients. More specifically, we consider a generic RVM structure of the form \eqref{Eq:eos.RVMgeneralized2}. From the QFT perspective, the two coefficients $\nueff$ and $\tilde{\nu}_{\rm eff}$ depend both on the number of bosons and fermions in a way which can be computed theoretically. Albeit we have presented the calculation for the scalar contribution only, the yield from the fermionic part will be shown in \hyperref[Chap:Fermions]{chapter\,\ref{Chap:Fermions}}. Nevertheless, note that despite the fact that the vacuum running is theoretically computable in QFT, which is perhaps the most remarkable observation of our published works \cite{Moreno-Pulido:2020anb,Moreno-Pulido:2022phq,Moreno-Pulido:2022upl,SolaPeracaula:2022hpd}, we have limited ourselves to the free theory and in practice a more realistic picture would arise after the introduction of interactions. Therefore, at present we cannot predict the precise quantitative evolution of the VED. But this does not preclude us, of course, from testing the phenomenological performance of the model. 

What we are going to do now is to advance some results regarding the model we will revisit in full detail in  \hyperref[Chap:PhenomenologyofRVM]{chapter\,\ref{Chap:PhenomenologyofRVM}}. While the general VED form \eqref{Eq:eos.RVMgeneralized2} was analyzed on pure phenomenological grounds in\,\cite{Sola:2015wwa,Sola:2016jky}, here we will minimize the number of parameters and shall consider the convenient situation $\tilde{\nu}_{\rm eff}=\nueff/2$, as in this way the VED adopts the form
\begin{equation}\label{Eq:eos.RRVM}
\rv(H) = \frac{3}{8\pi G_N}\left(c_{0} + \nueff\,H^2+ \tilde{\nu}_{\rm eff}\,\dot{H}\right)=\frac{3}{8\pi G_N}\left(c_{0} + \frac{\nueff}{12} {R}\right)\,,
\end{equation}
with ${R} = 12H^2 + 6\dot{H}$ the curvature scalar. This scenario was analyzed in\,\cite{SolaPeracaula:2021gxi} under different hypotheses, in particular it was assumed that the vacuum was interacting with cold dark matter, this will be explained in \hyperref[Chap:PhenomenologyofRVM]{chapter\,\ref{Chap:PhenomenologyofRVM}}. We restrict the number of assumptions to the minimum and just adapt to the precise situation that we have encountered in the QFT analysis presented in this chapter, in which matter is locally conserved (as in the standard $\CC$CDM) and the VED and gravitational coupling $G$ possess a mild cosmic evolution (as studied in the previous section). Interactions between matter and vacuum are not considered. This scenario is particularly well-behaved in the radiation dominated era since the relativistic matter density is not altered as compared to the standard model and hence cannot impinge on the BBN physics\,\cite{Sola:2015wwa,Sola:2016jky}. This is all the more true if we take into account that the vacuum energy density \eqref{Eq:eos.RRVM} remains also subdominant in the radiation epoch since $R\simeq 0$ in it.

In fact, in \cite{SolaPeracaula:2021gxi} it was shown that if there is a `DE threshold' $z_{*}$ near our time where the DE dynamics of the vacuum gets suddenly activated, this can be extremely helpful for solving the $\sigma_8$ tension within the RVM. At the same time, it was shown that if the gravitational coupling runs slowly (logarithmically) with the expansion, this can help to fix the $H_0$ tension. In \hyperref[Fig:eos.EosEntire]{Fig.\,\ref{Fig:eos.EosEntire}} we can see that a continuous (i.e. not abrupt) DE `threshold' window with low  $z_{*}={\cal O}(1)$ does indeed exist for the quantum vacuum, in the sense that for $z<z_{*}$ the vacuum gets progressively activated as DE ($\wv<-1/3$), whereas for $z>z_{*}$ the EoS of the quantum vacuum transmutes successively into that of dust matter and radiation. Additionally, in \hyperref[SubSect:runningG]{Sect.\,\ref{SubSect:runningG}} we have found that the dynamics of the vacuum is intertwined with that of the gravitational coupling through a log of the Hubble rate: $G=G(\ln H)$. Being these two crucial factors simultaneously present in our QFT approach, they combine constructively to relieve both tensions at a time. Incidentally, we note that the vacuum EoS for the current Universe \eqref{Eq:eos.EoSDeviation} in our QFT approach is similar to the EoS of the effective dark energy (DE) in a Brans-Dicke (BD) theory in the presence of a cosmological constant, see \hyperref[Chap:PhenomenologyofBD]{chapter\,\ref{Chap:PhenomenologyofBD}}.

As previously indicated, the RVM under consideration preserves local matter conservation. However, the VED evolves together with $G$ such that the Bianchi identity can be satisfied, just as explained in \hyperref[Sect:RenormalizedFriedmann]{Sect.\,\ref{Sect:RenormalizedFriedmann}}. On solving explicitly the model (details will be provided in \hyperref[Chap:PhenomenologyofRVM]{chapter\,\ref{Chap:PhenomenologyofRVM}}) we find the following evolution for the VED in linear order in the small parameter $\nueff$:
\begin{equation}\label{Eq:eos.VDEm}
\rv(a)=\left(\frac{\Omega_{\rm vac}^0}{ \Omega_{\rm m}^0}-\frac14\,\nueff\right) \rho_{\rm m}^0+\frac14\nueff\rho_{\rm m}^{0}a^{-3}+\mathcal{O}(\nueff^2)\,,
\end{equation}
where the current cosmological parameters $\Omega_i^0=(8\pi G_N) \rho_i^0/(3 H_0^2)$ satisfy $\Omega_{\rm vac}^0+ \Omega_{\rm m}^0=1$.
For $\nueff=0$ the VED is constant and $\rv=\frac{3 H_0^2}{8\pi G_N} \Omega_{\rm vac}^0=\rvo=$const. (i.e. we recover the $\CC$CDM behavior, as it should be), but for non-vanishing $\nueff$ the VED has a moderate dynamics since this parameter is small. In \hyperref[Fig:eos.DensitatLogDensitatsPaper]{Fig.\,\ref{Fig:eos.DensitatLogDensitatsPaper}} we plot the various energy densities for matter, radiation and vacuum (from \eqref{Eq:eos.VDEm}) using the best-fit values from Planck TT,TE,EE,+low E+lensing data\,\cite{aghanim2020planck}.

If $\nueff>0$ (as it follows from different phenomenological studies mentioned above), the EoS of the quantum vacuum satisfies $\wv>-1$ even at $z=0$ and also \eqref{Eq:eos.VDEm} predicts the VED to decrease with time. Hence it mimics quintessence without need of invoking \textit{ad hoc} scalar fields. This is quite revealing, as it suggests that such an effective quintessence behaviour may emerge from a fundamental QFT origin.
\begin{figure}[t]
\begin{center}
\includegraphics[scale=0.65]{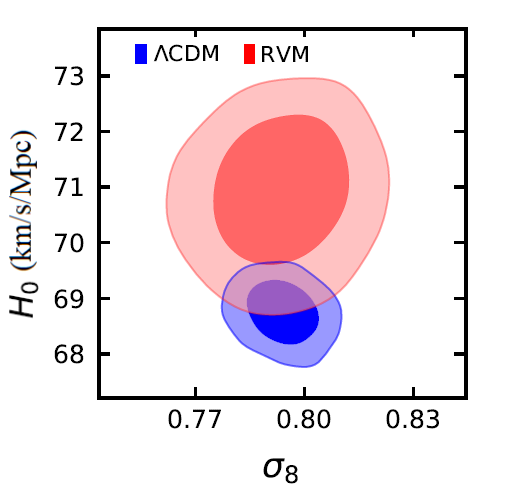}
\end{center}
\caption{Comparative contours at 1$\sigma$ and 2$\sigma$ c.l. in the $(\sigma_8, H_0)$-plane corresponding to the RVM and the $\Lambda$CDM for the data sets mentioned in the text. It can be seen that the RVM is quite effective in alleviating the $H_0$ tension and at the same time it reduces the $\sigma_8$ one.}
\label{Fig:rvm.CountourTypeII}
\end{figure}
Finally, in \hyperref[Fig:rvm.CountourTypeII]{Fig.\,\ref{Fig:rvm.CountourTypeII}} we illustrate an important phenomenological implication of the present framework, namely the possibility that the RVM could help in solving or at least alleviating the persistent $H_0$ and $\sigma_8$ tensions mentioned in the introduction. We refer the reader to the previous studies\,\cite{SolaPeracaula:2020vpg,SolaPeracaula:2021gxi} for a detailed definition and description of these data and the methodology used in the fitting analysis. These tensions are still the main focus of interest of many cosmologists\,\cite{verde2019tensions,DiValentino:2020zio,Abdalla:2022yfr,Perivolaropoulos:2021jda,DiValentino:2021izs,DiValentino:2020vvd,SolaPeracaula:2018xsi,Riess:2019cxk}. In that figure we show the $1\sigma$ and $2\sigma$ c.l. contours in the $(\sigma_8, H_0)$ plane both for the RVM and the $\CC$CDM. The used data sets in this analysis involve type Ia supernovae, baryonic acoustic oscillations, cosmic chronometers, large scale structure formation data (on $f\sigma_8$) and the cosmic microwave background observations from Planck 2018 (i.e. the data string SnIa+BAO+$H(z)$+LSS+CMB).
In the light of the results presented in \hyperref[Fig:rvm.CountourTypeII]{Fig.\,\ref{Fig:rvm.CountourTypeII}}, we can say that the comparative performances between the RVM and $\CC$CDM clearly show that the RVM may alleviate the two tensions $\sigma_8$ and $H_0$ at a time, which is remarkable. The value of $H_0$ tends to be higher in the case of the RVM as compared to the $\CC$CDM and as a result the current $5\sigma$ tension\cite{Riess:2021jrx} between the measurements of the local value of $H_0$ and from the early Universe (CMB) is rendered at the residual level of $\sim 1.6\sigma$. Similarly, the  $\sigma_8$ value is also reduced and the corresponding tension is brought to the inconspicuous level of $\sim 1.3\sigma$.  However, we believe that even from the short considerations highlighted in this section the reader can already get a flavor of the real potential of the RVM for improving the description of the cosmological data, {\it i.e.} when the cosmological vacuum becomes dynamical, and more specifically when it runs according to the QFT description presented here.

We remind the reader that all these disquisitions discussed in this section are going to be expanded in \hyperref[Chap:PhenomenologyofRVM]{chapter\,\ref{Chap:PhenomenologyofRVM}}. We have advanced some of the discussions here. However, there is a good reason for doing that. Our main point is to express that some of the features attributed to the vacuum fluid that were invoked in the aforementioned papers regarding the RVM (including\,\cite{SolaPeracaula:2021gxi}), were just contemplated from general arguments or for the sake of the fits. Now, we have seen how these emerge naturally from the QFT considerations presented here. Namely, a variable $G$, a late activation of the vacuum fluid as dynamical DE, and a mildly evolving VED which may behave as quintessence (or phantom) at the present time. Moreover, we have seen that only one parameter is needed to connect all these effects (although the possibility of considering a generalized RVM version may require more than one parameter\,\eqref{Eq:eos.RVMgeneralized2}). So that we have a strong motivation to study models incorporating these attributes jointly, as we expect to do in the near future\,\cite{CosmoTeam2023}.

\subsection{Implications for the early Universe: RVM-inflation}\label{SubSect:RVMInflation}

So far we have elaborated on the VED expression \eqref{Eq:QuantumVacuum.VEDscalesMandM0Final} in the low energy regime, in which we can neglect the ${\cal O}(H^4)$ terms of the form $\dot{H}^2, H\ddot{H}$ and  $H^2 \dot{H}$.  In such regime we know that the VDE can be put in the alternative form \eqref{Eq:QuantumVacuum.RVM2}, which fits in with the traditional RVM structure of the vacuum evolution and represents a small dynamical departure  with respect to the $\Lambda$CDM since $|\nueff|\ll1$. Rephrased in this fashion we can see that the obtained VED around our time represents a small variation with respect to the current value of the vacuum energy density, $\rho_{\rm vac}^0$. While the previous discussion obviously applies to the current Universe only, since we have neglected the ${\cal O}(H^4)$ terms on the RHS of Eq.\,\eqref{Eq:QuantumVacuum.VEDscalesMandM0Final}, we should emphasize that these terms can play a significant role in the early Universe. They are  generated from the functional differentiation of the $R^2$-term in the higher derivative part of the vacuum action (cf.  \hyperref[Appendix:Conventions]{chapter\,\ref{Appendix:Conventions}}), and therefore they play a similar role as in the case of Starobinsky's inflation\,\cite{Starobinsky:1980te,Starobinsky:1981vz,Vilenkin:1985md}. Notice that even though  all the terms  of the form $\dot{H}^2, H\ddot{H}$ and  $H^2 \dot{H}$ in Eq.\,\eqref{Eq:QuantumVacuum.VEDscalesMandM0Final} are denoted here as being of  ${\cal O}(H^4)$, none of them is really proportional to $ H^4$. As a result, they all vanish for $H$ strictly constant. In fact, Starobinsky's inflation is not triggered by an early epoch in which $H=$const. but by one in which $H$ decreases at constant rate $\dot{H}=$const, see e.g. \cite{Sola:2015rra} for a summarized discussion focusing on these well-known facts. The corresponding inflation period  is characterized by a final phase with rapid oscillations of the gravitational field, which is when the Universe leaves the inflationary phase and enters the radiation epoch after a reheating period. Prior to the oscillatory phase, hence within the inflationary period, $H$ decreases fast and $\dot{H}$ remains approximately constant (thence $\ddot{H}\simeq 0$)\cite{Sola:2015rra}. After the inflation period is accomplished we know that the Universe enters the radiation epoch, where $R=0$ and the higher order terms become irrelevant for the driving of the cosmic expansion.

After this brief summary of Starobisnky's inflation, we would like to note that once the vacuum energy density in cosmological spacetime is renormalized through the adiabatic procedure a definite prediction for a mechanism of early inflation emerges which is characterized by a short period where $H$=const. This constant must take, of course, a large value which we expect to lie around a characteristic GUT scale. It is nevertheless totally unrelated to the ground state  of a scalar field potential and hence does not require  any \textit{ad hoc} inflaton field. Such an alternative form of inflation,  based on the constancy of $H$ for a short lapse of time,  is called  `RVM-inflation'. To set off an inflationary phase with this mechanism we need powers of $H$ higher than  $H^2$, see\,\cite{Lima:2013dmf,Perico:2013mna,Sola:2015csa,Sola:2015rra,SolaPeracaula:2019kfm} for a phenomenological description. All that said, there are features of the RVM in the very early Universe which our analysis (strictly based on QFT) could not be sensitive to, and hence we would like to comment on them here. These features are connected with string theory contributions. As we have mentioned, the higher order terms of Eq.\,\eqref{Eq:QuantumVacuum.VEDscalesMandM0Final} seems to vanish for $H=$const.\,, and can not be source of such an inflation mechanism. Nevertheless, the effective generation of terms proportional to $H^4$ in the early Universe is perfectly possible from string-inspired mechanisms, see \cite{Basilakos:2019acj,Basilakos:2019mpe,Basilakos:2020qmu}, in which the $\sim H^4$ power is generated in the early Universe from the vacuum average of the (anomalous) gravitational Chern-Simons term  $\sim M_P\alpha' b(x) R_{\mu\nu\rho\sigma}(x)\, \widetilde R^{\mu\nu\rho\sigma}(x)$,  which is characteristic  of the  bosonic part of the low-energy effective action of the string gravitational multiplet.  Here $b(x)$ is the Kalb-Ramond axion field and  $\alpha'$ is the slope parameter ($M_{\rm s}=\sqrt{1/\alpha'}$ being the  string scale).  An effective metastable  vacuum is conceivable in this context since such state can be sustained until the Universe transits into the radiation phase, and this  occurs only after the gravitational anomalies are cancelled.  This must indeed happen because matter (relativistic and nonrelativistic particles) cannot coexist with gravitational anomalies.  These can actually be cancelled  by the chiral anomalies of matter itself, see\,\cite{Basilakos:2019acj,Basilakos:2019mpe,Basilakos:2020qmu} for details.  Before such thing occurs, a metastable de Sitter period remains temporally active and can bring about inflation through the (anomaly-generated)  $H^4$ term. The type of inflation produced by the $H^4$-term --- and, in general, by  higher order  (even) powers of $H$ --- is characteristic of  RVM-inflation. The latter  follows a different pattern as compared to Starobinsky's inflation, but graceful exit is still granted -- see\,\cite{Lima:2013dmf,Perico:2013mna,Sola:2015csa,SolaPeracaula:2019kfm} for details and particularly \cite{Sola:2015rra} for a comparison with Starobinsky's inflation\footnote{A detailed study of $H^4$-induced (and, in general, $H^{2n}$-induced) inflation and related considerations concerning cosmological horizons and entropy can be found in \cite{SolaPeracaula:2019kfm}.}.

It seems clear that the presence of the higher powers of the Hubble rate  in the early Universe can be very important from different perspectives. For example, as noted in \,\cite{Basilakos:2019acj,Basilakos:2019mpe},  they could help eschewing the possible trouble of string theories with the `swampland' criteria on the  impossibility to construct metastable de Sitter vacua in the string framework\,\cite{Vafa:2005ui,Agrawal:2018own,Obied:2018sgi}, which if so it would forbid the existence of de Sitter solutions in a low energy effective theory of quantum gravity. The existence of the $H^4$- terms does not depend on picking out a particular potential for the scalar field since, as we should recall here, no potential has been introduced at any time in our framework nor in that of\,\cite{Basilakos:2019acj,Basilakos:2019mpe,Basilakos:2020qmu}. Thus, the RVM string inflation approach could  provide a loophole to the swampland no-go criterion applied to fundamental scalar fields. But, of course, to fully establish it requires of a detailed investigation in the context of string-induced RVM\,\cite{Basilakos:2019acj,Basilakos:2019mpe,Basilakos:2020qmu},
which is certainly not the subject of the present work.  

Although is unclear that such a term proportional to $H^4$ term is present in our QFT framework, we can indeed establish 'RVM-Inflation' from first principles by recurring to even higher order terms, {\it i.e.} of adiabatic order 6. The ${\cal O}(T^{-6})$ terms accounted for here take now the lead, see Eq.\,\eqref{Eq:QuantumVacuum.OnshellHigher}. They originate from finite contributions unrelated to renormalization. Collecting the relevant terms from our $6th$-order  calculation, we find
\begin{equation}\label{Eq:eos.RVMinflation}
\rv^{\rm inf}(m)=\frac{\langle T_{00}^{\delta \phi}\rangle^{(6)}_{\rm ren}(m)}{a^2}=\frac{\txi}{80\pi^2 m^2}\, H^6+ f(\dot{H}, \ddot{H},\vardot{3}{H}...)\,,
\end{equation}
which we have labeled with a superindex `inf' because such an effective VED triggers inflation, as we shall see immediately. In computing the overall coefficient of $H^6$, we have defined the parameter
\begin{equation}\label{Eq:eos.xitilde}
 \txi=\left(\xi-\frac16\right)-\frac{2}{63}-360\left(\xi-\frac16\right)^3\,.
\end{equation}
Notice that in \eqref{Eq:eos.RVMinflation} we have only stood out the contributions from \eqref{Eq:QuantumVacuum.OnshellHigher} which can be responsible for fast inflation in a transient $H=$const. regime. The only relevant terms for such an inflationary interval are those carrying the power $H^6$ with some constant coefficient. The remaining terms, collected in the function $ f(\dot{H}, \ddot{H},\vardot{3}{H}...)$,  consist of different combinations of powers of $H$ accompanied in all cases with derivatives of $H$, and hence all these terms vanish for $H=$const. In other words, $f=0$ for $H=$const. in Eq.\,\eqref{Eq:eos.RVMinflation}. As a result, up to $6th$ adiabatic order the only terms which do not vanish for constant Hubble rate are the isolated powers $H^6$. The overall coefficient upon collecting all these terms is given by \eqref{Eq:eos.xitilde}.

For the current discussion on inflation  we may admit the presence of incoherent matter with density and pressure $(\rho_{\rm m},p_{\rm m})$ beyond our original field $\phi$, which is of course a most realistic assumption for this consideration.  The primeval, highly energetic, vacuum can then decay into relativistic particles of all species.  In the mentioned phenomenological approach, it is  considered a generalized RVM model of the form
\begin{equation}\label{Eq:eos.EffLambda}
\rv(H)=\frac{3}{\kappa^2}\left(c_0+\nu H^2+\tal \frac{H^{2p+2}}{H_I^{2p}}\right)\ \ \ \ \ \ \ \  \ \ \ \  (p=1,2,3,...)\,,
\end{equation}
where $\tal$ is another dimensionless coefficient. For  $\tal=0$  we recover the low-energy form of the RVM, {\it i.e.}  Eq.\,\eqref{Eq:QuantumVacuum.RVM2}, after we impose the boundary condition $\rv(H=H_0)=\rvo$  to determine the coefficient $c_0$.  At higher energies, however, the presence of higher powers of $H$ beyond $H^2$ can bring about inflation in the early Universe, and in this case $H_I$ defines (up to a coefficient) the scale of inflation, as we shall see. The effect of the higher powers of $H$ is negligible for the  current Universe.  For $\tal\neq 0$ the primeval vacuum can decay into matter (most likely relativistic) at high energies (when  $H\sim H_I$  very large),  thus  we can set  $\rho_{\rm m}=\rho_{\rm r}$ and $p_{\rm m}=p_{\rm r}=w_{\rm r}\rho_{\rm r}$  (with  $w_{\rm m}=w_{\rm r}=1/3$ in this case)  in \eqref{Eq:eos.FriedmanEqs}  and  \eqref{Eq:eos.DiffH}.  Additionally, since $f_6=0$ for $H=$const. we have  $\rveff=(4/3)\rv$ on the RHS of \eqref{Eq:eos.DiffH} for the inflationary period. Coefficients $c_0$ and $\nu$ are not important for the early Universe, and hence  we may just concentrate on the effect of $H^{2p+2}$ for the study of the RVM-inflationary mechanism. Once more the presence of only even powers  $H^{2p+2}\, (p=1,2,3,...)$ is related to the general covariance. Our QFT calculation has revealed that $p=2$ is singled out as the lowest possible power ($\sim H^6$) available for triggering inflation in the present framework.

Let us summarize the main traits of implementing RVM-inflation from our predicted VED form \eqref{Eq:eos.RVMinflation} in the very early Universe. First, we observe that we can take $\Pv\simeq-\rv$ for an inflationary regime in which  $H$ remains approximately constant since functions  $f_2,f_4$ and $f_6$  are essentially vanishing in Eq.\,\eqref{Eq:QuantumVacuum.VacuumPressureFullsplit}  during such an inflationary phase. In these conditions, taking $p=2$ in \eqref{Eq:eos.EffLambda} and neglecting the terms $c_0$ and $\nu H^2$ in front of the higher power $H^6$, we can actually solve for $H$ directly from \eqref{Eq:eos.DiffH}:
\begin{eqnarray}\label{Eq:eos.Hfunction}
H(a)=\frac{\tHI}{\left[1+\displaystyle{\left(\frac{a}{\astar}\right)^{8}}\right]^{1/4}}\,,
\end{eqnarray}
where we have traded cosmic time for the scale factor variable, $a$.
Next, using equations \eqref{Eq:eos.FriedmanEqs} we can solve for the explicit form of the radiation and vacuum energy densities\,\footnote{See Appendix B of Ref.\,\cite{SolaPeracaula:2019kfm}, where the analytic solution is given for arbitrary $\nu$, $\tal$ and $p$, but still with $c_0=0$. The analytic solution for $c_0\neq 0$ is only possible for the late Universe, where the high power  $H^{2p+2}\, (p\geq 1)$ is negligible\,\cite{SolaPeracaula:2019kfm}.}:
\begin{eqnarray}\label{Eq:eos.rhodensities}
\rho_{\rm r}(a)=\rI\,\frac{\displaystyle{\left(\frac{a}{\astar}\right)^{8}}}{\left[1+\displaystyle{\left(\frac{a}{\astar}\right)^{8}}\right]^{\frac{3}{2}}}\,\,, \ \ \ \ \ \ \ \ \
\rv(a)=\frac{\rI}{\left[1+\displaystyle{\left(\frac{a}{\astar}\right)^{8}}\right]^{\frac{3}{2}}}\,\,.
\end{eqnarray}
We are expressing the above results in terms of the  point $\astar$, which  defines  the transition between vacuum dominance and the radiation era, {\it i.e.} the point which satisfies $\rho_{\rm r}(\astar)=\rv(\astar)$. It can be estimated as $\astar\sim 10^{-29}$ within a typical GUT defined at the scale at $M_X\sim 10^{16}$ GeV\,\cite{SolaPeracaula:2019kfm}. Furthermore, we have defined
\begin{equation}\label{Eq:eos.defsHIrI}
\tHI=\frac{H_I}{\tal^{1/4}}\,, \ \ \ \ \ \ \ \ \ \ \ \  \rI=\frac{3}{\kappa^2}\,\tHI^2\,.
\end{equation}
Since $H(a=0)=\tHI$, it follows that this is the value of the Hubble rate in the very early Universe.  Similarly,  $\rv(0)=\rI$ is the VED at that initial point.  We can see that they are both finite. The model, therefore, presents no early singularities at all.  On comparing\,\eqref{Eq:eos.RVMinflation}  with the generic form \eqref{Eq:eos.EffLambda} -- with $p=2$ in our case -- and using the above definitions  we can easily identify
\begin{equation}\label{Eq:eos.HIeffective}
\tHI=\left(\frac{240\pi^2}{\txi}\right)^{1/4}\sqrt{\Mpl\, m}\,.
\end{equation}
Clearly, in order to have inflation  near a typical GUT scale we need   very massive particles with masses $m$ in the neighborhood of that scale. In addition, it  is  imperative, of course,  to have  $\txi>0$. For a single scalar field non-minimally coupled to gravity, we find numerically  from \eqref{Eq:eos.xitilde}  that this occurs   for $\xi\lesssim0.1023$.  Such a range excludes $\xi=1/6$, for which $\txi<0$, but it admits the minimal coupling situation $\xi=0$, and the negative values $\xi<0$.  In general,  we can expect to have many fields non-minimally coupled to gravity with different couplings and masses,  especially if we consider the matter content of GUT's.   We have checked that, in general, RVM-inflation can be made compatible with non-vanishing running of the VED at low energies, see Eq.\eqref{Eq:QuantumVacuum.RVM2}. However, a detailed account of RVM-inflation in the QFT context requires a  dedicated study and will be presented elsewhere.  In particular, we note that the fermionic contribution might be important as well, we will talk about this in \hyperref[Sect:RVMInfFermi]{Sect.\,\ref{Sect:RVMInfFermi}} of the next chapter. Although the generalization of the point-splitting and ARP methods for fermions have been amply discussed in the literature\,\cite{Christensen:1978yd,Landete:2013axa,Landete:2013lpa,delRio:2014cha,BarberoG:2018oqi},  its application to compute the scaling evolution of the  VED, and more specifically  within the off-shell ARP procedure that we have been using for scalar fields,  has been considered only very recently\,\cite{Samira2022}.  Overall, theoretical scenarios indeed exist for which RVM-inflation occurs along with $\nueff>0$ in Eq.\eqref{Eq:QuantumVacuum.nueffAprox},  thereby  being consistent with the sign picked up by the current phenomenological analyses\,\cite{Sola:2017znb,SolaPeracaula:2016qlq,SolaPeracaula:2021gxi}.
%
\begin{figure}[t]
\begin{center}
\includegraphics[scale=0.55]{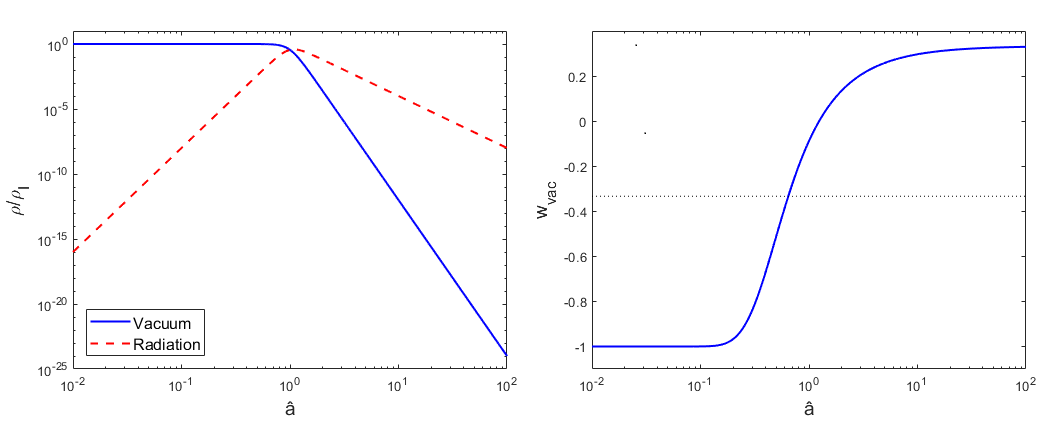}
\end{center}
\caption{Inflationary period. Left: It is shown the evolution of the energy densities\,\protect\eqref{Eq:eos.rhodensities} of vacuum and relativistic matter before and after the transition point $a_*\sim 10^{-29}$ from inflation to the early radiation epoch (see the text). The constant vacuum energy density during inflation decays into radiation and the standard FLRW regime starts. Right: The vacuum EoS is shown around the transition point. It is attached to -1 during the inflationary period, then it tends to mimic radiation after the reheating starts.}
\label{Fig:eos.JointInflation}
\end{figure}
%
From equations \eqref{Eq:eos.rhodensities} we learn that the VED is initially  constant and large, $ \rv(a)\simeq \rI$  for  $a\ll \astar$, and decreases very fast beyond that point.  On the other hand, the radiation energy is initially zero, $\rho_{\rm r}(0)=0$,  but  increases very fast in the beginning  ($\rho_{\rm r}\sim a^8$) owing to the vacuum decay into radiation. We can appraise this behavior in \hyperref[Fig:eos.JointInflation]{Fig.\,\ref{Fig:eos.JointInflation}}, where, on the left, we can see the transition from the pure vacuum state that brings about the inflationary phase, in which the VED remains approximately constant, into an incipient radiation epoch, which soon dominates the evolution of the Universe. At the beginning ($\hat{a}=0$) there is no radiation ($\rho_{\rm}(0)=0$) whilst the VED reaches its maximum value, {\it i.e.} $\rho_{\rm vac}(0)=\rho_I \propto \Mpl^2 H_I^2$. For $a\gg\astar$,  the frenzied growing  $\sim a^8$ of the radiation density turns into the dilution law $\rho_{\rm r}(a)\sim a^{-4}$  and hence we retrieve the standard  behavior of the radiation until our days.  The VED, on the other hand, remains negligible during the radiation epoch as compared to  $\rho_{\rm r}$ and therefore  it cannot perturb in any significant way the BBN processes which will occur  much later at  $a\sim 10^{-9}$. In the long run, after entering the matter-dominated epoch,  the VED  will  recover the (much more tranquil)  form \eqref{Eq:QuantumVacuum.RVM2}, which is characteristic of the current Universe and  evolves just  as a constant plus a mild component $ H^2\sim a^{-3}$ with a small coefficient. Obviously such form will appear in the solution only if we  keep the terms $c_0$ and $\nu$  in solving the equations, but in this section  we focus exclusively on the very early times when the  inflationary regime turns into the primeval radiation epoch.  Needless to say, the above solution is approximate and is in effect only when the terms carrying the time derivatives of $H$ are negligible as compared to $H^6$  during the short inflationary stage set off by $H\simeq$ const. This estimate is nonetheless sufficient to exhibit the main features of the early  phase of RVM-inflation.  The  terms with  time derivatives of $H$ are suppressed and cannot perturb significantly the inflationary period characterized by $H\simeq$const. nor can have any sizeable influence beyond the very early Universe  once the $H^2$ terms take over until the constant term $c_0$ becomes eventually dominant around our present time.

The EoS of the quantum vacuum in the early Universe follows from computing the corresponding vacuum pressure at that primeval stage up to 6{\it th} adiabatic order. The result is Eq.\,\eqref{Eq:QuantumVacuum.VacuumPressureFullsplit} which remember that adopts the form
\begin{equation}\label{Eq:eos.VacuumPressureFullsplit}
\Pv(M)=-\rv(M)+f_2(M,\dot{H})+ f_4(M,H,\dot{H},...,\vardot{3}{H})+f_6(\dot{H},...,\vardot{5}{H})+\cdots\,,
\end{equation}
in which the functions $f_2$, $f_4$ and $f_6$ involve  adiabatic contributions of second, fourth and sixth order, respectively,  and all of them carry at least one time derivative of $H$ as one may see from Eq.\,\eqref{Eq:eos.VacuumPressureFull}. Therefore,  all these functions vanish for $H=$const. ($\hat{a}\ll 1$) and we get  $\Pv=-\rv$  to a very good approximation.  The RVM inflationary period  is thus characterized by the traditional EoS of vacuum, $\wv=-1$. This can be appreciated in \hyperref[Fig:eos.JointInflation]{Fig.\,\ref{Fig:eos.JointInflation}}, on right.

All in all approximations made in Starobinsky and RVM inflation cases are similar since during the $\dot{H}\simeq$ const. regime all higher time derivatives are neglected -- cf. Ref.\,\cite{Sola:2015rra} for a comparative discussion of Starobinsky inflation versus RVM-inflation. The latter have been explored phenomenologically in the mentioned previous works (cf.\,\cite{Lima:2013dmf,Perico:2013mna,Sola:2015csa,Sola:2015rra,SolaPeracaula:2019kfm}), all of them were based on assuming the \textit{ad hoc} structure \eqref{Eq:eos.EffLambda}. Here, in contrast, we have been able to show that such a structure indeed emerges from QFT in curved spacetime and that $p = 2$ is the lowest possible value. In other words, we have proven from an explicit QFT calculation that RVM-inflation can be unleashed by the $\sim H^6$-term in the vacuum energy density. Thus, the two specific powers $H^2$ and $H^6$ are the ones which are picked out by the quantum effects of matter fields in the FLRW background. The former power affects the dynamics of the current Universe, whilst the latter is responsible for inflation in this context. Remember that only even powers are allowed by general covariance. The conspicuous absence of the power $H^4$ is not surprising: it is a built-in consequence of the subtraction procedure in the adiabatic renormalization of the EMT in four dimensions. At the end of the day, the RVM description has the ability to encompass the entire history of the Universe from inflation up to our days from first principles\,\footnote{We remind the reader that a related inflationary mechanism, based on the power $H^4$, is conceivable in the framework of\,\cite{Basilakos:2019acj,Basilakos:2019mpe,Mavromatos:2020kzj,Mavromatos:2021urx}. However, such a proposal is not based on the QFT action considered here but on gravitational-anomaly aspects which are specific of the bosonic part of the effective action of string theory. Because the leading power in this case is $H^4$ rather than $H^6$, such a stringy RVM-inflation is, in principle, distinguishable from the QFT one addressed here. These two independent options strengthen the general support for RVM inflation.}. A thorough account of the RVM mechanism of inflation will be presented in a devoted study.

\section{Discussion of the chapter}\label{Sect:eos.Discussion}

While in the previous chapter we focused more in the technical and mathematical aspects of our QFT calculation of dynamical VED, here we are centered in the possible cosmological consequences. The computation of the renormalized expression of Vacuum's pressure we performed at the beginning of the chapter, in \hyperref[Sect:RenormPressure]{Sect.\,\ref{Sect:RenormPressure}}, is a first step to scrutanize potential and crucial phenomenological consequences of the dynamical quanum vacuum. Before analyzing the whole form of Friedmann's equations in this case, we studied in \hyperref[SubSect:GeneralizedRVM]{Sect.\,\ref{SubSect:GeneralizedRVM}} an effective behaviour of VED at low energies in which the EoS is stuck to -1, as usual. We then derived the form of the effective VED in Eq.\,\eqref{Eq:eos.RVMgeneralized2} which generalizes the results of the previous chapter by the incorporation of linear corrections proportional to $\dot{H}$. 

In \hyperref[Sect:RenormalizedFriedmann]{Sect.\,\ref{Sect:RenormalizedFriedmann}}, the Friedmann equations, which incorporate the effects of renormalization, are presented. These equations describe the background evolution of the Cosmos, including higher-order derivative effects. At low energies, these higher-order effects can be neglected, and the Friedmann equations reduce to the standard ones, albeit with a mild evolution of both the vacuum energy density (VED) and the gravitational constant. This evolution is reflected in the conservation equations for the vacuum component, which, assuming the rest of the fluids to be covariantly conserved with the expansion, show that the Bianchi identity modifies the conservation equation for the quantum vacuum, resulting in an exchange of energy with the dynamically evolving background. Through this dynamical interplay with the VED, we find that the gravitational constant also evolves very mildly, following a logarithmic function of the Hubble rate: $G=G\left(\ln H\right)$

One of the main contributions of this chapter is the prediction of the EoS of the quantum vacuum, denoted as $\wv$, which has been computed using the quantum field theoretical approach introduced in the previous chapter. The computation of $\wv$ has been performed for the entire FLRW regime, and its potential phenomenological implications have been explored. Our study has shown that a proper renormalization of the quantum matter effects is necessary in order to understand the QFT vacuum in a curved background, and that the modification of the EoS of the quantum vacuum with respect to the classical result $\wv=-1$ is nontrivial and significant. The dynamical behavior of the quantum vacuum EoS, expressed as a function of redshift $\wv=\wv(z)$, represents a clear departure from the rigid value of $\wv=-1$ characteristic of the classical vacuum. Moreover, the EoS dynamics of the quantum vacuum are important because they carry a measurable imprint at present, behaving like quintessence with $\wv(z)\gtrsim-1$. Remarkably, no additional fields are needed to explain the cosmic acceleration, as it can be explained entirely by the quantum vacuum composed of fluctuations of the quantized matter fields. The evolution of the quantum vacuum is described as a cosmic chameleon, triggering inflation as the "true vacuum" with $\wv=-1$, adopting the EoS of matter ($\wv=1/3$ first, and $\wv=0$ later), and reappearing in our days as quintessence. In the late Universe, the quantum vacuum acts as dark energy, whose EoS is dynamically evolving.

In the last section, \hyperref[Sect:PhenoImplications]{Sect.\,\ref{Sect:PhenoImplications}}, we have also highlighted some of these important phenomenological applications of the RVM, which may help to improve the description of the overall cosmological data, and in particular to alleviate the $H_0$ and $\sigma_8$ tensions\,\cite{Abdalla:2022yfr}. First, we related our results with the most recent studies\,\cite{SolaPeracaula:2021gxi,CosmoTeam2023Universe} performed in the context of the RVM, and observe that different speculative features attributted to vacuum can be justified by the analysis done here. Secondly, we predict a new mechanism for inflation in \hyperref[SubSect:RVMInflation]{Sect.\,\ref{SubSect:RVMInflation}}, which is triggered by the aforementioned $\sim H^6$ terms. As a difference from the ${\cal O}(H^4)$ terms, which vanish for $H=$const. (as they all depend on time derivatives), the term $\sim H^6$ can perfectly bring about a short but fast period of $H=$const. inflation. We should stress, however, that the RVM inflationary mechanism is distinct from Starobinsky's inflation\,\cite{Starobinsky:1980te,Starobinsky:1981vz}, for which it is the time derivative of the Hubble rate which remains constant for a short stretch of time (viz. $\dot{H}=$ const.)  Noteworthy, there exists a stringy version of the RVM inflationary mechanism which can operate with $H^4$ terms, {\it i.e.} for which these terms appear without time derivatives and hence do not vanish for $H=$const.\,\cite{Mavromatos:2020kzj,Mavromatos:2021urx}. This is in contrast to the QFT version that we have described here, being characterized by $H^6$-inflation. This means that the stringy and non-stringy (QFT) mechanisms of RVM-inflation can, in principle, be distinguished. In both cases inflation is unleashed during a short time interval of the early Universe where $H=$const. and therefore requires that the effective behavior of the VED carries a higher (even) power $H^{N}\, (N=4,6,...)$ (beyond $H^2$). The existence of a high power of the Hubbe rate in the VED is the characteristic trademark of RVM-inflation. In the QFT case, it appears as a fundamental mechanism of inflation solely induced by the pure quantum matter effects on a classical gravitational background.

\blankpage

\chapter{Fermionic and bosonic contributions to the VED}\label{Chap:Fermions}

If we take quantum theory seriously, the most universal contribution to this vacuum energy density is the Zero-Point Energy (ZPE) of the massive quantum fields in the standard model of particle physics, and in fact in any QFT model. It has been repeteadly mentioned in earlier chapters that a naive calculation of this quantity leads to very large contributions proportional to the quartic power of the mass of the particles, $\rho_{ \rm ZPE}\sim m^4$, responsible of the so-called Cosmological Constant Problem (CCP)\cite{weinberg1989cosmological}. In fact, in \hyperref[Chap:QuantumVacuum]{Chap.\,\ref{Chap:QuantumVacuum}} and \hyperref[Chap:QuantumVacuum]{Chap.\,\ref{Chap:EoSVacuum}}, the adiabatic regularization prescription (ARP) was used to compute the renormalized VED, $\rv$, in the case of a non-minimally coupled real scalar field with gravity \cite{Moreno-Pulido:2022phq, SolaPeracaula:2022hpd}. We fully demonstrated how the final result of the running of VED in Curved spacetime \eqref{Eq:QuantumVacuum.VEDscalesMandM0Final} is free from the gigantic contribution coming from the ZPE. The overall contribution from bosons to the VED that we have found is well-behaved and can perfectly accommodate the measured value of the CC from observational cosmology without any fine-tuning.  Basically this is because the typical contribution expected for the VED in each epoch of the cosmic evolution is proportional to the tiny values of the $\beta$-function coefficients for bosons.

The main goal of this chapter, based on\,\cite{Samira2022}, is to extend the computations previously done with scalar fields and check if fermions also support the conclusions drawn from them, especially with regards to whether their corresponding vacuum fluctuations are free from the traditional fine-tuning problem, i.e., independent of the quartic powers of their masses. However, the extension proves to be rather non-trivial due to the involvement of Fermi-Dirac statistics and formal peculiarities associated with spinor calculus. Despite these complexities, it is reassuring to find that the many technicalities involved in the calculation do not alter the main conclusions derived from the scalar field calculation. Thus, the final results concerning the renormalized VED can be obtained by combining the contributions from an arbitrary number of quantized scalar and fermion fields. The formula obtained in this way is referred to as the bosonic and fermionic contributions to the VED. It is important to note that the theory does not have interactions between gauge fields and matter fields, so additional effects from such interactions are not necessary for the present study. However, the non-minimal coupling of the quantized scalar fields with the external (non-quantized) gravitational field is taken into account.  Therefore, we will omit the gauge fields in our considerations, with the understanding that a more complete calculation would be necessary in an interacting theory. All that said, the computation of the free field contribution from bosons and fermions in curved spacetime is already a formidable task, so for the sake of a stepwise and clearer presentation we will present the fermionic contributions here on equal footing to that of the scalar contribution in \hyperref[Chap:QuantumVacuum]{chapter\,\ref{Chap:QuantumVacuum}} and \hyperref[Chap:EoSVacuum]{chapter\,\ref{Chap:EoSVacuum}}. Additionally, we present in this chapter the equation of state (EoS) of the quantum vacuum with the contributions from boson and fermions. As the rest of the dissertation, all our calculations are performed assuming a spatially flat Friedmann-Lemaître-Robertson-Walker (FLRW) background.

The presentation of this chapter is structured in the following manner: In \hyperref[Sect:QuantizedFermion]{Sect.\,\ref{Sect:QuantizedFermion}}, we review the quantization of a Dirac fermion in a curved background, the corresponding Dirac equation and its spinor solutions
obtained from adiabatic expansion of the field modes. In \hyperref[Sect:ZPEFermions]{Sect.\,\ref{Sect:ZPEFermions}}, the off-shell adiabatic renormalization of the EMT for spin-1/2 fermions is addressed and we extract the renormalized ZPE, VED ($\rho_{\rm vac}$) and vacuum pressure ($P_{\rm vac}$) in
this context. Additionally, some remarks on the trace anomaly and its role in our approach are discussed and expand what is said in \hyperref[Sect:Trace]{Sect.\,\ref{Sect:Trace}}. \hyperref[Sect:CombinedBandF]{Sect.\,\ref{Sect:CombinedBandF}} contains the combined results from all the quantized matter fields. Specifically, we compute the renormalized VED for a system made of an arbitrary number of quantized scalar fields non-minimally coupled to curvature (with different masses and non-minimal couplings) and an arbitrary number of quantized spin-1/2 free fermion fields. In the same section we report on the corresponding running of the gravitational coupling $G = G(H)$, which goes hand in hand with the running of $\rho_{\rm vac}$ in order to preserve the Bianchi identity. We also discuss the mechanism of ‘RVM-inflation’ with the combined contribution from all these fields, and derive the equation of state (EoS) of the quantum vacuum for that system of quantized bosons and fermions fluctuating in it. As in the previous chapter, the vacuum EoS is no longer equal to $w_{\rm vac}=-1$, the reason being that the vacuum pressure and the VED are not exactly related in the usual manner (viz. $P_{\rm vac}=-\rho_{\rm vac}$) since $P_{\rm vac}$ and $\rho_{\rm vac}$ are independent functions of the Hubble rate $H$ and its time derivatives owing to the quantum effects. In the current Universe, there is still some remnant of these quantum effects which induce a small (but potentially measurable) departure making the quantum vacuum mimic quintessence. The conclusions are delivered in \hyperref[Sect:FermionsDiscussion]{Sect.\,\ref{Sect:FermionsDiscussion}} together with some additional discussion. Finally, \hyperref[Appendix:AdiabExpFermModes]{Appendix\,\ref{Appendix:AdiabExpFermModes}} and \hyperref[Appendix:AdiabExpFermEMT]{Appendix\,\ref{Appendix:AdiabExpFermEMT}} are rather bulky since they collect a number of cumbersome expressions related to the adiabatic expansion of the VEV of the EMT and the Fourier modes of the fermionic field (computed up to $6th$ order for the first time in the literature).

\section{Quantization of a fermionic field in curved spacetime}\label{Sect:QuantizedFermion}

Our goal is to extend the QFT results for the VED obtained for quantized scalar fields, which we have summarized in the previous section, to the case of  quantized spin-$1/2$ Dirac fermion fields and then combine the two types of contributions in closed form.  The calculation of the renormalized VED  for free spin-1/2 fermions is also nontrivial and rather cumbersome, and requires a devoted study, which we present here  (see also the appendices provided at the end for bulky complementary details).  While  the QFT treatment is analogous to the case of scalars,  the specific technicalities are  quite different and no less intricate, but fortunately the final result proves to be in consonance with the one previously derived for the scalars,  so it is perfectly possible to furnish a close form for the combined contribution to the VED involving an arbitrary number of non-interacting scalar and spin-1/2 fermion fields, cf. \hyperref[Sect:CombinedBandF]{Sect.\,\ref{Sect:CombinedBandF}}.

The study of the solutions of the Dirac equation in curved spacetime goes back to the works from many decades ago by Fock, Tetrode, Schr\"{o}dinger, McVittie, Bargmann,   Wheeler and others: see e.g. \cite{Parker:1980kw,Barut:1987mk,Villalba:1990bd,Finster:2009ah,Pollock:2010zz}, where the relevant historical references are given  and different aspects of spin-1/2 fermions in curved spacetime are studied, including  a detailed account  for the solutions in  FLRW spacetime -- see also the review \cite{Collas:2018jfx}, with a rather complete list of references.  On the other hand, the  subject of adiabatic regularization for fermions has been previously treated in the literature in different applications,  see e.g.  \cite{Christensen:1978yd} as well  as the more recent papers\,\cite{Landete:2013axa, Landete:2013lpa, delRio:2014cha, BarberoG:2018oqi} where  emphasis is made on exact solutions e.g. in de Sitter spacetime. The calculation of the renormalized VED in FLRW spacetime is,  however,  more complicated for it does not admit an exact solution.  Our strategy to circumvent this problem   is based on using an off-shell variant of the ARP framework\,\cite{Moreno-Pulido:2020anb, Moreno-Pulido:2022phq} which leads to the RVM  behavior of the vacuum energy\,\cite{Sola:2013gha,SolaPeracaula:2022hpd}. The RVM framework has proven rather successful in mitigating the cosmological tensions\,\cite{Perivolaropoulos:2021jda,Abdalla:2022yfr,Dainotti:2023yrk}, as shown in  different phenomenological analyses, such as\cite{SolaPeracaula:2021gxi} and \cite{Gomez-Valent:2018nib,Gomez-Valent:2017idt}.   On the theoretical side, attempts at computing the VED with other procedures has led to the traditional calamity with the quartic powers of the masses.  Here we will show that using the off-shell ARP to tackle the VED contribution from fermions generates a result which is free from these difficulties and fully along the lines of what has been obtained for the scalar fields in the previous sections and originally in\,\cite{Moreno-Pulido:2020anb, Moreno-Pulido:2022phq}.  Therefore, the combined contribution from fermions and scalar fields to the VED is compatible with a smooth running of the cosmological vacuum energy  and is consistent with the aforementioned phenomenological analysis of the RVM as a possible solution to the cosmological tensions.

Since it will be necessary a considerable amount of formalism to treat  fermions within the adiabatic approach, it is  convenient  to summarize first the necessary aspects of that formalism  before we can put forward our main results concerning their contribution to the vacuum energy density. It  will be useful to fix some notation as well.  Once more we perform the calculations  in FLRW spacetime with flat three-dimensional metric.
Consider a free Dirac spin-$1/2$ field, described by the  four-component spinor $\psi$. In our conventions,  the Dirac action in curved spacetime is given by
\begin{equation}\label{Eq:Fermions.FermionAction}
S_\psi(x)=-\int d^4x \sqrt{-g}\left[\frac{1}{2}i\left( \Bar{\psi}\underline{\gamma}^\mu \nabla_\mu \psi -\left(\nabla_\mu \Bar{\psi}\right)\underline{\gamma}^\mu \psi \right)+m_{\psi} \Bar{\psi} \psi\right]\,.
\end{equation}
In the above expression,  $m_{\psi}$ denotes the mass of the Dirac field and   $\Bar{\psi}\equiv \psi^\dagger \gamma^0$  the adjoint spinor.  Since we are in a curved background,  the partial derivative of a spinor $\partial_\mu\psi$  has been replaced with the corresponding covariant derivative $\nabla_\mu\psi$, which is defined below.  Moreover,  gamma matrices in curved spacetime are also needed, they are sometimes indicated (as above) with an underline to distinguish them from the Minkowski space gamma matrices. The former are  $\underline{\gamma}^\mu (x)$ (which are generally functions of the coordinates) whereas the latter are the constant matrices ${\gamma}^\alpha$ in flat spacetime. As it is well-known, to obtain a representation for the curved spacetime gamma matrices in terms of the Minkowskian gamma matrices we need  to introduce the local tetrad or  vierbein field (in $4$-dimensional spacetime)  $e^{\,\mu}_\alpha$. It is defined in each tangent space of the spacetime manifold and relates the curved spacetime metric with the Minkowskian one in the usual way:  $g^{\mu\nu}(x)=e^{\mu}_{\,\alpha}(x) e^{\nu}_{\beta}(x) \eta^{\alpha \beta}$,  where $\eta_{\alpha\beta}$ is the Lorentz metric in the local inertial frame specified by the normal coordinates at the given spacetime point.  The general relation between the  two sorts of gamma matrices is $\underline{\gamma}^\mu(x)=e^{\mu}_{\,\alpha}(x)\gamma^\alpha $.  Specifically, in a spatially flat FLRW spacetime the vierbein in conformal coordinates is  $e^{\mu}_{\alpha}={\rm diag}\left(1/a(\tau), 1/a(\tau), 1/a(\tau), 1/a(\tau)\right)$ where $a(\tau)$ is the scale factor as a function of the conformal time.  Whence the  gamma matrices in this background are  time-dependent and  related to the constant flat spacetime ones as follows:  $\underline{\gamma}^\mu(\tau)=\gamma^\mu /a(\tau)$.
This relation insures that they satisfy the following anti-commutation relations:
\begin{equation}\label{Eq:Fermions.DiracMatricesRel}
\left\{ \underline{\gamma}^\mu, \underline{\gamma}^\nu \right\} =-2g^{\mu\nu}\mathbb{I}_4\,,
\end{equation}
provided, of course,  the (constant) flat space gamma matrices satisfy $\left\{ {\gamma}^\alpha, {\gamma}^\beta \right\} =-2\eta^{\alpha\beta}\mathbb{I}_4$.
In order to obtain the equation of motion, {\it i.e.} the covariant Dirac equation in curved spacetime, one has to vary the covariant action \eqref{Eq:Fermions.FermionAction} with respect to the spinor field, giving
\begin{equation}\label{Eq:Fermions.CovDiracEq}
i \underline{\gamma}^\mu \nabla_\mu \psi+m_\psi \psi=i {e}^\mu_\alpha \gamma^\alpha \nabla_\mu \psi+m_\psi \psi=i \frac{1}{a}\left( \gamma^\alpha\partial_\alpha-  \gamma^\alpha\Gamma_\alpha\right) \psi+m_\psi \psi=0\,.
\end{equation}
The covariant derivative is defined through the spin connection, $\nabla_\mu \equiv \partial_\mu-\Gamma_\mu$.  The spinorial affine connection $\Gamma_\mu$  satisfies the equation\,\cite{Barut:1987mk}
\begin{equation}\label{Eq:Fermions.spinconnection}
\left[\Gamma_\nu, \underline{\gamma}^\mu(x)\right]=\frac{\partial \underline{\gamma}^\mu(x)}{\partial x^\nu}+\Gamma^\mu_{\nu\rho} \underline{\gamma}^\rho(x)\,,
\end{equation}
where $\Gamma^\mu_{\nu\rho}$ are the  Christoffel symbols.   The above equation is tantamount to require the vanishing of the covariant derivative of the curved space gamma matrices:  $\nabla_\nu \underline{\gamma}^\mu(x)=0$\,\cite{parker2009quantum}, i.e. the curved-space gamma matrices are defined to be covariantly constant over the spacetime manifold.
Using the Christoffel symbols in the conformally flat FLRW metric as given in \hyperref[Appendix:Conventions]{Appendix\,\ref{Appendix:Conventions}}, the explicit solution of Eq.\,\eqref{Eq:Fermions.spinconnection}  can be found, with the following result:  $\Gamma_0=0, \Gamma_j=-\left(\mathcal{H}/2\right)\gamma_j\gamma_0=-\left(a'/2a\right)\gamma_j\gamma_0$. Therefore,   $\gamma^\alpha\Gamma_\alpha=3(a'/2a)\gamma_0=-3(a'/2a)\gamma^0$. This expression can then  be inserted in Eq.\,\eqref{Eq:Fermions.CovDiracEq}.

In this way we have obtained an explicit form for the Dirac equation in  FLRW spacetime with spatially flat metric. We are now in position to attempt a solution by  expanding the quantized fermion field  in mode functions:
\begin{equation}\label{Eq:Fermions.psi(x)}
	\psi(x)=\int d^3k\sum_{\lambda=\pm 1}(B_{\vec{k},\lambda}u_{\vec{k},\lambda}\left(\tau,\vec{x}\right)+D^\dagger_{\vec{k},\lambda} v_{\vec{k},\lambda}\left(\tau, \vec{x}\right))\,.
\end{equation}
Here $B_{\vec{k},\lambda}$ and $D^\dagger_{\vec{k},\lambda}$ are creation and annihilation operators which satisfy the standard anticommutation relations
\begin{equation}\label{Eq:Fermions.AntiComRel}
\begin{split}
&\left\{D_{\vec{k},\lambda},D^\dagger_{\vec{q},\lambda^\prime}\right\}=\left\{B_{\vec{k},\lambda},B^\dagger_{\vec{q},\lambda^\prime}\right\}=\delta_{\lambda,\lambda^\prime}\delta^{(3)}\big(\vec{k}-\vec{q}\big)\,,\\
&\left\{D_{\vec{k},\lambda},D_{\vec{q},\lambda^\prime}\right\}=\left\{D^\dagger_{\vec{k},\lambda},D^\dagger_{\vec{q},\lambda^\prime}\right\}=\left\{B_{\vec{k},\lambda},B_{\vec{q},\lambda^\prime}\right\}=\left\{B^\dagger_{\vec{k},\lambda},B^\dagger_{\vec{q},\lambda^\prime}\right\}=0\,.
\end{split}
\end{equation}
The momentum expansion of the mode functions $u_{\vec{k},\lambda}$ and their charge conjugates $v_{\vec{k},\lambda}$ can be conveniently  written in terms of two $2$-component spinors $\xi_\lambda(\vec{k})$ and corresponding spinor modes $h^{\rm I}_k$ and $h^{\rm II}_k$:
\begin{equation}\label{Eq:Fermions.22componentspinors}
\begin{split}
&u_{\vec{k},\lambda}(\tau, \vec{x})=	\frac{e^{i\vec{k}\cdot\vec{x}}}{\sqrt{(2\pi a)^3}}\begin{pmatrix}
	h^{\rm I}_k( \tau )\xi_\lambda(\vec{k})\\
	h^{\rm II}_k(\tau)\frac{\vec\sigma.\vec{k}}{k}\xi_\lambda(\vec{k})
\end{pmatrix},\\
&v_{\vec{k},\lambda}(\tau , \vec{x})=	\frac{e^{-i\vec{k}\cdot\vec{x}}}{\sqrt{(2\pi a)^3}}\begin{pmatrix}
	-h^{\rm II*}_k ( \tau )\frac{\vec\sigma.\vec{k}}{k}\xi_{-\lambda}(\vec{k})\\
	-h^{\rm I*}_k (\tau) \xi_{-\lambda}(\vec{k})
\end{pmatrix}\,,
\end{split}
\end{equation}
with
\begin{equation}\label{Eq:Fermions.xidef}
\frac{\vec\sigma\cdot\vec{k}}{k}\xi_\lambda(\vec{k})=\lambda\xi_\lambda(\vec{k}), \hspace{0.3in}\lambda=\pm 1\,,\ \ \ \ \ \ \  \xi_\lambda^\dagger (\vec{k})\xi_\lambda(\vec{k})=1\,.
\end{equation}
Using this representation,  \eqref{Eq:Fermions.CovDiracEq} splits into a system of two coupled first order equations for each of the two types of spinor modes $h^{\rm I}_k$ and $h^{\rm II}_k$:
\begin{equation}\label{Eq:Fermions.h1h2eq}
h^{\rm I}_k=\frac{i a}{k}(\frac{1}{a}\partial_\tau+i m_\psi)h^{\rm II}_k(\tau),\hspace{0.5 in}h^{\rm II}_k=\frac{i a}{k}(\frac{1}{a}\partial_\tau-
i m_\psi)h^{\rm I}_k(\tau)\,.
\end{equation}
After straightforward calculation these equations can be rewritten as two second order decoupled equations:
\begin{equation}\label{Eq:Fermions.dceqofspinors}
\begin{split}
	&\bigg(\partial_\tau^2-i m_{\psi} a'+a^2 m_\psi^2+k^2
\bigg)h^{\rm I}_k(\tau)=0\rightarrow \big(\partial_\tau^2+\Omega_k^2(\tau)\big)h^{\rm I}_k(\tau)=0,\\
	&\bigg(\partial_\tau^2+i m_\psi a'+a^2 m_\psi^2+k^2
	\bigg)h^{\rm II}_k(\tau)=0\rightarrow \big(\partial_\tau^2+(\Omega_k^2(\tau))^*\big)h^{\rm II}_k(\tau)=0\,,
\end{split}	
\end{equation}
where 
\begin{equation}\label{Eq:Fermions.Omegak}
\Omega_k^2 \equiv\omega_k^2+a^2\Delta^2-i\sigma(\tau)\,,
\end{equation}
with
\begin{equation}\label{Eq:Fermions.adiabaticorder}
\begin{split}
&\omega_k (M) \equiv \sqrt{k^2+M^2 a^2},\\
&\sigma \equiv m_\psi a'=\sqrt{M^2+\Delta^2} \ a^\prime\,.
\end{split}
\end{equation}
The fact that \eqref{Eq:Fermions.dceqofspinors} only depends on the modulus of the momentum, $k$, justifies the notation used for the modes $h^{\rm I}_k, h^{\rm II}_k$, with no arrows.
Following the same prescription as in the case of scalar fields (cf. \hyperref[Sect:AdRegEMT]{Sect.\,\ref{Sect:AdRegEMT}}),  we have  introduced an off-shell scale $M$, which again will take the role of  renormalization scale. Correspondingly, we have defined  $\Delta^2\equiv m_\psi^2-M^2$  and once more assigned adiabaticity order $2$ to it.  We did not change the notation $\Delta$ as compared to the scalar case since the final formulas do not depend on $\Delta$ but on $M$ and the respective physical masses.  The argument of $\omega_k$ will be omitted  from now on, unless it takes a different value from $M$.  The normalization conditions for the mode functions involved in  $\psi$ are implemented through the Dirac scalar product:
\begin{equation}\label{Eq:Fermions.NormPrescription}
(u_{\vec{k},\lambda},u_{\vec{k}^\prime,\lambda^\prime})=\int d^3 x\, a^3 u_{\vec{k},\lambda}^\dagger u_{\vec{k}^\prime,\lambda^\prime} =\delta_{\lambda\lambda^\prime}\delta^3(\vec{k}-\vec{k}^\prime)
\end{equation}
and similarly for $ (v_{\vec{k},\lambda},v_{\vec{k}^\prime,\lambda^\prime})=\delta_{\lambda\lambda^\prime}\delta^3(\vec{k}-\vec{k}^\prime)$. It follows that
\begin{equation}\label{Eq:Fermions.normalizationrelation}
|h_{k}^{\rm I}|^2+|h_{k}^{\rm II}|^2=1\,.
\end{equation}
As mentioned in the previous section, the number of time derivatives of the cosmological scale factor $a(\tau)$ that appear in a term of the expansion is called adiabatic order of the term.

In order to solve the differential equations \eqref{Eq:Fermions.dceqofspinors}  we may follow a recursive process which preserves the adiabatic hierarchy, just as we did with the scalar fields.
Let us first  redefine $h^{\rm I}_k$ and the time variable as follow
\begin{equation}\label{Eq:Fermions.redefinefieldandtime}
h^{\rm I}_{k,1} \equiv\sqrt{\Omega_k}h^{\rm I}_k\hspace{0.5in}d\tau_1=\Omega_k d\tau.
\end{equation}
Substituting these relations into the equation for $h^{\rm I}_k$ in \eqref{Eq:Fermions.dceqofspinors} we find
\begin{equation}\label{Eq:Fermions.1storderofEom}
\frac{d^2}{d\tau^2_1}h^{\rm I}_{k,1}+\Omega_{k,1}^2 h^{\rm I}_{k,1}=0,\hspace{0.5in} \Omega_{k,1}^2 \equiv 1+\epsilon_2,\hspace{0.5 in}\epsilon_2\equiv-\Omega_{k}^{-1/2}\frac{d^2}{d\tau_1^2}\Omega_{k}^{1/2}\,.
\end{equation}
Since $\epsilon_2$ includes two derivatives, it contains terms of second and higher adiabatic order. We can ignore it to find the leading order solution
\begin{equation}\label{Eq:Fermions.hk1I}
h_{k,1}^{\rm I} \approx e^{-i\tau_1}\,,
\end{equation}
so that we get a first approximation
\begin{equation}\label{Eq:Fermions.FirstApprox}
h_{k}^{\rm I}\approx \frac{e^{-i\int^\tau \Omega_{k}d\tilde{\tau}}}{\sqrt{\Omega_{k}}}\,.
\end{equation}
Notice that $h_{k,1}^{\rm I}$ formally satisfies a differential equation with the same form as \eqref{Eq:Fermions.dceqofspinors} for $h_{k}^{\rm I}$. So that, we can repeat the process:
\begin{equation}\label{Eq:Fermions.hk2I}
h_{k,2}^{\rm I} \equiv \sqrt{\Omega_{k,1}} h_{k,1}^{\rm I}, \hspace{0.5 in} d\tau_2 \equiv \Omega_{k,1}d\tau_1\,.
\end{equation}
The corresponding differential equation for $h_{k,2}^{\rm I}$ is
 \begin{equation}\label{Eq:Fermions.Epsilon4}
\left(\frac{\partial^2}{\partial \tau_2^2}+\Omega_{k,2}^2\right)h_{k,2}^{\rm I}=0,\hspace{0.5in} \Omega_{k,2}^2 \equiv 1+\epsilon_4,\hspace{0.5 in}\epsilon_4\equiv-\Omega_{k,1}^{-1/2}\frac{d^2}{d\tau_2^2}\Omega_{k,1}^{1/2}\,.
\end{equation}
Once again, $\epsilon_4$ consists of terms of adiabatic order 4 and higher. We can approximate a solution of \eqref{Eq:Fermions.Epsilon4} by neglecting $\epsilon_4$:
\begin{equation}\label{Eq:Fermions.hk2Iaprox}
h_{k,2}^{\rm I} \approx e^{-i\tau_2}\,,
\end{equation}
whereby the approximation to  $h_{k}^{\rm I}$ can be further improved:
\begin{equation}\label{Eq:Fermions.2ndIterationh_kI}
h_{k}^{\rm I} \approx \frac{e^{-i \int^\tau \Omega_{k}\Omega_{k,1}\,d\tilde{\tau}} }{\sqrt{\Omega_{k}\Omega_{k,1}}}\,.
\end{equation}
By  iterating the procedure, we can obtain a better and better approximation to $h_{k}^{\rm I}$, and after $\ell>1$ steps we find
\begin{equation}\label{Eq:Fermions.GeneralSolution}
h_{k}^{\rm I} \approx \frac{e^{-i\int^\tau \Omega_{k}\cdots \Omega_{k,\ell-1}\,d\tilde{\tau}}}{\sqrt{\Omega_{k}\cdots \Omega_{k,\ell-1}}}\,,
\end{equation}
where, for $\ell \geq 1$,
\begin{equation}\label{Eq:Fermions.Omegakelle}
\Omega_{k,\ell}^2\equiv 1+\epsilon_{2\ell}, \hspace{0.5 in} d\tau_\ell \equiv \Omega_{k,\ell-1} d\tau_{\ell-1}, \hspace{0.5 in} \epsilon_{2\ell}\equiv -\Omega_{k,\ell-1}^{-1/2} \frac{d^2}{d\tau_\ell^2} \Omega_{k,\ell-1}^{1/2}\,.
\end{equation}
Now that the general method has been set up, let's find the $0th$ order solution for $h_{k}^{\rm I}$. From \eqref{Eq:Fermions.FirstApprox}, the most generic solution for $h_{k}^{\rm I}$ is
\begin{equation}\label{Eq:Fermions.hI(0)}
h^{\rm I}_k(\tau)\approx\frac{f_k^{(0)}}{\sqrt{\Omega_k}}e^{-i\int^\tau {\Omega_k\, d\tilde{\tau}}}=\frac{f_k^{(0)}}{(\omega^2_k+a^2\Delta^2-i\sigma )^{1/4}}e^{-i\int^\tau \sqrt{\omega_k^2+a^2\Delta^2-i\sigma}\,d\tilde{\tau}}\,,
\end{equation}
As for $h_{k}^{\rm II}$, by comparing both lines of $f_k^{(0)}$ (of adiabatic order $0$) accounts for the integration `constant' (strictly speaking, a function of the momentum but not of conformal time) in the exponential.
As for $h_k^{II}$, by comparing both lines of \eqref{Eq:Fermions.dceqofspinors} it is clear that it is possible to proceed in an analogous manner. So we obtain
\begin{equation}\label{Eq:Fermions.hII(0)}
h^{\rm II}_k(\tau)\approx\frac{g_k^{(0)}}{\sqrt{\Omega_k^*}}e^{-i\int^\tau {\Omega_k^*\, d\tilde{\tau}}}=\frac{g_k^{(0)}}{(\omega^2_k+a^2\Delta^2+i\sigma )^{1/4}}e^{-i\int^\tau \sqrt{\omega_k^2+a^2\Delta^2+i\sigma}\,d\tilde{\tau}}\,,
\end{equation}
where $g_k^{(0)}$ has the same paper as $f_k^{(0)}$.
To find the zeroth adiabatic order it is just enough to expand this solution and keep zero order terms. However, some extra caution is needed when dealing with the integrand in the exponential of \eqref{Eq:Fermions.hI(0)},  which may be expanded up to $1st$ order as
\begin{equation}\label{Eq:Fermions.Omegak01}
\Omega_k^{(0-1)}=\omega_k+\omega_k^{(1)}\,,
\end{equation}
where
\begin{equation}\label{Eq:Fermions.omegak(1)}
\omega_k^{(1)}\equiv -\frac{i a M }{2\omega_k}\frac{a^\prime}{a}\,.
\end{equation}
The reason is that the integration of the second term in the exponential factor is:
\begin{equation}\label{Eq:Fermions.ExpFactor}
e^{-i\int{\omega_k^{(1)}d\tau}}= \left( \frac{ \omega_k+aM }{k}\right)^{-1/2}=\left( \frac{ \omega_k-aM }{\omega_k+aM}\right)^{1/4}\,,
\end{equation}
so it yields a real term of adiabatic order zero, meaning that the expansion of $\Omega_k$ up to $1st$ order in the integral was mandatory. We have not included an explicit multiplicative factor related with the constant of integration\footnote{The same situation happens with indefinite integrals of higher order terms in the imaginary exponential of Eq.\,\eqref{Eq:Fermions.GeneralSolution}. They are written in an appropriate manner, contributing at bigger adiabatic orders. The final results, though, just depend on $f_k^{(0)}$ and not on the other higher order integrations constants, as dictated by the normalization condition \eqref{Eq:Fermions.normalizationrelation}. See \hyperref[Appendix:AdiabExpFermModes]{Appendix\,\ref{Appendix:AdiabExpFermModes}} for more details.} since it is already represented by $f_k^{(0)}$. We choose $f^{(0)}_k$ such that the above solution can be compatible with mode functions in Minkowskian spacetime, so we can write
\begin{equation}\label{Eq:Fermions.Zeroordersolution}
\begin{split}
&f_k^{(0)}=\sqrt{\frac{k}{2}},\hspace{0.5in}h^{\rm I(0)}_k(\tau)=\sqrt{\frac{\omega_k-aM}{2\omega_k}}e^{-i\int^\tau \omega_k \,d\tilde{\tau}},\\
&g_k^{(0)}=\sqrt{\frac{k}{2}},\hspace{0.5in}h^{\rm II(0)}_k(\tau)=\sqrt{\frac{\omega_k+aM}{2\omega_k}}e^{-i\int^\tau \omega_k \,d\tilde{\tau}}.
\end{split}
\end{equation}
Next we move on to the solution at $1st$ adiabatic order. As we have mentioned, the quantity $\epsilon_2$ defined in \eqref{Eq:Fermions.1storderofEom}, contains terms of adiabatic order two and higher, so it is not necessary to find the first order solution. It is enough to find the first order term from the denominator of \eqref{Eq:Fermions.hI(0)}.  So,
\begin{equation}
\begin{split}\label{Eq:Fermions.hI(1)}
h^{\rm I(0-1)}_k&\approx \left[\frac{1}{\sqrt{\omega_k}}\left(f^{(0)}_k+f^{(1)}_k \right)\left(1+\frac{iM a^\prime}{4\omega_k^2 }\right)e^{-\int^\tau \frac{M a^\prime}{2\omega_k}\,d\tilde{\tau}}\right] e^{-i\int^\tau \omega_k\, d\tilde{\tau}}\\
&=\sqrt{\frac{\omega_k-aM}{2\omega_k}}\left(1+i\frac{M a^\prime}{4\omega_k^2}+\sqrt{\frac{2}{k}}f^{(1)}_k\right)e^{-i\int^\tau {\omega_k\,d\tilde{\tau}}}\,.
\end{split}
\end{equation}
Similarly for the second spinor mode $h^{\rm II}_k$:
\begin{equation}
\begin{split}\label{Eq:Fermions.hII(1)}
h^{\rm II(0-1)}_k&\approx \left[\frac{1}{\sqrt{\omega_k}}\left(g^{(0)}_k+g^{(1)}_k \right)\left(1-\frac{iM a^\prime}{4\omega_k^2 }\right)e^{-\int^\tau \frac{M a^\prime}{2\omega_k}\,d\tilde{\tau}}\right] e^{i\int^\tau \omega_k\, d\tilde{\tau}}\\
&=\sqrt{\frac{\omega_k+aM}{2\omega_k}}\left(1-i\frac{M a^\prime}{4\omega_k^2}+\sqrt{\frac{2}{k}}g^{(1)}_k\right)e^{-i\int^\tau {\omega_k\,d\tilde{\tau}}}\,,
\end{split}
\end{equation}
where $f^{(1)}_k$ and $g^{(1)}_k$ come from integration constants, as mentioned in the footnote of the previous page. By imposing the normalization condition \eqref{Eq:Fermions.normalizationrelation}, which has to be satisfied at each adiabatic order, it is possible to see that these constants are purely imaginary, that is
\begin{equation} \label{Eq:Fermions.ConditionsI}
\operatorname{\mathbb{R}e}f_k^{(1)} =\operatorname{\mathbb{R}e} g_k^{(1)} = 0\,.
\end{equation}
To continue, we deal with the $2nd$ adiabatic order of the mode functions, i.e. $h_k^{\rm  I,II (2)}$. At this time, we have to include $\Omega^2_{k,1}=1+\epsilon_2$ in our considerations (this term contains $2nd$ order adiabatic terms and beyond).  Starting from Eq.  \eqref{Eq:Fermions.2ndIterationh_kI}, we have
\begin{equation}\label{Eq:Fermions.SecondOrderMode}
h^{\rm I}_k\approx\frac{f_k^{(0)}+f_k^{(1)}+f_k^{(2)}}{\sqrt{\Omega_k(1+\epsilon_2)^{1/2}}}e^{-i\int^\tau \Omega_k(1+\epsilon_2)^{1/2}\,d\tilde{\tau}}\,,
\end{equation}
where $\epsilon_2$ can be computed to be
\begin{equation}\label{Eq:Fermions.Epsilon2}
\begin{split}
&\epsilon_2 = \frac{5}{16\Omega_k^6}\left(2a a^\prime m_\psi^2-i m_\psi a^{\prime\prime}\right)^2-\frac{1}{4\Omega_k^4}\left(2{a^\prime}^2 m_\psi^2+2a a^{\prime \prime} m_\psi^2-i m_\psi a^{\prime\prime\prime} \right)\,.
\end{split}
\end{equation}
With this result, it is immediate to obtain an approximation for $\Omega_{k,1}$ valid up to the third adiabatic order:
\begin{equation}\label{Eq:Fermions.Omegak,1}
\begin{split}
\Omega_{k,1}=\left(1+\epsilon_2\right)^{1/2}&=1-\frac{a^2 M^2}{4\omega_k^4}\frac{a^{\prime\prime}}{a}-\frac{a^2 M^2}{4\omega_k^4}\left(\frac{a^\prime}{a}\right)^2+\frac{5}{8}\frac{a^4 M^4}{\omega_k^6}\left(\frac{a^\prime}{a}\right)^2+\frac{i a M}{8\omega_k^4}\frac{a^{\prime\prime\prime}}{a}\\
&-\frac{ia^3 M^3}{2\omega_k^6}\left(\frac{a^\prime}{a}\right)^3+\frac{15i a^5 M^5}{8\omega_k^8}\left(\frac{a^\prime}{a}\right)^3-\frac{9ia^3 M^3}{8\omega_k^6}\frac{a^\prime}{a}\frac{a^{\prime\prime}}{a}+\dots
\end{split}
\end{equation}
On the other hand, an expansion of the product $\Omega_k\Omega_{k,1}$ is necessary to improve the approximation of $h_k^{\rm I,II}$, as one can see from equation\,\eqref{Eq:Fermions.2ndIterationh_kI}. As earlier, if we wish to present a second order approximation of the modes we have to expand that product up to $3rd$ adiabatic order in the exponential. The expansion can be presented as follows:
and
\begin{equation}\label{Eq:Fermions.ExpansionofOmegakOmegak1I}
\Omega_k \Omega_{k,1}=\Omega_k  \left(1+\epsilon_2\right)^{1/2}=\omega_k+\omega_k^{(1)}+\omega_k^{(2)}+\omega_k^{(3)}+\dots
\end{equation}
where the dots represent the contributions of adiabatic order bigger than 3, and the indicated terms in the expansion read
\begin{equation}\label{Eq:Fermions.ExpansionofOmegakOmegak1}
\begin{split}
\omega_k^{(1)}&\equiv -\frac{i a M}{2\omega_k}\frac{a^\prime}{a}\,,\\
\omega_k^{(2)}&\equiv -\frac{a^2 M^2}{8\omega_k^3}\left(\frac{a^\prime}{a}\right)^2-\frac{a^2 M^2}{4\omega_k^3}\frac{a^{\prime\prime}}{a}+\frac{5a^4 M^4}{8\omega_k^5}\left(\frac{a^\prime}{a}\right)^2+\frac{a^2 \Delta^2}{2\omega_k}\,,\\
\omega_k^{(3)}&\equiv -\frac{5 i  a^3 M^3}{16\omega_k^5}\left(\frac{a^\prime}{a}\right)^3 -\frac{ i a^3 M^3}{\omega_k^5}\frac{a^\prime}{a}\frac{a^{\prime\prime}}{a}+\frac{i aM}{8\omega_k^3}\frac{a^{\prime\prime\prime}}{a}-\frac{i a\Delta^2}{4 M\omega_k}\frac{a^\prime}{a}+\frac{25 i a^5 M^5}{16\omega_k^7}\left(\frac{a^\prime}{a}\right)^3+\frac{i a^3 M\Delta^2}{4\omega_k^3}\frac{a^\prime}{a}\,.
\end{split}
\end{equation}
As noted before, $\omega_k^{(1)}$ and $\omega_k^{(3)}$ are purely imaginary, while $\omega_k$ and $\omega_k^{(2)}$ are real. Again, when integrated inside the exponential of equation \eqref{Eq:Fermions.2ndIterationh_kI} the former two give a real contribution, whereas the latter two become part of the phase of the mode and play the role of an effective frequency:
\begin{equation}
\begin{split}\label{Eq:Fermions.IntegralsExp}
&\exp\left(-i\int^\tau \Omega_k \left(1+\epsilon_2\right)^{1/2}d\tilde{\tau}\right) \approx \exp\left(-i\int^\tau \left(\omega_k^{(1)}+\omega_k^{(3)}\right)d\tilde{\tau}\right)\exp\left(-i\int^\tau \left(\omega_k+\omega_k^{(2)}\right)d\tilde{\tau}\right)\\
&=\left(\frac{\omega_k-a M}{\omega_k+a M}\right)^{1/4}\exp\left(-\frac{5 a^3 M^3}{16 \omega_k^5}\left(\frac{a^\prime}{a}\right)^2+\frac{a M}{8\omega_k^3}\frac{a^{\prime\prime}}{a}-\frac{a\Delta^2}{4M\omega_k}\right)\exp\left(-i\int^\tau \left(\omega_k+\omega_k^{(2)}\right)d\tilde{\tau}\right)\\
&\approx\left(\frac{\omega_k-a M}{\omega_k+a M}\right)^{1/4}\left(1-\frac{5 a^3 M^3}{16 \omega_k^5}\left(\frac{a^\prime}{a}\right)^2+\frac{a M}{8\omega_k^3}\frac{a^{\prime\prime}}{a}-\frac{a\Delta^2}{4M\omega_k}\right)\exp\left(-i\int^\tau \left(\omega_k+\omega_k^{(2)}\right)d\tilde{\tau}\right)\,.
\end{split}
\end{equation}
The last result holds good up to an arbitrary function of momentum (constant in conformal time) multiplying the whole result. We account for this arbitrary constant by introducing the functions $f_k^{(0)},f_k^{(1)},f_k^{(2)},\dots$ at each order.
An efficient strategy to compute the integrals involved in the above calculation (and many other ones of a similar sort, see \hyperref[Appendix:AdiabExpFermModes]{Appendix\,\ref{Appendix:AdiabExpFermModes}} for a  sample of them) is to set up an ansatz which respects the adiabaticity order of the calculation. The ansatz consists of a finite number of terms (in fact, a linear combination of them) taken each at the given adiabatic order and with coefficients (or `form factors') which must be determined. The terms of the ansatz are constructed out  of the derivatives of the scale factor and the parameter $\Delta^2$ (which we recall is of second adiabatic order). For instance, in order to compute the integral of $\omega_k^{(3)}$ in Eq.\,\eqref{Eq:Fermions.ExpansionofOmegakOmegak1}, we know that the result must be of second adiabatic order. Hence as a suitable ansatz we use a linear combination of second order adiabatic terms:
\begin{equation}\label{Eq:Fermions.Integralomega(3)}
-i\int^\tau \omega_k^{(3)}d\tilde{\tau}=Q_1\left(a,\omega_k\right)\left(\frac{a^\prime}{a}\right)^2+Q_2 \left(a,\omega_k\right)\frac{a^{\prime\prime}}{a}+Q_3\left(a,\omega_k\right)\Delta^2+\textrm{const.}
\end{equation}
where again  the term `const.' at the end means that it does not depend on the integration variable, $\tilde{\tau}$.
By taking derivatives with respect to (conformal) time of the last expression and comparing with $\omega_k^{(3)}$ one can identify the form factors $Q_1=-\frac{5a^3 M^3}{16\omega_k^5}$, $Q_2=\frac{aM}{8\omega_k^3}$ and $Q_3=-\frac{a}{4M\omega_k}$.
Using \eqref{Eq:Fermions.IntegralsExp} together with \eqref{Eq:Fermions.ExpansionofOmegakOmegak1} and \eqref{Eq:Fermions.SecondOrderMode}, the expansion of $h_{k}^{\rm I}$ up to $2nd$ order is
\begin{equation}\label{Eq:Fermions.hI(2)}
\begin{split}
h_k^{\rm I (0-2)}=\left(\frac{\omega_k-a M}{2\omega_k}\right)^{1/2}\Bigg(&1+\frac{i a^\prime M}{4\omega_k^2}+\sqrt{\frac{2}{k}}f_k^{(1)}\left(1+i\frac{a^\prime M}{4\omega_k^2}\right)+\sqrt{\frac{2}{k}}f_k^{(2)}+\frac{M a^{\prime \prime}}{8\omega_k^3}-\frac{5M^3{a^\prime}^2 a}{16\omega_k^5}\\
&-\frac{5a^2 {a^\prime}^2M^4}{16\omega_k^6}-\frac{{a^\prime}^2 M^2}{32\omega_k^4}+\frac{a a^{\prime\prime}M^2}{8\omega_k^4}-\frac{a\Delta^2}{4M\omega_k}-\frac{a^2 \Delta^2}{4\omega_k^2}\Bigg) e^{-i\int^\tau \left(\omega_k +\omega_k^{(2)}\right)d\tilde{\tau}}\,.
\end{split}
\end{equation}
In a similar way,
\begin{equation}\label{hII(2)}
\begin{split}
h_k^{\rm II (0-2)}=\left(\frac{\omega_k+a M}{2\omega_k}\right)^{1/2}\Bigg(&1-\frac{i a^\prime M}{4\omega_k^2}+\sqrt{\frac{2}{k}}g_k^{(1)}\left(1-i\frac{a^\prime M}{4\omega_k^2}\right)+\sqrt{\frac{2}{k}}g_k^{(2)}-\frac{M a^{\prime \prime}}{8\omega_k^3}+\frac{5M^3{a^\prime}^2 a}{16\omega_k^5}\\
&-\frac{5a^2 {a^\prime}^2M^4}{16\omega_k^6}-\frac{{a^\prime}^2 M^2}{32\omega_k^4}+\frac{a a^{\prime\prime}M^2}{8\omega_k^4}+\frac{a\Delta^2}{4M\omega_k}-\frac{a^2 \Delta^2}{4\omega_k^2}\Bigg) e^{-i\int^\tau \left(\omega_k +\omega_k^{(2)}\right)d\tilde{\tau}}\,.
\end{split}
\end{equation}

\begin{equation}\label{Eq:Fermions.ExpEpsilon2}
\begin{split}
e^{-i\int^\tau \Omega_k \left(1+\epsilon_2\right)^{1/2}d\tilde{\tau}}&\approx \left(\frac{\omega_k+a M}{\omega_k-a M}\right)^{1/4}\left(1-\frac{M a^{\prime \prime}}{8\omega_k^3}+\frac{5 M^3 \left(a^\prime \right)^2 a}{16 \omega_k^5}+\frac{a \Delta^2}{4M w_k}\right)\\
&\times e^{-i\int^\tau \left(\omega_k+\frac{1}{8  \omega_k^5}\left(-\left(a^\prime\right)^2M^2\omega_k^2-2a a^{\prime\prime}M^2\omega_k^2+a^2\left(5 \left(a^\prime\right)^2M^4+4\Delta^2 \omega_k^4\right)\right)\right)d\tilde{\tau}}.
\end{split}
\end{equation}
Using this last result together with \eqref{Eq:Fermions.ExpansionofOmegakOmegak1} and \eqref{Eq:Fermions.SecondOrderMode}, the expansion of $h_k^I$ up to $2nd$ order is
\begin{equation}\label{Eq:Fermions.hkI(2)}
\begin{split}
h_k^{I (0-2)}&=\left(\frac{\omega_k+a M}{2\omega_k}\right)^{1/2}\Bigg(1-\frac{i a^\prime M}{4\omega_k^2}+\sqrt{\frac{2}{k}}f_k^{(1)}\left(1-i\frac{a^\prime M}{4\omega_k^2}\right)+\sqrt{\frac{2}{k}}f_k^{(2)}-\frac{M a^{\prime \prime}}{8\omega_k^3}+\frac{5M^3\left(a^\prime\right)^2 a}{16\omega_k^5}\\
&\phantom{aaaaaaaaaaaaaaaaa}-\frac{5a^2 \left(a^\prime\right)^2M^4}{16\omega_k^6}-\frac{\left(a^\prime\right)^2 M^2}{32\omega_k^4}+\frac{a a^{\prime\prime}M^2}{8\omega_k^4}+\frac{a\Delta^2}{4M\omega_k}-\frac{a^2 \Delta^2}{4\omega_k^2}\Bigg)\\
&\times e^{-i\int^\tau \left(\omega_k +\frac{1}{8\omega_k^5}\left(-\left(a^\prime\right)^2M^2 \omega_k^2-2 a a^{\prime \prime}M^2 \omega_k^2+a^2\left(5 \left(a^\prime\right)^2M^4+4\Delta^2 \omega_k^4\right)\right)\right)d\tilde{\tau}}\,.
\end{split}
\end{equation}
In a similar way,
\begin{equation}\label{Eq:Fermions.hkII(2)}
\begin{split}
h_k^{II (0-2)}&=\left(\frac{\omega_k-a M}{2\omega_k}\right)^{1/2}\Bigg(1+\frac{i a^\prime M}{4\omega_k^2}+\sqrt{\frac{2}{k}}g_k^{(1)}\left(1+i\frac{a^\prime M}{4\omega_k^2}\right)+\sqrt{\frac{2}{k}}g_k^{(2)}+\frac{M a^{\prime \prime}}{8\omega_k^3}-\frac{5M^3\left(a^\prime\right)^2 a}{16\omega_k^5}\\
&\phantom{aaaaaaaaaaaaaaaaa}-\frac{5a^2 \left(a^\prime\right)^2M^4}{16\omega_k^6}-\frac{\left(a^\prime\right)^2 M^2}{32\omega_k^4}+\frac{a a^{\prime\prime}M^2}{8\omega_k^4}-\frac{a\Delta^2}{4M\omega_k}-\frac{a^2 \Delta^2}{4\omega_k^2}\Bigg)\\
&\times e^{-i\int^\tau \left(\omega_k +\frac{1}{8\omega_k^5}\left(-\left(a^\prime\right)^2M^2 \omega_k^2-2 a a^{\prime \prime}M^2 \omega_k^2+a^2\left(5 \left(a^\prime\right)^2M^4+4\Delta^2 \omega_k^4\right)\right)\right)d\tilde{\tau}}.
\end{split}
\end{equation}
The normalization condition fixes the following relations:
\begin{equation}\label{Eq:Fermions.ConditionsII}
\begin{split} 
&\left|f_k^{(1)}\right|^2=-\sqrt{2 k} \operatorname{\mathbb{R}e}f_k^{(2)},\\
&\left|g_k^{(1)} \right|^2=-\sqrt{2 k} \operatorname{\mathbb{R}e}g_k^{(2)}.
\end{split}
\end{equation}
So far, the expansion for the modes $h^I_k$ and $h^{II}_k$ up to $2nd$ order has been presented. One can continue with the procedure formerly described to reach higher orders, although of course the calculation becomes more and more involved. We should keep in mind, though, that the adiabatic expansion is an asymptotic expansion. While for renormalization purposes it is enough to stop the expansion at $4th$ adiabatic order (in $4$-dimensional spacetime), it is nonetheless necessary to reach up to $6th$ order to meet the finite terms $\sim H^6$ that are dominant in the early Universe and capable of triggering inflation in this framework (cf.\,\hyperref[Sect:RVMInfFermi]{Sect.\,\ref{Sect:RVMInfFermi}})\footnote{As explained in \hyperref[Chap:QuantumVacuum]{Chap.\,\ref{Chap:QuantumVacuum}}, owing to the renormalization prescription of the EMT -- see e.g. Eq. \eqref{Eq:QuantumVacuum.EMTRenormalizedDefinition} for the scalar case and its fermionic counterpart, Eq.\,\eqref{Eq:Fermions.RenormalizedEMTFermion} below -- the explicit $4th$ order powers $H^4$ just cancel out. As a result, the $6th$ order is the first non-vanishing contribution on-shell.} . We shall refrain from presenting these cumbersome formulas in the main text, see \hyperref[Appendix:AdiabExpFermModes]{Appendix\,\ref{Appendix:AdiabExpFermModes}}.

It is worth noticing that there is some residual freedom in the previous calculations since, we can not determine entirely the set of integration constants that appear during the calculations $f_k^{(1)},g_k^{(1)},f_k^{(2)},g_k^{(2)},\dots$
Because of the normalization condition \eqref{Eq:Fermions.normalizationrelation} of the mode functions, some restrictions such as \eqref{Eq:Fermions.ConditionsI} and \eqref{Eq:Fermions.ConditionsII} apply.
Fortunately, as commented in more detail in \hyperref[Appendix:AdiabExpFermModes]{Appendix\,\ref{Appendix:AdiabExpFermModes}}, the satisfaction of these restrictions is enough for the observables to be independent of this residual freedom. So that, is enough to set all of them to 0 to get, for instance, the desired values of the energy density and pressure.

\section{ZPE and VED for a fermionic field in FLRW spacetime}\label{Sect:ZPEFermions}

The computation of  the Fourier modes for a quantized fermion field through adiabatic expansion as explained  in the previous section is just the first step to compute the vacuum energy density (VED).  The next  step towards the VED  is to obtain the  ZPE  associated with Dirac fermions in curved spacetime. As it well known, traditional computations of ZPE suffer from the well-known headache of carrying highly unacceptable contributions proportional to the quartic powers of the masses, $\sim m^4$. This is so both for scalar and fermion fields, and it is already the case in flat, Minkowskian, spacetime, see e.g. \cite{Sola:2013gha,SolaPeracaula:2022hpd} for a detailed discussion and more references. In curved spacetime we have the same situation, in principle, but  in addition we encounter subleading,  curvature dependent,  contributions which do not exist in the flat case, as we shall see in a moment. To handle this issue, an adequate renormalization prescription is called for.

The derivation done here for fermions is closely related with the one performed for scalar fields in \hyperref[Chap:QuantumVacuum]{chapter\,\ref{Chap:QuantumVacuum}}. Once more the computation will be done up to $6th$ adiabatic order, since is the first non-vanishing order on-shell, {\it i.e.} when fixing the renormalization scale $M$ to the value of the mass of the fermion $m_\psi$. However, the off-shell computation at $4th$ order is already very useful as a means to determine the renormalization group running of the VED as a function of the scale $M$. This is one of the main new features of the ARP method proposed in \cite{Moreno-Pulido:2020anb,Moreno-Pulido:2022phq} which leads to the cosmic evolution of the VED. Next we consider the actual calculation for spinor fields.

To find out the ZPE, we start from the definition of EMT. In this case we have to evaluate the functional derivative
\begin{equation}\label{Eq:Fermions.Tmunudef}
      T_{\mu\nu}^{\psi}=-\frac{2}{\sqrt{-g}}\frac{\delta S_\psi}{\delta g^{\mu\nu}}\,,
\end{equation}
applied to the fermion action \eqref{Eq:Fermions.FermionAction}. Upon  a straightforward calculation we arrive at the following symmetric expression:
\begin{align}\label{Eq:Fermions.EMTFermion}
T_{\mu\nu}^{\psi}=\frac{i}{4}\bar{\psi}\left(\underline{\gamma}_\mu \nabla_\nu+\underline{\gamma}_\nu \nabla_\mu\right)\psi-\frac{i}{4}\left(\left(\nabla_\mu \bar{\psi} \right) \underline{\gamma}_\nu+ \left( \nabla_\nu \bar{\psi}\right)\underline{\gamma}_\mu\right)\psi\,,
\end{align}
in which the equation of motion \eqref{Eq:Fermions.CovDiracEq} and its hermitian conjugate have been used. We treat this spinor field as a field operator and upon using its expansion in Fourier modes and utilizing the anticommuting algebra of the creation and annihilation operators, Eq.\,\eqref{Eq:Fermions.AntiComRel}, we can compute the  VEV of the various components, which reflect  the contribution from the vacuum fluctuations of the quantized fermion fields. The method is exactly the same as the one used in previous chapters for scalar fields.
After significant work, we find that the VEV of the $00 th $ component of the EMT can be cast as follows:
\begin{equation}\label{Eq:Fermions.Tpsi00}
	\left<T^{\delta\psi}_{00}\right>=\frac{1}{2\pi^2 a}\int{dk k^2 \rho_k  }\,,
\end{equation}
where $\rho_k$ is a function of the previously defined mode functions (which can be computed through adiabatic expansion):
\begin{equation}\label{Eq:Fermions.producths}
\rho_k=\frac{i}{a}\left(h^{\rm I}_k h'^{\rm I*}_k+h^{\rm II}_k h'^{\rm II*}_k-h^{\rm I*}_k h'^{\rm I}_k-h^{\rm II*}_k h'^{\rm II}_k\right)\,.
\end{equation}
The explicit form of the adiabatic expansion of $\rho_k$ is rather cumbersome; the reader may find the final result of $\langle T_{00}^\psi \rangle$ in the \hyperref[Appendix:AdiabExpFermEMT]{Appendix\,\ref{Appendix:AdiabExpFermEMT}}. Let us note that for off-shell renormalization at a point $M$  it suffices to adiabatically  expand the solution up to $4th$ order  (as it was prescribed in Eq.\,\eqref{Eq:Fermions.RenormalizedEMTFermion} below). However,  we will provide the result up to $6th$ order so as to be sensitive to the on-shell result (occurring for  $M=m_\psi$)  and also because it is important for the inflationary mechanism in the early Universe (cf.\hyperref[Sect:RVMInfFermi]{Sec.\,\ref{Sect:RVMInfFermi}}).  Renormalization of the above expressions is indeed necessary since the VEV of the EMT is formally divergent. The UV-divergent contributions  appear up to $4th$ adiabatic order  (in $n=4$ spacetime dimensions), so that one has to subtract terms up to this order  to obtain a finite result.

\subsection{Divergence balance between  bosons and fermions in vacuum}\label{Sect:LeadingSubleading}
The unrenormalized  VEV  of the EMT can be split into two different parts, divergent (in the UV sense) and non-divergent.  Explicit calculation using the formulas of \hyperref[Appendix:AdiabExpFermEMT]{Appendix\,\ref{Appendix:AdiabExpFermEMT}}) shows that the divergent part  reads as follows:
\begin{align}\label{Eq:Fermions.Dpartofenergydensity}
	\begin{split}
		\left<T^{\delta\psi}_{00}\right>=\frac{1}{2\pi^2 a^2}\int_0^\infty dk k^2\left(-2\omega_k-\frac{a^2 \Delta^2}{\omega_k}+\frac{a^4\Delta^4}{4\omega_k^3}\right)
		+\frac{1}{2\pi^2 }\int_0^\infty dk k^2\left(\frac{ M^2}{4\omega_k^3}+\frac{\Delta^2}{4\omega_k^3}\right){\cal H}^2\,.
	\end{split}
\end{align}
As it is easy to see, there are terms diverging quartically, quadratically and logarithmically.
The non-divergent part contains the remaining terms, all of them being finite.
The above ZPE is, as warned,  an unrenormalized result at this point. However, before we proceed to  renormalize that expression in the next section, it may be instructive  to check if there is a chance for a cancellation between UV-divergent terms between fermions and bosons in the supersymmetric (SUSY) limit, if only  for the leading divergences.  In the on-shell case ($M=m$ and hence $\Delta^2=0$) the above equation \eqref{Eq:Fermions.Dpartofenergydensity} simplifies to
\begin{equation}\label{Eq:Fermions.FermUVpartOnshell}
 \left.\langle T_{00}^{\delta \psi}\rangle\right|_{(M=m)}= -\frac{1}{\pi^2 a ^2}\int dk k^2 \omega_k(m)+\frac{1}{8\pi^2 }\int_0^\infty dk k^2 \frac{ m^2}{\omega^3_k(m)} {\cal H}^2\,.
\end{equation}
It coincides with the Minkowskian result for $a=1$ (since ${\cal H}=0$).
Now, in a SUSY theory, in which the number of boson and fermion degrees of freedom (d.o.f.)  is perfectly balanced, we should expect that the leading (quartic) divergences cancel among the fermionic and bosonic contributions in the vacuum state\,\cite{Zumino:1974bg,Wess:1974tw} since in such case the  scalar and fermionic fields have the same mass $m$. Thus the quartically divergent contribution from the first term of \eqref{Eq:Fermions.FermUVpartOnshell} should be minus four times the corresponding result for one real scalar field\footnote{ In the chiral supermultiplets of SUSY theories  the number of d.o.f. from  bosons and from fermions is balanced, and they have the same mass.}.  We can check it is indeed so using the above formulas,  for in the on-shell limit and projecting  the UV-divergent terms of the first two adiabatic orders only,  we find that  the contribution from one real scalar field in  FLRW spacetime with spatially flat metric  is
\begin{equation}\label{Eq:Fermions.T002}
  \left.\langle T_{00}^{\delta \phi}\rangle^{ (0-2)}\right|_{(M=m)}  =\frac{1}{4\pi^2 a^2}\int dk k^2  \omega_k(m)
-\frac{3\left(\xi-\frac{1}{6}\right)}{4\pi^2 a^2}\int dk k^2\left(\frac{\mathcal{H}^2}{\omega_k(m)}+\frac{ a^2 m^2\mathcal{H}^2}{\omega_k^3(m)}\right)\,,
\end{equation}
We confirm that the first term (the quartically divergent one) of this expression  is of opposite sign to \eqref{Eq:Fermions.FermUVpartOnshell} and is a factor of $4$ smaller, as we indicated. So, in a SUSY theory,  where we would have $4$ real scalar d.o.f.  for each Dirac fermion, there would be an exact cancellation of the  leading UV-divergent terms.
In addition, we can see at once from \eqref{Eq:Fermions.T002}  that both the quadratic and logarithmic  divergences of  bosons  hinge on effects of the spacetime curvature since they are proportional to $\mathcal{H}^2$.  These terms, therefore,  vanish in Minkowski spacetime but are unavoidably present in the  FLRW background (except if $\xi=1/6$). On the other hand,  from the second term on the \textit{r.h.s.} of  Eq.  \eqref{Eq:Fermions.FermUVpartOnshell} it is clear that for fermions we only have subleading divergences of  logarithmic type, which  also hinge on curvature effects since they are again proportional to $\mathcal{H}^2$ and  would also vanish in Minkowski space.    Hence there is no possible cancellation of these subleading divergences between bosonic and fermionic d.o.f., in FLRW spacetime, even in the exact SUSY limit.  Of course, our framework is not placed in the context of supersymmetry, but it serves as a consistency check of our calculations.  See also the discussion in \cite{Maggiore:2010wr,Bilic:2011zm}.

Although it is possible to introduce a cutoff for a preliminary treatment of the subleading divergences (and maybe to speculate on its possible meaning)  it is not really necessary.  One simply has to implement appropriate renormalization since  renormalization is anyway necessary to deal meaningfully with the VED, as there is no way to cure the divergences from the combined contributions from bosons and fermions and it is not useful to be left with a ``physical'' cutoff.  Dealing with a cutoff is always ambiguous as it is generally not a covariant quantity.  Renormalization gets rid of cutoffs and one can preserve covariance, which is safer for a  physical interpretation of the final results. The adiabatic renormalization is ideal in this sense since the adiabatic expansion generates automatically a covariant result.

It is well-known that the renormalization program in QFT  requires the presence of a renormalization point, as well as a renormalization prescription. The renormalization point is a floating scale characteristic of the RG. As in the ordinary adiabatic procedure, to implement the renormalization of the EMT in $4$ spacetime dimensions we perform a subtraction of the first four adiabatic  orders, which are the only ones that can be UV-divergent\,\cite{birrell1984quantum,parker2009quantum,fulling1989aspects}. However, in contrast to the usual recipe, in which the subtraction is performed on the mass shell value $m$ of the quantum field,  we perform it at an arbitrary scale  $M$  since this enables  us to explore the RG evolution of the VED and ultimately connect it with its cosmic evolution. This is the specific feature of the  adiabatic  renormalization procedure (ARP)  for the VED that was proposed in \,\cite{Moreno-Pulido:2020anb, Moreno-Pulido:2022phq}  -- see also \cite{SolaPeracaula:2022hpd} for additional details and a comparison with other renormalization schemes. The resulting  renormalized VED ensuing from this procedure  is free from the usual troubles  with the quartic powers of the  masses and their inherent fine tuning problems.

Finally, let us note that dealing with the CCP in Minkowski spacetime  using,  for instance,  the MS scheme and assigning some value to the  't Hooft's mass unit $\mu$ in DR (as discussed so many times in the literature),  is entirely  meaningless. It is not only devoid of meaning in that a non-vanishing cosmological constant cannot be defined in Minkowski space without manifestly violating Einstein's equations; it is meaningless also on account of the fact that there is no sense in associating the scale $\mu$ with a cosmological variable, say $H$,  since, if  Einstein's equations are invoked,  the $\CC$ term as such in these equations cannot exist in Minkowski spacetime unless the VED is exactly $\rL+{\rm ZPE}=0$.   So there  is no cosmology whatsoever to do in  flat spacetime, despite some stubborn  attempts in the literature.  Persisting in this attitude leads to the nonsense of having to cope with $\sim m^4$ effects which must then be fine tuned among all the particles involved. This point has  been driven home repeatedly e.g.  in \cite{Sola:2013gha} and also recently in \cite{SolaPeracaula:2022hpd},  see also  \cite{Mottola:2022tcn}.  A realistic approach to the VED  within QFT in curved spacetime  must get rid of Minkowski space pseudo-argumentations. The approach that we present here is fully formulated in curved spacetime and the vacuum energy density evolves with the participation of the  curvature effects (powers of $H$) rather than with only powers of the masses, i.e. we pursue the successful renormalization program of \,\cite{Moreno-Pulido:2022phq,Moreno-Pulido:2020anb}. Therefore, when the background curvature vanishes, we consistently predict that the non-trivial effects which are responsible for the value of the vacuum energy density and the cosmological constant disappear (and hence we are left with no $\CC$ nor VED in the Universe).  Such is, of course, the situation in Minkowski space.  In practice, however,  we cannot reach that flat spacetime situation in our Universe since there exists four-dimensional curvature at all times during the indefinite process of expansion. But by the same token such an impossibility evinces the fact that the VED and its dynamical nature is a direct consequence of the expansion process (and of the spacetime curvature inherent to it). The expected size of the VED and of $\CC$ in our framework is indeed provided  by the magnitude of the spacetime curvature, which is of the typical value of the measured $\CC$. It is therefore not caused by the quartic power of the masses of the fields (which is the very root of the CC problem in most approaches). These powers  do not affect the running  of the VED in our framework. To put it in a nutshell: the renormalized VED in our framework is like a small quantum  `ripple' imprinted on the existing (classical) background curvature owing to the vacuum fluctuations of the quantized matter fields. In the absence of the background curvature, the ripple would disappear too since it is proportional to it through the coefficient $\nueff$, which encodes the quantum effects from the quantized matter fields.

Following the same approach as for scalar fields, in the next section we compute the quantum effects contributing to the VED from the quantized spin-1/2 fields and express them in renormalized form using the same subtraction scheme devised in \cite{Moreno-Pulido:2022phq,Moreno-Pulido:2020anb}.

\subsection{ZPE for fermions}\label{SubSect:RenZPEfermions}

Thus, following the same prescription \eqref{Eq:QuantumVacuum.EMTRenormalizedDefinition} as in the case of the scalar field, the renormalized form of the fermionic VEV of the EMT reads:
\begin{equation}\label{Eq:Fermions.RenormalizedEMTFermion}
\left<T_{\mu\nu}^{\delta\psi}\right>_{\rm ren}(M)\equiv\left<T^{\delta\psi}_{\mu\nu}\right>(m_\psi)-\left<T^{\delta\psi}_{\mu\nu}\right>^{(0-4)}(M)\,.
\end{equation}
Since our aim is to study the ZPE we will focus into the 00{\it th}-component of the former equation for the moment. Alternatively, it is written as \footnote{The subscript `Div' refers to the part of the EMT calculation comprising divergent integrals. These  appear only  up to the $4th$ adiabatic order. The subscript  `Non-Div', on the other hand, refers, of course, to the part of the EMT calculation involving finite integrals only. }
\begin{equation}\label{Eq:Fermions.T00ren}
\begin{split}
\left<T_{00}^{\delta\psi}\right>_{\rm ren}(M)&=\left<T_{00}^{\delta\psi}\right>_{\rm Div}(m_\psi)-\left<T_{00}^{\delta\psi}\right>_{\rm Div}(M)+\left<T_{00}^{\delta\psi}\right>_{\rm Non-Div}^{(0-4)}(m_\psi)-\left<T_{00}^{\delta\psi}\right>_{\rm Non-Div}^{(0-4)}(M)\\
&+\left<T_{00}^{\delta\psi}\right>^{(6)}(m_\psi)+\dots\\
&=\frac{1}{2\pi^2 a}\int_0^\infty dk k^2 \left(-\frac{2\omega_k (m_\psi)}{a}+\frac{2\omega_k (M)}{a}+\frac{a\Delta^2}{\omega_k(M)}-\frac{a^3\Delta^4}{4\omega_k^3 (M)}\right)\\
&+\frac{1}{2\pi^2 a}\int_0^\infty dk k^2 \left(\frac{a m_\psi^2}{4\omega_k^3(m_\psi)}-\frac{aM^2}{4\omega_k^3 (M)}-\frac{a\Delta^2}{4\omega_k^3(M)}\right)\left(\frac{a^\prime}{a}\right)^2\\
&+\left<T_{00}^{\delta\psi}\right>_{\rm Non-Div}^{(0-4)}(m_\psi)-\left<T_{00}^{\delta\psi}\right>_{\rm Non-Div}^{(0-4)}(M)+\left<T_{00}^{\delta\psi}\right>^{(6)}(m_\psi)+\dots\\
\end{split}
\end{equation}
where we have used the calculational results for the unrenormalized components of the VEV of the EMT recorded in \hyperref[Appendix:AdiabExpFermEMT]{Appendix\,\ref{Appendix:AdiabExpFermEMT}}
and we have introduced the notation $\omega_k(M) \equiv \sqrt{k^2+a^2 M^2}$ and  $\omega_k(m_\psi) \equiv \sqrt{k^2+a^2 m_\psi^2}$. The last line of \eqref{Eq:Fermions.T00ren} contains all the non-divergent terms, which constitute a perfectly finite contribution  and   is made of finite parts from the $4th$ order expansion and of the entire $6th$ order term, which is fully finite but rather cumbersome.
On the other hand, the first two lines in the last equality are a collection of terms that are individually divergent, but whose combination makes the integral convergent. In fact, by making use of simple algebraic manipulations at the level of the integrand one can show that explicitly. For instance, the rearrangement in the integrand
\begin{equation}\label{Eq:Fermions.Integrand}
dk k^2\left(\omega(m_\psi)-\omega(M)-\frac{a^2\Delta^2}{2\omega(M)}+\frac{a^4\Delta^4}{8\omega^3(M)}\right)= dk k^2a^6\Delta^6\frac{\omega(m_\psi)+3\omega(M)}{8\omega^3(M)(\omega(m_\psi)+\omega(M))^3}
\end{equation}
shows that terms seemingly diverging  as  $\sim k^4$ organize themselves to eventually  converge as $\sim 1/k^2$.  Needless to say, this is the consequence of the subtraction that has been operated. Similarly with the second integral in  \eqref{Eq:Fermions.T00ren}, whose individual terms are logarithmically divergent, but overall the integral  is once more convergent thanks to the involved subtraction.

The above renormalized result \eqref{Eq:Fermions.T00ren} would, of course,  vanish for $M=m_\psi$ if we were to stop the calculation at $4th$ adiabatic order, so in case that one wishes to obtain the renormalized on-shell result one has to either  compute the exact unrenormalized EMT on-shell  before subtracting the divergent adiabatic orders -- which is  possible but only in simpler cases such as in de Sitter space\,\cite{Landete:2013axa, Landete:2013lpa} -- or one has to face the calculation of the adiabatic expansion  up to $6th$-order at least. In the last case  the  term $\langle T_{00}^{\delta\psi}\rangle^{(6)}(m_\psi)$  indicated at the end of  Eq.\,\eqref{Eq:Fermions.T00ren} must be computed. This is what we have done here since an exact solution in the FLRW case is not possible.

The necessary work to reach up to $6th$ adiabatic  order for fermions  is  again significant, as it was previously  for  the scalar case.  The  unrenormalized components of the EMT up to the desired order are explicitly collected  in  \hyperref[Appendix:AdiabExpFermEMT]{Appendix\,\ref{Appendix:AdiabExpFermEMT}}. To subsequently obtain  the renormalized EMT  one has to  implement the subtraction  \eqref{Eq:Fermions.RenormalizedEMTFermion} and compute all the involved integrals. Despite the considerable amount of work involved,  the  final result to  the desired order  can nevertheless  be presented through a rather compact formula, as follows\footnote{We refer the reader to  \hyperref[Appendix:Conventions]{Appendix\ref{Appendix:Conventions}} for the computation/regularization of the involved integrals (depending on whether they are convergent or divergent) with the help of the master DR formula quoted there. Use of DR can be convenient since in certain cases the needed rearrangement of terms in the integrand to verify that the overall integral is actually convergent can be complicated.  Let us emphasize, however,  that DR is only used as an auxiliary regularization tool  for intermediate steps.   The final result has no memory of this intermediate step, see \hyperref[Appendix:Dimensional]{Appendix\,\ref{Appendix:Dimensional}} for an explicit nontrivial example.  To be sure,  no MS prescription is used for renormalization at any point of our calculation.  The crucial difference between the ARP and the  MS-like schemes is that the subtraction \eqref{Eq:Fermions.RenormalizedEMTFermion}  involves not just the UV-divergences  but also the  finite parts. }:
\begin{equation}\label{Eq:Fermions.EMT00}
	\begin{split}
\left<T^{\delta\psi}_{00}\right>_{\rm ren}^{(0-6)}(M,H)&=\frac{a^2}{32\pi^2}\left(3m_\psi^4-4m_\psi^2M^2+M^4-2m_\psi^4\ln \frac{m_\psi^2}{M^2}\right)\\
&+\frac{\mathcal{H}^2}{16\pi^2}\left(m_\psi^2-M^2-m_\psi^2\ln\frac{m_\psi^2}{M^2}\right)\\
&+\frac{1}{20160 \pi^2 a^4  m_\psi^2}\bigg( 204 \mathcal{H}^4 \mathcal{H}^\prime + 26 \left(\mathcal{H}^\prime\right)^3 - 30 \mathcal{H}^3\mathcal{H}^{\prime\prime} + 9 \left(\mathcal{H}^{\prime\prime}\right)^2 + 27 \mathcal{H}^2 \left( \mathcal{H}^{\prime}\right)^2 \\
&\phantom{aaaaaaaaaaaaaaa}- 72 \mathcal{H}^2 \mathcal{H}^{\prime\prime\prime}- 18 \mathcal{H}^\prime \mathcal{H}^{\prime\prime\prime} +
        \mathcal{H} (-78\mathcal{H}^\prime \mathcal{H}^{\prime\prime} + 18 \mathcal{H}^{\prime\prime\prime\prime})\bigg)\\
        &=\frac{a^2}{32\pi^2}\left(3m_\psi^4-4m_\psi^2M^2+M^4-2m_\psi^4\ln \frac{m_\psi^2}{M^2}\right)\\
        &+\frac{a^2 H^2}{16\pi^2}\left(m_\psi^2-M^2-m_\psi^2\ln\frac{m_\psi^2}{M^2}\right)\\
		&+\frac{a^2}{20160 \pi^2 m_\psi^2}\bigg( -31 H^6 - 108 H^4 \dot{H} - 46 \dot{H}^3 + 126 H^3 \ddot{H} + 9 \ddot{H}^2 -
         18 \dot{H} \vardot{3}{H}\\
        &\phantom{aaaaaaaaaaaaaa}+ 27 H^2 \left(7 \dot{H}^2 + 4 \vardot{3}{H} \right) + 6 H (23 \dot{H} \ddot{H} + 3 \vardot{4}{H})\bigg).
		\end{split}
\end{equation}
The final equality corresponds to the expression in terms of the cosmic time ($d()/dt\equiv\dot{()})$ with $H\equiv \dot{a}/a$. We point out that there is an explicit dependency on the Hubble function (and its derivatives) coming from $G_{\mu\nu}$. This justifies the notation  $\langle T_{00}^{\delta\psi} \rangle_{\rm ren}(M,H)$, with two arguments, where the dependence on the time derivatives of $H$ is omitted for simplicity.

We note that in the fermionic case there are no terms of ${\cal O}(H^4)$ in the evolution of the ZPE (and the VED, see next section), in stark contrast to the situation with scalars, see the last line of Eq.\,\eqref{Eq:QuantumVacuum.renormalized6th}, where we can recognize terms of the form $H^2\dot{H}, H\ddot{H}$ and $\dot{H}^2$ all of them of ${\cal O}(H^4)$.   We also remark what has been previously anticipated:  for $M=m_\psi$ (on-shell point) only the $6th$-order terms remain, which are the ones in the last two lines of Eq.\,\eqref{Eq:Fermions.EMT00}.  These terms are relevant for the RVM mechanism of inflation in the  very early Universe (cf. \,\hyperref[Sect:RVMInfFermi]{Sect.\,\ref{Sect:RVMInfFermi}}).  However, for the study of the renormalized theory at the point $M$ (generally different from the on-shell mass point  $m_\psi$) it is enough to consider the terms  up to $4th$ adiabatic order in Eq.\,\eqref{Eq:Fermions.EMT00}, see the next section.

So far, we have been able to provide the desired formula for the ZPE at the energy scale $M$ up to $6th$ adiabatic order,  cf. \eqref{Eq:Fermions.EMT00}. This is, however, not the end of the story, since a proper expression for the VED needs to take into account also the renormalized parameter $\rho_\Lambda$  in Einstein-Hilbert Action, as this parameter is part of the unrenormalized vacuum action and after renormalization it also runs with the scalae $M$, {\it i.e.} $\rL(M)$.  Both the ZPE and  $\rL(M)$ run with the scale and this will be crucial to study the properties of the renormalized VED. The running of the ZPE part between two different scales $M$ and $M_0$ can be illustrated by considering the difference of the respective ZPE values at these scales. From \eqref{Eq:Fermions.EMT00} we find
\begin{equation}\label{Eq:Fermions.TooTwoScales}
\begin{split}
\left<T^{\delta\psi}_{00}\right>_{\rm ren}(M,H)-\left<T^{\delta\psi}_{00}\right>_{\rm ren}(M_0,H)&=\frac{a^2}{32\pi^2}\left(M^4-M_0^4-4m_\psi^2 (M^2-M_0^2)+2m_\psi^4\ln \frac{M^2}{M_0^2}\right)\\
&+\frac{a^2 H^2}{16\pi^2}\left(-M^2+M_0^2+m_\psi^2\ln\frac{M^2}{M_0^2}\right)\,.
\end{split}
\end{equation}
The finite parts, and in particular the $6th$ order terms cancel of course in the above difference, but the latter will be essential in the on-shell case since the result would be zero without these higher order effects\,\footnote{Let us remark that the difference \eqref{Eq:Fermions.TooTwoScales} is an exact result, in the sense that it does not depend on the adiabaticity order we are working.  This is obvious from the renormalization prescription \eqref{Eq:Fermions.RenormalizedEMTFermion}, as all higher orders beyond the $4th$ one (not only the $6th$) cancel out in the subtraction, the reason being that these  adiabatic orders  are independent of the renormalization point $M$.  The latter is involved in the calculation of the EMT up to $4th$ order only (as these are the only adiabatic orders that are UV-divergent).}.
We should notice that, in contradistinction to the case with scalar fields, there are no contributions of ${\cal O}(H^4)$ such as $H^2\dot{H}$, $H\ddot{H}$ or $\dot{H}^2$ in the expression for the ZPE, as can be seen on comparing equations \eqref{Eq:QuantumVacuum.renormalized6th} and \eqref{Eq:Fermions.EMT00}. For this reason it is unnecessary to use the higher derivative (HD) tensor $\leftidx{^{(1)}}{\!H}_{\mu\nu}$ (cf. \hyperref[Appendix:Conventions]{Appendix\,\ref{Appendix:Conventions}}) as part of the renormalized Einstein's equations in the case of the fermion fields, again  in contrast to the situation with the scalar fields -- see \hyperref[Sect:RenormalizedVED]{Sect.\,\ref{Sect:RenormalizedVED}} for details.
Therefore, for fermions the subtraction at the two scales of the renormalized form of Einstein's equations can be done using the ordinary form of Einstein equations, {\it i.e.} Eq. \eqref{Eq:QuantumVacuum.EinsteinEq}, without higher order curvature terms, and  we find
\begin{equation}\label{Eq:Fermions.EinsteinSubstractionScales}
\left<T^{\delta\psi}_{\mu\nu}\right>_{\rm ren}(M,H)-\left<T^{\delta\psi}_{\mu\nu}\right>_{\rm ren}(M_0,H)=\left( \rho_\Lambda (M)-\rho_\Lambda (M_0) \right) g_{\mu\nu}+\left(\frac{1}{8\pi G(M)}-\frac{1}{8\pi G(M_0)}\right)G_{\mu\nu}\,.
\end{equation}
By comparison equations \eqref{Eq:Fermions.TooTwoScales} and \eqref{Eq:Fermions.EinsteinSubstractionScales}, and taking into account the tensorial structure of \eqref{Eq:Fermions.EinsteinSubstractionScales} and the explicit form of $G_{\mu\nu}$ in FLRW spacetime (cf. \hyperref[Appendix:Conventions]{Appendix\,\ref{Appendix:Conventions}}), we can perform the following identifications:
\begin{equation}\label{Eq:Fermions.RunningOfRhoLambda}
\begin{split}
\rho_\Lambda(M)-\rho_\Lambda(M_0)&=-\frac{1}{32\pi^2}\left(M^4-M_0^4-4m_\psi^2(M^2-M_0^2)+2m_\psi^4\ln \frac{M^2}{M_0^2}\right),\\
\frac{1}{8\pi G(M)}-\frac{1}{8\pi G(M_0)}&=\frac{1}{48\pi^2}\left(-M^2+M_0^2+m_\psi^2\ln\frac{M^2}{M_0^2}\right).
\end{split}
\end{equation}
\subsection{Renormalized  VED}\label{Sect:RenVEDfermions}

The same consideration of the scalar field case presented in \hyperref[Chap:QuantumVacuum]{chapter\,\ref{Chap:QuantumVacuum}} can be done here. So that, the VED associated to the fermionic field is
\begin{equation}\label{Eq:Fermions.DensityDefinition}
\rho_{\rm vac}^{\delta\psi} (M,H)=\frac{\left\langle T_{00}^{\rm vac}\right\rangle (M,H)}{a^2}=\rho_\Lambda (M)+\frac{\left\langle T_{00}^{\delta\psi} \right\rangle_{\rm ren} (M,H)}{a^2}\,.
\end{equation}
Now, if the subtraction of scales is done, we can write
\begin{equation}\label{Eq:Fermions.DensitiesDifference}
\begin{split}
\rho_{\rm vac}^{\delta\psi} (M,H)-\rho_{\rm vac}^{\delta\psi} (M_0,H)&=\frac{ \left\langle T_{00}^{\rm vac} \right\rangle (M,H)-\left\langle T_{00}^{\rm vac} \right\rangle (M_0,H)}{a^2}\\
&=\rho_\Lambda (M) - \rho_\Lambda (M_0)+\frac{\left\langle T_{00}^{\delta\psi} \right\rangle_{\rm ren} (M,H) -\left\langle T_{00}^{\delta\psi} \right\rangle_{\rm ren} (M_0,H)}{a^2}\\
&=\rho_\Lambda (M) - \rho_\Lambda (M_0)-\left( \rho_\Lambda (M) - \rho_\Lambda (M_0)\right)\\
&+\frac{3H^2}{8\pi}\left(\frac{1}{ G(M)}-\frac{1}{G(M_0)}\right)\\
&=\frac{H^2}{16\pi^2}\left(-M^2+M_0^2+m_\psi^2\ln\frac{M^2}{M_0^2}\right)\,,
\end{split}
\end{equation}
where in the last equality \eqref{Eq:Fermions.RunningOfRhoLambda} was used.  As expected, when written in terms of the ordinary Hubble function $H$ in cosmic time, the evolution of the VED  does not depend explicitly on the scale factor. For the sake of emphasizing the point, in the above equation we have explicitly indicated the cancellation of the terms carrying along the quartic powers of the masses, see the third equality in the above equation. As we can see, it is essential that the structure of the VED is obtained from the sum ``${\rm VED}=\rho_\Lambda+{\rm ZPE}$'', {\it i.e.} as we mentioned in a symbolic way in Eq.\,\,\eqref{Eq:QuantumVacuum.Symbolic}, since the mentioned cancellation occurs between  the renormalized expressions of $\rho_\Lambda$  and ${\rm ZPE}$ upon being subtracted at the two arbitrary scales $M$ and $M_0$. This means that the two values of the VED at these scales are related in a very smooth manner: in fact, they differ only by a term proportional to $H^2$, as it is obvious from \eqref{Eq:Fermions.DensitiesDifference}. Therefore, the evolution of the VED is well behaved, which means that, given its value at one scale, all other values at nearby scales are very close to it. The evolution is indeed slow and can be encoded into an effective contribution to the $\nueff$ parameter, as we did for bosons in Eq.\,\eqref{Eq:QuantumVacuum.nueffAprox}. The overall contribution from bosons and fermions to this parameter will be given in\,\hyperref[Sect:CombinedBandF]{Sect.\,\ref{Sect:CombinedBandF}}.

Even though Eq.\,\eqref{Eq:Fermions.DensitiesDifference} is formally correct, our job is not finished in the physical arena  yet.  Despite of the fact that such an equation describes the precise mathematical evolution of the VED with the renormalization scale, $M$, it is necessary to associate the latter with a suitable physical scale in order to extract useful phenomenological information out of it, exactly as we did for the scalar field in previous chapters. Again, the Hubble rate $H$  is a characteristic energy scale (in natural units) of the expanding Universe in the FLRW metric,  and hence proves to be a natural candidate for a representative physical scale in this context.  Whereby by following the same prescription used in \hyperref[Chap:QuantumVacuum]{Chap.\,\ref{Chap:QuantumVacuum}},  we set the renormalization energy scale to $M=H(t)$ (at the end of our calculations) in order  to track the physical evolution of the VED.  In other words, this prescription should allow us to explore the VED at different expansion history  times $H(t)$ in a physically meaningful way. In this way we obtain a well behaved evolution of the VED, which means that, given its value at one scale all other values at nearby scales are very close to it.  The dynamics of the VED is  slow and can be encoded into an effective contribution to the $\nueff$ parameter.  The combined contribution from bosons and fermions to this parameter  will be given in \hyperref[Sect:CombinedBandF]{Sect.\,\ref{Sect:CombinedBandF}}. Let us finally clarify the  sense of this scale setting, Namely, the full effective action does not depend on $M$, of course, but the renormalized VED indeed does since the effective action of vacuum is only a part of the full effective action. The scale dependence on $M$ from the other terms of the action, for example the terms carrying the running couplings of the RG-improved classical action, compensates for the $M$-dependence of the vacuum action.  Put another way,  only the full effective action (involving the classical part plus the nontrivial quantum vacuum effects) is scale- (i.e. RG-) independent.  This is of course the standard lore of the renormalization group (RG), see also \cite{SolaPeracaula:2022hpd} for an expanded discussion.  The choice of a particular scale helps of course in enhancing the physical significance of particular sectors of the full effective action. The procedure is of course akin to the usage of the RG in conventional gauge theories of strong and electroweak interactions, except that here one has to pick out an appropriate cosmological energy scale which is most adequate for the description of the Universe's expansion. The distinguished  scale $H$  appears to be the natural choice if the Universe where we live is indeed suitably described by the FLRW metric. In the next section we apply this approach to derive the  important RG equation of the VED itself.

\subsection{Renormalization group equation for the vacuum energy}\label{SubSect:RenGroupFermions}

One can also compute the $\beta$ function of the running vacuum associated to fermionic quantum fluctuations. Only the adiabatic terms below $4th$ order carry $M$-dependence by definition since the higher orders are finite and hence are not subtracted in the renormalization procedure. As it was noted before, in contrast to the scalar case the terms of $4th$ adiabatic order are not present for fermions. The computation follows the same strategy as for scalars. In this case we make use of equations \eqref{Eq:Fermions.TooTwoScales} and \eqref{Eq:Fermions.DensityDefinition}, and we find
\begin{equation}\label{Eq:Fermions.BetaFunctionFermion}
\begin{split}
    \beta_{\rv}^{\delta\psi}=&M\frac{\partial\rv^{\delta\psi}(M)}{\partial M}= \beta^{\delta\psi}_{\rL}+\frac{1}{8\pi^2}\left(M^2-m_\psi^2\right)^2-\frac{1}{8\pi^2}{H}^2\left(M^2-m_\psi^2\right)=-\frac{1}{8\pi^2}{H}^2\left(M^2-m_\psi^2\right)\,.
    \end{split}
\end{equation}
The second equality holds immediately after computing the $\beta$-function of the parameter $\rho_\Lambda$. From the first equation \eqref{Eq:Fermions.RunningOfRhoLambda} we find that
\begin{equation}\label{Eq:Fermions.BetaFunctionrLfermion}
 \beta_{\rL}^{\delta\psi}=M\frac{\partial\rL(M)}{\partial M}= -\frac{1}{8\pi^2}\left(M^2-m_\psi^2\right)^2
\end{equation}
and hence contains a term proportional to the quartic power of the particle mass; what's more, there  is an exact cancellation between the terms of the ZPE containing quartic powers of $M$ and $m_\psi$  and the expression of $\beta_{\rL}$.  The result \eqref{Eq:Fermions.BetaFunctionFermion} can also be consistently obtained directly from Eq.\,\eqref{Eq:Fermions.DensitiesDifference}. Notice that neither the parameter $\rL$ nor the ZPE have physical meaning in an isolated way, only the sum makes physical sense and defines the VED in the present context.  Let us compare the above results with those following from the contribution of one real scalar field $\phi$, see Eq.\,\eqref{Eq:QuantumVacuum.RGEVED1}:
\begin{equation}\label{Eq:Fermions.RGEVED1}
\beta_{\rv}^{\delta\phi}=\left(\xi-\frac{1}{6}\right)\frac{3 {H}^2 }{8 \pi^2}\left(M^2-m_\phi^2\right)+ {\cal O}(H^4)
\end{equation}
and
\begin{equation}\label{Eq:Fermions.BetaFunctionrL}
\beta_{\rL}^{\delta\phi} (M)=\frac{1}{2(4\pi)^2}(M^2-m_\phi^2)^2\,,
\end{equation}
 We can see that in both cases  the $\beta$-function of the VED  is proportional to $\beta_{\rv}\propto {H}^2 \left(M^2-m^2\right)$,  where $m=m_\phi$ or $m_\psi$, and therefore has a very smooth behavior thanks to the factor $H^2$.  In contrast, the $\beta$-function for the parameter $\rL$ in the gravitational action (which is often incorrectly identified as the VED in some  explicit QFT calculations of the vacuum energy in the literature) behaves in both cases as $\beta_{\rL}\propto \left(M^2-m^2\right)^2$ and hence  leads to undesired quartic contributions  $\sim m^4$  to the running. These are the problematic terms leading to fine tuning problems, but as can be seen these terms exactly cancel in  $\beta_{\rv}$ for the vacuum energy both for fermions and bosons in our renormalization scheme. Notice that there is a factor of $4$ between equations \eqref{Eq:Fermions.BetaFunctionrLfermion} and \eqref{Eq:Fermions.BetaFunctionrL} and have opposite sign. In a SUSY context,  Eq.\,\eqref{Eq:Fermions.BetaFunctionrL} should by multiplied b $4$to equalize bosonic and fermionic d.o.f in a given matter supermultiplet, all of whose members possess the same mass.  Then  $\beta_{\rL}^{\delta\phi}\rightarrow 4\beta_{\rL}^{\delta\phi}\equiv \beta_{\rL}^{\delta\phi\,( {\rm SUSY})} $, and  the sum of the two coefficients will indeed vanish in a supersymmetric context:
  \begin{equation}\label{Eq:Fermions.SUSYbetafunctions}
   \beta_{\rL}^{\delta\psi\,( {\rm SUSY})}+ \beta_{\rL}^{\delta\phi\,( {\rm SUSY})}=0\,.
  \end{equation}
But this is, of course, not a cancellation of the $\beta$-function coefficients for the VED of bosons and fermions in the SUSY limit, but only the cancellation of the contributions to the $\beta$-function coefficient for the formal parameter $\rL$ in the EH action \eqref{Eq:QuantumVacuum.EH}.
  This property is obviously connected with the discussion in \hyperref[Sect:LeadingSubleading]{Sec.\,\ref{Sect:LeadingSubleading}} about the balance of UV-divergences between fermions and bosons. In a SUSY theory the quartic divergences cancel prior to any renormalization process, as we have noticed, and the resulting $\beta$-function for the parameter $\rL$ is zero.  By the same token the running of the VED is freed from  $\sim m^4$ effects, which cancel among fermions and bosons in a SUSY context. The quartic powers  are independent of the curvature of spacetime.  However, the subleading divergences do depend on the background curvature and do not cancel at all, even in the exact SUSY limit \footnote{The SUSY considerations we have made here in passing only intend to clarify that in curved spacetime, irrespective of whether the quantized matter fields belong to a supersymmetric theory or not, the renormalization program is in any case mandatory to finally get rid of all the UV divergences. The calculations in this thesis, however,  do not presume any SUSY context at all, not even a SUSY-broken theory.  Our treatment of scalar and fermion fields is indeed completely general, in the sense that we are dealing with an arbitrary number of matter fields of both species without enforcing any balance between bosonic and fermionic d.o.f. -- see \hyperref[Sect:CombinedBandF]{Sec.\,\ref{Sect:CombinedBandF}} for more details.}.    The ``residual'' (finite) parts left in the renormalization process do not cancel either; they are actually  proportional to the curvature of the FLRW background, $R\sim H^2$. This fact translates into a correction to the physical vacuum energy density of order  $\sim m ^2 H^2$  both for bosons and fermions, which is far smaller than $m^4$. So the finite,  curvature dependent,  terms that remain after ARP renormalization are de facto the most important ones for our purposes since they lead to the RVM form of the VED! The renormalization of the formal parameter $\rL$, in contrast, has no physical imprint in the final result for the VED, except that the unwanted $m^4$ terms cancel against those involved in the ZPE, thus rendering the renormalized $V{\rm ED}=\rL+{\rm ZPE}$ free from quartic mass dependencies.

From the above RG equations we may write down the total contribution to the $\beta$-function of the VED from the matter fields.  Consider $N_{\rm f}$ species of fermion fields with  masses $m_{\psi,\ell}$ for each species  $\ell\in\{1,2,\dots , N_{\rm f}\}$, and similarly let $N_{\rm s}$ be the number of  scalar field species with  masses $m_{\phi, j}$, $j\in \{1,2,\dots, N_{\rm s}\}$. Some of these species may have the same mass, but this aspect is not relevant here, our formulas will include a summation over all contributions irrespective if some of them may be equal.
 The total $\beta$-function of the VED from an arbitrary number of  quantized matter fields can now be cast as follows:
  \begin{equation}\label{Eq:Fermions.Totalbetafunctions}
  \beta_{\rv}\equiv\sum_{j=1}^{N_{\rm s}} \beta_{\rv}^{\delta\phi_j}+ \sum_{\ell=1}^{N_{\rm f}}\beta_{\rv}^{\delta\psi_\ell}=\frac{3H^2}{8\pi^2}\left[\sum_{j=1}^{N_{\rm s}}\left(\xi_j-\frac{1}{6}\right) (M^2-m_{\phi_j}^2)-\frac{1}{3}\sum_{\ell=1}^{N_{\rm f}}(M^2-m_{\psi_\ell}^2)\right]+\mathcal{O}(H^4)\,.
  \end{equation}
The net outcome, therefore, is that the $\beta$-function of the vacuum energy density is free from undesirable contributions proportional to quartic mass powers of the quantized fields,  $\sim m^4$,  and hence these contributions do not appear in the renormalized theory.  This is of course an extremely welcome feature of our renormalization framework, which is, on inspection of the above equation,  fully shared by both  scalar and fermion fields.  Indeed, up to numerical factors fermions and scalar fields provide the same kind of leading contribution to the time evolution of the  cosmological vacuum energy. Overall we find that  the running of $\rv$ depends  only on quadratic terms in the fermion mass, namely $\sim m_{\psi_\ell}^2 H^2$, which are of the same type as in the case of bosons, namely $\sim m_{\phi_j}^2 H^2$, as discussed in \hyperref[Sect:RunningConnection]{Sect.\,\ref{Sect:RunningConnection}} and previously demonstrated in great detail in\,\cite{Moreno-Pulido:2020anb, Moreno-Pulido:2022phq}. These are actually very smooth owing to the presence of the $H^2$ factor. 

Integrating the RG equation corresponding to the $\beta$-function \eqref{Eq:Fermions.Totalbetafunctions} one finds  the expression for the evolution of the VED as a function of the  renormalization scale $M$ in the presence of any number of matter fields, see \hyperref[Sect:CombinedBandF]{Sec. \,\ref{Sect:CombinedBandF}}. In particular, integrating \eqref{Eq:Fermions.BetaFunctionFermion} for the case of one single fermion it is easy to verify that it leads exactly to \eqref{Eq:Fermions.DensitiesDifference}.

The kind of  much tempered behavior of the VED evolution that we have found here within our ARP renormalization program  is of the sort that was expected on the basis of semi-qualitative RG arguments and constitutes the characteristic running law of the so-called Running Vacuum Models (RVM), see \cite{Sola:2013gha,SolaPeracaula:2022hpd} and references therein. Thus, there is no need for fine-tuning in this scenario, since in such a renormalization procedure we have already gotten rid of the ugly contributions carried along by the quartic powers of the masses. In other words, the `problem' with the quartic powers of the masses does not appear in the physically renormalized theory. While the running of the formal parameter  $\rho_\Lambda$ with $M$ indeed carries $\sim m^4$ contributions, as it is obvious from the formulas above, this fact has no physical implication since $\rL$ is not itself a physical parameter (if taken in isolation) and  the unwanted terms carried by it exactly cancel out in the $\beta$-function for the  VED, as we have just proven.  As a result, the running of the VED is much softer, the `slope' is $\sim m^2 H^2$ rather than $\sim m^4$.   At variance with this result,  in the context of the MS renormalization approach, in which $\rho_\Lambda$ runs with the unphysical mass unit $\mu$ coming from dimensional regularization,  one is enforced to fine tune $\rho_\Lambda (\mu)$ against the large contribution proportional to $\sim m^4$ terms\,\cite{SolaPeracaula:2022hpd}.

\subsection{Renormalization of the fermionic vacuum pressure}\label{Sect:RenormFermVacuumPress}
Taking into account the perfect fluid form of the EMT associated with the vacuum,  the corresponding pressure is defined through the $ii$th-components. Any of them can be utilized owing to the assumed  homogeneity and isotropy. So, it is just enough to compute the $11$th-component\footnote{One can either compute the VEV of the  $T_{11}$  component, as we do here,  or use the  formula \eqref{Eq:eos.VacuumT11},  which allows to compute the vacuum pressure from the $00th$ component and the trace of the EMT. The result is the same, of course, owing to the isotropy of vacuum.}:
\begin{equation}\label{Eq:Fermions.ScalarFieldPressure2}
P_{\rm vac}(M)=\frac{\left\langle T_{11}^{\rm vac}\right\rangle}{a^2}=-\rho_{\Lambda}(M)+\frac{\left\langle T_{11}^{\rm \delta\psi} \right\rangle^{\rm ren}(M)}{a^2}\,.
\end{equation}  
From \eqref{Eq:Fermions.EMTFermion} and using once more the expansion of the spin-1/2 fermion fields in Fourier modes  (cf. \hyperref[Appendix:AdiabExpFermModes]{Appendix\,\ref{Appendix:AdiabExpFermModes}}) the result can be expressed  in the following way:
\begin{equation}\label{Eq:Fermions.T11psi}
    \begin{split}
             \left\langle T_{11}^{\delta\psi} \right\rangle=\frac{1}{2\pi^2a}\int_0^\infty dk k^2 P_k\,,
    \end{split}
\end{equation}
with
\begin{equation}\label{Eq:Fermions.Pk}
    P_k\equiv -\frac{2k}{3a}\left(h_k^I h_k^{II *}+h_k^{I*}h_k^{II}\right).
\end{equation}
Notice that there is a relation between $\rho_k$ and $P_k$
\begin{equation}\label{Eq:Fermions.Pkandrhok}
    P_k\equiv -\frac{2k}{3a}\left(h_{k}^{\rm I} h_k^{\rm II *}+h_k^{\rm I*}h_{k}^{\rm II}\right)
\end{equation}
and where the explicit expressions (in WKB-expanded form) for the fermion modes $h_k^{\rm I}$ and $h_k^{\rm II }$ can be found in the appendices.
Notice that there is a relation between $\rho_k$ and $P_k$,
\begin{equation}\label{Eq:Fermions.Relationpkrhok}
    P_k=-\frac{\rho^\prime_k}{3\mathcal{H}}\,,
\end{equation}
which follows from \eqref{Eq:Fermions.producths} using the mode equations \eqref{Eq:Fermions.h1h2eq}.
This relation  can be used as an alternative way to calculate $\langle T_{11}^{\delta\psi} \rangle$ from $\langle T_{00}^{\delta\psi} \rangle$:
\begin{equation}\label{Eq:Fermions.AlternativeExpression}
    \left\langle T_{11}^{\delta\psi} \right\rangle = -\frac{1}{3\mathcal{H}}\left(\left\langle T_{00}^{\delta\psi} \right\rangle^\prime+\mathcal{H}\left\langle T_{00}^{\delta\psi} \right\rangle\right)\,.
\end{equation}
For the sake of simplicity, the remaining discussions of this section will be restricted to the case of one single field. We shall retake the multifield case in \hyperref[Sect:CombinedBandF]{Sec.\,\ref{Sect:CombinedBandF}}.  Following the same steps and considerations made in the previous sections for the $00th$-component of the EMT, we reach the following expression for the renormalized value of the VEV of the $11th$-component of the EMT up to $6th$ adiabatic order:
\begin{equation}\label{Eq:Fermions.renEDsol}
\begin{split}
\left<T^{\delta\psi}_{11}\right>_{\rm ren}^{(0-6)}(M)&=-\frac{a^2}{32\pi^2}\big(M^4+3m_\psi^4-4m_\psi^2M^2-2m_\psi^4\ln\frac{m_\psi^2}{M^2}\big)\\
&+\frac{1}{48\pi^2}\left(M^2-m_\psi^2+m_\psi^2\ln\frac{m_\psi^2}{M^2}\right)\mathcal{H}^2+\frac{1}{24 \pi^2}\left(M^2-m_\psi^2+m_\psi^2\ln\frac{m_\psi^2}{M^2}\right) \mathcal{H}^\prime\\
&+\frac{1}{20160 \pi^2 a^4 m_\psi^2}\left(-245\mathcal{H}^2\left(\mathcal{H}^\prime\right)^2+8\left(\mathcal{H}^\prime\right)^3-98\mathcal{H}^3\mathcal{H}^{ \prime\prime}+35\left(\mathcal{H}^{\prime\prime}\right)^2-62\mathcal{H}^2\mathcal{H}^{\prime\prime\prime}\right. \\
&\left.+204\mathcal{H}^4\mathcal{H}^\prime -66\mathcal{H}\mathcal{H}^\prime\mathcal{H}^{\prime\prime}+56\mathcal{H}^\prime\mathcal{H}^{\prime\prime\prime}+42\mathcal{H}\mathcal{H}^{\prime\prime\prime\prime}-6\mathcal{H}^{\prime\prime\prime\prime\prime}\right)\\
&=-\frac{a^2}{32\pi^2}\big(M^4+3m_\psi^4-4m_\psi^2M^2-2m_\psi^4\ln\frac{m_\psi^2}{M^2}\big)\\
&+\frac{a^2}{16\pi^2}\left(M^2-m_\psi^2+m_\psi^2\ln\frac{m_\psi^2}{M^2}\right)H^2+\frac{a^2}{24 \pi^2}\left(M^2-m_\psi^2+m_\psi^2\ln\frac{m_\psi^2}{M^2}\right)\dot{H}\\
&+\frac{a^2}{20160 \pi^2  m_\psi^2}\left(31H^6+170H^4\dot{H}-45H^2\dot{H}^2-80\dot{H}^3-90H^3\ddot{H}-55\ddot{H}^2\right. \\
&\left.-150H^2\vardot{3}{H}-100\dot{H}\vardot{3}{H}-6H(65\dot{H}\ddot{H}+9\vardot{4}{H})-6\vardot{5}{H}\right)\,.
\end{split}
\end{equation}	
We may now  proceed to compute the vacuum EoS for the fermion fields up to the sixth adiabatic order. The best strategy is to compute first the pressure  through  Eq.\,\eqref{Eq:Fermions.renEDsol},  which can be inserted  into the relation \eqref{Eq:Fermions.ScalarFieldPressure2}.
Using next the VED expression \,\eqref{Eq:Fermions.DensityDefinition} for fermions -- with $\langle T^{\delta\psi}_{00}\rangle$ given by \eqref{Eq:Fermions.EMT00} -- the vacuum pressure can be seen to be equal to minus the VED plus some additional terms:
\begin{equation}\label{Eq:Fermions.PressureEnergyVacuum}
\begin{split}
P_{\rm vac}(M)&=-\rho_{\rm vac}(M)+\frac{1}{24 \pi^2}\left(M^2-m_\psi^2+m_\psi^2\ln\frac{m_\psi^2}{M^2}\right)\dot{H}\\
    	&+\frac{1}{20160 \pi^2  m_\psi^2}\left(62H^4\dot{H}+144H^2\dot{H}^2-126\dot{H}^3+36H^3\ddot{H}-46\ddot{H}^2-42H^2\vardot{3}{H}\right. \\
		&\left.\phantom{xxxxxxxxxxxx}-118\dot{H}\vardot{3}{H}-6H(42\dot{H}\ddot{H}+6\vardot{4}{H})-6\vardot{5}{H}\right)+\dots
\end{split}
\end{equation}
The additional terms represent a small (but worth noticing)  deviation from the classical vacuum EoS relation $P_{\rm vac}=-\rho_{\rm vac}$. The dominant vacuum  EoS is still the classical one up to a leading correction of ${\cal O}(\dot{H})$ (the second term on the \textit{r.h.s} of the above equation)  and several sorts of higher order corrections of  ${\cal O}(H^6)$ indicated in the last two lines.  The $\sim\dot{H}$  correction  in the first line of Eq.\,\eqref{Eq:Fermions.PressureEnergyVacuum} --  can obviously be relevant for the present Universe, and in particular it can modify the equation of state of the vacuum  it to depart from $-1$ at present.

The higher order terms in the last two lines, in contrast,  might be relevant only for the very early Universe, in principle. However, these terms involve  time derivatives and hence vanish for $H=$const. This fact will have implications for our discussion of RVM-inflation in \hyperref[Sect:RVMInfFermi]{Sect.\,\ref{Sect:RVMInfFermi}}), since inflation can be shown to exist in this framework for $H=$const. So at the end of the day, the higher order terms in the last two lines of Eq.\,\eqref{Eq:Fermions.PressureEnergyVacuum} become irrelevant both at low and high energies in this framework. The consequence is that the EoS of the quantum vacuum stays very close to $-1$  during inflation, in contrast to the vacuum EoS in subsequent eras of the cosmic evolution (cf. \hyperref[Sect:EoS-QVacuum]{Sect.\,\ref{Sect:EoS-QVacuum}}).

\subsection{Trace Anomaly}\label{Sect:TraceAnomFerm}

It is a very well known result that if a field theory has a classical action which is conformally invariant, then the trace of the classical EMT vanishes exactly. For this  it is necessary to work with a massless field, otherwise the presence of a mass breaks the symmetry since it introduces a fixed length scale. For instance, for a massless scalar field,
\begin{equation}\label{Eq:Fermions.LimTcl}
    \lim\limits_{\xi \to 1/6} \lim\limits_{m_\phi \to 0 } T_{\rm Cl.} \left(\phi\right)=0\,.
\end{equation}
 This follows immediately from \eqref{Eq:QuantumVacuum.KG} and \eqref{Eq:QuantumVacuum.EMTScalarField}. However, it is also true that this simple result does not hold when one takes into account the quantum effects from  the scalar field and constitutes the scalar part of the conformal anomaly,\cite{birrell1984quantum}.  This follows after a careful study of the diverging part of the vacuum effective action, $W_{\rm eff}^{\rm Div}$, in which $W_{\rm eff}$ was defined in \hyperref[Sect:EffectiveActionQFT]{Sect.\,\ref{Sect:EffectiveActionQFT}}. The part   $W_{\rm eff}^{\rm Div}$ is not conformally invariant for an arbitrary number of spacetime dimensions $n$ (although   $W_{\rm eff}$  is so in the massless limit), except for the case $n=4$. As a consequence, $W_{\rm eff}^{\rm Div}$ receives a finite payoff for $n\to 4$ owing to the existing pole $1/(n-4)$ in it. Correspondingly, the  VEV of the on-shell EMT receives a nontrivial contribution in the massless limit coming from the divergent part of the effective action, even in the case $\xi=1/6$:
\begin{equation}\label{Eq:Fermions.TraceLimitScalar}
   \lim\limits_{m_\phi\to 0}\lim\limits_{\xi \to 1/6}  \left\langle T^{\delta \phi} \right\rangle =-\lim\limits_{m_\phi\rightarrow 0} m_\phi^2 \left\langle \delta\phi^2\right\rangle\,.
\end{equation}

  The term $\left\langle \delta\phi^2\right\rangle$ contains some elements of $4th$ adiabatic order proportional to $1/m_\phi^2$, so that the corresponding limit results in a finite contribution.
 The same idea applies in the fermionic case,
\begin{equation}\label{Eq:Fermions.TraceLimitFermion}
     \lim_{m_\psi \to 0}\left\langle T^{\delta\psi}\right\rangle=-\lim_{m_\psi \to 0} m_\psi \left\langle \bar{\psi} {\psi}\right\rangle\,.
 \end{equation}
Here the term $\left\langle \bar{\psi} {\psi}\right\rangle$ contains $4th$ adiabatic order terms that are proportional to $1/m_\psi$ which make the limit non-trivial.
Technically speaking \eqref{Eq:Fermions.TraceLimitScalar} and \eqref{Eq:Fermions.TraceLimitFermion} are not yet what we call the {\it trace anomaly} or {\it conformal anomaly}.  This is due to the fact that the total effective action is conformally invariant and the corresponding EMT is traceless, so the part of the trace associated with the finite and divergent parts should be equal but with opposite sign in the conformal limit\,\cite{birrell1984quantum}. The anomaly is generated from the finite part, so its actual value for the scalar field case is
\begin{equation}\label{Eq:Fermions.AnomalyScalarField}
\begin{split}
\left\langle T^{\delta \phi} \right\rangle^{\rm  Anom.} &=-\lim\limits_{m_\phi\rightarrow 0}  \left\langle T^{\delta \phi} \right\rangle = \frac{1}{480\pi^2 a^4}\left(4\mathcal{H}^2\mathcal{H}^\prime -\mathcal{H}^{\prime\prime\prime}\right)\\
&=\frac{1}{2880\pi^2}\left(R^{\mu\nu}R_{\mu\nu}-\frac{R^2}{3}+\Box R\right)\,,
\end{split}
\end{equation}
where the conversion of the anomaly result into an invariant expression in the last step can be performed using the formulae of \hyperref[Appendix:Conventions]{Appendix\,\ref{Appendix:Conventions}}. This result was explicitly verified in the calculation of \hyperref[Sect:Trace]{Sect.\,\ref{Sect:Trace}}. We remark that  for an arbitrary curved background the expression for the conformal anomaly is more involved\,\cite{birrell1984quantum}. However, since the spatially flat FLRW spacetime  is conformally flat ({\it i.e.} conformal to Minkowski space)  the contribution from the Weyl tensor vanishes identically and hence also its square (entering the anomaly). Additional terms beyond $4th$ adiabatic order decouple when $m_\phi \rightarrow \infty$, satisfying the Appelquist-Carazzone decoupling theorem \cite{Appelquist:1974tg}. These terms are not finite in the massless limit, and hence do not take part of the anomaly. In practice we have derived the anomaly \eqref{Eq:Fermions.AnomalyScalarField} from the unrenormalized trace of the vacuum EMT for scalar fields, $\left\langle T^{ \phi} \right\rangle$, which is given in full detail in \hyperref[Sect:Trace]{Sect.\,\ref{Sect:Trace}}.
The corresponding conformal anomaly for fermions can be similarly extracted from the unrenormalized  $ \left\langle T^{ \delta\psi} \right\rangle$  and it is a bit cumbersome as well, so we shall spare details here.  We limit ourselves to provide the final result. Once more we can recognize the expression of the anomaly as a linear combination of finite terms of adiabatic order 4  which are independent of the mass scale. We find
\begin{equation}\label{Eq:Fermions.AnomalyFermionField}
\begin{split}
 \left\langle T^{\delta \psi} \right\rangle^{\rm  Anom.} &=-\lim\limits_{m_\psi\rightarrow 0}  \left\langle T^{
    \delta \psi} \right\rangle = \frac{1}{240\pi^2 a^4}\left(7\mathcal{H}^\prime\mathcal{H}^2 -3\mathcal{H}^{\prime\prime\prime}\right)\\
    &=\frac{11}{2880\pi^2}\left( R^{\mu\nu}R_{\mu\nu}-\frac{R^2}{3}+\frac{6}{11}\Box R\right)\,.
    \end{split}
\end{equation}
One natural question is related with the physical consequences of the conformal anomaly. It is well-known that it is a valuable theoretical concept  encoding essential information on the VEV of the renormalized EMT\cite{birrell1984quantum}, although it need not be itself part of the observable quantities of the renormalized theory.  There are some attempts in the literature to remove the anomaly by particular prescriptions or definitions of the renormalized EMT \cite{brown1978energy}.  This is also the case of the renormalization procedure employed in this chapter, as defined in \eqref{Eq:QuantumVacuum.EMTRenormalizedDefinition} and \eqref{Eq:Fermions.RenormalizedEMTFermion}, where the anomaly has no physical effects. The reason is that the on-mass-shell VEV of the EMT is subtracted at an arbitrary scale, $M$,  up to $4th$ adiabatic order. Since the anomaly is of $4th$ adiabatic order and it is independent of the mass of the fields and, of course, also from the arbitrary renormalization point, it gets cancelled exactly in our ARP renormalization procedure. Alternatively, one can think in terms of the effective action. Indeed, in \hyperref[Sect:EffectiveActionQFT]{Sect.\,\ref{Sect:EffectiveActionQFT}} we defined the renormalized effective lagrangian density off-shell at an arbitrary scale $M$,
\begin{equation}\label{Eq:Fermions.RenormalizedEffLagrangian}
L_{W}^{\rm Ren}(M)\equiv L_W (m)-L_W^{\rm Div}(M)
\end{equation}
and it was shown by expanding it through the DeWitt-Schwinger series that it eventually leads exactly to the same renormalized EMT defined by \eqref{Eq:QuantumVacuum.EMTRenormalizedDefinition}. This result was obtained explicitly for a scalar field $\phi$  and can be repeated for fermions, although we shall not provide details here. Now the anomaly is related with the divergent part of the effective Lagrangian, corresponding to the lowest adiabatic orders (up to  $4h$ order).  As a consequence  it gets once more exactly cancelled in \eqref{Eq:Fermions.RenormalizedEffLagrangian} analogously  to the subtraction of the EMT.

As previously indicated, the anomaly part of the trace is contained in the unrenormalized trace of the EMT (even though the anomaly itself is a finite part of it).  In our framework, however,  the anomaly cancels since the anomaly is independent of the mass scale and our  renormalized VEV of the EMT is defined through a subtraction of its value at two different scales, see equations \eqref{Eq:QuantumVacuum.EMTRenormalizedDefinition} and \eqref{Eq:Fermions.RenormalizedEMTFermion}. Thus the conformal anomaly is not involved in the renormalized expressions for the vacuum energy density and pressure in our framework.  Despite it  not having  physical consequences in our approach, the explicit calculation of the anomaly is certainly useful as a nontrivial cross-check of our intermediate results.

\section{Combined fermionic and bosonic contributions}\label{Sect:CombinedBandF}

Let us now determine the combined vacuum contributions from a multiplicity of non-interacting fermionic and bosonic degrees of freedom. We consider $N_{\rm f}$ species of fermion fields with  masses $m_{\psi,\ell}$  ($\ell\in\{1,2,\dots , N_{\rm f}\}$), and $N_{\rm s}$ scalar field species with  masses $m_{\phi, j}$ ($j\in \{1,2,\dots, N_{\rm s}\}$).

\subsection{Running vacuum from an arbitrary number of bosons and fermions}

The renormalized expression of the vacuum energy density is, in that case,
\begin{align}\label{Eq:Fermions.VEDMulticomponent}
    \rho_{\rm vac}(M,H)=\rho_\Lambda (M)+\frac{\sum_{j=1}^{N_{\rm s}} \left\langle T_{00}^{\delta \phi_j}\right\rangle_{\rm ren} (M,H)+\sum_{\ell=1}^{N_{\rm f}}\left\langle T_{00}^{\delta \psi_\ell}\right\rangle_{\rm ren} (M,H)}{a^2}\,.
\end{align}

Exactly as we did in \hyperref[SubSect:RenZPEfermions]{Sect.\,\ref{SubSect:RenZPEfermions}}, we need to  subtract Einstein's equations at two different
energy scales $M$ and $M_0$ in order to obtain the running of the couplings with the change of the scale.  We find:
\begin{equation}\label{Eq:Fermions.00Subtract}
\begin{split}
&\sum_{j=1}^{N_{\rm s}} \left(\left\langle T_{00}^{\delta \phi_j}\right\rangle_{\rm ren} (M,H)-\left\langle T_{00}^{\delta \phi_j}\right\rangle_{\rm ren} (M_0,H)\right)\\
&+\sum_{\ell=1}^{N_{\rm f}}\left(\left\langle T_{00}^{\delta \psi_\ell}\right\rangle_{\rm ren} (M,H)-\left\langle T_{00}^{\delta \psi_\ell}\right\rangle_{\rm ren} (M_0,H)\right)\\
&=\sum_{j=1}^{N_{\rm s}}\Bigg[\frac{a^2}{128\pi^2}\left(-M^4+M_0^4+4m_{\phi_j}^2 \left(M^2-M_0^2\right)-2m_{\phi_j}^4\ln\frac{M^2}{M_0^2}\right)\\
&\phantom{aaaaaaa}+\frac{3\left(\xi_j-\frac{1}{6}\right)a^2H^2}{16\pi^2}\left(M^2-M_0^2-m_{\phi_j}^2 \ln \frac{M^2}{M_0^2}\right)\\
&\phantom{aaaaaaa}+\frac{9\left(\xi_j-\frac{1}{6}\right)^2 a^2}{16\pi^2}\left(\dot{H}^2-2\ddot{H}H-6H^2\dot{H}\right)\ln \frac{M^2}{M_0^2}\Bigg]\\
&+\sum_{\ell=1}^{N_{\rm f}}\Bigg[\frac{a^2}{32\pi^2}\left(M^4-M_0^4-4m_{\psi_\ell}^2 \left(M^2-M_0^2\right)+2m_{\psi_\ell}^4\ln\frac{M^2}{M_0^2}\right)\\
&\phantom{aaaaaaa}+\frac{a^2H^2}{16\pi^2}\left(M_0^2-M^2+m_{\psi_\ell}^2 \ln \frac{M^2}{M_0^2}\right)\Bigg]\\
&=\left(\rho_\Lambda (M)-\rho_\Lambda (M_0)\right)g_{00}+\left(\frac{1}{8\pi G{(M)}}-\frac{1}{8\pi G(M_0)}\right)G_{00}+\left(a_1 (M)-a_1(M_0)\right)\leftidx{^{(1)}}{\!H}_{00}\,.
\end{split}
\end{equation}
Notice the appearance of the $00th$ component of $\leftidx{^{(1)}}{\!H}_{\mu\nu}$, which is a HD  tensor of  ${\cal O}(H^4)$,  hence of adiabatic order $4$, see \hyperref[Appendix:Conventions]{Appendix\,\ref{Appendix:Conventions}}.  Its presence in the generalized Einstein's GR equations is indispensable for renormalization purposes and  constitutes  a UV completion of the field equations.  No additional HD tensors are needed for conformally flat spacetimes\,\cite{birrell1984quantum}.  In our case, $\leftidx{^{(1)}}{\!H}_{\mu\nu}$  is necessary for the renormalization of the  short-distance effects produced by the quantum fluctuations of the scalar fields,  as these  involve  ${\cal O}(H^4)$  corrections.  However, as previously indicated  in \hyperref[Sect:ZPEFermions]{Sec.\,\ref{Sect:ZPEFermions}}, the renormalized EMT for fermions does not contain ${\cal O}(H^4)$ terms.  By using the expression of $\leftidx{^{(1)}}{\!H}_{00}$ in \hyperref[Appendix:Conventions]{Appendix\,\ref{Appendix:Conventions}} we can recognize the tensorial structure of the various terms, and from it we can pin down immediately the running of the couplings:
\begin{equation}\label{Eq:Fermions.rhoLsubtr}
\begin{split}
  \rho_\Lambda (M)-\rho_\Lambda (M_0)&=\frac{1}{128\pi^2}\left(-4N_{\rm f}+N_{\rm s}\right)\left(M^4-M_0^4\right)\\
  &+\frac{1}{32\pi^2}\left(4\sum_{\ell=1}^{N_{\rm f}}m_{\psi_\ell}^2 -\sum_{j=1}^{N_{\rm s}}m_{\phi_j}^2\right)\left(M^2-M_0^2\right)\\
    &+\frac{1}{64\pi^2}\left(-4\sum_{\ell=1}^{N_{\rm f}} m_{\psi_\ell}^4+\sum_{j=1}^{N_{\rm s}}m_{\phi_j}^4\right)\ln \frac{M^2}{M_0^2}\,,
\end{split}
\end{equation}
\begin{equation}\label{Eq:Fermions.Gsubtr}
\begin{split}
  \frac{1}{8\pi G (M)}-\frac{1}{8\pi G(M_0)}&=\frac{1}{48\pi^2}\left(-N_{\rm f}+3\sum_{j=1}^{N_{\rm s}}\left(\xi_j-\frac{1}{6}\right)\right)\left(M^2-M_0^2\right)\\
  &+\frac{1}{48\pi^2}\left(\sum_{\ell=1}^{N_{\rm f}}m_{\psi_\ell}^2 -3\sum_{j=1}^{N_{\rm s}}\left(\xi_j-\frac{1}{6}\right)m_{\phi_j}^2 \right)\ln \frac{M^2}{M_0^2}\,,
\end{split}
\end{equation}
\begin{equation}\label{Eq:Fermions.a1subtr}
\begin{split}
  a_1 (M)- a_1 (M_0)&=-\frac{1}{32\pi^2}\sum_{j=1}^{N_{\rm s}}\left(\xi_j-\frac{1}{6}\right)^2\ln\frac{M^2}{M_0^2}\,.
\end{split}
\end{equation}
From the above formulas we can now use  Eq.\,\eqref{Eq:Fermions.VEDMulticomponent} to find out the difference between the values of the VED at two different scales:
\begin{equation}\label{Eq:Fermions.rhovac}
\begin{split}
    \rho_{\rm vac} (M,H)-\rho_{\rm vac} (M_0,H_0)&=\frac{3}{16\pi^2}H^2\sum_{j=1}^{N_{\rm s}}\left(\xi_j-\frac{1}{6}\right)\left(M^2-m_{\phi_j}^2+m_{\phi_j}^2 \ln \frac{m_{\phi_j}^2}{M^2}\right)\\
    &-\frac{3}{16\pi^2}H_0^2\sum_{j=1}^{N_{\rm s}}\left(\xi_j-\frac{1}{6}\right)\left(M_0^2-m_{\phi_j}^2+m_{\phi_j}^2 \ln \frac{m_{\phi_j}^2}{M_0^2}\right)\\
    &+\frac{1}{16\pi^2}H^2\sum_{\ell=1}^{ N_{\rm f}}\left(-M^2+m_{\psi_\ell}^2-m_{\psi_\ell}^2 \ln \frac{m_{\psi_\ell}^2}{M^2}\right)\\
    &-\frac{1}{16\pi^2}H_0^2\sum_{\ell=1}^{ N_{\rm f}}\left(-M_0^2+m_{\psi_\ell}^2-m_{\psi_\ell}^2 \ln \frac{m_{\psi_\ell}^2}{M_0^2}\right)\\
    &+\frac{9}{16\pi^2}\left(2H\ddot{H}+6H^2\dot{H}-\dot{H}^2\right)\sum_{j=1}^{ N_{\rm s}}\left(\xi_j-\frac{1}{6}\right)^2\ln \frac{m_{\phi_j}^2}{M^2}\\
    &-\frac{9}{16\pi^2}\left(2H_0\ddot{H}_0+6H_0^2\dot{H}_0-\dot{H}_0^2\right)\sum_{j=1}^{ N_{\rm s}}\left(\xi_j-\frac{1}{6}\right)^2\ln \frac{m_{\phi_j}^2}{M_0^2}\\
    &+\frac{\sum_{j=1}^{ N_{\rm s}} \left\langle T_{00}^{\delta \phi_j}\right\rangle_{\rm ren}^{(6)} (M,H)+\sum_{\ell=1}^{ N_{\rm f}}\left\langle T_{00}^{ \psi_\ell}\right\rangle_{\rm ren}^{(6)} (M,H)}{a^2}\\
    &-\frac{\sum_{j=1}^{ N_{\rm s}} \left\langle T_{00}^{\delta \phi_j}\right\rangle_{\rm ren}^{(6)} (M_0,H_0)+\sum_{\ell=1}^{ N_{\rm f}}\left\langle T_{00}^{ \psi_\ell}\right\rangle_{\rm ren}^{(6)} (M_0,H_0)}{a^2}+\dots\\
\end{split}
\end{equation}
In the last line, the dots collectively represent all the terms of adiabatic 8 or beyond, which are not considered in our analysis. Notice that in the previous expression we have used the important relation \eqref{Eq:Fermions.rhoLsubtr}, which is essential to cancel the quartic mass contributions from the matter fields.

Following the same prescription that we used before to derive equations Eq.\,\eqref{Eq:QuantumVacuum.RVM2} and Eq.\,\eqref{Eq:QuantumVacuum.nueffAprox} for a single scalar field, we may implement now the scale settings $M=H$ and $M_0=H_0$ in order to compare the evolution of the VED between these two points, in the present case involving the full contributions from all the matter fields. For simplicity, let us call $\rv(H)\equiv\rho_{\rm vac} (H,H)$ and $\rv(H_0)\equiv\rho_{\rm vac} (H_0,H_0)$ when using the above expression \eqref{Eq:Fermions.rhovac}. The expansion history times $H$ and $H_0$ can be arbitrary, of course, but for obvious reasons we choose  $H_0=H(t_0)$ to be the value of the Hubble function at the present time, $t_0$, and $H=H(t)$ a value at a point in our past ($t<t_0$).  Therefore, the running of the VED between these two points can be expressed as follows:
\newpage
\begin{equation}\label{Eq:Fermions.DiffHHH0H0}
\begin{split}
    \rho_{\rm vac} (H)-\rho_{\rm vac} (H_0)&=\frac{3}{16\pi^2}H^2\sum_{j=1}^{N_{\rm s}}\left(\xi_j-\frac{1}{6}\right)\left(H^2-m_{\phi_j}^2+m_{\phi_j}^2 \ln \frac{m_{\phi_j}^2}{H^2}\right)\\
    &-\frac{3}{16\pi^2}H_0^2\sum_{j=1}^{N_{\rm s}}\left(\xi_j-\frac{1}{6}\right)\left(H_0^2-m_{\phi_j}^2+m_{\phi_j}^2 \ln \frac{m_{\phi_j}^2}{H_0^2}\right)\\
    &+\frac{1}{16\pi^2}H^2\sum_{\ell=1}^{ N_{\rm f}}\left(-H^2+m_{\psi_\ell}^2-m_{\psi_\ell}^2 \ln \frac{m_{\psi_\ell}^2}{H^2}\right)\\
    &-\frac{1}{16\pi^2}H_0^2\sum_{\ell=1}^{ N_{\rm f}}\left(-H_0^2+m_{\psi_\ell}^2-m_{\psi_\ell}^2 \ln \frac{m_{\psi_\ell}^2}{H_0^2}\right)\\
    &+\frac{9}{16\pi^2}\left(2H\ddot{H}+6H^2\dot{H}-\dot{H}^2\right)\sum_{j=1}^{ N_{\rm s}}\left(\xi_j-\frac{1}{6}\right)^2\ln \frac{m_{\phi_j}^2}{H^2}\\
    &-\frac{9}{16\pi^2}\left(2H_0\ddot{H}_0+6H_0^2\dot{H}_0-\dot{H}_0^2\right)\sum_{j=1}^{ N_{\rm s}}\left(\xi_j-\frac{1}{6}\right)^2\ln \frac{m_{\phi_j}^2}{H_0^2}\\
    &+\frac{\sum_{j=1}^{ N_{\rm s}} \left\langle T_{00}^{\delta \phi_j}\right\rangle_{\rm ren}^{(6)} (H,H)+\sum_{\ell=1}^{ N_{\rm f}}\left\langle T_{00}^{ \psi_\ell}\right\rangle_{\rm ren}^{(6)} (H,H)}{a^2}\\
    &-\frac{\sum_{j=1}^{ N_{\rm s}} \left\langle T_{00}^{\delta \phi_j}\right\rangle_{\rm ren}^{(6)} (H_0,H_0)+\sum_{\ell=1}^{ N_{\rm f}}\left\langle T_{00}^{ \psi_\ell}\right\rangle_{\rm ren}^{(6)} (H_0,H_0)}{a^2}\,.\\
\end{split}
\end{equation}
Obviously,  if the point $H$ is in the nearby past we can neglect all the ${\cal O}(H^4)$ terms generated in the above expression since they are much smaller than the ${\cal O}(H^2)$ contributions.  We will do this in the next section, where we  study in more detail the low energy regime, in particular the late time Universe where we live.  Let us however clarify that the ${\cal O}(H^2)$  terms are dominant not only for the late time Universe around our time, but in actual fact for the entire post-inflationary regime.
Finally,  we can extract the running of the gravitational constant from eq.\,\eqref{Eq:Fermions.Gsubtr}, with the following result:
\begin{equation}\label{Eq:Fermions.G(M)}
G(M)=\frac{G(M_0)}{1+\frac{G(M_0)}{2\pi}\left[ \left(\sum\limits_{j=1}^{N_{\rm s}}\left(\xi_j-\frac{1}{6}\right)-\frac{N_{\rm f}}{3}\right)(M^2-M_0^2)+\left(\sum\limits_{\ell=1}^{N_{\rm f}}\frac{m_{\psi_\ell}^2}{3} -\sum\limits_{j=1}^{N_{\rm s}}\left(\xi_j-\frac{1}{6}\right)m_{\phi_j}^2 \right) \ln \frac{M^2}{M_0^2}\right]} \,.
\end{equation}
We  discuss the running of $\rv$ and  $G$ in terms of $H$ in the next section.

\subsection{The low energy regime: evolution of $\rv$ and $G$  in the present Universe}\label{Sect:RVMpresentUniverse}

Of paramount importance is the evolution of the VED and of the gravitational coupling $G$  in the low energy regime, especially around our time.  Therefore, following our prescription, we evaluate  \eqref{Eq:Fermions.DiffHHH0H0} for the late Universe, when the dominant powers of $H$ are the $H^2$ ones. Such an expression  then boils down to
\begin{equation}\label{Eq:Fermions.LowEnergyRegime}
\rho_{\rm vac}(H)=\rho_{\rm vac}(H_0)+\frac{3\nu_{\rm eff}(H)}{8\pi}\mpl^2\left(H^2-H_0^2\right)+\mathcal{O}(H^4)\,,
\end{equation}
where the function $\nu_{\rm eff}(H)$ is defined as follows:
\begin{equation}\label{Eq:Fermions.nueff(H)}
\begin{split}
\nu_{\rm eff} (H)&=\frac{1}{2\pi}\Bigg[\sum_{j=1}^{N_{\rm s}}\left(\xi_j-\frac{1}{6}\right)\frac{m_{\phi_j}^2}{\mpl^2}\left(\ln\frac{m_{\phi_j}^2}{H_0^2}-1\right)-\frac{1}{3}\sum_{\ell=1}^{N_{\rm f}}\frac{m_{\psi_\ell}^2}{\mpl^2}\left(\ln\frac{m_{\psi_\ell}^2}{H_0^2}-1\right)\\
    &+\frac{H^2}{H^2-H_0^2}\ln\frac{H^2}{H_0^2}\left(\frac{1}{3}\sum_{\ell=1}^{N_{\rm f}}\frac{m_{\psi_\ell}^2}{\mpl^2 }-\sum_{j=1}^{N_{\rm s}}\left(\xi_j-\frac{1}{6}\right)\frac{m_{\phi_j}^2}{\mpl^2}\right)\Bigg]\,.
\end{split}
\end{equation}
It is indeed a function evolving with the Hubble rate,  but is almost constant since the dependence on $H$ is very mild, as we shall make manifest  in a moment.

Let us emphasize that the $\mathcal{O}(H^4)$ terms correcting the \textit{r.h.s.} of Eq.\,\eqref{Eq:Fermions.LowEnergyRegime} are completely irrelevant for the current Universe, and hence  they can be safely  ignored for the  FLRW regime, that is to say,  during the entire period  following the inflationary stage (cf.  next section). Therefore, equation \eqref{Eq:Fermions.LowEnergyRegime} should actually be relevant for the full cosmological evolution that is accessible (directly or indirectly) to our physical measurements and observations.

It is convenient to define the parameter
\begin{equation}\label{Eq:Fermions.epsilon}
 \epsilon \equiv \frac{1}{2\pi}\left(\sum_{j=1}^{N_{\rm s}}\left(\xi_j-\frac{1}{6}\right)\frac{m_{\phi_j}^2}{\mpl^2}-\frac{1}{3}\sum_{\ell=1}^{N_{\rm f}}\frac{m_{\psi_\ell}^2}{\mpl^2}\right)\,.
\end{equation}
This parameter is connected to the $\beta$-function \eqref{Eq:Fermions.Totalbetafunctions}  at low energies. Indeed, when we consider $M=H$ in the low energy regime, it is obvious that $H^2\ll m^2$ for any particle mass, and hence Eq.\,\eqref{Eq:Fermions.Totalbetafunctions} reduces to
 \begin{equation}\label{Eq:Fermions.epsiloncoeffbeta}
\beta_{\rv}=-\frac{3}{4\pi}\,\epsilon\,\mpl^2\,H^2\,,
  \end{equation}
within  a very good approximation. Quite obviously  we can see that  $\epsilon$ plays the role of coefficient of the low-energy $\beta$-function of the VED.  However, the eventual  coefficient that effectively  controls the final evolution of the VED is actually enhanced with respect to $\epsilon$ by a big logarithmic factor. To see this,  let us take the current limit ($H\to H_0$) of the function \eqref{Eq:Fermions.nueff(H)}:
\begin{equation}\label{Eq:Fermions.nueff0}
\nu_{\rm eff}^0 \equiv \lim\limits_{H\rightarrow H_0}\nu_{\rm eff}(H)=\frac{1}{2\pi}\left[\sum_{j=1}^{N_{\rm s}}\left(\xi_j-\frac{1}{6}\right)\frac{m_{\phi_j}^2}{\mpl^2}\left(\ln \frac{m_{\phi_j}^2}{H_0^2}-2\right)-\frac{1}{3}\sum_{\ell=1}^{N_{\rm f}}\frac{m_{\psi_\ell}^2}{\mpl^2}\left(\ln \frac{m_{\psi_\ell}^2}{H_0^2}-2\right)\right]\,.
\end{equation}
A simple rearrangement now shows that  we can rephrase \eqref{Eq:Fermions.nueff(H)} in terms of $\epsilon$ and $\nueff^0$:
\begin{equation}\label{Eq:Fermions.Approxnueff(H)}
      \nu_{\rm eff}(H)=\nu_{\rm eff}^0+ \left(1-\frac{H^2}{H^2-H_0^2}\ln \frac{H^2}{H_0^2}\right)\epsilon\,.
\end{equation}
This  formula is exact, but in practice some simplifications are perfectly possible.  For example, consider the big logarithms $\ln m_i^2/H_0^2$ (with $m_i$ any particle mass, boson or fermion)  involved in $\nueff$ but not in $\epsilon$.   For any known massive particle, we have  $\ln m_i^2/H_0^2\gg 1$, this being true even for the lightest neutrinos (recall that $H_0\sim 10^{-42}$ GeV$=10^{-30}$ meV). Typically $ \ln m_i^2/H_0^2={\cal O}(100)$ in all cases.  But as a matter of fact the only relevant contributions to $\nueff(H)$  come from the heavy massive particles  that belong to a GUT at a characteristic scale  $M_X\sim 10^{16}$GeV.  For these particles  (whether bosons or fermions, with masses $m_i\sim M_X$) we have   $m_i^2/\mpl^2\sim M_X^2/\mpl^2$ and this number is not so small since it may thrust the value of $\nueff$  up to  $\nueff\sim 10^{-3}$,  if one takes into account the large multiplicities of heavy fields existing in a typical GUT. This was first estimated  long ago in \cite{Sola:2007sv}.
Thus,  it is natural to expect $\left|\nu_{\rm eff}^0\right| \gg \left|\epsilon\right|$.It follows that we can safely neglect the term proportional to  $\epsilon$ in \eqref{Eq:Fermions.Approxnueff(H)}. It also means that we can neglect the very mild time-dependence of $\nu_{\rm eff}(H)$   and replace  it  with the constant coefficient $\nu_{\rm eff}^0$ in which $H$ is evaluated at the current time, $H=H_0$.   By the same token we can also ignore the numerical additive terms accompanying the big logarithms in \eqref{Eq:Fermions.nueff0}.  All in all, in very good approximation the evolution of the vacuum energy density can be described through  the formula
\begin{equation}\label{Eq:Fermions.VEDtotapprox}
\rho_{\rm vac}(H)= \rho_{\rm vac}(H_0)+\frac{3\nu_{\rm eff}}{8\pi}\mpl^2 (H^2-H_0^2)\,,
\end{equation}
with an effective $\nueff\simeq\nueff^0$ given by
\begin{equation}\label{Eq:Fermions.nueffapprox}
\nu_{\rm eff} =\frac{1}{2\pi}\left[\sum_{j=1}^{N_{\rm s}}\left(\xi_j-\frac{1}{6}\right)\frac{m_{\phi_j}^2}{\mpl^2}\,\ln \frac{m_{\phi_j}^2}{H_0^2}-\frac{1}{3}\sum_{\ell=1}^{N_{\rm f}}\frac{m_{\psi_\ell}^2}{\mpl^2}\,\ln \frac{m_{\psi_\ell}^2}{H_0^2}\right]\,.
\end{equation}
As it turns out, in practice  it all amounts to replace $\nueff(H)\to\nueff$ in \eqref{Eq:Fermions.LowEnergyRegime} since the result still retains a great degree of accuracy.
From the previous two equations, coefficient $\nueff$ is seen to play the role of $\beta$-function  for the running vacuum directly as a function of $H$.  If we compare \eqref{Eq:Fermions.nueffapprox} with \eqref{Eq:Fermions.epsilon} we can see that $\nueff$ and $\epsilon$ are roughly `proportional'  through a big log:
\begin{equation}\label{Eq:Fermions.relationepsilonnueff}
  \nueff\sim \epsilon \ln \frac{m_i^2}{H_0^2}\sim {\cal O}(100)\,\epsilon\,.
\end{equation}
Despite there being a summation over different masses, and hence such a proportionality not being strict,  the above relation is nevertheless true in order of magnitude.  The presence of the  big log factor in $\nueff$ makes the running of the VED faster than the tiny value of $\epsilon$ might convey at first sight.  On the other hand, as we shall see below, it is $\epsilon$ alone that controls the (much softer) running of the gravitational coupling $G$, which does not receive any enhancement from big log factors.

Equations \eqref{Eq:Fermions.VEDtotapprox} and \eqref{Eq:Fermions.nueffapprox} suffice to study the behavior of the VED near our time\footnote{It is apparent that for one single neutral scalar field and no fermion field the above expressions  we can reduce to the formulas for the scalar field presented earlier. For instance, compare \eqref{Eq:Fermions.nueffapprox} with \eqref{Eq:QuantumVacuum.nueffAprox}.}.
 These simplified  formulas have been previously used  de facto to fit the value of $\nueff$ from  the latest cosmological data, see e.g.\,\cite{SolaPeracaula:2021gxi} and references therein.  Here, however, we provide for the first time the full theoretical structure behind this parameter in the QFT context from the quantum effects induced by an arbitrary number of quantized matter fields.  The typical fitting value obtained in the mentioned reference is $\nueff\sim 10^{-3}$ and positive, which is well within the said expectations.  This phenomenological determination  picks up of course the net outcome from the various quantum matter fields involved in \eqref{Eq:Fermions.nueffapprox}, which at this point  cannot be discriminated in an individual way.
Finally, insofar as  the running gravitational constant is concerned, it can be written using the same renormalization scale as follows:
\begin{equation}\label{Eq:Fermions.GHExactllfields}
G(H)=\frac{G_N}{1+\frac{1}{2\pi}\left(\sum\limits_{j=1}^{N_{\rm s}}\left(\xi_j-\frac{1}{6}\right)-\frac{N_{\rm f}}{3}\right)\frac{H^2-H_0^2}{\mpl^2}-\epsilon\ln \frac{H^2}{H_0^2}}\,.
\end{equation}

{The former expression can be derived straightforwardly from eq.\,\eqref{Eq:Fermions.G(M)} by setting $M=H$ and $M_0=H_0$, with $H_0$ being the current value of the Hubble function, and we have defined  $G_N\equiv G(M_0)$  (the current value of the gravitational coupling). We follow  exactly the same recipe as for the VED.
In the low energy regime, where $H^2\ll \mpl^2$, we can approximate with high accuracy the former expression by just
\begin{equation}\label{Eq:Fermions.GHAllfields}
G(H)=\frac{G_N}{1-\epsilon\ln \frac{H^2}{H_0^2}}\,,
\end{equation}
where $\epsilon$ receives contributions from all the matter fields, see Eq.\,\eqref{Eq:Fermions.epsilon}. Recall from \eqref{Eq:Fermions.relationepsilonnueff} that $|\epsilon|\ll|\nueff|$, and also that  the running of  $G(H)$ is logarithmic, in contrast to the running of $\rv(H)$ which is quadratic in $H^2$ at low energies.  Therefore, the running of $G$ is much lesser than that of the VED. It may, however, be interesting to note that when $H$ approaches $\mpl$ the term $\sim H^2/\mpl^2$ in the denominator of the more accurate formula Eq.\,\eqref{Eq:Fermions.GHExactllfields} could be dominant over the logarithmic one.  If the multiplicity of matter  fields is large enough, such a term could make the gravitational coupling to evolve asymptotically free at very large energies when we approach the Planck scale.  Until that point,  $G$ increases at high energies for $\epsilon>0$. Beyond that point, decreases.

As an additional cross-check, we can see that the running of the vacuum energy density and of the  gravitational coupling  are compatible through the Bianchi identity. This can be translated into a local energy exchange between the vacuum fluid and the background gravitational field due to the quantum fluctuations. A deeper insight on the local (covariant) energy conservation and the Bianchi identity can be found in the previous works \cite{Moreno-Pulido:2022phq,Moreno-Pulido:2022upl}, where the reader may find a detailed derivation of the logarithmic evolution law, that is to say, equation\,\eqref{Eq:Fermions.GHAllfields}, in the simpler scenario of one real scalar field.
In fact, one finds that the $\beta$-function \eqref{Eq:Fermions.RGEVED1} for the VED running is crucially involved also in the local conservation law of the VED, which can be written in two alternative ways:
 \begin{equation}\label{Eq:Fermions.NonConserVED2}
\dot{\rho}_{\rm vac}+3H\left(\rv+P_{\rm vac}\right)=\frac{\dot{M}}{M}\,\beta_{\rv}=- \frac{\dot{G}}{G}\,\frac{3H^2}{8\pi G}\,.
\end{equation}
The first equality expresses the fact that the non-conservation of the VED is due to both the running of $\rv$ with $M$  (i.e. the fact that $\beta_{\rv}\neq 0$)  and also to the cosmic time dependence of $M$ (viz. $\dot{M}\neq 0$), whereas the second equality is a direct reflex of the Bianchi identity in Einstein's equations with variable $\rv$ and $G$, and hence provides a link between the time variation of the VED and that of the gravitational coupling $G$.
 The former equation does not depend on the number or nature of the fields involved, and holds as long as the  matter components are covariantly conserved on their own. The non-conservation of $\rv$, however, preserves the Bianchi identity thanks to the corresponding running of the gravitational coupling.  This does not preclude, however, that one can still formulate scenarios where matter can exchange energy with $\rv$, but we do not address this situation here.
 Taking the leading terms of  $\beta_{\rv}$ from the \textit{r.h.s.} of \eqref{Eq:Fermions.Totalbetafunctions} for the present Universe, and setting $M=H$,  we immediately obtain the following differential equation
\begin{equation}\label{Eq:Fermions.MixedConservationApprox3}
\frac{1}{G^2}\frac{dG}{dt}=\frac{2\epsilon}{G_N} \frac{\dot{H}}{H}\,,
\end{equation}
where $\epsilon$ is the full expression \eqref{Eq:Fermions.epsilon} involving the contributions from all the matter fields.
Its solution is precisely Eq.\,\eqref{Eq:Fermions.GHAllfields}, as can be readily checked.

It is also interesting to note from the above formulas that this framework predicts a (mild) cosmic time variation of the ``fundamental constants'', such as the gravitational coupling $G$ and $\rv$, as a function of  $H(t)$. Whence, it  implies a small evolution of these `constants' with the cosmological expansion. The possibility for such a variation is not new. It has long been  discussed in the literature\,\cite{Uzan:2010pm} and is still a hot matter of debate and of intensive test by different groups, see also \cite{Barontini:2021mvu} and the ample bibliography provided in them. Specific theoretical models accounting for such a possible variation are manifold, and in some cases they imply a time-dependence of the running couplings and masses in the particle and nuclear physics world, see e.g. \cite{Calmet:2001nu,Calmet:2017czo}. While most of the proposals are based on strict particle physics scenarios, in particular on GUT's, testing the evolution of the VED in curved spacetime  is a novel feature suggested in our framework, which was actually put forward on more phenomenological grounds sometime ago  in \cite{Fritzsch:2012qc,Fritzsch:2015lua,Fritzsch:2016ewd}. The QFT calculations presented in the current work provide indeed a solid theoretical support to these same ideas but from first principles. Recall the golden rule in this arena: when one ``fundamental constant'' varies, then all of them vary!

The formulas discussed above concern important epochs of the cosmological expansion such as the radiation-dominated epoch, matter-dominated epoch and the current epoch in which the vacuum energy resurfaces and became finally dominant over matter. During the entire FLRW regime the dominant power of the Hubble rate in the VED is $H^2$ or $\dot{H}$ (which are of the same adiabatic order). The terms with powers of $H$ (or of equal adiabatic order)  higher than $H^2$ (indicated by $\mathcal{O}(H^4)$ in \eqref{Eq:Fermions.LowEnergyRegime}) acquire real relevance much early on in the expansion history  since only during the most primitive stages of the Universe we encounter  a truly  high energy scenario. In the next section we extend the study of \hyperref[SubSect:RVMInflation]{chapter\,\ref{SubSect:RVMInflation}} by including the presence of several scalar and fermionic fields. 

\subsection{Inflation from running vacuum}\label{Sect:RVMInfFermi} 

It was noted in \hyperref[SubSect:RVMInflation]{Sect.\,\ref{SubSect:RVMInflation}} (see also \cite{Moreno-Pulido:2022phq}) that the quantum effects computed from the adiabatic expansion lead to higher powers of the Hubble rate and its derivatives, which are irrelevant for the current Universe but capable to bring about inflation in the very early Universe. They are characterized by a short period where $H$=const., provided this constant value is, of course, very large, namely around a characteristic GUT scale. The regime $H$=const. in our case  is totally  unrelated to the ground state  of a scalar field potential and therefore this new mechanism does not require  any \textit{ad hoc} inflaton field.  As said, it  is called  `RVM-inflation'.  Here we consider the contribution from the fermions fields and provide a formula for the dominant term of the energy density receiving contributions from an arbitrary number of non-minimally coupled scalar fields and also an arbitrary number of fermions fields.  The payoff from the latter stems from setting $H=$const in Eq.\,\eqref{Eq:Fermions.EMT00},  where we can see that all the time derivatives of the Hubble rate vanish except for a single term which is proportional to $H^6/m_\psi^2$.   The contribution from a non-minimally coupled scalar field was computed in \cite{Moreno-Pulido:2022phq} and here we just combine it with that of fermions assuming any number of both species.  Overall, we find that the total VED involving the contributions from bosons and fermions  at very high energies (hence relevant for triggering RVM-inflation in the very early Universe) can be put in the following fashion:
\begin{equation}\label{Eq:Fermions.RVMinflationCombined}
 \begin{split}
\rv^{\rm inf}=&C_{\rm inf}H^6+F(\dot{H}, \ddot{H},\vardot{3}{H}...),
\end{split}
\end{equation}
where
\begin{equation}\label{Eq:Fermions.Cinf}
C_{\rm inf}\equiv\frac{1}{80\pi^2}\left\{\sum_ j^{N_{\rm s}}\frac{1}{m_{\phi_j}^2}\left[\left(\xi_j-\frac16\right)-\frac{2}{63}-360\left(\xi_j-\frac16\right)^3\right] - \frac{31}{252}\sum_\ell^{N_{\rm f}} \frac{1}{m_{\psi_\ell}^2}\right\}\,.
\end{equation}
The terms collected in the function $ F(\dot{H}, \ddot{H},\vardot{3}{H}...)$ depend on different combinations of powers of $H$ involving derivatives of $H$ in all cases, and hence they all vanish for $H=$const. Thus   $F=0$ for $H=$const.  Overall we see that the dominant contribution is of the form $\rv^{\rm inf}\propto H^6$ with a complicated coefficient $C_{\rm inf}$ which depends on the number of scalar and fermions fields, their masses, multiplicities and also on the non-minimal couplings of the different scalars. In the case of fermions this contribution is seen to be negative-definite, whereas in the case of the scalars it can be positive.

 Let us note that during the inflationary period the EoS of the quantum vacuum is essentially $-1$, with very tiny deviations caused by terms which depend on the various time derivatives of the Hubble rate. To the extent that the condition $H$=const. is fulfilled these deviations are extremely small, see Eq.\,\eqref{Eq:Fermions.PressureEnergyVacuum}.   In the next section we shall see that, in contrast,  the EoS of the quantum vacuum in the present time can deviate from $-1$ by a small amount which is not as negligible as in the very early Universe and therefore could be detected and even mimic quintessence behavior.
 
The solution of the cosmological equations proceeds along the same lines as in \cite{Moreno-Pulido:2022phq}, except that now the fermionic contribution is also included but it only modifies the specific coefficient of $H^6$. Therefore, one finds again that a short period of inflation can be generated with $H\approx$ const. and subsequently the vacuum decays quickly into radiation\,\cite{Moreno-Pulido:2022phq}: 
\begin{equation}\label{Eq:Fermions.SolHaInflation}
H(a)=\frac{H_I}{\left(1+\hat{a}^8\right)},
\end{equation}
\begin{equation}\label{Eq:Fermions.rhodensities}
\rho_r(\hat{a})=\rI\,{\hat{a}^8}\left(1+\hat{a}^8\right)^{-\frac{3}{2}}\,\,, \ \ \ \ \ \ \ \ \
\rv(\hat{a})={\rI} \left(1+\hat{a}^{8}\right)^{-\frac{3}{2}}\,,
\end{equation}
 in which $\hat{a}\equiv a/\astar$,  with  $\astar$ is the transition point from the regime of vacuum dominance into one of radiation dominance, which can be estimated to be around $\astar\sim 10^{-30}$  (see Eq.\,\eqref{Eq:Fermions.astar} below). Moreover,  $H_I$ and $\rho_I$ are the value of $H$ and $\rho_{\rm vac}$, respectively, at the beginning of inflation, with $\rho_I= C_{\rm inf}H_I^6$. Applying the Friedmann equation, we find
\begin{equation}\label{Eq:Fermions.HIdef}
    H_I=\left(\frac{3}{8\pi G_I C_{\rm inf}}\right)^{1/4}\,,
\end{equation} 
\begin{equation}\label{Eq:Fermions.rhoIdef}
  \rho_I=\frac{3}{8\pi G_I}H_I^2=\frac{3}{8\pi G_I}\left(\frac{3}{8\pi G_I C_{\rm inf}}\right)^{1/2}=C_{\rm inf}^{-1/2}\left(\frac{3}{8\pi G_I }\right)^{3/2}\,,
\end{equation}
where $G_I\equiv G(H_I)$ is the gravitational coupling at $H=H_I$, the latter being value of the Hubble parameter at the inflationary era.  Needless to say, the difference between $G_I$ and the usual $G_N$ is not very important here since the running of $G$ is logarithmic, and hence the effect is very small as compared to the fast evolution of the $H^6$ term, so in practice we can neglect the running of $G$ for these considerations.   To trigger inflation in an effective way, we must have a positive coefficient $C_{\rm inf}>0$. In the light of Eq.\,\eqref{Eq:Fermions.Cinf} we can see that this is perfectly possible since the couplings $\xi_j$ and masses of the fields can take a variety of values that make this possible, as can be shown in a devoted study that will be presented elsewhere. 
 
The masses of the relevant fields involved must be very large,  say around a typical GUT scale, $M_X\sim 10^{16}$ GeV. This may not be obvious at first sight. A naive interpretation of the higher order terms of the VED, which are related with the $6th$ adiabatic order of the ZPE (see equation \eqref{Eq:Fermions.EMT00}), may give the erroneous impression that the relevant masses are to be the lightest possible ones, but this is by no means true since in such a situation the adiabatic expansion would break down. On the other hand, the analysis through the Friedmann equations reveals the correct dependency of the VED and of the Hubble function on the masses during the inflationary regime. From equation \eqref{Eq:Fermions.Cinf} it is obvious that $C_{\rm inf}^{-1/2}\propto m_{\phi,\psi}$, where the notation stands for a linear combination of the typical masses of the matter fields. The inflationary parameters \eqref{Eq:Fermions.HIdef}-\eqref{Eq:Fermions.rhoIdef}, therefore, depend on a positive power of $m_{\phi,\psi}$, and as a result the process of RVM-inflation is actually dominated by the heaviest masses, in contrast to naive expectations; namely, masses $m_{\phi,\psi}\sim M_X\sim 10^{16}$GeV of order of a typical GUT, as mentioned above. It follows that the same heavy masses which may generate a mild (but non-negligible) quadratic running $\sim H^2$ of the VED (with a coefficient $\nu_{\rm eff}\sim 10^{-3}$) in the late Universe  can also be responsible for driving fast inflation in the primeval stages of the cosmological evolution.  To see this feature more explicitly, let us recall that the differential equation driving the Hubble function in the presence of a high power $H^6$ in the VED\,\eqref{Eq:Fermions.RVMinflationCombined} reads \cite{Lima:2013dmf,Perico:2013mna,SolaPeracaula:2019kfm}
\begin{equation}\label{Eq:Fermions.Hdot2}
\dot H + \frac{3}{2} \, (1 + \omega_m) \, H^2 \, \Big( 1  - \frac{H^4}{H_I^4} \Big) =0\,,
\end{equation}
where $\omega_m=1/3$ is the EoS of matter in the relativistic epoch, and $H_I$ is given in \eqref{Eq:Fermions.HIdef}. We have neglected the influence of the term $H^2$ and also of the constant term in the very early Universe. It is obvious from \eqref{Eq:Fermions.Hdot2} that there is a constant solution $H=H_I$ to that equation, which is precisely the one which triggers the inflationary period.  From this observation one can then solve equation \eqref{Eq:Fermions.Hdot2} exactly to find Eq.\,\eqref{Eq:Fermions.SolHaInflation}. The latter shows clearly the departure of $H$ from $H_I$ when $\hat{a}>1$ (i.e. $a>a_*$). The inflationary phase actually occurs during the short period when the departure remains small, namely when $H$ remains approximately constant, $H\simeq H_I$. During such period the $F$-term on the \textit{r.h.s.} of Eq.\,\eqref{Eq:Fermions.RVMinflationCombined} just vanishes, $F(\dot{H}, \ddot{H},\vardot{3}{H}...)=0$, since all dependence on $H$ is through time derivatives.
From equations \eqref{Eq:Fermions.HIdef} and \eqref{Eq:Fermions.rhoIdef}, we find that the order of magnitude of the physical scales involved in RVM-inflation is the following:
\begin{equation}
 H_I\sim\left(M_X\,\mpl\right)^{1/2}\sim 10^{17}\,{\rm GeV}\,,\ \ \ \ \ \ \ \  \rho_I\sim M_X\,\mpl^3\sim\left(10^{18} {\rm GeV}\right)^4\,,
\end{equation} 
up to numerical coefficients and  multiplicity factors, of course. Thus,  if the masses of the relevant matter fields lie in the expected range for a GUT, the right order of magnitude for the relevant physical parameters at the inflationary epoch can be obtained.  Keeping in mind  that the mechanism of RVM-inflation can also be motivated in `stringy' scenarios\,\cite{Mavromatos:2020kzj,Mavromatos:2021urx,Mavromatos:2021sew,Basilakos:2020qmu,Basilakos:2019acj,Mavromatos:2022gtr},  it should be natural to expect RVM-infation in the range between the GUT scale and the Planck scale.  This is exactly what the above estimates suggest in order of magnitude.

One more observation is in order. One can easily check from \eqref{Eq:Fermions.rhodensities} that for $\hat{a}\gg 1$ (i.e.  $a\gg\astar$) we retrieve the standard decaying behavior of  radiation, $\rho_r(a)\sim a^{-4}$. This condition enforces the following relation between $\rho_r^0$ (the current value of radiation energy density) and $\rho_I$ (the energy density at the inflationary time):
\begin{equation}\label{Eq:Fermions.rhoI}
     \rho_{\rm I}\approx\rho_{\rm r}^0 \astar^{-4}\,.
\end{equation} 
Following the line of the previous estimations, it yields an equality point between vacuum energy and radiation around the value
\begin{equation}\label{Eq:Fermions.astar}
\astar=\left(\Omega_r ^0\,\frac{\rho_c^0}{\rho_I}\right)^{\frac{1}{4}}\simeq \left(10^{-4}\,\frac{10^{-47}}{10^{72}}\right)^{1/4}\sim 10^{-30}\,,
\end{equation}
where $\rho_c^0\sim 10^{-47}$ GeV$^4$ is the current critical density. In the meantime the vacuum energy becomes negligible and does not disturb primordial BBN, see  \cite{Moreno-Pulido:2022phq,Moreno-Pulido:2022upl}  for more details.  See also \cite{Lima:2013dmf,Perico:2013mna,Sola:2015rra,SolaPeracaula:2019kfm}   for interesting phenomenological applications prior to the QFT treatment of RVM-inflation,  first presented in\cite{Moreno-Pulido:2022phq}.

\subsection{Equation of state of the quantum vacuum}\label{Sect:EoS-QVacuum}

The quantum effects of the fields have an imprint on the vacuum equation of state, which is not exactly the traditional one $P_{\rm vac}=-\rho_{\rm vac}$.  From the expressions of the renormalized energy density and pressure of the vacuum that have been obtained in the previous sections and considering their generalization to an arbitrary number of fermions and scalars, we arrive at the following expression for the EoS of the quantum vacuum:
\begin{equation}\label{Eq:Fermions.EoScombined}
\begin{split}
    \omega_{\rm vac}(H)   &=-1+\frac{1}{8\pi^2\rho_{\rm vac}(H)}\Bigg\{\Bigg[\frac{1}{3}\sum_{\ell=1}^{N_{\rm f}} \left(H^2-m_{\psi_\ell}^2+m_{\psi_\ell}^2\ln \frac{m_{\psi_\ell}^2}{H^2}\right)\\
        &\phantom{xxxxxxxxxxxxxx}+\sum_{j=1}^{N_{\rm s}}\left(\xi_j-\frac{1}{6}\right)\left(m_{\phi_j}^2-H^2-m_{\phi_j}^2 \ln\frac{m_{\phi_j}^2}{H^2}\right)\Bigg]\dot{H}\\
        &\phantom{xxxxxxxxxxxxxx}-3\left[\sum_{j=1}^{N_{\rm s}}\left(\xi_j-\frac{1}{6}\right)^2\ln \frac{m_{\phi_j}^2}{H^2}\right]\left(6\dot{H}^2+3H\ddot{H}+\vardot{3}{H}\right)\Bigg\}+\mathcal{O}\left(H^6\right)\,.
    \end{split}
\end{equation}
Here  $\mathcal{O}\left(H^6\right)$ stands for the terms of adiabatic order $6$ or higher,  such as  $H^6/m^2$, $\ddot{H}^2/m^2,\dots$ ($m=m_{\psi_j}, m_{\phi_\ell}$ ).
All these higher order terms can be neglected during the postinflationary era. 
The above  relation shows that the quantum vacuum EoS is of dynamical nature. As we can see there is a deviation from the rigid value $-1$, which is the traditional EoS ascribed to the cosmological constant in the $\Lambda$CDM framework. The correction terms due to both bosonic and fermionic fields are small in the present era, in comparison with the constant term $-1$, but need not be negligible since the particle masses involved can be from a typical GUT, and hence one can estimate that the effective parameter  $\nu_{\rm eff}$  -- the parameter defined in Eq.\,\eqref{Eq:Fermions.nueffapprox} --could reach up to $ 10^{-3}$\cite{Sola:2007sv}.  Furthermore, if we focus only on the ${\cal O}(\dot{H})\sim {\cal O}(H^2) $ terms relevant for the current Universe and the radiation epoch we may also neglect the higher order adiabatic terms of ${\cal O}(H^4)$  in the last lines of Eq.\,\eqref{Eq:Fermions.EoScombined}.  Following the steps of \cite{Moreno-Pulido:2022upl}, the EoS can finally be written in a rather compact form as a function of the cosmological redshift:
\begin{equation} \label{Eq:Fermions.ApproximateEos}
\wv= -1+\frac{\left[\nu_{\rm eff}+\epsilon\left(1-\ln E^2 (z) \right)\right] \left[\Omega_{\rm m}^0 (1+z)^3+\frac{4\Omega_{\rm r}^0}{3}(1+z)^4\right]}{\Omega_{\rm vac}^0+\nu_{\rm eff}\left(-1+E^2 (z)\right)-\epsilon\left(1-E^2(z)+E^2(z) \ln E^2(z) \right)}+{\cal O}\left(\nueff^2\right)\,,
\end{equation}
where $\nu_{\rm eff}$  contains the combined effects from fermions and bosons, see Eq.\,\eqref{Eq:Fermions.nueff0}, and we have defined the normalized Hubble rate with respect to the present time ($H_0$):
\begin{equation}\label{Eq:Fermions.E2}
  E^2 (z)\equiv\frac{H(z)}{H_0}= \Omega_{\rm vac}^0+\Omega_{\rm m}^0 \left(1+z\right)^3+\Omega_{\rm r}^0 \left(1+z\right)^4\,.
\end{equation}
Here  $\Omega_{\rm vac}^0=\rho_{\rm vac}^0/\rho_{\rm c}^0 \approx 0.7$, $\Omega_{\rm m}^0=\rho_{\rm m}^0/\rho_{\rm c}^0 \approx 0.3$ and $\Omega_{\rm r}^0=\rho_{\rm r}^0/\rho_{\rm c}^0 \approx 10^{-4}$ are the current fractions of vacuum energy, dust-like matter and radiation, respectively.   The EoS  formula may be further simplified if we neglect the effect of  the small coefficient $\epsilon$  in Eq.\,\eqref{Eq:Fermions.Approxnueff(H)}.  This is justified since $\big|\nu_{\rm eff}\big| \gg \big| \epsilon \big|$ owing to the logarithmic extra terms $\ln m^2/H_0^2$ contained in $\nu_{\rm eff}^0$, which can typically be of $\mathcal{O}(100)$, see \eqref{Eq:Fermions.relationepsilonnueff}. Thus, to within a very good approximation, we can write
\begin{equation}\label{Eq:Fermions.ApproximateEos2}
\wv \simeq -1+\nu_{\rm eff} \,\frac{\Omega_{\rm m}^0 (1+z)^3+\frac{4\Omega_{\rm r}^0}{3}(1+z)^4}{\Omega_{\rm vac}^0+\nu_{\rm eff} \left(-1+E^2 (z)\right)}\,.
\end{equation}
Notice that the term proportional to $\nueff$ in the denominator cannot be neglected at large $z$ since it becomes dominant. In this case, the EoS takes on the form
\begin{equation}\label{Eq:Fermions.EosMDE}
\wv \approx -1+ \,\frac{\Omega_{\rm m}^0 (1+z)^3+\frac{4\Omega_{\rm r}^0}{3}(1+z)^4}{E^2 (z)}\ \ \  \ (z\gg 1)\,,
\end{equation}
where $\nueff$ has cancelled.
For example, for $z$ large enough but within the matter-dominated epoch the dominant term in equation \eqref{Eq:Fermions.E2} is the $\sim (1+z)^3$ one, and we can see from \eqref{Eq:Fermions.EosMDE} that the vacuum EoS then mimics matter since $\wv\simeq 0$. Similarly, at much larger values of $z$ already in the radiation-dominated epoch, where $\sim (1+z)^4$ is the dominant term, then $\wv\simeq 1/3$ and the vacuum imitates radiation. Such a `chameleonic' behavior of the quantum vacuum was first noticed in \cite{Moreno-Pulido:2022upl}, which is the base of the former chapter, and in fact the formula for the vacuum EoS that we have found here is a generalization for an arbitrary number of fermion and boson fields of the expression previously found in the previous chapter, see Eq. \eqref{Eq:eos.EoSChameleon}. Last but not least, the evolution of the vacuum EoS in the late Universe is no less remarkable and striking. From \eqref{Eq:Fermions.ApproximateEos2} we find  
\begin{equation}\label{Eq:Fermions.EoSDeviation}
\wv(z) \simeq   -1+\nueff \frac{\Omega_{\rm m}^0}{\Omega_{\rm vac}^0}(1+z)^3\ \ \ \ \ \ \ \ \ \  \ \  \ (z\lesssim 5)\,.
\end{equation}
In this approximation we recover once more the form \eqref{Eq:eos.EoSDeviation} of \hyperref[SubSect:QuantumVacuumEoSFLRW]{Sect.\,\ref{SubSect:QuantumVacuumEoSFLRW}}, but in this case with $\nueff$ involving the contributions from all the quantized matter fields.
Taking into account that the last fits of the RVM to the overall cosmological data favor a positive value of $\nueff>0$ and of order $10^{-3}$\,\cite{SolaPeracaula:2021gxi} (see also the previous phenomenological studies on the RVM reported in recent years\,\cite{Gomez-Valent:2014rxa,Sola:2015wwa,Sola:2016jky,SolaPeracaula:2016qlq,Sola:2017znb,SolaPeracaula:2017esw,Gomez-Valent:2018nib,Gomez-Valent:2017idt}), we learn that the deviation of the EoS from $-1$ in the present Universe is not completely negligible. The vacuum appears disguised as quintessence since $\wv\gtrsim -1$.

To summarize, the running vacuum mimics the EoS of the dominant component at a given time of the cosmic evolution, and at present is dynamical. This was first confirmed for scalar fields in \cite{Moreno-Pulido:2022upl}, see the former chapter.  In this chapter,  we have found  that the formal structure of the EoS does not change when we evaluate the fermionic contribution.  Therefore, we find that  $\wv$ is  $-1$ during inflation; stays  close to 1/3 in the radiation dominated epoch, and then close to 0 in the matter dominated epoch.  Furthermore, at present it mimics quintessence  since $\nueff$ is found to be positive in the phenomenological tests of the RVM and hence $\wv$  is slightly above $-1$. Therefore, the quantum EoS deviates from the traditional  value $\wv=-1$ of the classic vacuum.  The remarkable consequence is that we can have at present an effective quintessence behavior of the quantum vacuum without need of invoking {\rm ad hoc} scalar fields.  Finally, the vacuum  EoS  in the remote future will be as that in the very early (inflationary)  Universe, i.e. $\wv\to -1$, since the cosmic expansion asymptotes towards a new inflationary period.

\section{Discussion of the chapter}\label{Sect:FermionsDiscussion}

In this chapter, we have evaluated the contributions to the vacuum energy density (VED) from the quantized matter fields in a semiclassical gravity approach. By using the same regularization technique of Quantum Field Theory (QFT) in curved spacetime of \hyperref[Chap:QuantumVacuum]{chapter\,\ref{Chap:QuantumVacuum}} and \hyperref[Chap:EoSVacuum]{chapter\,\ref{Chap:EoSVacuum}} and making use of a specific (off-shell) subtraction prescription, we have been able to calculate the mode functions and the renormalized zero-point energy (ZPE)  from spin-1/2 quantum fields in a FLRW background up to sixth adiabatic order. 

Together with the contribution from the $\rL$ term in the Einstein-Hilbert action, we have obtained the properly renormalized VED. Since the corresponding computation for scalar fields had already been accounted for in the aforementioned works, we have put forward here the combined contribution to the VED  from an arbitrary number of quantized matter fields. We did not consider interactions among them, however, as the free field calculation in curved spacetime is already rather cumbersome in itself. One interesting difference between the expression of the ZPE of these two types of fields is that in the fermionic case the terms of fourth adiabatic order (viz. those involving  four time derivatives) are not present.  The final result  is that the overall VED of the quantized matter fields upon adiabatic renormalization appears to be a soft dynamical quantity with the cosmological evolution. This is a most remarkable outcome of the present study.  More specifically, the VED shows up in the form of an expansion in powers of the Hubble rate $H$ and its time derivatives, all these powers being of even adiabatic order, a property which is fully consistent (and expected) from the general covariance of the theory.  Such a series expansion appears to take the canonical form of the running vacuum model (RVM),  see \cite{Sola:2013gha,SolaPeracaula:2022hpd} and references therein.  This means, in particular, that the leading quantum effects obtained for the late Universe are of second adiabatic order, thus  $\sim H^2$ and $\sim\dot{H}$. Obviously, this may have consequences for the present Universe, and these consequences have been tested in previous phenomenological works. What is more, these quantum effects turn out to have a positive impact on a possible solution to the $\CC$CDM tensions, see e.g.\cite{SolaPeracaula:2021gxi,Gomez-Valent:2014rxa,Sola:2015wwa,Sola:2016jky,SolaPeracaula:2016qlq,Sola:2017znb,SolaPeracaula:2017esw,Gomez-Valent:2018nib,Gomez-Valent:2017idt}\footnote{In the upcoming study\,\cite{CosmoTeam2023Universe},  fully updated information  will be provided on the phenomenological performance of the RVM and its implications on the current $\CC$CDM tensions.}

In this chapter,  we have also discussed some of the theoretical difficulties in trying to renormalize the cosmological term, $\CC$, and its relation with the VED. To start with, it should be emphasized that these are two different concepts that can only be properly related in non-flat spacetime. If $\CC$ is taken to be the physically measured value of the cosmological term at present, $\CC_{\rm phys}$, then its relation with the current VED is $\rvo=\CC_{\rm phys}/(8\pi G_N)$. However, at a formal QFT level these quantities have to be derived from a gravitational action in curved spacetime and a lot more of care needs to be exercised. Leaving for the moment quantum gravity considerations for a better  future (viz. for when, hopefully, the quantum treatment of the gravitational field becomes possible),  the more pedestrian renormalization of $\rv$ in QFT in curved spacetime proves to be already quite helpful at present\cite{Moreno-Pulido:2020anb, Moreno-Pulido:2022phq}. It shows, for example -- and the present works attest once more for this fact --  that the renormalized VED in the FLRW background is a mild dynamical quantity evolving with the cosmic expansion, and hence that $\rvo$ is just its value at present. There is in fact no such thing as a rigid, everlasting,  cosmological constant in the context of  QFT in the FLRW background. In general, $\rv=\rv(H)$ is a function of the Hubble rate and its time derivatives.  What we call the `cosmological constant'  $\Lambda$  appears in our framework as the nearly sustained value of the renormalized quantity  $8\pi G(H)\rho_{\rm vac}(H)$  around (any)  given epoch $H$, where $G(H)$ is the renormalized gravitational coupling, which is also running, although very mildly (logarithmically) with $H$, i.e. $G=G(\ln H)$.  At present, $G(H_0)=G_N$ and $\rv(H_0)=\rvo$, and this defines $\CC_{\rm phys}=8\pi G_N\rvo$ in a precise way  in QFT in  FLRW spacetime (within our renormalization framework).

The longstanding and widespread confusion in the literature about the notion of cosmological constant, $\CC$, and that of  vacuum energy (density), $\rv$, has prevented from achieving a proper treatment of the renormalization of these quantities in cosmological spacetime.  In particular, the attempts to relate these concepts in the context of flat spacetime calculations  are meaningless and their repeated iteration has been highly  counterproducing\,\cite{SolaPeracaula:2022hpd}.

In the simplified scenario considered here, where only interactions with the gravitational background are allowed, the VED is the sum of two contributions, a parameter in the effective action, $\rho_\Lambda$, and the ZPE of the quantized fields. After renormalization, the VED depends on a  scale $M$, and the setting $M=H$ at the end of the calculation allows us to compare the VED at different epochs of the cosmic history, in a manner similar to the standard association made of the renormalization point with a characteristic energy scale of a given process in ordinary gauge theories. Thus the difference between the VED values  at any two points of the cosmological expansion, say $H(t_1)$ and $H(t_2)$,  provides a smooth running of the VED. Remarkably, such an evolution turns out to be free from the undesirable $\sim m^4$ contributions that emerge from the quantized matter fields in other frameworks. As a result there is no fine tuning involved in the evolution of the VED in the present calculation.
The VED, in fact, adopts the standard form of the RVM, which  in the late Universe reads $\rho_{\rm vac}(H_2)= \rho_{\rm vac} (H_1)+3\nu_{\rm eff}/(8\pi) \mpl^2 (H_2^2-H_1^2)$, where $H_1$ and $H_2$ can be, for example, the current value, $H_0$, and another value $H$ near our past. Finally, $\nu_{\rm eff}$ is a small parameter related to the $\beta$-function of the renormalization group running of the VED, whose value has been explicitly computed in this dissertation from the fluctuations of the quantized matter fields. Depending on the sign of $\nu_{\rm eff}$, the VED can mimic  quintessence or phantom-like behavior.

Much earlier in the cosmic history, the higher powers of $H$ (larger than $H^2$ and of even adiabatic  order to preserve covariance) took their turn  and could be relevant for the inflationary regime, in the sense that they had the capacity to trigger inflation through a mechanism that has been called `RVM-inflation'\cite{Moreno-Pulido:2020anb, Moreno-Pulido:2022phq,Moreno-Pulido:2022upl}. While the scalar field contribution to this inflationary mechanism had been computed in the previous references,  in this chapter we have accounted for the spin-1/2 fermionic contribution and combined the two types of effects for an arbitrary matter content. In both cases (scalar and fermion fields)  the sixth order adiabatic  terms ${\cal O}(H^6)$  had to be computed.
Finally, the renormalized vacuum fluid's pressure, $\Pv$, has been determined  using the same QFT techniques as for $\rv$. Equipped with these nontrivial results the equation of state (EoS) of the quantum vacuum can be computed from first principles.  The entire contribution from quantized matter fields (bosons and fermions) can be encoded in the effective $\nueff$ parameter. We find that the EoS function $\wv=\Pv/\rv$  deviates  from the traditional result  -1, a fact which is  worth emphasizing. This is true in most of the cosmological history, especially after inflation (which is the only period in our past where the vacuum EoS stayed very close to $-1$). It is no less noteworthy, as previously mentioned, that in the late Universe, and most particularly near our time, the vacuum EoS behaves as quintessence for $\nu_{\rm eff}>0$, the latter being the sign preferred by the existing phenomenological fits to the overall cosmological data -- see \cite{SolaPeracaula:2021gxi}, for example. For higher and higher redshifts during the FLRW regime, the vacuum EoS mimics the equation of state of the dominant matter component (relativistic or non-relativistic) at the corresponding epoch. Such a peculiar behavior of the running vacuum energy density  was referred to as ``chameleonic'' in\,\cite{Moreno-Pulido:2022upl}. The tracking of the EoS of matter by the vacuum ceases to exist in the late Universe, where the DE epoch breaks through and  $\wv$ behaves as effective quintessence, the reason being that the EoS is then in the process  to asymptote towards $-1$ in the remote future, exactly as it was in the primeval inflationary time. In fact, the inflationary process in the late Universe is eventually resumed, but very slowly.

Overall, by combining the results from an arbitrary number of  quantized matter fields we find that  the main cosmic running of $\rv$ depends on the  quadratic terms in the boson and fermion masses times the square of the Hubble function, i.e. $\sim m_\psi^2 H^2$ and $\sim m_\phi^2 H^2$. These effects are obviously much softer than the naively expected (hard) contributions $\sim  m_\psi^4$ and  $\sim m_\phi^4$.  As remarked, the soft terms have been profusely tested in phenomenological works on the RVM existing in the literature, see e.g.\,\cite{SolaPeracaula:2021gxi,Gomez-Valent:2014rxa,Sola:2015wwa,Sola:2016jky,SolaPeracaula:2016qlq,Sola:2017znb,SolaPeracaula:2017esw} --  and the upcoming \cite{CosmoTeam2023}. The QFT effects that we have computed here and in previous chapters, provide a solid theoretical underpinning of those phenomenological analyses. They even bring to light new relevant features, such as the dynamical character  of the EoS of the quantum vacuum, which is unprecedented in the literature to the best of our knowledge. In particular, they suggest that if in future cosmological observations we can collect clear signs that the EoS of the dark energy deviates from $-1$, such a feature could be explained by the running vacuum. It therefore opens the possibility that such observations (if confirmed)  may be accounted from  fundamental properties of QFT attributable to the fluctuations of the quantized matter fields in curved spacetime rather than to the existence of \textit{ad hoc} quintessence fields and the like.  This could be an extremely  interesting smoking gun of this approach. The EoS dynamics is prompted here by the virtual quantum effects produced by quantized fermion and boson fields, the same kind of effects which trigger a smooth evolution of the vacuum energy density in cosmological spacetime. As it turns out from the above considerations, in the RVM framework  the need for fundamental quintessence fields and also for inflaton fields subsides dramatically, for they can both be replaced by the running effect of the quantum vacuum.

On pure cosmological/observational grounds, the physical outcome of the theoretical framework presented here can be summarized in the following way.  The renormalization of the vacuum energy density in QFT in the FLRW background  leads to the  non-constancy of the `cosmological constant', $\CC$, in Einstein's equations  and predicts a slow time variation of the  vacuum energy density and of its  equation of state at the present time, which departs slightly from the traditional EoS value $-1$ for the vacuum.   This conclusion emerges from explicit QFT calculations in our approach and may point to a possible explanation for a wide range of cosmological problems that have been dealt with phenomenologically in the past in terms of \textit{ad hoc} quintessence or phantom fields. In our context, the currently measured cosmological `constant' is neither mimicked nor supplanted by any ersatz entity from the already too crammed black box of the dark energy.   We could simply phrase it in a nutshell as follows: the quantum vacuum shows up here as if it were a form of dynamical dark energy,  but it is (quantum) vacuum after all.  In fact, it is the same vacuum producing inflation in the early Universe by means of higher (even) powers of the Hubble rate $H^n$ beyond $n=2$,  and still leaving  a smoothly evolving cosmological term in the late Universe through the lowest possible power  compatible with general covariance, which is   $H^2$.  Formally, all these powers of the Hubble rate emerge as quantum effects  that are part of the effective action of vacuum. Today's physical cosmological term  appears  as  a quantity directly connected with  the  (properly renormalized)  vacuum energy density in QFT in curved spacetime:  $\Lambda_{\rm phys}=8\pi G\rv$.   As a running parameter,  it  is sensitive to the fluctuations  of the quantized matter fields.  Taking into account that in our framework  the scale  of renormalization is linked  with the cosmological expansion, represented by the Hubble rate, it turns out that $\Lambda_{\rm phys}$,  despite it appearing  as an approximately rigid term during  a typical  cosmic span around any given epoch, it is actually a physical observable in evolution during the entire cosmic history. 

Hence, the computations done here reaffirms the position at which we arrived at the end of \hyperref[Sect:EoS-QVacuum]{Sect.\,\ref{Sect:EoS-QVacuum}}. We claim that these effects are not disturbed by the presence of several fields, in this sense, they are very robust since do not rely in the specific kind of field. This could be an extremely interesting signature of this approach, which does not depend at all on \textit{ad hoc} quintessence fields to display a dynamical evolution at the current time. The EoS dynamics is prompted here by the virtual quantum effects produced by quantized fermion and boson fields, the same kind of effects which trigger a smooth and very mild evolution of the vacuum energy density in cosmological spacetime. Its time  variation,  which is ultimately of quantum origin,  is very small at present but it helps  in relieving the current tensions of the $\CC$CDM and  it might eventually provide an explanation for the cosmic acceleration observed in our Universe from first principles.

\blankpage

\chapter{Brans–Dicke cosmology with a \texorpdfstring{$\CC$}{L}-term: a possible solution to \texorpdfstring{$\CC$CDM}{LCDM} tensions}\label{Chap:PhenomenologyofBD}

The canonical picture of our Universe, formulated in the context of the General Relativity (GR) paradigm, is to assume that the cosmic acceleration is caused by a rigid cosmological constant (CC) term, $\CC$, in Einstein's equations, whose value has been pinned down  from a large set of cosmological observations, which by themselves also point to the existence of large amounts of dark matter (DM).   We call such an overall picture of the Universe the ``concordance (or standard) cosmological model'' \cite{peebles1993principles}, or simply GR-$\CC$CDM, where we append GR explicitly in the name because in this chapter we wish to study the $\CC$CDM model also from a different perspective to that of the GR paradigm. In particular, as a difference from the previous chapter, we wish to stick firmly to the $\CC$-term as the simplest provisional explanation for the cosmic acceleration. But we want to do it in the context of Brans Dicke theory of gravity, a fundamental change of the conceptual construct on gravitation which consists in a modification of General Relativity. It was conceived in the sixties by C. H. Brans and R. H. Dicke (``BD'' for short) \cite{brans1961mach, brans1962mach, dicke1962physical}. The main departure from GR was that $G$ was boldly assumed to be a dynamical variable rather than a constant of Nature. It actually traces back to early ideas in the thirties on the possibility of a time-evolving gravitational constant $G$ by Milne\,\cite{mccrea1935} and the suggestion by Dirac of the large number hypothesis \cite{dirac1937cosmological , dirac1938cosmological}, which led him also to propose the time evolution of $G$. Along similar lines, Jordan and Fierz speculated that the fine structure constant $\alpha_{\rm em}$ together with $G$ could be both space and time dependent\,\cite{jordan1937naturwiss, jordan1939ZPhys, jordan1955schwerkraft,Fierz:1956zz}. Finally, $G$ was formally associated to the existence of a dynamical scalar field $\psi\sim 1/G$ coupled to the curvature.  BD theory was the first historical attempt to extend GR in order to accommodate variations in the Newtonian coupling $G$. Subsequently these ideas were generalized in the form of scalar-tensor theories\,\cite{Bergmann:1968ve, Nordtvedt:1970uv , Wagoner:1970vr}, and thereafter further extended in multifarious ways  still compatible with the weak form of the Equivalence Principle\,\cite{Horndeski:1974wa,Fujii:2003pa}, see e.g. \cite{Sotiriou:2008rp,Capozziello:2011et,Clifton:2011jh,Will:2014kxa} for a review. We may call the  to the cosmological framework still encompassing all the phenomenological ingredients of the concordance $\CC$CDM  model, in particular a strict cosmological constant term $\CC$ and dark matter along with baryons, but now all of them ruled over by a different gravitational paradigm: BD-gravity, instead of GR. 

As it could not be in another way, we wish to explore the cosmological tensions (see \hyperref[Sect:BeyondSM]{Sect.\,\ref{Sect:BeyondSM}}) on $H_0$ and $\sigma_8$ in the context of the BD-$\CC$CDM model. The simultaneous solution/alleviation of the two tensions is not to easy to achieve since any new physics introduced to explain the severe $H_0$-tension should not aggravate the $\sigma_8$ one in a noticeable way. This ``golden rule'' to ameliorate the acute $H_0$-tension will be our guiding principle. We will see that  the BD-$\CC$CDM naturally implements the aforesaid golden rule for safely quenching the two tensions with a minimal number of extra parameters with respect the concodance model within GR.

Many alternative Dark Energy (DE) proposal have been addressed in the literature to improve the overall fit to the cosmological data and possibly to relax the main tensions, but most of them are still based on Einstein's GR. This may represent a fundamental limitation to the possible solution of the mentioned tensions, even more so if we take into account that these problems can be affected by assumptions on the basic parameters of gravitation. Recent analyses comparing different models of dark energy using similar data can be found e.g. in\,\cite{SolaPeracaula:2018wwm,Gomez-Valent:2020mqn}, and references therein. It is remarkable that the BD-$\CC$CDM can be rewritten as an effective GR framework (with $G$ and $\CC$ remain both constant), in which the combinations of the dynamical character of $G$ in the context of BD-gravity with a cosmological term $\CC$ can be seen as an effective dynamical dark energy (DDE) capable to overcome the mentioned tensions. This DDE possesses a time-evolving component of a very specific nature, as a constant term plus a small dynamical term  $\sim\nu H^2$ ($|\nu|\ll1$). This form is well-known in the literature and goes under the name of Running Vacuum Model (RVM) similarly to the description of vacuum energy obtained in \hyperref[Chap:QuantumVacuum]{chapter\,\ref{Chap:QuantumVacuum}} -- see also \cite{Sola:2013gha, Sola:2015rra, GomezValent:2017kzh} and references therein.

A first hint that Brans-Dicke gravity could lead to such a promising (BD-like) version of the RVM, was put forward in \cite{SolaPeracaula:2018dsw}. Further elaboration on this idea and a first comparison with data was subsequently given in  \cite{deCruzPerez:2018cjx}, and finally a more sophisticated study using a full Boltzmann code for the CMB part was presented  in\,\cite{SolaPeracaula:2019zsl,SolaPeracaula:2020vpg}. These two papers are the basis of this chapter, particularly the second one. We would also like to remark recent studies on testing  BD theories and on attempts at mitigating the tensions using particular potentials in the BD framework, see e.g. \cite{Umilta:2015cta,Ballardini:2016cvy,Rossi:2019lgt}. At the same time, there are different works trying to loosen the  GR-$\CC$CDM tensions by considering the possibility of a variable Newton's constant $G$. Usually these models are essentially GR-like, in the sense that {the basic term of the gravitational action still has the form of the Hilbert-Einstein term, plus a non-minimal coupling of curvature with a scalar field.}  This type of models and variations thereabout are  valuable and have been recently used to try to mitigate the $H_0$ or the $\sigma_8$ tensions\,\cite{Ballesteros:2020sik,Braglia:2020iik,Ballardini:2020iws,Bertini:2019xws,Rodrigues:2015hba}, but more difficult is to try to fulfill the mentioned golden rule -- which requires not to aggravate one of the two tensions when improving the other. In point of fact, in some cases the $\sigma_8$ one actually gets significantly worse. We further comment on these models in our section of discussion.

Here we present a comprehensive study of BD-$\CC$CDM cosmology vis-\`a-vis observations\,\cite{SolaPeracaula:2020vpg} using a significant amount of new and updated data, in particular we make use of the full Planck 2018 likelihood as well as of additional datasets (e.g. on Strong-Lensing and updated structure formation data) which prove quite revealing. Furthermore, we discuss here at length many analytical and numerical details of the calculation and the possible implications that the BD-$\CC$CDM model can have on the current cosmological observations, most particularly on the $\sigma_8$ and $H_0$ tensions. If a successful phenomenological description of the data and the loosening of the tensions could be reconfirmed with the advent of new and more precise data and new analyses in the future, it would be tantalizing to suggest the possibility that the underlying fundamental theory of gravity might actually be BD rather than GR. But there is still a long way to follow, of course.

The layout of this chapter reads as follows. In \hyperref[Sect:BDgravity]{Sect.\,\ref{Sect:BDgravity}} we introduce the basics of the Brans-Dicke (BD) model. In particular, we discuss the introduction of the cosmological term in this theory and the notion of effective gravitational coupling. \hyperref[Sect:Preview]{Section\,\ref{Sect:Preview}} is a preview of the following sections, and may help the reader to get a road map of the basic results of the chapter and above all an explanation of why BD-cosmology with a CC can be a natural and efficient solution to the $H_0$ and $\sigma_8$ tensions. \hyperref[Sect:EffectiveEoS]{Section\,\ref{Sect:EffectiveEoS}} shows how to parametrize BD-cosmology as a departure from GR-cosmology. We show that BD with a CC appears as GR with an effective equation of state (EoS) for the dark energy which behaves quintessence-like at a high confidence level. The remaining of the chapter presents the technical details and the numerical analysis, as well as complementary discussions. Thus, \hyperref[Sect:StructureFormationBD]{Section\,\ref{Sect:StructureFormationBD}} discusses the perturbations equations for BD-gravity in the linear regime (leaving for \hyperref[Appendix:CosmoPerturbationsSynchronous]{Appendix\,\ref{Appendix:CosmoPerturbationsSynchronous}} and \hyperref[Appendix:PerturbationTheoryNewtonian]{Appendix\,\ref{Appendix:PerturbationTheoryNewtonian}} an extended discussion with more technical details in different gauges); \hyperref[Sect:Mach]{Section\,\ref{Sect:Mach}} defines four scenarios for the BD-cosmology in the light of  Mach's Principle; \hyperref[Sect:MethodData]{Sect.\,\ref{Sect:MethodData}} carefully describes the data used from the various cosmological sources; \hyperref[Sect:NumericalAnalysis]{Section\,\ref{Sect:NumericalAnalysis}} presents the numerical analysis and results. In \hyperref[Sect:ConsiderationsBD]{Section\,\ref{Sect:ConsiderationsBD}} we perform a detailed discussion of the obtained results and we include a variety of extended considerations, in particular we assess the impact of massive neutrinos in the BD-$\CC$CDM framework. Finally, in \hyperref[Sect:DiscussionBD]{Sect.\,\ref{Sect:DiscussionBD}} we summarize the contents of the chapter. Four appendices at the end provide additional complementary material. In \hyperref[Appendix:Semi-Analytical]{Appendix\,\ref{Appendix:Semi-Analytical}} we compute semi-analytical solutions to the BD equations in different epochs, which are helpful to further understand the numerical results. We also recall the reader at this point why BD-cosmology mimics the Running Vacuum Model (RVM); in \hyperref[Appendix:FixedPoints]{Appendix\,\ref{Appendix:FixedPoints}} we compute the fixed points of the BD-cosmology with a cosmological constant. The aforementioned \hyperref[Appendix:CosmoPerturbationsSynchronous]{Appendix\,\ref{Appendix:CosmoPerturbationsSynchronous}} and \hyperref[Appendix:PerturbationTheoryNewtonian]{Appendix\,\ref{Appendix:PerturbationTheoryNewtonian}} provide the perturbations equations in the synchronous and conformal Newton gauges, respectively, and illustrate the correspondence between the two.

\section{BD-$\CC$CDM: Brans-Dicke gravity with a cosmological constant}\label{Sect:BDgravity}

Since the appearance of GR, more than one hundred years ago, many alternative theories of gravity have arisen, see e.g. \cite{Sotiriou:2008rp,Capozziello:2011et,Clifton:2011jh,Will:2014kxa} and references therein. The most important one, however, was proposed by Brans and Dicke almost sixty years ago\,\cite{brans1961mach}. This theoretical framework  contains an additional gravitational  \textit{d.o.f.}  as compared to GR, and as a consequence it is different from GR in a fundamental way, see the previously cited reviews. The new \textit{d.o.f.}  is a scalar field, $\psi$, coupled to the Ricci scalar of curvature, $R$.  BD-gravity is indeed the first historical attempt to extended GR to accommodate variations in the Newtonian coupling $G$. A generalization of it has led to a wide panoply of scalar-tensor theories since long ago\,\cite{Bergmann:1968ve, Nordtvedt:1970uv, Wagoner:1970vr, Horndeski:1974wa}.  The theory is also characterized by  a   (dimensionless) constant parameter, $\oD$, in front of the kinetic term of $\psi$.

\subsection{Action and field equations}\label{SubSect:ActionBD}

In our study we will consider the original BD-action extended with a cosmological constant density term, $\rL$, as it is essential to mimic the conventional $\CC$CDM model based on GR and reproduce its main successes. In this way we obtain what we have called the `BD-$\CC$CDM model' in the introduction of the chapter, {\it i.e.} the version of the  $\CC$CDM within the BD paradigm.
The BD action reads, in the Jordan frame, as follows\,\footnote{Our conventions may be found in \hyperref[Appendix:Conventions]{Appendix\,\ref{Appendix:Conventions}}}:
\begin{eqnarray}\label{Eq:BD.BDaction}
S_{\rm BD}=\int d^{4}x\sqrt{-g}\left[\frac{1}{16\pi}\left(R\psi-\frac{\oD}{\psi}g^{\mu\nu}\partial_{\nu}\psi\partial_{\mu}\psi\right)-\rL\right]+\int d^{4}x\sqrt{-g}\,{\cal L}_{\rm m}(\chi_i,g_{\mu\nu})\,. 
\end{eqnarray}
The (dimensionless) factor  in front of the kinetic term of $\psi$, {\it i.e.} $\oD$, will be referred to as the BD-parameter. While it is true that this parameter is not restricted to be a constant, throughout this chapter we will consider just the canonical option  $\oD=$const.
The last term of \eqref{Eq:BD.BDaction} stands for the matter action $S_{\rm m}$, which is constructed from the Lagrangian density of the matter fields, collectively denoted as $\chi_i$. There is no potential for the BD-field $\psi$ in the original BD-theory, but we admit the presence of a CC term associated to $\rL$. By not introducing any specific potential we keep the number of additional parameters to the minimum.

The Brans-Dicke field, $\psi$, has dimension $2$ in natural units ({\it i.e.} mass dimension squared), in contrast to the dimension $1$ of ordinary scalar fields. This is because we wish the effective value of $G$ at any time to be given directly by $1/\psi$. It goes without saying that  $\psi$ must be a field variable evolving very slowly with time.

The field equations of motion ensue after performing variation of the action \eqref{Eq:BD.BDaction}  with respect to both the metric and the scalar field $\psi$.  While the  first variation yields
\begin{equation}\label{Eq:BD.BDFieldEquation1}
\psi\,G_{\mu\nu}+\left(\Box\psi +\frac{\oD}{2\psi}\left(\nabla\psi\right)^2\right)\,g_{\mu\nu}-\nabla_{\mu}\nabla_{\nu}\psi-\frac{\oD}{\psi}\nabla_{\mu}\psi\nabla_{\nu}\psi=8\pi\left(\,T_{\mu\nu}-g_{\mu\nu}\rL\right)\,,
\end{equation}
the second variation gives the wave equation for $\psi$, which  depends on the curvature scalar $R$:
\begin{equation}\label{Eq:BD.BDFieldEquation2a}
\Box\psi-\frac{1}{2\psi}\left(\nabla\psi\right)^2+\frac{\psi}{2\oD}\,R=0\,.
\end{equation}
To simplify the notation, we have written  $(\nabla\psi)^2\equiv g^{\mu\nu}\nabla_{\mu}\psi\nabla_{\nu}\psi$.
In the first field equation, $G_{\mu\nu}=R_{\mu\nu}-(1/2)Rg_{\mu\nu}$ is the Einstein tensor, and on its RHS $T_{\mu \nu}=-(2/\sqrt{-g})\delta S_{\rm m}/\delta g^{\mu\nu}$ is the energy-momentum tensor of matter.
We can take the trace of Eq.\eqref{Eq:BD.BDFieldEquation1} to eliminate $R$ from \eqref{Eq:BD.BDFieldEquation2a}, what leads to a most compact result for the wave equation of $\psi$:
\begin{equation}\label{Eq:BD.BDFieldEquation2}
\Box\psi=\frac{8\pi}{2\oD+3}\,\left(T-4\rL\right)\,,
\end{equation}
%
\begin{equation} \label{Eq:BD.EMT}
\tilde{T}_{\mu\nu} =T_{\mu\nu}  -\rho_\CC g_{\mu\nu}=p\,g_{\mu\nu}+(\rho + p)u_\mu{u_\nu}\,,
\end{equation}
{with $\rho \equiv \rho_{\rm m} + \rho_\gamma+\rho_\nu + \rho_\Lambda$ and $p \equiv p_{\rm m} + p_\gamma + p_\nu + p_\Lambda$. The matter part $\rho_{\rm m} \equiv \rho_{\rm b} + \rho_{cdm}$, contains the pressureless contribution from baryons and cold dark matter. Photons are of course relativistic, so $p_\gamma=\rho_\gamma/3$. The functions $\rho_\nu$ and $p_\nu$ include the effect of massive and massless neutrinos, and therefore must be computed numerically.}

As in GR, we have included a constant vacuum energy density, $\rL$, in the BD-action \eqref{Eq:BD.BDaction}, with the usual equation of state $p_\Lambda=-\rL$. The quantum matter fields usually induce an additional, and very large, contribution to $\rL$. This is of course the origin of the Cosmological Constant Problem\,\cite{weinberg1989cosmological,Sola:2013gha,sahni2000case,peebles2003cosmological,padmanabhan2003cosmological,copeland2006dynamics,Amendola:2015ksp}\footnote{A recent proposal to alleviate the CCP within BD-gravity was made in \cite{SolaPeracaula:2018dsw}. Interestingly, the Higgs potential itself can be motivated in BD-gravity theories\,\cite{Sola:2016our,SolaPeracaula:2018dsw}.}.

Let us write down the field equations in the flat FLRW metric,  $ds^2=-dt^2 + a^2\delta_{ij}dx^idx^j$. Using the total density $\rho$ and pressure $p$ as indicated above, Eq.\,\eqref{Eq:BD.BDFieldEquation1} renders the two independent equations
\begin{equation}
3H^2 + 3H\fracdpsipsi -\frac{\wBD}{2}\left(\fracdpsipsi\right)^2 = \frac{8\pi}{\psi}\rho\label{Eq:BD.Friedmannequation}
\end{equation}
and
\begin{equation}
2\dot{H} + 3H^2 + \frac{\ddpsi}{\psi} + 2H\frac{\dpsi}{\psi} + \frac{\wBD}{2}\left(\frac{\dpsi}{\psi}\right)^2 = -\frac{8\pi}{\psi}p\,,\label{Eq:BD.pressureequation}
\end{equation}
whereas \eqref{Eq:BD.BDFieldEquation2} yields
\begin{equation}\label{Eq:BD.FieldeqPsi}
\ddpsi +3H\dpsi = \frac{8\pi}{2\wBD +3}(\rho - 3p)\,.
\end{equation}
Here dots stand for derivatives with respect to the cosmic time and $H=\dot{a}/a$ is the Hubble rate. For constant $\psi=1/G_N$, the first two 
equations reduce to the Friedmann and pressure equations of GR, and the third requires $\oD\to\infty$ for consistency (except in the pure radiation-dominated epoch, where $\rho - 3p=0$) .  The connection between GR and  the $\oD\to\infty$ limit is sometimes not as straightforward as one might  naively think\,\cite{Faraoni:1998yq,Faraoni:1999yp}. We will have due occasion in this chapter to appraise the significance of this important observation (cf. de BD scenarios described in \hyperref[Sect:Mach]{Sect.\,\ref{Sect:Mach}}).

By combining the above equations we expect to find a local covariant conservation law, similar to GR. This is because there is no interaction between matter and the BD-field. Although the details are more involved than in GR, the result can be obtained  upon  straightforward calculation of the covariant derivative on both sides of Eq.\,\eqref{Eq:BD.BDFieldEquation1} and using the Bianchi identity satisfied by $G_{\mu\nu}$ and the field equation of motion for $\psi$. The final result turns out to be the same:
\begin{equation}\label{Eq:BD.FullConservationLaw}
\dot{\rho} + 3H(\rho + p)=\sum_{N} \left[\dot{\rho}_N + 3H(\rho_N + p_N)\right] = 0\,,
\end{equation}
where the sum is over all components, {\it i.e.} baryons, dark matter, neutrinos, photons and vacuum.

Here we take the point of view that all of the matter components are separately conserved in the main periods of the cosmic evolution.  In particular, the vacuum component $\rL$ obviously does not contribute in the sum since it is assumed to be constant and $\rho_\CC + p_\CC=0$ .

Hereafter,  for convenience,  we will  use a dimensionless BD-field, $\varphi$, and the inverse of the BD-parameter, according to the following definitions:
\begin{equation}\label{Eq:BD.definitions}
\varphi(t) \equiv G_N\psi(t)\,,\qquad  \qquad\epsilon_{\rm BD} \equiv \frac{1}{\omega_{\rm BD}}\,.
\end{equation}
In this expression, $G_N = 1/\mPl^2$, with $\mPl$ the Planck mass as defined previously; $G_N$ gives the local value of the gravitational coupling, e.g. obtained from Cavendish-like  (torsion balance) experiments. Note that a nonvanishing value of $\eBD$ entails a deviation from GR. Being $\varphi(t)$  a dimensionless quantity,  we can recover GR by enforcing the simultaneous limits   $\epsilon_{\rm BD} \rightarrow 0$ \textit{and} $\varphi\to 1 $.  We emphasize that it is not enough to set $\epsilon_{\rm BD} \rightarrow 0$.  In this partial limit, we can only insure that $\varphi$ (and $\psi$, of course) does not evolve, but it does not fix its constant value. As we will see later on,  this feature can be important in our analysis. Using \eqref{Eq:BD.definitions}, we can see that \eqref{Eq:BD.BDaction} can be rewritten
\begin{equation}
\begin{split}
S_{\rm BD}&=\int d^{4}x\sqrt{-g}\left[\frac{1}{16\pi G_N}\left(R\varphi-\frac{\oD}{\varphi}g^{\mu\nu}\partial_{\nu}\varphi\partial_{\mu}\varphi - 2\Lambda\right)\right]\\
&+\int d^{4}x\sqrt{-g}\,{\cal L}_{\rm m}(\chi_i,g_{\mu\nu})\,, \label{Eq:BD.BDactionvarphi}
\end{split}
\end{equation}
where $\CC$ is the cosmological constant, which is related with the associated vacuum energy density as $\rL=\CC/(8\pi G_N)$.

\subsection{Cosmological constant and vacuum energy in BD theory}\label{SubSect:CosmoConstantBDTheory}

There are several ways to introduce the cosmological constant in the BD framework, so  a few comments are in order at this point, see e.g. the comprehensive exposition\,\cite{Fujii:2003pa} and the works\,\cite{mathiazhagan1984inflationary, La1989, Weinberg:1989mp, barrow1990extended} in the context of inflation. We can sum up the situation by mentioning  essentially three ways.
In one of them, the BD-action \eqref{Eq:BD.BDaction} is obtained upon promoting $G_N$ in the Einstein-Hilbert (EH) action with the $\rL$ term into a dynamical scalar field $1/\psi$,  adding the corresponding kinetic term and  keeping $\rho_\Lambda=const.$ The CC term $\Lambda$ is then related with the vacuum energy density through $\rho_\Lambda=\Lambda/(8\pi G_N )$. The EoS of the vacuum fluid is defined as $p_\Lambda=-\rho_\Lambda$. With these definitions, which we adopt throughout this chapter, the CC is not directly coupled to the BD-field. The latter, therefore, has a trivial (constant) potential, and the late-time acceleration source behaves as in the GR-$\Lambda$CDM model.

Alternatively, one could also adopt the EH action with CC and promote $G_N$ to a dynamical scalar, adding the corresponding kinetic term as before, but now keeping $\Lambda=const.$ instead of $\rho_\Lambda=const$. In this case, $\Lambda$ is linearly coupled to $\psi$. The potential energy density for the scalar field takes the form $\rho_\Lambda(\psi)= \psi \Lambda/(8\pi)$, so it is time-evolving. The coupling between the cosmological constant and the BD-field modifies the equations of motion. Thus, it may also alter the physics with respect to the option \eqref{Eq:BD.BDaction}, at least when the dynamics of the scalar field is not negligible \cite{Nariai:1969vh,endo1977cosmological, Uehara1982, lorenz1984exact, barrow1990extended, romero1992brans, Tretyakova:2011ch}.

A third possibility, of course, is to consider more general potentials, but they do not have a direct interpretation of a  CC term as in the  original GR action with a cosmological term, see e.g.\,\cite{Esposito-Farese:2000pbo, Alsing:2011er, Ozer:2017oik, Faraoni:2004pi}.  Let us briefly explain why. If one starts from a general potential for the BD-field in the action, say some  arbitrary function of the BD-field $U(\psi)$ (not carrying along any additive constant)  in place of the constant term $\rL$ in \eqref{Eq:BD.BDaction}, then it is not so straightforward to  generate a CC term and still remain in a pure BD framework. For if one assumes that $\psi$ develops a vacuum expectation value\,\footnote{We may assume, for the sake of the argument, that it is a classical ground state, since the BD-field is supposed to be part of the external gravitational field. Recall that we do not assume gravity  being a quantized theory here but just a set of background fields, in this case composed of the metric components $g_{\mu\nu}$ and $\psi$ (or equivalently $\varphi$).}, then the theory (when written in terms of the  fluctuating field around the ground state) would split into a non-minimal term coupled to curvature and a conventional EH term, so it would not be a pure BD theory.

This point of view is perfectly possible and has been considered by other authors, on which we shall comment in some detail in\,\hyperref[Sect:ConsiderationsBD]{Sect.\,\ref{Sect:ConsiderationsBD}}. However, proceeding in this way would lead us astray from our scientific leitmotif in this chapter (which is, of course, to remain fully within the BD paradigm).  Yet there is another option which preserves our BD philosophy, which is to perform a conformal transformation to the Einstein frame, where one can define a strict CC term.  Then we could impose that the effective potential in that frame, $V$,  is a constant. However, once this is done, the original potential in the (Jordan) BD frame, $U$, would no longer be constant since it would be proportional to $V$ times $\varphi^2$, with $\varphi$ defined as in \eqref{Eq:BD.definitions}.  Conversely, one may assume $U=\rL=const$. in the Jordan frame (as we actually did) and then the vacuum energy density will appear mildly time-evolving in the Einstein frame (for sufficiently large $\wBD$, of course).  This last option is actually  another way to understand why the vacuum energy density can be perceived as a slightly time-evolving quantity when the BD theory is viewed from the GR standpoint. It is also the reason behind the fact that the BD-$\CC$CDM framework mimics the so-called running vacuum model (RVM), see \hyperref[Sect:RVMconnection]{Sect.\,\ref{Sect:RVMconnection}} and references therein for details.

As indicated, in this chapter we opt for considering the definition provided in \eqref{Eq:BD.BDaction}, as we wish  to preserve the exact canonical form of the late-time acceleration source that is employed in the GR-$\Lambda$CDM model. At the same time we exploit the connection of the BD framework with the RVM and its well-known successful phenomenological properties, see e.g.\,\cite{SolaPeracaula:2016qlq, SolaPeracaula:2017esw, Gomez-Valent:2018nib, Gomez-Valent:2017idt, Sola:2017znb, Sola:2016jky, Sola:2016hnq, Sola:2015wwa}.  In addition, in \hyperref[Sect:EffectiveEoS]{Sect.\,\ref{Sect:EffectiveEoS}} we consider a direct parametrization of the departures of BD-$\Lambda$CDM from  GR-$\Lambda$CDM.

\subsection{Effective gravitational strength}\label{SubSect:EffGravStrength}
 
From \eqref{Eq:BD.BDactionvarphi} it follows that the quantity
\begin{equation}\label{Eq:BD.Gvarphi}
G(\varphi)=\frac{G_N}{\varphi}
\end{equation}
constitutes the effective gravitational coupling at the level of the BD-action. We will argue that $G(\varphi) $ is larger than $G_N$ because $\varphi<1$ (as it will follow from our analysis).  The gravitational field between two tests masses, however, is \textit{not} yet $G(\varphi) $ but the quantity  $G_{\rm eff}(\varphi)$  computed below.

{Let us remark that if one would like to rewrite the BD action in terms of a canonically normalized scalar field $\phi$ (of dimension $1$) having  a non-minimal coupling to curvature of the form  $\frac12\xi\phi^2 R$, it would suffice to redefine the BD-field as  $\psi=8\pi \xi\,\phi^2$   with $\xi=1/(4\oD)=\eBD/4$, and then the scalar part of the action takes on the usual form}
\begin{eqnarray}
S_{\rm BD}=\int d^{4}x\sqrt{-g}\left(\frac12\,\xi\phi^2 R-\frac12 g^{\mu\nu}\partial_{\nu}\phi\partial_{\mu}\phi-\rL\right)+\int d^{4}x\sqrt{-g}\,{\cal L}_{\rm m}(\chi_i,g_{\mu\nu})\,. \label{Eq:BD.BDaction2}
\end{eqnarray}
This alternative expression allows us to immediately connect with the usual parametrized post-Newtonian (PN) parameters, which restrict the deviation of the scalar-tensor theories of gravity with respect to GR\,\cite{Boisseau:2000pr, Clifton:2011jh,Will:2014kxa}.  Indeed, if we start from the generic scalar-tensor  action
\begin{eqnarray}
S_{\rm }=\int d^{4}x\sqrt{-g}\left(\frac12\, F(\phi) R-\frac12 g^{\mu\nu}\partial_{\nu}\phi\partial_{\mu}\phi-V(\phi)\right)+\int d^{4}x\sqrt{-g}\,{\cal L}_{\rm m}(\chi_i,g_{\mu\nu})\,, \label{Eq:BD.BDactionST}
\end{eqnarray}
we can easily recognize from \eqref{Eq:BD.BDaction2} that $F(\phi)=\xi\phi^2$, and that the potential $V(\phi)$ is just replaced by the  CC density $\rL$.  {In this way we can easily apply the well-known formulae of the PN formalism. We find the following values for the main PN parameters $\gamma^{\rm PN}$ and $\beta^{\rm PN}$ in our case (both being equal to $1$ in strict GR)}:
\begin{equation}\label{Eq:BD.gammaPPN}
1-\gamma^{\rm PN} =\frac{F^{\prime}(\phi)^2}{F(\phi)+2F^{\prime}(\phi)^2}= \frac{4\xi}{1+ 8\xi}=\frac{\eBD}{1+ 2\eBD}\simeq \eBD +   {\cal O}(\eBD^2)
\end{equation}
and
\begin{equation}\label{Eq:BD.betaPPN}
 1-\beta^{\rm PN}=-\frac14\, \frac{F(\phi) F^{\prime}(\phi)}{2F(\phi)+3F^{\prime}(\phi)^2}\,\frac{d \gamma^{\rm PN}}{d\phi}  =0\,,
\end{equation}
where {the primes refer here to derivatives with respect to $\phi$}. We are neglecting terms of $ {\cal O}(\eBD^2)$  and, in the second expression,  we use the fact that  $d\gamma_{\rm PN}/d\phi=0$ since $\oD=$ const. (hence $\eBD=$ const. too) in our case.  {Therefore, in BD-gravity, $\gamma^{\rm PN}$ deviates from $1$  an amount given precisely by $\eBD$ (in linear order), whereas $\beta^{\rm PN}$ undergoes no deviation at all. Furthermore, the effective gravitational strength between two test masses in the context of the scalar-tensor framework \eqref{Eq:BD.BDactionST} is well-known\,\cite{Boisseau:2000pr,Will:2014kxa,Clifton:2011jh}. In our case it  leads to the following result:}
\begin{equation}\label{Eq:BD.LocalGN}
G_{\rm eff}(\varphi) =\frac{1}{8\pi F(\phi)}\frac{2F(\phi) + 4F^{\prime}(\phi)^2}{2F(\phi)+3F^{\prime}(\phi)^2}=\frac{1}{8\pi\xi\phi^2} \frac{1+8\xi}{1+6\xi}= G(\varphi) \frac{2+4\eBD}{2+3\eBD}\,,
\end{equation}
where $G(\varphi)$ is the effective gravitational coupling in the BD action, as indicated in Eq. \eqref{Eq:BD.Gvarphi}.
Expanding linearly in $\eBD$, we find
\begin{equation}\label{Eq:BD.LocalGN2}
G_{\rm eff}(\varphi) =G(\varphi)\left[1+\frac12\,\eBD + {\cal O}(\eBD^2)\right] \,.
\end{equation}
{We confirm from the above two equations that the physical gravitational field undergone by two tests masses is not just determined by the effective  $G(\varphi)$ of the action but by  $G(\varphi)$ times a correction factor which depends on $\eBD$ and is larger (smaller) than $G(\varphi)$  for  $\eBD>0\,   (\eBD<0 )$.}

{From the exact formula \eqref{Eq:BD.LocalGN} we realize that if the local gravitational constraint ought to hold strictly, {\it i.e.} $G_{\rm eff}\to G_N$,  such formula would obviously enforce}
\begin{equation}\label{Eq:BD.LC}
\varphi=\frac{2+4\eBD}{2+3\eBD}\,.
\end{equation}
Due to Eq.\,\eqref{Eq:BD.gammaPPN}, the bound obtained from the Cassini mission\,\cite{Bertotti:2003rm}, $\gamma^{\rm PN}-1=(2.1\pm 2.3)\times 10^{-5}$, translates directly into a constraint on  $\eBD\simeq (-2.1\pm 2.3)\times 10^{-5}$ (in linear order), what implies $\oD\gtrsim 10^4$ for the BD-parameter. Thus, if considered together with the assumption $\varphi\simeq 1$ we would be left with very little margin for departures of $\Geff$ from $G_N$. However,  as previously  indicated,  we will not apply these local astrophysical bounds in most of our analysis since  we will assume that $\eBD$ may not be constrained in the cosmological domain and that the cosmological value of the gravitational coupling $G(\varphi)$  is different from $G_N$  owing to some possible variation of the BD-field $\varphi$ at the cosmological level. This can still be compatible with the local astrophysical constraints provided that we assume the existence of a screening mechanism in the local range which `renormalizes' the value of $\wBD$ and makes it appear much higher than its `bare' value (the latter being accessible only at the cosmological scales, where matter is much more diluted and uninfluential) --- cf. \hyperref[Sect:Mach]{Sect.\,\ref{Sect:Mach}} for details on the various possible BD scenarios.  We know that this possibility remains open and hence it must be explored\,\cite{Avilez:2013dxa}, see also \cite{Amendola:2015ksp,Clifton:2011jh}  and references therein.

Henceforth we shall stick to the original BD-form (\ref{Eq:BD.BDaction}) of the action  since the field $\psi$ (or, alternatively, its dimensionless companion $\varphi$)  is directly related to the effective gravitational coupling and the  $\oD$ parameter can be ascribed as being part of the kinetic term of $\psi$.  In contrast, the form (\ref{Eq:BD.BDaction2}) involves both $\phi$ and $\xi=1/(4\oD)$ in the definition of the gravitational strength.

\section{Why  BD-$\CC$CDM alleviates at a time the $H_0$ and $\sigma_8$ tensions?  Detailed preview and main results  of our analysis.}\label{Sect:Preview}

The field $\varphi$ and the parameter $\epsilon_{\rm BD}$ defined in the previous section, Eq.\,\eqref{Eq:BD.definitions}, are the fundamental new ingredients of BD-gravity as compared to GR  in the context of our analysis.  Any departure of $\varphi$ from one and/or of  $\epsilon_{\rm BD}$ from zero should reveal an extra effect of the BD-$\CC$CDM model as compared to the  conventional GR-$\CC$CDM one.   We devote this section to study the influence of  $\varphi$ and $\epsilon_{\rm BD}$ on the various observables we use in here to constrain the BD model.  This preliminary presentation will serve as a preview of the results presented in the rest of the chapter and will allow us to anticipate why BD-$\Lambda$CDM is able to alleviate so efficiently both of the $H_0$ and $\sigma_8$ tensions that are found in the context of the traditional  GR-$\Lambda$CDM framework.

Interestingly, many Horndeski theories \cite{Horndeski:1974wa} reduce in practice to BD at cosmological scales \cite{Avilez:2013dxa}, so the ability of  BD-$\Lambda$CDM to describe the wealth of current observations can also be extended to other, more general, scalar-tensor theories of gravity. Hence, it is crucial to clearly identify the reasons why BD-$\Lambda$CDM leads to such an improvement in the description of the data. Only later on (cf. Sec. \hyperref[Sect:NumericalAnalysis]{Sect.\,\ref{Sect:NumericalAnalysis}}) we will fit in detail the overall  data to the BD-$\Lambda$CDM model and will display the final numerical results. Here, in contrast, we will endeavour  to show why BD-gravity has specific clues to the solution which are not available to GR.

We can show this in two steps. First, we analyze what happens when we set $\epsilon_{\rm BD}=0$  at fixed values of $\varphi$ different from $1$.  From Eq.\,\eqref{Eq:BD.LocalGN} we can see that this scenario means to stick to the standard GR picture, but assuming that the effective Newtonian coupling can act at cosmological scales with a (constant) value $G_{\rm eff}=G(\varphi)$  different from the local one $G_N$.   In a second stage,  we  study the effect of the time dependence of $\varphi$  (triggered by a nonvanishing value of $\epsilon_{\rm BD}$), {\it i.e.} we will exploit the departure of $G_{\rm eff}$  in  Eq.\,\eqref{Eq:BD.LocalGN} from $G_N$ caused by $\eBD\neq0$ \textit{and} a variable $G(\varphi)$.  It will become clear  from this two-step procedure why BD-gravity has the double ability to reduce the two $\CC$CDM tensions in an harmonic way. On the one hand, a value of $\varphi$ below $1$ in the late Universe increases the value of $G_{\rm eff}$  and hence of $H_0$, so it should be  able to significantly  reduce the $H_0$-tension;  and on the other hand {the dynamics of $\varphi$ triggered by a finite (but negative) value of $\epsilon_{\rm BD}$ helps to suppress the structure formation processes in the Universe, since it enhances the Hubble friction and also leads to a decrease of the Poisson term in the perturbations equation.} {The upshot is that the $\sigma_8$-tension becomes reduced as well.} Let us note that the lack of use of LSS data may lead to a different conclusion, in particular to $\eBD>0$, see e.g. \cite{Yadav:2019fjx}. This reference, in addition, uses only an approximate treatment of the CMB data through distance priors.

\subsection{Role of $\varphi$, and the $H_0$-tension}\label{Sect:rolesvarphiH0}

Let us start, then, by studying how the observables change with $\varphi$ when $\epsilon_{\rm BD}=0$,  for fixed values of the current energy densities. In the context of BD-gravity, as well as in GR,  if the energy densities are fixed at present we can fully determine their cosmological evolution, since all the species are self-conserved, as discussed in \hyperref[Sect:BDgravity]{Sect.\,\ref{Sect:BDgravity}}. In BD-gravity, with $\epsilon_{\rm BD}=0$, the Hubble function takes the  form
\begin{equation}\label{Eq:BD.H1}
H^2(a) = \frac{8\pi G_N}{3\varphi}\rho(a)\,,
\end{equation}
where $\varphi=const$. We have just removed the time derivatives of the scalar field in the Friedmann equation \eqref{Eq:BD.Friedmannequation}. We define $H_0$ from the value of the previous expression at $a=1$ (current value). Recall from the previous section that $\rho$ is the sum of all the energy density contributions, namely $\rho \equiv \rho_{\rm m} + \rho_\gamma + \rho_\nu + \rho_\Lambda$. Therefore,
\begin{equation}\label{Eq:BD.H1b}
H_0^2 = \frac{8\pi G_N}{3\varphi}\rho^0=\frac{8\pi G_N}{3\varphi}\left( \rho_{\rm m}^0+\rho_\gamma^0+ \rho_\nu^{0}+\rL\right)\,.
\end{equation}
$\rho^0=\rho(a=1)$ is the total energy density at present, $\rho_\gamma^0$ is the corresponding density of photons and $\rho_\nu^{0}$ that of neutrinos, and finally $\rL$ is the original cosmological constant density in the BD-action \eqref{Eq:BD.BDaction}.
Using \eqref{Eq:BD.H1b} it proves now useful to rewrite \eqref{Eq:BD.H1} in the alternative way:
\begin{equation}\label{Eq:BD.H2}
H^2(a)= H_0^2\left[1+\tilde{\Omega}_{\rm m}(a^{-3}-1)+\tilde{\Omega}_\gamma(a^{-4}-1)+\tilde{\Omega}_\nu(a)-\tilde{\Omega}_\nu\right]\,,
\end{equation}
where we use the modified cosmological parameters (more appropriate for the BD theory):
\begin{equation}\label{Eq:BD.tildeOmegues}
\tilde{\Omega}_i\equiv\frac{\rho^{0}_i}{\rho^{0}}=\frac{\Omega_i}{\varphi}\,,\ \ \ \ \ \ \ \ \  \rho^{0}=\frac{3H_0^2}{8\pi G_N}\,\varphi=\rco\varphi\,.
\end{equation}
The tilde is to distinguish the modified $\tilde{\Omega}_i$ from the usual cosmological parameters $\Omega_i=\rho_i^{0}/\rco$ employed in the GR-$\Lambda$CDM model.  In addition,  $\tilde{\Omega}_{\rm m}=\tilde{\Omega}_{cdm}+\tilde{\Omega}_{\rm b}$ is the sum of the contributions from CDM and baryons; and $\tilde{\Omega}_\gamma$ and $\tilde{\Omega}_\nu$ are the current values for the photons and neutrinos. For convenience, we also define $\tilde{\Omega}_{\rm r}=\tilde{\Omega}_\gamma+\tilde{\Omega}_\nu$.  We remark that he current total energy density $ \rho^{0}$ is related to the usual critical density $\rco=3H_0^2/(8\pi G_N)$  through a factor of $\varphi$, as indicated above.
The modified parameters obviously satisfy the canonical sum rule
\begin{equation}\label{Eq:BD.SumRuleBD}
\tilde{\Omega}_{\rm m}+\tilde{\Omega}_{\rm r}+\tilde{\Omega}_\CC=1\,.
\end{equation}
The form of \eqref{Eq:BD.H2} is completely analogous to the one found in GR-$\Lambda$CDM since in BD-$\CC$CDM the $\tilde{\Omega}_i$'s represent the fraction of energy carried by the various species in the current Universe, as the $\Omega_i$'s do in GR, so from the physical point of view the $\tilde{\Omega}_i$'s in BD and the $\Omega_i$'s in GR contain the same information\footnote{For $|\eBD|\neq 0$ and small, the $\tilde{\Omega}_i$ parameters defined in Eq.\,\eqref{Eq:BD.tildeOmegues} receive a correction of ${\cal O}(\eBD)$, see\,\hyperref[Sect:eBDands8]{Sect.\,\ref{Sect:eBDands8}}.}. $H_0$ represents in both cases the current value of the Hubble function. Nevertheless, there is a very important (although maybe subtle) difference, namely: in BD-$\Lambda$CDM there does not exist a  one-to-one correspondence between $H_0$ and $\rho^{0}$. In contradistinction to GR-$\Lambda$CDM, in the BD version of the concordance model the value of $\varphi$ can modulate $H_0$ for a given amount of the total (critical) energy density. In other words, given a concrete value of $H_0$ there is a $100\%$ degeneracy between $\varphi$ and $\rho^{0}$. This degeneracy is broken by the data, of course. The question we want to address is precisely whether there is still room for an increase of $H_0$ with respect to the value found in the GR-$\Lambda$CDM model once the aforementioned degeneracy is broken by observations. We will see that this is actually the case by analyzing what is the impact that $\varphi$ has on each observable considered in our analyses.
\newline
\newline
{\bf Supernovae data}
\newline
\newline
\noindent
In the case  of Type Ia  Supernovae data (SNIa)  we fit observational points on their apparent magnitudes $m(z)= M +5\log_{10}[D_L(z)/10{\rm pc}]$, where $M$ is the absolute magnitude of the SNIa and $D_L(z)$ is the luminosity distance. In a spatially flat Universe the latter reads,
\begin{equation}\label{Eq:BD.LuminosityDistance}
D_L(z)=c(1+z)\int^z_{0}\frac{dz^\p}{H(z^\p)}\,,
\end{equation}
where we have momentarily kept $c$ explicitly for the sake of better understanding. Considering \eqref{Eq:BD.H2}, we can easily see that if we only use SNIa data, there is a full degeneracy between $M$ and $H_0$ in the computation of the apparent magnitudes, so it is not possible to obtain information about $\varphi$, since we cannot disentangle it from the absolute magnitude. As in GR-$\Lambda$CDM, we can only get constraints on the current fraction of matter energy in the Universe, {\it i.e.} $\tilde{\Omega}_{\rm m}$.
\newline
\newline
{\bf Baryon acoustic oscillations}
\newline
\newline
\noindent
The constraints obtained from the analysis in real or Fourier space of the baryon acoustic oscillations (BAO) are usually provided by galaxy surveys in terms of $D_A(z)/r_{\rm s}$ and $r_{\rm s} H(z)$, or in some cases by a combination of these two quantities when it is not possible to disentangle the line-of-sight and transverse information, through the so-called dilation scale $D_V$ (cf. \hyperref[Sect:MethodData]{Sect.\,\ref{Sect:MethodData}}),
\begin{equation}\label{Eq:BD.DVScale}
\frac{D_V(z)}{r_{\rm s}}=\frac{1}{r_{\rm s}}\left[D_{\rm M}^2(z)\frac{cz}{H(z)}\right]^{1/3}\,,
\end{equation}
$D_{\rm M}=(1+z)D_{A}(z)$ being the comoving angular diameter distance, $D_A(z)=D_L(z)/(1+z)^2$ the proper angular diameter distance, and
\begin{equation}\label{Eq:BD.rs}
r_{\rm s}=\int_{z_d}^{\infty}\frac{c_{\rm s}(z)}{H(z)}\,dz
\end{equation}
the comoving sound horizon at the baryon drag epoch $z_d$. In the above equation, $ c_{\rm s}(z)$ is  the sound speed of the photo-baryon plasma, which depends on the ratio $\rho_{\rm b}^{0}/\rho^{0}_\gamma$. The current temperature of photons (and hence also its associated current energy density $\rho^{0}_\gamma$) is already known with high precision thanks to the accurate measurement of the CMB monopole \cite{Fixsen:2009ug}. Because of \eqref{Eq:BD.H1} it is obvious that we cannot extract information on $\varphi$ from BAO data when $\epsilon_{\rm BD}=0$, since it cancels exactly in the ratio $D_A(z)/r_{\rm s}$ and the product $r_{\rm s} H(z)$. Thus, BAO data provide constraints on $\tilde{\Omega}_{\rm m}$ and $\rho_{\rm b}^{0}$, but not on $\varphi$.
\newline
\newline
{\bf Redshift-space distortions (RSD)}
\newline
\newline
\noindent
{The  LSS observable $f(z)\sigma_8(z)$, which is essentially determined from RSD measurements,  is of paramount importance to study the formation of structures in the Universe. In BD-gravity, the equation of matter field perturbations is different from that of GR and  is studied in detail in \hyperref[Sect:StructureFormationBD]{Sect.\,\ref{Sect:StructureFormationBD}}.  Here we wish to make some considerations which will help to have a rapid overview of why BD-gravity with a cosmological constant can also help to improve the description of structure formation as compared to GR. The exact differential equation for the linear density contrast of the matter perturbations,  $\delta_{\rm m}=\delta\rho_{\rm m}/\rho_{\rm m}$,  in this context can be computed  at deep subhorizon scales and is given by (cf. \hyperref[Sect:StructureFormationBD]{Sect.\,\ref{Sect:StructureFormationBD}}):}
\begin{equation}\label{Eq:BD.ExactPerturScaleFactor}
\delta_{\rm m}^\pp+\left(\frac{3}{a}+\frac{H^\prime(a)}{H(a)}\right)\delta_{\rm m}^\p-\frac{4\pi \Geff(a)}{H^2(a)}\frac{\rho_{\rm m}(a)}{a^2}\,\delta_{\rm m}=0\,,
\end{equation}
{where for the sake of convenience we express it in terms of the scale factor and hence prime denotes here  derivative {\it w.r.t.}  such variable:  $()^\prime \equiv d()/d a$. In the above equation, $\Geff(a)$ is the effective gravitational coupling \eqref{Eq:BD.LocalGN} with $\varphi=\varphi(a)$ expressed in terms of the scale factor:}
\begin{equation}\label{Eq:BD.LocalGNa}
\Geff(a) = G(\varphi(a)) \frac{2+4\eBD}{2+3\eBD}\,.
\end{equation}
It crucially controls the  Poisson term of the perturbations equation, {\it i.e.} the last term in \eqref{Eq:BD.ExactPerturScaleFactor}.
As we can see, it is $\Geff(\varphi)$ and not just $G(\varphi)$ the coupling involved in the structure formation data since it is $\Geff(\varphi)$ the gravitational coupling felt by the test masses in BD-gravity.
It is obvious that the above Eq.\,\eqref{Eq:BD.ExactPerturScaleFactor} boils down to the GR form for $\eBD=0$ and $\varphi=1$.

With the help of the above equations  we  wish first to assess the bearing that $\varphi$ can have on the LSS observable $f(z)\sigma_8(z)$ at  fixed values of the current energy densities and for $\epsilon_{\rm BD}=0$. Recall that when  $\epsilon_{\rm BD}=0$ the BD-field cannot evolve at all, so it just remains fixed at some value.  For this consideration, we therefore set $\psi=$const. in equations \eqref{Eq:BD.Friedmannequation} and \eqref{Eq:BD.pressureequation} and of course the BD-theory becomes just a GR version with an effective coupling $\Geff=G(\varphi)=$const.  which nevertheless need not be identical to $G_N$. In these conditions, it is easy to verify that Eq.\,\eqref{Eq:BD.ExactPerturScaleFactor} adopts the following simpler form, in which $\Geff$ drops from the final expression:
\begin{equation}\label{Eq:BD.DC}
\delta_{\rm m}^{\pp}+\frac{3}{2a} \left(1-w(a)\right) \delta^\p_{\rm m} -\frac{3}{2a^2}\frac{\rho_{\rm m}(a)}{\rho(a)}\delta_{\rm m}=0\,.
\end{equation}
In the above expression, $w(a)=p(a)/\rho(a)$ is the equation of state (EoS) of the total cosmological fluid, hence  $\rho(a)$ and $p(a)$ stand respectively for the total density and pressure of the fluids involved (cf. \hyperref[Sect:BDgravity]{Sect.\,\ref{Sect:BDgravity}}). In particular, during the epoch of structure formation the matter particles contribute a negligible contribution to the pressure and the dominant component is that of the  cosmological term: $p(a)\simeq p_\CC=-\rL$.

It is important to realize that $\varphi=$ const. does not play any role in \eqref{Eq:BD.DC}. This means that its constant value, whatever it may be,  does not affect the evolution of the density contrast, which is only determined by the fraction of matter, $\rho_{\rm m}(a)/\rho(a)$,  and the  EoS of the total cosmological fluid.  The equation that rules the evolution of the density contrast is exactly the same as in the GR-$\Lambda$CDM model. Matter inhomogeneities grow in the same way regardless of the constant value $\Geff$ that we consider. Matter tends to clump more efficiently for larger values of the gravitational strength, of course, but the Hubble friction also grows in this case, since such an increase in $\Geff$ also makes the Universe to expand faster.
Surprisingly, if $\epsilon_{\rm BD}=0$, {\it i.e.} if $\Geff=$const., both effects compensate each other. Thus, the BD growth rate $f(a)=a\delta_{\rm m}^\prime(a)/\delta_{\rm m}(a)$ does not change  {\it w.r.t.} the GR scenario either. But what happens with $\sigma_8(z)$? It is computed through the following expression:
\begin{equation}\label{Eq:BD.sigma8}
\sigma_8^2(z)=\frac{1}{2\pi^2}\int_0^\infty dk\,k^2\,P(k,z)\,W^2(kR_8)\,,
\end{equation}
in which  $P(k,z)$ is the power spectrum of matter fluctuations and $W(kR_8)$ is the top hat smoothing function in Fourier space, with $R_8=8h^{-1}$ Mpc. Even for $\epsilon_{\rm BD}=0$ one would naively expect \eqref{Eq:BD.sigma8} to be sensitive to the value of $\varphi$, since some relevant features of the power spectrum clearly are. For instance, the scale associated to the matter-radiation equality reads
\begin{equation}\label{Eq:BD.keq}
k_{eq}=a_{eq}H(a_{eq})=H_0\tilde{\Omega}_{\rm m}\sqrt{\frac{2}{\tilde{\Omega}_{\rm r}}}\,,
\end{equation}
and $H_0\propto\varphi^{-1/2}$, so the peak of $P(k,z)$ is shifted when we change $\varphi$. Also the window function itself depends on $H_0$ through $R_8$. Despite this, the integral \eqref{Eq:BD.sigma8} does not depend on the Hubble function for fixed energy densities at present, and hence neither on $\varphi$. To see this, let us  first rescale the wave number  as follows  $k=\bar{k}h$, and we obtain
\begin{equation}\label{Eq:BD.sigma8v2}
\sigma_8^2(z)=\frac{1}{2\pi^2}\int_0^\infty d\bar{k}\,\bar{k}^2\,\underbrace{P(k=\bar{k}h,z)\,h^3}_{\equiv \bar{P}(\bar{k},z)}\,W^2(\bar{k}\cdot 8\,{\rm Mpc^{-1}})\,.
\end{equation}
The only dependence on $h$ is now contained in $\bar{P}(\bar{k},z)$. We can write $P(k,z)=P_0k^{n_{\rm s}}T^2(k/k_{eq})\delta^2_{\rm m}(z)$, with $T(k/k_{eq})$ being the transfer function and
\begin{equation}\label{Eq:BD.PowerSpectrum}
P_0=A_{\rm s}\frac{8\pi^2}{25}\frac{k_*^{1-n_{\rm s}}}{(\tilde{\Omega}_{\rm m} H_0^2)^2}\,,
\end{equation}
with $A_{\rm s}$ and $n_{\rm s}$ being the amplitude and spectral index of the dimensionless primordial power spectrum, respectively, and $k_*$ the corresponding pivot scale. The last relation can be found using standard formulae, see e.g.\,\cite{Gorbunov:2011zzc,Amendola:2015ksp}. Taking into account all these expressions we obtain
\begin{equation}
\bar{P}(\bar{k},z)=A_{\rm s}\frac{8\pi^2}{25}\frac{\bar{k}^{1-n_{\rm s}}_*\bar{k}^{n_{\rm s}}}{(10^4\varsigma^2\tilde{\Omega}_{\rm m})^2} T^2(\bar{k}/\bar{k}_{eq})\delta^2_{\rm m}(z)\,,
\end{equation}
where we have used $H_0=100\, h\, \varsigma$  with $\varsigma\equiv 1\,{\rm km/s/Mpc}=2.1332\times 10^{-44}$ GeV (in natural units). We see that all factors of $h$ cancel out. Now it is obvious that $\sigma_8(z)$ is not sensitive to the value of $\varphi$. We have explicitly checked this with our own modified version of the Einstein-Boltzmann system solver \texttt{CLASS} \cite{Blas:2011rf}, in which we have implemented the BD-$\Lambda$CDM model (see \hyperref[Sect:MethodData]{Sect.\,\ref{Sect:MethodData}} for details). The product $f(z)\sigma_8(z)$ does not depend on $\varphi$ when $\epsilon_{\rm BD}=0$, so RSD data cannot constrain $\varphi$ either.
\newline
\newline
\newline
{\bf Strong-Lensing time delay distances, distance ladder determination of $H_0$, and cosmic chronometers}
\newline
\newline
\noindent
In this chapter, we will use the Strong-Lensing time delay angular diameter distances provided by the H0LICOW collaboration \cite{Wong:2019kwg}, see \hyperref[Sect:MethodData]{Sect.\,\ref{Sect:MethodData}}. Contrary to SNIa and BAO data, these distances are not relative, but absolute. This allows us to extract information not only on $\Om$, but on $H_0$ too.  Furthermore,  the data on $H(z)$ obtained from cosmic chronometers (CCH) give us information about these two parameters as well. Cosmic chronometers have been recently employed in the reconstruction of the expansion history of the Universe using Gaussian Processes and the so-called Weighted Function Regression method \cite{Yu:2017iju,Gomez-Valent:2018hwc,Haridasu:2018gqm}, which do not rely on any particular cosmological model. The extrapolated values of the Hubble parameter found in these analyses are closer to the best-fit GR-$\Lambda$CDM value reported by Planck \cite{aghanim2020planck}, around $H_0\sim (67.5-69.5)$ km/s/Mpc, but they are still compatible within  $\sim 1\sigma$ c.l. with the local determination obtained with the distance ladder technique \cite{Riess:2018uxu,Riess:2019cxk,Reid:2019tiq} and the Strong-Lensing time delay measurements by H0LICOW \cite{Wong:2019kwg}. The statistical weight of the CCH data is not as high as the one obtained from these two probes, so when combined with the latter, the resulting value for $H_0$ is still in very strong tension with Planck \cite{Gomez-Valent:2018hwc,Haridasu:2018gqm}. As mentioned before, $H_0^2\propto \rho^{0}/\varphi$ when $\epsilon_{\rm BD}=0$. Thus, we can alleviate in principle the $H_0$-tension by keeping the same values of the current energy densities of all the species as in the best-fit GR-$\Lambda$CDM model reported by Planck \cite{aghanim2020planck}, lowering the value of $\varphi$ down at cosmological scales, below $1$, and assuming some sort of screening mechanism acting on high enough density regions that allows us to evade the solar system constraints and keep unmodified the stellar physics needed to rely on CCH, SNIa, the H0LICOW data, and the local distance ladder measurement of $H_0$. By doing this we do not modify at all the SNIa, BAO and RSD observables {\it w.r.t.} the GR-$\Lambda$CDM, and we automatically improve the description of the H0LICOW data and the local determination of $H_0$, which are the observables that prefer higher values of the Hubble parameter.  Let us also mention that the fact that $\varphi<1$ throughout the cosmic history (which means $G>G_N$)  allows to have a larger value of $H$ (for similar values of the density parameters) at any time as compared to the GR-$\CC$CDM and hence a smaller value of the sound horizon distance $r_{\rm s}$, Eq.\,\eqref{Eq:BD.rs}, what makes the model to keep a good  fit to the BAO data. This is confirmed by the numerical analysis presented in Tables \hyperref[Table:BD.TableFitBaseline]{\ref{Table:BD.TableFitBaseline}}, \hyperref[Table:BD.TableFit+SL]{\ref{Table:BD.TableFit+SL}}, \hyperref[Table:BD.TableFitSpectrum]{\ref{Table:BD.TableFitSpectrum}}, \hyperref[Table:BD.TableFitAlternativeDataset]{\ref{Table:BD.TableFitAlternativeDataset}} and \hyperref[Table:BD.TableFitCassini]{\ref{Table:BD.TableFitCassini}} as compared to the conventional $\CC$CDM values, see \hyperref[Table:BD.TableFitBaseline]{\ref{Table:BD.TableFitBaseline}}, \hyperref[Table:BD.TableFit+SL]{\ref{Table:BD.TableFit+SL}}, \hyperref[Table:BD.TableFitSpectrum]{\ref{Table:BD.TableFitSpectrum}} and \hyperref[Table:BD.TableFitGR]{\ref{Table:BD.TableFitGR}}. While the claim existing in the literature  that models which predict smaller values of $r_{\rm s}$ are the preferred ones for solving the $H_0$-tension is probably correct,  we should point out that this sole fact is no guarantee of success, as one still needs in general a compensation mechanism at low energies which prevents $\sigma_8$ from increasing and hence worsening such tension. In the BD-$\CC$CDM such compensation mechanism is provided by a  negative value of $\eBD$ (as we will show later), and for this reason the two tensions can be smoothed at the same time in an harmonic way.

Overall, as we have seen from the above discussion,  according to the (long) string of supernovae, baryonic acoustic oscillations, cosmic chronometers, Strong-Lensing and local Hubble parameter data (SNIa+BAO+RSD+CCH+H0LICOW+$H_0$) it is possible to loosen the $H_0$-tension, and this is already very remarkable, but we still have to see whether this is compatible with the very precise measurements of the photon temperature fluctuations in the CMB map or not. More specifically, we have to check whether it is possible to describe these anisotropies while keeping the current energy densities close to the best-fit GR-$\Lambda$CDM model from Planck, compatible with $\varphi<1$.
\newline
\newline
\begin{figure}[t!]
\begin{center}
\includegraphics[width=6in, height=4in]{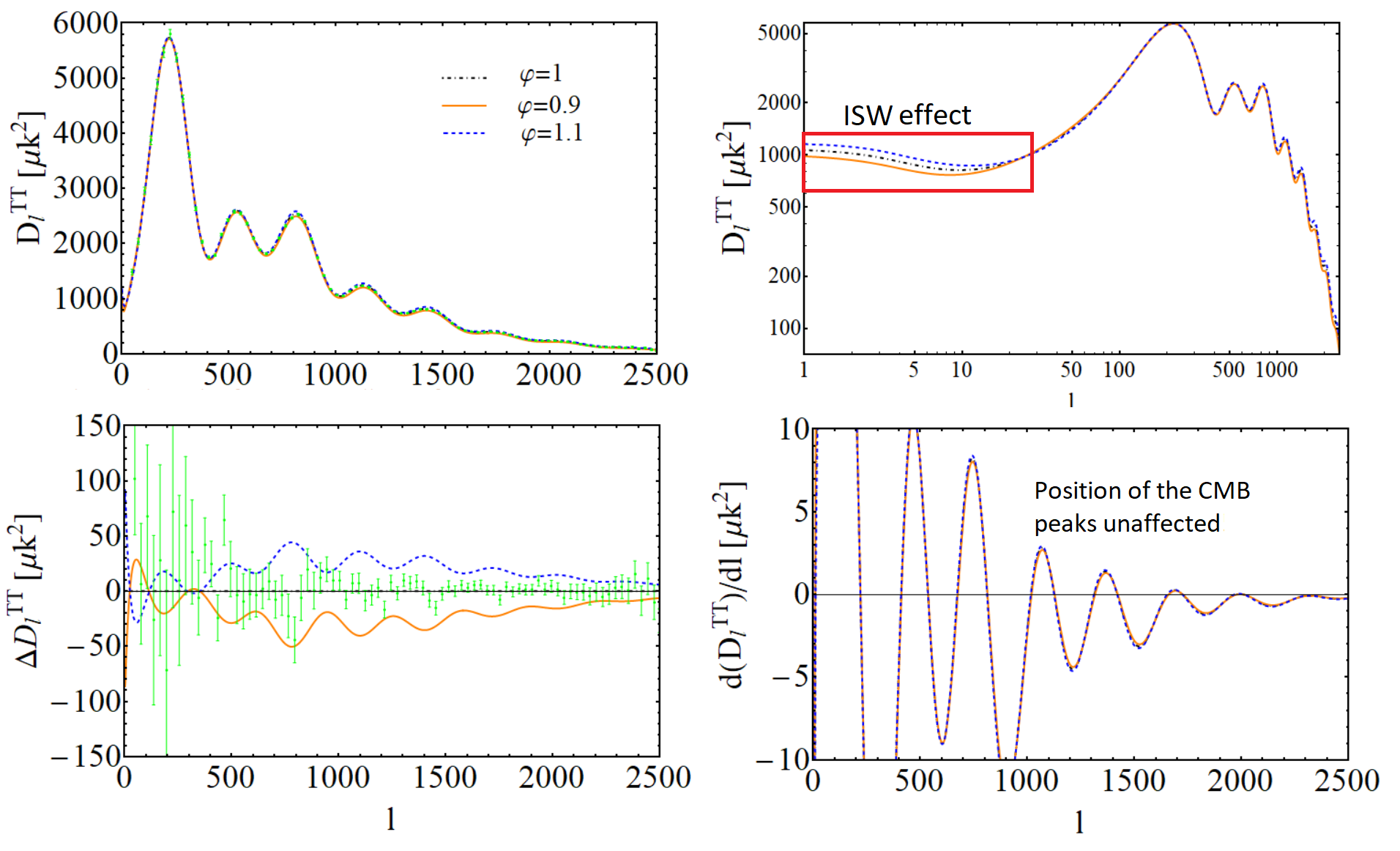}
\caption{{\it Left-upper plot:} Theoretical curves of the CMB temperature anisotropies obtained by fixing the current energy densities, $\tau$ and the parameters of the primordial power spectrum to the GR-$\Lambda$CDM best-fit values obtained from the analysis of TTTEEE+lowE data by Planck 2018 \protect\cite{aghanim2020planck} (which we will refer to as the $\CC$CDM baseline configuration, denoted by a dot-dashed black line), against the BD-$\Lambda$CDM model  keeping $\epsilon_{\rm BD}=0$ for two different constant values  $\varphi=0.9,1.1$  (orange and dashed blue lines, respectively); {\it Right-upper plot:} The same, but using a logscale in the $x$-axis to better appreciate the integrated Sachs-Wolfe effect at low multipoles; {\it Left-bottom plot:} Absolute differences of the data points and theoretical curves for $\varphi=0.9,1.1$ {\it w.r.t.} the $\varphi=1$ case; {\it Right-bottom plot:} Derivative of the functions plotted in the upper plots, to check the effect of $\varphi$ on the position of the peaks, which corresponds to the location of the zeros in this plot. No shift is observed, as expected (cf. the explanations in the main text).}\label{Fig:BD.Clsvarphi}
\end{center}
\end{figure}
\newline
\newline
{\bf CMB temperature anisotropies}
\newline
\newline
\noindent We expect the peak positions of the CMB temperature (TT, in short) power spectrum to remain unaltered under changes of $\varphi$ (when $\epsilon_{\rm BD}=0$), since they are always related with an angle, which is basically a ratio of cosmological distances ({a transverse distance to the line of sight divided by the angular diameter distance}).  If $\varphi=$const.,  such constant cancels  in the ratio, so there is no dependence on $\varphi$. In the right-bottom plot of \hyperref[Fig:BD.Clsvarphi]{Fig.\,\ref{Fig:BD.Clsvarphi}} we show the derivative of the $\mathcal{D}_l^{TT}$'s for three alternative values of $\varphi$. It is clear that the location of the zeros does not depend on the value of this parameter. Hence $\varphi$ does not shift the peaks of the TT CMB spectrum when we consider it to be a constant throughout the entire cosmic history, as expected. Nevertheless, there are two things that affect its shape and both are due to the impact that $\varphi$ has on the Bardeen potentials. We have seen before that the matter density contrast is not affected by $\varphi$ when it is constant, but taking a look on the Poisson equation in the BD model (cf. Appendix D), we can convince ourselves that $\varphi$ does directly affect the value of $\Psi$ and $\Phi$, since both functions are proportional to $\rho_{\rm m}\delta_{\rm m}/\varphi$ at subhorizon scales, see Eqs.\,\eqref{Eq:NewtGauge.PhiplusPsi}-\eqref{Eq:NewtGauge.Poisson3}. This dependence modifies two basic CMB observables:
\begin{itemize}
\item The CMB lensing, at low scales (large multipoles, $500\lesssim l\lesssim 2000$). In the left-bottom plot of \hyperref[Fig:BD.Clsvarphi]{Fig.\,\ref{Fig:BD.Clsvarphi}} we show the difference of the TT CMB spectra {\it w.r.t.} the GR-$\Lambda$CDM model. A variation of $\varphi$ changes the amount of CMB lensing, which in turn modifies the shape of the spectrum mostly from the third peak onwards. In that multipole range also the Silk damping plays an important role and leaves a signature \cite{Silk1968}.
\item The integrated Sachs-Wolfe effect \cite{Sachs:1967er}, at large scales (low multipoles, $l\lesssim 30$). Values of $\varphi<1$ (which, recall, lead to higher values of $H_0$) suppress the power of the $\mathcal{D}_l^{TT}$'s in that range. This is a welcome feature, since it could help us to explain the low multipole ``anomaly'' that is found in the context of the GR-$\Lambda$CDM model (see e.g. Fig.\,1 of \cite{aghanim2020planck}, and \cite{Das:2013sca}). Later on, we further discuss and confirm the alleviation  of this intriguing anomaly in light of the final fitting results, see \,\hyperref[Sect:LowMultipoles]{Sect.\,\ref{Sect:LowMultipoles}} and \hyperref[Fig:BD.Cls-ISW]{Fig.\,\ref{Fig:BD.Cls-ISW}}. This is obviously an additional bonus of the BD-$\CC$CDM framework.
\end{itemize}

\begin{figure}[t!]
\begin{center}
\includegraphics[width=6in, height=2.4in]{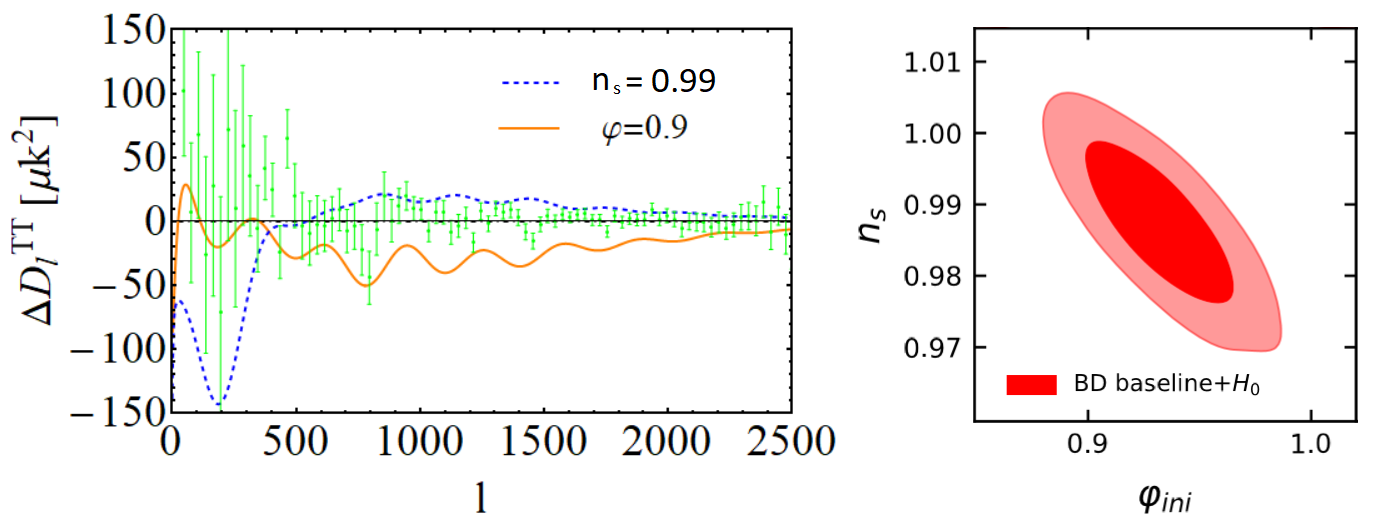}
\caption{{\it Left plot:} Differences between the CMB temperature spectrum obtained using the original $\CC$CDM baseline configuration (as used in \hyperref[Fig:BD.Clsvarphi]{Fig.\,\ref{Fig:BD.Clsvarphi}}) and those obtained using $\varphi=0.9$ (with the baseline $n_{\rm s}=0.9649$) and $n_{\rm s}=0.99$ (with the baseline $\varphi=1$). This is to show that the effects induced by a lowering of $\varphi$ can be compensated by an increase of the spectral index; {\it Right plot:} We also show the $1\sigma$ and $2\sigma$ c.l. regions in the $(\varphi_{\rm ini},n_{\rm s})$-plane, obtained using the baseline dataset, together with the Gaussian prior on $H_0$ from \protect\cite{Reid:2019tiq}, see \protect\hyperref[Sect:MethodData]{Sect-\,\ref{Sect:MethodData}} for details. Here one can clearly see the  anticorrelation between these two parameters.}\label{Fig:nsVSvarphi}
\end{center}
\end{figure}

Even though in \hyperref[Fig:BD.Clsvarphi]{Fig.\,\ref{Fig:BD.Clsvarphi}} we are considering large ($\sim 10\%$) relative deviations of $\varphi$ with respect to $1$, the induced deviations of the $\mathcal{D}_l^{TT}$'s are fully contained in the observational error bars at low multipoles, and they are not extremely big at large ones. The latter are only $1-2\sigma$ away from most of the data points.  We emphasize that we are only varying $\varphi$ here, so there is still plenty of room to correct these deviations by modifying the value of other parameters entering the model. To do that it would be great if we could still keep the values of the current energy densities as in the concordance model, since this would ensue the automatic fulfillment of the constraints imposed by the datasets discussed before. But is this possible? In \hyperref[Fig:nsVSvarphi]{Fig.\,\ref{Fig:nsVSvarphi}} we can see that e.g. an increase of $n_{\rm s}$ can compensate for the decrease of $\varphi$ pretty well. This is why in the BD model we obtain higher best-fit values of the spectral index {\it w.r.t.} the GR-$\Lambda$CDM, and a clear anti-correlation between these two parameters ({cf. \hyperref[Fig:nsVSvarphi]{Fig.\,\ref{Fig:nsVSvarphi}}, tables \hyperref[Table:BD.TableFitBaseline]{\ref{Table:BD.TableFitBaseline}}, \hyperref[Table:BD.TableFit+SL]{\ref{Table:BD.TableFit+SL}}, \hyperref[Table:BD.TableFitSpectrum]{\ref{Table:BD.TableFitSpectrum}}, \hyperref[Table:BD.TableFitCassini]{\ref{Table:BD.TableFitCassini}} and \hyperref[Sect:NumericalAnalysis]{Sect.\,\ref{Sect:NumericalAnalysis}} for details}). Small variations in other parameters can also help to improve the description of the data, of course, but the role of $n_{\rm s}$ seems to be important. In \hyperref[Fig:BD.pkns]{Fig.\,\ref{Fig:BD.pkns}} we can appreciate the change in the matter power spectrum induced by different values of $n_{\rm s}$. There is a modification in the range of scales that can be observationally accessed to with the analysis of RSD, but these differences are negative at $k\lesssim 0.07\,h{\rm\,Mpc^{-1}}$ and positive at larger values of the wave number (lower scales), so there can be a compensation when $\sigma_8$ is computed through \eqref{Eq:BD.sigma8}, leaving the value of the latter stable. Moreover, we will see below that $\epsilon_{\rm BD}$ can also help to decrease the value of $P(k)$ at $k\gtrsim 0.02\,h{\rm\,Mpc^{-1}}$, so the correct shape for the power spectrum is therefore guaranteed.

The upshot of this section is worth emphasizing: as it turns out, the sole fact of considering a cosmological Newtonian coupling about $\sim 10\%$ larger than the one measured locally can allow us to fit very well all the cosmological datasets, loosening the $H_0$-tension and keeping standard values of $\sigma_8$. It has become common in the literature to divide the theoretical proposals  able to decrease  the $H_0$-tension into two different classes depending on the stage of the Universe's expansion at which new physics are introduced \cite{Knox:2019rjx}: pre- and post-recombination solutions. The one we are suggesting here cannot be identified with any of these two categories, since it modifies the strength of gravity at cosmological scales not only before the decoupling of the CMB photons or the late-time Universe, but during the whole cosmological history, relying on a screening mechanism able to generate $\Geff=G=G_N$ in high density ({nonrelativistic}) environments where nonlinear processes become important, as e.g. in our own solar system\footnote{The study of these screening mechanisms, see e.g. \cite{Tsujikawa:2008uc,Amendola:2015ksp,Clifton:2011jh,Li:2020uaz} and references therein, can be the subject of  future work, but here we  remark that e.g. chameleon \cite{Khoury:2003aq}, symmetron \cite{Hinterbichler:2010es} or Vainshtein mechanisms (see \cite{Kimura:2011dc} and references therein), do not screen the value of $\varphi$ during the radiation-dominated epoch. This is important to loose the $H_0$-tension in the BD-$\Lambda$CDM framework through the increase of $H(z)$ at both, the early and late Universe.}. That there is indeed a  change of the gravity  strength  throughout the entire cosmological history in our study follows from the fact that $\eBD\neq0$ in the BD framework, and this is exactly the feature that we are going to exploit in the next section, a feature which adds up to the mere change of the global strength of the gravity interaction, which is still possible for $\eBD=0$ in the BD context, and that we have explored in the previous section.

\begin{figure}[t!]
\begin{center}
\includegraphics[width=6.5in, height=2.3in]{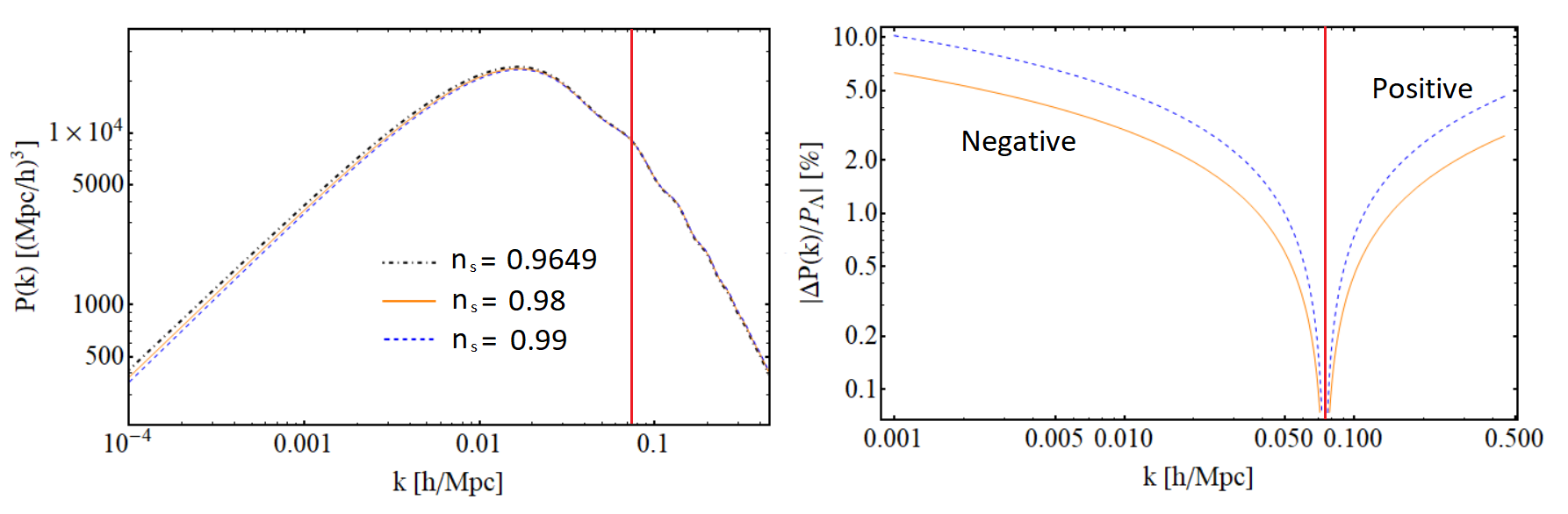}
\caption{{\it Left plot:} Here we compare the linear matter power spectrum obtained for the $\CC$CDM baseline configuration (dot-dashed black line) and two alternative values of the spectral index $n_{\rm s}$. The vertical red line indicates the value of the wave number at which there is the break in the right plot; {\it Right plot:} The absolute relative differences in $P(k)$ {\it w.r.t.} the baseline $\CC$CDM model. The ``positive'' (``negative'') region is the one in which $P(k)$ is larger (lower) than in the baseline setup. Both zones are delimited by a vertical red line. See the comments in the text.}\label{Fig:BD.pkns}
\end{center}
\end{figure}

\subsection{Role of $\epsilon_{\rm BD}$, and the $\sigma_8$-tension}\label{Sect:eBDands8}

Next we study the effect of $\epsilon_{\rm BD}\neq0$. We know that when we introduce the matter bispectrum information from BOSS \cite{Gil-Marin:2016wya} (which definitely prefers a lower amount of structure in the Universe than the data reported in \cite{Alam:2016hwk}) in our fitting analyses, we find a stronger signal for negative values of  $\epsilon_{\rm BD}$ (cf. \hyperref[Sect:NumericalAnalysis]{Sect.\,\ref{Sect:NumericalAnalysis}}). When $\epsilon_{\rm BD}\ne 0$ equation \eqref{Eq:BD.DC} is not valid anymore, since it was derived under the assumption of $\varphi=$const.  {We start once more from the exact perturbations equation for the matter density contrast  in the linear regime within the BD-gravity, {\it i.e.} Eq.\,\eqref{Eq:BD.ExactPerturScaleFactor}.  Let us consider its form within the approximation $|\epsilon_{\rm BD}|\ll 1$ (as it is preferred by the data, see e.g. \hyperref[Table:BD.TableFitBaseline]{Table\,\ref{Table:BD.TableFitBaseline}}). We come back to the BD-field equations  \eqref{Eq:BD.Friedmannequation} and \eqref{Eq:BD.pressureequation} and apply such approximation. In this way Eq.\,\eqref{Eq:BD.ExactPerturScaleFactor} can be expanded linearly in $\eBD$ as follows (primes are still denoting derivatives with respect to the scale factor)}:
\begin{equation}\label{Eq:BD.DC2}
\delta_{\rm m}^{\pp}+\left[\frac{3}{2a}\left(1-w(a)\right)+\frac{\mathcal{F}^\p}{2}-\frac{\varphi^\p}{2\varphi}\right] \delta_{\rm m}^\p-\frac{3}{2a^2}\frac{\rho_{\rm m}(a)}{\rho(a)}\delta_{\rm m}\left(1+\frac{\epsilon_{\rm BD}}{2}-\mathcal{F}\right)=0\,,
\end{equation}
where we have defined
\begin{equation}\label{Eq:BD.F}
\mathcal{F}=\mathcal{F}\left(\frac{\varphi^\p}{\varphi}\right)=-a\frac{\varphi^\p}{\varphi}+\frac{\omega_{\rm BD}}{6}a^2\left(\frac{\varphi^\p}{\varphi}\right)^2\,.
\end{equation}
In particular, we have expanded the effective gravitational coupling  \eqref{Eq:BD.LocalGNa} linearly as in \eqref{Eq:BD.LocalGN2}.
As we can show easily, the expression \eqref{Eq:BD.F} can be treated as a perturbation, since it is proportional to $\epsilon_{\rm BD}$. To prove this, let us borrow the solution for the matter-dominated epoch (MDE) derived from the analysis of fixed points presented in \hyperref[Appendix:FixedPoints]{Appendix\,\ref{Appendix:FixedPoints}}. Because the behavior of the BD-field towards the attractor at the MDE is governed by a power law of the form $\varphi\sim a^{\eBD}$ (cf. Eq.\,\eqref{Eq:FixedPoints.psiMDE}), we obtain
\begin{eqnarray}\label{Eq:BD.ximatter}
a\frac{\varphi^\p}{\varphi} &=& \epsilon_{\rm BD}+\mathcal{O}(\epsilon^2_{\rm BD})\,,\\
\mathcal{F}&=&-\frac{5}{6}\epsilon_{\rm BD}+\mathcal{O}(\epsilon^2_{\rm BD})\simeq -\frac{5}{6}\epsilon_{\rm BD}\,;\,\ \ \ \ \ \ \ \ \mathcal{F}^\p=\mathcal{O}(\epsilon^2_{\rm BD})\simeq 0\,.
\end{eqnarray}
This proves our contention that the function $\mathcal{F}$ in \eqref{Eq:BD.F} is of order $\eBD$ and its effects can be treated as a perturbation to the above formulas.
Incidentally, the relative change of $\varphi$ does not depend on $\varphi$ itself.
%
%
From the definition of $\mathcal{F}$ we can now refine the old Friedmann's equation \eqref{Eq:BD.H1} or \eqref{Eq:BD.H2} (only valids for $\eBD=0$) as follows. Starting from Eq.\,\eqref{Eq:BD.Friedmannequation}, it is easy to see that it can be cast in the Friedmann-like form:
\begin{equation}\label{Eq:BD.FriedmannWithF}
H^2(a)=\frac{8\pi G_N}{3\varphi (a) (1-\mathcal{F}(a))}\,\rho(a).
\end{equation}
Despite $\mathcal{F}(a)$ evolves with the expansion, as shown by \eqref{Eq:BD.F}, it is of order $\eBD$ and evolves very slowly. In this sense, Eq.\,\eqref{Eq:BD.FriedmannWithF} behaves approximately as an ${\cal O}(\eBD)$ correction to Friedmann's equation \eqref{Eq:BD.H1}.
Setting $a=1$, the value of the current Hubble parameter satisfies
\begin{equation}\label{Eq:BD.FriedmannWithFAtPresent}
H_0^2=\frac{8\pi G_N}{3\varphi_0 (1-\mathcal{F}_0)} \rho^0,
\end{equation}
where $\varphi_0 \equiv \varphi (a=1)$ and $\mathcal{F}_0 \equiv \mathcal{F}(a=1)$. The above equation implies that $\rho^0=\rho_c^0 \varphi_0 (1-\mathcal{F}_0 )$. We may now rewrite \eqref{Eq:BD.FriedmannWithF} in the suggestive form:
\begin{equation}\label{Eq:BD.H2withF}
H^2(a)= H_0^2\left[\hat{\Omega}_{\rm m}(a)a^{-3}+\hat{\Omega}_{\gamma}(a)a^{-4}+\hat{\Omega}_\nu(a)+\hat{\Omega}_\CC(a)\right]\,,
\end{equation}
provided we introduce the new `hatted' parameters $\hat{\Omega}_i(a)$, which are actually slowly varying functions of the scale factor:
\begin{figure}[t!]
\begin{center}
\includegraphics[width=4.2in, height=2.8in]{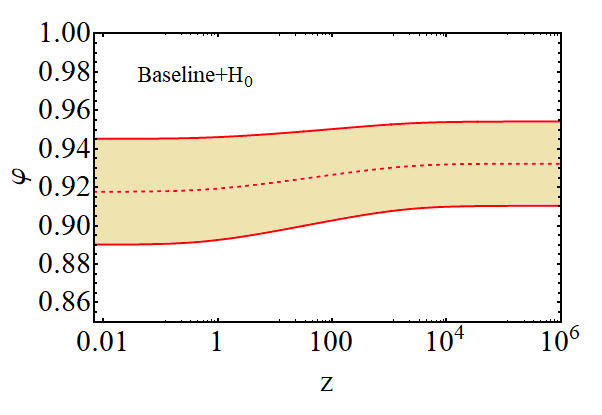}
\caption{Exact numerical analysis of the evolution of $\varphi$ as a function of the redshift across the entire cosmic history, starting from the radiation-dominated epoch up to our time. We use here the values of the BD-$\CC$CDM Baseline+$H_0$ dataset indicated in the figure itself (cf. \hyperref[Sect:MethodData]{Sect.\,\ref{Sect:MethodData}} and \hyperref[Sect:NumericalAnalysis]{Sect.\,\ref{Sect:NumericalAnalysis}}). In particular, $\epsilon_{\rm BD}=-0.00199^{+0.00142}_{-0.00147}$. The band around the central (dotted) curve shows the computed $1\sigma$ uncertainty from the Markov chains of our statistical analysis.}\label{Fig:BD.varphi}
\end{center}
\end{figure}
\begin{equation}\label{Eq:BD.hatOmega}
\hat{\Omega}_i (a)=\frac{\Omega_i}{\varphi (a)}\frac{1}{1-\mathcal{F}(a)}\simeq \frac{\Omega_i}{\varphi (a)}\left(1+\mathcal{F}(a)\right) \,,
\end{equation}
with $\Omega_i = \rho^0_i /\rho_c^0$ as previously. These functions also satisfy, exactly, the canonical sum rule at present: %
$\sum_i\hat{\Omega}_i(a=1)=1$.
For $\eBD=0$, the hatted parameters reduce to the old tilded ones \eqref{Eq:BD.tildeOmegues}, $\hat{\Omega}_i=\tilde{\Omega}_i$, and for typical values of $|\epsilon_{\rm BD}|\sim \mathcal{O}(10^{-3})$ the two sets of parameter differ by $\mathcal{O}(\epsilon_{\rm BD})$ only:
\begin{equation}\label{Eq:BD.hatOmega2}
\hat{\Omega}_i (a)=\frac{\Omega_i}{\varphi}+\mathcal{O}(\epsilon_{\rm BD})=\tilde{\Omega}_i+\mathcal{O}(\epsilon_{\rm BD}) \,.
\end{equation}
From \eqref{Eq:BD.DC2} we obviously recover  the previous Eq.\,\eqref{Eq:BD.DC} in the limit $\epsilon_{\rm BD}\to 0$, and it is easy to see that for non-null values of $\epsilon_{\rm BD}$ the density contrast acquires a dependence on the ratio $\varphi^\p/\varphi$ and its derivative, so it is sensitive to the relative change of $\varphi$ with the expansion.  Its time evolution is now possible by virtue of the third BD-field equation \eqref{Eq:BD.FieldeqPsi}, which can be expanded linearly in $\eBD$ in a similar way. After some calculations, we find
\begin{equation}\label{Eq:BD.FieldeqPsiapprox}
\varphi^\pp+\frac{1}{2a}\left(5-3w(a)\right)\varphi^\p=\frac{3\eBD}{2a^2}(1-3w(a))\varphi\,.
\end{equation}
{For $\eBD=0$ we recover the solution $\varphi=$const. In the radiation dominated epoch (RDE), $w\simeq 1/3$,  the RHS vanishes and in this case $\varphi$ need not be constant.  It is easy to see that the exact solution of this equation in that epoch is
\begin{equation}\label{Eq:BD.varphiRDE}
\varphi(a)=\varphi^{(0)}+\frac{\varphi^{(1)}}{a}\,,
\end{equation}
for arbitrary constants $\varphi^{(i)}$. The variation during the RDE is therefore very small since the dominant solution is a constant and the variation  comes only through a decaying mode $1/a\sim t^{-1/2}$ ($t$ is the cosmic time). For the MDE (for which $w=0$) there is some evolution, once more with a decaying mode but then through a sustained logarithmic term:
\begin{equation}\label{Eq:BD.varphiMDE}
\varphi(a)\sim \varphi^{(0)}\left(1+\eBD\ln a\right)+ \varphi^{(1)}a^{-3/2}\rightarrow \varphi^{(0)} \left(1+\eBD\ln a\right) \,,
\end{equation}
where coefficient $\varphi^{(0)}$ is to be adjusted from the boundary conditions between epochs\footnote{Approximate solutions to the BD-field equations for the main cosmological variables  in the different epochs are discussed in \hyperref[Appendix:Semi-Analytical]{Appendix\,\ref{Appendix:Semi-Analytical}}.}. The dynamics of $\varphi$ for $\eBD\neq0$ is actually mild in all epochs since $\eBD$ on the RHS of Eq.\,\eqref{Eq:BD.FieldeqPsiapprox} is small.
However mild it might be, the  dynamics of $\varphi$ modifies both the friction and Poisson terms in Eq. \eqref{Eq:BD.DC2}, and it is therefore  of pivotal importance to understand what are the changes that are induced by positive and negative values of $\epsilon_{\rm BD}$ on these terms during the relevant epochs of the structure formation history. An exact (numerical) solution is displayed in \hyperref[Fig:BD.varphi]{Fig.\,\ref{Fig:BD.varphi}}, where we can see that $\varphi$ remains within the approximate interval  $0.918\lesssim\varphi\lesssim 0.932$ for the entire cosmic history (starting from the RDE up to our time).  This plot has been obtained from the overall numerical fit performed to the observational data used in this analysis within one of the BD-$\CC$CDM baseline datasets considered (cf. \hyperref[Sect:NumericalAnalysis]{Sect.\,\ref{Sect:NumericalAnalysis}}). The error band around the main curve includes the $1\sigma$-error computed from our statistical analysis. Two very important things are to be noted at this point: on the one hand the variation of $\varphi$  is indeed small, and on the other hand $\varphi<1$, and hence $G(\varphi)>G_N$ for the whole cosmic span.}

\begin{figure}[t!]
\begin{center}
\includegraphics[width=6.2in, height=2.3in]{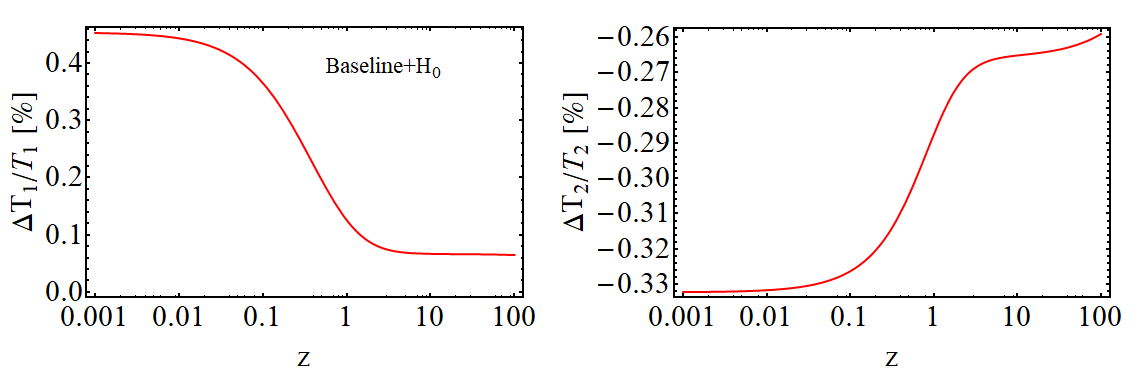}
\caption{{\it Left plot:} Relative difference between the friction term $T_1$ of Eq. \protect\eqref{Eq:BD.DC2}, $\delta_{\rm m}^\pp+T_1\delta_{\rm m}^\p-T_2\delta_{\rm m}=0$, using the best-fit values of the BD-$\CC$CDM model obtained with the Baseline+$H_0$ dataset, {\it i.e.} with $\epsilon_{\rm BD}=-0.00199$, and the case with $\epsilon_{\rm BD}=0$ (cf. \hyperref[Table:BD.TableFitBaseline]{Table\,\ref{Table:BD.TableFitBaseline}} for the values of the other parameters); {\it Right plot:} The same, but for the last term of Eq. \protect\eqref{Eq:BD.DC2}, $T_2$ (Poisson term). We can clearly appreciate that negatives values of $\epsilon_{\rm BD}$ produce a higher Hubble friction, {\it i.e.} a higher $T_1$, and a lower $T_2$, {\it w.r.t.} the $\epsilon_{\rm BD}=0$ case. Both things lead to a decrease of $\delta_{\rm m}$. The two plots have been obtained with our modified version of \texttt{CLASS}. See the main text for further details.}\label{Fig:BD.termsdeltaEq}
\end{center}
\end{figure}

We can start considering what is the influence of the scalar field dynamics on the perturbations during the pure MDE. Using the relations \eqref{Eq:BD.ximatter} and \eqref{Eq:BD.F} in \eqref{Eq:BD.DC2} we find ({setting $w=0$, $\rho\simeq \rho_{\rm m}$ and neglecting $\rL$ in the MDE}):
\begin{equation}\label{Eq:BD.DCmatter}
\delta_{\rm m}^{\pp}+\frac{\delta_{\rm m}^\p}{2a}(3-\epsilon_{\rm BD})-\frac{3}{2a^2}\delta_{\rm m}\left(1+\frac{4\epsilon_{\rm BD}}{3}\right)=0\,.
\end{equation}
{If $\epsilon_{\rm BD}<0$ the Poisson term (the last in the above equation) decreases and, on top of that, the Hubble friction increases {\it w.r.t.} the case with $\varphi=$const. (or the GR-$\CC$CDM model, if we consider the same energy densities). Both effects help to slow down the structure formation in the Universe. Of course, if $\epsilon_{\rm BD}>0$ the opposite happens. This is confirmed by solving explicitly Eq.\,\eqref{Eq:BD.DCmatter}.  Despite an exact solution to Eq.\,\eqref{Eq:BD.DCmatter} can be found, it suffices to quote it at $\mathcal{O}(\epsilon_{\rm BD})$ and neglect the $\mathcal{O}(\epsilon^2_{\rm BD})$ corrections.  The growing and  decaying modes at leading order read  $\delta^+_{\rm m}(a)\sim a^{1+\eBD}$ and  $\delta^-_{\rm m}(a)\sim a^{-\frac12(3+\eBD)}$, respectively. The latter just fades soon into oblivion  and the former explains why negative values of $\epsilon_{\rm BD}$ are favored by the data on RSD, since $\eBD<0$ obviously slows down the rate of structure formation and hence acts as an effective (positive) contribution to the vacuum energy density\footnote{This feature was already noticed in the preliminary treatment of Ref.\,\cite{deCruzPerez:2018cjx} for the BD theory itself, and it was actually pointed out as a general feature of the class of Running Vacuum Models (RVM), which helps to cure the $\sigma_8$-tension\,\cite{Gomez-Valent:2018nib,Gomez-Valent:2017idt}. This is remarkable, since the RVM's turn out to mimic BD-gravity, as first noticed in \cite{SolaPeracaula:2018dsw}. For a summary, see \hyperref[Sect:RVMconnection]{Appendix\,\ref{Sect:RVMconnection}}.}. The preference for negative values of $\eBD$ is especially clear when the RSD data include the matter bispectrum information, which tends to accentuate the slowing down of the growth function, as noted repeatedly in a variety of previous works\,\cite{SolaPeracaula:2016qlq, SolaPeracaula:2017esw, Sola:2017znb, Sola:2016jky}. We may clearly appraise this feature also in the present study, see e.g. Tables \hyperref[Table:BD.TableFitBaseline]{\ref{Table:BD.TableFitBaseline}} and \hyperref[Table:BD.TableFit+SL]{\ref{Table:BD.TableFit+SL}} (with spectrum {\it and} bispectrum) and \hyperref[Table:BD.TableFitSpectrum]{Table\,\ref{Table:BD.TableFitSpectrum}} (with spectrum but no bispectrum), where the $\sigma_8$ value is in general well-behaved ($\sigma_8\simeq 0.8$) in the BD-$\CC$CDM framework when $\eBD<0$, but it is clearly reduced (at a level $\sigma_8\simeq 0.78-0.79$)  in the presence of bispectrum data.  And in both cases the value of $H_0$  is in the range of $70-71$ km/s/Mpc. Most models trying to explain both tensions usually increase $\sigma_8$ substantially  ($0.82-0.85$).}

We can also study the pure vacuum-dominated epoch (VDE) in the same way. In this case $\varphi\sim a^{2\eBD}$ (cf. \hyperref[Appendix:FixedPoints]{Appendix\,\ref{Appendix:FixedPoints}}), and hence
\begin{equation}\label{Eq:BD.xivacuum}
a\frac{\varphi^\p}{\varphi} = 2\epsilon_{\rm BD}+\mathcal{O}(\epsilon^2_{\rm BD})\simeq 2\epsilon_{\rm BD}\,,
\end{equation}
again with $\mathcal{F}^\p=\mathcal{O}(\epsilon^2_{\rm BD})\simeq 0$. The Poisson term can be neglected in this case since $\rho_{\rm m}\ll\rho\simeq \rL$, and hence,
\begin{equation}\label{Eq:BD.DCvacuum}
\delta_{\rm m}^{\pp}+\frac{\delta_{\rm m}^\p}{a}(3-\epsilon_{\rm BD})=0\,.
\end{equation}
When the vacuum energy density rules the expansion of the Universe, there is a stable constant mode solution $\delta_{\rm m}=$const. and a decaying mode that decreases faster than in the GR scenario if $\epsilon_{\rm BD}<0$, again due to the fact that the friction term is in this case larger than in the standard picture, specifically the latter reads $\delta^-_{\rm m}(a)\sim a^{-2+\epsilon_{\rm BD}}$ in the ${\cal  O}(\eBD)$ approximation.
\begin{figure}[t!]
\begin{center}

\includegraphics[width=6in, height=2.1in]{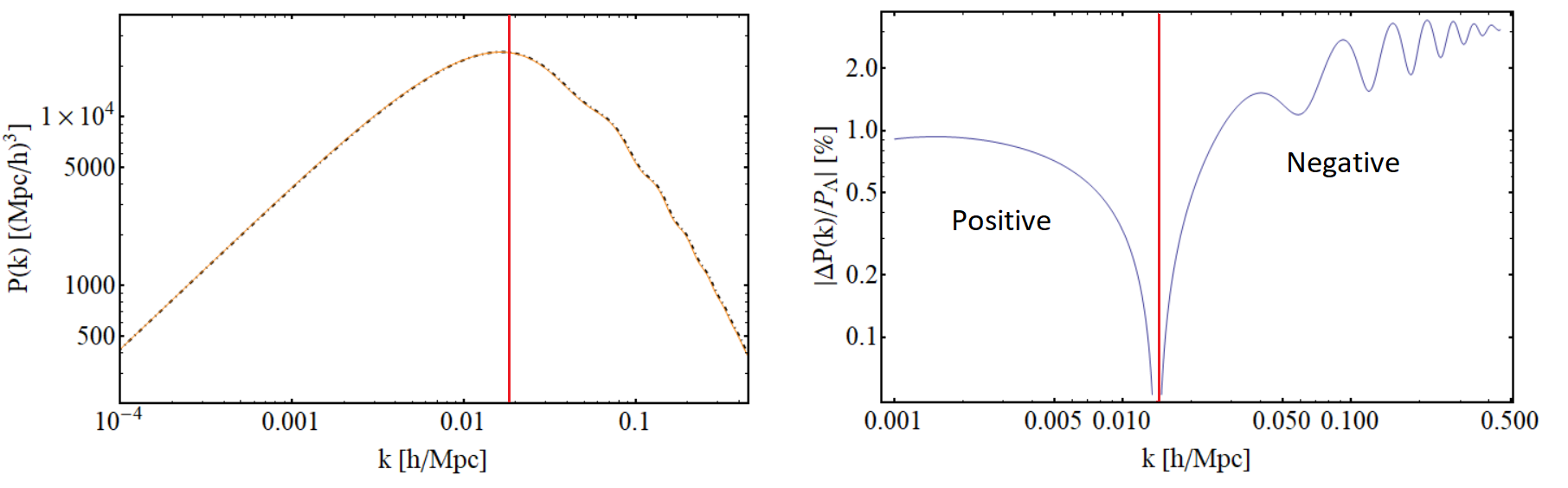}
\caption{As in \hyperref[Fig:BD.pkns]{Fig.\,\ref{Fig:BD.pkns}}, but comparing now the $\CC$CDM baseline configuration defined there with the case $\epsilon_{\rm BD}=-0.003$ (and $\varphi_{\rm ini}=1$) of the BD-$\CC$CDM model.}\label{Fig:BD.pkepsilon}
\end{center}
\end{figure}
The analytical study of the transition between the matter and vacuum-dominated epochs is more difficult, but with what we have already seen it is obvious that the amount of structure generated also in this period of the cosmic expansion will be lower than in the $\varphi=$const. case if $\epsilon_{\rm BD}$ takes a negative value. In \hyperref[Fig:BD.termsdeltaEq]{Fig.\,\ref{Fig:BD.termsdeltaEq}} we show this explicitly.

From this analysis it should be clear that if $\epsilon_{\rm BD}<0$ there is a decrease of the matter density contrast for fixed energy densities when compared with the GR-$\Lambda$CDM scenario, and also with the BD scenario with $\varphi=const$. In \hyperref[Fig:BD.pkepsilon]{Fig.\ref{Fig:BD.pkepsilon}} we can see this feature directly in the matter power spectrum, which is  seen to be suppressed with respect to the case $\epsilon_{\rm BD}=0$ {for those scales that are relevant for the RSD, {\it i.e.} within the range of wave numbers  $0.01 h {\rm Mpc^{-1}}\lesssim k\lesssim 0.1 h {\rm Mpc^{-1}}$  (corresponding to distance scales roughly between a few dozen to a few hundred Mpc)}. However, we still don't know whether these negative values of $\epsilon_{\rm BD}$ can also be accommodated by the other datasets.

\begin{figure}[t!]
\begin{center}
\includegraphics[width=6in, height=4in]{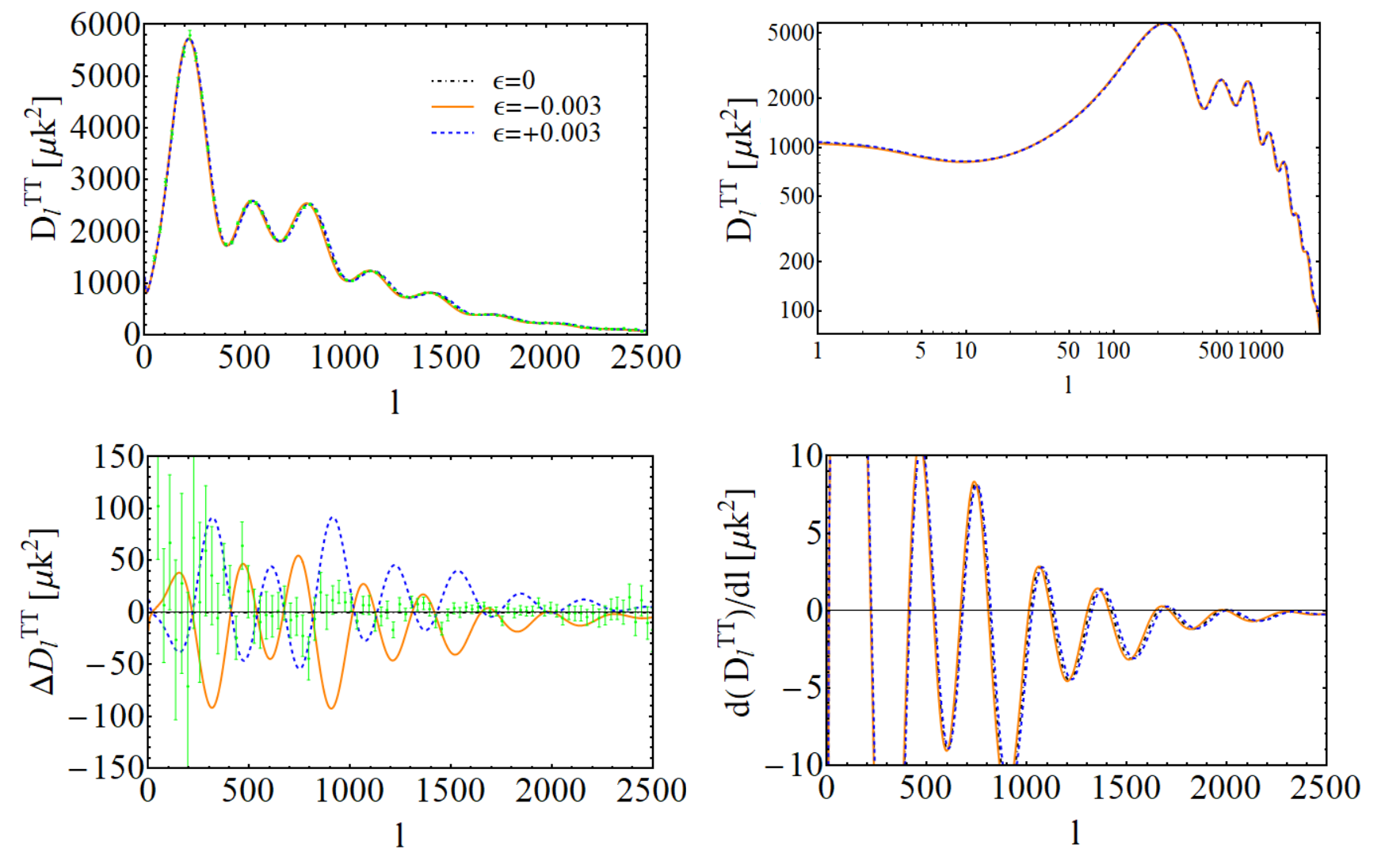}
\caption{As in \hyperref[Fig:BD.Clsvarphi]{Fig.\,\ref{Fig:BD.Clsvarphi}}, but comparing the GR-$\CC$CDM baseline scenario (equivalent to the BD-$\CC$CDM one for  $\epsilon_{\rm BD}=0$, $\varphi=1$) with the case $\epsilon_{\rm BD}=\pm 0.003$, using as initial condition $\varphi_{\rm ini}=1$.}\label{Fig:BD.Cls-epsilon}
\end{center}\end{figure}

To check this, let us recall from our discussion above (see also \hyperref[Appendix:Semi-Analytical]{Appendix\,\ref{Appendix:Semi-Analytical}} and \hyperref[Appendix:FixedPoints]{Appendix\,\ref{Appendix:FixedPoints}} for more details) that the evolution of the BD-field takes place basically during the MDE.  In the RDE the scalar field is essentially frozen once the decaying mode becomes irrelevant, and although in the late-time Universe $\varphi$ evolves faster than in the MDE (compare \eqref{Eq:BD.ximatter} and \eqref{Eq:BD.xivacuum}) it remains almost constant in the redshift range $z\lesssim \mathcal{O}(1)$, in which all the non-CMB data points lie, particularly the LSS data.  {Let us note that the typical values of $\eBD$  fitted from the overall set of data used in our analysis (cf. \hyperref[Sect:MethodData]{Sect.\,\ref{Sect:MethodData}}) place that parameter  in the ballpark  of $|\epsilon_{\rm BD}|\sim \mathcal{O}(10^{-3})$ (cf.Tables \hyperref[Table:BD.TableFitBaseline]{\ref{Table:BD.TableFitBaseline}},\hyperref[Table:BD.TableFit+SL]{\ref{Table:BD.TableFit+SL}} and \hyperref[Table:BD.TableFitSpectrum]{\ref{Table:BD.TableFitSpectrum}}, for example).  Schematically, we can think that the field takes a value $\varphi_{\rm ini}$ during the RDE, then if $\epsilon_{\rm BD}<0$ it decreases an amount $\Delta\varphi_{\rm ini}<0$  with respect to its original value $\varphi_{\rm ini}$, during the MDE, and  finally $\varphi^{0}\approx\varphi_{\rm ini}+\Delta\varphi_{\rm ini}$, with $\varphi(z)\approx\varphi^0$ at $z\lesssim\mathcal{O}(1)$}. Thus, Eq. \eqref{Eq:BD.H1} still applies in good approximation during the RDE and the late-time Universe, but with two different values of the cosmological Newton's coupling in the two (widely separated) cosmological eras. Taking these facts into account it is easy to understand why SNIa data are not able to tell us anything about $\epsilon_{\rm BD}$ when used alone, since again $\varphi^{0}$ is fully degenerated with the absolute magnitude parameter $M$ of the supernovae. This does not mean, though, that SNIa data cannot tell us anything about the $H_0$-tension when they are considered together with other datasets that do provide constraints on $H_0$, since the constraints that SNIa impose on $\Om$ help to break degeneracies present in the other datasets and to tighten the allowed region of parameter space. H0LICOW and CCH data, for instance, will allow us to put constraints again on $\Om$ and also on $H_0$ (hence on $\rho^{0}/\varphi^{0}$). BAO will constrain $\rho_{\rm b}^{0}$, $\Om$, and now also $\Delta\varphi_{\rm ini}/\varphi_{\rm ini}$, which is proportional to $\epsilon_{\rm BD}$. For instance, $r_{\rm s} H(z)\propto \left(1-\frac{\Delta\varphi_{\rm ini}}{2\varphi_{\rm ini}}\right)$.

The BD effect caused on the temperature spectrum of CMB anisotropies is presented in the fourfold plot in \hyperref[Fig:BD.Cls-epsilon]{Fig.\,\ref{Fig:BD.Cls-epsilon}}. Due to the fact that now we have $\epsilon_{\rm BD}\ne 0$, $\Delta\varphi\ne 0$,  a small shift in the location of the peaks is naturally generated. In the right-bottom plot of such figure one can see that negative values of $\epsilon_{\rm BD}$ move the peaks slightly towards lower multipoles, and the other way around for positive values of this parameter. It is easy to understand why.  To start with, let us remark that we have produced all the curves of this plot using the same initial condition $\varphi_{\rm ini}=1$ and the same $\rho^0_{\rm b}$, $\rho^0_{cdm}$ and $\rho_\Lambda$ and, hence, fixing in the same way the complete evolution of the energy densities for the different plots in that figure. This means that the differences in the Hubble function can only be due to differences in the evolution of the BD scalar field. The modified expansion histories produce changes in the value of $\theta_*=r_{\rm s}/D_{A,rec}$ (with $D_{A,rec}$ being the angular diameter distance to the last scattering surface), so also in the location of the peaks. If $\epsilon_{\rm BD}<0$,  $\varphi$ decreases with the expansion, so its value at recombination and at present is lower than when $\epsilon_{\rm BD}=0$, and correspondingly $G(\varphi)$ will be higher. Because of this, the value of the Hubble function will be larger, too, and the cosmological distances lower, so the relation between $\theta_*(\epsilon_{\rm BD}\ne 0)$ and  $\theta_*(\epsilon_{\rm BD}= 0)$ can be written as follows:
\begin{equation}
\theta_*(\epsilon_{\rm BD}\ne 0)=\frac{r_{\rm s}(\epsilon_{\rm BD}\ne 0)}{D_{A,rec}(\epsilon_{\rm BD}\ne 0)}=\frac{X\cdot r_{\rm s}(\epsilon_{\rm BD}= 0)}{Y\cdot D_{A,rec}(\epsilon_{\rm BD}=0)}= \frac{X}{Y}\,\theta_*(\epsilon_{\rm BD}=0)\,,
\end{equation}
{where the rescaling factors satisfy  $0<X,Y<1$ for $\epsilon_{\rm BD}<0$. As we have already mentioned before, most of the variation of $\varphi$ occurs during the MDE, so the largest length reduction will be in the cosmic stretch from recombination to the present time, and thereby  $Y<X$}. Thus, if $\epsilon_{\rm BD}< 0$ we find $\theta_*(\epsilon_{\rm BD}< 0)>\theta_*(\epsilon_{\rm BD}=0)$ and the peaks of the TT CMB spectrum shifts towards lower multipoles. Analogously, if $\epsilon_{\rm BD}$ is positive $X,Y>1$, with $Y>X$, so $\theta_*(\epsilon_{\rm BD}> 0)<\theta_*(\epsilon_{\rm BD}=0)$ and the peaks move to larger multipoles. It turns out, however, that these shifts, and also the changes in the amplitude of the peaks, can be compensated by small changes in the baryon and DM energy densities, as we will show in \hyperref[Sect:NumericalAnalysis]{Sect.\,\ref{Sect:NumericalAnalysis}}.

At this point we would like to recall why in the GR-$\Lambda$CDM concordance model it is not possible to reconcile the local measurements of $H_0$ with its CMB-inferred value. In the concordance model, which we assume spatially flat,  the current value of the cold dark matter density is basically fixed by the amplitude of the first peak of the CMB temperature anisotropies, and $\rho_{\rm b}^0$ by the relative amplitude of the second and third peaks with respect to the first one. {As a result, even if the cosmological term plays no role in the early Universe, one finds that  in order to explain the precise location of the CMB peaks the value of $\rL$ obtained from the matching of the predicted  $\theta_*$ with the measured peak positions determines $\rL$ so precisely that it leaves little margin. This causes a problem since such narrow range of values is not in the right range to explain the value of the current Hubble parameter measured with the cosmic distance ladder technique \cite{Riess:2018uxu,Riess:2019cxk,Reid:2019tiq} and the Strong-Lensing time delay angular diameter distances from H0LICOW \cite{Wong:2019kwg}. In other words,  despite the concordance model fits in a remarkably successful way the CMB and BAO, and also the SNIa data, there is an irreducible internal discordance in the parameter needs to explain with precision both the physics of the early and of the late-time Universe. This is of course the very expression of the  $H_0$-tension, to which we have to add the $\sigma_8$ one.}

Cosmographic analyses based on BAO and SNIa data calibrated with the GR-$\Lambda$CDM Planck preferred value of $r_{\rm s}$ also lead to low estimates of $H_0$. This is the so-called inverse cosmic distance ladder approach, adopted for instance in \cite{Aubourg:2014yra,Bernal:2016gxb,Feeney:2018mkj,Macaulay:2018fxi}. This has motivated cosmologists to look for alternative theoretical scenarios (for instance, the generic class of  EDE  proposals)  able to increase the expansion rate of the Universe before the decoupling of the CMB photons and, hence, to lower $r_{\rm s}$ down.  {This, in principle, demands an increase of the Hubble function at present in order not to spoil the good fit to the BAO and CMB observables.} {Nevertheless, not all the models passing the BAO and CMB constraints and predicting a larger value of $H_0$ satisfy the `golden rule' mentioned in the Introduction of the chapter, since they can lead e.g. to a worsening of the $\sigma_8$-tension}.  As an example, we can mention some early DE models, e.g. those discussed in \cite{Poulin:2018cxd}. In these scenarios there is a very relevant DE component which accounts for the $\sim 7\%$ of the total matter-energy content of the Universe at redshifts $\sim 3000-5000$, before recombination. This allows of course to enhance the expansion rate and reduce $r_{\rm s}$. After such epoch, the DE decays into radiation. In order not to alter the position of the CMB peaks and BAO relative distances, an increase of the DM energy density is needed. According to \cite{Hill:2020osr}, this leads to an excess of density power and an increase of $\sigma_8$ which is not welcome by LSS measurements, including RSD, Weak-Lensing and galaxy clustering data.  Another example is the interesting modified gravity model analyzed in \cite{Ballesteros:2020sik}, based on changing the cosmological value of $G$ also in the pre-recombination era, thus mimicking an increase of the effective number of relativistic degrees of freedom in such epoch. The additional component gets eventually diluted at a rate faster than radiation in the MDE and it is not clear if an effect is left at present\footnote{{In stark contrast to the model of \cite{Ballesteros:2020sik}, in BD-$\Lambda$CDM cosmology the behavior of the effective $\rho_{\rm BD}$  (acting as a kind of additional DE component during the late Universe) mimics pressureless matter during the MDE epoch and modifies the effective EoS of the DE at present, see the next \hyperref[Sect:EffectiveEoS]{Sect.\,\ref{Sect:EffectiveEoS}} for details.}}. This model also fits the CMB and BAO data well and loosens at some extent the $H_0$-tension, but violates the golden rule of the tension solver, as it spoils the structure formation owing to the very large values of $\sigma_8\sim 0.84-0.85$ that are predicted (see the discussion in\hyperref[Sect:ConsiderationsBD]{\,\,Sect.\,\ref{Sect:ConsiderationsBD}} for more details).

Our study shows that a value of the cosmological gravitational coupling about $\sim 10\%$ larger than $G_N$ can ameliorate in a significant way the $H_0$-tension, while keeping the values of all the current energy densities very similar to those found in the GR-$\Lambda$CDM model. If, apart from that, we also allow for a very slow running (increase) of the cosmological $G$ triggered by negative values of order $\epsilon_{\rm BD}\sim -\mathcal{O}(10^{-3})$, we can  mitigate at the same time the $\sigma_8$-tension when only the CMB TT+lowE anisotropies are considered. When the CMB polarizations and lensing are also included in the analysis, then $\sigma_8$ is kept at the GR-$\Lambda$CDM levels, and the sign of $\epsilon_{\rm BD}$ is not conclusive.  {In all situations we can preserve the golden rule}. We  discuss in detail the numerical  results of our analysis  in \hyperref[Sect:NumericalAnalysis]{Sect.\,\ref{Sect:NumericalAnalysis}}.

\section{Effective equation of state of the dark energy in the BD-$\CC$CDM model}\label{Sect:EffectiveEoS}
\noindent
Our aim in this section is to write down the Brans-Dicke cosmological equations \eqref{Eq:BD.Friedmannequation}-\eqref{Eq:BD.FieldeqPsi} in the context of what we may call the ``effective GR-picture''. This means to rewrite them  in such a way that they can be thought of as an effective model within the frame of GR, thus providing a parametrized departure from GR at the background level. We will see that the main outcome of this task, at least qualitatively, is that the BD-$\CC$CDM model (despite it having a constant vacuum energy density $\rL$) appears as one in the GR class, but with a dynamical DE rather than a CC. The dynamics of such an effective form of DE is a function of the BD-field $\varphi$. We wish to compute its effective EoS.  In order to proceed, the first step is to  rewrite Eq.\, \eqref{Eq:BD.Friedmannequation} \`a la  Friedmann:
\begin{equation}\label{Eq:BD.BDFriedmann}
3H^2 = 8\pi{G_N}(\rho + \rvphi)\,,
\end{equation}
where $\rho$ is the total energy density as defined previously (coincident with that of the GR-$\CC$CDM model),  and  $\rvphi$ is the additional ingredient that is needed, which reads
\begin{equation}\label{Eq:BD.rhoBD}
\rvphi\equiv \frac{3}{8\pi{G_N}}\left(H^2\dvphi - H\dot{\varphi} + \frac{\omega_{\rm BD}}{6}\frac{\dot{\varphi}^2}{\varphi}\right).
\end{equation}
Remember the definition $\varphi(t) \equiv G_N\psi(t)$ made in \eqref{Eq:BD.definitions}, and we have now introduced
\begin{equation}\label{Eq:BD.Deltaphi}
\dvphi(t)\equiv 1-\varphi(t)\,,
\end{equation}
which tracks the small departure of $\varphi$ from one and hence of $G(\varphi)$ from $G_N$  (cf. \hyperref[Sect:BDgravity]{Sect.\,\ref{Sect:BDgravity}}). Note that $\varphi=\varphi(t)$ evolves in general with the expansion, but very slowly  since $\eBD$ is presumably fairly small.

From the above  Eq.\,\eqref{Eq:BD.rhoBD} it is pretty clear that we have absorbed all the terms beyond the $\Lambda$CDM model into the expression of $\rvphi$. While it is true that we define this quantity as if it were an energy density, it is important to bear in mind that it is not associated to any kind of particle, it is just a way to encapsulate those terms that are not present in the standard model.  This quantity, however, satisfies a local conservation law as if it were a real energy density, as we shall see in a moment.  From the generalized Friedmann equation \eqref{Eq:BD.BDFriedmann} and the explicit expression for $\rvphi$ given above we can write down the generalized cosmic sum rule verified by the BD-$\CC$CDM model in the effective GR-picture:
\begin{equation}\label{Eq:BD.SumRuleBDexact}
\Omega_{\rm m}+\Omega_{\rm r}+\Omega_\CC+\Omega_\varphi=1\,,
\end{equation}
where the $\Omega_i$ are the usual (current) cosmological parameters of the concordance $\CC$CDM, whereas $\Omega_\varphi$ is the additional one that parametrizes the departure of the  BD-$\CC$CDM model from the  GR-$\CC$CDM  in the context of the GR-picture, and reads
\begin{equation}\label{Eq:BD.OmegaBD}
\Omega_{\varphi}=\frac{\rho_{\varphi}^0}{\rco}.
\end{equation}
Notice that the above sum rule is exact and it is different from that in Eq.\,\eqref{Eq:BD.SumRuleBD} since the latter is only approximate for the case when $\eBD=0$ or very small.  These are two different pictures of the same BD-$\CC$CDM model. The modified cosmological parameters (\ref{Eq:BD.tildeOmegues}) depend on $\varphi$ whereas here the $\varphi$-dependence has been fully concentrated on $\Omega_\varphi$.  It is interesting to write down the exact equation \eqref{Eq:BD.SumRuleBDexact} in the form
\begin{equation}\label{Eq:BD.OMegaBD}
\Omega_{\rm m}+\Omega_{\rm r}+\Omega_\CC=1-\Omega_{\varphi}= 1-\Delta\varphi_0+\frac{\dot{\varphi}_0}{H_0}-\frac{\wBD}{6}\frac{\dot{\varphi}_0^2}{H_0^2\varphi_0}\,.
\end{equation}
in which $\varphi_0=\varphi(z=0)$ and $\dot{\varphi}_0=\dot{\varphi}(z=0)$. For $\eBD\simeq 0$ we know that $\varphi\simeq$const. and we can neglect the time derivative terms and then we find the approximate form $\Omega_{\rm m}+\Omega_{\rm r}+\Omega_\CC=1-\Omega_{\varphi}\simeq  1-\Delta\varphi_0$.
This equation suggests that a value of $\Delta\varphi_0\neq 0$ would emulate the presence of  a small  fictitious spatial curvature in the GR-picture. See e.g.\cite{Park:2019emi,Khadka:2020vlh,Cao:2020jgu} and references therein for the study of a variety models explicitly involving spatial curvature.

The second step in the process of constructing the GR-picture of the BD theory  is to express \eqref{Eq:BD.pressureequation} as in the usual pressure equation for GR, and this forces us to define a new pressure quantity $p_{\varphi}$ associated to $\rvphi$. We find
\begin{equation}\label{Eq:BD.PressureEq}
2\dot{H} + 3H^2 = -8\pi{G_N}(p + \pvphi),
\end{equation}
with
\begin{equation} \label{Eq:BD.pBD}
p_{\varphi}\equiv \frac{1}{8\pi{G_N}}\left(-3H^2\dvphi -2\dot{H}\dvphi  + \ddot{\varphi} + 2H\dot{\varphi} + \frac{\omega_{\rm BD}}{2}\frac{\dot{\varphi}^2}{\varphi}\right).
\end{equation}
{On the face of the above definitions \eqref{Eq:BD.rhoBD} and \eqref{Eq:BD.pBD}, we can now interpret the BD theory as an effective theory within the frame of General Relativity, which deviates from it an amount $\dvphi$. Indeed, for $\dvphi=0$ we have $\rvphi=\pvphi=0$ and we recover GR.  Mind that $\dvphi=0$ means not only that $\varphi=$const (hence that $\eBD=0$, equivalently $\wBD \to\infty$), but also that that constant is exactly $\varphi=1$.  In such case $\Geff$ is also constant and $\Geff=G_N$ exactly. The price that we have to pay for such a GR-like description of the BD model is the appearance of the fictitious BD-fluid with energy density $\rvphi$ and pressure $\pvphi$, which complies with the following conservation equation throughout the expansion of the Universe}\,\footnote{{The new `fluid' that one has to add to GR to effectively mimic BD plays a momentous role to explain the $H_0$ and $\sigma_8$-tensions. In a way it mimics the effect of the `early DE' models mentioned in the previous section, except that the BD-fluid persists for the entire cosmic history and is instrumental both in the early as well as in the current Universe so as to preserve the golden rule of the tension solver: namely, it either smoothes the two tensions of GR or improves one of them without detriment of the other. }}:
\begin{equation}
\dot{\rho}_{\varphi}+3H(\rho_{\varphi}+p_{\varphi})= 0.
\end{equation}
{One can check that this equation holds after a straightforward calculation, which makes use of the three  BD-field equations \eqref{Eq:BD.Friedmannequation}-\eqref{Eq:BD.FieldeqPsi}.}
Although at first sight the above conservation equation can be surprising actually it is not, since it is a direct consequence of the Bianchi identity. Let us now assume that the effective BD-fluid can be described by an equation of state like $\pvphi = w_{\varphi}\rvphi$, so
\begin{equation}\label{BDEoS0}
w_{\varphi}=\frac{\pvphi}{\rvphi}=\frac{-3H^2\dvphi -2\dot{H}\dvphi  + \ddot{\varphi} + 2H\dot{\varphi} + \frac{\omega_{\rm BD}}{2}\frac{\dot{\varphi}^2}{\varphi}}{3H^2\dvphi - 3H\dot{\varphi} + \frac{\omega_{\rm BD}}{2}\frac{\dot{\varphi}^2}{\varphi}}.
\end{equation}
The contribution from those terms containing derivatives of the BD-field are subdominant for the whole cosmic history.  We have verified this fact numerically, see \hyperref[Fig:BD.varphi]{Fig.\,\ref{Fig:BD.varphi}}. While the variation of $\varphi$ between the two opposite ends of the cosmic history is of $\sim 1.5\%$ and is significant for our analysis, the instantaneous variation is actually negligible. Thus,  $H\dot{\varphi}$ and  $\ddot{\varphi}$  are both much smaller than $\dot{H}\dvphi$,  and  in this limit we can approximate \eqref{BDEoS0} very accurately as
\begin{equation}\label{Eq:BD.BDEoS}
w_{\varphi}(t) \simeq -1 - \frac{2}{3}\frac{\dot{H}}{H^2}\,\ \ \ \ \ \ ({\rm for}\ H\dot{\varphi}, \ddot{\varphi}\ll \dot{H}\dvphi)\,.
\end{equation}
This EoS turns  out to be the standard total EoS of the  $\Lambda$CDM, which boils down to the EoS corresponding to the different epochs of the cosmic evolution ({\it i.e.} $w=1/3, 0, -1$ for RDE, MDE and VDE).  This means that the EoS of the  BD-fluid mimics these epochs.   We can go a step further and define not just the BD-fluid but the combined system of the BD-fluid and the vacuum energy density represented by the density $\rL$ associated to the cosmological constant.  {We define the following effective EoS for such combined fluid:}
\begin{figure}[t!]
\begin{center}
\includegraphics[width=6.8in, height=2.8in]{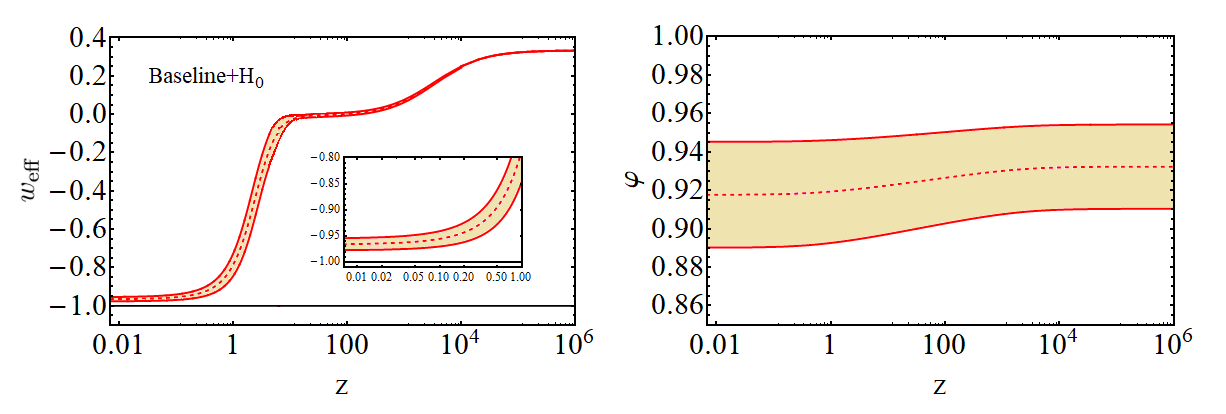}
\caption{{\it Left plot:} Effective equation of state of the DE in the BD-$\CC$CDM model as a function of the redshift. The inner plot magnifies the region around our time. We can see that the BD model mimics quintessence with a significance of more than $3\sigma$; {\it Right plot:} It shows once more \hyperref[Fig:BD.varphi]{Fig.\,\ref{Fig:BD.varphi}} in order to ease the comparison of the EoS evolution, which is associated to the evolution of the BD-field $\varphi$ -- cf. Eq.\eqref{Eq:BD.effEoS}. The shadowed bands in these plots correspond to the 1$\sigma$ regions.}\label{Fig:BD.weffvarphi}
\end{center}
\end{figure}
%
\begin{equation}\label{Eq:BD.effEoS}
\weff\equiv \frac{p_\Lambda + p_{\varphi}}{\rho_\Lambda + \rho_{\varphi}} = -1 + \frac{p_{\varphi} + \rho_{\varphi}}{\rho_\Lambda+ \rho_{\varphi}}=-1+\frac{-2\dot{H}\dvphi+f_1(\varphi,\dot{\varphi},\ddot{\varphi})}{\CC+ 3 H^2\,\dvphi+f_2(\varphi,\dot{\varphi})}\,,
\end{equation}
where the two functions
\begin{equation}\label{Eq:BD.f1f2}
  f_1(\varphi,\dot{\varphi},\ddot{\varphi})=\ddot{\varphi}-H\dot{\varphi}+\wBD \frac{\dot{\varphi}^2}{\varphi}\,,\ \ \ \ \ \ \  f_2(\varphi,\dot{\varphi})= -3H\dot{\varphi}+\frac{\wBD}{2}  \frac{\dot{\varphi}^2}{\varphi}
\end{equation}
{involve differentiations with respect to the slowly varying field $\varphi$ and as before they are negligible,  in absolute value,  as compared to $\dot{H}\Delta\varphi$ and  ${H^2}\Delta\varphi$. The effective EoS \eqref{Eq:BD.effEoS} is a time-evolving quantity which mimics dynamical DE at low redshifts.  At very high redshifts $z\gg1$, well beyond the DE dominated epoch, we can neglect $\CC$ in the denominator of the EoS and the dominant term is $ 3 H^2\,\dvphi$. Similarly, in the numerator the dominant term is always $-2\dot{H}\dvphi$.  Therefore, at high redshifts the effective EoS \eqref{Eq:BD.effEoS} behaves as \eqref{Eq:BD.BDEoS} since $\dvphi$ cancels out: $\weff(z)\simeq w_{\varphi}(z)\,\ ( z\gg1)$, which means that it just reproduces the standard EoS of the GR-$\CC$CDM.}
The exact EoS \eqref{Eq:BD.effEoS}, however,  must be computed numerically, and it is displayed in \hyperref[Fig:BD.weffvarphi]{Fig.\ref{Fig:BD.weffvarphi}}, together with the numerical plot of $\varphi$ (which we have already shown in \hyperref[Fig:BD.varphi]{Fig.\ref{Fig:BD.varphi}}). We have used the Baseline+$H_0$ dataset defined in \hyperref[Sect:MethodData]{Sect.\,\ref{Sect:MethodData}}.  For a semi-qualitative discussion of the combined EoS it will suffice an analytical approximation, as we did before with $w_{\varphi}$. The most relevant part of $\weff (z)$ as to the possibility of disentangling the dynamical DE effects triggered by the underlying BD model is near the present time ($z<1$). Thus, neglecting the contribution from the functions $f_{1,2}$, but now keeping the $\CC$-term in the denominator of  \eqref{Eq:BD.effEoS} we can use the Hubble function of the concordance model and we find the following result at linear order in $\dvphi$:
\begin{equation}\label{Eq:BD.BDEoSz0}
\weff(z)\simeq-1-\frac{2\dot{H}\dvphi}{\CC}\simeq -1+\dvphi\,\frac{\Omo}{\OLo}\,(1+z)^3\,,
\end{equation}
where $\Omo$ and $\OLo$ are the current values of the cosmological parameters, which satisfy $\Omo+\OLo=1$  for spatially flat Universe.
{As has been stated before, the previous approximate formula is valid only for $z<1$, but it shows very clearly that for $\dvphi>0$ (resp. $<0$) we meet quintessence-like (resp. phantom-like) behavior.  As we have repeatedly emphasized, our analysis points to $\eBD<0$ and hence $\varphi$ decreases with the expansion, remaining smaller than one. From Eq.\,\eqref{Eq:BD.Deltaphi} this means $\dvphi>0$  and therefore we find that the effective GR behavior of the BD-$\CC$CDM is quintessence-like.}  We can be more precise at this point.
We have numerically computed the value of the exact function \eqref{Eq:BD.effEoS} at $z=0$, taking into account the contribution from all the terms, in particular the slowly varying functions \eqref{Eq:BD.f1f2}, see Tables \hyperref[Table:BD.TableFitBaseline]{\ref{Table:BD.TableFitBaseline}}, \hyperref[Table:BD.TableFit+SL]{\ref{Table:BD.TableFit+SL}}, \hyperref[Table:BD.TableFitSpectrum]{\ref{Table:BD.TableFitSpectrum}} and \hyperref[Table:BD.TableFitAlternativeDataset]{\ref{Table:BD.TableFitAlternativeDataset}}. The results obtained from three of the most prominent datasets defined in \hyperref[Sect:MethodData]{Sect.\,\ref{Sect:MethodData}} read as follows:
\begin{align}\label{Eq:BD.weffpresent}
&{\bf Baseline}:\quad &\weff(0)=& -0.983^{+0.015}_{-0.014}\\
&{\bf Baseline+H_0}:\quad &\weff(0)=& -0.966^{+0.012}_{-0.011}\\
&{\bf Baseline+H_0+SL}:\quad &\weff(0) =& -0.962\pm 0.011.
\end{align}
As can be seen, there is a non-negligible departure from the constant EoS value $-1$  of the  GR-$\Lambda$CDM, which reaches the $\sim 3\sigma$ c.l. when the prior on $H_0$ from the local distance ladder measurement by SH0ES \cite{Reid:2019tiq} is included in the analysis, and $\sim 3.5\sigma$ c.l. when also the angular diameter distances from H0LICOW \cite{Wong:2019kwg} are taken into account. The effective quintessence EoS $\weff(0)>-1$ is one of the ingredients that allows the BD-$\CC$CDM model to significantly loosen the $H_0$-tension, since it is a direct consequence of having $\varphi<1$ (or, equivalently, $G>G_N$) (cf. \hyperref[Sect:Preview]{Sect.\,\ref{Sect:Preview}} for details).

We have obtained the above results from the equations of motion once a metric was assumed; however, it is possible to obtain all the expressions listed in this section starting from the BD action \eqref{Eq:BD.BDaction} itself and then considering the FLRW metric.  {To show this, let us use the dimensionless field $\varphi=G_N\psi$ and the variable $\dvphi$ defined in \eqref{Eq:BD.Deltaphi}. }First of all we split the whole action in three pieces
\begin{equation}\label{Eq:BD.BDactionDecomp}
S_{\rm BD}[\varphi]=S_{EH}+S_{GR}[\varphi]+S_{\rm m},
\end{equation}
where
\begin{equation}\label{Eq:BD.EH}
S_{EH}\equiv\int d^4x \sqrt{-g}\left[\frac{R}{16 \pi G_N}-\rho_\Lambda \right],
\end{equation}
is the usual Einstein-Hilbert action, whereas
\begin{equation}\label{Eq:BD.GRPictureAction}
S_{GR}[\varphi] \equiv \int d^4x \sqrt{-g}\frac{1}{16\pi G_N}\left[ -R\dvphi-\frac{\omega_{\rm BD}}{\varphi}g^{\mu \nu}\partial_\nu \varphi \partial_\mu \varphi \right],
\end{equation}
is the action parametrizing the deviation of the BD theory from the  GR-picture expressed  in terms of the scalar field $\varphi$. As expected, for $\varphi=1$ that action vanishes identically. Finally,
\begin{equation}\label{Eq:BD.MatterAction}
S_{\rm m}\equiv\int d^4x \sqrt{-g}\mathcal{L}_{\rm m} (\chi_i,g_{\mu \nu})
\end{equation}
is the action for the matter fields. Since there is no interaction involving $\varphi$ with other components, the BD-field $\varphi$ is covariantly conserved, as remarked in \hyperref[Sect:BDgravity]{Sect.\,\ref{Sect:BDgravity}}. In order to compute the energy-momentum tensor and find out the effective density and pressure of the BD-field, we apply the usual definition of that tensor in curved spacetime:
\begin{equation} \label{Eq:BD.energy-momentum}
T_{\mu \nu}^{\rm BD}=-\frac{2}{\sqrt{g}}\frac{\delta S_{GR}[\varphi]}{\delta g^{\mu \nu}}\,.
\end{equation}
After some calculations we arrive at
\begin{equation}\label{Eq:BD.BDEMT}
\begin{split}
T_{\mu \nu}^{\rm BD}&=\frac{R_{\mu \nu}}{8\pi G_N}\dvphi-\frac{\nabla_\nu \nabla_\mu\varphi}{8\pi G_N}+\frac{g_{\mu \nu}\Box \varphi}{8\pi G_N}+\frac{\omega_{\rm BD}}{8\pi \varphi}\partial_\nu \varphi \partial_\mu \varphi\\
&-\frac{g_{\mu \nu}}{16\pi G_N}\left(R\dvphi+\frac{\omega_{\rm BD}}{\varphi}g^{\alpha \beta}\partial_\alpha \varphi  \partial_\beta \varphi \right).
\end {split}
\end{equation}
Since $\varphi$ has no interactions it behaves as any free scalar field, so its energy-momentum tensor must adopt the perfect fluid form at the background level:
\begin{equation}\label{Eq:BD.TmunuBD}
T_{\mu \nu}^{\rm BD}=\pvphi g_{\mu \nu}+(\rvphi+\pvphi)u_\mu u_\nu.
\end{equation}
Now we can compare this form with \eqref{Eq:BD.BDEMT}. It is straightforward to obtain the energy density as well as the corresponding pressure, we only need to compute $\rvphi=T_{00}^{\rm BD}$ and $\pvphi = (T^{\rm BD} + \rvphi)/3$, being $T^{\rm BD} =g^{\mu\nu}T^{\rm BD}_{\mu\nu}$ the trace of the tensor. {Using at this point the spatially flat FLRW metric one can work out  the explicit result for $\rvphi$ and $\rvphi$  and reconfirm that it acquires the form previously indicated in the equations \eqref{Eq:BD.rhoBD} and \eqref{Eq:BD.pBD}. This  provides perhaps a more formal derivation of these formulas and serves as a cross-check of them.}

\section{Structure formation  in the linear regime. Perturbations equations}\label{Sect:StructureFormationBD}

In order to perform a complete analysis of the model, we need to study the evolution of the perturbed cosmological quantities in the context of BD theory. For a review of the standard model perturbations equations, see e.g. \cite{Ma:1995ey,Liddle:2000cg,Lyth:2009zz}.
We assume a FLRW metric written in conformal time, denoted by $\eta$, in which the line element is $ds^2=a^2(\eta)[-d\eta^2+(\delta_{ij}+h_{ij})dx^idx^j]$. Here $h_{ij}$ is a perturbation on the spatial part of the metric which can be expressed in momentum space as follows,
\begin{equation}\label{Eq:BD.MainhFourier}
h_{ij}(\eta,\vec{x})=\int d^3k\, e^{-i\vec{k}\cdot\vec{x}}\left[\hat{k}_i\hat{k}_j h(\eta,\vec{k})+\left(\hat{k}_i\hat{k}_j-\frac{\delta_{ij}}{3}\right)6\xi(\eta,\vec {k})\right].
\end{equation}
As we see, in momentum space the trace $h\equiv \delta^{ij}h_{ij}$ decouples from the traceless part of the perturbation, $\xi$. Now, we are going to list the perturbations equations at late stages of expansion in momentum space at deep subhorizon scales, that is, we assume $\mathcal{H}^2\ll k^2$, with  $\mathcal{H}\equiv a^\prime/a$. Although primes were used previously for derivatives with respect to the scale factor, they will henceforth stand for derivatives with respect to the conformal time within the main text ({except in \hyperref[Appendix:FixedPoints]{Appendix\,\ref{Appendix:FixedPoints}}): $()^\prime \equiv d()/d \eta$. For example, it is easy to see that $\mathcal{H}= a H$.
One may work with the standard differential equation for the density contrast at deep subhorizon scales,
\begin{equation} \label{Eq:BD.DensityConstrastLCDM}
\delta_{\rm m}^{\prime \prime}+\mathcal{H}\delta_{\rm m}^\prime-4\pi{G_N}\bar{\rho}_{\rm m} a^2 \delta_{\rm m}=0,
\end{equation}
where $\delta_{\rm m} \equiv \delta \rho_{\rm m} / \bar{\rho}_{\rm m}$ is the density contrast, the bar over $\bar{\rho}_{\rm m}$ indicates that is a background quantity and the evolution of $\mathcal{H}$ and $\bar{\rho}_{\rm m}$ is the one expected by the background equations of the BD theory in Section \hyperref[Sect:BDgravity]{Sect.\,\ref{Sect:BDgravity}}. This expression is just the corresponding one for the $\Lambda$CDM, completely neglecting any possible perturbation in the BD-field, namely $\delta \varphi$. However, it is possible to see that a second order differential equation for the density contrast can be written, even if the perturbation in $\varphi$ is not neglected. This is done in detail in the \hyperref[Appendix:CosmoPerturbationsSynchronous]{Appendix\,\ref{Appendix:CosmoPerturbationsSynchronous}} and \hyperref[Appendix:PerturbationTheoryNewtonian]{Appendix\,\ref{Appendix:PerturbationTheoryNewtonian}} for the Synchronous as well as for the Newtonian gauges, respectively. In this section, we present the main perturbations equations in the case of the Synchronous gauge and discuss the interpretation of the result.

If $\vec{v}_{\rm m}$ is the physical 3-velocity of matter  (which is much smaller than 1 and can be treated as a perturbation), then we can define its divergence, $\theta_{\rm m}\equiv \vec{\nabla} \cdot (\vec{v}_{\rm m})$. At deep subhorizon scales it is possible to see that the equation governing its evolution is
\begin{equation}\label{Eq:BD.ThetamEq}
\theta_{\rm m}^\prime+\mathcal{H}\theta_{\rm m}=0.
\end{equation}
{Since $da^{-1}/d\eta=- \mathcal{H}/a$,  we arrive to a decaying solution  $\theta_{\rm m}  \propto a^{-1}$. A common assumption is to set $\theta_{\rm m} \sim 0$ in the last stages of the Universe, which is what we will do in our analysis. This allows us to simplify the equations. Another simplification occurs if we take into account that we are basically interested in computing the matter perturbations only at deep subhorizon scales, namely for $k^2\gg \mathcal{H}^2$, which allows us to neglect some terms as well (cf. \hyperref[Appendix:CosmoPerturbationsSynchronous]{Appendix\,\ref{Appendix:CosmoPerturbationsSynchronous}}). Altogether we are led to the following set of perturbations equations in the synchronous gauge:}
\begin{equation}\label{Eq:BD.Mainsimpli0}
\delta_{\rm m}^\prime=-\frac{h^\prime}{2}\,.
\end{equation}
\begin{equation}\label{Eq:BD.Mainsimpli1}
k^2\delta \varphi+\frac{h^\prime}{2}\bar{\varphi}^\prime =\frac{8 \pi G_N}{3+2\omega_{\rm BD}}a^2 \bar{\rho}_{\rm m}\delta_{\rm m}\,,
\end{equation}
\begin{equation}\label{Eq:BD.Mainsimpli2}
\bar{\varphi}(\mathcal{H}h^\prime-2\xi k^2)+k^2\delta\varphi+\frac{h^\prime}{2}\bar{\varphi}^\prime=8\pi G_N a^2 \bar{\rho}_{\rm m}\delta_{\rm m}\,,
\end{equation}
\begin{equation}\label{Eq:BD.Mainsimpli3}
2k^2\delta \varphi+\bar{\varphi}^\prime h^\prime+\bar{\varphi}\left(h^{\prime \prime}+2h^\prime \mathcal{H}-2k^2 \xi\right)=0\,.
\end{equation}
{Combining these four equations simultaneously (cf. \hyperref[Appendix:CosmoPerturbationsSynchronous]{Appendix\,\ref{Appendix:CosmoPerturbationsSynchronous}} for more details) and  without doing any further approximation, one finally obtains the following compact equation for the matter density contrast of the BD theory at deep subhorizon scales:}
\begin{equation}\label{Eq:BD.ExactPerturConfTime}
\delta_{\rm m}^{\prime\prime}+\mathcal{H}\delta_{\rm m}^\prime-\frac{4\pi G_N a^2}{\bar{\varphi}}\bar{\rho}_{\rm m}\delta_{\rm m}\left(\frac{4+2\omega_{\rm BD}}{3+2\omega_{\rm BD}}\right)=0\,.
\end{equation}
{In other words},
\begin{equation}\label{Eq:BD.ExactPerturConfTime2}
\delta_{\rm m}^{\prime\prime}+\mathcal{H}\delta_{\rm m}^\prime- 4\pi \Geff(\bar{\varphi}) a^2\,\bar{\rho}_{\rm m}\delta_{\rm m}=0\,.
\end{equation}
The quantity
\begin{equation}\label{Eq:BD.MainGeffective}
 \Geff(\bar{\varphi})=\frac{G_N}{\bar{\varphi}}\left(\frac{4+2\omega_{\rm BD}}{3+2\omega_{\rm BD}}\right)=\frac{G_N}{\bar{\varphi}}\left(\frac{2+4\eBD}{2+3\eBD}\right)
\end{equation}
is precisely the effective coupling previously introduced in Eq.\,\eqref{Eq:BD.LocalGNa}; it modifies the Poisson term of the perturbations equation with respect to that of the standard model, Eq.\,\eqref{Eq:BD.DensityConstrastLCDM}.  There is, in addition, a modification in the friction term between the two models, which is of course associated to the change in ${\cal H}$.

The argument of $\Geff$ in \eqref{Eq:BD.MainGeffective} is not $\varphi$ but the background value $\bar{\varphi}$ since the latter is what remains in first order of perturbations from the consistent splitting of the field into the background value and the perturbation: $\varphi=\bar{\varphi}+\delta\varphi$. Notice that there is no dependence left on the perturbation $\delta\varphi$. As we can see from \eqref{Eq:BD.MainGeffective}, the very same effective coupling that rules the attraction of two tests masses in BD-gravity is the coupling strength that governs the formation of structure in this theory, as it could be expected. But this does not necessarily mean that the effective gravitational strength governing the process of structure formation is the same as for two tests masses on Earth. We shall elaborate further on this point in the next section. At the moment we note that if we compare the above perturbations equation with the standard model one \eqref{Eq:BD.DensityConstrastLCDM}, the former reduce to the latter in the limit $\wBD\to\infty$ ({\it i.e.} $\eBD\to 0$) \textit{and} $\bar{\varphi}=1$.

The form of \eqref{Eq:BD.ExactPerturConfTime2} in terms of the scale factor variable rather than in conformal time was given previously in \hyperref[Sect:rolesvarphiH0]{Sect.\,\ref{Sect:rolesvarphiH0}} when we considered a preview of the implications of BD-gravity on structure formation data\footnote{{Recall, however, that prime in Eq.\,\eqref{Eq:BD.ExactPerturConfTime2} stands for differentiation {\it w.r.t.}  conformal time whereas in Eq.\,\eqref{Eq:BD.ExactPerturScaleFactor} denotes differentiation  {\it w.r.t.} the scale factor. These equations perfectly agree and represent the same perturbations equation for the matter density field in BD-gravity in the respective variables. They are also in accordance with the perturbations equation obeyed by the matter density field within the context of scalar-tensor theories with the general action \eqref{Eq:BD.BDactionST} (see \cite{Boisseau:2000pr}) of which the form (\ref{Eq:BD.BDaction2}) is a  particular case.}}. The transformation of derivatives between the two variables can be easily performed with the help of the chain rule $d/d\eta=a{\cal H} d/da$.

\section{Different BD scenarios and Mach's Principle}\label{Sect:Mach}
As previously indicated, the  relation \eqref{Eq:BD.LocalGNa}, which appears now in the cosmological context in the manner \eqref{Eq:BD.MainGeffective}, follows from the computation of the gravitational field felt by two test point-like (or spherical) masses in interaction in BD-gravity within the weak-field limit\cite{brans1961mach, dicke1962physical}, see also \cite{Fujii:2003pa} and references therein. Such relation shows in a manifest way the integration of Mach's principle within the BD context, as it postulates a link between the measured local value of the gravitational strength, $G_N$, as measured at the Earth surface, and its cosmological value, $\Geff(\varphi)$, which depends on $\varphi$ and $\wBD$. In particular, $\varphi$  may be sensitive to the mean energy densities and pressures of all the matter and energy fields that constitute the Universe. {If there is no mechanism screening the BD-field on Earth, $\Geff(\bar{\varphi})(z=0)=G_N$.} {However, one can still fulfill this condition  if  Eq.\,\eqref{Eq:BD.MainGeffective} constraints the current value of the cosmological BD-field $\bar{\varphi}$ }
\begin{equation}\label{Eq:BD.constraintvarphi}
\bar{\varphi}(z=0)=\frac{4+2\omega_{\rm BD}}{3+2\omega_{\rm BD}}=\frac{2+4\eBD}{2+3\eBD}\simeq 1+\frac12\,\eBD+{\cal O}(\eBD^2)\,.
\end{equation}
That is to say, such constraint permits to reconcile $\Geff(\bar{\varphi})(z=0)$ with $G_N$ by still keeping  $\bar{\varphi}(z=0)\ne 1$ and $\epsilon_{\rm BD}\ne 0$. Hence the BD-field can be dynamical and there can be a departure of $G(\bar{\varphi})$ from $G_N$ even at present. This constraint, however, is much weaker than the one following from taking the more radical approach in which $\Geff(\varphi)$ and  $G_N$  are enforced to coincide upon imposing the double condition $\eBD\to0$ ({\it i.e.} $\wBD\to\infty$) \textit{and} $\bar{\varphi}=1$. It is this last setup which anchors the BD theory to remain exactly (or very approximately) close to GR at all scales.

However, if we take seriously the stringent constraint imposed  by the Cassini probe on the post-Newtonian parameter $\gamma^{PN}$\cite{Bertotti:2003rm}, which leads to a very large  value of $\wBD\gtrsim 10^4$ (equivalently, a very small value of $\eBD=1/\wBD$), as we discussed in \,\hyperref[Sect:BDgravity]{Sect.\,\ref{Sect:BDgravity}},  $\bar{\varphi}$ must remain almost constant throughout the cosmic expansion, thus essentially equal to $\bar{\varphi}(z=0)$. However, the Cassini limit leaves $\bar{\varphi}(z=0)$ unconstrained, so this constant  is not restricted to be in principle equal to $1$.  In this case the relation \eqref{Eq:BD.constraintvarphi} may or may not apply; there is in fact no especial reason for it {(it will depend on the effectiveness of the screening mechanism on Earth)}. If it does, {\it i.e.} if there is no screening, $\Geff$ is forced to be very close to $G_N$ {$\forall{z}$};
if it does not, $\bar{\varphi}$ can freely take {(almost constant)} values  which do not push $\Geff$ to stay so glued to $G_N$.  It is  interesting to see  the  extent  to which the cosmological constraints can compete with the local ones given the current status of precision they can both attain.

So, as it turns out, we find that one of the two interpretations  leads to values of $\Geff$ very close to $G_N$ {$\forall{z}$} on account of the fact that  we are  imposing  very large values of  $\oD$ {and assuming \eqref{Eq:BD.constraintvarphi},} 
whereas the other achieves the same aim (viz. $\Geff $  can stay very close to $G_N$) for intermediate values of $\wBD$ provided they are  linked to $\bar{\varphi}(z=0)$ through the constraint  \eqref{Eq:BD.constraintvarphi}. This last option, as indicated, is not likely {since this would imply the existence of a screening at the scales probed by Cassini that may become ineffective on Earth, where the densities are higher}. Finally, we may as well have a situation where the cosmological $\Geff $  remains different from  $G_N$ even if the Cassini limit is enforced. For this to occur we need an (essentially constant) value of  $\bar{\varphi}\neq 1$ (different from the one associated to that constraint) at the cosmological level. This can still be compatible with the local constraints provided $\varphi$ is screened {on Earth} at $z=0$.

{A more open-minded and general approach, which we are going to study in this chapter, is to take the last mentioned option but without the Cassini limit. This means that $\eBD$ is not forced to be so small and hence  $\varphi$ can still have some appreciable dynamics. We  assume that the pure BD model applies
from the very large cosmological domains to those at which the matter perturbations remain linear.  Equation \eqref{Eq:BD.MainGeffective} predicts the cosmological value of $\Geff$ once $\wBD$ and the initial value $\varphi_{\rm ini}$ are fitted to the data. We can dispense with the Cassini constraint (which affects $\wBD$ only) because we assume that some kind of screening mechanism
acts at very low (astrophysical) scales, namely in the nonlinear domain, without altering the pure BD model at the cosmological level.
 To construct a concrete screening mechanism would imply to specify some microscopic interaction properties of $\varphi$ with matter, but these do not affect the analysis at the cosmological level, where there are no place with high densities of material particles. But once such mechanism is constructed (even if not being the primary focus of our work) the value of the BD-field $\varphi$  is ambient dependent, so to speak, since $\varphi$  becomes sensitive to the presence of large densities of matter.   This possibility is well-known in the literature through the chameleon mechanism\,\cite{Khoury:2003aq} and in the case of the BD-field was previously considered in \cite{Avilez:2013dxa} without letting the Brans-Dicke parameter $\omega_{\rm BD}$ to acquire negative values, and using  datasets which now can be considered a bit obsolete. Here we do, instead,  allow negative values for $\omega_{\rm BD}$ (we have seen in \hyperref[Sect:Preview]{Sect.\,\ref{Sect:Preview}} the considerable advantages involved in this possibility), and moreover we are using a much more complete set of observations from all panels of data taking.  In this scenario we cannot make use of \eqref{Eq:BD.MainGeffective} to connect the locally measured value of the gravitational strength $G_N$ with the BD-field at cosmological scales. We just do not need to know how the theory exactly connects these two values. We reiterate once more: we will not focus on the screening mechanism itself here but rather on the properties of the BD-field in the Universe {in the large}, {\it i.e.} at the cosmological level. As it is explained in \cite{Avilez:2013dxa} -- see also \cite{Clifton:2011jh} --  many scalar-tensor theories of gravity belonging to the Horndeski class\,\cite{Horndeski:1974wa} could lead to such kind of BD behavior at cosmological level and, hence, it deserves a dedicated and updated analysis, which is currently lacking in the literature.}

{To summarize, the following interpretations of the BD-gravity framework considered here are, in principle, possible in the light of the current observational data:}

\begin{itemize}

\item {\bf BD-Scenario I}: \textit{Rigid Scenario for both the Local and Cosmological domains}. In it, we have $\Geff(\varphi (z))\simeq G(\varphi (z))\simeq G_N$, these three couplings being so close that in practice BD-gravity is indistinguishable from GR.  In this case, the BD-gravity framework is assumed to hold on equal footing with all the scales of the Universe, local and cosmological.  There are no screening effects from matter. In this context, one interprets that the limit from the Cassini probe\,\cite{Bertotti:2003rm}  leads to a very large value of $\wBD$, which enforces $\varphi$ to become essentially rigid, and one assumes that such constant value is very close to $1$ owing to the relation \eqref{Eq:BD.constraintvarphi}.  Such a rigid scenario is, however, unwarranted. It is possible, although  is not necessary since, strictly speaking,  there is no direct connection between the Cassini bound on  $\wBD$ and the value that $\varphi$ can take. Thus, in this scenario the relation \eqref{Eq:BD.constraintvarphi} is just assumed. In point of fact, bounds on  $\wBD$  can only affect the time evolution of $\varphi$, they do not constraint its value.

\item {\bf BD-Scenario II (Main)}: {\it  Locally  Constrained but Cosmologically Unconstrained Scenario}. It is our main scenario.  It assumes a constrained  situation in the local domain, caused by the presence of chameleonic forces,  but permits an unconstrained picture for the entire cosmological range. In other words, the Cassini limit that holds for the post-Newtonian parameter $\gamma^{PN}\, $in the local astrophysical level (and hence on $\wBD$) is assumed to reflect just the presence of screening effects of matter in that nonlinear domain. These effects are acting on $\varphi$  and produce the illusion that  $\wBD$  has a very large value (as if `dressed' or `renormalized' by the chamaleonic forces). One expects that the `intrusive' effects of matter are only possible in high density (hence nonrelativistic) environments, and in their presence we cannot actually know the real (`naked') value of $\wBD$  through local experiments alone. We assume that the screening disappears as soon as we leave the astrophysical scales and plunge into the cosmological ones; then, and only then, we can measure the naked or ``bare' value of $\wBD$ (stripped from such effects). We may assume that the screening ceases already at the LSS scales where linear structure formation occurs, see e.g. \cite{Tsujikawa:2008uc} for examples of potentials which can help to realize this mechanism. The bare value of $\wBD$ can then be fitted to the overall data, and in particular to the LSS formation data. Since $\wBD$ does no longer appear that big (nor it has any a priori sign) the BD-field $\varphi$ can evolve in an appreciable way at the cosmological level: it increases with the expansion if $\wBD>0$, and decreases with the expansion if $\wBD<0$.  In this context, its  initial value, $\varphi_{\rm ini}$, becomes a relevant cosmological parameter, which must be taken into account as a fitting parameter on equal footing with $\wBD$ and all of the conventional parameters entering the fit. Using the large wealth of cosmological data, these parameters can be fixed at the cosmological level without detriment of the observed physics at the local domain, provided there is a screening mechanism insuring that $\Geff(\varphi)(0)$  stays sufficiently  close  to $G_N$ in the local neighbourhood. The numerical results for this important scenario are presented in Tables \hyperref[Table:BD.TableFitBaseline]{\ref{Table:BD.TableFitBaseline}}, \hyperref[Table:BD.TableFit+SL]{\ref{Table:BD.TableFit+SL}}, \hyperref[Table:BD.TableFitSpectrum]{\ref{Table:BD.TableFitSpectrum}},  \hyperref[Table:BD.TableFitAlternativeDataset]{\ref{Table:BD.TableFitAlternativeDataset}} and \hyperref[Table:BD.TableFitS8]{\ref{Table:BD.TableFitS8}}.

\item  {\bf BD-Scenario III}:  {\it Cassini-constrained Scenario}. A more restricted version of scenario II  can appear if the `bare value' of $\wBD$ is as large as in the Cassini bound.  In such case $\wBD$ is, obviously,  perceived large in both domains,  local  and cosmological. Even so, and despite of the fact that $\varphi$ varies very slowly in this case, one can still exploit the dependence of the fit on the initial value of the BD-field, $\varphi_{\rm ini}$, and use it as a relevant cosmological parameter.  In practice, this situation has only one additional degree of freedom as compared to Scenario I (and one less than in Scenario II), but it is worth exploring -- see our results in \hyperref[Table:BD.TableFitCassini]{Table\,\ref{Table:BD.TableFitCassini}}. As these numerical results show, the Cassini bound still leaves considerable freedom to the BD-$\CC$CDM model for improving the $H_0$ tension (without aggravating the $\sigma_8$ one) since the value of $\varphi$ is still an active degree of freedom, despite its time evolution is now more crippled.

\item  {\bf BD-Scenario IV}: {\it Variable-$\wBD(\varphi)$  Scenario}. Here one admits that the parameter $\wBD$ is actually a function of the BD-field, $\wBD=\wBD(\varphi)$, which can be modeled and adapted to the constraints of the  local and cosmological domains, or even combined with the screening effects of the local Universe.  We have said from the very beginning that we would assume $\wBD=$const. throughout our analysis, and in fact we shall stick to that hypothesis; so here we mention the variable $\wBD$ scenario  only for completeness. In any case, if a function $\wBD(\varphi)$ exists such that it takes very large values in the local Universe while it takes much more moderate values in the cosmological scales, that sort of scenario would be in the main tantamount to Scenario II insofar as concerns its cosmological implications.

\end{itemize}

{In our analysis we basically choose BD- Scenarios II and III (the latter being a particular case of the former), which represent the most tolerant  point of view within the canonical  $\wBD=$const. option. Scenario II offers the most powerful framework amenable to provide a cure for the tensions afflicting the conventional $\CC$CDM model based on  GR.  Thus, we assume that we can  measure the cosmological  value of the gravity strength in BD theory -- {\it i.e.} the value given in Eq.\,\eqref{Eq:BD.MainGeffective} -- {by using only cosmological data. We combine the information from the LSS processes involving linear structure formation with the background information obtained from low, intermediate, and very high redshift probes, including BAO, CMB, and the distance ladder measurement of $H_0$.}
The values of  $\varphi_{\rm ini}$ and $\wBD$ are fitted to the data, and with them we obtain not only $\varphi(z=0)$  but we determine the effective cosmological gravity strength at all epochs  from \eqref{Eq:BD.MainGeffective}. The cosmological value of the gravity coupling  can be considered as the `naked' or `bare' value of the gravitational interaction, stripped from screening effects of matter, in the same way as $\wBD$ measured at cosmological scales is the bare value freed of these effects.  Even though $\Geff$ can be different from $G_N,$  we do not object to that since it can be ascribed to screening forces caused by the huge amounts of clustered matter in the astrophysical environments.  For this reason we do not adopt the local constraints for our cosmological analysis presented in this chapter, {\it i.e.} we adhere to Scenario II as our main scenario.  Remarkably enough, we shall see that Scenario III still possesses a large fraction of the potentialities inherent to Scenario II, notwithstanding the Cassini bound.  In this sense Scenarios II and III are both extremely interesting. A  smoking gun of such overarching possible picture  is the possible detection of the dynamical dark energy EoS encoded in the BD theory within the GR-picture (cf. \hyperref[Sect:EffectiveEoS]{Sect.\,\ref{Sect:EffectiveEoS}}), which reveals itself in the form of effective quintessence,  as well as through the large smoothing achieved of  the main tensions afflicting the conventional $\CC$CDM. From here on, we present the bulk of our analysis and detailed results after we have already discussed to a great extent their possible implications.}

\section{Data and methodology}\label{Sect:MethodData}

We fit the BD-$\Lambda$CDM together with the concordance GR-$\Lambda$CDM model and the GR-XCDM (based on the XCDM parametrization of the DE \cite{Turner:1998ex}) to the wealth of cosmological data compiled from distant type Ia supernovae (SNIa), baryonic acoustic oscillations (BAO), a set of measurements of the Hubble function at different redshifts, the Large-Scale Structure (LSS) formation data encoded in $f(z_i)\sigma_8(z_i)$, and the CMB temperature and low-$l$ polarization data from the Planck satellite. The joint combination of all these individual datasets will constitute our \textbf{Baseline Data} configuration. Moreover, we also study the repercussion of some alternative data, by adding them to the aforementioned baseline setup. These additional datasets are: a prior on the value of $H_0$ (or alternatively an effective calibration prior on M) provided by the SH0ES collaboration; the CMB high-$\ell$ polarization and lensing data from Planck; the Strong-Lensing (SL) time delay angular diameter distances from H0LICOW; and, finally, Weak-Lensing (WL) data from KiDS.  The following is a description of the data points included in our datasets, which we used in our analyses, along with the corresponding references\footnote{For more general explanations of the datasets, the reader may refer to  \hyperref[Appendix:Description]{Appendix\,\ref{Appendix:Description}}.}:
\newline
\newline
\textbf{CMB}: The baseline dataset contains the full Planck 2018 TT+lowE likelihood\,\cite{aghanim2020planck}. In order to study the influence of the CMB high-$\ell$ polarizations and lensing we consider two alternative (non-baseline) datasets, in which we substitute the TT+lowE likelihood by: (i) the TTTEEE+lowE likelihood, which incorporates the information of high multipole polarizations; (ii) the full TT-TEEE+lowE+lensing likelihood, in which we also incorporate the Planck 2018 lensing data. In Tables \hyperref[Table:BD.TableFitAlternativeDataset]{\ref{Table:BD.TableFitAlternativeDataset}} and \hyperref[Table:BD.TableFitGR]{\ref{Table:BD.TableFitGR}} these scenarios are denoted as B+$H_0$+pol and B+$H_0$+pol+lens, respectively.
\newline
\newline
\textbf{SNIa}: We use the full Pantheon likelihood, which incorporates the information from 1048 SNIa\,\cite{Pan-STARRS1:2017jku}. In addition, we also include the 207 SNIa from the DES survey\,\cite{DES:2018paw}. These two SNIa samples are uncorrelated, but the correlations between the points within each sample are non-null and have been duly incorporated in our analyses through the corresponding covariance matrices.
\newline
\newline
\textbf{BAO}: We use data on both, isotropic and anisotropic BAO analyses. We provide the detailed list of data points and corresponding references in \hyperref[Table:BD.BAO]{Table\,\ref{Table:BD.BAO}}. A few comments are in order about the use of some of the BAO data points considered in this chapter. Regarding the Ly$\alpha$-forest data, we opt to use the auto-correlation information from\,\cite{deSainteAgathe:2019voe}. Excluding the Ly$\alpha$ cross-correlation data allows us to avoid double counting issues between the latter and the eBOSS data from \cite{Gil-Marin:2018cgo}, due to the partial (although small) overlap in the list of quasars employed in these two analyses. It is also important to remark that we consider in our baseline dataset the BOSS data reported in \cite{Gil-Marin:2016wya}, which contains information from the spectrum (SP) and the bispectrum (BP). The bispectrum information could capture some details otherwise missed when only the spectrum is considered, so it is worth to use it\footnote{See also Ref.\cite{SolaPeracaula:2018wwm} for additional comments on the significance of the bispectrum data as well as its potential implications on the possible detection of dynamical dark energy.}. Therefore, we study the possible significance of the bispectrum component in the data by carrying out an explicit comparison of the results obtained with the baseline configuration to those obtained by substituting the data points from \cite{Gil-Marin:2016wya} with those from \cite{Alam:2016hwk}, which only incorporate the SP information. The results are provided in \hyperref[Table:BD.TableFitBaseline]{Table\,\ref{Table:BD.TableFitBaseline}} and \hyperref[Table:BD.TableFitSpectrum]{Table\,\ref{Table:BD.TableFitSpectrum}}, respectively. In Tables \hyperref[Table:BD.TableFit+SL]{\ref{Table:BD.TableFit+SL}}, \hyperref[Table:BD.TableFitAlternativeDataset]{\ref{Table:BD.TableFitAlternativeDataset}}, \hyperref[Table:BD.TableFitGR]{\ref{Table:BD.TableFitGR}}, \hyperref[Table:BD.TableFitXCDM]{\ref{Table:BD.TableFitXCDM}}, we use the SP+BSP combination \cite{Gil-Marin:2016wya}. In \hyperref[Table:BD.TableFitCassini]{Table\,\ref{Table:BD.TableFitCassini}} we employ both SP and SP+BSP.
%
\begin{table}[t]
\begin{center}
\resizebox{14.5cm}{!}{
\begin{tabular}{| c | c |c | c |c|}
\multicolumn{1}{c}{Survey} &  \multicolumn{1}{c}{$z$} &  \multicolumn{1}{c}{Observable} &\multicolumn{1}{c}{Measurement} & \multicolumn{1}{c}{{\small References}}
\\\hline
6dFGS+SDSS MGS & $0.122$ & $D_V(r_{s,fid}/r_{\rm s})$[Mpc] & $539\pm17$[Mpc] &\cite{Carter:2018vce}
\\\hline
 WiggleZ & $0.44$ & $D_V(r_{\rm s,fid}/r_{\rm s})$[Mpc] & $1716.4\pm 83.1$[Mpc] &\cite{Kazin:2014qga} \tabularnewline
\cline{2-4} & $0.60$ & $D_V(r_{\rm s,fid}/r_{\rm s})$[Mpc] & $2220.8\pm 100.6$[Mpc]&\tabularnewline
\cline{2-4} & $0.73$ & $D_V(r_{\rm s,fid}/r_{\rm s})$[Mpc] &$2516.1\pm 86.1$[Mpc] &
\\\hline

DR12 BOSS (BSP)& $0.32$ & $Hr_{\rm s}/(10^{3}km/s)$ & $11.549\pm0.385$   &\cite{Gil-Marin:2016wya}\\ \cline{3-4}
 &  & $D_A/r_{\rm s}$ & $6.5986\pm0.1337$ &\tabularnewline \cline{3-4}
 \cline{2-2}& $0.57$ & $Hr_{\rm s}/(10^{3}km/s)$  & $14.021\pm0.225$ &\\ \cline{3-4}
 &  & $D_A/r_{\rm s}$ & $9.389\pm0.1030$ &\\\hline

DR12 BOSS (SP) & $0.38$ & $D_{\rm M}(r_{\rm s}/r_{s,fid})$[Mpc] & $1518\pm22$   &\cite{Alam:2016hwk} \\ \cline{3-4}
 &  & $H(r_{s,fid}/r_{\rm s})$[km/s/Mpc] & $81.5\pm1.9$ & \tabularnewline \cline{3-4}
 \cline{2-2}& $0.51$ & $D_{\rm M}(r_{\rm s}/r_{s,fid})$[Mpc] & $1977\pm27$ & \\ \cline{3-4}
 &  & $H(r_{s,fid}/r_{\rm s})$[km/s/Mpc] & $90.4\pm1.9$ & \\ \cline{3-4}
 \cline{2-2}& $0.61$ & $D_{\rm M}(r_{\rm s}/r_{s,fid})$[Mpc]  & $2283\pm32$ & \\ \cline{3-4}
 &  & $H(r_{s,fid}/r_{\rm s})$[km/s/Mpc] & $97.3\pm2.1$ & \\\hline

DES & $0.81$ & $D_A/r_{\rm s}$ & $10.75\pm0.43$ &\cite{Abbott:2017wcz}
\\\hline
eBOSS DR14 & $1.19$ & $Hr_{\rm s}/(10^{3}km/s)$ & $19.6782\pm1.5867  $ &\cite{Gil-Marin:2018cgo}\\ \cline{3-4}
 &  & $D_A/r_{\rm s}$ & $12.6621\pm0.9876$ &\tabularnewline \cline{3-4}
 \cline{2-2}& $1.50$ & $Hr_{\rm s}/(10^{3}km/s)$  & $19.8637\pm2.7187$ &\\ \cline{3-4}
 &  & $D_A/r_{\rm s}$ & $12.4349\pm1.0429$ &\\ \cline{3-4}
 \cline{2-2}& $1.83$ & $Hr_{\rm s}/(10^{3}km/s)$  & $26.7928\pm3.5632$ &\\ \cline{3-4}
 &  & $D_A/r_{\rm s}$ & $13.1305\pm1.0465$ &\\\hline

Ly$\alpha$-F eBOSS DR14 & $2.34$ & $D_H/r_{\rm s}$ & $8.86\pm0.29$   &\cite{deSainteAgathe:2019voe}
\\ \cline{3-4} &  & $D_{\rm M}/r_{\rm s}$ & $37.41\pm 1.86$
&\\\hline
\end{tabular}}
\caption{Published values of BAO data, see the quoted references and text in \protect\hyperref[Sect:MethodData]{Sect.\,\ref{Sect:MethodData}}. Although we include in this table the values of $D_H/r_{\rm s}=c/(r_{\rm s}H)$ and $D_{\rm M}/r_{\rm s}$ for the Ly$\alpha$-forest auto-correlation data from \protect\cite{deSainteAgathe:2019voe}, we have performed the fitting analysis with the full likelihood. The fiducial values of the comoving sound horizon appearing in the table are $r_{\rm s,fid} = 147.5$ Mpc for \protect\cite{Carter:2018vce}, $r_{s,fid} = 148.6$ Mpc for \protect\cite{Kazin:2014qga}, and $r_{\rm s,fid} = 147.78$ Mpc for \protect\cite{Alam:2016hwk}.}\label{Table:BD.BAO}
\end{center}
\end{table}
%
\newline
\newline
\textbf{Cosmic Chronometers}: We use the 31 data points on $H(z_i)$, at different redshifts, from \cite{Jimenez:2003iv,Simon:2004tf,Stern:2009ep,Moresco:2012jh,Zhang:2012mp,Moresco:2015cya,Moresco:2016mzx,Ratsimbazafy:2017vga}. All of them have been obtained making use of the differential age technique applied to passively evolving galaxies \cite{Jimenez:2001gg}, which provides cosmology-independent constraints on the Hubble function, but are still subject to systematics coming from the choice of the stellar population synthesis technique, and also the potential contamination of young stellar components in the quiescent galaxies \cite{Lopez-Corredoira:2017zfl,Lopez-Corredoira:2018tmn,Moresco:2018xdr}. For this reason we consider a more conservative dataset that takes into account these additional uncertainties. To be concrete, we use the processed sample presented in Table 2 of \cite{Gomez-Valent:2018gvm}. See therein for further details.
\newline
%
\begin{table}[t]
\begin{center}
\resizebox{10cm}{!}{
\begin{tabular}{| c | c |c | c |}
\multicolumn{1}{c}{Survey} &  \multicolumn{1}{c}{$z$} &  \multicolumn{1}{c}{$f(z)\sigma_8(z)$} & \multicolumn{1}{c}{{\small References}}
\\\hline
6dFGS+2MTF & $0.03$ & $0.404^{+0.082}_{-0.081}$ & \cite{Qin:2019axr}
\\\hline
SDSS-DR7 & $0.10$ & $0.376\pm 0.038$ & \cite{Shi:2017qpr}
\\\hline
GAMA & $0.18$ & $0.29\pm 0.10$ & \cite{Simpson:2015yfa}
\\ \cline{2-4}& $0.38$ & $0.44\pm0.06$ & \cite{Blake:2013nif}
\\\hline
DR12 BOSS (BSP)& $0.32$ & $0.427\pm 0.056$  & \cite{Gil-Marin:2016wya}\\ \cline{2-3}
 & $0.57$ & $0.426\pm 0.029$ & \\\hline
 WiggleZ & $0.22$ & $0.42\pm 0.07$ & \cite{Blake:2011rj} \tabularnewline
\cline{2-3} & $0.41$ & $0.45\pm0.04$ & \tabularnewline
\cline{2-3} & $0.60$ & $0.43\pm0.04$ & \tabularnewline
\cline{2-3} & $0.78$ & $0.38\pm0.04$ &
\\\hline

DR12 BOSS (SP) & $0.38$ & $0.497\pm 0.045$ & \cite{Alam:2016hwk}\tabularnewline
\cline{2-3} & $0.51$ & $0.458\pm0.038$ & \tabularnewline
\cline{2-3} & $0.61$ & $0.436\pm0.034$ &
\\\hline

VIPERS & $0.60$ & $0.49\pm 0.12$ & \cite{Mohammad:2018mdy}
\\ \cline{2-3}& $0.86$ & $0.46\pm0.09$ &
\\\hline
VVDS & $0.77$ & $0.49\pm0.18$ & \cite{Guzzo:2008ac},\cite{Song:2008qt}
\\\hline
FastSound & $1.36$ & $0.482\pm0.116$ & \cite{Okumura:2015lvp}
\\\hline
eBOSS DR14 & $1.19$ & $0.4736\pm 0.0992$ & \cite{Gil-Marin:2018cgo} \tabularnewline
\cline{2-3} & $1.50$ & $0.3436\pm0.1104$ & \tabularnewline
\cline{2-3} & $1.83$ & $0.4998\pm0.1111$ &

\\\hline
 \end{tabular}}
\caption{Published values of $f(z)\sigma_8(z)$, see the quoted references and text in \hyperref[Sect:MethodData]{Sect.\,\ref{Sect:MethodData}}.}\label{Table:BD.fs8}
\end{center}
\end{table}
%
\newline
\newline
\textbf{LSS}: In this chapter the LSS dataset contains the data points on the product of the ordinary growth rate $f(z_i)$ with $\sigma_8(z_i)$ at different effective redshifts. They are all listed in \hyperref[Table:BD.fs8]{Table\,\ref{Table:BD.fs8}}, together with the references of interest. In order to correct the potential bias introduced by the particular choice of a fiducial model in the original observational analyses we apply the rescaling correction explained in \cite{Macaulay:2013swa}. See also Sec. II.2 of \cite{Nesseris:2017vor}. The internal correlations between the BAO and RSD data from \cite{Gil-Marin:2016wya}, \cite{Alam:2016hwk} and \cite{Gil-Marin:2018cgo} have been duly taken into account through the corresponding covariance matrices provided in these three references.
\newline
\newline
\textbf{Prior on $H_0$}: We include as a prior in almost all the non-baseline datasets the value of the Hubble parameter measured by the SH0ES collaboration, $H_0= (73.5\pm 1.4)$ km/s/Mpc \cite{Reid:2019tiq}. It is obtained with the cosmic distance ladder method using an improved calibration of the Cepheid period-luminosity relation. It is based on distances obtained from detached eclipsing binaries located in the Large Magellanic Cloud, masers in the galaxy NGC $4258$ and Milky Way parallaxes. This measurement is in $4.1\sigma$ tension with the value obtained by the Planck team under the TTTEEE+lowE+lensing dataset, and using the GR-$\Lambda$CDM model, $H_0= (67.36\pm 0.54)$ km/s/Mpc \cite{aghanim2020planck}. In only one alternative dataset we opt to use instead the SH0ES effective calibration prior on the absolute magnitude  $M$ of the SNIa, as provided in \cite{Camarena:2019rmj}:  $M=-19.2191\pm 0.0405$. This case is denoted ``$M$'' in our tables. It is obtained from the calibration of `nearby' SNIa (at $z\lesssim 0.01$) with Cepheids\,\cite{Riess:2019cxk}.
It has been recently argued in \cite{Camarena:2019moy,Camarena:2019rmj} (and later on also in \cite{Dhawan:2020xmp,Benevento:2020fev}) that in cosmological studies it is better to use this SH0ES constraint rather than the direct prior on $H_0$ when combined with low-redshift SNIa data to avoid double counting issues. We show that the results obtained with these two methods are compatible and lead to completely consistent results (see the discussion in \hyperref[Sect:NumericalAnalysis]{Sect.\,\ref{Sect:NumericalAnalysis}}, and also \hyperref[Table:BD.TableFitBaseline]{Table\,\ref{Table:BD.TableFitBaseline}} and \hyperref[Table:BD.TableFitAlternativeDataset]{Table\,\ref{Table:BD.TableFitAlternativeDataset}} ).
\newline
\newline

\textbf{SL}: In one of the non-baseline datasets we use, in combination with the SH0ES prior on $H_0$, the data extracted from the six gravitational lensed quasars of variable luminosity reported by the H0LICOW team\footnote{\url{http://shsuyu.github.io/H0LiCOW/site/}}: B1608+656, RX51131-1231, HE0435-1223, SDSS 1206+4332, WFI 2033-4723 and PG 1115+080.  As explained in \hyperref[Sect:Preview]{Sect.\,\ref{Sect:Preview}}, the fact of being absolute distances (instead of relative, as in the SNIa and BAO datasets) {allows them to directly constrain the Hubble parameter in the context of the GR-$\Lambda$CDM as follows:  $H_0=73.3^{+1.7}_{-1.8}$ km/s/Mpc}. Interestingly, the non-detection of Strong-Lensing time delay variations can be used to put an upper bound on $\dot{G}/G$ at the redshift and location of the lens \cite{Giani:2020fpz}. These constraints, though, cannot be applied to our model since we assume that the field is screened in these dense regions (see Scenarios II and III in \hyperref[Sect:Mach]{Sect.\,\ref{Sect:Mach}}). Moreover, they are still too weak -- $\dot{G}/G\lesssim 10^{-2}\,yr^{-1}$ \cite{Giani:2020fpz} -- to have an impact on our results.
\newline
\newline
\textbf{WL}: We used data from the Kilo-Degree Survey (KiDS)\,\cite{Hildebrandt:2016iqg,Joudaki:2017zdt,Kohlinger:2018sxx,Wright:2020ppw} as an alternative dataset. Nonlinear effects for small angular scales cannot be calculated for the BD-$\Lambda$CDM model using the Halofit module \cite{Takahashi:2012em}, which only implements the GR-$\Lambda$CDM model and minimal extensions. However, we found that without information from the small scales, predictability is significantly reduced and $S_8$ remains largely unconstrained. In \hyperref[Table:BD.TableFitAlternativeDataset]{Table\,\ref{Table:BD.TableFitAlternativeDataset}} and \hyperref[Table:BD.TableFitGR]{Table\,\ref{Table:BD.TableFitGR}} we show the results obtained using the full KiDS likelihood \cite{Hildebrandt:2016iqg,Kohlinger:2018sxx}\footnote{See \url{http://kids.strw.leidenuniv.nl/sciencedata.php} for more details.}, {\it i.e.} including all the scales (also the small ones). The results, though, should be interpreted with caution.
\newline
\newline
We study the performance of the BD-$\CC$CDM, GR-$\CC$CDM and GR-XCDM models under different datasets. In the following we briefly summarize the composition of each of them:
\begin{itemize}

\item {\bf Baseline (B)}: Here we include the Planck 2018 TT+lowE CMB data, together with SNIa +BAO+$H(z_i)$+LSS (see \hyperref[Table:BD.TableFitBaseline]{Table\,\ref{Table:BD.TableFitBaseline}} and \hyperref[Table:BD.TableFitXCDM]{Table\,\ref{Table:BD.TableFitXCDM}}). It is important to remark that for the BOSS BAO+LSS data we consider \cite{Gil-Marin:2016wya}, which includes the information from the spectrum (SP) as well as from the bispectrum (BSP). We construct some other datasets using this baseline configuration as the main building block. See the other items, below.

\item {\bf Baseline{\boldmath+$H_0$}} ({\bf B{\boldmath+$H_0$}}): Here we add the SH0ES prior on the $H_0$ parameter from \cite{Reid:2019tiq} to the baseline dataset (see again \hyperref[Table:BD.TableFitBaseline]{Table\,\ref{Table:BD.TableFitBaseline}} and \hyperref[Table:BD.TableFitXCDM]{\ref{Table:BD.TableFitXCDM}}).

\item {\bf Baseline{\boldmath+$H_0$+}SL}: The inclusion of the Strong-Lensing (SL) data from \cite{Wong:2019kwg} exacerbates more the $H_0$-tension in the context of the GR-$\CC$CDM model (see e.g. \cite{verde2019tensions} and \hyperref[Sect:ConsiderationsBD]{Sect.\,\ref{Sect:ConsiderationsBD}}), so it is interesting to also study the ability of the BD-$\CC$CDM to fit the SL data when they are combined with the previous B+$H_0$ dataset, and compare the results with those obtained with the GR-$\CC$CDM. The corresponding fitting results are displayed in \hyperref[Table:BD.TableFit+SL]{Table\,\ref{Table:BD.TableFit+SL}}.

\item  {\bf Spectrum}: In this dataset we replace the SP+BSP data from \cite{Gil-Marin:2016wya} used in the Baseline dataset (see the first item, above) by the data from \cite{Alam:2016hwk}, which only contains the spectrum (SP) information ({\it i.e.} the usual matter power spectrum).

\item {\bf Spectrum{\boldmath+$H_0$}}: As in the preceding item, but including the $H_0$ prior from SH0ES \cite{Reid:2019tiq}.

\end{itemize}
The aforementioned datasets are all based on the BD-Scenario II (cf. \hyperref[Sect:Mach]{Sect.\,\ref{Sect:Mach}}) and can be considered as the main ones (cf. Tables \hyperref[Table:BD.TableFitBaseline]{\ref{Table:BD.TableFitBaseline}}, \hyperref[Table:BD.TableFit+SL]{\ref{Table:BD.TableFit+SL}}, \hyperref[Table:BD.TableFitSpectrum]{\ref{Table:BD.TableFitSpectrum}} and \hyperref[Table:BD.TableFitXCDM]{\ref{Table:BD.TableFitXCDM}}), nevertheless we also consider a bunch of alternative datasets (also based on the BD-Scenario II). We present the corresponding results in \hyperref[Table:BD.TableFitAlternativeDataset]{Table\,\ref{Table:BD.TableFitAlternativeDataset}} and the first five columns of \hyperref[Table:BD.TableFitGR]{Table\,\ref{Table:BD.TableFitGR}} for the BD-$\CC$CDM and GR-$\CC$CDM models, respectively.

\begin{itemize}

\item {\bf B{\boldmath+$M$}}: In this scenario we replace the prior on $H_0$ \cite{Reid:2019tiq} employed in the B+$H_0$ dataset with the effective SH0ES calibration prior on the absolute magnitude of SNIa $M$ provided in \cite{Camarena:2019rmj}.

\item {\bf B{\boldmath+$H_0$+}pol}: Here we add the CMB high-$\ell$ polarization data from Planck 2018 \cite{aghanim2020planck} to the B+$H_0$ dataset described before, {\it i.e.} we consider the Planck 2018 TTTEEE+lowE likelihood for the CMB.

\item {\bf B{\boldmath+$H_0$+}pol+lens}: In addition to the datasets considered in the above case we also include the CMB lensing data from Planck 2018 \cite{aghanim2020planck}, {\it i.e.} we use the Planck 2018 TTTEEE+lowE+ lensing likelihood.

\item {\bf B{\boldmath+$H_0$+}WL}: In this alternative case we consider the Weak-Lensing (WL) data from KiDS \cite{Hildebrandt:2016iqg,Kohlinger:2018sxx}, together with the B+$H_0$ dataset.

\item  {\bf CMB+BAO+SNIa}: By considering only this data combination, we study the performance of the the BD-$\CC$CDM and the GR-$\CC$CDM models under a more limited dataset, obtained upon the removal of the data that trigger the $H_0$ and $\sigma_8$ tensions, {\it i.e.} the prior on $H_0$ from SH0ES and the LSS data. The use of the BAO+SNIa data helps to break the strong degeneracies found in parameter space when only the CMB is considered. Here we use the TT+lowE Planck 2018 likelihood \cite{aghanim2020planck}.

\end{itemize}

Finally, in \hyperref[Table:BD.TableFitCassini]{Table\,\ref{Table:BD.TableFitCassini}} we present the results obtained for the BD-$\CC$CDM in the context of the Cassini-constrained scenario, or Scenario III (see \hyperref[Sect:Mach]{Sect.\,\ref{Sect:Mach}} for the details). The corresponding results for the GR-$\CC$CDM are shown in the third column of \hyperref[Table:BD.TableFitBaseline]{Table\,\ref{Table:BD.TableFitBaseline}}, and the last three rows of \hyperref[Table:BD.TableFitGR]{Table\,\ref{Table:BD.TableFitGR}}. In all these datasets we include the Cassini bound\,\cite{Bertotti:2003rm} (see \hyperref[Sect:BDgravity]{Sect.\,\ref{Sect:BDgravity}} for details). The main purpose of this scenario is to test the ability of the BD-$\CC$CDM to fit the observational data with $\eBD\simeq 0$ {\it and} $\varphi\neq 1$.

\begin{itemize}

\item {\bf B{\boldmath+$H_0$+}Cassini}: It contains the very same datasets as in the Baseline+$H_0$ case, but here we also include the Cassini constraint.

\item {\bf B{\boldmath+$H_0$+}Cassini (No LSS)}: Here we study the impact of the LSS data in the context of Scenario III, by removing them from the previous B+$H_0$+Cassini dataset.

\item {\bf Dataset \cite{Ballesteros:2020sik}}: To ease the comparison with the results obtained in \cite{Ballesteros:2020sik}, here we use exactly the same dataset as in that reference, namely: the Planck 2018 TTTEEE+lowE+lensing likelihood \cite{aghanim2020planck}, the BAO data from \cite{Beutler:2011hx,Ross:2014qpa,Alam:2016hwk}, and the SH0ES prior from \cite{Riess:2019cxk}, $H_0=74.03\pm 1.42$ km/s/Mpc.

\item {\bf Dataset\cite{Ballesteros:2020sik}+LSS}: Here we consider an extension of the previous scenario by adding the LSS data on top of the data from \cite{Ballesteros:2020sik}.

\end{itemize}

We believe that all the datasets and scenarios studied in this chapter cover a wide range of possibilities and show in great detail which is the phenomenological performance of the BD-$\CC$CDM, GR-$\CC$CDM and GR-XCDM models.

The speed of gravitational waves at $z\approx 0$, $c_{gw}$, has been recently constrained to be extremely close to the speed of light, $|c_{gw}/c-1|\lesssim 5\cdot 10^{-16}$ \cite{TheLIGOScientific:2017qsa}. In the BD-$\CC$CDM model $c_{gw}=c\,$ $\forall{z}$, so this constraint is automatically fulfilled, see Appendix C.6 and references \cite{Creminelli:2017sry,Ezquiaga:2017ekz} for further details. We have also checked that the BD-$\CC$CDM respects the bound on $G(\varphi)$ at the Big Bang Nucleosynthesis (BBN) epoch, $|G(\varphi_{\rm BBN})/G_N-1|\lesssim 0.1$ \cite{Uzan:2010pm}, since $G(\varphi_{\rm BBN})\simeq G(\varphi_{\rm ini})$ and our best-fit values satisfy $G(\varphi_{\rm ini})>0.9 G_N$ regardless of the dataset under consideration, see the fitting results in Tables \hyperref[Table:BD.TableFitBaseline]{\ref{Table:BD.TableFitBaseline}}, \hyperref[Table:BD.TableFit+SL]{\ref{Table:BD.TableFit+SL}}, \hyperref[Table:BD.TableFitSpectrum]{\ref{Table:BD.TableFitSpectrum}}, \hyperref[Table:BD.TableFitAlternativeDataset]{\ref{Table:BD.TableFitAlternativeDataset}} and \hyperref[Table:BD.TableFitCassini]{\ref{Table:BD.TableFitCassini}}.

To obtain the posterior distributions and corresponding constraints for the various dataset combinations described above we have run the Monte Carlo sampler \texttt{Montepython}\footnote{\url{http://baudren.github.io/montepython.html}} \cite{Audren:2012wb} together with the Einstein-Boltzmann system solver \texttt{CLASS}\footnote{\url{http://lesgourg.github.io/class\%_public/class.html}} \cite{Blas:2011rf}. We have duly modified the latter to implement the background and linear perturbations equations of the BD-$\Lambda$CDM model. {We use adiabatic initial conditions for all matter species. Let us note that the initial perturbation of the BD-field and its time derivative can be set to zero, as the DM velocity divergence when the synchronous gauge is employed, see \hyperref[Sect:IC]{Sect.\,\ref{Sect:IC}} for a brief discussion.} To get the contour plots and one-dimensional posterior distributions of the parameters entering the models we have used the \texttt{Python} package \texttt{GetDist}\footnote{https://getdist.readthedocs.io/en/latest/} \cite{Lewis:2019xzd}, and to compute the full Bayesian evidences for the different models and dataset combinations, we have employed the code \texttt{MCEvidence}\footnote{https://github.com/yabebalFantaye/MCEvidence} \cite{Heavens:2017afc}. The Deviance Information Criterion (DIC)\,\cite{Spiegelhalter:2002yvw} has been computed with our own numerical code. The results are displayed in Tables \hyperref[Table:BD.TableFitBaseline]{\ref{Table:BD.TableFitBaseline}}, \hyperref[Table:BD.TableFit+SL]{\ref{Table:BD.TableFit+SL}}, \hyperref[Table:BD.TableFitSpectrum]{\ref{Table:BD.TableFitSpectrum}}, \hyperref[Table:BD.TableFitAlternativeDataset]{\ref{Table:BD.TableFitAlternativeDataset}}, \hyperref[Table:BD.TableFitGR]{\ref{Table:BD.TableFitGR}}, \hyperref[Table:BD.TableFitXCDM]{\ref{Table:BD.TableFitXCDM}}, \hyperref[Table:BD.TableFitS8]{\ref{Table:BD.TableFitS8}}, \hyperref[Table:BD.TableFitCassini]{\ref{Table:BD.TableFitCassini}}, and also in \hyperref[Fig:BD.BayesRatio]{Fig.\,\ref{Fig:BD.BayesRatio}} Figs. 10-11. They are discussed in the next section.

\section{Numerical analysis. Results}\label{Sect:NumericalAnalysis}

In the following we put the models under consideration to the test, using the various datasets described in \hyperref[Sect:MethodData]{Sect.\,\ref{Sect:MethodData}}. We will review some statistical concepts of our analysis, however the reader may refer to \hyperref[Appendix:Bayesian]{chapter\,\ref{Appendix:Bayesian}} for additional explanations. Please note that the notation used in the appendix may differ from that used in this section. Here we perform the statistical analysis of the models in terms of a joint likelihood function, which is the product of the individual likelihoods for each data source and includes the corresponding covariance matrices. For a fairer comparison with the GR-$\CC$CDM  we use standard information criteria in which the presence of extra parameters in a given model is conveniently penalized so as to achieve a balanced comparison with the model having less parameters.  More concretely, we employ the full Bayesian evidence to duly quantify the fitting ability of the BD-$\CC$CDM model as compared to its GR analogue. Given a dataset $\mathcal{D}$, the probability of a certain model $M_i$ to be the best one among a given set of models $\{M\}$ reads,
\begin{figure}[t!]
\begin{center}
\includegraphics[width=4.5in, height=3.5in]{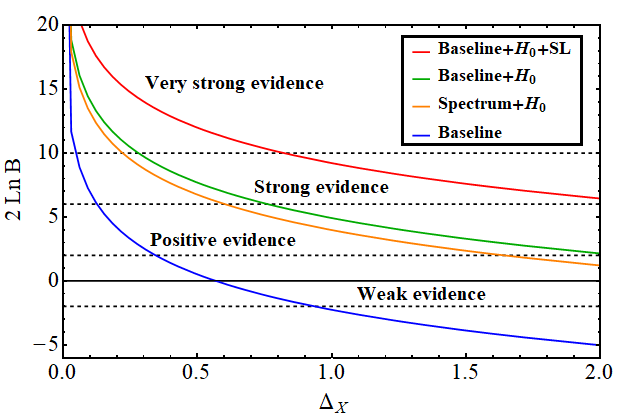}
\caption{The full Bayesian evidence curves for the BD-$\CC$CDM as compared to the GR-$\CC$CDM, using different datasets and as a function of $\Delta_X=\Delta\epsilon_{\rm BD}/10^{-2}=\Delta\varphi_{\rm ini}/0.2$, with $\Delta\epsilon_{\rm BD}$ and $\Delta\varphi_{\rm ini}$ being the (flat) prior ranges for $\epsilon_{\rm BD}$ and $\varphi_{\rm ini}$, respectively. The curves are computed using the exact evidence formula, Eq. \protect\eqref{Eq:BD.evidence}. The marked evidence ranges conform with the conventional Jeffreys' scale, see the main text in \protect\hyperref[Sect:NumericalAnalysis]{Sect.\,\ref{Sect:NumericalAnalysis}}.}\label{Fig:BD.BayesRatio}
\end{center}
\end{figure}
%
\begin{equation}\label{Eq:BD.BayesTheorem}
P(M_i|\mathcal{D})=\frac{P(M_i)\mathcal{E}(\mathcal{D}|M_i)}{P(\mathcal{D})}\,,
\end{equation}
{where $P(M_i)$ is the prior probability of the model $M_i$, $P(\mathcal{D})$ the probability of having the dataset $\mathcal{D}$, and the normalization condition $\sum_{j\in\{M\}}P(M_j)=1$ is assumed. The quantity $\mathcal{E}(\mathcal{D}|M_i)$ is the so-called marginal likelihood or evidence\,\cite{Amendola:2015ksp}. If the model $M_i$ has $n$ parameters contained in the vector $\vec{p}^{M_i}=(p^{M_i}_1, p^{M_i}_2,...,p^{M_i}_n)$, the evidence takes the following form:}

\renewcommand{\arraystretch}{1.1}
\begin{table}[t!]
\begin{center}
\resizebox{1\textwidth}{!}{

\begin{tabular}{|c  |c | c |  c | c | c  |}
 \multicolumn{1}{c}{} & \multicolumn{2}{c}{Baseline} & \multicolumn{2}{c}{Baseline+$H_0$}
\\\hline
{\scriptsize Parameter} & {\scriptsize GR-$\Lambda$CDM}  & {\scriptsize BD-$\Lambda$CDM} & {\scriptsize GR-$\Lambda$CDM}  &  {\scriptsize BD-$\Lambda$CDM}
\\\hline
{\scriptsize $H_0$ (km/s/Mpc)}  & {\scriptsize $68.20^{+0.41}_{-0.40}$} & {\scriptsize $68.86^{+1.15}_{-1.24}$} & {\scriptsize $68.57^{+0.36}_{-0.42}$}  & {\scriptsize $70.83^{+0.92}_{-0.95}$}
\\\hline
{\scriptsize$\omega_{\rm b}$} & {\scriptsize $0.02227^{+0.00019}_{-0.00018}$}  & {\scriptsize $0.02251^{+0.00026}_{-0.00027}$} & {\scriptsize $0.02238\pm 0.00019$}  &  {\scriptsize $0.02275^{+0.00024}_{-0.00026}$}
\\\hline
{\scriptsize$\omega_{cdm}$} & {\scriptsize $0.11763^{+0.00090}_{-0.00092}$}  & {\scriptsize $0.11598^{+0.00159}_{-0.00152}$} & {\scriptsize $0.11699^{+0.00092}_{-0.00083}$}  &  {\scriptsize $0.11574^{+0.00164}_{-0.00158}$}
\\\hline
{\scriptsize$\tau$} & {{\scriptsize$0.050^{+0.004}_{-0.008}$}} & {{\scriptsize$0.052^{+0.006}_{-0.008}$}} & {{\scriptsize$0.051^{+0.005}_{-0.008}$}}  &   {{\scriptsize$0.053^{+0.006}_{-0.008}$}}
\\\hline
{\scriptsize$n_{\rm s}$} & {{\scriptsize$0.9683^{+0.0039}_{-0.0038}$}}  & {{\scriptsize$0.9775^{+0.0084}_{-0.0086}$}} & {{\scriptsize$0.9703^{+0.0038}_{-0.0036}$}} &   {{\scriptsize$0.9873^{+0.0076}_{-0.0075}$}}
\\\hline
{\scriptsize$\sigma_8$}  & {{\scriptsize$0.797^{+0.005}_{-0.006}$}}  & {{\scriptsize$0.785\pm 0.013$}} & {{\scriptsize$0.796^{+0.006}_{-0.007}$}}  &   {{\scriptsize$0.789\pm 0.013$}}
\\\hline
{\scriptsize$r_{\rm s}$ (Mpc)}  & {{\scriptsize$147.83^{+0.29}_{-0.30}$}}  & {{\scriptsize$145.89^{+2.26}_{-2.49}$}} & {{\scriptsize$147.88\pm 0.31$}}  &   {{\scriptsize$142.46^{+1.84}_{-1.86}$}}
\\\hline
{\scriptsize $\epsilon_{\rm BD}$} & - & {{\scriptsize $-0.00184^{+0.00140}_{-0.00142}$}} & - &   {{\scriptsize $-0.00199^{+0.00142}_{-0.00147}$}}
\\\hline
{\scriptsize$\varphi_{\rm ini}$} & - & {{\scriptsize $0.974^{+0.027}_{-0.031}$}} & - &    {{\scriptsize $0.932^{+0.022}_{-0.023}$}}
\\\hline
{\scriptsize$\varphi(0)$} & - & {{\scriptsize $0.960^{+0.032}_{-0.037}$}} & - &    {{\scriptsize $0.918^{+0.027}_{-0.029}$}}
\\\hline
{\scriptsize$\weff(0)$} & - & {{\scriptsize $-0.983^{+0.015}_{-0.014}$}} & - &    {{\scriptsize $-0.966^{+0.012}_{-0.011}$}}
\\\hline
{\tiny $\dot{G}(0)/G(0) (10^{-13}yr^{-1})$} & - & {{\scriptsize $2.022^{+1.585}_{-1.518}$}} & - &    {{\scriptsize $2.256^{+1.658}_{-1.621}$}}
\\\hline
{\scriptsize$\chi^2_{\rm min}$} & {\scriptsize 2271.98}  & {\scriptsize 2271.82} & {\scriptsize 2285.50}  &  {\scriptsize 2276.04}
\\\hline
{\scriptsize$2\ln B$} & {\scriptsize -}  & {\scriptsize -2.26} & {\scriptsize -}  &  {\scriptsize +4.92}
\\\hline
{\scriptsize$\Delta {\rm DIC}$} & {\scriptsize -}  & {\scriptsize -0.54} & {\scriptsize -}  &  {\scriptsize +4.90}
\\\hline
\end{tabular}}
\end{center}
\caption{The mean fit values and $68.3\%$ confidence limits for the considered models using our baseline dataset in the first block, {\it i.e.} SNIa+$H(z)$+BAO+LSS+CMB TT data, and baseline+$H_0$ in the second one (cf. \protect\hyperref[Sect:MethodData]{Sect.\,\ref{Sect:MethodData}} for details). These results have been obtained within our main BD scenario (Scenario II of \,\protect\hyperref[Sect:Mach]{Sect.\,\ref{Sect:Mach}}). In all cases a massive neutrino of $0.06$ eV has been included. First we display the fitting results for the six conventional parameters, namely: $H_0$, the reduced density parameters for baryons ($\omega_{b}=\Omega_{\rm b} h^2$) and CDM ($\omega_{cdm}=\Omega_{cdm} h^2$), the reionization optical depth $\tau$, the spectral index $n_{\rm s}$ of the primordial power-law power spectrum, and, for convenience, instead of the amplitude $A_{\rm s}$ of such spectrum we list the values of $\sigma_8$. We also include the sound horizon at the baryon drag epoch, $r_{\rm s}$, obtained as a derived parameter. Right after we list the values of the free parameters that characterize the BD model: $\epsilon_{\rm BD}$ \protect\eqref{Eq:BD.definitions} and the initial condition for the BD-field, $\varphi_{\rm ini}$. We also include the values of the BD-field, the (exact) effective EoS parameter \protect\eqref{Eq:BD.effEoS}, and the ratio between the derivative and the value of Newton's coupling, all computed at $z=0$. Finally, we report the values of the minimum of the $\chi^2$-function, $\chi^2_{\rm min}$, the exact Bayes ratios (computed under the conditions explained in the main text of Sec. 8), and the DIC. It is also worth to remark that the baseline dataset employed here includes the contribution not only of the spectrum, but also the bispectrum information from BOSS \protect\cite{Gil-Marin:2016wya}, see \hyperref[Sect:MethodData]{Sect.\,\ref{Sect:MethodData}} for details.}\label{Table:BD.TableFitBaseline}
\end{table}
\begin{equation}\label{Eq:BD.evidence}
\mathcal{E}(\mathcal{D}|M_i)=\int \mathcal{L}(\mathcal{D}|\vec{p}^{M_i},M_i)\pi(\vec{p}^{M_i}) d^np^{M_i}\,,
\end{equation}
{with $\mathcal{L}(\mathcal{D}|\vec{p}^{M_i},M_i)$ the likelihood and $\pi(\vec{p}^{M_i})$ the prior of the parameters entering the model $M_i$. The evidence is larger for those models that have more overlapping volume between the likelihood and the prior distributions, but penalizes the use of additional parameters having a non-null impact on the likelihood. Hence, the evidence constitutes a good way of quantifying the performance of the model by implementing in practice the Occam razor principle. We can compare the fitting performance of BD-$\CC$CDM and GR-$\Lambda$CDM models by assuming equal prior probability for both of them, {\it i.e.} $P({\rm BD-}\CC{\rm CDM})=P({\rm GR-}\CC{\rm CDM})$ (``Principle of Insufficient Reason''). The ratio of their associated probabilities can then be directly written as the ratio of their corresponding evidences, {\it i.e.}}

\renewcommand{\arraystretch}{1.1}
\begin{table}[t!]
\begin{center}
\resizebox{0.65\textwidth}{!}{

\begin{tabular}{|c  |c | c |  c |}
 \multicolumn{1}{c}{} & \multicolumn{2}{c}{Baseline+$H_0$+SL}
\\\hline
{\scriptsize Parameter} & {\scriptsize GR-$\CC$CDM}  & {\scriptsize BD-$\CC$CDM}
\\\hline
{\scriptsize $H_0$ (km/s/Mpc)}  & {\scriptsize $68.74^{+0.37}_{-0.40}$} & {\scriptsize $71.30^{+0.80}_{-0.84}$}
\\\hline
{\scriptsize$\omega_{\rm b}$} & {\scriptsize $0.02242^{+0.00018}_{-0.00019}$}  & {\scriptsize $0.02281\pm 0.00025$}
\\\hline
{\scriptsize$\omega_{cdm}$} & {\scriptsize $0.11666^{+0.00087}_{-0.00086}$}  & {\scriptsize $0.11560^{+0.00158}_{-0.00169}$}
\\\hline
{\scriptsize$\tau$} & {{\scriptsize$0.051^{+0.005}_{-0.008}$}} & {{\scriptsize$0.053^{+0.006}_{-0.008}$}}
\\\hline
{\scriptsize$n_{\rm s}$} & {{\scriptsize$0.9708^{+0.0036}_{-0.0038}$}}  & {{\scriptsize$0.9901^{+0.0075}_{-0.0070}$}}
\\\hline
{\scriptsize$\sigma_8$}  & {{\scriptsize$0.795^{+0.006}_{-0.007}$}}  & {{\scriptsize$0.789\pm 0.013$}}
\\\hline
{\scriptsize$r_{\rm s}$ (Mpc)}  & {{\scriptsize$147.93^{+0.30}_{-0.31}$}}  & {{\scriptsize$141.68^{+1.69}_{-1.73}$}}
\\\hline
{\scriptsize $\epsilon_{\rm BD}$}  & {\scriptsize -}   & {{\scriptsize$-0.00208^{+0.00151}_{-0.00140}$}}
\\\hline
{\scriptsize$\varphi_{\rm ini}$}  & {\scriptsize -}   & {{\scriptsize$0.923^{+0.019}_{-0.021}$}}
\\\hline
{\scriptsize$\varphi(0)$}  & {\scriptsize -}   & {{\scriptsize$0.908^{+0.026}_{-0.028}$}}
\\\hline
{\scriptsize$\weff(0)$}  & {\scriptsize -}   & {{\scriptsize$-0.962\pm0.011$}}
\\\hline
{\tiny $\dot{G}(0)/G(0) (10^{-13}yr^{-1})$} & {\scriptsize -}   & {{\scriptsize$2.375^{+1.612}_{-1.721}$}}
\\\hline
{\scriptsize$\chi^2_{\rm min}$} & {\scriptsize 2320.40}  & {\scriptsize 2305.80}
\\\hline
{\scriptsize$2\ln B$} & {\scriptsize -}  & {\scriptsize $+9.22$}
\\\hline
{\scriptsize$\Delta {\rm DIC}$} & {\scriptsize -}  & {\scriptsize $+9.15$}
\\\hline
\end{tabular}}
\end{center}
\caption{\scriptsize Fitting results as in \hyperref[Table:BD.TableFitBaseline]{Table\,\ref{Table:BD.TableFitBaseline}}, but adding the Strong-Lensing data, {\it i.e.} we use here the dataset Baseline+$H_0$+SL, for both the GR-$\CC$CDM and the BD-$\CC$CDM.}\label{Table:BD.TableFit+SL}
\end{table}


\renewcommand{\arraystretch}{1.1}
\begin{table}[t!]
\begin{center}
\resizebox{1\textwidth}{!}{

\begin{tabular}{|c  |c | c |  c | c | c  |}
 \multicolumn{1}{c}{} & \multicolumn{2}{c}{Spectrum} & \multicolumn{2}{c}{Spectrum+$H_0$}
\\\hline
{\scriptsize Parameter} & {\scriptsize GR-$\Lambda$CDM}  & {\scriptsize BD-$\Lambda$CDM} & {\scriptsize GR-$\Lambda$CDM}  &  {\scriptsize BD-$\Lambda$CDM}
\\\hline
{\scriptsize $H_0$ (km/s/Mpc)}  & {\scriptsize $68.00^{+0.47}_{-0.48}$} & {\scriptsize $68.86^{+1.27}_{-1.34}$} & {\scriptsize $68.61^{+0.46}_{-0.49}$}  & {\scriptsize $70.94^{+1.00}_{-0.98}$}
\\\hline
{\scriptsize$\omega_{\rm b}$} & {\scriptsize $0.02223^{+0.00020}_{-0.00021}$}  & {\scriptsize $0.02241\pm 0.00027$} & {\scriptsize $0.02239^{+0.00020}_{-0.00019}$}  &  {\scriptsize $0.02264^{+0.00026}_{-0.00025}$}
\\\hline
{\scriptsize$\omega_{cdm}$} & {\scriptsize $0.11809^{+0.00112}_{-0.00095}$}  & {\scriptsize $0.11743^{+0.00168}_{-0.00170}$} & {\scriptsize $0.11695\pm 0.00104$}  &  {\scriptsize $0.11702^{+0.00169}_{-0.00167}$}
\\\hline
{\scriptsize$\tau$} & {{\scriptsize$0.051^{+0.005}_{-0.008}$}} & {{\scriptsize$0.053^{+0.006}_{-0.008}$}} & {{\scriptsize$0.053^{+0.006}_{-0.008}$}}  &   {{\scriptsize$0.054^{+0.007}_{-0.008}$}}
\\\hline
{\scriptsize$n_{\rm s}$} & {{\scriptsize$0.9673^{+0.0039}_{-0.0044}$}}  & {{\scriptsize$0.9742^{+0.0086}_{-0.0090}$}} & {{\scriptsize$0.9705^{+0.0040}_{-0.0041}$}} &   {{\scriptsize$0.9845^{+0.0076}_{-0.0077}$}}
\\\hline
{\scriptsize$\sigma_8$}  & {{\scriptsize$0.800^{+0.006}_{-0.007}$}}  & {{\scriptsize$0.798\pm 0.014$}} & {{\scriptsize$0.798\pm 0.007$}}  &   {{\scriptsize$0.801^{+0.015}_{-0.014}$}}
\\\hline
{\scriptsize$r_{\rm s}$ (Mpc)}  & {{\scriptsize$147.75^{+0.31}_{-0.35}$}}  & {{\scriptsize$145.82^{+2.33}_{-2.52}$}} & {{\scriptsize$147.88^{+0.33}_{-0.32}$}}  &   {{\scriptsize$142.55^{+1.71}_{-1.96}$}}
\\\hline
{\scriptsize $\epsilon_{\rm BD}$} & - & {{\scriptsize $-0.00079^{+0.00158}_{-0.00157}$}} & - &   {{\scriptsize $-0.00081^{+0.00162}_{-0.00165}$}}
\\\hline
{\scriptsize$\varphi_{\rm ini}$} & - & {{\scriptsize $0.976^{+0.028}_{-0.032}$}} & - &    {{\scriptsize $0.935^{+0.020}_{-0.024}$}}
\\\hline
{\scriptsize$\varphi(0)$} & - & {{\scriptsize $0.970^{+0.034}_{-0.038}$}} & - &    {{\scriptsize $0.929^{+0.028}_{-0.030}$}}
\\\hline
{\scriptsize$\weff(0)$} & - & {{\scriptsize $-0.987^{+0.016}_{-0.014}$}} & - &    {{\scriptsize $-0.971^{+0.013}_{-0.011}$}}
\\\hline
{\tiny $\dot{G}(0)/G(0) (10^{-13}yr^{-1})$} & - & {{\scriptsize $0.864^{+1.711}_{-1.734}$}} & - &    {{\scriptsize $0.913^{+1.895}_{-1.791}$}}
\\\hline
{\scriptsize$\chi^2_{\rm min}$} & {\scriptsize 2269.04}  & {\scriptsize 2268.28} & {\scriptsize 2283.66}  &  {\scriptsize 2274.64}
\\\hline
{\scriptsize$2\ln B$} & {\scriptsize -}  & {\scriptsize $-2.94$} & {\scriptsize -}  &  {\scriptsize $+3.98$}
\\\hline
{\scriptsize$\Delta {\rm DIC}$} & {\scriptsize -}  & {\scriptsize $-3.36$} & {\scriptsize -}  &  {\scriptsize $+4.76$}
\\\hline
\end{tabular}}
\end{center}
\caption{As in \hyperref[Table:BD.TableFitBaseline]{\ref{Table:BD.TableFitBaseline}}, but replacing the BOSS BAO+LSS data from \protect\cite{Gil-Marin:2016wya} with those from \protect\cite{Alam:2016hwk}, which only includes the spectrum information. See the discussion of the results in \protect\hyperref[Sect:NumericalAnalysis]{Sect.\,\ref {Sect:NumericalAnalysis}}.}\label{Table:BD.TableFitSpectrum}
\end{table}


%
\begin{equation}\label{Eq:BD.BayesRatio}
\frac{P({\rm BD-}\CC{\rm CDM}|\mathcal{D})}{P({\rm GR-}\CC{\rm CDM}|\mathcal{D})}=\frac{\mathcal{E}(\mathcal{D}|{\rm BD-}\CC{\rm CDM})}{\mathcal{E}(\mathcal{D}|{\rm GR-}\CC{\rm CDM})} \equiv B\,,
\end{equation}
where $B$ is the so-called Bayes ratio (or Bayes factor) and is the quantity we are interested in. Notice that when $B>1$ this means that data prefer the BD-$\CC$CDM model over the GR version, but of course depending on how much larger than 1 it is we will have different levels of statistical significance for such preference. It is common to adopt in the literature the so-called Jeffreys' scale to categorize the level of evidence that one can infer from the computed value of the Bayes ratio. Jeffrey's scale actually is usually written not directly in terms of $B$, but in terms of $2\ln B$. The latter is sometimes estimated with a simple Schwarz (or Bayesian) information criterion $\Delta$BIC \cite{Schwarz1978,KassRaftery1995}, although $2\ln B$ is a much more rigorous, sophisticated (and difficult to compute) statistics than just the usual  $\Delta$BIC estimates based on using the minimum value of $\chi^2$, the number of points and the number  of independent fitting parameters. If   $2\ln B$ lies below $2$ in absolute value, then we can conclude that the evidence in favor of BD-$\CC$CDM (against GR-$\CC$CDM)  is at most only {\it weak}, and in all cases {\it not conclusive}; if $2<2\ln B<6$ the evidence is said to be {\it positive}; if, instead, $6<2\ln B<10$, then it is considered to be {\it strong}, whereas if $2\ln B>10$ one is entitled to speak of {\it very strong} evidence in favor of the BD-$\CC$CDM over the GR-$\CC$CDM model. For more technical details related with the evidence and the Bayes ratio we refer the reader to \cite{KassRaftery1995,Burnham2002,Amendola:2015ksp}. Notice that the computation of \eqref{Eq:BD.BayesRatio} is not easy in general; in fact, it can be rather cumbersome since we usually work with models with a high number of parameters, so the multiple integrals that we need to compute become quite demanding from the computational point of view. We have calculated the evidences numerically, of course, processing the Markov chains obtained from the Monte Carlo analyses carried out with \texttt{CLASS}+\texttt{MontePython} \cite{Blas:2011rf,Audren:2012wb} with the numerical code \texttt{MCEvidence} \cite{Heavens:2017afc}, which is publicly available (cf.  \hyperref[Sect:MethodData]{Sect.\,\ref{Sect:MethodData}}). We report the values obtained for $2\ln B$ \eqref{Eq:BD.BayesRatio} in Tables \hyperref[Table:BD.TableFitBaseline]{\ref{Table:BD.TableFitBaseline}}, \hyperref[Table:BD.TableFit+SL]{\ref{Table:BD.TableFit+SL}}, \hyperref[Table:BD.TableFitSpectrum]{\ref{Table:BD.TableFitSpectrum}}, \hyperref[Table:BD.TableFitAlternativeDataset]{\ref{Table:BD.TableFitAlternativeDataset}}, \hyperref[Table:BD.TableFitXCDM]{\ref{Table:BD.TableFitXCDM}} and \hyperref[Table:BD.TableFitCassini]{\ref{Table:BD.TableFitCassini}}. \hyperref[Table:BD.TableFitBaseline]{Table\,\ref{Table:BD.TableFitBaseline}} contains the fitting results for the BD- and GR$-\CC$CDM models obtained with the Baseline and Baseline+$H_0$ datasets. In \hyperref[Table:BD.TableFit+SL]{Table\,\ref{Table:BD.TableFit+SL}} we present the results for the Baseline+$H_0$+SL dataset. In \hyperref[Table:BD.TableFitSpectrum]{Table\,\ref{Table:BD.TableFitSpectrum}} we show the output of the fitting analysis for the same models and using the same data as in \hyperref[Table:BD.TableFitBaseline]{Table\,\ref{Table:BD.TableFitBaseline}}, but changing the BOSS data from \cite{Gil-Marin:2016wya}, which contain both  the mater spectrum and bispectrum information, by the BOSS data from \cite{Alam:2016hwk}, which only incorporate the spectrum part ({\it i.e.} the usual matter power spectrum). In \hyperref[Table:BD.TableFitAlternativeDataset]{Table\,\ref{Table:BD.TableFitAlternativeDataset}} we plug the results obtained for the BD-$\CC$CDM with the alternative datasets described in \hyperref[Sect:MethodData]{Sect.\,\ref{Sect:MethodData}}, and in \hyperref[Table:BD.TableFitGR]{Table\,\ref{Table:BD.TableFitGR}} we show the corresponding results for the GR-$\CC$CDM. Next, in \hyperref[Table:BD.TableFitXCDM]{Table\,\ref{Table:BD.TableFitXCDM}} we present the results with the Baseline and Baseline+$H_0$ data configurations obtained using the GR-XCDM parametrization. In \hyperref[Table:BD.TableFitS8]{Table\,\ref{Table:BD.TableFitS8}} we display the values of the parameters $\sigma_8$ and $S_8$ for the GR- and BD-$\CC$CDM models, as well as the parameter $\tilde{S}_8 = S_8/\sqrt{\varphi(0)}$ for the BD. Finally, in \hyperref[Table:BD.TableFitCassini]{Table\,\ref{Table:BD.TableFitCassini}} we present the fitting results for the BD-$\CC$CDM model, considering the (Cassini-constrained) Scenario III described in \hyperref[Sect:Mach]{Sect.\,\ref{Sect:Mach}}.

\renewcommand{\arraystretch}{1.1}
\begin{table}[t!]
\begin{center}
\resizebox{1\textwidth}{!}{
\begin{tabular}{|c  |c | c |  c | c | c | c | c | c| c|}
 \multicolumn{1}{c}{} & \multicolumn{2}{c}{} & \multicolumn{1}{c}{} & \multicolumn{1}{c}{} & \multicolumn{1}{c}{} & \multicolumn{1}{c}{} & \multicolumn{1}{c}{}
\\\hline
Datasets & {\small $H_0$}  & {\small $\omega_{\rm m}$} & {\small $\sigma_8$ } & {\small $r_{\rm s}$ (Mpc) }  & {\small $\epsilon_{\rm BD}\cdot 10^{3}$ }  & {\small $\weff(0)$}  & {\small $2\ln B$}
\\\hline
{\small B+$M$}  & {\small $71.19^{+0.92}_{-1.02}$} & {\small $0.1390^{+0.0014}_{-0.0015}$} & {\small $0.788^{+0.012}_{-0.013}$} & {\small $141.87^{+2.06}_{-1.81}$} & {\small $-2.16^{+1.42}_{-1.36}$} & {\small $-0.963^{+0.012}_{-0.011}$} & {\small $+10.38$}
\\\hline
{\small B+$H_0$+pol}  & {\small $69.85^{+0.81}_{-0.85}$} & {\small $0.1409^{+0.0012}_{-0.0011}$} & {\small $0.801\pm 0.011$} & {\small $144.72^{+1.51}_{-1.83}$} & {\small $-0.30^{+1.20}_{-1.23}$} &  {\small $-0.985^{+0.012}_{-0.009}$} & {\small $-1.44$ }
\\\hline
{\small B+$H_0$+pol+lens}  & {\small $69.74^{+0.82}_{-0.77}$} & {\small $0.1416^{+0.0011}_{-0.0010}$} & {\small $0.808\pm 0.009$} & {\small $144.66^{+1.56}_{-1.61}$} & {\small $0.00^{+1.11}_{-1.07} $} &  {\small $-0.986\pm 0.010$} & {\small $-1.98$ }
\\\hline
{\small B+$H_0$+WL}  & {\small $70.69^{+0.91}_{-0.90}$} & {\small $0.1398^{+0.0015}_{-0.0013}$} & {\small $0.794^{+0.011}_{-0.012}$} & {\small $142.76^{+1.79}_{-1.86}$} & {\small $-1.42^{+1.29}_{-1.37}$} & {\small $-0.970^{+0.011}_{-0.010}$} & {\small $+4.34$ }
\\\hline
{\small CMB+BAO+SNIa}  & {\small $68.63^{+1.44}_{-1.50}$} & {\small $0.1425\pm 0.0019$} & {\small $0.818^{+0.017}_{-0.018}$} & {\small $146.62^{+2.92}_{-2.93}$} & {\small $1.14^{+1.84}_{-1.68}$} &  {\small $-0.999^{+0.020}_{-0.017}$} & {\small $-3.00$ }
\\\hline
\end{tabular}}
\end{center}
\caption{Fitting results for the BD-$\Lambda$CDM model obtained with some alternative datasets and in all cases within the main BD Scenario II. Due to the lack of space, we employ some abbreviations, namely: {\bf B} for the Baseline dataset described in \hyperref[Sect:MethodData]{Sect.\,\ref{Sect:MethodData}}; {\bf pol} for the Planck 2018 (TE+EE) high-$\ell$  polarization data; and {\bf lens} for the CMB lensing. {The $\omega_{\rm m}$ parameter contains the contribution of baryons and dark matter.  In the last row, CMB refers to the TT+lowE Planck 2018 likelihood (cf. \hyperref[Sect:MethodData]{Sect.\,\ref{Sect:MethodData}} for details on the data). $H_0$ is given in km/s/Mpc. For a discussion of the results, see \hyperref[Sect:NumericalAnalysis]{Sect.\,\ref{Sect:NumericalAnalysis}}.}}\label{Table:BD.TableFitAlternativeDataset}
\end{table}


The evidence \eqref{Eq:BD.evidence} clearly depends on the priors for the parameters entering the model. For the 6 parameters in common in the BD- and GR-$\Lambda$CDM models, namely, $(\omega_{\rm b}=\Omega_{\rm b} h^2,\omega_{cdm}=\Omega_{cdm}h^2,H_0,\tau,A_{\rm s},n_{\rm s})$, if we use the same flat priors in both models they cancel exactly in the computation of the Bayes ratio \eqref{Eq:BD.BayesRatio}. Thus, the latter does not depend on the priors for these parameters if their ranges are big enough so as to not alter the shape of the likelihood severely. The Bayes ratio is, though, sensitive to the priors for the two additional parameters introduced in the BD-$\CC$CDM model in our Scenario II (cf. \hyperref[Sect:Mach]{Sect.\,\ref{Sect:Mach}}), {\it i.e.} $\varphi_{\rm ini}$ and $\epsilon_{\rm BD}$, since they are not canceled in \eqref{Eq:BD.BayesRatio}. We study the dependence of the evidence on these priors in \hyperref[Fig:BD.BayesRatio]{Fig.\ref{Fig:BD.BayesRatio}}, where we plot $2\ln B$ obtained for the BD-$\CC$CDM model from different datasets, and as a function of a quantity that we call $\Delta_X$, defined as
\begin{equation}\label{Eq:BD.DeltaX}
\Delta_X\equiv \frac{\Delta\epsilon_{\rm BD}}{10^{-2}}=\frac{\Delta\varphi_{\rm ini}}{0.2}\,,
\end{equation}
with $\Delta\epsilon_{\rm BD}$ and $\Delta\varphi_{\rm ini}$ being the flat prior ranges of $\epsilon_{\rm BD}$ and $\varphi_{\rm ini}$, centered at $\epsilon_{\rm BD}=0$ and $\varphi_{\rm ini}=1$, respectively. $\Delta_X$ will be equal to one when $\Delta\epsilon_{\rm BD}=10^{-2}$ and $\Delta\varphi_{\rm ini}=0.2$, which are natural values for these prior ranges. The former implies $\oD>100$ and the latter could be associated to the range $0.9<\varphi_{\rm ini}<1.1$. We do not expect $\oD\lesssim 100$ since this would imply an exceedingly large departure from GR, even at cosmological scales, where this lower bound was already set using the first releases of WMAP CMB data almost twenty years ago, see e.g. \cite{Nagata:2003qn,Acquaviva:2004ti}. Regarding the prior range $0.9<\varphi_{\rm ini}<1.1$, it is also quite natural, since this is necessary to satisfy the BBN bounds \cite{Uzan:2010pm}. In all tables we report the values of $2\ln B$ obtained by setting the natural value $\Delta_X=1$, and in \hyperref[Fig:BD.BayesRatio]{Fig.\ref{Fig:BD.BayesRatio}}, as mentioned before, we also show how this quantity changes with the prior width, in terms of the variable $\Delta_X$ \eqref{Eq:BD.DeltaX}.

In the Cassini-constrained Scenario III (cf. again \hyperref[Sect:Mach]{Sect.\,\ref{Sect:Mach}}), we also allow variations of $\varphi_{\rm ini}$ and $\epsilon_{\rm BD}$ in our Monte Carlo runs, of course, but the natural prior range for $\eBD$ is now much smaller than in Scenario II, since now we expect it to be more constrained by the local observations. It is more natural to take in this case a prior range $\Delta\eBD=5\cdot 10^{-5}$ (still larger than Cassini's bound), and this is what we do in all the analyses of this scenario. See the comments in \hyperref[Sect:ConsiderationsBD]{Sect.\,\ref{Sect:ConsiderationsBD}}, and \hyperref[Table:BD.TableFitCassini]{Table\,\ref{Table:BD.TableFitCassini}}.

In Tables \hyperref[Table:BD.TableFitBaseline]{\ref{Table:BD.TableFitBaseline}}, \hyperref[Table:BD.TableFit+SL]{\ref{Table:BD.TableFit+SL}} and \hyperref[Table:BD.TableFitSpectrum]{\ref{Table:BD.TableFitSpectrum}} apart from the Bayes ratio, we also include the Deviance Information Criterion\,\cite{Spiegelhalter:2002yvw}, which is strictly speaking an approximation of the exact Bayesian approach that works fine when the posterior distributions are sufficiently Gaussian. The DIC is defined as
\begin{equation}\label{Eq:BD.DIC}
{\rm DIC}=\chi^2(\hat{\theta})+2p_D\,.
\end{equation}
Here $p_D=\overline{\chi^2}-\chi^2(\hat{\theta})$ is the effective number of parameters of the model and $\overline{\chi^2}$ the mean of the overall $\chi^2$ distribution. DIC is particularly suitable for us, since we can easily compute all the quantities involved directly from the Markov chains generated with \texttt{MontePython}. To compare the ability of the BD- and GR-$\Lambda$CDM models to fit the data, one has to compute the respective differences of DIC values between the first and second models. They are denoted $\Delta$DIC in our tables, and this quantity is the analogous to $2\ln B$.

\subsection{Comparing with the XCDM parametrization}

\renewcommand{\arraystretch}{1.1}
\begin{table}[t!]
\begin{center}
\resizebox{1\textwidth}{!}{
\begin{tabular}{|c  |c | c |  c | c | c |}
 \multicolumn{1}{c}{} & \multicolumn{2}{c}{} & \multicolumn{1}{c}{} & \multicolumn{1}{c}{}
\\\hline
{\scriptsize Datasets} & {\small $H_0$}  & {\small $\omega_{\rm m}$} & {\small $\sigma_8$ } & {\small $r_{\rm s}$ (Mpc) }
\\\hline
{\small B+$M$} & {\small $68.64^{+0.39}_{-0.38}$} & {\small $0.1398\pm 0.0009$} & {\small $0.796^{+0.005}_{-0.006}$} & {\small $147.87^{+0.29}_{-0.30}$}
\\\hline
{\small B+$H_0$+pol}  & {\small $68.50^{+0.33}_{-0.36}$} & {\small $0.1408^{+0.0007}_{-0.0008}$} & {\small $0.799\pm 0.006$} & {\small $147.53\pm 0.022$}
\\\hline
{\small B+$H_0$+pol+lens}  & {\small $68.38^{+0.35}_{-0.33}$} & {\small $0.1411\pm 0.0007$} & {\small $0.803\pm0.006$} & {\small $147.44^{+0.22}_{-0.21}$}
\\\hline
{\small B+$H_0$+WL}  & {\small $68.64\pm 0.37$} & {\small $0.1399^{+0.0008}_{-0.0009}$} & {\small $0.795^{+0.005}_{-0.007}$} & {\small $147.89^{+0.31}_{-0.29}$}
\\\hline
{\small CMB+BAO+SNIa}  & {\small $67.91^{+0.39}_{-0.41}$} & {\small $0.1413\pm 0.0009$} & {\small $0.805^{+0.006}_{-0.007}$} & {\small $147.66^{+0.31}_{-0.29}$}
\\\hline\hline
{\small B+$H_0$ (No LSS)}  & {\small $68.38^{+0.42}_{-0.37}$} & {\small $0.1405\pm 0.0009$} & {\small $0.802^{+0.007}_{-0.008}$} & {\small $147.76\pm 0.30$}
\\\hline
{\small Dataset \cite{Ballesteros:2020sik}}  & {\small $68.17^{+0.43}_{-0.44}$} & {\small $0.1416\pm 0.0009$} & {\small $0.810^{+0.006}_{-0.007}$} & {\small $147.35^{+0.22}_{-0.24}$}
\\\hline
{\small Dataset \cite{Ballesteros:2020sik} + LSS (SP)}  & {\small $68.36\pm 0.42$}  & {\small $0.1412\pm 0.0009$} & {\small $0.806\pm 0.006$} & {\small $147.43\pm $0.23}
\\\hline
\end{tabular}}
\end{center}
\caption{Different fitting results for the GR-$\Lambda$CDM model. The first five rows correspond to the different non-baseline datasets explored for the  BD-$\CC$CDM  in \hyperref[Table:BD.TableFitAlternativeDataset]{Table\,\ref{Table:BD.TableFitAlternativeDataset}}. The last three rows correspond  to other scenarios tested with the BD-$\CC$CDM model in \hyperref[Table:BD.TableFitCassini]{Table\,\ref{Table:BD.TableFitCassini}}, see \,\hyperref[Sect:ConsiderationsBD]{Sect.\,\ref{Sect:ConsiderationsBD}} for more details. $H_0$ is given in km/s/Mpc.}\label{Table:BD.TableFitGR}
\end{table}

As mentioned, in our numerical analysis of the data we also wish to consider the effect of a simple but powerful DDE parametrization, which is the traditional XCDM\cite{Turner:1998ex}. In this very simple framework, the DE is self-conserved and is associated to some unspecified entity or fluid (called X) which exists together with ordinary baryonic and cold dark matter, but it does not have any interaction with them.  The energy density of X is simply given by $\rho_X(a) = \rho_{X0}a^{-3(1+w_0)}$,   $\rho_{X0}=\rho_\Lambda$ being the current DE density value and $w_0$ the (constant) EoS parameter of such fluid. More complex parametrizations of the EoS can be considered, for instance the CPL one\,\cite{Chevallier:2000qy,Linder:2002et}, in which there is a  time evolution of the EoS itself. However, we have previously shown its incapability to improve the XCDM performance in solving the two tensions, see\,\cite{SolaPeracaula:2018wwm}. Thus, in this chapter we prefer to stay as closer as possible to the standard cosmological model and we will limit ourselves to analyze the XCDM only. By setting $w_0=-1$ we retrieve the $\Lambda$CDM model with constant $\rho_\Lambda$. For $w_0 \gtrsim-1$ the XCDM mimics quintessence, whereas for $w_0 \lesssim -1$ it mimics phantom DE. The fitting results generated from the XCDM on our datasets are used in our analysis as a figure of merit or benchmark to compare with the corresponding fitting efficiency of both the BD-$\CC$CDM and the GR-$\CC$CDM models. In the next section, we comment on the comparison.
%



\setlength{\tabcolsep}{0.7em}
\renewcommand{\arraystretch}{1.1}
\begin{table}[t!]
\begin{center}
\resizebox{0.7\textwidth}{!}{

\begin{tabular}{|c  |c | c |  c |}
 \multicolumn{1}{c}{} & \multicolumn{2}{c}{GR-XCDM}
\\\hline
{\normalsize Parameter} & {\normalsize Baseline}  & {\normalsize Baseline+$H_0$}
\\\hline
{\normalsize  $H_0$ (km/s/Mpc) }  & {\normalsize  $67.34^{+0.63}_{-0.66}$ } & {\normalsize $68.40^{+0.60}_{-0.62}$}
\\\hline
{\normalsize$\omega_{\rm b}$} & {\normalsize   $0.02235^{+0.00021}_{-0.00020}$}  & {\normalsize $0.02239^{+0.00019}_{-0.00020}$}
\\\hline
{\normalsize$\omega_{cdm}$} & {\normalsize $0.11649^{+0.00108}_{-0.00111}$}  & {\normalsize $0.11671^{+0.00117}_{-0.00109}$}
\\\hline
{\normalsize$\tau$} & {{\normalsize$0.053^{+0.006}_{-0.008}$}} & {{\normalsize$0.051^{+0.005}_{-0.008}$}}
\\\hline
{\normalsize$n_{\rm s}$} & {{\normalsize$0.9709\pm 0.0043$}}  & {{\normalsize$0.9707^{+0.0042}_{-0.0043}$}}
\\\hline
{\normalsize$\sigma_8$}  & {{\normalsize$0.782^{+0.011}_{-0.010}$}}  & {{\normalsize$0.792\pm 0.011$}}
\\\hline
{\normalsize$r_{\rm s}$ (Mpc)}  & {{\normalsize$148.05^{+0.32}_{-0.34}$}}  & {{\normalsize$147.95^{+0.33}_{-0.34}$}}
\\\hline
{\normalsize$w_0$}  & {{\normalsize$-0.956\pm0.026$}}  & {{\normalsize$-0.991^{+0.026}_{-0.024}$}}
\\\hline
{\normalsize$\chi^2_{\rm min}$} & {\normalsize 2269.88}  & {\normalsize 2285.22}
\\\hline
{\normalsize$2\ln B$} & {\normalsize $-2.23$}  & {\normalsize $-5.21$}
\\\hline
\end{tabular}}
\end{center}
\caption{As in \hyperref[Table:BD.TableFitBaseline]{Table\,\ref{Table:BD.TableFitBaseline}}, but for the XCDM parametrization (within GR). Motivated by previous works (see e.g. \protect\cite{SolaPeracaula:2018wwm}), we have used the (flat) prior $-1.1<w_0<-0.9$.}\label{Table:BD.TableFitXCDM}
\end{table}

\vspace{0.3cm}

\section{Extended considerations}\label{Sect:ConsiderationsBD}

In this chapter, we have dealt with Brans-Dicke  (BD) theory in  extenso. Our main goal was to assess if BD-gravity can help to smooth out the main two tensions besetting the usual concordance $\CC$CDM model (based on GR): i) the $H_0$-tension (the most acute existing discordance at present), and ii) the $\sigma_8$-tension, which despite not being so sharp it often occurs that the (many) models in the market dealing with the former tend to seriously  aggravate the latter.  As we have explained in the Introduction,  the  `golden rule' to be preserved by the tension solver should be to find a clue on how to tackle the main discrepancy (on the local $H_0$ parameter)  while at the same time to curb the $\sigma_8$ one, or at least not to worsen it. We have found that BD-gravity could be a key paradigm capable of such achievement.  Specifically, we have considered the original BD model with the only addition of a cosmological constant (CC), and we have performed a comprehensive analysis  in the light of a rich and updated set of observations. These involve a large variety of experimental inputs of various kinds, such as  the long chain SNIa+$H(z)$+BAO+LSS+CMB of data sources, which we have considered  at different levels and combinations; and tested with the inclusion of other potentially important factors such as the influence of the bispectrum component in the structure formation data (apart from the ordinary power spectrum); and also assessed the impact of  gravitational lensing data of different sorts (Weak and Strong-Lensing).

Although  BD-gravity is fundamentally different from GR, we have found very useful to try to pick out possible measurable signs of the new gravitational paradigm by considering the two frameworks in the (spatially flat) FLRW metric and compare the  versions of the $\CC$CDM model resulting in each case, which we have called BD-$\CC$CDM and GR-$\CC$CDM, respectively.  We have parametrized the departure of the former from the latter at the background level (cf. \,\hyperref[Sect:EffectiveEoS]{Sect.\,\ref{Sect:EffectiveEoS}}) and we have seen that BD-$\CC$CDM can appear in the form of a dynamical dark energy (DDE) version of the GR-$\CC$CDM, in which the vacuum energy density is evolving through a non-trivial EoS  (cf. \hyperref[Fig:BD.weffvarphi]{Fig.\ref{Fig:BD.weffvarphi}}).  We have called it  the `GR-picture' of the BD theory.  The resulting effective behavior is  $\CC$CDM-like with, however, a mild time-evolving  quasi-vacuum component. In fact, such behavior is not `pure vacuum' -- which is why we call it quasi-vacuum --  despite of the fact that the original BD-$\CC$CDM theory possesses a rigid cosmological constant.  Specifically, using the numerical fitting results of our analysis we find that such EoS  shows up in effective quintessence-like form at more than $3\sigma$ c.l. (this is perfectly appreciable at naked eye in \hyperref[Fig:BD.weffvarphi]{Fig.\ref{Fig:BD.weffvarphi}}). Our fit to the data demonstrates that such an effective  representation of BD-gravity can be competitive with the concordance model with a rigid $\CC$-term. It may actually create the fiction that the DE is dynamical when viewed within the GR framework, whilst it is  just a rigid CC in the underlying  BD action.  The practical outcome is that the BD approach with a CC definitely  helps to smooth out some of the tensions afflicting the $\CC$CDM in a manner very similar to the Running Vacuum Model, see e.g.\,\cite{SolaPeracaula:2016qlq,SolaPeracaula:2017esw,Gomez-Valent:2018nib,Gomez-Valent:2017idt,Sola:2017znb,Sola:2016hnq}, and this success might ultimately reveal the signs of the  BD theory. We conclude that finding  traces of vacuum dynamics, accompanied with apparent deviations from the standard matter conservation law\,\cite{Fritzsch:2012qc} could be the `smoking gun' pointing to the possibility that the gravity theory sitting behind these effects is not GR but BD.

\renewcommand{\arraystretch}{1.1}
\begin{table}[t!]
\begin{center}
\resizebox{1\textwidth}{!}{

\begin{tabular}{|c  |c | c |  c |c|c|c|}
 \multicolumn{1}{c}{} & \multicolumn{2}{c}{}
\\\hline
{\normalsize Scenarios} & {\normalsize ${\sigma}_{8(GR)}$} & {\normalsize ${\sigma}_{8(BD)}$} & {\normalsize ${S}_{8(GR)}$} & {\normalsize ${S}_{8(BD)}$}  & {\normalsize $\tilde{S}_{8(BD)}$}
\\\hline
{\normalsize Baseline } & {\normalsize $0.797^{+0.005}_{-0.006}$} & {\normalsize $0.785\pm 0.013$} & {\normalsize $0.800^{+0.010}_{-0.011}$} & {\normalsize $0.777^{+0.019}_{-0.020}$} & {\normalsize $0.793\pm 0.012$}
\\\hline
{\normalsize Baseline+$H_0$} & {\normalsize $0.796^{+0.006}_{-0.007}$} & {\normalsize $0.789\pm 0.013$} & {\normalsize $0.793^{+0.011}_{-0.010}$} & {\normalsize $0.758\pm 0.018$} & {\normalsize $0.792\pm 0.013$}
\\\hline
{\normalsize Baseline+$H_0$+SL} & {\normalsize $0.795^{+0.006}_{-0.007}$} & {\normalsize $0.789\pm 0.013$}& {{\normalsize$0.789^{+0.011}_{-0.010}$}} & {\normalsize $0.753^{+0.017}_{-0.018}$}  & {{\normalsize$0.791^{+0.012}_{-0.013}$}}
\\\hline
{\normalsize Spectrum}  & {\normalsize $0.800^{+0.006}_{-0.007}$}& {\normalsize $0.798\pm 0.014$} &{\normalsize $0.807\pm 0.013$} & {\normalsize $0.793^{+0.021}_{-0.022}$} & {\normalsize $0.805\pm 0.014$}
\\\hline
{\normalsize Spectrum+$H_0$} & {\normalsize $0.798\pm 0.007$} & {\normalsize $0.801^{+0.015}_{-0.014}$} & {{\normalsize$0.794^{+0.012}_{-0.013}$}} & {\normalsize $0.773^{+0.019}_{-0.021}$} & {{\normalsize$0.802^{+0.013}_{-0.014}$}}
\\\hline
\end{tabular}}
\end{center}
\caption{Fitted values for the $\sigma_{8(M)}$ (here M stands for GR or BD) obtained under different dataset configurations. We also include the derived values of $S_{8(M)}$ = $\sigma_8\sqrt{\Omega_{\rm m}/0.3}$ and the renormalized $\tilde{S}_8$ = $\sigma_8\sqrt{\tilde{\Omega}_{\rm m}/0.3}$, with $\tilde{\Omega}_{\rm m}$ defined as in \hyperref[Sect:rolesvarphiH0]{Sect.\,\ref{Sect:rolesvarphiH0}}. These results correspond to the main Scenario II of the BD-$\CC$CDM model.}\label{Table:BD.TableFitS8}
\end{table}


\subsection{Alleviating the $H_0$-tension}\label{Sect:H0-tension}

Our analysis of the BD theory with a CC, taken at face value,  suggests that the reason for the enhancement of $H_0$ in the BD model is because the effective gravitational coupling acting at cosmological scales, $\Geff\sim G_N/\varphi$, is higher than the one measured on Earth (see \hyperref[Fig:BD.varphi]{Fig.\ref{Fig:BD.varphi}}). This possibility allows the best-fit current energy densities of all the species to remain compatible at $\lesssim 1\sigma$ c.l. with the ones obtained in the GR-$\CC$CDM model (cf. e.g. Tables\,\hyperref[Table:BD.TableFitBaseline]{\ref{Table:BD.TableFitBaseline}}, \hyperref[Table:BD.TableFit+SL]{\ref{Table:BD.TableFit+SL}} and \hyperref[Table:BD.TableFitSpectrum]{\ref{Table:BD.TableFitSpectrum}}).  
Thus, since $\varphi<1$  we find that the increase of the Hubble parameter is basically due to the increase of  the effective $G$, and there is no need for a strong modification of the energy densities of the various species filling the Universe. This is a welcome feature since  the measurable cosmological mass parameter in the BD-$\CC$CDM model is, for sufficiently small $\eBD$,  not the usual $\Omega_{\rm m}$, but precisely the tilded one, related to it through $\tilde{\Omega}_{\rm m}=\Omega_{\rm m}/\varphi$.  The latter   is about   $\sim  8-9\%$  bigger than its standard model counterpart  ($\tilde{\Omega}_{\rm m}> \Omega_{\rm m}$)   as it follows from \hyperref[Fig:BD.varphi]{Fig.\ref{Fig:BD.varphi}}, where we can read off the current value  $\varphi(z=0)\simeq 0.918$.  Now, because at the background level  it is possible to write an approximate  Friedmann's equation \eqref{Eq:BD.H2}  in terms of the tilded parameters, these are indeed the ones that are actually  measured from SNIa and BAO observations in the BD context\footnote{Recall that for $\eBD\neq0$ the tilded  parameters $\tilde{\Omega}_i$  (which were originally defined for $\eBD=0$) receive a correction and become the hatted parameters $\hat{\Omega}_i$ introduced in Eq.\,\eqref{Eq:BD.hatOmega}.  However, the difference between them is of ${\cal O}(\eBD)$, see Eq.\,\eqref{Eq:BD.hatOmega2}, and since $|\epsilon_{\rm BD}|\sim \mathcal{O}(10^{-3})$ it can be ignored.}. The differences, however, as we have just pointed out, are not to be attributed to a change in the physical energy content of matter but to the fact that $\varphi<1$  throughout the entire cosmic history, as clearly shown in \hyperref[Fig:BD.varphi]{Fig.\ref{Fig:BD.varphi}}.  Obviously, the measurement of the parameters $\tilde{\Omega}_i$  can be performed  through the very same data  and procedures well accounted for  in the context of the GR-$\CC$CDM framework.  This explanation is perfectly consistent with the fact that when the Friedmann's law is expressed in terms of the effetive $G$, as indicated  in Eq.\,\eqref{Eq:BD.H1}, the local value of the Hubble parameter $H_0$ becomes bigger owing to $\Geff=G_N/\varphi$ being bigger than $G_N$.  Thus,  when we compare the early and local measurements of $H_0$ we do not meet  any anomaly in this approach.

We also recall at this point that there is no correction from $\wBD$ on the effective coupling $\Geff$, Eq.\,\eqref{Eq:BD.MainGeffective},  in the local domain. This is because in our context  $\wBD$ appears as being very large owing to the assumed screening of the BD-field caused by the clustered matter (cf. \hyperref[Sect:Mach]{Sect.\,\ref{Sect:Mach}}). From \hyperref[Fig:BD.varphi]{Fig.\ref{Fig:BD.varphi}} and \hyperref[Table:BD.TableFitBaseline]{Table\,\ref{Table:BD.TableFitBaseline}} we find that the BD model leads to a value of $H_0$ a factor  $\Geff^{1/2}(z=0)/G_N^{1/2}\sim 1/\varphi^{1/2}(z=0)=1/\sqrt{0.918}$, {\it i.e.} $\sim 4.5\%$, bigger than the one inferred from the CMB in the GR-$\CC$CDM model, in which $\Geff=G_N\,(\forall{z})$. It is reassuring to realize  that such a `renormalization factor'  can  enhance the low Planck 2018  CMB measurement of the  Hubble parameter (viz. $H_0=67.4\pm 0.5$ km/s/Mpc\,\cite{aghanim2020planck})  up to the range of  $70-71$km/s/Mpc (cf. e.g. Tables\,\hyperref[Table:BD.TableFitBaseline]{\ref{Table:BD.TableFitBaseline}}, \hyperref[Table:BD.TableFit+SL]{\ref{Table:BD.TableFit+SL}}, \hyperref[Table:BD.TableFitSpectrum]{\ref{Table:BD.TableFitSpectrum}} and \hyperref[Table:BD.TableFitAlternativeDataset]{\ref{Table:BD.TableFitAlternativeDataset}}), hence  much closer to the local measurements. For example, SH0ES yields $ H_0= (73.5\pm 1.4)$ km/s/Mpc\,\cite{Reid:2019tiq}; and when the latter is combined  with Strong-Lensing data from the H$0$LICOW collab.\cite{Wong:2019kwg} it leads to  $ H_0= (73.42\pm 1.09)$ km/s/Mpc. This combined value is $5\sigma$ at odds with the Planck 2018 measurement, a serious tension.

\renewcommand{\arraystretch}{1.1}
\begin{table}[t!]
\begin{center}
\resizebox{1\textwidth}{!}{
\begin{tabular}{|c  |c | c |  c | c | c  |}
 \multicolumn{1}{c}{} & \multicolumn{4}{c}{BD-$\Lambda$CDM (Scenario III: Cassini-constrained)}
\\\hline
{\scriptsize Parameter} & {\scriptsize B+$H_0$ (No LSS)}  & {\scriptsize B+$H_0$ } & {\scriptsize Dataset \cite{Ballesteros:2020sik}}  &  {\scriptsize Dataset \cite{Ballesteros:2020sik} + LSS (SP)}
\\\hline
{\scriptsize $H_0$ (km/s/Mpc)}  & {\scriptsize $70.99^{+0.94}_{-0.97}$} & {\scriptsize $70.80^{+0.81}_{-0.91}$} & {\scriptsize $70.01^{+0.86}_{-0.92}$}  & {\scriptsize $70.03^{+0.90}_{-0.88}$}
\\\hline
{\scriptsize$\omega_{\rm b}$} & {\scriptsize $0.02257\pm 0.00021$}  & {\scriptsize $0.02256^{+0.00019}_{-0.00020}$} & {\scriptsize $0.02271\pm 0.00016$}  &  {\scriptsize $0.02272^{+0.00015}_{-0.00016}$}
\\\hline
{\scriptsize$\omega_{cdm}$} & {\scriptsize $0.11839^{+0.00093}_{-0.00094}$}  & {\scriptsize $0.11748\pm 0.00089$} & {\scriptsize $0.11885^{+0.00092}_{-0.00095}$}  &  {\scriptsize $0.11827^{+0.00089}_{-0.00093}$}
\\\hline
{\scriptsize$\tau$} & {\scriptsize $0.057^{+0.007}_{-0.008}$} & {\scriptsize$0.050^{+0.004}_{-0.008}$} & {\scriptsize$0.061^{+0.007}_{-0.008}$}  &   {\scriptsize$0.058^{+0.006}_{-0.008}$}
\\\hline
{\scriptsize$n_{\rm s}$} & {\scriptsize $0.9824^{+0.0057}_{-0.0058}$}  & {\scriptsize$0.9811^{+0.0051}_{-0.0052}$} & {\scriptsize$0.9783^{+0.0052}_{-0.0059}$} &   {\scriptsize$0.9701^{+0.0056}_{-0.0054}$}
\\\hline
{\scriptsize$\sigma_8$}  & {\scriptsize $0.815^{+0.008}_{-0.009}$}  & {\scriptsize$0.804^{+0.006}_{-0.007}$} & {\scriptsize$0.817\pm 0.007$}  &   {\scriptsize$0.812^{+0.006}_{-0.007}$}
\\\hline
{\scriptsize$r_{\rm s}$ (Mpc)}  & {\scriptsize $142.14^{+1.91}_{-1.72}$}  & {\scriptsize$143.31^{+1.72}_{-1.63}$} & {\scriptsize$143.58^{+1.62}_{-1.55}$}  &   {\scriptsize$144.10^{+1.62}_{-1.52}$}
\\\hline
{\scriptsize $\epsilon_{\rm BD}$} & {\scriptsize $-0.00002\pm 0.00002$} & {\scriptsize $-0.00002\pm 0.00002$} & {\scriptsize $-0.00002\pm 0.00002$} &  {\scriptsize $-0.00002\pm 0.00002$}
\\\hline
{\scriptsize$\varphi_{\rm ini}$} & {\scriptsize $0.933\pm 0.021$} & {\scriptsize $0.944\pm 0.020$} &{\scriptsize $0.955^{+0.018}_{-0.019}$}&    {\scriptsize $0.960^{+0.020}_{-0.018}$}
\\\hline
{\scriptsize$\varphi(0)$} & {\scriptsize $0.933^{+0.020}_{-0.021}$} & {\scriptsize $0.944^{+0.019}_{-0.020}$} &{\scriptsize $0.955^{+0.018}_{-0.019}$}     &{\scriptsize $0.960^{+0.020}_{-0.017}$}
\\\hline
{\scriptsize$\weff(0)$} & {\scriptsize $-0.972\pm 0.009$} & {\scriptsize $-0.977\pm 0.008$} & {\scriptsize $-0.981^{+0.008}_{-0.007}$} &    {\scriptsize $-0.983\pm 0.008$}
\\\hline
{\tiny $\dot{G}(0)/G(0) (10^{-13}yr^{-1})$} & {\scriptsize $0.025^{+0.025}_{-0.026}$} & {\scriptsize $0.026^{+0.027}_{-0.028}$} & {\scriptsize $0.023^{+0.026}_{-0.027}$} &   {\scriptsize $0.020\pm 0.026$}
\\\hline
{\scriptsize$\chi^2_{\rm min}$} & {\scriptsize 2256.14}  & {\scriptsize 2278.34} & {\scriptsize 2797.44}  &  {\scriptsize 2812.68}
\\\hline
{\scriptsize$2\ln B$} & {\scriptsize +9.03}  & {\scriptsize $+5.21$} & {\scriptsize +3.45}  &  {\scriptsize +2.21}
\\\hline
\end{tabular}}
\end{center}
\caption{Fitting results for the BD-$\CC$CDM, in the context of the BD-Scenario III explained in \protect\hyperref[Sect:Mach]{Sect.\,\ref{Sect:Mach}} under different datasets.  As characteristic of Scenario III, in all of these datasets the Cassini constraint on the post-Newtonian parameter $\gamma^{PN}$ has been imposed \protect\cite{Bertotti:2003rm}. In the first two fitting columns we use the Baseline+$H_0$ dataset described in \protect\hyperref[Sect:MethodData]{Sect.\,\ref{Sect:MethodData}}. However, we exclude the LSS data in the first column while it is kept in the second. In the third and fourth fitting columns we report on the results obtained using the very same dataset as in Ref.\,\protect\cite{Ballesteros:2020sik}, just to ease the comparison between the BD-$\CC$CDM and the variable $G$ model studied in that reference (cf. their Table 1). This dataset includes the Planck 2018 TTTEEE+lensing likelihood \protect\cite{aghanim2020planck}, BAO data from \protect\cite{Beutler:2011hx,Ross:2014qpa,Alam:2016hwk} and the SH0ES prior from \protect\cite{Riess:2019cxk}. In the last fitting column, however, we add the LSS data to the previous set but with no bispectrum (cf. \hyperref[Table:BD.fs8]{Table\,\ref{Table:BD.fs8}} and \protect\hyperref[Sect:MethodData]{Sect.\,\ref{Sect:MethodData}}). The corresponding results for the  GR-$\CC$CDM can be found in \hyperref[Table:BD.TableFitBaseline]{Table\,\ref{Table:BD.TableFitBaseline}} and \hyperref[Table:BD.TableFitGR]{Table\,\ref{Table:BD.TableFitGR}}.}\label{Table:BD.TableFitCassini}
\end{table}


\begin{figure}[t!]
\begin{center}
\includegraphics[width=6.5in, height=6.5in]{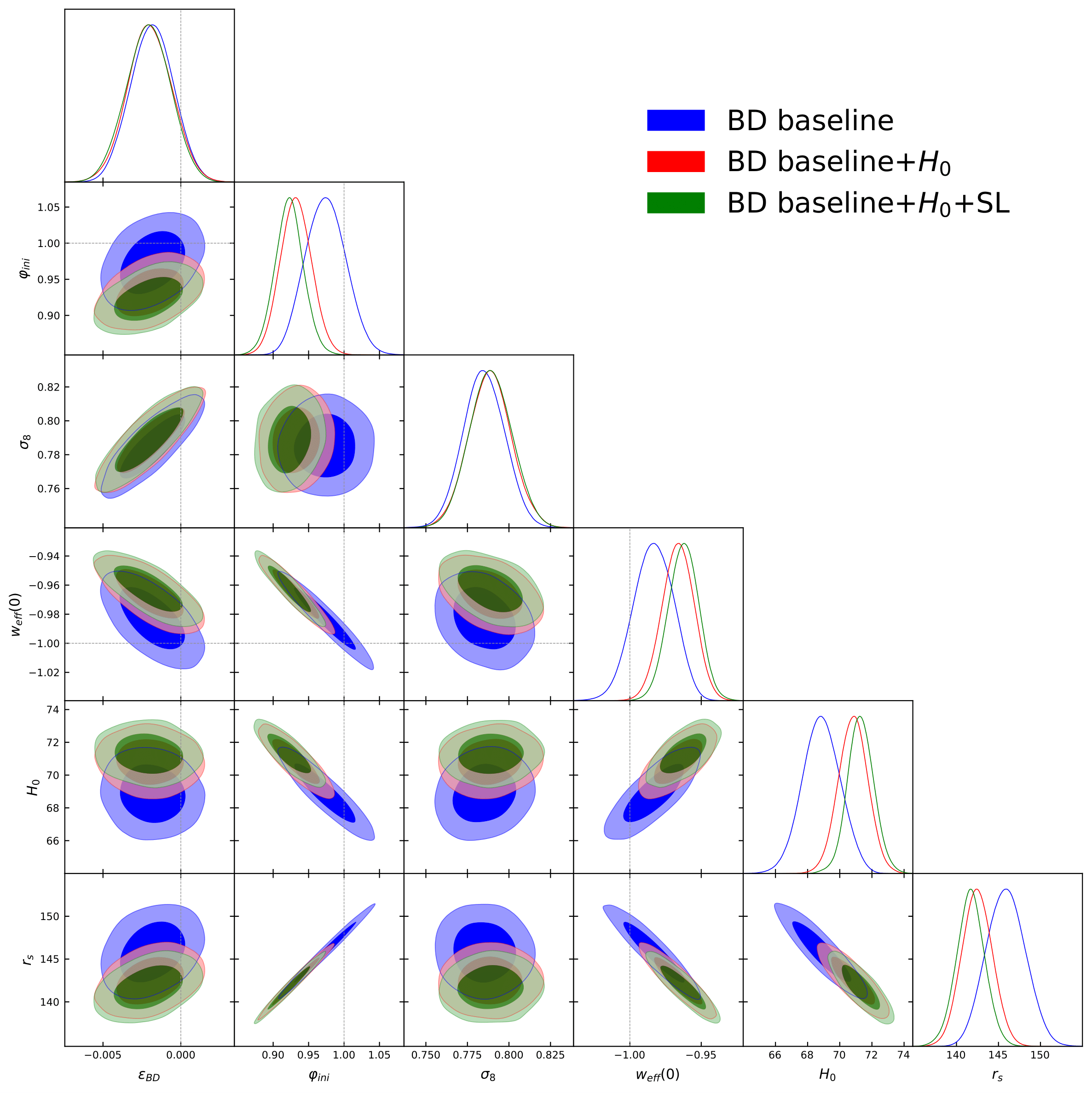}
\caption{Triangular matrix containing the two-dimensional marginalized distributions for some relevant combinations of parameters in the BD-$\CC$CDM model (at $1\sigma$ and $2\sigma$ c.l.), together with the corresponding one-dimensional marginalized likelihoods for each of them. $H_0$ is expressed in km/s/Mpc, and $r_{\rm s}$ in Mpc. See \hyperref[Table:BD.TableFitBaseline]{Table\,\ref{Table:BD.TableFitBaseline}} and \hyperref[Table:BD.TableFit+SL]{Table\,\ref{Table:BD.TableFit+SL}} for the numerical fitting results.}\label{Fig:BD.triangular}
\end{center}
\end{figure}

On the other hand, if we compare e.g. the fitting value predicted within the BD-$\CC$CDM model from \hyperref[Table:BD.TableFitBaseline]{Table\,\ref{Table:BD.TableFitBaseline}} (namely $H_0=70.83^{+0.92}_{-0.95}$ km/s/Mpc) with the aforementioned  SH0ES determination, we can see that the difference is of only $1.58\sigma$. If we next compare our fitting result from Table\,4  ($H_0= 71.30^{+0.80}_{-0.84}$ km/s/Mpc), which incorporates the H$0$LICOW Strong-Lensing data in the fit as well, with the combined SH0ES  and H$0$LICOW result (viz. the one which is in $5\sigma$ tension with the CMB value)  we obtain once more an inconspicuous tension of only  $1.55\sigma$.  In either case  it is far away from any perturbing discrepancy.  In fact, no discrepancy which is not reaching a significance of at least  $3\sigma$ can be considered sufficiently worrisome.

\subsection{Alleviating the $\sigma_8$-tension}\label{sec:s8-tension}

Furthermore, the smoothing of the tension applies to the $\sigma_8$ parameter as well, with the result that it essentially disappears within a similar level of inconspicuousness. In fact, values such as $\sigma_8=0.789\pm 0.013$ and $\tilde{S}_8=0.792\pm 0.013$ (obtained within the Baseline$+H_0$ dataset, see \hyperref[Table:BD.TableFitS8]{Table\,\ref{Table:BD.TableFitS8}}) are in good agreement with weak gravitational lensing observations derived from shear data (cf. the WL data block mentioned in Sec. \hyperref[Sect:MethodData]{Sect.\,\ref{Sect:MethodData}}). Let us take  the value by Joudaki et al. 2018 of the combined observable $S_8 = 0.742\pm 0.035$, for example,  obtained by KiDS-450, 2dFLenS and BOSS collaborations from a joint analysis of weak gravitational lensing tomography and overlapping redshift-space galaxy clustering\,\cite{Joudaki:2017zdt}.  These observations can be compared  with our prediction for ${S}_8=\sigma_8\sqrt{\Omega_{\rm m}/0.3}$  and  with the `renormalized' form of that quantity within the BD-$\CC$CDM model, namely $\tilde{S}_8=\sigma_8\sqrt{\tilde{\Omega}_{\rm m}/0.3}$,  which depends on the modified cosmological parameter $\tilde{\Omega}_{\rm m}$, which is slightly higher,  recall  our Eq.\,\eqref{Eq:BD.tildeOmegues}\footnote{Although we could use the hatted parameter $\hat{S}_8=\sigma_8\sqrt{\hat{\Omega}_{\rm m}/0.3}$, instead of  $\tilde{S}_8$, we have already pointed out  that the difference between $\hat{\Omega}_{\rm m}$ and $\tilde{\Omega}_{\rm m}$ is negligible for $|\epsilon_{\rm BD}|\sim \mathcal{O}(10^{-3})$, and so is between $\hat{S}_8$ and  $\tilde{S}_8$.}. Both ${S}_8$ and $\tilde{S}_8$ are displayed together in \hyperref[Table:BD.TableFitS8]{Table\,\ref{Table:BD.TableFitS8}} for the main scenarios, also in company with $\sigma_8$ values for the GR and BD models.  Differences of the mentioned experimental measurements with respect to e.g. our prediction for the Baseline$+H_0$ dataset,   are at the level of  $0.5\sigma-1.3\sigma$ depending on whether we use $S_8$ or $\tilde{S}_8$, whence statistically irrelevant in any case. More details can be appraised on some of  these observables and their correlation with $H_0$ in Figs. 10 and 11, on which we shall further comment later on.

\begin{figure}[t!]
\begin{center}
\includegraphics[width=4.3in, height=3.8in]{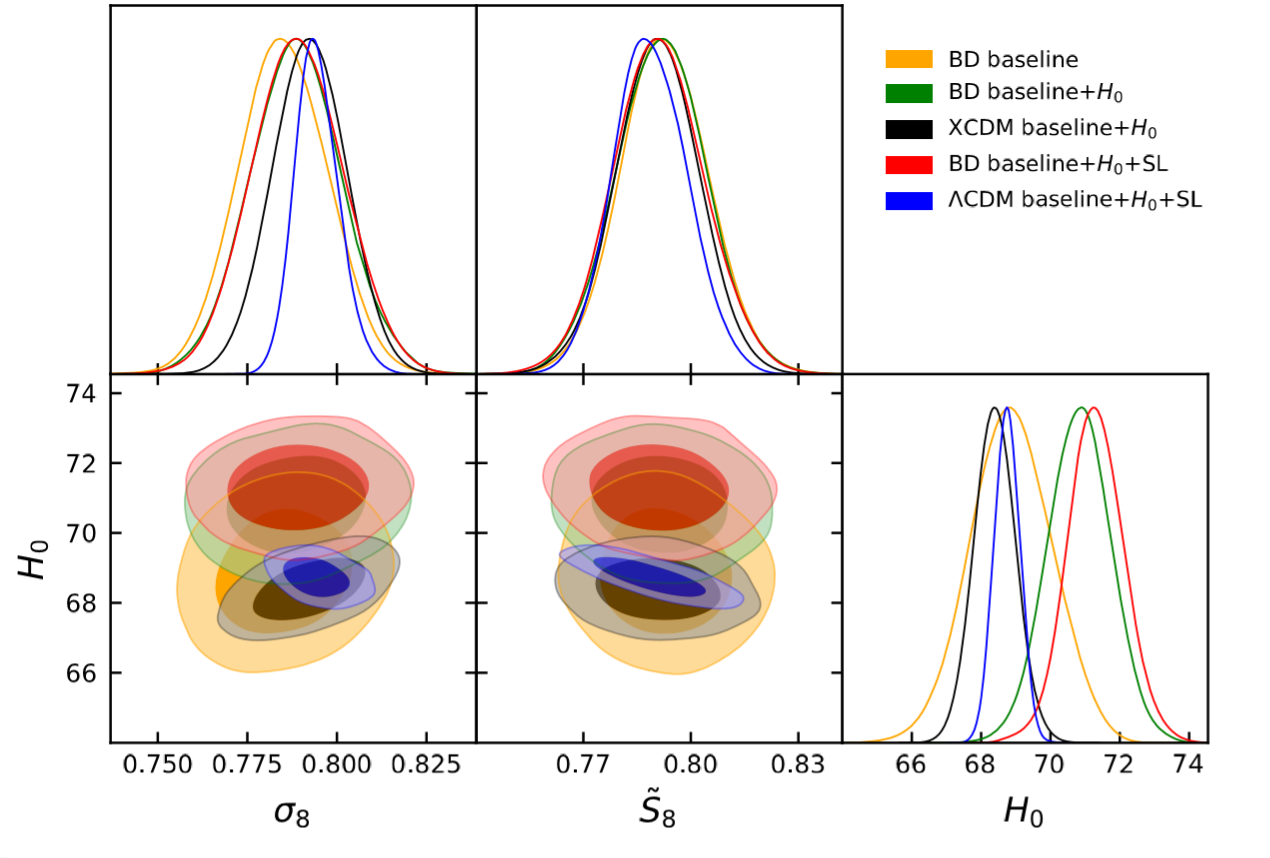}
\caption{Constraints obtained for $\sigma_8$ and $\tilde{S}_8$ versus $H_0$ (in km/s/Mpc) from the fitting analyses of the GR- and BD-$\CC$CDM models, and the GR-XCDM parametrization. We show both, the contour lines in the corresponding planes of parameter space, and the associated marginalized one-dimensional posteriors. The centering of the parameters in the ranges $\sigma_8<0.80$ and $H_0\gtrsim71$ km/s/Mpc is a clear sign of the smoothening of the $\sigma_8$-tension and, more conspicuously, of the $H_0$ one within the BD-$\CC$CDM model. We can also see that while a simple XCDM parametrization for the DE can help to diminish $\sigma_8$ as compared to the concordance model, it is however unable to improve the $H_0$ tension, which is kept at a similar level as within the concordance model.}\label{Fig:BD.H0S8}
\end{center}
\end{figure}

\subsection{Comparing different scenarios}\label{sec:Others}

We have also tested the performance of the BD- and GR-$\CC$CDM models when we include the CMB high-$\ell$ (TE+EE) polarization data from Planck 2018, with and without the CMB lensing, in combination with the baseline dataset and the SH0ES prior on $H_0$ (cf. \hyperref[Table:BD.TableFitAlternativeDataset]{Table\,\ref{Table:BD.TableFitAlternativeDataset}} and \hyperref[Table:BD.TableFitGR]{Table\,\ref{Table:BD.TableFitGR}}). The values of the Hubble parameter  in these cases are a little bit lower than when we consider only the temperature and low-$l$ polarization (TT+lowE) CMB data, but the tension is nevertheless significantly reduced, being now of only $\sim 2.2\sigma$ c.l., whereas in the GR-$\CC$CDM model it is kept at the $\sim 3.5\sigma$ level. The values of $\sigma_8$ are still low, $\sim 0.80-0.81$. The information criteria in these cases, though, have no preference for any of the two models, they are not conclusive.

We also examine the results that are obtained when we do not include in our fitting analyses any of the data sources that trigger the tensions. We consider here the CMB, BAO and SNIa datasets (denoted as CMB+BAO+SNIa in \hyperref[Table:BD.TableFitAlternativeDataset]{Table\,\ref{Table:BD.TableFitAlternativeDataset}} and \hyperref[Table:BD.TableFitGR]{Table\,\ref{Table:BD.TableFitGR}}), but exclude the use of the SH0ES prior on $H_0$ and the LSS data.  As expected,  the evidence for the BD model decreases since now we do not give to it the chance of showing its power.  Even though the description of the data improves,  it is not enough to compensate for the penalty received owing to the use of the two additional parameters $(\epsilon_{\rm BD},\varphi_{\rm ini})$,  and in this case we read  $2\ln B=-3.00$ (from \hyperref[Table:BD.TableFitAlternativeDataset]{Table\,\ref{Table:BD.TableFitAlternativeDataset}}). Thus, there is here a marginal preference for the GR scenario, but as previously mentioned, this is completely normal, since we are removing precisely the data sources whose correct description demands for new physics. Even so, the $H_0$-tension is again remarkably reduced from  $3.8\sigma$  in  GR-$\CC$CDM to only  $2.4\sigma$  in the BD-$\CC$CDM. The respective values of $\sigma_8$ remain compatible with $0.80$  within  $\sim 1\sigma$.

It is also interesting to compare the results that we have obtained within the BD framework  with other approaches in the literature,  in which the variation of $G$ is dealt with as a small departure from GR, namely in a context where the action still contains a large mass scale $M$ near the Planck mass  $m_{\rm Pl}$, together with some scalar field which parametrizes the deviations from it. This is of course fundamentally different from the BD paradigm but it bears relation owing to the variation of the effective $G$, and it has also been used to try to smooth out the tensions. However, as already mentioned (see \hyperref[Sect:Preview]{Sect.\,\ref{Sect:Preview}}), it is not easy at all for a given model to fulfill the `golden rule', {\it i.e.} to loosen the two tensions at a time, or just to alleviate one of them without worsening the other.  Different proposals have appeared in the market trying to curb the $H_0$-tension, e.g. the so-called early dark energy models\,\cite{Poulin:2018cxd,Chudaykin:2020acu,Braglia:2020bym}, and the model with variable $G$ recently considered in \cite{Ballesteros:2020sik,Braglia:2020iik,Ballardini:2020iws}. Although the physical mechanism of the EDE and the aforementioned models with variable $G$ is of course very different, their aim is pretty similar. They reduce the sound horizon $r_{\rm s}$ at recombination in order to force the increase of the Hubble function in the late Universe. This allows them to generate larger values of $H_0$ in order to keep a good fit to the CMB and BAO data, but this happens only at the expense of increasing the tension in $\sigma_8$, since they do not have any compensation mechanism able to keep the structure formation in the late Universe at low enough levels.  Some of these models  appear not to be particularly disfavored notwithstanding. But this is simply because they did not use LSS data in their fits, {\it i.e.} they did not put their models to the test of structure formation and for this reason they have more margin to adjust the remaining  observables without getting any statistical punishment.  So the fact that the significant increase of $\sigma_8$  that they find is not statistically penalized is precisely because they do not use  LSS data,  such as e.g. those on the observable $f(z)\sigma_8(z)$ displayed in our \hyperref[Table:BD.fs8]{Table\,\ref{Table:BD.fs8}}.  In this respect, EDE cosmologies are an example; they seem to be unable to alleviate the $H_0$-tension when LSS data are taken into account, as shown in \cite{Hill:2020osr}.

\subsection{Imposing the Cassini constraint}\label{sec:Cassini}

To further illustrate the capability of the BD-$\CC$CDM model to fit the data under more severe conditions, in  \hyperref[Table:BD.TableFitCassini]{Table\,\ref{Table:BD.TableFitCassini}} we consider four possible settings to fit our Cassini-constrained  BD-Scenario III defined in \,\hyperref[Sect:Mach]{Sect.\,\ref{Sect:Mach}}. Recall that this BD scenario involves the stringent Cassini bound on the post-Newtonian parameter $\gamma^{PN}$ \cite{Bertotti:2003rm}, which we have discussed in Sec.\,\ref{Sect:BDgravity}. The first two fitting columns of \hyperref[Table:BD.TableFitCassini]{Table\,\ref{Table:BD.TableFitCassini}} correspond to our usual Baseline$+H_0$ dataset,  in one case (first fitting column in that table) we omit the LSS data, whereas in the second column we restore it. In this way we can check the effect of the structure formation data on the goodness of the fit. The comparison between the results presented in these two columns shows, first and foremost, that the Cassini bound does not have a drastically nullifying effect, namely it does not render the BD-$\CC$CDM model irrelevant to the extent of  making it indistinguishable from GR-$\CC$CDM, not at all, since the quality of the fits is still fairly high (confer the Bayes factors in the last row). In truth, the fit quality is still comparable to that of the Baseline$+H_0$ scenario (cf. \hyperref[Table:BD.TableFitBaseline]{Table\,\ref{Table:BD.TableFitBaseline}}). However, the description of the LSS data is naturally poorer since $\eBD$ is smaller and the model cannot handle so well  the features of the structure formation epoch, thus yielding slightly higher values of $\sigma_8$. Second, the fact that the scenario without LSS furnishes a higher Bayes factor  just exemplifies the aforementioned circumstance that when cosmological models are tested without using this kind of data the results may in fact not be sufficiently reliable. When the LSS data enter the fit (third fitting colum in that table), we observe, interestingly enough, that the BD-$\CC$CDM model is still able to keep $H_0$  in the safe range, it does improve the value of $\sigma_8$ as well  ({\it i.e.} it becomes lower) and, on top of that, it carries a (`smoking gun')  signal of almost $2.9\sigma$ c.l. --  encoded in the value of $\weff$ --  pointing to quintessence-like behavior. Overall it is quite encouraging since it shows that the Cassini bound does not exceedingly hamper the BD-$\CC$CDM model capabilities. Such bound constraints the time evolution of $\varphi$ (because $|\eBD|$ is forced to be much smaller) but it does not preclude $\varphi$ from choosing a suitable value in compensation (cf. BD-Scenario III in \,\hyperref[Sect:Mach]{Sect.\,\ref{Sect:Mach}}).

In the last two columns of \hyperref[Table:BD.TableFitCassini]{Table\,\ref{Table:BD.TableFitCassini}}, we can further  check what are the changes in the previous fitting results when we use a  more restricted dataset,  e.g. the one used in Ref.\,\cite{Ballesteros:2020sik}, in which the Cassini bound is also implemented. These authors study a model which represents a modification of GR through an effective $G\sim 1/M^2$, with $M$ a mass very near  the Planck mass, $\mPl$, which is allowed to change slowly through a scalar field $\phi$ as follows: $M^2\to M^2+\beta\phi ^2$, where $\beta$ is a small (dimensionless) parameter. The authors assume that the Cassini bound on the post-Newtonian parameter $\gamma^{PN}$ \cite{Bertotti:2003rm} is in force (see \,\hyperref[Sect:MethodData]{Sect.\,\ref{Sect:MethodData}} for details). However, they do not consider LSS data (only CMB and BAO). We may compare the results they obtain within that variable $G$ model (cf. their Table 1) with those we obtain within the BD-Scenario III under the very same dataset as these authors. The results are displayed in the third fitting column of \hyperref[Table:BD.TableFitCassini]{Table\,\ref{Table:BD.TableFitCassini}}.  We obtain  $H_0=(70.01^{+0.86}_{-0.92})$ km/s/Mpc and $\sigma_8=0.817\pm 0.007$, whereas they obtain $H_0=(69.2^{+0.62}_{-0.75})$ km/s/Mpc and $\sigma_8=0.843^{+0.015}_{-0.024}$. Clearly, the BD-$\CC$CDM is able to produce larger central values of $H_0$ and lower values of $\sigma_8$, even under the Cassini bound, although the differences are  within errors. The value of $2\ln\,B$ lies around $+3.5$ and hence points to a mild positive evidence in favor of the BD model. This is consistent with the associated deviation we find  of $G(0)$ from $G_N$ at $2.43\sigma$ c.l., and with a signal of effective quintessence at $2.53\sigma$ c.l. within the GR-picture.

Let us  now consider what is obtained if we add up the LSS data to this same BD-Scenario III, still with the restricted dataset od Ref.\,\cite{Ballesteros:2020sik}. As expected, the inclusion of the structure formation data
pushes the value of $\sigma_8=0.812^{+0.006}_{-0.007}$ down as compared to their absence ($\sigma_8=0.817\pm 0.007$). This is the most remarkable difference between the two cases, as one cannot appreciate significant changes in the other parameters. Something very similar happens when we compare the values of $\sigma_8$ of the first two fitting columns of \hyperref[Table:BD.TableFitCassini]{Table\,\ref{Table:BD.TableFitCassini}}, in which we consider the B+$H_0$ dataset without LSS data (first fitting column) and with LSS data (second fitting column). The relative improvement {\it w.r.t.} the model of \cite{Ballesteros:2020sik} is therefore greater in the presence of LSS data, whose use has been omitted in that reference. In that variable $G$ model one finds a larger value of $H(z)$ at recombination thanks to the larger values of $G$ in that epoch, but at present $G$ is forced to be almost equal to $G_N$ (being the differences not relevant for cosmology). This means that: (i) in that model $G$ decreases with time, which leads to a kind of effective $\epsilon_{\rm BD}>0$; (ii) the model cannot increase $H_0$ with a large value of $G(z=0)$. Both facts do limit significantly  the effectiveness of the model in loosening the tensions. In the BD-$\CC$CDM model under consideration, instead, we find that $G$ has to be $\sim 8-9\%$ larger than $G_N$ not only in the pre-recombination Universe, but also at present, and this allows to reduce significally the $H_0$-tension. Moreover, we find that a mild increase of the cosmological $G$ with the expansion leads also to an alleviation of the $\sigma_8$-tension.

Under all of the datasets studied in \hyperref[Table:BD.TableFitCassini]{Table\,\ref{Table:BD.TableFitCassini}} we obtain central values of $|\epsilon_{\rm BD}|\sim \mathcal{O}(10^{-5})$, which are compatible with 0 at $1\sigma$. Notice that this value is just of order of the Cassini bound on $\eBD$, as could be expected. Notwithstanding, and remarkably enough, the stringent bound imposed by the Cassini constraint, which enforces $\epsilon_{\rm BD}$ to remain two orders of magnitude lower than in the main Baseline  scenarios, is nevertheless insufficient to wipe out the  positive effects from the BD-$\CC$CDM model. They are still capable to emerge with a sizeable part of the genuine BD signal. This is, as anticipated in \hyperref[Sect:Mach]{Sect.\,\eqref{Sect:Mach}}, mainly due to the fact that the Cassini bound cannot restrict the value of the BD field  $\varphi$, only its time evolution.

\subsection{More on alleviating simultaneously the two tensions}\label{Sect:TwoTensions}

A few additional comments on our results concerning the parameters $H_0$  and $\sigma_8$ are now in order. Their overall impact can be better assessed by examining  the triangular matrix of fitted contours involving all the main parameters, as shown in \hyperref[Fig:BD.triangular]{Fig.\ref{Fig:BD.triangular}},
in which we offer the numerical results of several superimposed analyses based on different datasets, all of them within Baseline scenarios. We project the contour lines in the corresponding planes of parameter space, and  show the associated marginalized one-dimensional posteriors.   The fitted value of the EoS parameter at $z=0$  shown there, $\weff(0)$, is to be understood, of course, as a derived parameter from the prime ones of the fit, but we include this information along with the other parameters in order to further display the significance of the obtained signal:  $\gtrsim 3\sigma$  quintessence-like behavior.  Such signal, therefore, mimics  `GR+ DDE' and  hints at something beyond pure GR-$\CC$CDM. What we find in our study is that such time-evolving DE behavior is actually of quasi-vacuum type and appears as a kind of  signature of the underlying BD theory\footnote{As we recall in \hyperref[Sect:RVMconnection]{subsection\,\ref{Sect:RVMconnection}} of \hyperref[Appendix:Semi-Analytical]{Appendix\,\ref{Appendix:Semi-Analytical}}, such  kind of  dynamical behavior of the vacuum is characteristic of the Running Vacuum Model (RVM),  a version of the   $\CC$CDM in which the vacuum energy density is not just a constant  but involves also a dynamical term $\sim H^2$.  The description of the BD-$\CC$CDM model in the GR-picture mimics a  behavior of this sort \,\cite{SolaPeracaula:2018dsw,deCruzPerez:2018cjx}.}.

Figure\,10 provides  a truly panoramic and graphical view of our main fitting results,  and  from where we can comfortably  judge the impact of the BD framework for describing the overall cosmological data. It is fair to say that it appears at a level highly competitive with GR --  in fact, superior to it.  In all datasets involving the local $H_0$ input in the fit analyses, the improvement is substantial and manifest. Let us stand out only three of the entries in that graphical matrix:  i) for the parameters $(\sigma_8,H_0)$,  all the contours in the main dataset scenarios are centered around values of $\sigma_8<0.80$ and $H_0\gtrsim71$ km/s/Mpc, which are the coveted ranges for every model aiming at smoothing the two tensions at a time;  ii) for the pair $(H_0,r_{\rm s})$, the contours are centered around the same range of relevant  $H_0$ values as before, and also around values of the comoving sound horizon (at the baryon drag epoch) $r_{\rm s}\lesssim142$ Mpc, these being significantly smaller  than those of the concordance $\CC$CDM  (cf. \hyperref[Table:BD.TableFitGR]{Table\,\ref{Table:BD.TableFitGR}}) and hence consistent with larger values of the expansion rate at that epoch;  iii) and for  $(\weff(0), H_0)$,  the relevant  range $H_0\gtrsim71$ km/s/Mpc is once more picked out, together with an effective quintessence signal $\weff(0)>-1$ at more than $3\sigma$ (specifically $3.45\sigma$ for the scenario of \hyperref[Table:BD.TableFit+SL]{Table\,\ref{Table:BD.TableFit+SL}}, in which strong lensing data are included in the fit).

It is also interesting to focus once more our attention on \hyperref[Fig:BD.H0S8]{Fig.\ref{Fig:BD.H0S8}}, where we provide devoted contours involving  both  $\tilde{S}_8$ and $\sigma_8$  versus $H_0$.  On top of the observations we have previously  made on these observables, we  can compare here our basic dataset scenarios for the BD-$\CC$CDM model with the yield of a simple  XCDM parametrization of the DDE. In previous studies we had already shown that such parametrization can help to deal with the $\sigma_8$ tension\,\cite{SolaPeracaula:2018wwm}. Nonetheless, as we can see here, it proves completely impotent for solving  or minimally helping to alleviate the $H_0$-tension since the values predicted for this parameter stay as low as in the concordance model. This shows, once more,  that in order to address a possible solution to the two tensions simultaneously, it is not enough to have just some form of dynamics in the DE sector; one really needs a truly specific one,  e.g. the one provided (in an effective way) by the BD-$\CC$CDM model.

\subsection{Predicted relative variation of the effective gravitational strength}\label{Sect:VariationG}

In the context of the BD framework it is imperative, in fact mandatory,  to discuss the current values of the relative variation of the effective gravitational strength, viz. of $\dot{G}(0)/G(0)$, which follow from our fitting analyses (see the main \hyperref[Table:BD.TableFitBaseline]{Table\,\ref{Table:BD.TableFitBaseline}}, \hyperref[Table:BD.TableFit+SL]{Table\,\ref{Table:BD.TableFit+SL}} and \hyperref[Table:BD.TableFitSpectrum]{Table\,\ref{Table:BD.TableFitSpectrum}}). The possible time evolution of that quantity hinges directly on $\eBD$, of course, since the latter is the parameter that controls the (cosmological) evolution of the gravitational coupling in the BD theory. It is easy to see from Eq.\,\eqref{Eq:BD.definitions} that $\dot{G}(0)/G(0)=-\dot{\varphi}(0)/\varphi(0)\simeq -\eBD H_0$, where we use the fact that $\varphi\sim a^{\eBD}$ in the matter-dominated epoch (cf. \hyperref[Appendix:FixedPoints]{Appendix\,\ref{Appendix:FixedPoints}}). Recalling that $H_0\simeq 7\times 10^{-11}$ yr$^{-1}$ (for $h\simeq 0.70$), we find  $\dot{G}(0)/G(0)\simeq -\eBD\cdot 10^{-10}\,yr^{-1}$. Under our main BD-Scenario II (cf. \,\hyperref[Sect:Mach]{Sect.\,\ref{Sect:Mach}}) we obtain values for $\dot{G}(0)/G(0)$ of order $\mathcal{O}(10^{-13})\,yr^{-1}$ (and positive), just because $\epsilon_{\rm BD}\sim \mathcal{O}(10^{-3})$ (and negative). Being  $\dot{G}(0)/G(0)>0$ it means that the effective gravitational coupling obtained by our global cosmological fit increases with the expansion, and hence it was smaller in the past. This suggests that the sign $\eBD<0$, which is directly picked out by the data, prefers a kind of asymptotically free behavior for the gravitational coupling since the epochs in the past are more energetic,  in fact characterized by larger values of $H$ (with natural dimension of energy).  The central values show a mild time variation at present, at a level of
 $1.3\sigma$,  both for $\eBD$ and $\dot{G}(0)/G(0)$, when the bispectrum data from BOSS is also included in the analysis (cf. \hyperref[Table:BD.TableFitBaseline]{Table\,\ref{Table:BD.TableFitBaseline}} and \hyperref[Table:BD.TableFit+SL]{Table\,\ref{Table:BD.TableFit+SL}}). Such departure goes below $1\sigma$ level when only the spectrum is considered (see \hyperref[Table:BD.TableFitSpectrum]{Table\,\ref{Table:BD.TableFitSpectrum}}). In the context of the BD-Scenario III, in which $\eBD$ is very tightly constrained by the Cassini bound \cite{Bertotti:2003rm}, namely at a level of $\mathcal{O}(10^{-5})$, we find $\dot{G}(0)/G(0)\sim 10^{-15}\,yr^{-1}$, which is compatible with 0 at $1\sigma$. All that said, we should emphasize once more  that the fitting values that we obtain for $\dot{G}(0)/G(0)$ refer to the cosmological time variation of $G$ and, therefore, cannot be directly compared with constraints existing in the literature based on strict local gravity measurements, such as e.g. those from the lunar laser ranging experiment -- $\dot{G}(0)/G(0)=(2\pm 7)\cdot 10^{-13}yr^{-1}$ \cite{Muller:2007zzb} -- (see e.g. the review \cite{Uzan:2010pm} for a detailed presentation of many other local constraints on $\dot{G}(0)/G(0)$). Even though this bound turns out to be preserved within our analysis, it is not in force at the cosmological level provided  an screening mechanism acting at these scales is assumed, as in our case. Thus, the local measurements have no bearing a priori on the BD-$\CC$CDM cosmology. The opposite may not be true, for despite  the fact that the values reported in our tables are model-dependent, they prove to be quite efficient and show that the cosmological observations can compete in precision with the local measurements.

\begin{figure}[t!]
\begin{center}
\includegraphics[width=4.in, height=3in]{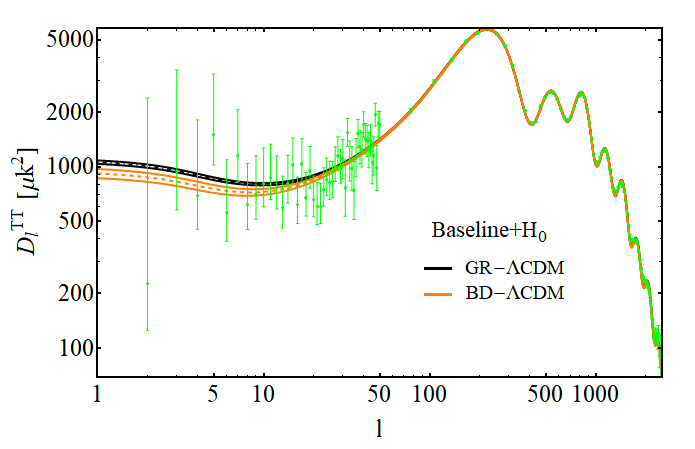}
\caption{CMB temperature power spectrum for the GR-$\CC$CDM (in black) and BD-$\CC$CDM (in orange), obtained from the fitting results within the Baseline+$H_0$ dataset (cf. \hyperref[Table:BD.TableFitBaseline]{Table\,\ref{Table:BD.TableFitBaseline}} and Sec. \protect\ref{Sect:MethodData}). We plot the central curves together with the corresponding $1\sigma$ bands. In the inner plot we zoom in the multipole range $l\in [0,30]$ and include the Planck 2018 \protect\cite{aghanim2020planck} data error bars (in green). At low multipoles ($l\lesssim 30$) the BD-$\CC$CDM model produces less power than the concordance GR-$\CC$CDM model owing to the suppression of the Integrated Sachs-Wolfe effect that we have discussed in \protect\hyperref[Sect:Preview]{Sect.\,\ref{Sect:Preview}}. The differences are $\gtrsim 1\sigma$, and allow to soften a well-known low-multipole CMB anomaly. See the main text for further discussion.}\label{Fig:BD.Cls-ISW}
\end{center}
\end{figure}

 \begin{table}[h!]
\begin{center}
\begin{tabular}{|c  |c | c |  c | c | c  |}
 \multicolumn{1}{c}{} & \multicolumn{2}{c}{Baseline} & \multicolumn{2}{c}{Baseline+$H_0$}
\\\hline
{\scriptsize Parameter} & {\scriptsize GR-$\Lambda$CDM}  & {\scriptsize BD-$\Lambda$CDM} & {\scriptsize GR-$\Lambda$CDM}  &  {\scriptsize BD-$\Lambda$CDM}
\\\hline
{\scriptsize $H_0$ (km/s/Mpc)}  & {\scriptsize $67.75^{+0.46}_{-0.48}$} & {\scriptsize $68.86^{+1.26}_{-1.22}$} & {\scriptsize $68.35^{+0.49}_{-0.46}$}  & {\scriptsize $70.81^{+0.95}_{-0.92}$}
\\\hline
{\scriptsize$\omega_{\rm b}$} & {\scriptsize $0.02231^{+0.00018}_{-0.00020}$}  & {\scriptsize $0.02214\pm 0.00031$} & {\scriptsize $0.02239\pm 0.00019$}  &  {\scriptsize $0.02239^{+0.00029}_{-0.00032}$}
\\\hline
{\scriptsize$\omega_{cdm}$} & {\scriptsize $0.11655^{+0.00110}_{-0.00111}$}  & {\scriptsize $0.11860^{+0.00194}_{-0.00197}$} & {\scriptsize $0.11625^{+0.00099}_{-0.00106}$}  &  {\scriptsize $0.11821^{+0.00214}_{-0.00200}$}
\\\hline
{\scriptsize$\tau$} & {{\scriptsize$0.053^{+0.006}_{-0.008}$}} & {{\scriptsize$0.054^{+0.006}_{-0.008}$}} & {{\scriptsize$0.053^{+0.006}_{-0.008}$}}  &   {{\scriptsize$0.054^{+0.006}_{-0.009}$}}
\\\hline
{\scriptsize$n_{\rm s}$} & {{\scriptsize$0.9706^{+0.0041}_{-0.0040}$}}  & {{\scriptsize$0.9691^{+0.0104}_{-0.0093}$}} & {{\scriptsize$0.9717^{+0.0041}_{-0.0042}$}} &   {{\scriptsize$0.9791^{+0.0097}_{-0.0090}$}}
\\\hline
{\scriptsize$\sigma_8$}  & {{\scriptsize$0.770^{+0.016}_{-0.018}$}}  & {{\scriptsize$0.759^{+0.0184}_{-0.0163}$}} & {{\scriptsize$0.780^{+0.018}_{-0.015}$}}  &   {{\scriptsize$0.762^{+0.021}_{-0.018}$}}
\\\hline
{\scriptsize$r_{\rm s}$ (Mpc)}  & {{\scriptsize$148.03\pm0.33$}}  & {{\scriptsize$146.16^{+2.40}_{-2.58}$}} & {{\scriptsize$148.04^{+0.32}_{-0.34}$}}  &   {{\scriptsize$142.81^{+1.91}_{-2.03}$}}
\\\hline
{\scriptsize$m_{\nu}$ (eV)}  & {{\scriptsize$0.161^{+0.058}_{-0.059}$}}  & {{\scriptsize$0.409^{+0.139}_{-0.198}$}} & {{\scriptsize$0.118^{+0.053}_{-0.058}$}}  &   {{\scriptsize$0.409^{+0.183}_{-0.228}$}}
\\\hline
{\scriptsize $\epsilon_{\rm BD}$} & - & {{\scriptsize $0.00459^{+0.00316}_{-0.00319}$}} & - &   {{\scriptsize $0.00433^{+0.00395}_{-0.00336}$}}
\\\hline
{\scriptsize$\varphi_{\rm ini}$} & - & {{\scriptsize $0.979^{+0.028}_{-0.032}$}} & - &    {{\scriptsize $0.938^{+0.023}_{-0.024}$}}
\\\hline
{\scriptsize$\varphi(0)$} & - & {{\scriptsize $1.016^{+0.041}_{-0.049}$}} & - &    {{\scriptsize $0.972^{+0.042}_{-0.043}$}}
\\\hline
{\scriptsize$\omega_{eff}(0)$} & - & {{\scriptsize $-1.005^{+0.021}_{-0.017}$}} & - &    {{\scriptsize $-0.986^{+0.017}_{-0.016}$}}
\\\hline
{\tiny $\dot{G}(0)/G(0) (10^{-13}yr^{-1})$} & - & {{\scriptsize $-5.048^{+3.485}_{-3.434}$}} & - &    {{\scriptsize $-4.915^{+3.749}_{-4.501}$}}
\\\hline
{\scriptsize$\chi^2_{\rm min}$} & {\scriptsize 2270.54}  & {\scriptsize 2268.44} & {\scriptsize 2285.02}  &  {\scriptsize 2274.76}
\\\hline
{\scriptsize$2\ln B$} & {\scriptsize -}  & {\scriptsize -0.12} & {\scriptsize -}  &  {\scriptsize +9.34}
\\\hline
{\scriptsize$\Delta {\rm DIC}$} & {\scriptsize -}  & {\scriptsize -0.81} & {\scriptsize -}  &  {\scriptsize +8.22}
\\\hline
\end{tabular}
\caption{As in \hyperref[Table:BD.TableFitBaseline]{Table\,\ref{Table:BD.TableFitBaseline}}, but here we allow the variation of the mass of the massive neutrino ($m_\nu$) in the Monte Carlo routine, instead of setting it to $0.06$ eV. The other two neutrinos remain massless. We have used the same conservative prior range for $m_\nu\in [0,1]$ eV in both, the GR- and BD-$\Lambda$CDM models.}\label{Table:BD.TableFitNeutrino}
\end{center}
\end{table}

\subsection{One more bonus: suppressing the power at low multipoles }\label{Sect:LowMultipoles}

An additional bonus from the  BD cosmology is worth mentioning before we close this lengthy study. It is found in the description of the CMB temperature anisotropies. As we have discussed in \hyperref[Sect:Preview]{Sect.\,\ref{Sect:Preview}}, the BD-$\CC$CDM model is, in principle, able to suppress the power at low multipoles ($l\lesssim 30$), thereby softening one of the so-called CMB anomalies that are encountered in the context of the GR-$\CC$CDM model. This is basically due to the low values of $\varphi<1$ preferred by the data, which in turn produce a suppression  of the Integrated Sachs-Wolfe (ISW) effect\,\cite{Sachs:1967er,Das:2013sca}. We have confirmed that this suppression actually occurs for the best-fit values of the parameters in our analysis, cf. \hyperref[Fig:BD.Cls-ISW]{Fig.\ref{Fig:BD.Cls-ISW}}. The aforementioned anomaly is not very severe, since the  power at low multipoles is affected by a large cosmic variance and cannot be measured very precisely. Nevertheless, it is a subtle anomaly which has been there unaccounted for a long time and could not be improved in a consistent way: that is to say, usually models ameliorating the low tail of the spectrum do spoil the high part of it. However, here  the suppression of power with respect to the GR-$\CC$CDM at $l\lesssim 30$ is fully consistent and is another very welcome feature of the BD-$\CC$CDM model, which is not easy at all to attain.  It is interesting to mention that the ISW effect can be probed by cross correlating the CMB temperature maps with the LSS data, e.g. with foreground galaxies number counts, especially if using  future surveys which should have much smaller uncertainties.  This can be a useful probe for DE and a possible distinctive signature for DDE theories\,\cite{Pogosian:2004wa,Zucca:2019ohv}\,\footnote{We thank L. Pogosian for interesting comments along these lines.}.  The upshot of our investigation is that once more  the `golden rule' mentioned in the Introduction is preserved here and the curing effects from the BD-$\CC$CDM stay aligned: the  three tensions of the  GR-$\CC$CDM ($H_0$, $\sigma_8$ and  the exceeding CMB power at low multipoles) can be improved at a time.

\subsection{The effect of massive neutrinos }\label{Sect:MassiveNeutrino}

Finally, it is worth assessing the impact of leaving the sum of the three neutrino masses ($M_\nu\equiv\sum m_\nu$) as a free parameter, rather than fixing it. We model this scenario in two different ways:

\begin{itemize}

\item Considering one massive ($m_\nu$) and two massless neutrinos, so $M_\nu=m_\nu$. Here we try to mimic the physics encountered when the neutrino masses follow the normal hierarchy, in which one of the neutrinos is much heavier than the other two, for reasonable values of $M_\nu$.
    
\item Considering three massive neutrinos of equal mass $m_\nu$, such that $M_\nu=3m_\nu$. This is the degenerated case, utilized also in the analysis by Planck 2018 \cite{aghanim2020planck} (cf. Sec. 7.5.1 therein).
    
\end{itemize}

In both cases, the BD-$\CC$CDM has one additional parameter {\it w.r.t.} the same model with fixed mass $M_\nu=m_\nu=0.06$ eV that we have discussed in the previous sections, and three more than the vanilla $\Lambda$CDM model: $(\eBD,\varphi_{\rm ini}, M_\nu)$. Upon taking into account the constraints obtained from neutrino oscillation experiments on the mass-squared splittings through the corresponding likelihoods we would perhaps find more precise bounds on the sum of the neutrino masses\,\cite{Loureiro:2018pdz}, but here we want to carry out a more qualitative analysis to study how massive neutrinos may impact on our results, considering only cosmological data. The results of the first mass scenario are shown in \hyperref[Table:BD.TableFitNeutrino]{Table\,\ref{Table:BD.TableFitNeutrino}}. Those for the second scenario are not tabulated, but are commented below.

As can be seen from \hyperref[Table:BD.TableFitNeutrino]{Table\,\ref{Table:BD.TableFitNeutrino}}, the effect of having a neutrino with an adjustable mass picked out by the fitting process is non-negligible. It produces a significant lowering of the value of $\sigma_8$ while preserving the value of $H_0$ at a level comparable to the previous tables in which the neutrino had a fixed mass of $0.06$ eV, therefore preserving what we have called the `golden rule'.  Let us note, however, that the sign of $\eBD$ has changed now with respect to the situation with a fixed light mass, it is no longer negative but positive and implies a value of the BD parameter of $\wBD\simeq230$.  Last but not least, the fitted value of the neutrino mass is $m_\nu=0.409$ eV, which is significantly higher than the upper bound placed by Planck 2018 under the combination TTTEEE+lowE+lensing+BAO: $M_\nu\equiv\sum m_\nu<0.120$ eV ($95\%$ c.l.)\,\cite{aghanim2020planck}.

Similar results are obtained with the second mass scenario mentioned above.  Let us summarize them. For the Baseline+$H_0$ dataset, the common fitted mass value obtained for the three neutrinos is $m_\nu=0.120^{+0.054}_{-0.068}$ eV, which means that $M_\nu\simeq 0.360$ eV, slightly lower than in the first scenario. The corresponding BD parameter reads comparable, $\wBD\simeq 261$, and the values for $\sigma_8$ and $H_0$ are also very close to the previous case, so the two scenarios share similar advantages. The fact that the Planck 2018 upper limit for $M_\nu$ is overshooted in both neutrino mass scenarios does not necessarily exclude them, as the limits on  $M_\nu$ are model-dependent, see e.g. \cite{Loureiro:2018pdz} and \cite{Ballardini:2020iws}. In particular, Planck 2018 obviously used  GR-$\CC$CDM. We can check in \hyperref[Table:BD.TableFitNeutrino]{Table\,\ref{Table:BD.TableFitNeutrino}} that the neutrino mass values for BD-$\CC$CDM are substantially different. The results that we have obtain for GR-$\CC$CDM are fully compatible with those from Planck 2018.

We conclude that the influence of neutrino masses on the BD-$\CC$CDM fitting results is potentially significant, but it cannot be settled at this point. The subject obviously deserves further consideration in the future.

\section{Discussion of the chapter}\label{Sect:DiscussionBD}

To summarize, we have presented a rather comprehensive work on the current status of the Brans-Dicke theory with a cosmological constant in the light of the modern observations. Such framework constitutes a new version of the concordance $\CC$CDM model in the context of a distinct gravity paradigm, in which the gravitational constant is no longer a fundamental constant of Nature but a dynamical field. We have called this framework BD-$\CC$CDM model to distinguish it from the conventional one, the GR-$\CC$CDM model, based on General Relativity.  This chapter is based in our paper \cite{SolaPeracaula:2020vpg} which, at the same time, is an expanded and fully updated analysis of our previous and much shorter presentation\,\cite{SolaPeracaula:2019zsl}, in which we have  replaced the Planck 2015 data by the Planck 2018 data, and we have included additional sets of modern cosmological observations. We reconfirm the results of \,\cite{SolaPeracaula:2019zsl} and provide now a bunch of new results which fit in with the conclusions of our previous work and reinforce its theses.  To wit:  in the light of the figures and tables that we have presented in the current study we may assert that the BD-$\CC$CDM model fares better not only as compared to the GR-$\CC$CDM  with a rigid cosmological constant (cf. Tables \hyperref[Table:BD.TableFitBaseline]{\ref{Table:BD.TableFitBaseline}}, \hyperref[Table:BD.TableFit+SL]{\ref{Table:BD.TableFit+SL}}, \hyperref[Table:BD.TableFitSpectrum]{\ref{Table:BD.TableFitSpectrum}}, \hyperref[Table:BD.TableFitAlternativeDataset]{\ref{Table:BD.TableFitAlternativeDataset}}, \hyperref[Table:BD.TableFitGR]{\ref{Table:BD.TableFitGR}} and \hyperref[Table:BD.TableFitBaseline]{\ref{Table:BD.TableFitCassini}}) but also when the CC term is replaced with a dynamical parametrization of the DE, such as the traditional XCDM, which acts as a benchmark (cf. \hyperref[Table:BD.TableFitXCDM]{Table\,\ref{Table:BD.TableFitXCDM}}).  We find that the GR-XCDM is completely unable to enhance the value of $H_0$ beyond that of the concordance model.  In particular, in \hyperref[Table:BD.TableFitBaseline]{Table\,\ref{Table:BD.TableFitBaseline}} and \hyperref[Table:BD.TableFit+SL]{Table\,\ref{Table:BD.TableFit+SL}}  we can see that the information criteria (Bayes factor and Deviance Information Criterion) do favor  significantly and consistently the BD-$\CC$CDM model as compared to GR-$\CC$CDM and  GR-XCDM. There is a very good resonance between the Bayesian evidence criterion and the DIC differences, which definitely uphold the BD framework at a level of $+5$ units for the Baseline+$H_0$ dataset scenario, meaning that the degree of support of BD versus GR is in between \textit{positive} to \textit{strong} (cf. Sec.\,\ref{Sect:NumericalAnalysis}). This support is further enhanced up to more than $+9$ units, hence  in between \textit{strong} to \textit{very strong}, for the case  when we include the Strong-Lensing data in the fit (see \hyperref[Table:BD.TableFit+SL]{Table\,\ref{Table:BD.TableFit+SL}}). The exact Bayesian evidence curves computed in \hyperref[Fig:BD.BayesRatio]{Fig.\,\ref{Fig:BD.BayesRatio}} reconfirm these results in a graphical way. The pure baseline dataset scenario (in which the local $H_0$ value is not included) shows weak evidence; however, as soon as the local $H_0$ value is fitted along with the remaining parameters the evidence increases rapidly and steadily,  reaching the status of positive, strong and almost very strong depending on the datasets.

Another dataset scenario which is particularly favored in our analysis is the one based on considering the effective calibration prior on the absolute magnitude $M$ of the nearer SNIa data in the distance ladder (as defined in \hyperref[Sect:MethodData]{Sect.\,\ref{Sect:MethodData}}), instead of the local value $H_0$ from SH0ES. The results for the Baseline dataset in combination with $M$ (denoted B+$M$) can be read off from \hyperref[Table:BD.TableFitAlternativeDataset]{Table\,\ref{Table:BD.TableFitAlternativeDataset}} (first row).  We can see it yields a tantalizing overall output, with values of $H_0$ and  $\sigma_8$ in the correct ranges for solving the two tensions, and fully compatible with the results obtained using the prior on $H_0$, as expected. In addition, the corresponding Bayes factor for this scenario points to a remarkably high value $2\ln B>+10$, thereby carrying a very strong Bayesian evidence,  in fact comparable to the Baseline+$H_0$+SL scenario of \hyperref[Table:BD.TableFit+SL]{Table\,\ref{Table:BD.TableFit+SL}}.  In all of the mentioned cases in our summary the information criteria  definitely endorse the BD-cosmology versus the GR one.

Finally, we have also assessed the influence of the neutrino masses in the context of the BD-$\CC$CDM model, see \hyperref[Table:BD.TableFitNeutrino]{Table\,\ref{Table:BD.TableFitNeutrino}}. We have found that massive neutrinos can help to further reduce the predicted value of $\sigma_8$ to the level of what is precisely needed to describe the weak gravitational lensing observations derived from direct shear data. This would completely dissolve the $\sigma_8$-tension without detriment of the positive results obtained to loosen the $H_0$-tension, {\it i.e.} by preserving the `golden rule'.  Such a conclusion is, however, provisional as it requires a devoted study of the neutrino sector (extending the analysis of \hyperref[Sect:MassiveNeutrino]{Sect.\,\ref{Sect:MassiveNeutrino}}).

Overall, the statistical support in favor of the BD-$\CC$CDM model against the concordance  GR-$\CC$CDM model is rather significant. It is not only that the $H_0$ and $\sigma_8$ tensions are simultaneously dwarfed to a level where they are both rendered inessential ($\lesssim 1.5\sigma$), but also that all tested BD scenarios involving the local $H_0$ value provide a much better global fit than the concordance model on the basis  of a rich and updated set of modern observations from all the main cosmological data sources available at present. If we take into account that the BD-$\CC$CDM framework is not just some \textit{ad hoc} phenomenological toy-model, or some last-minute smart parametrization just concocted to solve or mitigate the two tensions, but the next-to-leading fundamental theory candidate directly competing with GR, it may give us a sense of the potential significance of these results.

\chapter{Running vacuum against the \texorpdfstring{$H_0$}{Ho} and \texorpdfstring{$\sigma_8$}{s8} tensions}\label{Chap:PhenomenologyofRVM}

Despite Einstein's original formulation\,\cite{Einstein:1917ce}, in which the cosmological term $\CC$ is treated as a strict constant in the gravitational field equations, the idea that $\CC$ (and its associated vacuum energy density $\rv$) can be a dynamical quantity should be most natural in the context of an expanding Universe. This point of view has led to the notion of dynamical dark energy (DDE) in its multifarious forms\,\cite{padmanabhan2003cosmological,peebles2003cosmological,copeland2006dynamics,Amendola:2015ksp}.  Herein, however, we stick to the notion of dynamical vacuum energy (DVE) as the ultimate cause of DDE. Despite the fact that $\rv$ has long been associated with the so-called cosmological constant problem\,\cite{weinberg1989cosmological,Sola:2013gha,Sola:2014tta,Sola:2016zeg}, which involves severe fine-tuning of the parameters, such a conundrum actually underlies all of the DE models known up to date, with no exception\,\cite{Sola:2013gha,Sola:2014tta,Sola:2016zeg}. Our calculations of $\rv$ in the context of quantum field theory (QFT) in curved spacetime have brought new light into this problem (see \hyperref[Chap:QuantumVacuum]{chapter\,\ref{Chap:QuantumVacuum}} and \hyperref[Chap:EoSVacuum]{chapter\,\ref{Chap:EoSVacuum}}) and suggest that if  the vacuum energy density (VED) is renormalized using an appropriate regularization procedure, it evolves in a mild way as a series of powers of the Hubble rate $H$ and its cosmic time derivatives: $\rv(H,\dot{H},...)$, denoted $\rv(H)$ for short. This fact was long foreseen from general renormalization group arguments which led to the notion of running vacuum models (RVM's), see the reviews \cite{Sola:2013gha,Sola:2014tta,Sola:2016zeg,Sola:2015rra} and references therein.
\renewcommand{\arraystretch}{1.2}
\begin{table*}[t!]
\begin{center}
\resizebox{1\textwidth}{!}{
\begin{tabular}{|c  |c | c |  c | c | c  |c  |}
 \multicolumn{1}{c}{} & \multicolumn{4}{c}{Baseline}
\\\hline
 Parameter &  GR-$\Lambda$CDM  &  type I RRVM &  type I RRVM$_{\rm thr.}$  &   type II RRVM &   BD-$\Lambda$CDM
\\\hline
 $H_0$(km/s/Mpc)  &  $68.37^{+0.38}_{-0.41}$ &  $68.17^{+0.50}_{-0.48}$ &  $67.63^{+0.42}_{-0.43}$  &  $69.02^{+1.16}_{-1.21}$ &  $69.30^{+1.38}_{-1.33}$
\\\hline
$\omega_{\rm b}$ &  $0.02230^{+0.00019}_{-0.00018}$  &  $0.02239^{+0.00023}_{-0.00024}$ &  $0.02231^{+0.00020}_{-0.00019}$  &   $0.02245^{+0.00025}_{-0.00027}$ &  $0.02248\pm 0.00025$
\\\hline
$\omega_{\rm dm}$ &  $0.11725^{+0.00094}_{-0.00084}$  &  $0.11731^{+0.00092}_{-0.00087}$ &  $0.12461^{+0.00201}_{-0.00210}$  &   $0.11653^{+0.00158}_{-0.00160}$ &    $0.11629^{+0.00148}_{-0.00151}$
\\\hline
$\nu_{\rm eff}$ & {-}  &  $0.00024^{+0.00039}_{-0.00040}$ &  $0.02369^{+0.00625}_{-0.00563}$  &   $0.00029\pm 0.00047$ & {-}
\\\hline
$\epsilon_{\rm BD}$ & {-}  & {-} & {-}  &  {-} &  $-0.00109\pm ^{+0.00135}_{-0.00141}$
\\\hline
$\varphi_{\rm ini}$ & {-}  & {-} & {-}  &   $0.980^{+0.031}_{-0.027}$ &  $0.972^{+0.030}_{-0.037}$
\\\hline
$\varphi_0$ & {-}  & {-} & {-}  &   $0.973^{+0.036}_{-0.033}$ &  $0.963^{+0.036}_{-0.041}$
\\\hline
$\tau_{\rm reio}$ & {$0.049^{+0.008}_{-0.007}$} & {$0.051^{+0.008}_{-0.009}$} & {$0.058^{+0.007}_{-0.009}$}  &   {$0.051\pm 0.008$} & {$0.051\pm 0.008$}
\\\hline
$n_{\rm s}$ & {$0.9698^{+0.0039}_{-0.0036}$}  & {$0.9716^{+0.0044}_{-0.0047}$} & {$0.9703\pm 0.038$} &   {$0.9762^{+0.0081}_{-0.0091}$} &
\\\hline
$\sigma_8$  & {$0.796\pm 0.007$}  & {$0.789^{+0.013}_{-0.014}$} & {$0.768^{+0.010}_{-0.009}$}  &   {$0.791^{+0.013}_{-0.012}$} & {$0.790^{+0.013}_{-0.012}$}
\\\hline
$S_8$  & {$0.796\pm 0.011$}  & {$0.791^{+0.014}_{-0.013}$} & {$0.797^{+0.012}_{-0.011}$}  &   {$0.781^{+0.021}_{-0.020}$} & {$0.777^{+0.021}_{-0.022}$}
\\\hline
$r_{\rm s}$ (Mpc)  & {$147.90^{+0.30}_{-0.31}$}  & {$147.99^{+0.35}_{-0.36}$} & {$147.81\pm 0.30$}  &   {$146.30^{+2.39}_{-2.30}$} & {$145.72^{+2.44}_{-2.90}$}
\\\hline
$\chi^2_{\rm min}$ & { 2290.20} & { 2289.72} & { 2272.44}  &   { 2288.74} & { 2289.40}
\\\hline
$\Delta{\rm DIC}$  & {-}  & { -2.70} & { +13.82}  &   { -4.59} & { -3.53}
\\\hline
\end{tabular}}
\end{center}
\caption{The mean values and 68.3\% confidence limits for the models under study using our Baseline dataset, which is almost the same as the one employed in \protect\cite{SolaPeracaula:2020vpg,SolaPeracaula:2019zsl} (see previous chapter and \hyperref[Appendix:Description]{Appendix\,\ref{Appendix:Description}}), with few changes: (i) for the eBOSS survey we have replaced the data from \protect\cite{Gil-Marin:2018cgo} with the one from \protect\cite{Neveux:2020voa}; (ii) the LyF data have been updated, replacing \protect\cite{deSainteAgathe:2019voe} with \protect\cite{duMasdesBourboux:2020pck}; (iii) finally, we have replaced the two $f\sigma_8$ data points \protect\cite{Qin:2019axr,Huang:2021quh} with the one provided in \protect\cite{Said:2020epb}. We display the fitting values for the usual parameters, to wit: $H_0$, the reduced density parameter for baryons ($w_{\rm b} = \Omega^0_{\rm b}{h^2}$) and CDM ($w_{\rm dm} = \Omega^0_{\rm dm}{h^2}$), with $\Omega_i^0=8\pi G_N\rho^0_i/3H_0^2$ and $h$ the reduced Hubble constant, the reionization optical depth $\tau_{\rm reio}$, the spectral index $n_{\rm s}$ and the current matter density rms fluctuations within spheres of radius $8h^{-1}$~Mpc, {\it i.e.}  $\sigma_8$. We include also a couple of useful derived parameters, namely: the sound horizon at the baryon drag epoch $r_{\rm s}$ and $S_8\equiv \sigma_8\sqrt{\Omega^0_{\rm m}/0.3}$. For all the RRVM's we show $\nu_{\rm eff}$, and for the type II {and BD-$\Lambda$CDM}\protect\cite{SolaPeracaula:2020vpg,SolaPeracaula:2019zsl} we also report the initial and current values of $\varphi$, $\varphi_{\rm ini}$ and $\varphi_0$, respectively. {The parameter $\epsilon_{\rm BD}\equiv1/\omega_{\rm BD}$ (inverse of the Brans-Dicke parameter\,\protect\cite{brans1961mach}) controls the dynamics of the scalar field\protect\cite{SolaPeracaula:2020vpg,SolaPeracaula:2019zsl}.}
We provide the corresponding values of $\chi^2_{\rm min}$ and $\Delta$DIC.}\label{Table:PhenomenologyRVM.Fit1}
\end{table*}
%
For the current Universe, the leading VED term is constant but the next-to-leading one is dynamical, specifically it evolves as a power  $\sim H^2$ with a small coefficient $|\nu|\ll1$. For the early Universe, terms of order $\sim H^4$ or higher appear and these can trigger inflation\,\cite{Lima:2013dmf,Sola:2015rra,Sola:2015csa,SolaPeracaula:2019kfm}. It is remarkable that the fourth power $H^4$ can be motivated within  the context of string theory calculations at low energy (meaning near the Planck scale)\,\cite{Basilakos:2019acj,Basilakos:2020qmu}, what reveals a distinctive mechanism of inflation different from that of Starobinsky inflation\,\cite{Starobinsky:1980te}, for example. See\,\cite{Moreno-Pulido:2020anb,Mavromatos:2020kzj} for a detailed discussion. Here, however, we will concentrate on the post-inflationary Universe, where only the leading power $\sim H^2$ is involved in the dynamics of  $\rv$. A variety of phenomenological analyses have supported this possibility in recent years\,\cite{SolaPeracaula:2016qlq,SolaPeracaula:2017esw,Gomez-Valent:2018nib,Gomez-Valent:2017idt,Sola:2017znb,Sola:2016jky,Sola:2015wwa,Sola:2016hnq,Gomez-Valent:2014rxa,Basilakos:2014tha,Basilakos:2009wi,SolaPeracaula:2020vpg,SolaPeracaula:2019zsl,Perico:2016kbu,Geng:2017apd}.

In this chapter, based on our paper\,\cite{SolaPeracaula:2021gxi}, we present a devoted study of the class of RVM's based on a large and updated string SNIa+BAO+$H(z)$+LSS+CMB of modern cosmological observations, in which for the first time the CMB part involves the full Planck 2018 likelihood. We also test the potential dependence of the results on the threshold redshift $\zstar$  at which the DVE becomes  activated in the recent past. We find that different RVM's prove very helpful to alleviate the persisting tensions between the concordance $\CC$CDM model and the structure formation data (the so-called $\sigma_8$ tension) and the mismatch between the local values of the Hubble parameter and those derived from the  CMB\,\cite{aghanim2020planck} (the $H_0$ tension). These tensions are well described in the literature, see e.g. the reviews\,\cite{verde2019tensions,DiValentino:2020zio,DiValentino:2020vvd}. Many models in the market try to address them, see e.g. ref.\,\cite{DiValentino:2019jae} and the long list of references therein.

In the current (fully updated) study we find significant signals of DVE (using  $\zstar\simeq 1$) at $\sim 3.6\sigma$ c.l., which can be enhanced up to  $\sim4.0\sigma$.  Finally, we show that the  RVM's provide an overall fit to the cosmological data which is comparable or significantly better than in the $\CC$CDM case, as confirmed by calculating the relative  Deviance Information Criterion (DIC) differences obtained form the Monte Carlo chains of our numerical analysis.

\section{Running vacuum Universe}\label{Sect:RunningVacuumUniverse}

As indicated, the total vacuum part of the energy-momentum tensor,  $T_{\mu \nu}^{\rm vac}$,  can be appropriately  renormalized into a finite quantity which depends on the Hubble rate $H$ and its time derivatives\, \cite{Moreno-Pulido:2020anb}.  The corresponding  $00$-component  defines the vacuum energy density (VED),  $\rv(H)$.  Let us denote by  $\rvo\equiv\rv(H_0)=\CC/(8\pi G_N)$ ($G_N$ being Newton's constant)  the current value of the latter, with $H_0$  today's value of the Hubble parameter and $\CC$ the measured  cosmological constant term.  We define two types of DVE scenarios.  In type I scenario the vacuum is in interaction with matter, whereas in type II  matter is conserved at the expense of an exchange between the vacuum and a slowly evolving gravitational coupling $G (H)$.   The combined cosmological `running' of these quantities  insures the accomplishment of the Bianchi identity (and the  local conservation law).
\renewcommand{\arraystretch}{1.2}
\begin{table*}[t!]
\begin{center}
\resizebox{1\textwidth}{!}{
\begin{tabular}{|c  |c | c |  c | c | c  |c  |}
 \multicolumn{1}{c}{} & \multicolumn{4}{c}{Baseline + $H_0$}
\\\hline
 Parameter &  GR-$\Lambda$CDM  &  type I RRVM &  type I RRVM$_{\rm thr.}$  &   type II RRVM &   BD-$\Lambda$CDM
\\\hline
 $H_0$ (km/s/Mpc)  &  $68.75^{+0.41}_{-0.36}$ &  $68.77^{+0.49}_{-0.48}$ &  $68.14^{+0.43}_{-0.41}$  &  $70.93^{+0.93}_{-0.87}$  &  $71.23^{+1.01}_{-1.02}$
\\\hline
$\omega_{\rm b}$ &  $0.02240^{+0.00019}_{-0.00021}$  &  $0.02238^{+0.00021}_{-0.00023}$ &  $0.02243^{+0.00019}_{-0.00018}$  &   $0.02269^{+0.00025}_{-0.00024} $  &  $0.02267^{+0.00026}_{-0.00023} $
\\\hline
$\omega_{\rm dm}$ &  $0.11658^{+0.00080}_{-0.00083}$  &  $0.11661^{+0.00084}_{-0.00085}$ &  $0.12299^{+0.00197}_{-0.00203}$  &   $0.11602^{+0.00162}_{-0.00163}$  &  $0.11601^{+0.00161}_{-0.00157}$
\\\hline
$\nu_{\rm eff}$ & {-}  &  $-0.00005^{+0.00040}_{-0.00038}$ &  $0.02089^{+0.00553}_{-0.00593}$ &   $0.00038^{+0.00041}_{-0.00044}$  & {-}
\\\hline
$\epsilon_{\rm BD}$ & {-}  & {-} & {-}  &  {-} &  $-0.00130\pm ^{+0.00136}_{-0.00140}$
\\\hline
$\varphi_{\rm ini}$ & {-}  & {-} & {-}  &   $0.938^{+0.018}_{-0.024}$  &  $0.928^{+0.024}_{-0.026}$
\\\hline
$\varphi_0$ & {-}  & {-} & {-}  &   $0.930^{+0.022}_{-0.029}$  &  $0.919^{+0.028}_{-0.033}$
\\\hline
$\tau_{\rm reio}$ & {$0.050^{+0.008}_{-0.007}$} & {$0.049^{+0.009}_{-0.008}$} & {$0.058^{+0.008}_{-0.009}$}  &   {$0.052\pm 0.008$}  & {$0.052\pm 0.008$}
\\\hline
$n_{\rm s}$ & {$0.9718^{+0.0035}_{-0.0038}$}  & {$0.9714\pm 0.0046$} & {$0.9723^{+0.0040}_{-0.0039}$} &   {$0.9868^{+0.0072}_{-0.0074}$}  & {$0.9859^{+0.0073}_{-0.0072}$}
\\\hline
$\sigma_8$  & {$0.794\pm 0.007$}  & {$0.795\pm 0.013$} & {$0.770\pm 0.010$}  &   {$0.794^{+0.013}_{-0.012}$}  & {$0.792^{+0.013}_{-0.012}$}
\\\hline
$S_8$  & {$0.788^{+0.010}_{-0.011}$}  & {$0.789\pm 0.013$} & {$0.789\pm 0.011$} &   {$0.761^{+0.018}_{-0.017}$}  & {$0.758^{+0.019}_{-0.018}$}
\\\hline
$r_{\rm s}$ (Mpc)  & {$147.97^{+0.29}_{-0.31}$}  & {$147.94^{+0.35}_{-0.36}$} & {$147.88^{+0.33}_{-0.29}$}  &   {$143.00^{+1.54}_{-1.96}$}  & {$142.24^{+1.99}_{-2.12}$}
\\\hline
$\chi^2_{\rm min}$  & { 2302.14}  & { 2301.90} & { 2288.82}  &   { 2296.38}  & { 2295.36}
\\\hline
$\Delta{\rm DIC}$ & {-}  & { -2.36} & { +10.88}  &   { +5.52}  & { +6.25}
\\\hline
\end{tabular}}
\end{center}
\caption{Same as in \hyperref[Table:PhenomenologyRVM.Fit1]{Table\,\ref{Table:PhenomenologyRVM.Fit1}}, but also considering the prior on $H_0=(73.5\pm 1.4)$ km/s/Mpc from SH0ES \protect\cite{Riess:2019cxk,Reid:2019tiq}.}\label{Table:PhenomenologyRVM.Fit2}
\end{table*}

Let us therefore consider a generic cosmological framework described by the spatially flat Friedmann-Lema\^\i tre-Robertson-Walker  (FLRW) metric. The vacuum energy density in the RVM can be written in the form \,\cite{Sola:2013gha,Sola:2014tta,Sola:2016zeg,Sola:2015rra}:
\begin{equation}\label{Eq:PhenomenologyRVM.RVMvacuumdadensity}
\rv(H) = \frac{3}{8\pi{G}_N}\left(c_{0} + \nu{H^2+\tilde{\nu}\dot{H}}\right)+{\cal O}(H^4)\,,
\end{equation}
in which the ${\cal O}(H^4)$ terms will be neglected for the physics of  the post-inflationary epoch.
The above generic structure can be motivated from the aforementioned explicit QFT calculations on a FLRW background\,\cite{Moreno-Pulido:2020anb}, see also \hyperref[SubSect:GeneralizedRVM]{Sect.\,\ref{SubSect:GeneralizedRVM}} of this dissertation.
The additive constant $c_0$ is fixed by the boundary condition $\rho_{\rm vac}(H_0)=\rvo$.  Notice that the two dynamical components $H^2$ and $\dot{H}$ are dimensionally homogeneous and, in principle, independent.
Their associated  (dimensionless) coefficients $\nu$ and $\tilde{\nu}$ encode the dynamics of the vacuum at low energy  and we naturally expect $|\nu,\tilde{\nu}|\ll1$. An estimate of $\nu$ in QFT indicates that it is of order $10^{-3}$ at most \cite{Sola:2007sv}. In the calculation of  \cite{Moreno-Pulido:2020anb} these coefficients are expected to be of order $\sim M_X^2/\mpl^2\ll 1$, where  $\mpl\simeq 1.22\times 10^{19}$ GeV is the Planck mass and $M_X$  is of order of a typical Grand Unified Theory (GUT) scale, times a multiplicity factor accounting for the number of heavy particles in the GUT.
We will be particularly interested in the RVM density obtained from the choice $\tilde{\nu}=\nu/2$.  As a result, $\rv(H) ={3}/(8\pi G_N)\left[c_0 + \nu\left({H^2+\frac12\dot{H}}\right)\right]$. We will call this form of the VED the  `RRVM' since it realizes the generic RVM density \eqref{Eq:PhenomenologyRVM.RVMvacuumdadensity} through the Ricci scalar $R = 12H^2 + 6\dot{H}$, namely
\begin{equation}\label{Eq:PhenomenologyRVM.RRVM}
\rv(H) =\frac{3}{8\pi{G_N}}\left(c_0 + \frac{\nu}{12}R\right)\equiv \rv(R)\,.
\end{equation}
Such a RRVM implementation has the advantage that it gives a safe path to the early epochs of the cosmological evolution since in the radiation dominated era we have $R/H^2\ll 1$, and hence we do not generate any conflict with the BBN nor with any other feature of the modern Universe.  Of course, early on the RVM has its own mechanism for inflation (as we have already explained in previous chapters), but we shall not address these aspects here, see \cite{Sola:2013gha,Sola:2014tta,Sola:2016zeg,Sola:2015rra,Lima:2013dmf, Sola:2015csa,SolaPeracaula:2019kfm,Mavromatos:2020kzj}.

\subsection{Type I RRVM}\label{SubSect:TypeIRRVM}

Friedmann's equation and the acceleration equation relate $H^2$ and $\dot{H}$ with the energy densities and pressures for the different species involved, and read
\begin{align}
3H^2 &= 8\pi{G_N}\left(\rho_{\rm m} + \rho_{\rm{ncdm}} + \rho_\gamma + \rv(H)\right),\label{Eq:PhenomenologyRVM.FriedmannEquation} \\
3H^2 + 2\dot{H} &= -8\pi{G_N}\left(p_{\rm{ncdm}} + p_\gamma + p_{\rm vac}(H)\right)\label{Eq:PhenomenologyRVM.AccelerationEquation}\,.
\end{align}
The total nonrelativistic matter density is the sum of the cold dark matter (CDM) component and the baryonic one: $\rho_{\rm m} = \rho_{\rm dm} + \rho_{\rm b}$.  The contributions of massive and massless neutrinos are included in $\rho_{\rm ncdm}$  (`${\rm ncdm}$' means non-CDM). Therefore the total (relativistic and nonrelativistic)  matter density is $\rho_t=\rho_{\rm m}+ \rho_\gamma+\rho_{\rm ncdm}$. Similarly, the total matter pressure reads  $p_t=p_{\rm{ncdm}} + p_\gamma$  (with  $p_\gamma=(1/3)\rho_\gamma$).
We note that  there is a transfer of energy from the relativistic neutrinos to the nonrelativistic ones along the whole cosmic history, and hence it is not possible (in an accurate analysis) to make a clear-cut separation between the two.  Our procedure adapts to our own modified version of the system solver \texttt{CLASS}\,\cite{Blas:2011rf}.  The latter solves the coupled system of Einstein's and Boltzmann's differential equations  for any value of the scale factor and, in particular, provides the functions
$\rho_{\rm h} = \rho_{\rm ncdm} - 3p_{\rm ncdm}$ and $\rho_\nu = 3p_{\rm ncdm}$ for the nonrelativistic and relativistic neutrinos, respectively.
This allows to compute the combination $R/12=H^2 + (1/2)\dot{H}$  appearing in \eqref{Eq:PhenomenologyRVM.RRVM}  in terms of the energy densities and pressures using \eqref{Eq:PhenomenologyRVM.FriedmannEquation} and \eqref{Eq:PhenomenologyRVM.AccelerationEquation}:
\begin{equation}\label{Eq:PhenomenologyRVM.combination}
R = 8\pi{G_N}\left(\rho_{\rm m} + 4\rv + \rho_{\rm h}\right)\,.
\end{equation}
Notice that the photon contribution cancels exactly in this expression and hence $\rv$ from \eqref{Eq:PhenomenologyRVM.RRVM} remains much smaller than the photon density in the radiation epoch, entailing no alteration of the thermal history.  While neutrinos do not behave as pure radiation for the aforementioned reasons,
one can check numerically (using \texttt{CLASS}) that the ratio $r \equiv{\rho_{\rm h}}/{\rho_{\rm m}}$ is very small throughout the entire cosmic history up to our time (remaining always below $10^{-3}$). Thus, we can neglect it in \eqref{Eq:PhenomenologyRVM.RRVM} and we can solve for the vacuum density as a function of the scale factor $a$ as follows:
\begin{equation}\label{Eq:PhenomenologyRVM.VacuumDens}
\rv(a) = \rvo + \frac{\nu}{4(1-\nu)}(\rho_{\rm m}(a) - \rho^0_{\rm m})\,,
\end{equation}
where `$0$' (used as subscript or superscript) always refers to current quantities.  For $a=1$ (today's Universe)  we confirm the correct normalization:  $\rv(a=1) = \rvo$. Needless to say,  $\rho_{\rm m}(a)$ is not just $\sim a^{-3}$ since the vacuum is exchanging energy with matter here.  This is obvious from the fact that the CDM exchanges energy with the vacuum (making it dynamical):
\begin{equation}\label{Eq:PhenomenologyRVM.LocalConsLaw}
\dot{\rho}_{\rm dm} + 3H\rho_{\rm dm} = -\dot{\rho}_{\rm vac}\,.
\end{equation}
Baryons do not interact with the vacuum, which implies $\dot{\rho}_{\rm b} + 3H\rho_{\rm b} =0$, and as a result the total matter contribution ($\rho_{\rm m}$)  satisfies the same local conservation law \eqref{Eq:PhenomenologyRVM.LocalConsLaw} as CDM:
$\dot{\rho}_{\rm m} + 3H\rho_{\rm m} = -\dot{\rho}_{\rm vac}$.  Using it with \eqref{Eq:PhenomenologyRVM.VacuumDens} we find
$\dot{\rho}_{\rm m} + 3H\xi\rho_{\rm m} = 0$, where we have defined  $\xi \equiv \frac{1 -\nu}{1 - \frac{3}{4}\nu}$.
Since $\nu$ is small, it is convenient to encode the deviations with respect to the standard model in terms of the effective parameter $\nueff\equiv\nu/4$:
\begin{equation}\label{Eq:PhenomenologyRVM.xi}
\xi = 1-\nu_{\rm eff} + \mathcal{O}\left(\nu_{\rm eff}^2\right)\,.
\end{equation}
It is straightforward to find the expression for the matter densities:
\begin{equation}\label{Eq:PhenomenologyRVM.MassDensities}
\rho_{\rm m}(a) = \rho^0_{\rm m}{a^{-3\xi}}\,, \ \ \ \rho_{\rm dm}(a) = \rho^{0}_{\rm m}{a^{-3\xi}}  - \rho^0_{\rm b}{a^{-3}} \,.
\end{equation}
They recover the $\CC$CDM form for $\xi=1$ ($\nueff=0$).  The small departure is precisely what gives allowance for a mild dynamical vacuum evolution:
\begin{align}  \label{Eq:PhenomenologyRVM.Vacdensity}
\rv(a) &= \rvo + \left(\frac{1}{\xi} -1\right)\rho^0_{\rm m}\left(a^{-3\xi} -1\right)\,.
\end{align}
The vacuum  becomes rigid only for $\xi=1$ ($\nueff=0$).

\subsection{Type II RRVM}\label{SubSect:TypeIIRRVM}

For type II models matter is conserved (no exchange with vacuum), but the vacuum can still evolve provided the gravitational coupling also evolves (very mildly) with the expansion: $G=G(H)$. Following the notation of \cite{SolaPeracaula:2020vpg,SolaPeracaula:2019zsl}, let us define (just for convenience) an auxiliary variable  $\varphi=G_N/G $ -- in the manner of  a Brans-Dicke {(BD)} field\,{\cite{brans1961mach}} (see the previous chapter), without being really so. Notice that $\varphi\neq 1$ in the cosmological domain, but remains very close to it, see \hyperref[Table:PhenomenologyRVM.Fit1]{Table\,\ref{Table:PhenomenologyRVM.Fit1}} and \hyperref[Table:PhenomenologyRVM.Fit2]{Table\,\ref{Table:PhenomenologyRVM.Fit2}}. 
{For convenience, in the last column of \hyperref[Table:PhenomenologyRVM.Fit1]{Table\,\ref{Table:PhenomenologyRVM.Fit1}} and \hyperref[Table:PhenomenologyRVM.Fit2]{Table\,\ref{Table:PhenomenologyRVM.Fit2}} (and \hyperref[Fig:PhenomenologyRVM.XCDMEvolution]{Fig.\,\ref{Fig:PhenomenologyRVM.XCDMEvolution}}) we include the updated results of \cite{SolaPeracaula:2020vpg,SolaPeracaula:2019zsl} (BD model with a cosmological constant) with the data changes indicated in the caption of \hyperref[Table:PhenomenologyRVM.Fit1]{Table\,\ref{Table:PhenomenologyRVM.Fit1}}.}

Friedman's equation for type-II model takes the form
\begin{equation}\label{Eq:PhenomenologyRVM.fried}
3H^2=\frac{8\pi G_N}{\varphi}\left[\rho_t+C_0+\frac{3\nu}{16\pi G_N}(2H^2+\dot{H})\right]\,,
\end{equation}
with $C_0=3c_0/(8\pi G_N)$.
The Bianchi identity dictates the correlation between the dynamics of $\varphi$ and that of $\rv$ \footnote{{For type-II models the running of $G$ is triggered by that of $\rv$ via the Bianchi identity \eqref{Eq:PhenomenologyRVM.Bianchi}. If matter is self-conserved,
such running is unavoidable from the existence of the quantum effects  $\sim H^2$ (and/or $\dot{H}$) inducing the running of $\rv$, see \cite{Moreno-Pulido:2020anb}.
This does not exclude other microscopic mechanisms, but for type-II the running of $G$  is necessary to comply with general covariance. In the BD case of the previous chapter\,\cite{SolaPeracaula:2020vpg,SolaPeracaula:2019zsl}, instead,
 $\varphi$ is an explicit field ingredient of the classical action.}}:
\begin{equation}\label{Eq:PhenomenologyRVM.Bianchi}
\frac{\dot{\varphi}}{\varphi}=\frac{\dot{\rho}_{\rm vac}}{\rho_t+\rv}\,,
\end{equation}
where $\rho_t $ is as before the total matter energy density and $\rv$ adopts exactly the same form as in \eqref{Eq:PhenomenologyRVM.RRVM}. Using these equations one can show that the approximate behavior of the VED in the present time is (recall that $|\nueff|\ll1$):
\begin{equation}\label{Eq:PhenomenologyRVM.VDEm}
\rv(a)=C_0(1+4\nueff)+\nueff\rho_{\rm m}^{0}a^{-3}+\mathcal{O}(\nueff^2)\,.
\end{equation}
Again, for $\nueff=0$ the VED is constant, but otherwise it shows a moderate dynamics of ${\cal O}(\nueff)$ as in the type I case \eqref{Eq:PhenomenologyRVM.Vacdensity}.
Here, however, the exact solution must be found numerically.  One can also show that the behavior of $\rv(a)$ in the radiation dominated epoch is also of the form \eqref{Eq:PhenomenologyRVM.VDEm}, except that the constant additive term can be completely neglected.  It follows that  $\rv(a)\ll\rho_{\rm r}(a)=\rho_{\rm r}^0 a^{-4}$  for $a\ll 1$ and hence the VED for the type II model does not perturb the normal thermal history (as in the type I model). Finally, one finds $\varphi(a)\propto a^{-\epsilon}\approx 1-\epsilon\ln\,a$ in the current epoch (with $0<\epsilon\ll 1$ of order $\nueff$), thus confirming the very mild (logarithmic) evolution of $G$.
%
\begin{figure*}
\centering
\includegraphics[angle=0,width=0.9\linewidth]{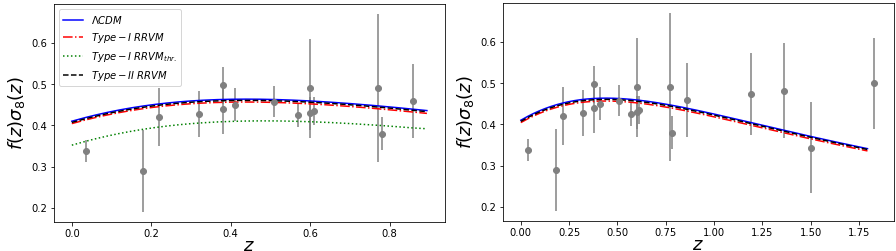}
\caption{
Theoretical curves of $f(z)\sigma_8 (z)$ for the various models and the data points employed in our analysis, in two different redshift windows. To generate this plot we have used the central values of the cosmological parameters shown in \hyperref[Table:PhenomenologyRVM.Fit1]{Table\,\ref{Table:PhenomenologyRVM.Fit1}}. The type I running vacuum model with threshold redshift $\zstar\simeq 1$  has a most visible and favorable impact on solving the $\sigma_8$ tension.
}\label{Fig:PhenomenologyRVM.fs8Evolution}
\end{figure*}
%

\section{Threshold redshift scenario for type I models}\label{Sect:ThresholdRedshift}

One possibility that has been explored in the literature in different type of models is to admit that the dynamics of vacuum is relatively recent (see e.g. \cite{Salvatelli:2014zta,Martinelli:2019dau}).  This means to study the consequences of keeping deactivated the interaction between the vacuum energy density and the CDM for most of cosmic history until the late Universe when the DE becomes apparent. We denote the threshold value of the scale factor when the activation takes places by $\astar$. According to this scenario the VED was constant prior to $a=\astar$ and it just started to evolve for $a>\astar$.  While $\rv$ is a continuous function, its derivative is not since we mimic such situation through a Heaviside step function  $\Theta(a-\astar)$.  If we would have a microscopic description of the phenomenon it should not be necessary to assume such a sudden (finite) discontinuity. However, a  $\Theta$-function description will be enough for our purposes.  Therefore, we assume that in the range   $a<\astar$  (hence for $z> \zstar$) we have
\begin{align}\label{Eq:PhenomenologyRVM.Vacdensitystar}
&\rho_{\rm dm}(a) = \rho_{\rm dm}(\astar)\left(\frac{a}{\astar}\right)^{-3},  \ \ \ \ \ \ \ \ \ \  \nonumber\\
&\rv (a) = \rv(\astar)=\text{const.} \ \ \ \ \ \ \ \ \ \ \ \ \ \   (a<\astar)\,,
\end{align}
where $\rho_{\rm dm}(\astar)$ and  $\rv(\astar)$ are computed from \eqref{Eq:PhenomenologyRVM.MassDensities} and \eqref{Eq:PhenomenologyRVM.Vacdensity}, respectively.
In the complementary range, instead,  {\it i.e.} for  $a>\astar$   ($0<z<\zstar$) near our time,  the original equations \eqref{Eq:PhenomenologyRVM.MassDensities} and \eqref{Eq:PhenomenologyRVM.Vacdensity} hold good.

Notice that the above threshold procedure is motivated specially within type I models in order to preserve the canonical evolution law for the matter energy density when the redshift is sufficiently high. In fact, the threshold redshift value need not be very large and as we shall see in the next section,  if fixed by optimization it turns out to be of order  $\zstar\simeq 1$.  Above it ($z>\zstar$) the matter density evolves as in the $\CC$CDM and in addition $\rv$ remains constant. Its dynamics is only triggered at (and below) $\zstar$.  An important consequence of  such threshold is that the cosmological physics during the  CMB epoch (at $z\simeq 1000$) is exactly as in the $\CC$CDM.  For type II models there is still some evolution of the VED at the CMB epoch, but the matter density follows the same law as in the $\CC$CDM case. For this reason we will not investigate here the threshold scenario for type II models.

\section{Cosmological perturbations}\label{Sect:CosmPerturbations}

So far so good for the background cosmological equations in the presence of dynamical vacuum. However,  an accurate description of the  large scale structure  (LSS) formation data is also of paramount importance, all the more if we take into account that one of the aforementioned $\CC$CDM tensions (the $\sigma_8$ one) stems from it.  Allowing for some evolution of the vacuum can be the clue to solve the $\sigma_8$ tension since such dynamics affects nontrivially the cosmological perturbations\,\cite{Gomez-Valent:2018nib,Gomez-Valent:2017idt}.
We consider the perturbed, spatially flat,  FLRW metric   $ds^2=-dt^2+(\delta_{ij}+h_{ij})dx^idx^j$, in which $h_{ij}$ stands for the metric fluctuations. These fluctuations are coupled to the  matter density perturbations $\delta_{\rm m}=\delta\rho_{\rm m}/\rho_{\rm m}$.
We shall refrain from providing details of this rather technical part, which will be deferred for an expanded presentation  elsewhere.  However, the reader can check e.g.  \cite{Gomez-Valent:2017idt,SolaPeracaula:2017esw,Gomez-Valent:2018nib,Gomez-Valent:2017idt,Sola:2017znb,Sola:2016jky,Sola:2015wwa,Sola:2016hnq,Gomez-Valent:2014rxa,Basilakos:2014tha,Basilakos:2009wi,SolaPeracaula:2020vpg,SolaPeracaula:2019zsl} for the basic discussion of the RVM perturbations equations. The difference is that here we have implemented the full perturbations analysis  in the context of the Einstein-Boltzmann code \texttt{CLASS}\,\cite{Blas:2011rf}  (in the synchronous gauge\,\cite{Ma:1995ey}).   Let us nonetheless mention a few basic perturbations equations which have a more direct bearing on the actual fitting analysis presented in our tables and figures.  Since baryons do not interact with the time-evolving VED the perturbed conservation equations are not directly affected. However, the corresponding equation for CDM is modified in the following way:
\begin{equation}\label{Eq:PhenomenologyRVM.perturbCDM}
\dot{\delta}_{\rm dm}+\frac{\dot{h}}{2}-\frac{\dot{\rho}_{\rm vac}}{\rho_{\rm dm}}\delta_{\rm dm}=0\,,
\end{equation}
with $h=h_{ii}$ denoting the trace of $h_{ij}$. We remark that the term $\dot{\rho}_{\rm vac}$ is nonvanishing for these models and affects the fluctuations of CDM in a way which obviously produces a departure from the $\CC$CDM. The above equation is, of course, coupled with the metric fluctuations and the combined system must be solved numerically.

The analysis of the linear LSS regime is performed with the help of the weighted linear growth $f(z)\sigma_8(z)$, where $f(z)$ is the growth factor and $\sigma_8(z)$ is the rms mass fluctuation amplitude on scales of $R_8=8\,h^{-1}$ Mpc at redshift $z$.   The quantity $\sigma_8(z)$ is directly provided by \texttt{CLASS} and the calculation of $f(a)$ (with $z=a^{-1}-1$ in our normalization) can be obtained as follows. If $\vec{k}$ denotes the comoving wave vector and $\vec{k}/a$  the physical one,  at  subhorizon scales its modulus (square) satisfies  $k^2/a^2\gg  H^2$. If, in addition, we are in the linear regime the matter density contrast can  be written as $\delta_{\rm m}(a,\vec{k}) = D(a)F(\vec{k})$\,\cite{Dodelson:2003ft,Amendola:2015ksp},
where the dependence on $\vec{k}$ factors out. The properties of $F(\vec{k})$ are determined by the initial conditions and $D(a)$ is called the growth function.  The relation between the matter power spectrum and the density contrast reads
$P_{\rm m}(a,\vec{k}) = C\langle\delta_{\rm m}(a,\vec{k})\delta^{*}_{\rm m}(a,\vec{k})\rangle  \equiv D^2(a)P(\vec{k})$,
where $C$ is a constant and $P(\vec{k}) =C\langle{F(\vec{k})}F^{*}(\vec{k})\rangle$ is the primordial power spectrum (determined from the theory of inflation).  Since neither $F(\vec{k})$ nor  $P(\vec{k})$  depend on $a$, the linear growth $f(a) = d\ln \delta_{\rm m}(a,\vec{k})/d\ln a$ is given by $f(a)={d\ln D(a)}/{d\ln a}$, and ultimately by
\begin{equation}\label{Eq:PhenomenologyRVM.f(a)}
f(a) =\frac{d\ln P^{1/2}_{\rm m}(a,\vec{k})}{d\ln a} = \frac{a}{2P_{\rm m}(a,\vec{k})}\frac{dP_{\rm m}(a,\vec{k})}{da}\,.
\end{equation}
It follows that we may extract the  (observationally measured) linear growth function  $f(a)$ directly from the matter power spectrum $P_{\rm m}(a,\vec{k})$, which is computed numerically by \texttt{CLASS}  for all  values of $a$ and $\vec{k}$ (assuming   adiabatic initial conditions). This allows us to compare theory and observation for the important  LSS part.
%
\begin{figure*}
\centering
\includegraphics[angle=0,width=0.9\linewidth]{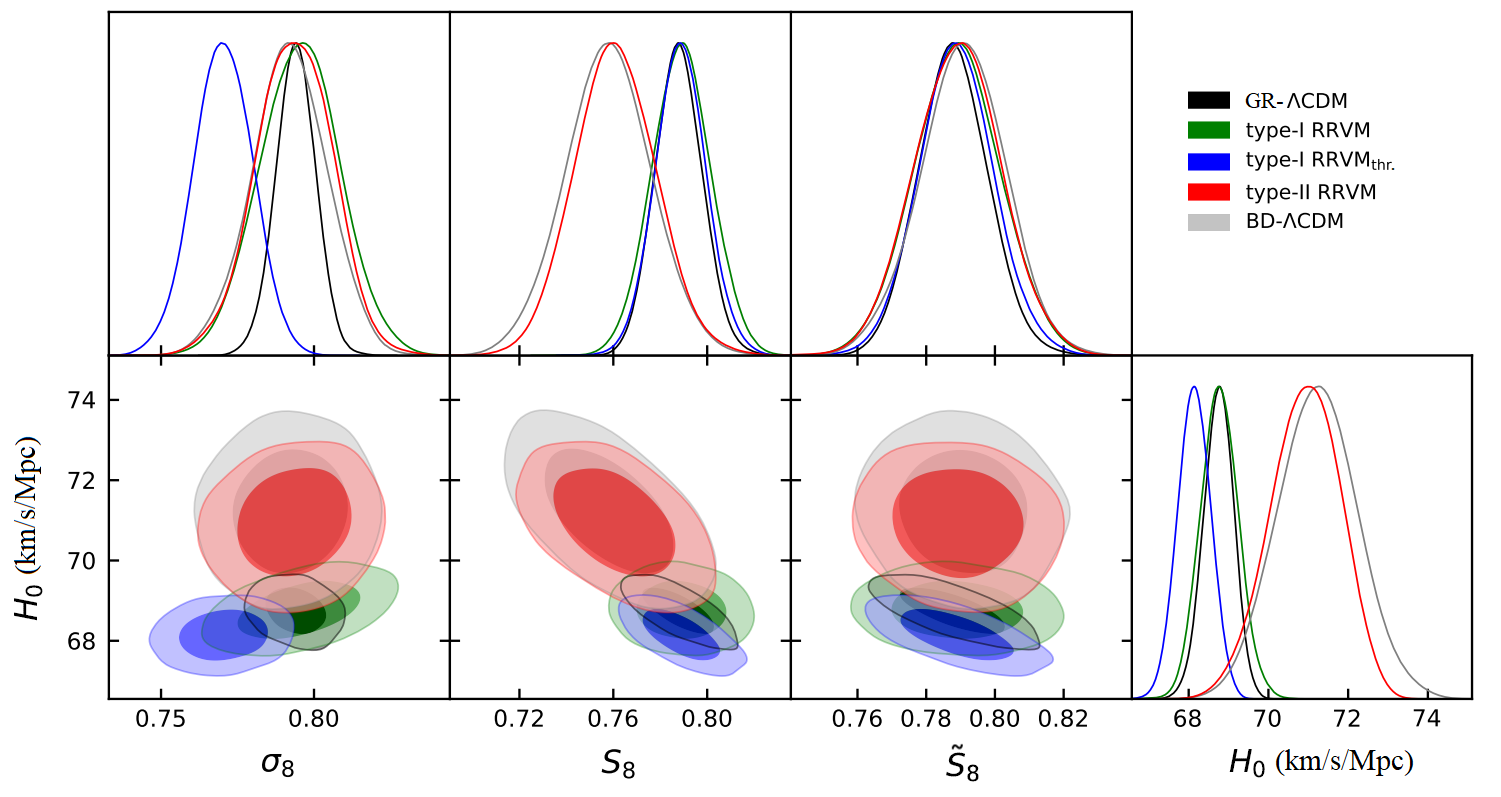}
\caption{
$1\sigma$ and $2\sigma$ contours in the $H_0$-$\sigma_8,S_8,\tilde{S}_8$ planes and the corresponding one-dimensional posteriors for the {GR- and BD- $\Lambda$CDM and the RRVM's} obtained from the fitting analyses with our Baseline+$H_0$ data set. The type II model manifestly alleviates the $H_0$ tension without spoiling the $\sigma_8$ one (even if phrased through the alternative parameters $S_8$ or $\tilde{S}_8$, see text), whereas the type I model with threshold redshift $\zstar\simeq 1$ can fully solve the latter (see also \hyperref[Fig:PhenomenologyRVM.fs8Evolution]{Fig.\,\ref{Fig:PhenomenologyRVM.fs8Evolution}}) but cannot address the former.
}\label{Fig:PhenomenologyRVM.XCDMEvolution}%
\end{figure*}
%

\section{Fitting results and discussion of the chapter}\label{Sect:FittingResults}

To compare the  RRVM's (types I and II) with the $\CC$CDM, we have defined a joint likelihood function ${\cal L}$. The overall fitting results are reported in \hyperref[Table:PhenomenologyRVM.Fit1]{Table\,\ref{Table:PhenomenologyRVM.Fit1}} and \hyperref[Table:PhenomenologyRVM.Fit2]{Table\,\ref{Table:PhenomenologyRVM.Fit2}}. The used data sets are the same as those described  in detail in  the previous chapter for BD models, except the updated values pointed out in the caption of \hyperref[Table:PhenomenologyRVM.Fit1]{Table\,\ref{Table:PhenomenologyRVM.Fit1}}.  Assuming Gaussian errors, the total $\chi^2$ to be minimized in our case is given by
\begin{equation}\label{Eq:PhenomenologyRVM.ChiSq}
\chi^2_{\rm tot}=\chi^2_{\rm SNIa}+\chi^2_{\rm BAO}+\chi^2_{ H}+\chi^2_{\rm f\sigma_8}+\chi^2_{\rm CMB}\,.
\end{equation}
The above $\chi^2$ terms are defined in the standard way from the data including the covariance matrices\,\cite{Amendola:2015ksp}.  In particular, the $\chi^2_{H}$ part may contain or not the local $H_0$ value measured by Riess et al.\,\cite{Riess:2019cxk,Reid:2019tiq} depending on the setup indicated in the tables (apart from the cosmic chronometer data employed also in \cite{SolaPeracaula:2020vpg,SolaPeracaula:2019zsl}). The local determination of $H_0$ (which is around $4\sigma$ away from the corresponding Planck 2018 value based on the CMB) is the origin of the so-called $H_0$ tension\,\cite{verde2019tensions,DiValentino:2020zio,DiValentino:2020vvd}.
Taking into account that the RRVM's of type I and II have one and two more parameters, respectively, as compared to the $\CC$CDM, a fairer  model comparison is achieved by computing the differences between the Deviance Information Criterion \cite{Spiegelhalter:2002yvw}, of the $\CC$CDM model and the RRVM's: $\Delta{\rm DIC}={\rm DIC}_{\rm \CC CDM}-{\rm DIC}_{\rm RRVM}$.  These differences will be (and in fact are) positive if the RRVM's fit better the overall data than the $\CC$CDM. The DIC is defined as
\begin{equation}
{\rm DIC}=\chi^2(\overline{\theta})+2p_D\,.
\end{equation}
Here $p_D=\overline{\chi^2}-\chi^2(\overline{\theta})$ is the effective number of parameters of the model, and $\overline{\chi^2}$ and $\overline{\theta}$ the mean of the overall $\chi^2$ distribution and the parameters, respectively. The DIC is a  good approximation to the exact Bayesian approach and works optimal if the posterior distributions are sufficiently Gaussian.  To obtain the posterior distributions and corresponding constraints for the various dataset combinations we have used  the Monte Carlo cosmological parameter inference code \texttt{Montepython}\cite{Audren:2012wb} in combination with the mentioned Einstein-Boltzmann code \texttt{CLASS}\,\cite{Blas:2011rf}.

The value of DIC can be computed directly from the Markov chains generated with \texttt{MontePython}.
For values $+5<\Delta{\rm DIC} <+10$ we would conclude strong evidence of the RRVM's as compared to the $\CC$CDM, and for $\Delta{\rm DIC}>+10$ the evidence is very strong. Such is the case when we use a threshold  redshift $\zstar\simeq 1$ in type I RRVM (cf. \hyperref[Table:PhenomenologyRVM.Fit1]{Table\,\ref{Table:PhenomenologyRVM.Fit1}} and \hyperref[Table:PhenomenologyRVM.Fit2]{Table\,\ref{Table:PhenomenologyRVM.Fit2}}). In contrast, when the threshold is removed we find only moderate evidence against it ($-3<\Delta{\rm DIC} <-2$), although the fitting performance keeps on being slightly better (smaller $\chi^2_{\rm min}$) than the GR-$\Lambda$CDM, similar to e.g. coupled dark energy \cite{Gomez-Valent:2020mqn}. Quite obviously, the effect of the threshold can be very important and indicates that a mild dynamics of the vacuum is very much welcome, especially if it is activated at around the very epoch when the vacuum dominance appears, namely at around $z\simeq 1$. To be more precise, the vacuum dominance in the $\CC$CDM starts at around $z\simeq 0.3$.  Therefore, these results suggest that if the vacuum starts to be slightly dynamical at an earlier point  which is `close' (in redshift terms) to the transition from deceleration to acceleration ($z\simeq 0.7$), then the impact on the description of the overall SNIa+BAO+$H(z)$+LSS+CMB data becomes extraordinarily significant on statistical terms. Before the transition point, physics can remain basically unaltered with respect to the standard $\Lambda$CDM model, but the vacuum dynamics allows to suppress an exceeding amount of LSS in the Universe, leading to a better description of the $f(z)\sigma_8(z)$ data set. It is not just that the total  $\chi^2_{\rm min}$ is 13 to 18 units smaller as compared to the $\CC$CDM in the presence of the threshold $\zstar$ (cf. \hyperref[Table:PhenomenologyRVM.Fit1]{Table\,\ref{Table:PhenomenologyRVM.Fit1}} and \hyperref[Table:PhenomenologyRVM.Fit2]{Table\,\ref{Table:PhenomenologyRVM.Fit2}}), but the fact that the information criteria (which take into account the penalty to be paid by the RRVM's for having more parameters) still decides very strongly in its favor. In the absence of the $H_0$ prior \cite{Riess:2019cxk,Reid:2019tiq}, type II RRVM performs a bit better than the GR-$\Lambda$CDM (cf. \hyperref[Table:PhenomenologyRVM.Fit1]{Table\,\ref{Table:PhenomenologyRVM.Fit1}}), but the improvement is not sufficient. Occam's razor penalizes the model for having two additional parameters than GR-$\Lambda$CDM and leads to a moderately negative evidence against it. When we include the prior, however, we get a strong evidence in its favor ($\Delta {\rm DIC}\gtrsim +5$, cf.\hyperref[Table:PhenomenologyRVM.Fit2]{Table\,\ref{Table:PhenomenologyRVM.Fit2}}), since this model can accommodate higher values of the Hubble parameter and hence loosen the $H_0$ tension. This is similar to what we found in \cite{SolaPeracaula:2020vpg,SolaPeracaula:2019zsl} for Brans-Dicke cosmology with $\CC\neq 0$.

Finally, we want to remark a few things about the RRVM's under study, in connection with the cosmological tensions, cf. \hyperref[Table:PhenomenologyRVM.Fit1]{Table\,\ref{Table:PhenomenologyRVM.Fit1}} and \hyperref[Table:PhenomenologyRVM.Fit2]{Table\,\ref{Table:PhenomenologyRVM.Fit2}}, and the contours in \hyperref[Fig:PhenomenologyRVM.XCDMEvolution]{Fig.\,\ref{Fig:PhenomenologyRVM.XCDMEvolution}}: (i) the only model capable of alleviating the $H_0$ tension is RRVM of type II; (ii) the values of $S_8$ in all RRVM's are perfectly compatible with recent weak lensing and galaxy clustering measurements \cite{Heymans:2020gsg}. For type II a related observable analogous to (but different from) $S_8$ is possible: $\tilde{S}_8\equiv S_8/\sqrt{\varphi_0}$. It is connected with the time variation of $G=G_N/\varphi$ and can be viewed also as a rescaling $\Omega_{\rm m}^0\to \Omega_{\rm m}^0/\varphi_0$ in the effective Friedmann's equation for type II models, see the previous chapter, where we do similar definitions for BD models. We show the corresponding countours in \hyperref[Fig:PhenomenologyRVM.XCDMEvolution]{Fig.\,\ref{Fig:PhenomenologyRVM.XCDMEvolution}}; (iii) Quite remarkable is the fact that the value of $\sigma_8$ is significantly lower in the type I RRVM$_{\rm thr.}$ to the point that the $\sigma_8$ tension can be fully accounted for. We have checked that this feature is shared by the more general RVM class \eqref{Eq:PhenomenologyRVM.RVMvacuumdadensity} using the same threshold redshift.

We find significant evidence that a  mild dynamics of the cosmic vacuum would be helpful to describe the overall cosmological observations as compared to the standard cosmological model with a rigid $\CC$-term.  For type I models the level of evidence is very strongly supported by the DIC criterion provided there exists a threshold redshift $\zstar\simeq 1$ where the vacuum dynamics is triggered. With such dynamics the $\sigma_8$ tension is rendered virtually nonexistent ($\lesssim 0.4\sigma$) \cite{Heymans:2020gsg}. The $H_0$ tension, however,  can only be improved within the type II model  with variable $G$ {and also with the related BD-$\CC$CDM model already studied in \hyperref[Chap:PhenomenologyofBD]{chapter\,\ref{Chap:PhenomenologyofBD}}. For both the two tensions can be dealt with at a time, the $H_0$ remaining at $\sim 1.6\sigma$ \cite{Riess:2019cxk,Reid:2019tiq} and the $\sigma_8$  one at $\sim 1.3\sigma$ (or at only $\sim 0.4\sigma$ if stated in terms of $S_8$)\,\cite{Heymans:2020gsg}. The simultaneous alleviation of the two tensions is remarkable and is highly supported by the DIC criterion.

\blankpage

\chapter{Conclusions}\label{Conclusions}

In this dissertation, we present the most relevant works and results from the PhD period. In chapters \hyperref[Chap:QuantumVacuum]{\ref{Chap:QuantumVacuum}}, \hyperref[Chap:EoSVacuum]{\ref{Chap:EoSVacuum}}, and \hyperref[Chap:Fermions]{\ref{Chap:Fermions}}, we report our theoretical calculations concerning the running of the vacuum energy density (VED) in the context of Quantum Field Theory (QFT) and obtain significant results that were previously absent from the literature. Unfortunately, the widespread confusion in the literature between the cosmological constant, $\CC$, and the VED, $\rv$, has hindered a proper treatment of the renormalization of these quantities in cosmological spacetime. Perhaps the most pernicious practice has been the repeated attempts to relate these concepts in the context of flat spacetime calculations, which is meaningless\,\cite{SolaPeracaula:2022hpd,Mottola:2022tcn}. As we indicated, if we consider $\CC$ as the physically measured value, then its relation with the current $\rv$ is straightforward: $\rvo=\CC/(8\pi G_N)$. However, more care needs to be taken at a formal level where these quantities are derived from a gravitational action in curved spacetime. Even though a quantum theory of gravity is currently inaccessible, we show that studying $\rv$ in QFT in curved spacetime can be quite helpful. The most noticeable result is that the VED acquires a dynamical nature, resembling the Running Vacuum Models (RVM), which have been present in the literature for two decades (see \,\cite{Sola:2013gha,SolaPeracaula:2022hpd} and references therein). This family of models were originally justified by generic renormalization group arguments and state that the VED evolves with the cosmological expansion as a series of powers of the Hubble function. Here we present the first rigorous derivation within QFT in curved spacetime of this evolution  

On the other hand, in chapters \hyperref[Chap:PhenomenologyofBD]{\ref{Chap:PhenomenologyofBD}} and \hyperref[Chap:PhenomenologyofRVM]{\ref{Chap:PhenomenologyofRVM}}, we have presented our fits and analysis of two different kinds of extensions of the cosmological standard model: the family of Ricci Running Vacuum Models (RRVM) and Brans-Dicke $\Lambda$CDM (BD-$\Lambda$CDM). These models represent small but noticeable deviations from the $\Lambda$CDM model. We have studied them in detail, both at the background and perturbation levels, and have confronted them against a robust dataset composed of a variety of cosmological probes.

Let us start summarizing the main conclusions obtained in the chapters dedicated to the QFT in Curved spacetime computations (cf. chapters \hyperref[Chap:QuantumVacuum]{\ref{Chap:QuantumVacuum}}, \hyperref[Chap:EoSVacuum]{\ref{Chap:EoSVacuum}} and \hyperref[Chap:Fermions]{\ref{Chap:Fermions}}):

\begin{itemize}

\item[$\bullet$] We presented a novel formalism based on the well-known formalism of adiabatic regularization in curved spacetime for an expanding Universe. This method consists of an adiabatic expansion of the interested quantities, such as the mode functions satisfying the wave equation, which yields an asymptotic series that can be used to approximate physical observables. In traditional references, the computations are performed on-shell, but here we work off-shell, defining a renormalization scale we call $M$. While this method was originally presented for computing the running couplings in an FLRW evolving background\,\cite{Ferreiro:2018oxx}, we go further and systemize and extend its use for studying the evolution of the vacuum energy-momentum tensor (EMT) along the cosmological history in great detail in our papers\,\cite{Moreno-Pulido:2020anb,Moreno-Pulido:2022phq} (see also\,\cite{SolaPeracaula:2022hpd} for a review) and shown in \hyperref[Chap:QuantumVacuum]{chapter\,\ref{Chap:QuantumVacuum}}.

\item[$\bullet$] Prior to compute the physical quantities of interest, one had to get rid of the manifest divergences that appear in the form of diverging integrals.  This is solved after performing a renormalization procedure based on the subtraction of the given UV-divergent quantity at an off -shell scale, M, which regularizes the integrals. Thanks to this method, we were able to renormalize the value of the so-called zero-point energy (ZPE), which receives contributions from several free fields with bosonic or fermionic nature. Not only was the ZPE renormalized, but we were also able to properly renormalize the running couplings of the theory, in particular, $\rho_\Lambda(M)$ of the Einstein-Hilbert action and $G$, the gravitational coupling. As for our definition of VED, in our simplified framework, $\rho_{\rm vac}$ receives contributions from two different origins: the aforementioned ZPE, coming from the various quantum fields present in the standard model of particle physics (or even beyond), and $\rho_\Lambda(M)$, which has to be interpreted as a geometrical contribution to the VED.

\item[$\bullet$] The renormalized results obtained in our analysis are robust. Indeed, in order to strengthen our conclusion, which we have first presented on the basis of the WKB expansion of the Fourier modes of the field, we have subsequently corroborated it from the perspective of the effective action formalism. It means that we have solved the curved spacetime Feynman propagator of the nonminimally coupled scalar field to gravity using the adiabatic method and computed the effective action using the heat-kernel expansion. Since that expansion has also been performed off-shell (i.e. at the arbitrary scale $M$ rather than at the physical mass $m$), it was necessary to compute the corresponding corrections induced on the DeWitt-Schwinger coefficients. With the help of the effective action we have rederived the renormalized EMT and obtained the same results as with the adiabatic method. As a bonus we have extracted the renormalization group equations for the couplings and also for the VED itself.

\item[$\bullet$] Because of inappropriate renormalization schemes and computational procedures,  the computation of the VED has been persistently  affected  by severe fine tuning problems owing to the presence of $m^4$ terms in these schemes. Only an approach based on the effective action and/or an appropriate renormalization of the energy-momentum tensor (EMT) has the capability to capture all the essential physical features. We show there that the appropriate calculation and renormalization of the zero-point energy (ZPE) in curved spacetime, together with the $\rL$ parameter in the action, produces a dynamical quantity, $\rho_{\rm vac}(M)$, whose running is free from undesired $m^4$ contributions, where $m$ is the mass of a field involved in the computation. Indeed, the VED is given schematically by $\rL+{\rm ZPE}$, and upon renormalizing the energy-momentum tensor directly, or indirectly through its relation with the effective action of vacuum (cf. Sect. \ref{Sect:EffectiveActionQFT}), the two procedures converge to the same result. We find that although both $\rL$ and ${\rm ZPE}$ bring forth quartic powers $\sim m^4$, they do not affect the running of the VED. Formally, this is because the $\beta$-function of the VED does not depend on $m^4$ terms, as we have seen in \hyperref[Sect:appendixAbisbis1]{Appendix\,\ref{Sect:appendixAbisbis1}}. Thus, there is no need for fine-tuning if our desire is to compute the evolution of $\rho_{\rm vac}$ with time. While this is not equivalent to solving the problem of the cosmological constant, it is remarkable enough that we can free our results from these undesirable contributions. Notice that a naive application of other methods such as the MS-scheme renormalization has several problems with fine-tuning when dealing with this quartic power.

\item[$\bullet$] Scale dependencies (explicit and implicit) obviously cancel out in the full effective action of the theory, which is renormalization group invariant, and involves in particular the vacuum action as well as many other scale-dependent terms, such as the classical Lagrangian with the corresponding running couplings, which contribute also in a crucial way to warrant the overall scale-independence of the full effective action. Now except for the effective action itself, which is a rather formal object, in cosmology we cannot play with RG-invariant quantities which are more common in particle physics (such as scattering cross-sections, decay rates, etc. and in general different kinds of Green’s functions related to observable quantities). Thus, we must content ourselves with using different parts of the full effective action that remain scale dependent. The VED is one of these parts and hence it appears as one of the scale-dependent quantities upon renormalization. Therefore, after applying the renormalization procedure, it is necessary to fix the off-shell scale to a suitable, relevant physical scale if we want to obtain meaningful results. This is in no way different from the case of ordinary gauge theories. The only difference is that we are in a cosmological context and one has to make a choice and test its effectiveness. In our case we have proposed the Hubble function, $M=H$, as the most natural quantity for this role, which has dimension of energy in natural units and it therefore can be considered a characteristic energy scale in an expanding Friedmann-Lemaître-Robertson-Walker (FLRW) spacetime.

\item[$\bullet$] As indicated in the previous point, the most natural quantity selected for this role in FLRW spacetime is the Hubble function, $H$, which traces the characteristic energy scale of the expanding Universe. Therefore, we set $M=H$ and end up with a smooth, dynamical vacuum energy density parametrized as $\rho_{\rm vac}(H)$, where $H$ designates, generically, the dependency of $\rho_{\rm vac}$ on $H$ and its derivatives. The low-energy regime, near our time, is characterized by a mild evolution proportional to the difference $H^2-H_0^2$, where $H_0$ is the value of the Hubble function at the present time, and $H$ is a close value in the recent past. In other words, if $a_1$ and $a_2$ are two values of the scale factor close to its current value, then $\rho_{\rm vac}(a_1)-\rho_{\rm vac}(a_2)\sim \nu_{\rm eff}\mpl^2 (H^2(a_1)-H^2(a_2))$, up to a great degree of approximation. Here, $|\nu_{\rm eff}|$ is an effective parameter proportional to the quadratic powers of the masses of the particles. The detailed structure of $\nueff$ depends on all the quantum matter fields involved in the calculation. We have provided an analytic expression for $\nueff$ obtained within our QFT formalism, which depends on the masses of all quantized fields and the non-minimal couplings to gravity. However, its numerical value must ultimately be determined experimentally by confronting the model with cosmological data. The $H^2$ term is nevertheless sufficient to describe the dynamics of the vacuum in the current Universe, while the higher-order components can play a role in the early Universe, and in particular for describing inflation. This path also leads to a mildly logarithmic running of the gravitational coupling $G$ (or equivalently, a running $\Mpl^2$).  

\item[$\bullet$] The law governing the vacuum energy density that we have rigorously derived was proposed a long time ago for the Running Vacuum Models (RVM). These models have the same low-energy regime for the vacuum energy density as presented here. Previously, they were motivated by generic Renormalization Group arguments, but now we have derived the dynamic expression for the first time within the context of Quantum Field Theory, using the adiabatic regularization procedure and a renormalization scheme with all detail. Furthermore, we have observed that deviations from the more traditional law, $\rho_{\rm vac}(H)=\rho_{\rm vac} (H_0)+3\nu/(8\pi G_N)(H^2-H_0^2)$ are possible, such as replacing $H^2$ with a linear combination of $H^2$ and $\dot{H}$. This fact motivates the study of alternative models originated from the RVM, such as the Ricci-RVM (RRVM) of \hyperref[Chap:PhenomenologyofRVM]{chapter\,\ref{Chap:PhenomenologyofRVM}}.

\item[$\bullet$] The fundamental approach described here does not require invoking ad-hoc phantom or quintessence fields, or even particular forms for the effective potentials, which are absent in our simple model. The vacuum-to-vacuum fluctuations of quantum fields in curved spacetime are sufficient to describe the dynamical behavior of the vacuum fluid, after appropriate renormalization following our adiabatic procedure. Let us also note that, apart from the QFT framework shown in this dissertation which supports the theoretical structure of the RVM, an intriguing 'stringy' version of the RVM is also possible and with interesting phenomenological implications as well\,\cite{Basilakos:2019acj,Basilakos:2020qmu,Mavromatos:2020kzj,Mavromatos:2021urx}.

\item[$\bullet$] An analogous procedure applies to the vacuum pressure as well. This allows us to compute the equation of state $w_{\rm vac}\equiv P_{\rm vac}/\rho_{\rm vac}$, as we originally presented in our paper\,\cite{Moreno-Pulido:2022upl}, of the quantum vacuum from first principles, which does not remain fixed at -1 as in the traditional cosmological constant of the $\Lambda$CDM model. Instead, it appears to mimic the dominant component of the Universe throughout cosmic evolution, approaching -1 only in the late period of expansion. This surprising result is obtained from the direct calculation performed in QFT and to the best of our knowledge is unprecedented in the literature. Interestingly, the effective picture of the Brans-Dicke-$\Lambda$CDM model presented in \hyperref[Chap:PhenomenologyofBD]{Chap.\,\ref{Chap:PhenomenologyofBD}} exhibits a similar behavior. The reason is that one can show that the  Brans-Dicke-CDM model with a cosmological constant mimics the RVM.

\item[$\bullet$] The higher-order terms in the adiabatic expansion correspond to higher powers of the Hubble function and/or a higher number of time derivatives. While these terms do not affect in a significant way any fit of the RVM to the modern cosmological data, such as the SNIa + $H(z)$ + BAO + LSS + CMB observations, they may play a crucial role in the early Universe. Specifically, a simple but interesting mechanism for inflation relying on the higher-order terms in the expansion of $\rho_{\rm vac}(H)$ has been described. These terms are proportional to $H^6$ and can lead to an enhancement in the magnitude of $\rho_{\rm vac}(H)$ at high redshifts, in the remote past. This simple mechanism predicts a graceful end to inflation followed by a reheating process that fills the Universe with radiation. Although many details still need to be addressed and a dedicated study is necessary, this possibility is worth considering.

\item[$\bullet$] The conclusions presented here originated from our calculations with a single real scalar field\,\cite{Moreno-Pulido:2020anb,Moreno-Pulido:2022phq,SolaPeracaula:2022hpd,Moreno-Pulido:2022upl}, which are summarized in chapters \hyperref[Chap:QuantumVacuum]{\ref{Chap:QuantumVacuum}} and \hyperref[Chap:EoSVacuum]{\ref{Chap:EoSVacuum}}. However, in chapter \hyperref[Chap:Fermions]{\ref{Chap:Fermions}}, based on our paper\,\cite{Samira2022}, we analyze a more general scenario where multiple free spin-1/2 and scalar fields are present. Despite this generalization, we arrived at the same conclusions, providing strong support for our previous results. This raises the question of whether the characteristic running law of the vacuum energy density from QFT is universal.

\end{itemize}

Now, we will list our conclusions with respect the phenomenological part of this dissertation (cf. chapter \hyperref[Chap:PhenomenologyofBD]{\ref{Chap:PhenomenologyofBD}} and \hyperref[Chap:PhenomenologyofRVM]{\ref{Chap:PhenomenologyofRVM}}).

\begin{itemize} 

\item[$\bullet$] In chapter \hyperref[Chap:PhenomenologyofBD]{\ref{Chap:PhenomenologyofBD}}, we dedicate an extensive study to the Brans-Dicke model, or Brans-Dicke-$\Lambda$CDM (BD-$\Lambda$CDM), as we call it in our papers\,\cite{SolaPeracaula:2019zsl,SolaPeracaula:2020vpg}. This model is a simple but non-trivial extension of the $\Lambda$CDM by allowing the gravitational constant $G$ to vary with time. This variation is parametrized using a BD-field, such that $G=G_N/\varphi$, where $G_N\approx 6.67 \times 10^{–11} m^3 kg^{–1} s^{-2}$ is the local gravitational constant measured by Cavendish-like experiments at the surface of our planet. The action contains a kinetic term for the field and the traditional cosmological constant $\rho_\Lambda$, but we did not assume any potential for the BD-field.

\item[$\bullet$] We studied different scenarios depending on how we interpret the Cassini constraint and how we fix the initial condition of the BD-field. When the initial value of the BD-field is considered as a free parameter, our fits indicate that the preferred value of the BD-field (or the gravitational constant) is lower than 1 (or higher than $G_N$), and the mean value of the $H_0$ parameter is higher than that of the $\Lambda$CDM in each possible dataset. In particular, when including the SH0ES' prior on the value of $H_0$, the mean value of this parameter falls in the range of 70-72 km/s/Mpc, depending on the dataset, thereby reducing the tension to smaller values.

\item[$\bullet$] On the other hand, the dynamics of the field can play an important role in the $\sigma_8$ tension. The variation of the field with time is encoded in the parameter $\epsilon_{\rm BD}$. In the $\Lambda$CDM model, this parameter is exactly 0, while in Brans-Dicke models, it may have a small but significant value. In our analysis, we concluded that a non-zero $\epsilon_{\rm BD}$ has a relevant impact on structure formation, lowering the mean value of $\sigma_8$ with respect to the $\Lambda$CDM model and increasing the errors under different datasets, thus reducing the associated tension. The combination of both effects, the freeing of the initial value and its dynamical nature, may ameliorate both tensions at the same time.

\item[$\bullet$] In \hyperref[Chap:PhenomenologyofRVM]{chapter\,\ref{Chap:PhenomenologyofRVM}}, we present a detailed analysis of RRVM models, or Ricci-RVM, emerging from some effective approaches to the traditional RVM that have been mentioned previously. They consist in a slight modification of the classical law of the RVM, where the Hubble parameter is replaced by the Ricci scalar, $\rho_{\rm vac}(R)$. This model can be motivated by an effective approach (see \hyperref[SubSect:GeneralizedRVM]{Sect.\,\ref{SubSect:GeneralizedRVM}}) or by purely phenomenological motivations. We test two different scenarios: one where there is an interaction between matter and vacuum (type I), and another where matter is conserved, but the vacuum can evolve by exchanging energy with the background gravitational field (type II). In the former, we even consider the possibility of this interaction being activated recently at a low redshift, which we call the threshold. These features appear to be very useful in reducing cosmological tensions.

\item[$\bullet$] The concept of a step function threshold may appear entirely ad hoc and without any fundamental reason beyond the scope of pure phenomenology. However, as discussed in chapter \ref{Chap:EoSVacuum}, this threshold could be interpreted as the point at which the equation of state (EoS) of the vacuum fluid drops below $-1/3$ at recent redshifts of around $z\sim 1-10$ (as shown in Figure \ref{Fig:eos.EosEntire} in chapter \ref{Chap:EoSVacuum}). This indicates that vacuum behaving as a dynamical dark energy is a relatively recent event in cosmological history. While this is not precisely the same as the step function used in our analysis, it could be a fundamental reason why our approach appears to work well in addressing the $\sigma_8$ tension. Nevertheless, it remains essential to fit a model that incorporates the extra features of quantum vacuum encountered in this work, and we plan to do so soon \,\cite{CosmoTeam2023}.

\item[$\bullet$] The effective parameter that controls the running of the VED, $\nu_{\rm eff}$, is found to be positive based on the fitting results. In the type I scenario with a threshold, this parameter is different from 0 at the $\sim 3 \sigma$ level, while in the type II scenario, there is still a signal at the $\sim \sigma$ level. This suggests that the VED decreases slowly with expansion, and thus the RVM model behaves similarly to quintessence. This result is consistent with previous studies, and readers can find a comprehensive list of references in the introduction of \hyperref[Chap:PhenomenologyofRVM]{chapter \ref{Chap:PhenomenologyofRVM}}.

\item[$\bullet$] Type I models with a recent threshold can perfectly handle the $\sigma_8$ tension and reduce it to a harmless discrepancy below $0.4\sigma$. For type II models, the tension is reduced at the $1.3\sigma$ level in terms of $\sigma_8$ and $\sim 0.5\sigma$ in terms of $\tilde{S}_8$. Moreover, it shares similarities with the Brans-Dicke model presented earlier in this dissertation. It should be noted that its effect in reducing the $H_0$ tension is noticeable, remaining at the $1.6\sigma$ level. The alleviation of both tensions at the same time is a significant signal and is supported by the DIC criterion.

\item[$\bullet$] In summary, RRVM and BD offer promising extensions of the $\Lambda$CDM that can accommodate the general fit of cosmological data and hold potential for addressing the tensions that plague modern cosmology\,\cite{Abdalla:2022yfr}. However, our work is not yet complete. Further analyses are required to test our predictions against the latest cosmological data. Fortunately, in this era of precision cosmology, performing these tests will not take long, with vast amounts of data coming from independent surveys about to appear in the next years.

\end{itemize}

To conclude, let us reflect once more on the Cosmological Constant Problem (CCP). It poses a significant challenge in reconciling the predictions of general Quantum Field Theory (QFT) and string theory with the measurements from cosmological probes,\cite{weinberg1989cosmological,witten2001cosmological}. Given its magnitude and significance, it can be argued that the CCP is one of the most important (if not the most important) questions facing theoretical physics today, and an urgent need exists for a fundamental-level explanation.

The heart of our findings is that the QFT calculations carried out in this study indicate that the vacuum has dynamic properties. If the CCP is reframed as the necessity to determine the specific value of the VED at any given time based on first principles, then we are bound to fail.  We have emphasized that while the renormalization program is extremely useful in providing us with a law for the evolution of the vacuum energy density at any point in cosmic history, it is important to remember that we require observational data that cannot be obtained from the formalism. This additional ingredient is crucial for renormalization theory to make predictions at other points in time. 

Despite years of continued attempts, this is an insurmountable challenge, it seems that QFT does not provide any insight in this task of predicting the value of VED without any experimental input. Instead, within the QFT framework, we encounter a humble yet astounding and satisfactory result: the CCP in QFT is no longer the issue of comprehending how the enormous contributions of $\sim m^4$ from quantum fields cancel each other out. Rather, once we measure the precise value of the current VED, denoted as $\rvo\equiv \rv(H_0)$, which is unknown to us a priori, it will no longer be disturbed by the enormous $\sim m^4$ effects from any kind of quantum field we come across in the Universe. 

Additionally, QFT informs us that there is no definite cosmological constant viewed as an everlasting fundamental entity of Nature. In other words,$\rvo=\rv(H_0)$ should \textit{not} be interpreted as a fixed cosmological constant term, but merely as the value of the vacuum energy density at present ($H=H_0$). At any cosmic time characterized by the cosmological energy scale $H(t)$, there is a VED that evolves smoothly, known as $\rv(H)$. The dynamics are smooth such that $\CC=8\pi G(H) \rv(H)$ changes very gradually for a long period, creating the illusion of an almost constant quantity that the standard model explains as the cosmological constant.

However, we should remark once again that the CCP has many faces. While bringing some light to the problem of fine-tuning is a step forward, it is still different from solving the problem in its entirety. We must conclude that the so-called `old cosmological constant problem'\,\cite{weinberg1989cosmological}, formulated as the problem of explaining the value of the cosmological constant rather than its running, unfortunately persists.

Throughout the years of dedicated study on the dynamical vacuum energy, we have found promising signs from both a theoretical and phenomenological perspective. The theoretical framework presented here reinforces the position of the RVM family and related models. Despite enduring admonitions and severe criticisms over time, the quantum vacuum may very well be the ultimate \textit{raison d'être} for dark energy within the fundamental framework of QFT in curved spacetime. It could provide a clue to solving cosmological tensions and the ultimate explanation for cosmic acceleration.

\begin{appendices}

\blankpage


\chapter{Conventions, notations and useful formulas}\label{Appendix:Conventions}

All along this text we will use natural units (with some exceptions that will be indicated), therefore $\hbar=c=1$ and $G_N=1/m_{\rm Pl.}^2$, where $G_N$ is the Gravitational Constant measured on Earth by Cavendish-like torsion experiments and $\mpl \simeq 1.22\times 10^{19}$ GeV stands for Planck's Mass. As for the conventions on geometrical quantities used throughout this work, they read as follows:
\begin{itemize}
\item[$\bullet$] Signature of the metric $g_{\mu\nu}$, $(-, +,+,+ )$\,.
\item[$\bullet$] Riemann tensor, $R^\lambda_{\,\,\,\,\mu \nu \sigma} = \partial_\nu \, \Gamma^\lambda_{\,\,\mu\sigma} + \Gamma^\rho_{\,\, \mu\sigma} \, \Gamma^\lambda_{\,\, \rho\nu} - (\nu \leftrightarrow \sigma)$\,.
\item[$\bullet$] Ricci tensor, $R_{\mu\nu} = R^\lambda_{\,\,\,\,\mu \lambda \nu}$\,.
\item[$\bullet$] Ricci scalar, $R = g^{\mu\nu} R_{\mu\nu}$\,.
\end{itemize}
Overall, these correspond to the $(+, +, +)$ conventions in the classification by Misner-Thorn-Wheeler\,\cite{Misner:1973prb}. Our Framework is a Friedmann-Lemaître-Robertson-Walker (FLRW) background with null spatial curvature. As usual, the Einstein tensor is defined through $G_{\mu\nu}=R_{\mu\nu}-\frac12\,R g_{\mu\nu}$ and the Einstein field equations read $G_{\mu\nu}+\CC g_{\mu\nu}=8\pi G_N\, T_{\mu\nu}$. We assume spatially flat three-dimensional geometry along the different works summarized in the dissertation.

We denote with primes the derivatives with respect to conformal time ($\tau$) and with dots the derivatives with respect to cosmic time ($t$). Thus, ${\cal H}=a H$,  $a'=a \mathcal{H}=a^2 H$ and $a''=a^3(2 H^2+\dot{H})$. The Christoffel symbols associated to the conformally flat metric $ds^2=a^2(\eta)\eta_{\mu\nu}dx^\mu dx^\nu$, with $\eta_{\mu\nu}={\rm diag} (-1, +1, +1, +1)$, are the following:
\begin{equation}\label{Eq:Conventions.Christoffel}
\Gamma_{00}^{0}=\mathcal{H},\qquad \Gamma_{ij}^0=\mathcal{H}\delta_{ij}, \qquad \Gamma_{j0}^i=\mathcal{H}\delta_j^i\,.
\end{equation}
We can also derive the following useful relations to convert the derivatives of the Hubble rate with respect to conformal time into derivatives with respect to cosmic time, which are repeatedly used in the calculations quoted in the main text:
\begin{equation}\label{Eq:Conventions.ConformalHvsCosmicH}
\begin{split}
\mathcal{H}^\prime=&a^2(H^2+\dot{H})\,,\\
\mathcal{H}^{\prime\prime}=&a^3\left(2H^3+4 H\dot{H}+\ddot{H}\right)\,,\\
\mathcal{H}^{\prime\prime\prime}=&a^4\left(6H^4+18 H^2\dot{H}+ 4\dot{H}^2+ 7 H\ddot{H}+\vardot{3}{H}\right)\,,\\
\mathcal{H}^{\prime\prime\prime\prime}=&a^5\left(24H^5+96 H^3\dot{H}+ 52 H\dot{H}^2+ 46 H^2\ddot{H}+15\dot{H}\ddot{H}\ddot{H}+11 H\vardot{3}{H}+\vardot{4}{H}\right)\,,\\
\mathcal{H}^{\prime\prime\prime\prime\prime}=&a^6 \left( 120 H^6 + 600 H^4 \dot{H} + 548 H^2 \dot{H}^2 + 52 \dot{H}^3 +
326 H^3 \ddot{H} + 271 H \dot{H} \ddot{H} + 15 \ddot{H}^2 \right.\\
&\left. \phantom{aa}+101 H^2 \vardot{3}{H} + 26 \dot{H} \vardot{3}{H} + 16 H \vardot{4}{H} + \vardot{5}{H} \right)\,.
\end{split}
\end{equation}
For convenience we quote the Ricci scalar and the non-vanishing components of the curvature tensors in alternative forms:
\begin{equation}\label{Eq:Conventions.R}
R={6}\frac{a^{\prime\prime}}{a^3}=\frac{6}{a^2}\,(\mathcal{H}^\prime+\mathcal{H}^2)=6\,\left(\frac{\dot{a}^2}{a^2}+\frac{\ddot{a}}{a}\right)=6(2H^2+\dot{H})
\end{equation}
and
\begin{equation}\label{Eq:Conventions.R00G00}
 \ \ \ R_{00}=-3\mathcal{H}^\prime=-3a^2(H^2+\dot{H})\,, \qquad G_{00}=3\mathcal{H}^2=3a^2H^2\,.
\end{equation}
For reference we also quote the well-known definitions of Euler's density $E$ and the square of the Weyl tensor ($C^2$):
\begin{equation}\label{Eq:Conventions.EC2}
E= R^{\alpha\beta\gamma\delta}R_{\alpha\beta\gamma\delta}-4R^{\alpha\beta}R_{\alpha\beta}+R^2\,, \qquad \ \
C^2=R^{\alpha\beta\gamma\delta}R_{\alpha\beta\gamma\delta}-2R^{\alpha\beta}R_{\alpha\beta}+\frac{1}{3}R^2\,.
\end{equation}
It follows that
\begin{equation}\label{Eq:Conventions.R2R4}
R^{\alpha\beta\gamma\delta}R_{\alpha\beta\gamma\delta}= 2 C^2-E+\frac{1}{3}R^2, \qquad \ \
R^{\alpha\beta}R_{\alpha\beta}=\frac12\, (C^2-E)+\frac{1}{3}R^2\,.
\end{equation}
From the density $E$ one defines the Gauss-Bonnet term,
\begin{equation}\label{Eq:Conventions.GaussBonnet}
\GB= \int d^nx \sqrt{-g} E\,,
\end{equation}
which is a topological invariant in $n=4$ (not so in other dimensions). Such a topological invariance implies that the metric functional variation of $\GB$ vanishes identically in four dimensions:
\begin{equation}\label{Eq:Conventions.Topological}
 \frac{\delta \GB}{\delta g^{\mu\nu}}=0\ \ \ \ \ \ \ (n=4)\,.
\end{equation}
From the basic HD terms one may construct the higher derivative (HD) part of the vacuum action, henceforth $n=4$:
\begin{equation}\label{Eq:Conventions.HDaction}
S_{\rm HD}= \int d^4x \sqrt{-g} \left (\alpha_1 C^2+\alpha_2 R^2+\alpha_3 E+\alpha_4 \Box R\right)\equiv \int d^4x \sqrt{-g} L_{\rm HD} \,.
\end{equation}
The purely geometric terms in \eqref{Eq:Conventions.HDaction} are generated by the quantum matter contributions, and hence these HD terms are necessary for the renormalization procedure. The bare couplings $\alpha_i$ become renormalized couplings $\alpha_i$(M) which run with the renormalization scale $M$.
That HD gravitational action (with effective Lagrangian $ L_{\rm HD}$) is to be added to the EH action plus matter, cf. Eq.\,\eqref{Eq:QuantumVacuum.EH}, in order to have a well-defined and renormalizable semiclassical theory of quantum fields in curved spacetime. The total action of gravity plus matter therefore reads
\begin{equation}\label{Eq:Conventions.Totalaction}
S_{\rm tot}= S_{\rm EH}+S_{\rm HD}+S_{\rm m} \,.
\end{equation}
In it, the total vacuum action is the sum of the first two pieces, whereas the last piece is the matter action.
By functionally differentiating the $R^2$ and $R_{\alpha\beta}R^{\alpha\beta}$ terms with respect to the metric, we obtain two (conserved) higher order curvature tensors (of adiabatic order $4$), namely
\begin{equation}\label{Eq:Conventions.H1munu}
\leftidx{^{(1)}}{\!H}_{\mu\nu}=\frac{1}{\sqrt{-g}}\frac{\delta}{\delta g^{\mu\nu}}\int d^4 x \sqrt{-g}\,R^2 =-2\nabla_\mu\nabla_\nu R+2g_{\mu\nu}\Box R-\frac{1}{2}g_{\mu\nu}R^2+2RR_{\mu\nu}.
\end{equation}
and
\begin{equation}\label{Eq:Conventions.H2munu}
\begin{split}
\leftidx{^{(2)}}{\!H}_{\mu\nu}&=\frac{1}{\sqrt{-g}}\frac{\delta}{\delta g^{\mu\nu}}\int dx^4 \sqrt{-g}R_{\alpha\beta}R^{\alpha\beta}\\
&=2{R^\alpha}_\mu R_{\alpha \nu}-2g_{\mu\beta}\nabla_\alpha\nabla_\nu R^{\alpha\beta}+\Box R_{\mu\nu}+\frac{1}{2}g_{\mu\nu}\Box R-\frac{1}{2}g_{\mu\nu}R^{\alpha\beta}R_{\alpha \beta}\,.
\end{split}
\end{equation}
One can also define $H_{\mu\nu}=\frac{1}{\sqrt{-g}}\frac{\delta}{\delta g^{\mu\nu}}\int dx^4 \sqrt{-g}R^{\alpha\beta\gamma\delta}R_{\alpha\beta\gamma\delta}$. However, because of the topological property \eqref{Eq:Conventions.Topological} in $n=4$, one can easily show that the new HD tensor can be written in terms of the previously defined ones as follows: $H_{\mu\nu}=4\leftidx{^{(2)}}{\!H}_{\mu\nu}-\leftidx{^{(1)}}{\!H}_{\mu\nu}$. Using this property to compute the functional derivative of the Weyl tensor squared defined in
\eqref{Eq:Conventions.EC2} we find
\begin{equation}\label{Eq:Conventions.deltaC2}
\frac{1}{\sqrt{-g}}\frac{\delta C^2}{\delta g^{\mu\nu}}=H_{\mu\nu}-2\leftidx{^{(2)}}{\!H}_{\mu\nu}+\frac13 \leftidx{^{(1)}}{\!H}_{\mu\nu}= 2 \leftidx{^{(2)}}{\!H}_{\mu\nu}-\frac23 \leftidx{^{(1)}}{\!H}_{\mu\nu}\,.
\end{equation}
The previous relation implies that for conformally flat spacetimes (like FLRW), for which the Weyl tensor vanishes identically, the basic two HD tensors $\leftidx{^{(2)}}{\!H}_{\mu\nu}$ and $ \leftidx{^{(1)}}{\!H}_{\mu\nu}$ are not independent:
\begin{equation}\label{Eq:Conventions.ConfFlat}
\leftidx{^{(2)}}{\!H}_{\mu\nu}=\frac13\,\leftidx{^{(1)}}{\!H}_{\mu\nu}\,.
\end{equation}
We remark that the two HD tensors $,\leftidx{^{(1)}}{\!H}_{\mu\nu}$ and $\leftidx{^{(2)}}{\!H}_{\mu\nu}$, are conserved tensors, namely they satisfy the local conservation laws
\begin{equation}\label{Eq:Conventions.conservH1H2}
\nabla^\mu\, \leftidx{^{(1)}}{\!H}_{\mu\nu}=0\,, \ \ \ \ \ \ \ \ \ \ \ \ \ \ \ \nabla^\mu\, \leftidx{^{(2)}}{\!H}_{\mu\nu}=0\,.
\end{equation}
These laws are fulfilled identically and independently of each other, even if the background geometry is non-conformally flat and the relation \eqref{Eq:Conventions.ConfFlat} is not satisfied. This should not be surprising for the following reason. Tensors $\leftidx{^{(1)}}{\!H}_{\mu\nu}$ and $\leftidx{^{(2)}}{\!H}_{\mu\nu}$ represent the most general modification of the \textit{l.h.s.} of Einstein's equations in the presence of HD terms. In fact, the metric variation of the total action \eqref{Eq:Conventions.Totalaction} produces the generalized Einstein's equations:
\begin{equation}\label{Eq:Conventions.MostGenEinstEqs}
 G_{\mu \nu}+b_1(M) \leftidx{^{(1)}}{\!H}_{\mu\nu}+ b_2(M) \leftidx{^{(2)}}{\!H}_{\mu\nu}=8\pi G(M) \left\langle T^{\rm tot}_{\mu\nu}\right\rangle_{\rm ren}(M)\,.
\end{equation}
One would expects that $\leftidx{^{(1)}}{\!H}_{\mu\nu}$ and $\leftidx{^{(2)}}{\!H}_{\mu\nu}$ should not perturb the consistency between the Bianchi identity $\nabla^{\mu}G_{\mu\nu}=0$ satisfied by the Einstein tensor and the local conservation law $\nabla^{\mu}T^{\rm tot}_{\mu\nu}=0$ (where the EMT $T^{\rm tot}_{\mu\nu}$ involves all forms of energy, matter and vacuum, whether interacting or not). One can verify, of course, by explicit calculations from the above definitions that the two local conservation laws \eqref{Eq:Conventions.conservH1H2} are indeed satisfied. This fact insures that acting with $\nabla^\mu$ on both sides of \eqref{Eq:Conventions.MostGenEinstEqs} gives consistently zero. For the explicit derivation of the relations \eqref{Eq:Conventions.conservH1H2}, the following standard relation can be used:
\begin{equation}\label{Eq:Conventions.CommutationNablas}
 \left(\nabla_\nu \nabla_\mu-\nabla_\mu \nabla_\nu\right)v_\alpha =R^\sigma_{\textrm{	}\alpha \mu \nu}v_\sigma\,,
\end{equation}
which holds for any covariant vector field $v_\alpha$. In particular, for $v_\alpha=\nabla_\alpha\phi$ we find
\begin{equation}\label{Eq:Conventions.CommutationNablas2}
\left(\nabla_\nu\nabla_\mu-\nabla_\mu\nabla_\nu\right)\nabla_\alpha \phi=R^\sigma_{\
\alpha\mu\nu}\nabla_\sigma \phi\,.
\end{equation}
It shows that in curved spacetime the successive action of three $\nabla_\mu$ operators cannot be performed by commuting the last two being applied, while of course $\nabla_\nu\nabla_\mu\phi=\nabla_\mu\nabla_\nu\phi$ (because the Christoffel symbols are symmetric). The relation \eqref{Eq:Conventions.CommutationNablas2} can be used to derive the rule for commuting the nabla and box operators, which we need as well in the text:
\begin{equation}\label{Eq:Conventions.CommutationNablaBox}
\nabla_\mu \Box \phi-\Box\nabla_\mu \phi=-R_{\mu\nu}\nabla^\nu \phi\,.
\end{equation}
Additional formulas which are used in the main text involving the above HD tensors in the specific context of the FLRW metric are the following.
The  $00th$ and $11th$-components of the $\leftidx{^{(1)}}{\!H}_{\mu\nu}$ tensor in the conformally flat metric reads
\begin{equation}\label{Eq:Conventions.H100}
\leftidx{^{(1)}}{\!H}_{00}=\frac{-18}{a^2}\left(\mathcal{H}^{\prime 2}-2\mathcal{H}^{\prime \prime}\mathcal{H}+3 \mathcal{H}^4 \right)= -18 a^2\left(\dot{H}^2-2H\ddot{H}-6H^2\dot{H}\right)\,,
\end{equation}
\begin{equation}\label{Eq:Conventions.H111}
\leftidx{^{(1)}}{\!H}_{11}=-a ^2\left(108 H^2 \dot{H}+54\dot{H}^2+72H\ddot{H}+12\vardot{3}{H}\right)\,.
\end{equation}
We will also need the invariants
\begin{equation}\label{Eq:Conventions.RmunusquareBoxR}
 R^{\mu\nu} R_{\mu\nu}=\frac{12}{a^4}\left({\cH^\prime}^2+\cH^\prime\cH^2+\cH^4\right)\,,\qquad\Box R=-\frac{6}{a^4}\left(\cH^{\prime\prime\prime}-6\cH'\cH^2\right)\,,
\end{equation}
which hold good for flat three-dimensional FLRW spacetime.

For gamma matrices (in flat spacetime), the standard Dirac basis is chosen for our calculations with spin-1/2 fermions:
\begin{equation}\label{Eq:Conventions.GammaMatrices}
\gamma^0=\begin{pmatrix}
I & 0\\
0 & -I
\end{pmatrix}\qquad \gamma^k =\begin{pmatrix}
0 & \sigma_k\\
-\sigma_k & 0
\end{pmatrix}\,,
\end{equation}
where $\sigma_k$ ($k=1,2,3$)  are the usual Pauli matrices. In terms of the above $\gamma^\alpha$, the curved spacetime $\gamma$-matrices read $\underline{\gamma}^\mu(x)=e^{\mu}_{\,\alpha}(x)\gamma^\alpha $, where $e^{\mu}_{\,\alpha}(x) $ is the vierbein (cf. \hyperref[Sect:QuantizedFermion]{Sect.\,\ref{Sect:QuantizedFermion}}).

\section{Master Integrals}\label{Sect:MasterInt}

Integrals over $3$-dimensional momentum appear quite often in our calculations. For our purposes it will suffice to focus on integrals of the form
\begin{equation}\label{Eq:Conventions.MasterIntegral3}
I_3(p,Q)\equiv\int \frac{d^3 k}{(2\pi)^3}\frac{1}{\omega^p_k(Q)}=\frac{1}{2\pi^2}\int dk k^2 \frac{1}{\omega^p_k(Q)}=\frac{1}{2\pi^2}\int dk k^2\frac{1}{(k^2+Q^2)^{p/2}}\,,
\end{equation}
where $k\equiv|\bk|$, $\omega_k(Q)=\sqrt{k^2+Q^2}$ and $Q$ is an arbitrary scale. In $n-1$ spatial dimensions,
\begin{equation}\label{Eq:Conventions.DRFormula}
\begin{split}
I_{n-1}(p,Q)&\equiv \int \frac{\mu^{3-(n-1)}d^{n-1} k}{(2\pi)^{(n-1)}}\frac{1}{(k^2+Q^2)^{p/2}}=\frac{\mu^{3-(n-1)}}{(4\pi)^{(n-1)/2}}\frac{\Gamma\left(\frac{p-(n-1)}{2} \right)}{\Gamma\left( \frac{p}{2}\right)}\left(Q^2\right)^{\frac{(n-1)-p}{2}}\\
&=\frac{1}{(4\pi)^{3/2}}\frac{\Gamma\left(\frac{p-3}{2} +\epsilon\right)}{\Gamma\left( \frac{p}{2}\right)}\left(Q^2\right)^{\frac{3-p}{2}}\left(\frac{Q^2}{4\pi \mu^2}\right)^{-\epsilon}\,.
\end{split}
\end{equation}
Here $\Gamma(x)$ is Euler's $\Gamma$ function, which satisfies the functional relation $\Gamma(x+1)= x\,\Gamma (x)$. The scale $\mu$ (with natural dimension one) has been introduced such that the new integration measure $d^{n-1} k\to \mu^{2\epsilon} d^{n-1} k$ has the same dimension as $d^3k$, where $\epsilon\equiv\frac {3-(n-1)}{2}=\frac {4-n}{2}$. Of course, the limit $\epsilon\to 0$ ({\it i.e.} $n-1\to 3$) at the end of the calculation is understood. Such a limit is trivial for $p> 3$, but not so for $p\leq 3$ since in the last case poles $\sim\frac{1}{\epsilon}$ appear in the result of \eqref{Eq:Conventions.DRFormula}, which can be used to regularize the UV-divergent terms appearing in many of the integrals appearing in our calculation, see e.g. Eq.\,\eqref{Eq:QuantumVacuum.EMTFluctuations}. The limit $\epsilon\to 0$ also generates finite parts which must be carefully included. Despite the fact that the adiabatic subtraction procedure  provides overall UV-convergent integrals, as explained in detail in the main text, one can also use dimensionally regularized integrals to track the poles found in intermediate results. The following properties of the $\Gamma$ function are useful:
\begin{equation}\label{Eq:Conventions.Gammaepsilon}
\Gamma(\epsilon)=\frac{1}{\epsilon} - \gamma_E +{\cal O}(\epsilon)\,,\ \ \ \ \ \Gamma(-1+\epsilon)=-\frac{1}{\epsilon} - 1+\gamma_E +{\cal O}(\epsilon)\,,
\end{equation}
where $\gamma_E$ is Euler's constant. Using the functional definition of $\Gamma$ mentioned above, one can easily extend these formulas to parameterize the divergent behavior of $\Gamma$ around any negative integer.

A simpler version of \eqref{Eq:Conventions.DRFormula}, that may be used for explicitly convergent integrals is
\begin{equation}
I_3 \equiv \int_0^\infty \frac{k^2}{\omega_k^p (Q)}dk=\int_0^\infty \frac{k^2}{\left( k^2+Q^2 \right)^{p/2}}dk=\frac{1}{Q^{p-3}}\frac{(p-5)!!}{(p-2)!!}\,,
\end{equation}
valid for odd numbers $p \geq 5$, and !! represents the double factorial, defined recursively as $N!!=N\times(N-2)!!$ and $1!!=0!!=1$.

With respect \hyperref[Chap:QuantumVacuum]{Chap.\,\ref{Chap:QuantumVacuum}} and \hyperref[Chap:Fermions]{Chap.\,\ref{Chap:Fermions}}, the following observation is in order at this point. It is important to clarify that, in our renormalization scheme, the auxiliary 't Hooft's mass unit $\mu$ used in the above formulae plays no role and cancels out completely at the level of the final results. This is so in all the computations presented in this dissertation. The appearance of $\mu$ in intermediate steps is related to have used (optionally) dimensional regularization in some parts of our calculation. Use of DR, however, is not essential at all and it can be totally circumvented. This is shown in the calculations given in \hyperref[Appendix:Dimensional]{Appendix \,\ref{Appendix:Dimensional}}, where the regularization of the EMT is performed using DR after the results have already been obtained using the subtraction prescription in the main text. Similarly, use of DR in the effective action approach of \hyperref[Sect:EffectiveActionQFT]{Sect.\,\ref{Sect:EffectiveActionQFT}} is only for convenience, we have rederived the same results using the scale subtraction procedure, {\it i.e.} the one we have employed in \hyperref[Sect:RenormZPE]{Sect.\,\ref{Sect:RenormZPE}} and \hyperref[Sect:RenormalizedVED]{Sect.\,\ref{Sect:RenormalizedVED}}. The same is true for \hyperref[Chap:Fermions]{Chap.\,\ref{Chap:Fermions}}: The presented results do not depend on the mathematical procedure used for solving/regularize integrals and it is ultimately a matter of choice. We emphasize, however, that we did not use the MS scheme of renormalization at any point in our study of the VED, although this is of course independent of using DR as an intermediate regularization technique, if desired. In contrast, the subtracting scale $M$ remains always in our results as it is inherent to our renormalization method, no matter whether we decide to use DR in intermediate steps for regularization or just proceed to rearrange the terms of the integrands of our subtracted integrals to show by explicit calculation that the result is overall convergent.

\blankpage

\chapter{Combining adiabatic and dimensional regularization in chapter 2}\label{Appendix:Dimensional}

In this appendix, we sketch the calculation of the regularized EMT by using dimensional regularization (DR). Let us nonetheless emphasize that while we will use minimal subtraction of poles as a regularization procedure, we do \textit{not} intend to renormalize the theory with this prescription. If we would do that the renormalized vacuum energy would still exhibit the unwanted $\sim m^4$ contributions. In the following, we show that after the ARP has been performed,  the divergent integrals appearing in the intermediate calculations can be regularized through DR and then we can recover exactly the same result \eqref{Eq:QuantumVacuum.VEDscalesMandM0Final} for the renormalized VED.

\section{Dimensionally regularized ZPE in FLRW spacetime}\label{Sect:DimRegZPEinFLRW}

Next we summarize how to obtain the same expression for the renormalized VED as the one we have found in \hyperref[Sect:TotalVED]{Sect.\,\ref{Sect:TotalVED}}, but now using DR in the intermediate steps to regularize the divergent integrals.
Our common starting point is Eq.\,\eqref{Eq:QuantumVacuum.DecompositionEMT},
\begin{equation}\label{Eq:Dimensional.DR_EMT}
\langle T_{00}^{\delta \phi}\rangle (M)= \langle T_{00}^{\delta \phi}\rangle_{Div}(M)+\langle T_{00}^{\delta \phi}\rangle_{Non-Div}(M),
\end{equation}
where the divergent and non-divergent contributions are the same ones as in equations \eqref{Eq:QuantumVacuum.DivergentPart} and \eqref{Eq:QuantumVacuum.Non-DivergentPart}, respectively. The order of adiabaticity of these expressions, therefore is the same as in the calculation presented in the main text, and we shall take this fact for granted hereafter. We should remind the reader that in these expressions the WKB expansion of the modes has been performed off-shell, {\it i.e.} at an arbitrary mass scale $M$ which is generally different from the physical mass, $m$. However, at this point we take a different route for the rest of the calculation, namely we compute the divergent parts with the help of the DR formula \eqref{Eq:Conventions.DRFormula}. Next we expand in $\epsilon$ before taking the limit $\epsilon\to 0$ and leave only the $\epsilon$ dependence at the poles located at $\epsilon=0$ ({\it i.e.} $N=3$). The final result is
\begin{equation}\label{Eq:Dimensional.DivergentPartExplicitelyDR}
\begin{split}
\langle T_{00}^{\delta \phi}\rangle_{Div}(M)&=-\frac{M^4 a^2}{64\pi^2}\left[\frac{1}{\epsilon}+\frac{3}{2}-\gamma_E+\ln 4\pi +\ln \frac{\mu^2}{M^2}\right]\\
&-\frac{3M^2 \mathcal{H}^2}{16\pi^2}\left(\xi-\frac{1}{6} \right)\left[\frac{1}{\epsilon}-1-\gamma_E+\ln 4\pi +\ln\frac{\mu^2}{M^2}\right]\\
&-\frac{9}{16\pi^2 a^2}\left(\xi-\frac{1}{6}\right)^2(2\mathcal{H}^{\prime \prime}-\mathcal{H}^{\prime 2}-3\mathcal{H}^4)\left[\frac{1}{\epsilon}-\gamma_E+\ln 4\pi +\ln \frac{\mu^2}{M^2}\right]\\
&-\frac{\Delta^2 a^2 M^2}{32\pi^2}\left[\frac{1}{\epsilon}+1-\gamma_E+\ln 4\pi +\ln \frac{\mu^2}{M^2}\right]
-\frac{\Delta^4 a^2}{64\pi^2}\left[\frac{1}{\epsilon}-\gamma_E+\ln 4\pi +\ln \frac{\mu^2}{M^2}\right]\\
&-\left(\xi-\frac{1}{6}\right)\frac{3\Delta^2 \mathcal{H}^2}{16\pi^2}\left[\frac{1}{\epsilon}-\gamma_E+\ln 4\pi +\ln \frac{\mu^2}{M^2}\right]
\end{split}
\end{equation}
This equation can be conveniently split into a UV-divergent part involving the poles at $\epsilon=0$ and a finite part. Defining
\begin{equation}\label{Eq:Dimensional.Depsilon}
 D_\epsilon=\frac{1}{\epsilon}-\gamma_E+\ln 4\pi
\end{equation}
and recalling that $\Delta^2=m^2-M^2$, we obtain
\begin{equation}\label{Eq:Dimensional.split1}
\begin{split}
\langle T_{00}^{\delta \phi}\rangle_{Div}(M)&=-\frac{m^4 a^2}{64\pi^2}D_\epsilon-\frac{3 m^2\cH^2}{16\pi^2} \left(\xi-
\frac16\right) D_\epsilon
-\frac{9}{16\pi^2 a^2}\left(\xi-\frac16\right)^2 (2\mathcal{H}^{\prime \prime}-\mathcal{H}^{\prime 2}-3\mathcal{H}^4)D_\epsilon\\
&+\langle T_{00}^{\delta \phi}\rangle_{\rm FR}(M)\,.
\end{split}
\end{equation}
The UV-divergent part, in the first line, depends only on the physical mass of the particle, $m$, whereas the finite remainder (denoted with the label FR) depends both on the mass and on the renormalization point $M$:
\begin{equation}\label{Eq:Dimensional.FRi}
\begin{split}
\langle T_{00}^{\delta \phi}\rangle_{\rm FR}(M)&=-\frac{M^4 a^2}{64\pi^2}\left[\frac{3}{2}+\ln\frac{\mu^2}{M^2}\right]-\frac{3M^2 \mathcal{H}^2}{16\pi^2}\left(\xi-\frac{1}{6}\right)\left[ -1+\ln\frac{\mu^2}{M^2} \right]\\
&-\frac{\Delta^2 a^2 M^2}{32\pi^2}\left[1+\ln \frac{\mu^2}{M^2}\right]-\frac{\Delta^4 a^2 }{64\pi^2}\ln \frac{\mu^2}{M^2}
-\left(\xi-\frac{1}{6}\right)\frac{3\Delta^2 \mathcal{H}^2}{16\pi^2}\ln \frac{\mu^2}{M^2}\\
&-\frac{9}{16\pi^2 a^2}\left(\xi-\frac{1}{6} \right)^2(2\mathcal{H}^{\prime \prime}\mathcal{H}-\mathcal{H}^{\prime 2}-3\mathcal{H}^4)\ln\frac{\mu^2}{M^2}\\
&=\frac{a^2}{128\pi^2}\left(M^4-4m^2M^2-2m^4\ln \frac{\mu^{2}}{M^2}\right)\\
&+\frac{3}{16\pi^2}\left(\xi-\frac{1}{6}\right) \cH^2\left(M^2 -m^2\ln \frac{\mu^{2}}{M^2}\right)\\
&-\frac{9}{16\pi^2 a^2}\left(\xi-\frac{1}{6} \right)^2(2\mathcal{H}^{\prime \prime}\mathcal{H}-\mathcal{H}^{\prime 2}-3\mathcal{H}^4)\ln\frac{\mu^2}{M^2}\,,
\end{split}
\end{equation}
where in the second equality we have used once more $\Delta^2=m^2-M^2$. At this stage, the DR procedure carries a dependence on the artificial mass scale $\mu$. However, in our case $\mu$ will play no role since we are not just aiming at a conventional renormalization based on minimal subtraction, so $\mu$ serves only as an auxiliary variable which will eventually disappear from the renormalized result. We should emphasize that the relevant renormalization scale in our calculation is not $\mu$ but $M$. DR is used here only as a technique to display explicitly the divergences of the EMT and to enable their subtraction with the conventional counterterm procedure.

\section{Counterterms}\label{Sect:Counterterms}

While the calculation can be fully carried out without any use of DR, provided one defines a properly subtracted EMT from the beginning with the ARP (cf. \hyperref[Sect:RenormZPE]{Sect.\,\ref{Sect:RenormZPE}}), we follow now the more conventional approach. Thus we remove the unphysical divergences of the EMT by generating counterterms from the coupling constants present in the extended gravitational action with the HD terms. The modified Einstein's equations read formally as in Eq.\,(\ref{Eq:QuantumVacuum.MEEs}) but carrying the bare couplings, {\it i.e.} couplings which are formally UV-divergent and scale independent:
\begin{equation}\label{Eq:Dimensional.EinsteinField}
\frac{1}{8\pi G_N}G_{\mu \nu}+\rho_\Lambda g_{\mu \nu}+a_1 H_{\mu \nu}^{(1)}=\langle T_{\mu \nu}^{\delta \phi} \rangle +T_{\mu \nu}^{\phi_{\rm b}}\,.
\end{equation}
We will focus on the $00$-component of this equation since we are interested in the ZPE.

Following the standard renormalization procedure, we split each of the bare couplings on the \textit{l.h.s} of the above equation into the renormalized term (which depends on the renormalization point $M$), and a counterterm (which does not depend on $M$):
\begin{equation}\label{Eq:Dimensional.splitcounters}
\begin{split}
G_N^{-1}&= G_N^{-1}(M)+\delta G_N^{-1},\\
\rho_\Lambda&=\rho_\Lambda(M)+\delta\rho_\Lambda,\\
a_1&=a_1(M)+\delta a_1.
\end{split}
\end{equation}
%
We define the counterterms such that we can subtract the universal terms $\gamma_E$ and $4\pi$ of the DR procedure alongside with the poles, as it is conventional in the modified MS (or $\overline{\rm MS}$)\,\cite{donoghue2014dynamics,Manohar:1996cq}. That is why we have defined the quantity $D_\epsilon$ in Eq.\,\eqref{Eq:Dimensional.Depsilon}.
As we can see, three `primitive divergences' appear in the unrenormalized form of the EMT, which are proportional to $\sim m^4$, $\sim m^2 (\xi-1/6)$ and $(\xi-1/6)^2$, respectively. These can be cancelled by the corresponding counterterms generated from the bare couplings in Eq.\,\eqref{Eq:QuantumVacuum.splitcounters}, {\it i.e.} the counterterms can now be precisely used to cancel the three divergent quantities proportional to $D_\epsilon$ in Eq.\,\eqref{Eq:Dimensional.split1}.  Using the $00$-components of the geometric tensors given in \hyperref[Appendix:Conventions]{Appendix\,\ref{Appendix:Conventions}}, they are readily found to be
\begin{equation}\label{Eq:Dimensional.counters}
\begin{split}
\delta G_N^{-1} &=-\frac{m^2}{2\pi}\,\left(\xi-\frac{1}{6}\right)\, D_\epsilon\,,\\
\delta\rho_\Lambda &=+\frac{m^4}{64\pi^2}\, D_\epsilon\,,\\
\delta a_1 &= -\frac{1}{32\pi^2}\,\left(\xi-\frac{1}{6}\right)^2\, D_\epsilon\,.
\end{split}
\end{equation}
We confirm that they depend on the physical mass $m$ and not on the renormalization point $M$.
The renormalized Einstein equation resulting from cancelling the poles with the counterterms take on the same form as in Eq.\,\eqref{Eq:QuantumVacuum.MEEs}, in which the couplings are now the renormalized ones and explicitly depend on the mass scale $M$. The $00$-component reads
\begin{equation}\label{Eq:Dimensional.Einsteinequationmuind}
\frac{1}{8\pi G_N (M) } G_{00}+\rho_\Lambda (M) g_{00}+a_1 (M) H_{00}^{(1)}= \langle \widetilde{{T}_{00}^{\delta \phi}}\rangle (M)+T_{00}^{\phi_{\rm b}},
\end{equation}
where the tilded quantity
\begin{equation}\label{Eq:Dimensional.T00RenDR}
 \langle \widetilde{{T}_{00}^{\delta \phi}}\rangle (M)= \langle T_{00}^{\delta \phi}\rangle_{\rm FR}(M)+\langle T_{00}^{\delta \phi}\rangle_{Non-Div}(M)\,.
\end{equation}
is the finite part left of the EMT \eqref{Eq:Dimensional.DR_EMT} after removing the poles. Being finite we might be tempted to call it (provisionally) the renormalized ZPE, but in fact is not our final renormalized expression.  Some further insight on it can be achieved by considering the term labelled FR, which is given by \eqref{Eq:Dimensional.FRi}.
If we apply the limit $a=1$ (so that $\cH$ and all its derivatives vanish) and project the result on-shell ($M=m$, hence $\Delta=0$), the whole expression \eqref{Eq:Dimensional.FRi} shrinks to just one of the equivalent forms
\begin{equation}\label{Eq:Dimensional.ZPEflat}
 \left. \langle T_{00}^{\delta \phi}\rangle_{\rm FR}(m)\right|_{\rm Minkowski}=-\frac{m^4}{64\pi^2}\left[\frac{3}{2}+\ln\frac{\mu^2}{m^2}\right]=\frac{m^4}{128\pi^2}\left[-3-2\ln\frac{\mu^2}{m^2}\right]=\frac{m^4}{64\pi^2}\left[\ln\frac{m^2}{\mu^2}-\frac{3}{2}\right]\,.\phantom{XX}\
\end{equation}
This is nothing but the standard (one-loop) ZPE in flat spacetime, namely it is the renormalized form of the UV-divergent integral \eqref{Eq:QuantumVacuum.Minkowski} within the $\overline{\rm MS}$. As we can see, Eq.\,\eqref{Eq:Dimensional.ZPEflat} brings a explicit dependence on $\mu$ and above all it grows as the quartic power of the mass of the field. Because the total VED is the sum of \eqref{Eq:Dimensional.ZPEflat} plus the renormalized $\rL$ -- cf. Eq.\,\eqref{Eq:QuantumVacuum.EMTvacuum} -- we are led to face a huge contribution from the quartic term $\sim m^4$ (for virtually every known particle, except a very light neutrino), which amounts to a large fine-tuning between these two quantities. This is odd, in fact unacceptable. As discussed in detail in\,\cite{Sola:2013gha,Sola:2014tta,Sola:2011qr}, the flat space formula carries indeed the core of the cosmological constant problem\,\cite{weinberg1989cosmological} and the curved spacetime calculation just inherits it at this point, but it does not aggravate it further. Thus, not surprisingly the subtraction of this part leaves a well-behaved result (cf. \hyperref[Sect:SubtractMinkowski]{Sect.\,\ref{Sect:SubtractMinkowski}}). However, let us continue with our renormalization procedure and evade this conundrum within the present context.

\section{Renormalized ZPE and absence of \texorpdfstring{$\sim m^4$}{m4} contributions}\label{SubSect:RenZPEabscencem4}

The problem stems from the tilded definition of the renormalized EMT given in \eqref{Eq:Dimensional.T00RenDR}, which is just a variant of the $\overline{\rm MS}$-renormalized one, although carrying off-shell $\Delta^2$-corrections. However, a well-defined expression can be obtained if we call back anew our definition of renormalized EMT as in \eqref{Eq:QuantumVacuum.EMTRenormalizedDefinition} of the main text. The prescription amounts to take the on-shell value (at the physical mass $m$) and subtract from it the terms up to 4th adiabatic order at some arbitrary mass scale $M$. This provides automatically an overall finite result, as we have proven in the main text without using DR. Taking into account that in this alternative procedure we have already removed the poles appearing in the intermediate steps with the help of DR, it suffices to perform the aforementioned subtraction directly with the finite expression \eqref{Eq:Dimensional.T00RenDR}:
\begin{eqnarray}\label{Eq:Dimensional.EMTRenormalized2}
\langle T_{00}^{\delta \phi}\rangle_{\rm Ren}(M)&=& \langle \widetilde{{T}_{00}^{\delta \phi}}\rangle(m) - \langle \widetilde{{T}_{00}^{\delta \phi}}\rangle (M)\nonumber\\
&=&\langle T_{00}^{\delta \phi}\rangle_{\rm FR}(m)-\langle T_{00}^{\delta \phi}\rangle_{\rm FR}(M)+\langle T_{00}^{\delta \phi}\rangle_{\rm Non-div}(m)-\langle T_{00}^{\delta \phi}\rangle_{\rm Non-div}(M)\nonumber\\
&=&\langle T_{00}^{\delta \phi}\rangle_{\rm FR}(m)-\langle T_{00}^{\delta \phi}\rangle_{\rm FR}(M)-\left(\xi-\frac{1}{6}\right)\frac{3\Delta^2 \cH^2}{8\pi^2}+\dots
\end{eqnarray}
Upon some simple rearrangements, it finally yields
\begin{equation}\label{Eq:Dimensional.RenormalizedFinal}
\begin{split}
\langle T_{00}^{\delta \phi}\rangle_{\rm Ren}(M)&=\frac{a^2}{128\pi^2 }\left(-M^4+4m^2M^2-3m^4+2m^4 \ln \frac{m^2}{M^2}\right)\nonumber\\
&-\left(\xi-\frac{1}{6}\right)\frac{3 \mathcal{H}^2 }{16 \pi^2 }\left(m^2-M^2-m^2\ln \frac{m^2}{M^2} \right)\\
&+\left(\xi-\frac{1}{6}\right)^2 \frac{9\left(2 \mathcal{H}^{\prime \prime} \mathcal{H}- \mathcal{H}^{\prime 2}- 3 \mathcal{H}^{4}\right)}{16\pi^2 a^2}\ln \frac{m^2}{M^2}+\dots \\
\end{split}
\end{equation}
The $\mu$-dependence has cancelled at this point, and as we can see this equation turns out to be exactly the same one as in Eq.\,\eqref{Eq:QuantumVacuum.ExplicitRenormalized}. Therefore, from this point onwards we can reproduce the same renormalized VED \eqref{Eq:QuantumVacuum.VEDscalesMandM0Final}, just starting from \eqref{Eq:QuantumVacuum.Totalrhovac} and subtracting its value at the two scales $M$ and $M_0$. Once more the result is that the VED at the scale $M$ can be related with its value at another scale $M_0$ without receiving any contribution from the quartic values of the mass scales or of the mass of the particle. Thus, on using this renormalization procedure we can get rid of the dependence on the quartic powers of the masses as well as on the spurious DR parameter $\mu$.

The lesson we can learn is the following. While the mere $\overline{\rm MS}$ renormalization of the VED (based on using DR together with the subtraction of the poles by the counterterms) leaves a result which is explicitly dependent both on the artifical DR scale $\mu$ and on the quartic powers of the masses\,\cite{Kohri:2016lsj,Kohri:2017iyl}, the extended ARP technique\,\cite{Ferreiro:2018oxx} allows to relate the renormalized quantities at different scales. With detailed calculations, which we have presented here through two different approaches (one of them not using DR at all), we have shown that we can avert the mentioned problems associated to a mere removal of the poles by the counterterms. The common final result emerging from the two procedures is an expression for the running of the renormalized EMT in a FLRW background as a function of the Hubble rate, thus allowing to trace the VED evolution throughout the cosmic history. The result we have obtained is indeed much closer in spirit to the renormalization group approach of the RVM, cf.\,\cite{Sola:2013gha,Sola:2014tta,Sola:2011qr,Sola:2015rra} and references therein -- particularly\,\cite{Shapiro:2000dz,Shapiro:2003kv,Sola:2007sv,Shapiro:2009dh} -- in which such mild evolution of the vacuum energy density in terms of (even) powers of the Hubble rate was predicted on very general grounds. Here we have provided for the first time a detailed account from explicit QFT calculations under an appropriate renormalization scheme leading to a possible physical interpretation of the results. The outcome is Eq.\,\eqref{Eq:QuantumVacuum.VEDscalesMandM0Final}.

We should perhaps repeat once more that such relation is \textit{not} a prediction of the value of the CC and in general of the VED, as this is out of the scope of renormalization theory. Every renormalization calculation needs a set of renormalization conditions. Behind these renormalization conditions there is a set of physical (and hopefully known) inputs and from these observational inputs we can predict other physical results. In the present instance, this means that given the VED at one scale (entailing a physical input) we can predict its value at another scale. What is, however, distinctive in the kind of calculation we have presented here is the fact that the connection between the renormalized values of the VED at different points appears smooth enough, {\it i.e.} it does not involve $\sim m^4$ terms, which are usually very large for ordinary particle masses in the standard model of particle physics (let alone in GUT's) and this suggests that no fine tuning is actually involved.

The unsatisfactory status of the $m^4$ terms in cosmology is very similar to the hierarchy problem associated to the $m^2$ terms in ordinary gauge theories\,\cite{Veltman:1980mj}, but even worse in magnitude. In stark contrast to the usual situation with these terms, in the approach we have outlined we do not need to call for special cancellations (fine-tuning) among the various $m^4$ contributions from different particles, such as e.g. when using the Pauli sum rules -- see \,\cite{Visser:2016mtr} for a detailed discussion --, nor to invoke the existence of emergent scales or very small dimensionless parameters suppressing the undesired effects, see e.g.\,\cite{Bjorken:2001yv,Bjorken:2001pe,Bjorken:2010qx,Jegerlehner:2018zxm,Jegerlehner:2013cta,Sola:2016our,Bass:2020nrg} for a variety of contexts of this sort. The problem is fixed here automatically by the renormalization process itself that we have used.

\blankpage

\chapter{Running vacuum and gravitational coupling in the RVM presented in chapter 2}\label{Appendix:Abis}

In this appendix, we provide calculational details on the formulas for the running vacuum and gravitational coupling introduced in the main text, and their interrelationship. Recall that $ \rv(M)$ is an abridged notation for $\rv=\rv(M,H, \dot{H},\ddot{H},...)$, {\it i.e.} the vacuum energy density (VED), which is a function not only of the scale $M$ but also of the Hubble rate and its time derivatives.  The value of $H=H(t)$ defines an expansion history time $t$. When compared with our current cosmic time, $t_0$, the difference $t_0-t$ defines our lookback time to the events occurring around the expansion history epoch $H$. It is advisable to make the original shorthanded notation a bit more explicit for the kind of discussion in this appendix. Rather than denoting the renormalized value of the VED at the scale $M$ for a fixed expansion rate $H$ (and corresponding time derivatives) by just $\rv(M)$ , we will use $\rv(M,H)$. The second argument denotes generically all the dependency in $H,\dot{H},\ddot{H},\dots$ The values of $M$ and $H$ are independent, of course, but a selected choice of the renormalization point $M$ near $H$ corresponds to choose the RG scale around the characteristic energy scale of FLRW spacetime at a given moment, and hence it should have more physical significance. This is actually in analogy with the standard practice in ordinary gauge theories, where the choice of RG scale is usually made near the typical energy of the process. For the FLRW Universe, the natural choice for the process of expansion is $M=H$ and we will see it is consistent. In what follows we derive the `low energy' form of the VED along these lines. Subsequently we will focus on the running gravitational coupling $G(M)$ and its relation with the running $\rv(M)$.

\section{Running VED}\label{Sect:Abis1}

The expression for VED at the scale $M$ for a given expansion history time $H$  is provided by our renormalization procedure and it is given by Eq.\,\eqref{Eq:QuantumVacuum.RenVDEexplicit}. This expression contains the contributions from all the possible adiabatic orders up to the limit of the asymptotic expansion. Suppose, however, that we consider the renormalized VED at a given expansion history time $H$ for different values of the renormalization scale, say $M$ and $M_0$. The difference of renormalized VED values at these scales at a fixed $H$ can be computed in an exact way, see Eq.\,\eqref{Eq:QuantumVacuum.VEDscalesMandM0Final}. The exactness of such formula stems from the fact that the renormalization scale dependence of the EMT ({\it i.e.} the $M$-dependence) can only be carried by the terms that are originally divergent (those up to $4th$ adiabatic order). The renormalized EMT at the scales $M$ and $M_0$ at fixed cosmic time (hence at fixed $H$) is obtained upon subtracting the corresponding on-shell value at these respective scales, as explained in \hyperref[Sect:RenormZPE]{Sect.\,\ref{Sect:RenormZPE}}. Therefore, the difference of renormalized VED values at $M$ and $M_0$ is free from all of the finite contributions from $6th$ adiabatic order and higher. Only ${\cal O}\left(H^2\right)$ and ${\cal O}\left(H^4\right)$ ({\it i.e.} second and fourth adiabatic orders, respectively) remain, as it is manifest in Eq.\,\eqref{Eq:QuantumVacuum.VEDscalesMandM0Final}. However, despite of the fact that such result is exact, we wish to focus on lookback times accessible to observations, hence with values of $H$ which are moderate enough for the ${\cal O}\left(H^4\right)$ terms to be negligible. The desired difference between $\rv(M,H)$ and $\rv(M_0,H)$ within our lookback observational range therefore reads
\begin{equation}\label{Eq:RVM.RenVDECurrentUniv}
\begin{split}
&\rv(M,H)-\rv(M_0,H)= \rho_\Lambda (M)- \rho_\Lambda (M_0)\\
&+\frac{1}{128\pi^2 }\left(-M^4+M_0^4+4m^2(M^2-M_0^2)-2m^4 \ln \frac{M^2}{M_0^2}\right)\\
&+\left(\xi-\frac{1}{6}\right)\frac{3 \mathcal{H}^2 }{16 \pi^2 a^2}\left(M^2-M_0^2-m^2\ln \frac{M^2}{M_0^2} \right)+\cdots\\
&= \left(\xi-\frac{1}{6}\right)\frac{3 {H}^2 }{16 \pi^2}\left(M^2-M_0^2-m^2\ln \frac{M^2}{M_0^2} \right)+\cdots
\end{split}
\end{equation}
where the dots denote the kind of neglected contributions mentioned above. As we know, the first two terms in the above formula cancel against each other thanks to the relation \eqref{Eq:QuantumVacuum.SubtractionrL}. The obtained result is given, of course, by the ${\cal O}\left(H^2\right)$ part of Eq.\,\eqref{Eq:QuantumVacuum.VEDscalesMandM0Final}.

Similarly, from Eq.\,\eqref{Eq:QuantumVacuum.RenVDEexplicit} we may also find the difference between the values of the VED corresponding to two accessible lookback times for a given renormalization point $M$:
\begin{equation}\label{Eq:RVM.DiffHH0Mfix}
\begin{split}
\rv(M,H)-\rv(M,H_0)=3\frac{\left(\xi-\frac{1}{6}\right)}{16\pi^2}(H^2-H_0^2)\left(M^2-m^2+m^2\ln \frac{m^2}{M^2}\right)+\cdots
\end{split}
\end{equation}
In this expression we have disregarded not only the ${\cal O}\left(H^4\right)$ terms but also the higher order ones (which are indeed present here, in contrast to Eq.\,\eqref{Eq:RVM.RenVDECurrentUniv}), as they entail no significant contribution at present. They can be important only for the early Universe, e.g. during the inflationary regime, see \hyperref[SubSect:RVMInflation]{Sect.\,\ref{SubSect:RVMInflation}}.

We may as well compute the scaling evolution of the VED when we change both the cosmic times and the renormalization points. Keeping our focus on cosmic epochs $H$ and $H_0$ accessible to our observations, the result immediately follows from Eq.\,\eqref{Eq:QuantumVacuum.RenVDEexplicit} upon neglecting the ${\cal O}\left(H^4\right)$ terms and higher:
\begin{equation}\label{Eq:RVM.DiffHH0MM0}
\begin{split}
\rv(M,H)-\rv(M_0,H_0)&
=\frac{3\left(\xi-\frac{1}{6}\right)}{16\pi^2}\left[H^2\left(M^2-m^2+m^2\ln\frac{m^2}{M^2}\right)\right.\\
&\left.-H_0^2\left(M_0^2-m^2+m^2\ln\frac{m^2}{M_0^2}\right)\right]+\cdots\,,
\end{split}
\end{equation}
where again the first two terms on the RHS of Eq.\,\eqref{Eq:RVM.RenVDECurrentUniv} are involved here, but cancel each other for the aforementioned reasons.
Finally, let us consider what should be the physical (measurable) difference between the VED values at different epochs of the cosmic evolution within our observational range. According to our prescription, choosing the renormalization point $M$ near $H$ (and hence bringing the RG scale near the characteristic energy scale of the FLRW spacetime at the given epoch) ought to be the most suited physical choice in consonance with the usual practice based on selecting the RG scale choice near the typical energy of the process in particle physics.  As indicated, in our case the `process' is nothing but the cosmic expansion of the Universe at a given epoch. Thus, to compute the scaling evolution of the VED in the span mediating in between the two cosmic epochs $H$ and $H_0$, follows directly from Eq.\,\eqref{Eq:RVM.DiffHH0MM0} upon picking out the renormalization points $M$ and $M_0$ at precisely the values of the Hubble rate in those epochs, respectively: $M=H$ and $M_0=H_0$. Defining for convenience $\rv(H)\equiv\rv(M=H,H)$ and similarly $\rv(H_0)\equiv \rv(H_0,H_0)$, and neglecting as always the higher order terms ${\cal O}\left(H^4\right)$ , we find
\begin{equation}\label{Eq:RVM.DiffVEDphys}
\begin{split}
\rv(H)-\rv(H_0)&=\frac{3\left(\xi-\frac{1}{6}\right)}{16\pi^2}\bigg[H^2\left(H^2-m^2+m^2\ln\frac{m^2}{H^2}\right)\\
&\phantom{xxxxxxxxxx}-H_0^2\left(H_0^2-m^2+m^2\ln\frac{m^2}{H_0^2}\right)\Bigg]+\cdots\\
& \simeq \frac{3\left(\xi-\frac16\right)m^2 }{16\pi^2}\left[-\left(H^2-H_0^2\right)+H^2\ln\frac{m^2}{H^2}-H_0^2\ln\frac{m^2}{H_0^2}\right]\\
&=\frac{3\left(\xi-\frac16\right)m^2 }{16\pi^2}\left[-1+\ln \frac{m^2}{H^2}-\frac{H_0^2}{H^2-H_0^2}\ln \frac{H^2}{H_0^2}\right] \left(H^2-H_0^2\right)\,.
\end{split}
\end{equation}
The previous formula shows that there is in effect a `running' or change of the VED from $H_0$ to $H$. Notice that if $m$ is an ordinary particle mass (e.g. within the standard model of particle physics) the running would be very small. Suppose, however, that $m$ is a particle mass near some GUT scale, then it is natural to measure its value in units of the Planck mass $\mpl$ and factor out the ratio $m/\mpl$. We do this in defining the effective running parameter
\begin{equation}\label{Eq:RVM.nueff2}
\nueff(H)\equiv\frac{1}{2\pi}\,\left(\xi-\frac16\right)\,\frac{m^2}{\mpl^2}\left(-1+\ln \frac{m^2}{H^{2}}-\frac{H_0^2}{H^2-H_0^2}\ln \frac{H^2}{H_0^2}\right)\,.
\end{equation}
The running VED formula \eqref{Eq:RVM.DiffVEDphys} can now be written in a rather compact form as follows:
\begin{equation}\label{Eq:RVM.RVMcanonical}
\rv(H)\simeq \rvo+\frac{3\nueff(H)}{8\pi}\,(H^2-H_0^2)\,\mpl^2=\rvo+\frac{3\nueff(H)}{8\pi G_N}\,(H^2-H_0^2)\,,
\end{equation}
where $\rv(H_0)$ is identified with today's VED value, $\rvo$, and $G_N$ is assumed to be the currently measured value of the gravitational constant.
As a matter of fact, $\nueff(H)$ in \eqref{Eq:RVM.nueff2} is not a parameter, of course, since it is a function of $H$. However, it varies very slowly with the Hubble rate. The last term of \eqref{Eq:RVM.nueff2} is logarithmic and becomes quickly suppressed for increasingly large values of $H$ above $H_0$, whereas the second term furnishes (on account of $\ln \frac{m^{2}}{H^2}\gg1$) the dominant contribution to the effective running parameter:
\begin{equation}\label{Eq:RVM.nueffAprox2}
\nueff(H)\simeq\frac{1}{2\pi}\,\left(\xi-\frac{1}{6}\right)\,\frac{m^2}{\mpl^2}\ln\frac{m^2}{H^2}\,.
\end{equation}
In the approximation $H=H_0$ (valid to within a few percent in the accessible part of the expansion history for large $m$), it just renders Eq.\,\eqref{Eq:QuantumVacuum.nueffAprox} in the main text, and Eq.\,\eqref{Eq:RVM.RVMcanonical} is nothing but the canonical form of the VED for the running vacuum model (RVM), as given in the main text in Eq.\,\eqref{Eq:QuantumVacuum.RVM2}. In it, the running parameter is treated essentially as a constant. 

In actual fact, for a large stretch of the recent Universe we can just set $H=H_0$ in Eq.\,\eqref{Eq:RVM.nueffAprox2} since it differs less than $7\%$ through the entire period from now up to the decoupling time. Even if $\xi$ is not known, the ratio $m^2/\mpl^2\ll1$ in the prefactor of \eqref{Eq:RVM.nueff2} is very small, so we can expect that $\nueff$ is essentially a tiny quantity with a very mild variation with $H$. From the foregoing, it follows that it can be treated to a good approximation as a small parameter within the observable Universe. Notice, however, that for  $m$ large, say of order of a GUT scale $M_X\sim 10^{16}$ GeV, we have $m^2/\mpl^2\sim 10^{-6}$, which is not hopelessly small. Since $\xi$ is, in principle, arbitrary and we have in general a large multiplicity of heavy scalar particles in a typical GUT, the effective value of $\nueff$ can not be excluded to be in the small but sizeable range $10^{-4}-10^{-3}$\,\cite{Sola:2007sv}. This theoretical expectation is actually corroborated by the phenomenological analysis. The RVM has been fitted to the data and the obtained results for $\nueff $ lie in expected ballpark of $\sim 10^{-3}$,  see e.g.\,\cite{Sola:2017znb,SolaPeracaula:2016qlq,SolaPeracaula:2017esw} and\,\cite{SolaPeracaula:2021gxi}.

\section{Time versus scaling evolution of the VED}\label{Sect:appendixAbisbis1}

To further illustrate the meaning and consistency of the above formulas for the running VED, it is interesting to compare the scaling versus time evolution laws of the VED at the differential level. The former is, of course, determined by the $\beta$-function \eqref{Eq:QuantumVacuum.RGEVED1} of the VED, whereas the latter can be computed as follows. For two given expansion epochs $H$ and $H_0$, the time evolution is determined by Eq.\,\eqref{Eq:RVM.DiffVEDphys}, or equivalently by \eqref{Eq:RVM.nueff2} and \eqref{Eq:RVM.RVMcanonical}. However, we would like this result for an infinitesimal change of $H$ around $H=H_0$, which means to compute the derivative of $\rv(H)$ with respect to $H$ at $H=H_0$, or, for convenience, the logarithmic derivative $d\rv(H)/d\ln H=H d\rv(H)/dH$. The result follows from Eq.\,\eqref{Eq:RVM.DiffVEDphys}.
We find
\begin{equation}\label{Eq:RVM.DiffLimiting}
\begin{split}
H_0\frac{d\rv(H_0)}{dH}=
\left(\xi-\frac{1}{6}\right)\frac{3 H_0^2 m^2}{8 \pi^2 }\left(-2+\ln\frac{m^2}{H_0^2}\right)\,.
\end{split}
\end{equation}
This equation can be written in an approximate way as follows:
\begin{equation}\label{Eq:RVM.ApproxDiffLimiting}
\begin{split}
H_0\frac{d\rv(H_0)}{dH}\simeq \left(\xi-\frac{1}{6}\right)\frac{3 H_0^2 m^2}{8 \pi^2 }\ln\frac{m^2}{H_0^2}=\frac{3\nueff }{4\pi} \,\mpl^2 H_0^2\,.
\end{split}
\end{equation}
where in the last step we used $\ln\frac{m^2}{H^2}\gg 1$ and took the approximate expression \eqref{Eq:QuantumVacuum.nueffAprox} for the coefficient $\nueff$.
It is also instructive to obtain the same result using the chain rule to compute the total derivative with respect to $M$ and set $M=H_0$ at the end of the calculation, as in this way the role of the $\beta$-function for the VED, $\beta_{\rv}$, becomes manifest:
\begin{equation}\label{Eq:RVM.DerivTotal}
\begin{split}
M\frac{d\rv(M,H)}{dM}=&M\left(\frac{\partial\rv}{\partial M}+\frac{\partial\rv}{\partial H}\frac{\partial H}{\partial M}\right)=\beta_{\rv}+M\frac{\partial\rv}{\partial H}\frac{\partial H}{\partial M}\,,
\end{split}
\end{equation}
or, more explicitly, for $H=H_0$:
\begin{equation}\label{Eq:RVM.DerivTotal2}
\begin{split}
M\frac{d\rv(M,H_0)}{dM}&=\left(\xi-\frac{1}{6}\right)\frac{3 H_0^2 }{8 \pi^2 }\left(M^2-m^2\right)\\
&+M\left(\xi-\frac16\right)\frac{3H_0}{8\pi^2}\left(M^2-m^2+m^2\ln\frac{m^2}{M^2}\right)\frac{\partial H_0}{\partial M}\, ,
\end{split}
\end{equation}
where both for the VED expression \eqref{Eq:QuantumVacuum.RenVDEexplicit} and for its $\beta$-function \eqref{Eq:QuantumVacuum.RGEVED1} we used the relevant ${\cal O}(H^2)$ terms only, and of course we borrowed the $\beta$-function for the renormalized coupling $\rL(M)$, as given in Eq.\,\eqref{Eq:QuantumVacuum.BetaFunctionrL}. Setting at this point $M=H_0$ and dropping the terms higher than ${\cal O}(H^2)$ we strike once more Eq.\,\eqref{Eq:RVM.DiffLimiting}. From the latter we can see that the sign of the total variation of the VED is given by the sign of $\xi-\frac16$ and hence also by the sign of $\nueff$ , see \eqref{Eq:RVM.ApproxDiffLimiting}. This is in full accordance with the result \eqref{Eq:QuantumVacuum.RVM2}. In particular, we can see from the last derivation that the total evolution of the VED is dominated by the variation of $\rv$ with $H$ rather than with the scale $M$ (before setting $M=H$). When we set $M=H$ at the end, the dominant term is that one carrying $\partial\rv/\partial H$ in \eqref{Eq:RVM.DerivTotal}, whose sign is the same as that of $\nueff$. It is therefore the total, rather than just the partial, derivative with respect to $M$ what matters for the study of the physical evolution of the VED. This is a consequence of the time dependence of $M$ in cosmology, in contrast to the situation in ordinary gauge theories. Writing the leading term of Eq.\,\eqref{Eq:QuantumVacuum.RGEVED1}  at low energies as
\begin{equation}\label{Eq:RVM.BetaFuncoeff}
\begin{split}
\beta_{\rv}=\frac{3\,\bv}{4\pi}\, H_0^2 m^2\,\ \ \ \ \ \ \ \ \ \ \ \ \ \ \bv\equiv -\frac{1}{2\pi}\left(\xi-\frac16\right)\,,
\end{split}
\end{equation}
we find from \eqref{Eq:RVM.nueffAprox2} the approximate formula relating $\nueff$ with the coefficient $ \bv$ of the $\beta$-function for the running VED:
\begin{equation}\label{Eq:RVM.DTotalversusBeta}
\nueff=- \bv\,\frac{m^2}{\mpl^2}\,\ln\frac{m^2}{H_0^2}\,.
\end{equation}
The sign between $\nueff$ and $\bv$ reflects the mentioned antagonism between the variations of the VED with $H$ and with $M$ before the latter is set equal to the former. Recall that $H_0$ here may be the current value of the Hubble rate, but for that matter it can represent any point of the expansion history at low energies. As noted in the previous section, $\nueff$ remains essentially constant since the change of $\ln\frac{m^2}{H^2}$ is bound within a few percent from now until recombination.

Finally, it is also instructive to derive once more the above result using a third alternative procedure, as in this way we may crosscheck our formulas from different perspectives. In particular, let us remember that Eq.\,\eqref{Eq:eos.NonConserVED1} provides direct and precious information about the time evolution of the VED and it involves the influence from the vacuum pressure, which as we know is not exactly equal to minus the vacuum density in this QFT framework ({\it i.e.} the EoS of the quantum vacuum is not exactly equal to $-1$, see \hyperref[Sect:RenormPressure]{Sect.\,\ref{Sect:RenormPressure}}). We want to show that we can test the consistency of this feature as well.  Upon inserting \eqref{Eq:eos.EoSaprox} into the term $\rv+\Pv$ of Eq.\,\eqref{Eq:eos.NonConserVED2} we find
\begin{equation}\label{Eq:RVM.ThirdTest}
\begin{split}
\dot{\rho}_{\rm vac}=\frac{\dot{M}}{M}\beta_{\rho_{\rm vac}}-3H(\rho_{\rm vac}+P_{\rm vac})=\frac{\dot{M}}{M}\beta_{\rho_{\rm vac}}-3H\frac{\xi-\frac16}{8\pi^2}\dot{H}m^2\left(1-\ln\frac{m^2}{H^2}\right)+\cdots
\end{split}
\end{equation}
where the dots just denote that we are not considering terms higher than ${\cal O}(H^2)$, which is our usual assumption for the present considerations. Using once more the $\beta$-function \eqref{Eq:QuantumVacuum.RGEVED1} of the VED to the same consistent order and setting $M=H_0$, we find
\begin{equation}\label{Eq:RVM.ThirdTest2}
\begin{split}
\dot{\rho}_{\rm vac}(t_0)=\left(\xi-\frac16\right)\frac{3m^2}{8\pi^2}H_0\dot{H_0}\left(-2+\ln\frac{m^2}{H_0^2}\right)\,.
\end{split}
\end{equation}
This equation gives the rate of change of the VED at $t=t_0$ (corresponding to $H=H_0$), which may refer to the present time or for that matter any other cosmic time. For $\nueff>0$ (respectively for $\nueff<0$) and taking into account that $\dot{H}<0$ at all (post-inflationary) times, together with the fact that $\ln \frac{m^{2}}{H^2}\gg1$, it is not difficult to see that the VED increases (resp. decreases) towards the past and decreases (resp. increases) towards the future. We can actually show that the above equation coincides with Eq.\,\eqref{Eq:RVM.DiffLimiting}. Indeed, bearing in mind that $\dot{\rho}_{\rm vac}(t)=\dot{H}\frac{d\rv}{d H}$ we find
\begin{equation}\label{Eq:RVM.ThirdTest3}
\begin{split}
H \frac{d\rv}{d H}=\frac{H}{\dot{H}} \dot{\rho}_{\rm vac}(t)=\left(\xi-\frac{1}{6}\right)\frac{3 H^2 m^2}{8 \pi^2 }\left(-2+\ln\frac{m^2}{H_0^2}\right)\,,
\end{split}
\end{equation}
which for $t=t_0$ exactly matches Eq.\,\eqref{Eq:RVM.DiffLimiting} (q.e.d.) Thus the three approaches do converge to the same result, which can be summarized as follows: for $\nueff>0$ the VED is larger in the past and behaves effectively as quintessence, whereas for $\nueff<0$ the VED is smaller in the past (equivalently, it increases towards the future) and hence it behaves effectively as phantom DE.  This is exactly the same message encoded in Eq.\,\eqref{Eq:QuantumVacuum.RVM2}. The fact that the quantum vacuum can mimic both quintessence and phantom DE shows that it may not be necessary to introduce \textit{ad hoc} scalar fields in the classical action to generate dynamical DE, since the latter could just be caused by the fact that the quantum vacuum is in permanent cosmic evolution!

\section{Running $G$}\label{Sect:appendixAbis2}

Here we elaborate further on the derivation of Eq.\,\eqref{Eq:eos.GNHfinal} in the main text. We take up the discussion from Eq.\,\eqref{Eq:eos.MixedConservationApprox1}, in which we have disregarded the HD contributions present in Eq.\,\eqref{Eq:eos.MixedConservation}, as they are negligible for the present Universe (and for that matter, also at all times after inflation). We can admit the concurrence of relativistic and non-relativistic ordinary matter components $(\rho_{\rm m},p_{\rm m})$ apart from the background scalar field $(\rho_\phi,p_\phi)$. If the former are locally conserved ($\dot{\rho}_{\rm m}+3H(\rho_{\rm m}+p_{\rm m})=0$) it is not difficult to see that the structure of \eqref{Eq:eos.MixedConservationApprox1} remains unaltered:
\begin{equation}\label{Eq:RVM.ModifiedConsLaw}
\dot{G}(M)\left(\rho_{\rm m}+\rho_\phi+\rv(H)\right)+G(H)\dot{\rho}_{\rm vac}(H)+3HG(H)\left(\rv(H)+P_{\rm vac}(H)\right)=0\,,
\end{equation}
where we have set $M=H$, according to the prescription discussed in the previous section.
Using the Friedmann equation we can get rid of the total energy density in the above expression no matter the number of components involved in it: $\rho_{\rm m}+\rho_\phi+\rv=3H^2/(8\pi G)$. In addition, we insert the expression $\rv(H)$ from \eqref{Eq:QuantumVacuum.RVM2} in the above equation, and also $\rv(H)+P_{\rm vac}(H)$ from \eqref{Eq:eos.EoSaprox}. As always we neglect of course the higher order terms ${\cal O}\left(H^4\right)$ generated in intermediate calculations, which are irrelevant after the inflationary epoch. All in all, Eq.\,\eqref{Eq:RVM.ModifiedConsLaw} can be rewritten
\begin{equation}\label{Eq:RVM.ModifiedConsLaw2}
\begin{split}
& \frac{3H^2 \dot{G}}{8\pi G}+3 G\frac{d}{dt}\left[ \rvo+\frac{1}{\kappa^2}\nueff (H)(H^2-H_0^2)\right]+3HG\frac{\left(\xi-\frac{1}{6}\right)}{8\pi^2}\dot{H}m^2\left(1-\ln \frac{m^2}{H^2}\right)\\
&=\frac{3H^2 \dot{G}}{8\pi G}+3\frac{G}{\kappa^2}\Bigg(\dot{\nu}_{\rm eff}(H) (H^2-H_0^2)+2H\dot{H}\nueff (H)+H\frac{\left(\xi-\frac{1}{6}\right)}{\pi}\dot{H}\frac{m^2}{m^2_{\rm Pl}}\left(1-\ln \frac{m^2}{H^2}\right)\Bigg)\\
&=\frac{3H^2 \dot{G}}{8\pi G}+3\frac{G}{\kappa^2}\left(\dot{\nu}_{\rm eff} (H) (H^2-H_0^2)-2H_0^2\frac{H\dot{H}}{H^2-H_0^2}\frac{\left(\xi-\frac{1}{6}\right)}{2\pi}\frac{m^2}{\mpl^2}\ln \frac{H^2}{H_0^2}\right)=0\,,
\end{split}
\end{equation}
where we recall that $\kappa^2= 8\pi G_N$ is constant, whereas $G=G(H(t))$ is the function that we wish to determine by solving the above differential equation.
Notice that to compute $\dot{\nu}_{\rm eff}(H)=\frac{d}{dt}{\nu}_{\rm eff}(H)$ we use the exact expression \eqref{Eq:RVM.nueff2} rather than just a constant approximation. At this point it is important to keep all terms since, in general, expressions that are neglected are not necessarily negligible after being differentiated. We find that there is a partial cancellation between the last two terms of \eqref{Eq:RVM.ModifiedConsLaw2} in the second line, which can be pinpointed if we use the explicit form of \eqref{Eq:RVM.nueff2}. The intermediate result at this point, prior to calculating the time derivative of the last term of \eqref{Eq:RVM.nueff2}, reads as follows:
\begin{equation}\label{Eq:RVM.MixedConservationAppox2}
\begin{split}
&\frac{3H^2 \dot{G}}{8\pi G}+3\frac{G}{\kappa^2}\left[\frac{\left(\xi-\frac{1}{6}\right)}{2\pi}\frac{m^2}{m_{\rm Pl}^2}(H^2-H_0^2)\left(-2\frac{\dot{H}}{H}-H_0^2\frac{d}{dt}\left(\frac{\ln\frac{H^2}{H_0^2}}{H^2-H_0^2}\right)\right)\right.\\
&\left.\phantom{aaaaaaaaaaaa} -\frac{H\dot{H}\left(\xi-\frac{1}{6}\right)}{\pi}\frac{m^2}{m_{\rm Pl}^2}H_0^2\frac{\ln\frac{H^2}{H_0^2}}{H^2-H_0^2}\right]=0\,.\\
\end{split}
\end{equation}
Computing the pending derivative it is possible to produce further simplifications among the various terms until reaching the following beautifully simple and compact expression:
\begin{equation}\label{Eq:RVM.MixedConservationApprox3}
\frac{d{G}}{G^2}=\frac{\left(\xi-\frac{1}{6}\right)}{\pi}m^2\frac{dH}{H}\,,
\end{equation}
in which we have replaced the time derivatives $\dot{G}=dG/dt$ and $\dot{H}=dH/dt$ by just the differentials $dG$ and $dH$ since $dt$ cancels on both sides.
Finally, integrating by simple quadrature Eq.\,\eqref{Eq:RVM.MixedConservationApprox3} from the present time $(H_0,G(H_0)=G_N)$ up to an arbitrary moment around the present $(H,G(H))$ we meet after some elementary algebra the final result
\begin{equation}\label{Eq:RVM.runGH}
G(H)=\frac{G_N}{1-\frac{\left(\xi-\frac{1}{6}\right)}{2\pi}\frac{m^2}{m^2_{\rm Pl}}\ln \frac{H^2}{H_0^2}}=\frac{G_N}{1+\bv\frac{m^2}{m^2_{\rm Pl}}\ln \frac{H^2}{H_0^2}}\,,
\end{equation}
in which $G_N$ defines the local gravity value usually associated to the inverse Planck mass squared: $ G(H_0)=G_N=1/m^2_{\rm Pl}$ (in natural units). In this way we have proven \eqref{Eq:eos.GNHfinal} (q.e.d.) The obtained formula is our QFT prediction for the physical running of the gravitational coupling with the cosmic expansion. The presence of the coefficient $\bv$ in it -- cf. Eq.\,\eqref{Eq:RVM.BetaFuncoeff} -- denotes its connection with the scaling evolution of the VED.

\chapter{Adiabatic expansion of the fermionic modes of chapter 4}\label{Appendix:AdiabExpFermModes}

In the main text (cf. \hyperref[Sect:QuantizedFermion]{Sect.\,\ref{Sect:QuantizedFermion}}) we have presented an iterative procedure which allows us to determine the two types of field modes $h_k^{\rm I}$ and $h_k^{\rm II}$ which are necessary to construct the $2$-component spinor fields. They are functions of both momentum ($k$) and conformal time ($\tau$) and have the following structure:

\begin{equation}\label{Eq:FermMode.ansatz2}
\begin{split}
&	h^{\rm I}_k(\tau)=\sqrt{\frac{\omega_k-aM}{2\omega_k}}F(\tau)\, e^{-i\int^\tau  W_k (\tilde{\tau}) d\tilde{\tau}},\\
&	h^{\rm II}_k(\tau)=\sqrt{\frac{\omega_k+aM}{2\omega_k}}G(\tau)\,e^{-i\int^\tau  W_k (\tilde{\tau}) d\tilde{\tau}}\,,
\end{split}
\end{equation}
where
\begin{equation}\label{Eq:FermMode.ExpansionF}
F \equiv1+F^{(1)}+F^{(2)}+F^{(3)}+\cdots,
\end{equation}
\begin{equation}\label{Eq:FermMode.ExpomegakansionG}
G \equiv 1+G^{(1)}+G^{(2)}+G^{(3)}+\cdots,
\end{equation}
\begin{equation}\label{Eq:FermMode.ExpansionW}
W_k \equiv \omega_k+\omega_k^{(2)}+\omega_k^{(4)}+\omega_k^{(6)}+\cdots
\end{equation}
Here $W_k$ is a real function playing an analogous role to the effective frequency introduced (with the same notation) in the scalar field case, Eq.\,\eqref{Eq:QuantumVacuum.WKBIteration}. The superscript $(n=1,2,3,...)$ indicates that the corresponding quantity is of adiabatic order $n$. The modes \eqref{Eq:FermMode.ansatz2} are constrained to satisfy the normalization condition
\begin{equation}\label{Eq:FermMode.NormalizationConditionMode}
    |h_{k}^{\rm I}|^2+|h_{k}^{\rm II}|^2=1\,.
\end{equation}
Some of the notation is similar to that of\,\cite{Landete:2013axa, Landete:2013lpa, delRio:2014cha}, although we use conformal metric and different conventions, and moreover we deal with  FLRW spacetime rather than with de Sitter (where the EMT takes a simpler form).  In addition, as explained in the main text, we perform the renormalization at the arbitrary scale $M$  (not at the on-shell point). This is important in order to test the scaling dependence of the renormalized VED, which is the main aim of our calculation.

In what follows we use the notation  $\omega_k  \equiv \sqrt{k^2+M^2 a^2}$.  Recall that unless it is explicitly noted the mass scale involved is the off-shell point $M$.  When the subtraction \eqref{Eq:Fermions.RenormalizedEMTFermion} is implemented within our  renormalization procedure, one just sets $M=m_\psi$ in the subtracted part. The mass $m_\psi$ can be conveniently expanded in even adiabatic orders as
\begin{equation}\label{Eq:FermMode.expansionm}
m_\psi=\sqrt{M^2+ \Delta^2}=M+\frac{\Delta^2}{2M}-\frac{\Delta^4}{8M^3}+\frac{\Delta^6}{16 M^5}+\dots
\end{equation}
After completing  $\ell \geq 1$ steps in the process described in \hyperref[Sect:CombinedBandF]{Sect.\,\ref{Sect:CombinedBandF}} we end up with expression \eqref{Eq:Fermions.GeneralSolution}, which depends on the following expression:
\begin{equation}\label{Eq:FermMode.GeneralSolutionProduct}
    \Omega_k \Omega_{k,1}\cdots \Omega_{k,\ell-1} = \omega_k+\omega_k^{(1)}+\omega_k^{(2)}+\dots+\omega_k^{(2\ell-1)}+\dots
\end{equation}
where $\omega_k^{(j)}$ represents a function of adiabatic order $j$.
Functions $W_k (\tau,M)$, $F(\tau,M)$ and $G(\tau,M)$ in the ansatz \eqref{Eq:FermMode.ansatz2} can be estimated with the help of this product. However the following clarifications may be necessary to better understand this process, together with the explanations already given in the main text, see \hyperref[Sect:QuantizedFermion]{Sec.\,\ref{Sect:QuantizedFermion}}:
\begin{itemize}

\item For $\ell=1$ we only have performed one iterative step, and at this point we need to deal with the square root of
\begin{equation}\label{Eq:FermMode.ExampleOmegak}
\begin{split}
\Omega_k^2=\omega_k^2-i\sigma+a^2\Delta^2=\omega_k^2-iMa^\prime+a^2\Delta^2-\frac{ia\Delta^2}{2M}\frac{a^\prime}{a}+\dots
\end{split}	
\end{equation}
The dots ``..." in \eqref{Eq:FermMode.ExampleOmegak} account for terms of adiabatic order 4 and beyond. The square root of the previous result yields
\begin{equation}\label{Eq:FermMode.ExampleOmegakSquareRoot}
\begin{split}
	\Omega_k = &\omega_k+\omega_k^{(1)}+\frac{a^2\omega_k}{8} \left[ \frac{M^2}{\omega_k^4}\left(\frac{a^\prime}{a}\right)^2+\frac{4\Delta^2}{\omega_k^2} \right]\\
	&-\frac{iaM}{16\omega_k}\frac{a^\prime}{a}\left[\frac{4\Delta^2}{M^2}-\frac{4a^2\Delta^2}{\omega_k^2}-\frac{a^2 M^2 }{\omega_k^4}\left(\frac{a^\prime}{a}\right)^2\right]+\dots
\end{split}
\end{equation}
with
\begin{equation}\label{Eq:FermMode.omegak(1)}
\omega_k^{(1)}\equiv-\frac{iaM}{2\omega_k}\frac{a^\prime}{a}\,.
\end{equation}
From the \textit{r.h.s} of \eqref{Eq:FermMode.ExampleOmegakSquareRoot}, the first two terms, $\omega_k$ and $\omega_k^{(1)}$, are used in the first order approximation of the modes (see \,\eqref{Eq:Fermions.hI(1)} and previous equations in the main text).

Now suppose that we proceed with a further step in the iterative process, $\ell=2$. We have to deal with the square root of
\begin{equation}\label{Eq:FermMode.ExampleOmegak1}
			\Omega_k^2 \Omega_{k,1}^2 =\left(\omega_k^2-iMa^\prime+a^2\Delta^2-\frac{ia\Delta^2 }{2M}\frac{a^\prime}{a}+\dots\right)\left( 1+\epsilon_2\right)\,.\\	
\end{equation}
The introduction of
\begin{equation}\label{Eq:FermMode.Epsilon2expansion}
		\epsilon_2=\epsilon_2^{(2)}+\epsilon_2^{(3)}+\dots\,,
\end{equation}
whose expression can be seen in \eqref{Eq:Fermions.Epsilon2}, does not alter the $0th$  nor the $1st$ orders of \eqref{Eq:FermMode.ExampleOmegak} and \eqref{Eq:FermMode.ExampleOmegakSquareRoot} since $\epsilon_2$ is a sum of terms of adiabatic order $2$ and higher:
\begin{equation}\label{Eq:FermMode.ExampleOmegak1SquareRoot}
\begin{split}
\Omega_k\Omega_{k,1}=&\omega_k+\omega_k^{(1)}+\frac{\omega_k}{2} \epsilon_2^{(2)}+\frac{a^2 \Delta^2}{2\omega_k}+\frac{a^2 M^2}{8\omega_k^3}\left(\frac{a^\prime}{a}\right)^2+\frac{\omega_k}{2}\epsilon_2^{(3)}-\frac{iaM}{4\omega_k}\epsilon_2^{(2)}\frac{a^\prime}{a}\\
&-\frac{ia\Delta^2}{4M\omega_k}\frac{a^\prime}{a}+\frac{ia^3 M\Delta^2 }{4\omega_k^3}\frac{a^\prime}{a}+\frac{ia^3M^3}{16\omega^5_k}\left(\frac{a^\prime}{a}\right)^3+\dots
\end{split}
\end{equation}
Nonetheless, the $2nd, 3rd,\dots$ adiabatic orders of \eqref{Eq:FermMode.ExampleOmegak1SquareRoot} do not coincide with the ones of  \eqref{Eq:FermMode.ExampleOmegakSquareRoot}. Similarly, by going to the next iterative step, $\ell=3$, implies working with the square root of the product
\begin{equation}\label{Eq:FermMode.ExampleOmegak2}
		\Omega_k^2\Omega_{k,1}^2\Omega_{k,2}^2=\left(\omega_k^2-iMa^\prime+a^2\Delta^2+\dots\right)\left( 1+					\epsilon_2\right)\left(1+\epsilon_4\right)\\	
\end{equation}
with	
\begin{equation}\label{Eq:FermMode.Epsilon4expansion}
		\epsilon_4=\epsilon_4^{(4)}+\epsilon_4^{(5)}+\dots
\end{equation}
The introduction of $\epsilon_4$ does not alter neither the 0{\it th}, 1{\it st}, 2{\it nd} nor the 3{\it rd} adiabatic orders of \eqref{Eq:FermMode.ExampleOmegak1} or \eqref{Eq:FermMode.ExampleOmegak1SquareRoot}, since $\epsilon_4$ is a sum of terms of adiabatic order $4$ and higher. By the same token, the 4{\it th} and 5{\it th} adiabatic orders and beyond in \eqref{Eq:FermMode.ExampleOmegak1} do not coincide with the ones in \eqref{Eq:FermMode.ExampleOmegak2}. Similar considerations apply to the square root of these quantities.

We can sum up this by saying that for each iterative step we can compute two consecutive adiabatic orders of \eqref{Eq:FermMode.GeneralSolutionProduct} that will not be altered by the subsequent steps. Then, after $\ell$ steps, the $0,1,\dots, 2\ell-1$ adiabatic orders of the product \eqref{Eq:FermMode.GeneralSolutionProduct} are trustworthy for the estimation of the modes.

\item The \textit{r.h.s.}  of \eqref{Eq:FermMode.GeneralSolutionProduct} has a pure imaginary part conformed by the odd orders, precisely those which do no take part in the computation of $W_k$:
    \begin{equation}\label{Eq:FermMode.GeneralSolutionExponentialAdiabOrder}
        \begin{split}
            &-i\int^\tau \left(\omega_k+\omega_k^{(1)}+\omega_k^{(2)}+\dots+\omega_k^{(2\ell-1)}\right)d\tilde{\tau}\\
            &=-i\int^\tau \left(W_k^{\left(0-2(\ell-1)\right)}\right)d\tilde{\tau}-i\int^\tau \left(\omega_k^{(1)}+\omega_k^{(3)}+\dots+\omega_k^{(2\ell-1)}\right)d\tilde{\tau}\,.
        \end{split}
    \end{equation}
However, because of the factor $-i$ in front of the integral, the imaginary terms in the integrand become real and are then necessary for the computation of $F$ and $G$ in \eqref{Eq:FermMode.ansatz2}.

\item We did not specify the limits of integration in \eqref{Eq:Fermions.GeneralSolution} for the following reasons. On the one hand, even terms take part in the pure imaginary exponential of \eqref{Eq:FermMode.ansatz2}. Now because the imaginary exponential does not appear in the final result of the relevant quantities that we compute in the main text (since they cancel in the products of a function times its complex conjugate) one might   wrongly be led to conclude that $W_k$ does not influence the calculation of the EMT. This would, however, be incorrect since the derivatives of the modes $h_k^{\rm I,II}$ are present in these calculations. On the other hand, after integrating the odd terms without specifying the limits in the integral, there exists some residual freedom in the form of a set of functions of the momentum only (i.e. not depending on time). These are called $f_k^{(0)},g_k^{(0)},f_k^{(1)}, g_k^{(1)},\dots$ where the superscript indicates the adiabatic order. If our goal is to compute an adiabatic expansion of the modes up to 6{\it th} order  we need 7 arbitrary constants for $h^{\rm I}$, namely $f_k^{(0)},\dots,f_k^{(6)}$. Similarly for $h^{\rm II}$,  which we denote $g_k^{(0)},\dots,g_k^{(6)}$.
	
\item  From Eqs. \eqref{Eq:Fermions.h1h2eq} \eqref{Eq:Fermions.dceqofspinors}, it is clear that $h_k^{\rm I}(\tau,M)=h_{k}^{\rm II}(\tau,-M)$ so $F(\tau,M)=G(\tau,-M)$ and $f_k^{(n)}(M)=g_k^{(n)}(-M)$.

\end{itemize}

With this considerations in mind let's us put forward the adiabatic expansion of $W_k$ explicitly.
As said, $W_k$ can be specified though the even terms of \eqref{Eq:FermMode.GeneralSolutionProduct}. We are interested to compute at least up to 6{\it th} adiabatic order (that means, at least, $\ell = 4$ steps). Therefore we find:
\begin{equation}\label{Eq:FermMode.WkAdiabaticOrdersExpansion}
W_k^{(0-6)}(\tau)=\omega_k+\omega_k^{(2)}+\omega_k^{(4)}+\omega_k^{(6)}\,,
\end{equation}
with
\begin{equation}\label{Eq:FermMode.Wk0th}
\omega_k^{(0)}=\omega_k\,,
\end{equation}
\begin{equation}\label{Eq:FermMode.Wk2nd}
\omega_k^{(2)}=\frac{a^2\Delta^2}{2\omega_k}-\frac{ a^2 M^2}{8 \omega_k^3}\left(\frac{a^\prime}{a}\right)^2+\frac{5 a^4 M^4}{8\omega_k^5}\left(\frac{a^\prime}{a}\right)^2-\frac{a^2 M^2}{4\omega_k^3}\frac{a^{\prime\prime} }{a}\,,
\end{equation}
\begin{equation}\label{Eq:FermMode.Wk4th}
\begin{split}
\omega_k^{(4)}&=-\frac{a^4\Delta^4}{8\omega_k^3}+\left(-\frac{25 a^6 M^4}{16\omega_k^7}+\frac{23 a^4 M^2}{16\omega_k^5}-\frac{a^2}{8\omega_k^3}\right)\Delta^2 \left(\frac{a^\prime}{a}\right)^2
+\left(\frac{3a^4M^2}{8\omega_k^5}-\frac{ a^2}{4\omega_k^3}\right)\Delta^2  \frac{a^{\prime \prime}}{a}\\
&-\left(\frac{1105a^8 M^8}{128\omega_k^{11}}-\frac{267a^6 M^6}{64\omega_k^9}+\frac{21 a^4 M^4}{128\omega_k^7}\right)\left(\frac{a^\prime}{a}\right)^4+\left(\frac{221a^6 M^6}{32\omega_k^9}-\frac{57 a^4 M^4}{32\omega_k^7}\right)\left(\frac{a^\prime}{a}\right)^2 \frac{a^{\prime\prime}}{a}\\
&-\left(\frac{19a^4 M^4 }{32\omega_k^7}-\frac{a^2 M^2}{32\omega_k^5}\right)\left(\frac{a^{\prime\prime}}{a}\right)^2-\left(\frac{7a^4M^4}{8\omega_k^7}-\frac{a^2 M^2}{16\omega_k^5}\right)\frac{a^{\prime}}{a}\frac{a^{\prime\prime\prime}}{a}+\frac{a^2 M^2}{16\omega_k^5}\frac{a^{\prime\prime\prime\prime}}{a}\,,
\end{split}
\end{equation}
\begin{equation*}
\begin{split}
\omega_k^{(6)}&=\frac{a^6 \Delta^6}{16\omega_k^5}+\left(\frac{175a^8 M^4}{64\omega_k^9}-\frac{215 a^6 M^2 }{64\omega_k^7}+\frac{13a^4}{16\omega_k^5}\right) \Delta^4  \left(\frac{a^\prime}{a}\right)^2-\left(\frac{15a^6 M^2 }{32\omega_k^7}-\frac{3a^4}{8\omega_k^5}\right)\Delta^4 \frac{a^{\prime\prime}}{a}\\
&+\left(\frac{12155 a^{10} M^8}{256 \omega_k^{13}}-\frac{6823a^8 M^6}{128\omega_k^{11}}+\frac{3351 a^6 M^4 }{256\omega_k^9}-\frac{21 a^4 M^2}{64\omega_k^7}\right)\Delta^2 \left(\frac{a^{\prime}}{a}\right)^4\\
&+\left(\frac{133 a^6  M^4 }{64\omega_k^9}-\frac{81 a^4  M^2}{64\omega_k^7}+\frac{a^2}{32\omega_k^5}\right)\Delta^2 \left(\frac{a^{\prime\prime}}{a}\right)^2\\
\end{split}
\end{equation*}
 \begin{equation}\label{Eq:FermMode.Wk6th}
\begin{split}
\phantom{xxxxxx}&-\left(\frac{1989a^8 M^6}{64\omega_k^{11}}-\frac{1725 a^6 M^4}{64\omega_k^9}+\frac{57a^4 M^2}{16\omega_k^7}\right) \Delta^2 \left(\frac{a^{\prime}}{a}\right)^2 \frac{a^{\prime\prime}}{a}\\
&+\left(\frac{49a^6 M^4}{16\omega_k^9}-\frac{61 a^4  M^2 }{32\omega_k^7}+\frac{a^2}{16\omega_k^5}\right)\Delta^2 \frac{a^\prime}{a} \frac{a^{\prime\prime\prime}}{a}-\left(\frac{5a^4   M^2 }{32\omega_k^7}-\frac{a^2}{16\omega_k^5}\right)\Delta^2 \frac{a^{\prime\prime\prime\prime}}{a}\\
&+\left(\frac{414125 a^{12} M^{12}}{1024 \omega_k^{17}} - \frac{
 338935 a^{10} M^{10}}{1024 \omega_k^{15}} + \frac{56271 a^8 M^8 }{1024 \omega_k^{13}} - \frac{869 a^6 M^6}{1024 \omega_k^{11}}\right)\left(\frac{a^\prime}{a}\right)^6\\
& -\left(\frac{248475 a^{10} M^{10}}{512 \omega_k^{15}}-\frac{73457 a^8 M^8 }{256 \omega_k^{13}}
+ \frac{12087 a^6 M^6 }{512 \omega_k^{11}}\right)\left(\frac{a^{\prime}}{a}\right)^4 \frac{a^{\prime\prime}}{a} \\
&+ \left(\frac{34503 a^8 M^8}{256 \omega_k^{13}} - \frac{3225 a^6 M^6 }{64 \omega_k^{11}}+ \frac{249 a^4 M^4 }{256 \omega_k^9}\right)\left(\frac{a^{\prime}}{a}\right)^2\left(\frac{a^{\prime\prime}}{a}\right)^2\\
 &- \left(\frac{631 a^6 M^6 }{128 \omega_k^{11}} -\frac{109 a^4 M^4 }{128 \omega_k^9}\right)\left(\frac{a^{\prime\prime}}{a}\right)^3 + \left(\frac{
 1055 a^8 M^8 }{16 \omega_k^{13}} -\frac{3171 a^6 M^6}{128 \omega_k^{11}}+\frac{69 a^4 M^4}{128 \omega_k^9}\right) \left(\frac{a^\prime}{a}\right)^3 \frac{a^{\prime\prime\prime}}{a} \\
 &-\left(\frac{ 1391 a^6 M^6}{64 \omega_k^{11}}-\frac{245 a^4 M^4 }{64 \omega_k^9}\right)\frac{a^{\prime}}{a} \frac{a^{\prime\prime}}{a}\frac{ a^{\prime\prime\prime}}{a} +\left(\frac{69 a^4 M^4}{128 \omega_k^9} -\frac{a^2 M^2}{128 \omega_k^7}\right)\left(\frac{a^{\prime\prime\prime}}{a}\right)^2 \\
 &-\left(\frac{815 a^6 M^6}{128 \omega_k^{11}}-\frac{149  a^4 M^4 }{128 \omega_k^9}\right)\left(\frac{a^\prime}{a}\right)^2 \frac{a^{\prime\prime\prime\prime}}{a}+\left(\frac{ 55 a^4 M^4}{64 \omega_k^9} -\frac{a^2 M^2}{64 \omega_k^7}\right)\frac{a^{\prime\prime}}{a} \frac{a^{\prime\prime\prime\prime}}{a}\\
 & +\left(\frac{ 27 a^4 M^4 }{64 \omega_k^9}-\frac{ a^2 M^2}{64 \omega_k^7}\right)\frac{a^\prime}{a} \frac{a^{ \prime\prime\prime\prime\prime}}{a} -\frac{ a^2 M^2 }{64 \omega_k^7}\frac{a^{\prime\prime\prime\prime\prime\prime}}{a}\,.
\end{split}
\end{equation}
The odd terms in the expansion \eqref{Eq:FermMode.GeneralSolutionProduct} yield a real exponential contribution in the integrals involved in \eqref{Eq:FermMode.ansatz2} and hence do not contribute to the expansion of $W_k$ in \eqref{Eq:FermMode.ExpansionW}, but are nevertheless necessary to compute the amplitude of the modes. Notice that after computing the integral, the adiabatic order decreases by one unit, so in order to estimate the amplitude up to 6{\it th} order is mandatory to compute up to $\omega_k^{(7)}$. The corresponding integrals for these terms are listed below:
\begin{equation}\label{Eq:FermMode.Integral1st}
		-i\int{\omega_k^{(1)}d\tau}=\int^\tau \left[-\frac{ M a^\prime}{2\omega_k}\right]d\tilde{\tau}=\log\left(\frac{\omega_k-aM}{\omega_k+aM}\right)^{1/4}\,,
\end{equation}
\begin{equation}\label{Eq:FermMod.Integral3rd}
	\begin{split}
		-i\int{\omega_k^{(3)}d\tau}=&\Delta^2\int^\tau \left[\frac{ a^2 a^\prime  M}{4\omega_k^3}-\frac{i a^\prime }{4M\omega_k}\right]d\tilde{\tau}\\
		&+\int^\tau \left[\frac{25  a^2 M^5 {a^\prime}^3}{16\omega_k^7}-\frac{5   M^3 {a^\prime}^3}{16\omega_k^5}-\frac{ a M^3 a^\prime 				a^{\prime\prime}}{\omega_k^5}+\frac{Ma^{\prime\prime\prime}}{8\omega_k^3}\right]d\tilde{\tau}\\
		&=-\frac{a\Delta^2}{4M\omega_k}+\frac{aM}{8\omega_k^3}\frac{a^{\prime\prime}}{a}-\frac{5a^3 M^3 }{16\omega_k^5}\left(\frac{a^\prime}{a}\right)^2\,,
\end{split}
\end{equation}
\begin{equation}\label{Eq:FermMode.Integral5th}
	\begin{split}
	-i\int^\tau \omega_k^{(5)}d\tilde{\tau}	&=\Delta^4\int^\tau \left[\frac{-3  a^4  M a^\prime}{16\omega_k^5}+\frac{a^2  a^\prime}{8 M\omega_k^3}+\frac{ a^\prime}{16 M^3 \omega_k}\right]d\tilde{\tau}\phantom{aaaaaaaaaaaaaaaaaaaaaaaaaaaaaaaaaaaaaa}\\
&+\Delta^2\int^\tau \bigg[-\frac{175 a^4  M^5 {a^\prime}^3}{32\omega_k^9}+\frac{75  a^2  M^3 {a^\prime}^3}{16 \omega_k^7}-\frac{15  M {a^\prime}^3}{32  \omega_k^5}+\frac{5a^3  M^3 a^\prime a^{\prime \prime}}{2 \omega_k^7}-\frac{3 a  M a^\prime a^{\prime\prime}}{2\omega_k^5}\\
&\phantom{aaaaaaaaa}-\frac{3 a^2 M a^{\prime\prime\prime}}{16\omega_k^5}+\frac{a^{\prime\prime\prime}}{16M\omega_k^3}\Bigg]d\tilde{\tau}\\
\phantom{aaaaaaaaa}&+\int^\tau\Bigg[-\frac{12155  a^4 M^9 {a^\prime}^5}{256 \omega_k^{13}}+\frac{3453 a^2 M^7 {a^\prime}^5}{128 \omega_k^{11}}-\frac{399  M^5{a^\prime}^5}{256 \omega_k^9}+\frac{1547  a^3 M^7 {a^\prime}^3 a^{\prime\prime}}{32\omega_k^{11}}\\
&\phantom{aaaaaaa}-\frac{543  a M^5 {a^\prime}^3 a^{\prime\prime}}{32\omega_k^9}-\frac{575  a^2 M^5 a^\prime {a^{\prime\prime}}^2}{64\omega_k^9}+\frac{89 M^3 a^\prime {a^{\prime\prime}}^2}{64\omega_k^7}-\frac{417 a^2 M^5 {a^\prime}^2 a^{\prime\prime\prime}}{64\omega_k^9}\\
&\phantom{aaaaaaa}+\frac{63  M^3 {a^\prime}^2 a^{\prime\prime\prime}}{64\omega_k^7}+\frac{33  a M^3 a^{\prime\prime} a^{\prime\prime\prime}}{32\omega_k^7}+\frac{19 a M^3 a^\prime a^{\prime\prime\prime\prime}}{32\omega_k^7}-\frac{M {a^{\prime\prime\prime\prime\prime}}}{32 \omega_k^5}\Bigg]d\tilde{\tau}\\
&=\left(\frac{a^3}{16 M \omega_k^3}+\frac{a}{16M^3 \omega_k}\right)\Delta^4+\left(-\frac{3a^3 M}{16\omega_k^5}+\frac{a}{16M \omega_k^3}\right)\Delta^2 \frac{a^{\prime\prime}}{a}\\
&+\left(\frac{25 a^5 M^3}{32 \omega_k^7}-\frac{15 a^3 M }{32 \omega_k^5}\right)\Delta^2 \left(\frac{a^\prime}{a}\right)^2+\left(-\frac{221a^5M^5}{64\omega_k^9}+\frac{35a^3 M^3}{64\omega_k^7}\right) \frac{a^{\prime \prime}}{a} \left(\frac{a^\prime}{a}\right)^2\\
&+\left(\frac{1105 a^7 M^7 }{256\omega_k^{11}}-\frac{399 a^5 M^5}{256 \omega_k^9}\right)  \left(\frac{a^\prime}{a}\right)^4+\frac{19a^3 M^3 }{64\omega_k^7}\left(\frac{a^{\prime\prime}}{a}\right)^2+\frac{7 a^3  M^3}{16 \omega_k^7}\frac{a^\prime}{a} \frac{a^{\prime\prime  \prime}}{a}-\frac{a M}{32\omega_k^5}\frac{a^{\prime\prime\prime\prime}}{a}\,,
	\end{split}
\end{equation}
\begin{equation*}
\begin{split}
-i \int^\tau \omega_k^{(7)}d \tilde{\tau}=&\Delta^6 \int^\tau \left[\frac{5  a^6 M a^\prime}{32\omega_k^7}-\frac{3 a^4  a^\prime}{32 M \omega_k^5}-\frac{ a^2 a^\prime}{32 M^3 \omega_k^3}-\frac{  a^\prime}{32 M^5 \omega_k}\right]d\tilde{\tau}\\
		&+\Delta^4 \int^\tau \Bigg[\frac{1575  a^6  M^5 {a^\prime}^3}{128 \omega_k^{11}}-\frac{1925  a^4  M^3 {a^\prime}^3}{128 \omega_k^9}		+	\frac{525  a^2  M {a^\prime}^3}{128 \omega_k^7}-\frac{15   {a^\prime}^3 }{128 M \omega_k^5}-\frac{35  a^5  M^3 a^\prime a^{\prime \prime}}{8 \omega_k^9} \\
		&  \phantom{aaaaaaaaa} + \frac{15  a^3  M a^\prime a^{\prime\prime}}{4\omega_k^7}-\frac{3 a  a^\prime a^{\prime				\prime}}{8 M \omega_k^5}+\frac{15  a^4  M a^{\prime\prime\prime}}{64 \omega_k^7}-\frac{3 a^2  a^{\prime\prime \prime}}{32 M \omega_k^5}-\frac{  a^{\prime \prime  \prime}}{64 M^3 \omega_k^3}\Bigg]d\tilde{\tau}\\
		&+\Delta^2\int^\tau \Bigg[\frac{158015  a^6  M^9 {a^\prime}^5}{512 \omega_k^{15}}-\frac{185361  a^4  M^7 {a^\prime}^5}{512\omega_k^{13}}+\frac{51933  a^2  M^5 {a^\prime}^5}{512 \omega_k^{11}}\\
		&\phantom{aaaaaaaaa}-\frac{1995   M^3 {a^\prime}^5}{512 \omega_k^9}-\frac{17017 a^5  M^7 {a^\prime}^3 a^{\prime \prime}}{64\omega_k^{13}}+\frac{3929 a^3  M^5 {a^\prime}^3 a^{\prime\prime}}{16 \omega_k^{11}}\\
  &\phantom{aaaaaaaaa}-\frac{2715 a  M^3 {a^\prime}^3 a^{\prime \prime}}{64 \omega_k^9}+\frac{5175  a^4  M^5 a^\prime {a^{\prime\prime}}^2}{128 \omega_k^{11}}-\frac{1749  a^2  M^3 a^\prime {a^{\prime\prime}}^2}{64\omega_k^9}\\
		&\phantom{aaaaaaaaa}+\frac{267   M a^\prime {a^{\prime\prime}}^2}{128 \omega_k^{7}}+\frac{3753  a^4  M^5 {a^\prime}^2 a^{\prime\prime\prime}}{128\omega_k^{11}}-\frac{1263  a^2  M^3 {a^\prime}^2 a^{\prime\prime \prime}}{64\omega_k^9}\\
  &\phantom{aaaaaaaaa}+\frac{189   M {a^\prime}^2 a^{\prime\prime\prime}}{128\omega_k^7}-\frac{231  a^3  M^3 a^{\prime\prime}a^{\prime\prime\prime}}{64\omega_k^9}+\frac{99  a  M a^{\prime\prime}a^{\prime \prime \prime}}{64 \omega_k^7}-\frac{133 a^3  M^3 a^\prime a^{\prime\prime\prime\prime}}{64\omega_k^9}\\
		&\phantom{aaaaaaaaa}+\frac{57  a  M a^\prime a^{\prime\prime\prime\prime}}{64 \omega_k^7}+\frac{5  a^2  		M {a^{\prime\prime\prime\prime\prime}}}{64\omega_k^7}-\frac{  {a^{\prime\prime\prime\prime\prime}}}{64 M \omega_k^5}\Bigg]d\tilde{\tau}\\
	&+\int^\tau \Bigg[\frac{7040125  a^6 M^{13}{a^\prime}^7}{2048 \omega_k^{19}}-\frac{6664175  a^4 M^{11}{a^\prime}^7}{2048\omega_k^{17}}+\frac{1429935  a^2 M^9 {a^\prime}^7}{2048 \omega_k^{15}}-\frac{39325  M^7 {a^\prime}^7}{2048 \omega_k^{13}}\\
		&\phantom{aaaaaa}-\frac{1242375  a^5 M^{11}{a^\prime}^5 a^{\prime\prime}}{256 \omega_k^{17}}+\frac{449881  a^3 M^9 {a^\prime}^5 a^{\prime\prime}}{128 \omega_k^{15}}-\frac{112779  a M^7 {a^\prime}^5 a^{\prime\prime}}{256 \omega_k^{13}}\\
  &\phantom{aaaaaa}+\frac{945489  a^4 M^9 {a^\prime}^3 {a^{\prime\prime}}^2}{512\omega_k^{15}}-\frac{30273  a^2 M^7 {a^\prime}^3 {a^{\prime\prime}}^2}{32\omega_k^{13}}+\frac{24435  M^5 {a^\prime}^3 {{a^{\prime\prime}}^2}}{512 \omega_k^{11}}\\
  &\phantom{aaaaaa}-\frac{10361  a^3 M^7 a^\prime {a^{\prime\prime}}^3}{64\omega_k^{13}}+\frac{1639  a M^5 a^\prime {a^{\prime\prime}}^3}{32\omega_k^{11}}-\frac{90425  a^3 M^7 {a^\prime}^2 a^{\prime\prime}a^{\prime\prime\prime}}{256 \omega_k^{13}}\\
		&\phantom{aaaaaa}+\frac{687335  a^4 M^9 {a^\prime}^4 a^{\prime\prime\prime}}{1024 \omega_k^{15}}-\frac{173583  a^2 M^7 {a^\prime}^4 a^{\prime\prime\prime}}{512 \omega_k^{13}}+\frac{17259  M^5 {a^\prime}^4 a^{\prime\prime\prime}}{1024 \omega_k^{11}}\\
		&\phantom{aaaaaa}+\frac{27923  a M^5 {a^\prime}^2a^{\prime\prime}a^{\prime\prime\prime}}{256\omega_k^{11}}+\frac{4675  a^2 M^5 {a^{\prime \prime}}^2a^{\prime\prime\prime}}{256 \omega_k^{11}}-\frac{649  M^3 {a^{\prime \prime}}^2 a^{\prime\prime\prime}}{256\omega_k^9}\\
		&\phantom{aaaaaa}+\frac{3403  a^2 M^5 a^\prime {a^{\prime \prime \prime}}^2}{256\omega_k^{11}}-\frac{447  M^3 a^\prime {a^{\prime\prime\prime}}^2}{256\omega_k^9}-\frac{17405  a^3 M^7 {a^\prime}^3 a^{\prime\prime\prime\prime}}{256\omega_k^{13}}\\
		&\phantom{aaaaaa}+\frac{5215  a M^5 {a^\prime}^3 a^{\prime\prime			\prime\prime}}{256 \omega_k^{11}}+\frac{2701  a^2 M^5 a^\prime a^{\prime\prime}a^{\prime\prime\prime\prime}}{128 \omega_k^{11}}-\frac{349  M^3 a^\prime a^{\prime\prime}a^{\prime\prime\prime\prime}}{128 \omega_k^9}\\
        &\phantom{aaaaaa}-\frac{31 a M^3 a^{\prime\prime\prime}a^{\prime\prime\prime\prime}}{32\omega_k^9}+\frac{1301 a^2 M^5 {a^\prime}^2 {a^{\prime\prime\prime\prime\prime}}}{256 \omega_k^{11}}-\frac{159 M^3 {a^\prime}^2 {a^{\prime\prime\prime\prime\prime}}}{256 \omega_k^9}\\
        &\phantom{aaaaaa}-\frac{41 a M^3 a^{\prime\prime}a^{\prime\prime\prime\prime\prime}}{64 \omega_k^9}-\frac{17 a M^3 a^\prime a^{\prime\prime\prime\prime\prime\prime}}{64 \omega_k^9}+\frac{M a^{\prime\prime\prime\prime\prime\prime\prime}}{128\omega_k^{7}}\Bigg]d\tilde{\tau}\\
		\end{split}
\end{equation*}
\begin{equation}\label{Eq:FermMode.Integral7th}
	\begin{split}          
		&\phantom{aaaaaaaaaaaaa}=-\left(\frac{a^5}{32M \omega_k^5}+\frac{a^3}{48M^3\omega_k^3}+\frac{a}{32M^5 \omega_k}\right)\Delta^6\\
        &\phantom{aaaaaaaaaaaaa}-\left(\frac{175a^7 M^3}{128 \omega_k^9}-\frac{75 a^5 M}{64\omega_k^7}+\frac{15 a^3 }{128 M \omega_k^5}\right)\Delta^4 \left(\frac{a^\prime}{a}\right)^2\\
		&\phantom{aaaaaaaaaaaaa}+\left(\frac{15 a^5 M}{64\omega_k^7}-\frac{3a^3}{32 M \omega_k^5}-\frac{a}{64 M^3 \omega_k^3}\right)\Delta^4 \frac{a^{\prime \prime}}{a}-\left(\frac{133 a^5 M^3}{128\omega_k^9}-\frac{57a^3 M}{128\omega_k^7}\right)\Delta^2  \left(\frac{ a^{\prime\prime}}{a}\right)^2\\
        &\phantom{aaaaaaaaaaaaa}-\left(\frac{12155 a^9 M^7}{512\omega_k^{13}}-\frac{11326 a^7 M^5 }{512 \omega_k^{11}}+\frac{1995 a^5 M^3}{512 \omega_k^9}\right)\Delta^2  \left(\frac{a^\prime}{a}\right)^4\\
		&\phantom{aaaaaaaaaaaaa}+\left(\frac{1989 a^7 M^5}{128\omega_k^{11}}-\frac{1350 a^5 M^3 }{128 \omega_k^9}+\frac{105 a^3 M}{128\omega_k^7}\right)\Delta^2 \frac{a^{\prime\prime}}{a}\left(\frac{a^\prime}{a}\right)^2\\
        &\phantom{aaaaaaaaaaaaa}-\left(\frac{49 a^5 M^3}{32 \omega_k^9}-\frac{21 a^3 M}{32\omega_k^7}\right) \Delta^2 \frac{a^{\prime\prime\prime}}{a}\frac{a^{\prime}}{a}+\left(\frac{5a^3 M}{64 \omega_k^7}-\frac{a}{64 M \omega_k^5}\right)\Delta^2 \frac{a^{\prime\prime\prime\prime}}{a}\\
		&\phantom{aaaaaaaaaaaaa}-\left(\frac{414125a^{11}M^{11}}{2048 \omega_k^{17}}-\frac{459355 a^9 M^9 }{3072\omega_k^{15}}+\frac{39325 a^7 M^7}{2048 \omega_k^{13}}\right) \left(\frac{a^\prime}{a}\right)^6-\frac{55 a^3 M^3  }{128\omega_k^9}\frac{a^{\prime\prime\prime\prime}}{a}\frac{a^{\prime\prime}}{a}\\
        &\phantom{aaaaaaaaaaaaa}-\left(\frac{34503a^7 M^7}{512\omega_k^{13}}-\frac{11037 a^5 M^5  }{512 \omega_k^{11}}\right)\left(\frac{a^{ \prime \prime}}{a}\frac{a^\prime}{a}\right)^2-\frac{27a^3 M^3  }{128\omega_k^9}\frac{a^{\prime\prime\prime\prime\prime}}{a}\frac{a^\prime}{a}+\frac{a M  }{128\omega_k^7}\frac{a^{\prime\prime\prime\prime\prime\prime}}{a}\\
        &\phantom{aaaaaaaaaaaaa}-\left(\frac{1055 a^7 M^7}{32\omega_k^{13}}-\frac{330 a^5 M^5}{32\omega_k^{11}}\right) \frac{a^{\prime\prime\prime}}{a}\left(\frac{a^\prime}{a}\right)^3+\left(\frac{815 a^5 M^5}{256\omega_k^{11}}-\frac{105 a^3 M^3}{256\omega_k^9}\right)\frac{a^{\prime\prime\prime\prime}}{a}\left(\frac{a^\prime}{a}\right)^2\\
		&\phantom{aaaaaaaaaaaaa}+\left(\frac{248475 a^9 M^{9}}{1024\omega_k^{15}}-\frac{64863 a^7 M^7 }{512 \omega_k^{13}}+\frac{6699 a^5 M^5}{1024 \omega_k^{11}}\right)\frac{a^{\prime\prime}}{a}\left(\frac{a^{\prime}}{a}\right)^4-\frac{69a^3 M^3}{256\omega_k^9}\left(\frac{a^{\prime\prime\prime}}{a}\right)^2\\	
		&\phantom{aaaaaaaaaaaaa}+\left(\frac{631 a^5 M^5}{256\omega_k^{11}}-\frac{271 a^3 M^3}{768 \omega_k^9}\right)\left(\frac{a^{\prime\prime}}{a}\right)^3+\left(\frac{1391 a^5 M^5}{128\omega_k^{11}}-\frac{189 a^3 M^3}{128\omega_k^9}\right)\frac{a^{\prime\prime\prime}}{a}\frac{a^{\prime\prime}}{a}\frac{a^{\prime}}{a}\,.
	\end{split}
\end{equation}
 Use of Mathematica\,\cite{Mathematica} has been helpful to work out the above integrals. The computational strategy consists in using a sufficiently general ansatz for the structure of the result that is compatible with the adiabaticity order of the calculation, and then solve for the coefficients (form factors)  of the ansatz from pure algebraic manipulations. The procedure has been illustrated with a specific example in equation \eqref{Eq:Fermions.Integrand} of the main text.

Finally, by applying the normalization condition \eqref{Eq:FermMode.NormalizationConditionMode} for the modes at each order, there exists a constraint to fix  the residual freedom mentioned earlier, which is parametrized by the time independent factors $f_k$ and $g_k$ (only depending on the momentum $k$):
\begin{equation}\label{Eq:FermMode.NormalizationConditions}
\begin{split}
&\mathbb{R}e f^{(1)}_k=0\,,\\
&\left( \mathbb{I}m f^{(1)}_k \right)^2+\sqrt{2k}\mathbb{R}e f^{(2)}_k=0\,,\\
&2\mathbb{I}m f^{(2)}_k\mathbb{I}m f^{(1)}_k+\sqrt{2k}\mathbb{R}e f^{(3)}_k=0\,,\\
&\left|f^{(2)}_k\right|^2+2\mathbb{I}m f^{(1)}_k\mathbb{I}m f^{(3)}_k+\sqrt{2k}\mathbb{R}e f^{(4)}_k=0\,,\\
&2\mathbb{I}m f^{(1)}_k\mathbb{I}m f^{(4)}_k+2\mathbb{I}m f^{(2)}_k\mathbb{I}m f^{(3)}_k+2\mathbb{R}e f^{(2)}_k\mathbb{R}e f^{(3)}_k+\sqrt{2k}\mathbb{R}e f^{(5)}_k=0\,,\\
&2\mathbb{I}m f^{(1)}_k\mathbb{I}m f^{(5)}_k+2\mathbb{I}m f^{(2)}_k\mathbb{I}m f^{(4)}_k+\left|f^{(3)}_k\right|^2+2\mathbb{R}e f^{(2)}_k\mathbb{R}e f^{(4)}_k+\sqrt{2k}\mathbb{R}e f^{(6)}_k=0\,.
\end{split}
\end{equation}
Notice that, imposing the former conditions is equivalent to claim that
\begin{equation}\label{Eq:FermMode.NormalizationConditionII}
	\Bigg| 1+\sqrt{\frac{2}{k}}\left(f_k^{(1)}+f_k^{(2)}+f_k^{(3)}+f_k^{(4)}+f_k^{(5)}+f_k^{(6)}\right) \Bigg|^2\approx 1\,,
\end{equation}
where the departure from 1 are just terms of adiabatic order 7 or bigger. Similarly for the functions $g_k$. It can be shown that the observables made out of quadratic terms in the modes $h_k^{\rm I},h_{k}^{\rm II}$ (such as e.g. the EMT), depend on the particular values of $f_k$ in the form \eqref{Eq:FermMode.NormalizationConditionII}. It follows that they are not actual degrees of freedom if they satisfy the conditions \eqref{Eq:FermMode.NormalizationConditions}. It is then safe to set particular values for the functions as long as quantities are computed up to $6th$ adiabatic order. The simplest solution satisfying the normalization conditions \eqref{Eq:FermMode.NormalizationConditions} is $f_k^{(1)}=f_k^{(2)}=f_k^{(3)}=f_k^{(4)}=f_k^{(5)}=f_k^{(6)}=0$ and it is the chosen option for the formulas shown in the rest of this Appendix.

Equipped with these results we are able to calculate the different orders of $F\left(\tau,M\right)$ up to $6th$ order. A comparison between the general equation\,\eqref{Eq:Fermions.GeneralSolution} and the ansatz\,\eqref{Eq:FermMode.ansatz2}, the different orders of $F$ are conformed by the rightful combinations of terms of the denominator $\sqrt{\Omega_k\Omega_{k,1}\Omega_{k,2}\Omega_{k,3}}$ and the real factors of the  exponential $\exp\left(-i\int^\tau \Omega_k\Omega_{k,1}\Omega_{k,2}\Omega_{k,3} d\tilde{\tau}\right)$.
Now the different orders of $F$ are\,\footnote{For $h^{\rm II}_k$, one can make use of the relation $G^{(n)}(M)=F^{(n)}(-M)$.}:
\begin{equation}\label{Eq:FermMode.F1st}
    \begin{split}
		&F^{(1)}(M)=\frac{i a M}{4\omega_k^2}\frac{a^\prime}{a}\,,
    \end{split}
\end{equation}
\begin{equation}\label{Eq:FermMode.F2nd}
    \begin{split}
		F^{(2)}(M)&=\left(-\frac{a^2}{4\omega_k^2}-\frac{a }{4 M \omega_k}\right)\Delta^2-\left(\frac{5 a^4 M^4 }{16 \omega_k^6}+ \frac{5 a^3 M^3 }{16 \omega_k^5 }
		+\frac{M^2 a^2}{32 \omega_k^4}\right) \left(\frac{a^ \prime}{a}\right)^2\\
		&+ \left(\frac{a^2 M^2 }{8\omega_k^4} + \frac{a M}{8\omega_k^3}\right) \frac{a^{\prime\prime}}{a}\,,
    \end{split}
\end{equation}
\begin{equation}\label{Eq:FermMode.F3rd}
    \begin{split}
		F^{(3)}(M)&=i\left(-\frac{5 M a^3 }{16 \omega_k ^4}-\frac{a^2}{16 \omega_k ^3}+\frac{a}{8 M \omega_k ^2}\right)\frac{a^\prime}{a}\Delta^2\\
		&+i\left(-\frac{65 M^5 			a^5}{64 \omega_k^8}-\frac{5 M^4 a^4}{64 \omega_k^7}+\frac{21 M^3 a^3}{128 \omega_k^6}\right)\left(\frac{a^\prime}{a}\right)^3\\
		&+i\left(\frac{19 M^3 a^3}{32 \omega_k^6}+\frac{M^2 a^2}{32 \omega_k^5}\right)\frac{a'a''}{a^2}-i\frac{a M}{16 \omega_k ^4}\frac{a^{\prime\prime\prime}}{a}\,,
    \end{split}
\end{equation}
\begin{equation}\label{Eq:FermMode.F4th}
    \begin{split}
		F^{(4)}(M)&=\left(\frac{5a^4}{32 \omega_k^4}+\frac{a^3}{8M\omega_k^3}+\frac{a^2}{32 M^2 \omega_k^2}+\frac{a}{16 M^3 \omega_k}\right)\Delta^4\\
		&+\left(\frac{65a^6M^4}{64\omega_k^8}+\frac{15a^5 M^3}{16\omega_k^7}-\frac{61 a^4 M^2}{128\omega_k^6}-\frac{59 a^3 M}{128\omega_k^5}-\frac{a^2}{32\omega_k^4}			\right)\left(\frac{a^\prime}{a}\right)^2\Delta^2\\
		&+\left(-\frac{9a^4 M^2}{32\omega_k^6}-\frac{a^3 M}{4\omega_k^5}+\frac{3a^2}{32\omega_k^4}+\frac{a}{16M\omega_k^3}\right)\frac{a^{\prime \prime}}{a}\Delta^2\\
		&+\left(\frac{2285 a^8 M^8}{512\omega_k^{12}}+\frac{565a^7 M^7}{128 \omega_k^{11}}-\frac{349 a^6 M^6}{256\omega_k^{10}}-\frac{793 a^5 M^5}{512\omega_k^9}-				\frac{85 a^4 M^4}{2048 \omega_k^8}\right)\left(\frac{a^\prime}{a}\right)^4\\
		&+\left( -\frac{457 a^6 M^6}{128\omega_k^{10}}-\frac{113 a^5 M^5}{32\omega_k^9}+\frac{113 a^4 M^4 }{256\omega_k^8}+\frac{139 a^3M^3}{256 \omega_k^7}\right)				\left(\frac{a^\prime}{a}\right)^2\frac{a^{\prime\prime}}{a}\\
		&+\left(\frac{41a^4 M^4}{128\omega_k^8}+\frac{5a^3 M^3}{16\omega_k^7}-\frac{a^2 M^2}{128\omega_k^6}\right)\left(\frac{a^{\prime\prime}}{a}\right)^2+					\left(\frac{7 a^4 M^4 }{16 \omega_k ^8}+\frac{7 a^3 M^3 }{16 \omega_k^7}+\frac{a^2 M^2}{64 \omega_k^6}\right)\frac{a^\prime}{a}\frac{a^{\prime\prime\prime}}{a}\\
		&-\left(\frac{a^2 M^2}{32\omega_k^6}+\frac{a M}{32\omega_k^5}\right)\frac{a^{\prime\prime\prime\prime}}{a}\,,
    \end{split}
\end{equation}
\begin{equation}\label{Eq:FermMode.F5th}
    \begin{split}
		F^{(5)}(M)&=i\left(\frac{45a^5 M}{128\omega_k^6}+\frac{3a^4}{32\omega_k^5}-\frac{19a^3}{128 M \omega_k^4}-\frac{a^2}{64M^2 \omega_k^3}-\frac{a}						{32M^3\omega_k^2}\right)\frac{a^\prime}{a}\Delta^4\\
		&+i\left(\frac{1105 a^7 M^5}{256 \omega_k^{10}}+\frac{35a^6 M^4}{64\omega_k^9}-\frac{1563 a^5M^3}{512\omega_k^8}-\frac{101 a^4 M^2}{512\omega_k^7}+\frac{63 			a^3 M}{256\omega_k^6}\right)\left(\frac{a^\prime}{a}\right)^3\Delta^2\\
		&+i\left(-\frac{247a^5 M^3}{128\omega_k^8}-\frac{15 a^4 M^2}{64\omega_k^7}+\frac{113 a^3 M}{128\omega_k^6}+\frac{a^2}{32\omega_k^5}\right)\frac{a^\prime}{a}				\frac{a^{\prime\prime}}{a}\Delta^2\\
		&+i\left(\frac{9 a^3 M}{64\omega_k^6}+\frac{a^2}{64\omega_k^5}-\frac{a}{32M \omega_k^4}\right)\frac{a^{\prime\prime\prime}}{a}\Delta^2\\
		&+\left(\frac{57125 a^9 M^9}{2048\omega_k^{14}}+\frac{715 a^8 M^8}{512\omega_k^{13}}-\frac{7657 a^7 M^7}{512 \omega_k^{12}}-\frac{903 a^6 M^6}							{2048\omega_k^{11}}+\frac{6511a^5 M^5}{8192\omega_k^{10}}\right)\left(\frac{a^\prime}{a}\right)^5\\
		&+i\left(-\frac{14167 a^7 M^7}{512\omega_k^{12}}-\frac{301 a^6 M^6}{256\omega_k^{11}}+\frac{9273 a^5 M^5}{1024\omega_k^{10}}+\frac{161 a^4 M^4}{1024\omega_k^9}			\right)\frac{a^{\prime\prime}}{a}\left(\frac{a^\prime}{a}\right)^3\\
		&+i\left(\frac{2525 a^5 M^5}{512\omega_k^{10}}+\frac{19 a^4 M^4}{128\omega_k^9}-\frac{361 a^3 M^3}{512\omega_k^8}\right)\frac{a^\prime}{a}								\left(\frac{a^{\prime\prime}}{a}\right)^2\\
		&+i\left(\frac{933 a^5 M^5}{256\omega_k^{10}}+\frac{33 a^4 M^4}{256\omega_k^9}-\frac{255 a^3M^3}{512\omega_k^8}\right)\left(\frac{a^\prime}{a}							\right)^2\frac{a^{\prime\prime\prime}}{a}\\
		&+i\left(-\frac{69 a^3 M^3}{128\omega_k^8}-\frac{M^2 a^2}{128\omega_k^7}\right)\frac{a^{\prime\prime}}{a}\frac{a^{\prime\prime\prime}}{a}-i\left(\frac{41a^3 M^3}{128\omega_k^8}+\frac{M^2 a^2}{128\omega_k^7}\right)\frac{a^{\prime}}{a}\frac{a^{\prime\prime\prime\prime}}{a}+i\frac{a M}{64\omega_k^6}\frac{a^{\prime\prime\prime\prime\prime}}{a}\,,
		\end{split}
\end{equation}

\begin{equation}\label{Eq:FermMode.F6th}
    \begin{split}
		F^{(6)}(M)&=\left(-\frac{15 a^6 }{128\omega_k^6}-\frac{11 a^5}{128 M \omega_k^5}-\frac{3 a^4}{128M^2 \omega_k^4}-\frac{5a^3}{128 M^3 \omega_k^3}-\frac{a^2}{64 			M^4 \omega_k^2}-\frac{a}{32M^5 \omega_k}\right)\Delta^6\\
		&+\left(-\frac{1105 a^8 M^4}{512\omega_k^{10}}-\frac{965 a^7 M^3}{512\omega_k^9}+\frac{1713 a^6 M^2}{1024\omega_k^8}+\frac{715 a^5 M}{512\omega_k^7}-\frac{149 			a^4}{1024\omega_k^6}-\frac{57 a^3}{512 M \omega_k^5}\right)\left(\frac{a^\prime}{a}\right)^2\Delta^4\\
		&+\left(\frac{117 a^6 M^2}{256\omega_k^8}+\frac{97 a^5 M}{256\omega_k^7}-\frac{55 a^4}{256\omega_k^6}-\frac{33 a^3 }{256 M\omega_k^5}-\frac{a^2}{128 M^2 				\omega_k^4}-\frac{a}{64 M^3 \omega_k^3}\right)\frac{a^{\prime\prime}}{a}\Delta^4\\
		&+\left(-\frac{57125 a^{10} M^8}{2048 \omega_k^{14}}-\frac{54265 a^9 M^7}{2048\omega_k^{13}}+\frac{24479 a^8 M^6}{1024\omega_k^{12}}+\frac{47405 a^7 M^5}				{2048\omega_k^{11}}-\frac{28887 a^6 M^4}{8192 \omega_k^{10}}\right.\\
		&\left. \phantom{aaa}-\frac{31635 a^5  M^3}{8192 \omega_k^9}-\frac{85 a^4 M^2 }{1024 \omega_k^8}\right)\left(\frac{a^\prime}{a}\right)^4\Delta^2\\
        &+\left(\frac{9597 a^8 M^6}{512 \omega_k^{12}}+\frac{9045 a^7 M^5}{512 \omega_k^{11}}-\frac{11985 a^6 M^4}{1024 \omega_k^{10}}-\frac{5619 a^5 M^3}{512\omega_k^9}+\frac{765 a^4 M^2}{1024\omega_k^8}\right.\\
        &\phantom{aaa}\left.+\frac{417 a^3 M}{512\omega_k^7}\right)\left(\frac{a^\prime}{a}\right)^2\frac{a^{\prime\prime}}{a}\Delta^2+\left(\frac{13 a^4 M^2}{128 \omega_k^8}+\frac{3a^3 M}{32\omega_k^7}-\frac{3a^2}{128\omega_k^6}-\frac{a}{64M \omega_k^5}\right)\frac{a^{\prime\prime\prime\prime}}{a}\Delta^2\\
        &+\left(-\frac{697 a^6 M^4}{512\omega_k^{10}}-\frac{641 a^5 M^3}{512\omega_k^9}+\frac{301 a^4 M^2}{512 \omega_k^8}+\frac{241 a^3 M}{512 \omega_k^7}-\frac{a^2}{128 \omega_k^6}\right)\left(\frac{a^{\prime\prime}}{a}\right)^2\Delta^2\\
        &+\left(-\frac{119 a^6 M^4}{64\omega_k^{10}}-\frac{7 a^5 M^3}{4\omega_k^9}+\frac{183 a^4 M^2}{256\omega_k^8}+\frac{167 a^3 M}{256 \omega_k^7}+\frac{a^2}{64\omega_k^6}\right)\frac{a^\prime}{a}\frac{a^{\prime\prime\prime}}{a}\Delta^2\\
		&+\left(-\frac{1690275 a^{12}M^{12}}{8192\omega_k^{18}}-\frac{1678975 a^{11}M^{11}}{8192 \omega_k^{17}}+\frac{2377685 a^{10} M^{10}}{16384 \omega_k^{16}}+				\frac{1231405 a^9 M^9}{8192 \omega_k^{15}}\right.\\
		&\phantom{aaa}\left.-\frac{519009a^8 M^8}{32768 \omega_k^{14}}-\frac{627179 a^7 M^7}{32768 \omega_k^{13}}-\frac{13989 a^6 M^6}{65536 \omega_k^{12}}\right)\left(\frac{a^\prime}{a}\right)^6+					\left(\frac{a^2 M^2}{128\omega_k^8}+\frac{aM}{128\omega_k^7}\right)\frac{a^{\prime\prime\prime\prime\prime\prime}}{a}\\
        &+\left(\frac{1014165a^{10}M^{10}}{4096 \omega_k^{16}}+\frac{1007385 a^9 M^9}{4096 \omega_k^{15}}-\frac{250133 a^8 M^8}{2048\omega_k^{14}}-\frac{521273 a^7M^7}{4096 \omega_k^{13}}\right.\\
        &\phantom{aaa}\left.+\frac{74799 a^6 M^6}{16384\omega_k^{12}}+\frac{106819 a^5 M^5}{16384 \omega_k^{11}}\right)\frac{a^{\prime\prime}}{a}\left(\frac{a^\prime}{a}\right)^{4}-\left(\frac{27 a^4 M^4}{128\omega_k^{10}}+\frac{27 a^3 M^3}{128\omega_k^9}+\frac{a^2 M^2}{256\omega_k^8}\right)\frac{a^\prime}{a}\frac{a^{\prime\prime\prime\prime\prime}}{a}\\
		&-\left(\frac{141309 a^8 M^8}{2048\omega_k^{14}}+\frac{140205 a^7 M^7}{2048\omega_k^{13}}-\frac{85737 a^6 M^6}{4096 \omega_k^{12}}-\frac{44397 a^5 M^5}{2048\omega_k^{11}}-\frac{441 a^4 M^4}{4096 \omega_k^{10}}\right)\left(\frac{a^\prime}{a} \frac{a^{\prime\prime}}{a}\right)^2\\
		&+\left(\frac{2643 a^6 M^6}{1024\omega_k^{12}}+\frac{2603 a^5 M^5}{1024\omega_k^{11}}-\frac{403 a^4 M^4}{1024\omega_k^{10}}-\frac{363 a^3 M^3}{1024\omega_k^9}\right)\left(\frac{a^{\prime\prime}}{a}\right)^3\\
		&+\left(-\frac{8545 a^8 M^8}{256 \omega_k^{14}}-\frac{4255 a^7 M^7}{128\omega_k^{13}}+\frac{9721 a^6 M^6}{1024\omega_k^{12}}+\frac{10541 a^5 M^5}						{1024\omega_k^{11}}+\frac{277 a^4 M^4}{2048 \omega_k^{10}}\right)\frac{a^{\prime\prime\prime}}{a}\left(\frac{a^\prime}{a}\right)^3\\
		&+\left(\frac{353 a^6 M^6}{32\omega_k^{12}}+\frac{1405 a^5 M^5}{128\omega_k^{11}}-\frac{697 a^4 M^4}{512\omega_k^{10}}-\frac{755a^3 M^3}{512\omega_k^9}\right)			\frac{a^\prime}{a}\frac{a^ {\prime\prime}}{a}\frac{a^{\prime\prime\prime}}{a}\\
  		&+\left(-\frac{69 a^4 M^4}{256 \omega_k^{10}}-\frac{69 a^3 M^3}{256 \omega_k^{9}}-\frac{a^2 M^2}{512\omega_k^8}\right)\left(\frac{a^{\prime\prime\prime}}{a}\right)^2+\left(-\frac{113 a^4 M^4}{256\omega_k^{10}}-\frac{7a^3M^3}{16\omega_k^9}+\frac{a^2 M^2}{256\omega_k^8} \right)\frac{a^{\prime\prime\prime\prime}}{a}\frac{a^{\prime\prime}}{a}\\
		&+\left(\frac{1645a^6 M^6}{512\omega_k^{12}}+\frac{205 a^5 M^5}{64\omega_k^{11}}-\frac{349 a^4 M^4}{1024\omega_k^{10}}-\frac{419 a^3 M^3}{1024\omega_k^9}				\right)\frac{a^{\prime\prime\prime\prime}}{a}\left(\frac{a^\prime}{a}\right)^2\,.
	\end{split}
\end{equation}

\chapter{ Adiabatic expansion of $\left\langle T_{\mu\nu} \right\rangle$  for spin-1/2 fields in chapter 4}\label{Appendix:AdiabExpFermEMT}

The unrenormalized components of the VEV of the EMT,  $\langle T_{\mu\nu}^{\delta \psi} \rangle$,  for spin-1/2 fermions can be obtained through the adiabatic expansion, which we compute up to $6th$ order. For the $00$ component we have
\begin{equation}\label{Eq:FermMode.ExpansionT00}
    \left\langle T_{00}^{\delta\psi} \right\rangle =\left\langle T_{00}^{\delta\psi} \right\rangle^{(0)}+\left\langle T_{00}^{\delta\psi} \right\rangle^{(2)}+\left\langle T_{00} ^{\delta\psi}\right\rangle^{(4)}+\left\langle T_{00}^{\delta\psi} \right\rangle^{(6)}+\cdots
\end{equation}
The various terms in the expansion \eqref{Eq:FermMode.ExpansionT00} read as follows:
\begin{equation}\label{Eq:FermMode.ExpansionT00order0}
\left\langle T_{00}^{\delta\psi} \right\rangle^{(0)}=\frac{1}{2\pi^2 a}\int_0^\infty dk k^2\left[-\frac{2\omega_k}{a}\right]\,,
\end{equation}
\begin{equation}\label{Eq:FermMode.ExpansionT00order2}
\left\langle T_{00}^{\delta\psi} \right\rangle^{(2)}=\frac{1}{2\pi^2 a}\int_0^\infty dk k^2\left[-\frac{a\Delta^2}{\omega_k}-\frac{a^3 M^4}{4\omega_k^5}\left(\frac{a^\prime}{a}\right)^2+\frac{a M^2}{4\omega_k^3}\left(\frac{a^\prime}{a}\right)^2\right]\,,
\end{equation}
\begin{equation}\label{Eq:FermMode.ExpansionT00order4}
\begin{split}
\left\langle T_{00}^{\delta\psi} \right\rangle^{(4)}&=\frac{1}{2\pi^2 a}\int_0^\infty dk k^2\Bigg[\frac{a^3 \Delta^4}{4\omega_k^3}+\left(\frac{5a^5M^4 \Delta^2}{8\omega_k^7}-\frac{7a^3 M^2 \Delta^2}{8 \omega_k^5}+\frac{a\Delta^2}{4\omega_k^3}\right)\left(\frac{a^\prime}{a}\right)^2\\
&+\left(\frac{105 a^7 M^8}{64\omega_k^{11}}-\frac{63a^5 M^6}{32\omega_k^9}+\frac{21 a^3 M^4}{64\omega_k^7}\right)\left(\frac{a^\prime}{a}\right)^4+\left(-\frac{7a^5 M^6}{8\omega_k^9}+\frac{7 a^3 M^4}{8\omega_k^7}\right)\frac{a^{\prime\prime}}{a}\left(\frac{a^\prime}{a}\right)^2 \\
&+\left(-\frac{a^3 M^4}{16\omega_k^7}+\frac{a M^2}{16\omega_k^5}\right)\left(\frac{a^{\prime\prime}}{a}\right)^2+\left(\frac{a^3 M^4}{8\omega_k^7}-\frac{a M^2}{8\omega_k^5}\right)\frac{a^\prime}{a}\frac{a^{\prime\prime\prime}}{a}
\Bigg]\,,
\end{split}
\end{equation}

\begin{equation}\label{Eq:FermMode.ExpansionT00order6}
\begin{split}
\left\langle T_{00}^{\delta\psi} \right\rangle^{(6)}&=\frac{1}{2\pi^2 a}\int_0^\infty dk k^2\Bigg[-\frac{a^5 \Delta^6}{8\omega_k^5}+\left(\frac{35 a^7  M^4}{32\omega_k^9}+\frac{55 a^5 M^2}{32\omega_k^7}-\frac{5a^3 }{8\omega_k^5}\right)\left(\frac{a^\prime}{a}\right)^2\Delta^4\\
&+\left(-\frac{1155 a^9 M^8}{128\omega_k^{13}}+\frac{987 a^7 M^6}{64\omega_k^{11}}-\frac{903 a^5 M^4}{128\omega_k^9}+\frac{21a^3 M^2}{32\omega_k^7}\right)\left(\frac{a^\prime}{a}\right)^4\Delta^2\\
&+\left(\frac{63 a^7 M^6}{16\omega_k^{11}}-\frac{91 a^5 M^4}{16\omega_k^9}+\frac{7a^3 M^2}{4\omega_k^7}\right)\frac{a^{\prime\prime}}{a}\left(\frac{a^\prime}{a}\right)^2\Delta^2\\
&+\left(\frac{7a^5 M^4 }{32\omega_k^9}-\frac{9 a^3 M^2}{32\omega_k^7}+\frac{a}{16\omega_k^5}\right)\left(\frac{a^{\prime\prime}}{a}\right)^2\Delta^2\\
&+\left(-\frac{7a^5 M^4 }{16\omega_k^9}+\frac{9a^3M^2}{16\omega_k^7}-\frac{a}{8\omega_k^5}\right)\frac{a^\prime}{a}\frac{a^{\prime\prime\prime}}{a}\Delta^2\\
&+\bigg(-\frac{25025 a^{11} M^{12}}{512\omega_k^{17}}+\frac{39039 a^9 M^{10}}{512 \omega_k^{15}}-\frac{14883 a^7 M^8}{512\omega_k^{13}}+\frac{869 a^5 M^6}{512\omega_k^{11}}\bigg)\left(\frac{a^\prime}{a}\right)^6\\
&+\left(\frac{3003 a^9 M^{10}}{64\omega_k^{15}}-\frac{2013 a^7 M^8}{32\omega_k^{13}}+\frac{1023 a^5 M^6}{64\omega_k^{11}}\right)\frac{a^{\prime\prime}}{a}\left(\frac{a^\prime}{a}\right)^4\\
&+\left(-\frac{5a^5 M^6}{16\omega_k^{11}}+\frac{5a^3 M^4}{16\omega_k^9} \right)\left(\frac{a^{\prime\prime}}{a}\right)^3\\
&+\left(-\frac{891a^7M^8}{128\omega_k^{13}}+\frac{501a^5 M^6 }{64\omega_k^{11}}-\frac{111 a^3 M^4}{128\omega_k^9}\right)\left(\frac{a^{\prime\prime}}{a} \frac{a^{\prime}}{a}\right)^2\\
&+\left(-\frac{429a^7 M^8}{64\omega_k^{13}}+\frac{249a^5 M^6}{32\omega_k^{11}}-\frac{69a^3 M^4}{64\omega_k^9}\right)\left( \frac{a^\prime}{a} \right)^3\frac{a^{\prime\prime\prime}}{a}\\
&+\left(\frac{15 a^5 M^6}{16\omega_k^{11}}-\frac{15 a^3M^4}{16\omega_k^9}\right)\frac{a^\prime}{a}\frac{a^{\prime\prime}}{a}\frac{a^{\prime\prime\prime}}{a}+\left(-\frac{a^3 M^4}{64\omega_k^9}+\frac{a M^2}{64\omega_k^7}\right)\left(\frac{a^{\prime\prime\prime}}{a}\right)^2\\
&+\left(\frac{9a^5 M^6}{16\omega_k^{11}}-\frac{9a^3 M^4}{16 \omega_k^9}\right)\frac{a^{\prime\prime\prime\prime}}{a}\left(\frac{a^\prime}{a}\right)^{2}\\
&+\left(\frac{a^3 M^4}{32\omega_k^9}-\frac{a M^2}{32\omega_k^7}\right)\frac{a^{\prime\prime}}{a}\frac{a^{\prime\prime\prime\prime}}{a}+\left(-\frac{a^3M^4}{32\omega_k^9}+\frac{aM^2}{32\omega_k^7}\right)\frac{a^\prime}{a}\frac{a^{\prime\prime\prime\prime\prime}}{a}\Bigg]\,.
\end{split}
\end{equation}
\newpage
To obtain the component $\langle T_{11}^\psi \rangle$ a similar expansion as in \eqref{Eq:FermMode.ExpansionT00} holds.  However, it  is possible to use the following relation with the previously computed  $ \langle T_{00}^\psi \rangle$ component:
\begin{equation}\label{Eq:FermMode.ExpansionT11}
\left\langle T_{11}^{\delta\psi}\right\rangle=-\frac{1}{3\mathcal{H}}\left(\left\langle T_{00}^{\delta\psi}\right\rangle^\prime+\mathcal{H}\left\langle T_{00}^{\delta\psi}\right\rangle \right)\,.
\end{equation}
As a result we find\footnote{We  refer the reader to \hyperref[Sect:MasterInt]{Appendix\,\ref{Sect:MasterInt}} for the explicit computation of the integrals below.}:
\begin{equation}\label{Eq:FermMode.ExpansionT11order0}
\left\langle T_{11}^{\delta\psi} \right\rangle^{(0)}=\frac{1}{2\pi^2 a} \int_0^\infty dk k^2\left[\frac{2a M^2}{3 \omega_k}-\frac{2\omega_k}{3a}\right]\,,
\end{equation}
\begin{equation}\label{Eq:FermMode.ExpansionT11order2}
\begin{split}
\left\langle T_{11}^{\delta\psi} \right\rangle^{(2)}&=\frac{1}{2\pi^2 a}\int_0^\infty dk k^2\Bigg[\left(-\frac{a^3 M^2}{3\omega_k^3}+\frac{a}{3\omega_k}\right)\Delta^2+\left(-\frac{5a^5M^6}{12\omega_k^7}+\frac{a^3 M^4}{3\omega_k^5}+\frac{a M^2}{12\omega_k^3}\right)\left(\frac{a^\prime}{a}\right)^2\\
&+\left(\frac{a^3 M^4}{6\omega_k^5}-\frac{a M^2}{6\omega_k^3}\right)\frac{a^{\prime\prime}}{a}\Bigg]\,,
\end{split}
\end{equation}
\begin{equation}\label{Eq:FermMode.ExpansionT11order4}
\begin{split}
\left\langle T_{11}^{\delta\psi} \right\rangle^{(4)}&=\frac{1}{2\pi^2 a}\int_0^\infty dk k^2\Bigg[\left(\frac{a^5 M^2}{4\omega_k^5}-\frac{a^3}{4\omega_k^3}\right)\Delta^4+\left(-\frac{5a^5 M^4}{12\omega_k^7}+\frac{7a^3 M^2}{12\omega_k^5}-\frac{a }{6\omega_k^3}\right)\frac{a^{\prime\prime}}{a}\Delta^2\\
&+\left(\frac{35 a^7 M^6}{24\omega_k^9}-\frac{25 a^5 M^4}{12\omega_k^7}+\frac{13a^3M^2}{24\omega_k^5}+\frac{a}{12\omega_k^3}\right) \left(\frac{a^\prime}{a}\right)^2 \Delta^2\\
&+\left(\frac{385 a^9 M^{10}}{64\omega_k^{13}}-\frac{483a^7 M^8}{64\omega_k^{11}}+\frac{91a^5 M^6}{64\omega_k^9}+\frac{7a^3 M^4}{64\omega_k^7}\right)\left(\frac{a^\prime}{a}\right)^4\\
&+\left(-\frac{77a^7M^8}{16\omega_k^{11}}+\frac{21a^5 M^6}{4\omega_k^9}-\frac{7a^3 M^4}{16\omega_k^7}\right)\frac{a^{\prime\prime}}{a}\left(\frac{a^\prime}{a}\right)^2+\left(\frac{7a^5 M^6}{16\omega_k^9}-\frac{11a^3 M^4}{24\omega_k^7}+\frac{a M^2}{48\omega_k^5}\right)\left(\frac{a^{\prime\prime}}{a}\right)^2\\
&+\left(\frac{7a^5 M^6}{12\omega_k^9}-\frac{13a^3 M^4}{24\omega_k^7}-\frac{a M^2}{24\omega_k^5}\right)\frac{a^\prime}{a}\frac{a^{\prime\prime\prime}}{a}+\left(-\frac{a^3 M^4}{24\omega_k^7}+\frac{a M^2}{24\omega_k^5}\right)\frac{a^{\prime\prime\prime\prime}}{a}\Bigg]\,,
\end{split}
\end{equation}
\begin{equation}\label{Eq:FermMode.ExpansionT11order6}
\begin{split}
\left\langle T_{11}^{\delta\psi} \right\rangle^{(6)} & = \frac{1}{2\pi^2 a}\int_0^\infty dk k^2\Bigg[\left(-\frac{5a^7 M^2}{24\omega_k^7}+\frac{5a^5}{24\omega_k^5}\right)\Delta^6+\left( \frac{35 a^7 M^4}{48\omega_k^9}-\frac{55a^5 M^2}{48\omega_k^7}+\frac{5a^3}{12 \omega_k^5}\right) \frac{a^{\prime\prime}}{a} \Delta^4 \\
&+\left(-\frac{105 a^9 M^6}{32\omega_k^{11}}+\frac{35 a^7 M^4}{6\omega_k^9}-\frac{265 a^5 M^2}{96\omega_k^7}+\frac{5a^3 }{24\omega_k^5}\right)\left(\frac{a^\prime}{a}\right)^2\Delta^4\\
&+\left(-\frac{5005 a^{11}M^{10}}{128\omega_k^{15}}+\frac{9163 a^9 M^8}{128\omega_k^{13}}-\frac{4683 a^7 M^6}{128\omega_k^{11}}+\frac{497 a^5 M^4}{128\omega_k^9}+\frac{7a^3 M^2}{32\omega_k^7}\right)\left(\frac{a^\prime}{a}\right)^4\Delta^2\\
&+\left(\frac{847a^9 M^8}{32\omega_k^{13}}-\frac{343 a^7 M^6}{8\omega_k^{11}}+\frac{553 a^5 M^4}{32\omega_k^9}-\frac{7a^3M^2}{8\omega_k^7}\right)\frac{a^{\prime\prime}}{a}\left(\frac{a^\prime}{a}\right)^2\Delta^2\\
&+\left(-\frac{63 a^7 M^6}{32\omega_k^{11}}+\frac{35a^5M^4}{12\omega_k^9}-\frac{31a^3 M^2}{32\omega_k^7}+\frac{a}{48\omega_k^5}\right)\left(\frac{a^{\prime\prime}}{a}\right)^2\Delta^2\\
&+\left(-\frac{21 a^7 M^6}{8\omega_k^{11}}+\frac{175a^5 M^4}{48\omega_k^9}-\frac{47 a^3 M^2}{48\omega_k^7}-\frac{a}{24\omega_k^5}\right)\frac{a^\prime}{a}\frac{a^{\prime\prime\prime}}{a}\Delta^2 \\
&+\left(\frac{7a^5 M^4}{48\omega_k^9}-\frac{3a^3 M^2}{16\omega_k^7}+\frac{a}{24\omega_k^5}\right)\frac{a^{\prime\prime\prime\prime}}{a}\Delta^2\\
&-\left(\frac{425425a^{13} M^{14}}{1536\omega_k^{19}}-\frac{355355 a^{11}M^{12}}{768\omega_k^{17}}+\frac{25883 a^9 M^{10}}{128\omega_k^{15}}-\frac{12221 a^7 M^8}{768\omega_k^{13}}-\frac{869a^5 M^6}{1536 \omega_k^{11}}\right)\left(\frac{a^\prime}{a}\right)^6\\
&+\left(\frac{85085 a^{11} M^{12}}{256\omega_k^{17}}-\frac{124839 a^9 M^{10}}{256\omega_k^{15}}+\frac{40623 a^7 M^8}{256\omega_k^{13}}-\frac{869 a^5 M^6}{256\omega_k^{11}}\right)\frac{a^{\prime\prime}}{a}\left(\frac{a^\prime}{a}\right)^4\\
&+\left(-\frac{11869 a^9 M^{10}}{128\omega_k^{15}}+\frac{15301 a^7 M^8}{128\omega_k^{13}}-\frac{3395 a^5 M^6}{128\omega_k^{11}}-\frac{37 a^3 M^4}{128\omega_k^{9}}\right)\left(\frac{a^\prime}{a} \frac{a^{\prime\prime}}{a}\right)^2\\
&+\left(\frac{671 a^7 M^8}{192\omega_k^{13}}-\frac{391 a^5 M^6}{96 \omega_k^{11}}+\frac{37 a^3 M^4}{64\omega_k^{9}}\right)\left(\frac{a^{\prime\prime}}{a}\right)^3\\
&+\left(-\frac{715 a^9 M^{10}}{16 \omega_k^{15}}+\frac{3597a^7 M^8}{64\omega_k^{13}}-\frac{357 a^5 M^6}{32\omega_k^{11}}-\frac{23 a^3 M^4}{64\omega_k^9}\right)\frac{a^{\prime\prime\prime}}{a}\left(\frac{a^\prime}{a}\right)^3\\
&+\left(\frac{473a^7 M^8}{32\omega_k^{13}}-\frac{263 a^5 M^6}{16\omega_k^{11}}+\frac{53 a^3 M^4}{32\omega_k^9}\right)\frac{a^{\prime\prime\prime}}{a}\frac{a^{\prime\prime}}{a}\frac{a^\prime}{a}\phantom{............}\\
&+\left(-\frac{23 a^5 M^6}{64\omega_k^{11}}+\frac{17 a^3  M^4}{48\omega_k^9}+\frac{a M^2}{192\omega_k^7}\right)\left(\frac{a^{\prime\prime\prime}}{a}\right)^2\phantom{............}\\
&+\left(\frac{275 a^7 M^8}{64\omega_k^{13}}-\frac{149a^5 M^6}{32\omega_k^{11}}+\frac{23 a^3 M^4}{64\omega_k^9}\right)\left(\frac{a^{\prime}}{a}\right)^2\frac{a^{\prime\prime\prime\prime}}{a}\phantom{............}\\
&+\left(-\frac{19 a^5 M^6}{32\omega_k^{11}}+\frac{29 a^3 M^4}{48\omega_k^9}-\frac{a M^2}{96\omega_k^7}\right)\frac{a^{\prime\prime}}{a}\frac{a^{\prime\prime\prime\prime}}{a}\phantom{............}\\
&+\left(-\frac{9a^5 M^6}{32\omega_k^{11}}+\frac{13a^3 M^4}{48\omega_k^9}+\frac{a M^2}{96\omega_k^7}\right)\frac{a^{\prime}}{a}\frac{a^{\prime\prime\prime\prime\prime}}{a}+\left(\frac{a^3 M^4}{96\omega_k^9}-\frac{a M^2}{96\omega_k^7}\right)\frac{a^{\prime\prime\prime\prime\prime\prime}}{a}\Bigg]\,.\phantom{............}
\end{split}
\end{equation}

\newpage

\chapter{Semi-analytical solutions in different epochs of chapter 5}\label{Appendix:Semi-Analytical}

Our aim in this appendix is to find semi-analytical solutions for the BD equations in the various epochs of the cosmic history, by using a perturbative approach. We express the BD-field and the scale factor up to linear order in $\epsilon_{\rm BD}$ as follows\footnote{The use of the superindex $(n)$ indicates the order in the expansion in terms of $\epsilon_{\rm BD}$. It has nothing to do to with the adiabatic orders we presented in previous chapters.},
\begin{equation}\label{Eq:SemiAnalytical.ExpansionOfvarphi}
\varphi(t)=\varphi^{(0)}+\epsilon_{\rm BD}\varphi^{(1)} (t)+\mathcal{O}(\epsilon_{\rm BD}^2), \\
\end{equation}
\begin{equation}\label{Eq:SemiAnalytical.ExpansionOfa}
 a(t)=a^{(0)}(t)+\epsilon_{\rm BD} a^{(1)}(t)+\mathcal{O}(\epsilon_{\rm BD}^2)\,, 
\end{equation}
with the functions with superscript $(0)$ denoting the solutions of the background equations in standard GR with a constant Newtonian coupling that can be in general different from $G_N$, and the functions with superscript $(1)$ denoting the first-order corrections induced by a non-null $\epsilon_{\rm BD}$. Neglecting the higher-order terms is a very good approximation for all the relevant epochs of the expansion history due to the small values of $\epsilon_{\rm BD}$ allowed by the data. Plugging these expressions in Eqs. \eqref{Eq:BD.Friedmannequation}-\eqref{Eq:BD.FullConservationLaw} we can solve the system and obtain the dominant energy density at each epoch and the Hubble function, which of course can also be written as
\begin{equation}\label{Eq:SemiAnalytical.ExpansionOfrho}
\rho_N(t) = \rho_N^{(0)}(t)+\epsilon_{\rm BD} \rho_N^{(1)}(t)+\mathcal{O}(\epsilon_{\rm BD}^2)\,,\\
\end{equation}
\begin{equation}\label{Eq:SemiAnalytical.ExpansionOfH}
H(t) = H^{(0)}(t)+\epsilon_{\rm BD} H^{(1)}(t)   +\mathcal{O}(\epsilon_{\rm BD}^2)\,,
\end{equation}
respectively, where $N=R,M,\Lambda$ denotes the solution at the radiation, matter, and $\Lambda$-dominated epochs\footnote{For the sake of simplicity and to ease the obtention of analytical expressions, in this appendix we consider three massless neutrinos.}. We make use of the following relations,
\begin{equation}\label{Eq:SemiAnalytical.KGFirstOrder}
\begin{split}
&\frac{1}{2\omega_{\rm BD}+3}=\frac{\epsilon_{\rm BD}}{2}+\mathcal{O}(\epsilon_{\rm BD}^2)\,, \qquad
\frac{\dot{\varphi}}{\varphi}=\epsilon_{\rm BD} \frac{\dot{\varphi}^{(1)}}{\varphi^{(0)}}+\mathcal{O}(\epsilon_{\rm BD}^2)\,,\\
&\frac{\omega_{\rm BD}}{2}\left(\frac{\dot{\varphi}}{\varphi} \right)^2=\frac{\epsilon_{\rm BD}}{2}\left(\frac{\dot{\varphi}^{(1)}}{\varphi^{(0)}}\right)^2+\mathcal{O}(\epsilon_{\rm BD}^2)\,.
\end{split}
\end{equation}
The Klein-Gordon equation is already of first order in $\epsilon_{\rm BD}$,
\begin{equation}\label{Eq:SemiAnalytical.KGFirstOrderII}
\ddot{\varphi}^{(1)}+3H^{(0)}\dot{\varphi}^{(1)}= 4\pi G_N(\rho^{(0)}_N-3p^{(0)}_N)\,, 
\end{equation}
and this allows us to find $\varphi^{(1)}$ without knowing $H^{(1)}$ nor the linear corrections for the energy densities and pressures. The Friedmann equation leads to
\begin{equation}\label{Eq:SemiAnalytical.FriedmannZeroOrder}
H^{(0)}=\left(\frac{8\pi{G_N}}{3\varphi^{(0)}}\rho_N^{(0)}\right)^{1/2}\,,
\end{equation}
at zeroth order, and
\begin{equation}\label{Eq:SemiAnalytical.FriedmannFirstOrder}
6H^{(0)}H^{(1)}+3H^{(0)}\frac{\dot{\varphi}^{(1)}}{\varphi^{(0)}}-\frac{1}{2}\left(\frac{\dot{\varphi}^{(1)}}{\varphi^{(0)}}\right)^2=3\left(H^{(0)}\right)^2 \left(\frac{\rho^{(1)}_N}{\rho_N^{(0)}} - \frac{\varphi^{(1)}}{\varphi^{(0)}}\right)\,, 
\end{equation}
at first order, which let us to compute $a^{(1)}(t)$ once we get $\rho^{(1)}_N(t,a^{(1)}(t))$ from the conservation equation and substitute it in the {\it r.h.s}. $H^{(1)}$ is then trivially obtained using the computed scale factor. Alternatively, one can also combine \eqref{Eq:SemiAnalytical.FriedmannFirstOrder} with the pressure equation to obtain a differential equation for $H^{(1)}$ and directly solve it. Proceeding in this way one obtains, though, an additional integration constant that must be fixed using the Friedmann and conservation equations.

\section{Radiation dominated epoch (RDE)}\label{Sect:RDEAnalytical}
In the RDE the trace of the energy-momentum tensor is negligible when compared with the total energy density in the Universe, so we can set the right-hand side of \eqref{Eq:SemiAnalytical.KGFirstOrder} to zero. The leading order of the scale factor and the Hubble function take the following form, respectively: $a^{(0)}(t)=A t^{1/2}$,  $H^{(0)}(t)=1/(2t)$, with $A \equiv (32\pi{G_N}\rho^{0}_{\rm r}/3\varphi^{(0)})^{1/2}$ and $\rho^{0}_{\rm r}$ the current value of the radiation energy density. Using these relations we can find the BD-field as well as the scale factor at first order. They read,
\begin{equation}\label{Eq:SemiAnalytical.RDEVarphiExp}
\varphi (t) =\varphi^{(0)}+\epsilon_{\rm BD}\left(C_{R1} + C_{R2}\,t^{-1/2}\right)+\mathcal{O}(\epsilon_{\rm BD}^2)\,,
\end{equation}
and
\begin{equation}\label{Eq:SemiAnalytical.RDEScaleFactorExp}
a(t)= At^{1/2}\left( 1  + \epsilon_{\rm BD}\left[\frac{C_{R2}^2 \ln(t)}{24(\varphi^{(0)})^2{t}}-\frac{C_{R1}}{4\varphi^{(0)}}+\frac{C_{R3}}{t} \right] +\mathcal{O}(\epsilon_{\rm BD}^2) \right)\,,
\end{equation}
where $C_{R1}$, $C_{R2}$ and $C_{R3}$ are integration constants. {Let us note that in the RDE the evolution of the BD-field $\varphi$ is essentially frozen, since there is no growing mode. The term evolving as $\sim t^{-1/2}$ is the decaying mode, and after some time we are eventually left with a constant contribution}.

Finally, it is easy to find the corresponding Hubble function
\begin{equation}\label{Eq:SemiAnalytical.RDEHubbleFunctionExp}
H(t) = \frac{1}{2t}\left(1  + \epsilon_{\rm BD}\left[ -\frac{C_{R2}^2}{12(\varphi^{(0)})^2}\frac{\ln(t)}{t}+\frac{1}{t}\left(\frac{C_{R2}^2}{12(\varphi^{(0)})^2}-2C_{R3} \right) \right] +\mathcal{O}(\epsilon_{\rm BD}^2)\right)\,.
\end{equation}
At late enough times it is natural to consider that decaying mode is already negligible, and this allows us to simplify a lot the expressions, by setting $C_{R2}=0$. The scalar field remains then constant in very good approximation when radiation rules the expansion of the Universe, and the other cosmological functions are the same as in GR ($a\sim t^{1/2}$,  $H\sim \frac{1}{2t}$), but with an effective gravitational coupling $G=G_N/\varphi$.

\section{Matter dominated epoch (MDE)}\label{Sect:MDEAnalytical}

When nonrelativistic matter is dominant in the Universe the scalar field evolves as
\begin{equation}\label{Eq:SemiAnalytical.varphiMDEApp}
\varphi(t) = \varphi^{(0)} + \epsilon_{\rm BD}\left[\frac{C_{M1}}{t}+\frac{2\varphi^{(0)}}{3}\ln(t) + C_{M2}\right] +\mathcal{O}(\epsilon_{\rm BD}^2)\,,
\end{equation}
where $C_{M1}$ and $C_{M2}$ are integration constants. At leading order the scale factor and the Hubble function take the following form, $a^{(0)}(t) = Bt^{2/3}$ and $H^{(0)}(t) = 2/3t$ respectively, with $B \equiv \left(6\pi{G_N}\rho^0_{\rm m}/\varphi^{(0)}\right)^{1/3}$ and $\rho^0_{\rm m}$ the current value of the matter energy density. If we neglect the decaying mode in \eqref{Eq:SemiAnalytical.varphiMDEApp} we find $\varphi(a)=\varphi^{(0)}(1+\epsilon_{\rm BD}\ln a+\mathcal{O}(\epsilon^2_{\rm BD}))$. This solution is also found from the analysis of fixed points of \hyperref[Appendix:FixedPoints]{Appendix\,\ref{Appendix:FixedPoints}}, at leading order in $\epsilon_{\rm BD}$. The scale factor reads in this case,
\begin{equation}\label{Eq:SemiAnalytical.ScaleFactorMDEApp}
a(t) = Bt^{2/3}\left(1 + \epsilon_{\rm BD}\left[ -\left(\frac{C_{M1}}{\varphi^{(0)}}\right)^2\frac{1}{8t^2}-\frac{C_{M2}}{3\varphi^{(0)}}-\frac{1}{18}-\frac{2}{9}\ln(t)+\frac{C_{M3}}{t} \right] +\mathcal{O}(\epsilon_{\rm BD}^2) \right)\,,
\end{equation}
where $C_{M3}$ is another integration factor, and the Hubble function,
\begin{equation}\label{Eq:SemiAnalytical.HubbleFunctionMDEApp}
H(t) = \frac{2}{3t}\left(1 +\epsilon_{\rm BD}\left[\left(\frac{C_{M1}}{\varphi^{(0)}}\right)^2\frac{3}{8t^2}-\frac{1}{3}-\frac{3C_{M3}}{2t} \right]+\mathcal{O}(\epsilon_{\rm BD}^2)\right)\,.
\end{equation}
{Once more, after we neglect the contribution from the decaying modes, the usual cosmological functions are as is GR in the MDE ($a\sim t^{2/3}$, $H\sim \frac{2}{3t}$). However, in contrast to the RDE there is some mild evolution (a logarithmic one with the cosmic time) of the BD-field. Since $\varphi^{(0)}$ is obviously positive, it follows from  Eq.\,\eqref{Eq:SemiAnalytical.varphiMDEApp} that the sign of such evolution (implying growing or decreasing behavior) is entirely defined by the sign of the BD-parameter $\eBD$.  Our fit to the overall data clearly shows that $\eBD<0$ (cf. \hyperref[Sect:NumericalAnalysis]{Sect.\,\ref{Sect:NumericalAnalysis}})  and hence $\varphi$ decreases with the expansion during the MDE, which means that the effective gravitation coupling $G=G_N/\varphi$ increases with the expansion.}

\section{$\Lambda$-dominated  or VDE}\label{Sect:VDEAnalytical}
Here we assume that the energy density of the Universe is completely dominated by a vacuum fluid, with constant energy density $\rho_\Lambda$.  {This period occurs in the very early Universe during inflation, and in the very late one when matter has diluted significantly and a new period of inflation occurs}. The usual solution for the Hubble function in that epoch is $H^{(0)}(t)=H_\Lambda$, where $H_\Lambda$ is a constant, fixed by \eqref{Eq:SemiAnalytical.FriedmannZeroOrder}. Taking this into account we find
\begin{equation}\label{Eq:SemiAnalytical.varphi(t)Inflation}
\varphi(t)=\varphi^{(0)}+\epsilon_{\rm BD}\left(2H_\Lambda \varphi^{(0)} t+C_{\Lambda 1} e^{-3H_\Lambda t}+C_{\Lambda 2} \right)+\mathcal{O}(\epsilon_{\rm BD}^2), 
\end{equation}
and for the scale factor
\begin{equation}\label{Eq:SemiAnalytical.ScaleFactor(t)Inflation}
\begin{split}
&a(t)=C_{\Lambda 4}e^{H_\Lambda t}\left(1+\epsilon_{\rm BD}\left[-\frac{H_\Lambda^2 t^2}{2}-\left(\frac{C_{\Lambda 1}}{\varphi^{(0)}}\right)^2\frac{e^{-6H_\Lambda t}}{8}+C_{\Lambda 3}H_\Lambda t\left(\frac{C_{\Lambda 2}}{2\varphi^{(0)}}+\frac{2}{3} \right) \right]\right)+\mathcal{O}(\epsilon_{\rm BD}^2),
\end{split}
\end{equation}
where $C_{\Lambda 1},$ $C_{\Lambda 2}$, $C_{\Lambda 3}$, and $C_{\Lambda 4}$ are integration constants. Note that by performing the limit $\epsilon_{\rm BD} \rightarrow 0$ we recover the usual GR expressions, as in all the previous formulas. The Hubble function at first order takes the form
\begin{equation}\label{Eq:SemiAnalytical.H(t)Inflation}
\begin{split}
&H(t)=H_\Lambda\left(1 +\epsilon_{\rm BD}\left[-\frac{2}3+\frac{3}{4}\left(\frac{C_{\Lambda 1}}{\varphi^{(0)}}\right)^2e^{-6H_\Lambda t}-H_\Lambda t-\frac{C_{\Lambda 2}}{2\varphi^{(0)}} \right]+\mathcal{O}(\epsilon_{\rm BD}^2)\right).
\end{split}
\end{equation}
The term accompanied by the constant $C_{\Lambda 1}$ in \eqref{Eq:SemiAnalytical.varphi(t)Inflation} is a decaying mode, which we considered to be negligible already in the RDE. Thus, we can safely remove it, and $C_{\Lambda 2}$ can be fixed by the condition $\varphi(t_*)=\varphi_*$, at some $t_*$ deeply in the VDE. The scalar field evolves then as $\varphi(t) \sim 2\epsilon_{\rm BD} H_\Lambda t \sim 2\epsilon_{\rm BD}\ln a$, which is the behavior we obtain also from the analysis of fixed points (cf. \hyperref[Appendix:FixedPoints]{Appendix\,\ref{Appendix:FixedPoints}}). We can also see that during this epoch the BD-field decreases with the expansion because $\eBD<0$, as indicated before.

\section{Mixture of matter and vacuum energy}\label{Sect:MixtureMatterVacuumAnalytical}

In a Universe with a non-negligible amount of vacuum and matter energy densities, it is also possible to obtain an analytical expression for the BD scalar field at leading order in $\epsilon_{\rm BD}$. Unfortunately, this is not the case for the scale factor and the Hubble function, so we will present here only the formula for $\varphi$. In the $\Lambda$CDM the scale factor is given by
\begin{equation}\label{Eq:SemiAnalytical.ScaleFactor(t)Mixture}
a^{(0)}(t)=\left(\frac{\tilde{\Omega}_{\rm m}}{\tilde{\Omega}_{\Lambda}}\right)^{1/3} \sinh^{2/3} \left(\frac{3}{2}\sqrt{\tilde{\Omega}_{\Lambda}}H_0 t \right)\,,
\end{equation}
so
\begin{equation}\label{Eq:SemiAnalytical.H(t)Mixture}
H^{(0)}(t)=H_0\sqrt{\tilde{\Omega}_\CC} \coth \left(\frac{3}{2}\sqrt{\tilde{\Omega}_{\Lambda}}H_0 t\right)\,.
\end{equation}
Solving the Klein-Gordon equation, we obtain
\begin{equation}\label{Eq:SemiAnalytical.varphi1VDE}
\varphi^{(1)}(t)=\sqrt{\tilde{\Omega}_\Lambda}H_0 t  \coth \left(\frac{3}{2}\sqrt{\tilde{\Omega}_\Lambda}H_0 t \right)+\frac{2}{3}\ln\left( \sinh \left(\frac{3}{2}\sqrt{\tilde{\Omega}_\Lambda}H_0 t \right) \right)\,.
\end{equation}
One can easily check that in the limits $H_0t\ll 1$ and $H_0 t\gg 1$ we recover the behavior that we have found in previous sections for the matter and $\Lambda$-dominated Universes, respectively. {When we substitute the previous expression in Eq.\,\eqref{Eq:SemiAnalytical.ExpansionOfvarphi}, we confirm once more that $\varphi$ decreases with the expansion (since $\eBD<0$ and $\dot{\varphi}^{(1)}(t)>0$ $\forall{t}$). This is the period when the Universe is composed of a  mixture of matter and vacuum energy at comparable proportions, and corresponds to the current Universe.  Thus  $G=G_N/\varphi$ increases with the expansion in the present Universe as it was also the case in the preceding MDE period, which is of course an important feature that helps to solve the $H_0$-tension, as explained in different parts of the chapter, and in particular in the preview \hyperref[Sect:Preview]{Sect.\,\ref{Sect:Preview}}.}


\section{Connection of the BD-$\CC$CDM model with the  Running Vacuum Model}\label{Sect:RVMconnection}

Analytical solutions to the system \eqref{Eq:BD.Friedmannequation}-\eqref{Eq:BD.FieldeqPsi} are not known and for this reason our actual analysis proceeds numerically. However, as we have seen in the previous sections it is possible to search for  approximate solutions in the different epochs, which can help to better understand the numerical results and the qualitative behavior of the BD-$\CC$CDM model. Actually, a first attempt in this direction trying to show that BD-$\CC$CDM  can mimic the Running Vacuum Model (RVM) was done in \cite{SolaPeracaula:2018dsw,deCruzPerez:2018cjx} and we refer the reader to these references for details. Here we just summarize the results and adapt them to the current notation.  It is based on searching for solutions in the MDE in the form of a power-law ansatz in which the BD-field $\varphi$ evolves very slowly:
\begin{equation}\label{Eq:SemiAnalytical.powerlaw}
\varphi(a) = \varphi_0\,a^{-\epsilon}\  \qquad (|\epsilon|\ll1)\,.
\end{equation}
Obviously $\epsilon$ must be a very small parameter in absolute value since $G(a)\equiv G(\varphi(a))$ cannot depart too much from $G_N$.  On comparing with the analysis of fixed points given in \hyperref[Appendix:FixedPoints]{Appendix\,\ref{Appendix:FixedPoints}} -- cf. Eq.\,\eqref{Eq:FixedPoints.psiMDE} -- we can anticipate that $\epsilon\propto -\eBD$, although we do not expect perfect identification since \eqref{Eq:SemiAnalytical.powerlaw} is a mere ansatz solution in the MDE whereas \eqref{Eq:FixedPoints.psiMDE} is an exact phase trajectory in that epoch. For $\epsilon>0$ (hence $\eBD<0$), the effective  coupling increases with the expansion and hence is asymptotically free since  $G(a)$  is smaller in the past, which is the epoch when the
Hubble rate (with natural dimension of energy) is bigger. For $\epsilon<0$ ($\eBD>0$), instead, $G(a)$  decreases with the expansion.

Using the power-law ansatz \eqref{Eq:SemiAnalytical.powerlaw} we find
\begin{equation}\label{Eq:SemiAnalytical.derivatives}
 \frac{\dot\varphi}{\varphi}= -\epsilon {H}\,,\ \ \ \ \ \ \ \ \ \ \ \
 \frac{\ddot\varphi}{\varphi} = -\epsilon\dot{H} + \epsilon^2{H^2}. 
\end{equation}
Plugging these relations into the system of equations \eqref{Eq:BD.Friedmannequation}-\eqref{Eq:BD.FieldeqPsi} and after some calculation it is possible to arrive at the following pair of Friedmann-like equations to ${\cal O}(\epsilon)$\cite{SolaPeracaula:2018dsw,deCruzPerez:2018cjx}:
\begin{equation}\label{Eq:SemiAnalytical.EffectiveFriedmann}
   H^2=\frac{8\pi G}{3}\left(\rho^0_{\rm m} a^{-3+\epsilon}+\rDE(H)\right)
\end{equation}
and
\begin{equation}\label{Eq:SemiAnalytical.currentacceleration}
\frac{\ddot{a}}{a}=-\frac{4\pi G}{3}\,\left(\rho_{\rm m}^0 a^{-3+\epsilon}+\rDE(H)+3p_{\Lambda}\right)\,,
\end{equation}
with $G=G_N/\varphi_0$. The first equation emulates an effective Friedmann's equation with time-evolving cosmological term, in which the DE appears as dynamical:
\begin{equation}\label{Eq:SemiAnalytical.rLeff}
  \rDE(H)=\rL+\frac{3\,\nu_{\rm eff}}{8\pi G} H^2\,.
\end{equation}
Here
\begin{equation}\label{Eq:SemiAnalytical.nueff}
\ \nu_{\rm eff}\equiv\epsilon\left(1+\frac16\,\oD\epsilon\right) 
\end{equation}
is the coefficient controlling the dynamical character of the dark energy \eqref{Eq:SemiAnalytical.rLeff}. The structure of this dynamical dark energy (DDE) is reminiscent of the Running Vacuum Model (RVM), see \cite{Sola:2013gha,Sola:2015rra,GomezValent:2017kzh} and references therein. In the language used in this chapter, the analogue of the above RVM form of the Friedmann equation can be derived from Eq.\,\eqref{Eq:BD.FriedmannWithF} upon taking into account that the function ${\cal F}$ is of ${\cal O}(\eBD)$. In fact, ${\cal F}$ is the precise analogue of $\nu_{\rm eff}$, for if we set $\epsilon\to -\eBD$  in \eqref{Eq:SemiAnalytical.nueff} it boils down to the value quoted in  Eq.\,\eqref{Eq:BD.F}. The two languages are similar, but not identical, for the reasons explained above.

Notice from \eqref{Eq:SemiAnalytical.EffectiveFriedmann} that,  to ${\cal O}(\epsilon)$:
\begin{equation}\label{Eq:SemiAnalytical.SumRule}
\Omo +\OLo =1-\nu_{\rm eff}\,,
\end{equation}
so the usual sum rule of GR is slightly violated by the BD model when parametrized as a deviation with respect to GR.
Only for $\epsilon=0$ we have $\nu_{\rm eff}=0$ and then  we recover the usual cosmic sum rule\footnote{It is interesting to note that the presence of $\nueff\neq0$ emulates a fictitious spatial curvature. This is  the analog, in RVM language, of the similar situation  noted  in \hyperref[Sect:EffectiveEoS]{Sect.\,\ref{Sect:EffectiveEoS}} when we defined the GR-picture of the BD model. A persistent irreducible value of this parameter in future observations might serve also as a hint of the underlying BD physics.}.
The parameter $\nu_{\rm eff}$ becomes associated to the dynamics of the DE. Worth noticing, the above expression adopts the form of the RVM, see \cite{Sola:2013gha,Sola:2015rra,GomezValent:2017kzh} and references therein, in particular \cite{Shapiro:2000dz,Sola:2007sv,Shapiro:2009dh} -- where the running parameter is usually denoted $\nu$ and is associated to the $\beta$-function of the running vacuum.  Recently, the parameter $\nu$ (and in general the structure of the RVM) has been elucidated from direct calculations in QFT in curved spacetime within GR\,\cite{Moreno-Pulido:2020anb}. For additional discussions on the running of the CC term, see e.g. \cite{Babic:2004ev,Ward:2010qs,Antipin:2017pbt}. The RVM has been shown to be phenomenologically promising to alleviate some of the existing tensions within the $\CC$CDM, particularly the $\sigma_8$-tension\,\,\cite{SolaPeracaula:2016qlq, SolaPeracaula:2017esw, Gomez-Valent:2018nib, Gomez-Valent:2017idt, Sola:2017znb, Sola:2016jky , Sola:2016hnq , Sola:2015wwa , Geng:2017apd , Rezaei:2019xwo, Geng:2020mga}. It is therefore not surprising that the mimicking of the RVM by the BD-$\CC$CDM model enjoys of the same virtues. In actual fact, the particular RVM form obtained in BD-gravity (we may call it ``BD-RVM'' for short)  is even more successful since it can cure both tensions, the $H_0$ and $\sigma_8$ one.  The reason why the BD-RVM can cure also the $H_0$-tension is because we need the evolution of the effective gravitational coupling $\Geff$ to achieve that, as we have seen in the preview \hyperref[Sect:Preview]{Sect.\,\ref{Sect:Preview}}, whereas the $\sigma_8$-tension can be cured with $\nueff$, which is associated to $\epsilon\propto-\eBD$  (the second ingredient characteristic of BD-gravity), and hence the two key elements are there to make a successful phenomenological performance.

On the other hand, from \eqref{Eq:SemiAnalytical.currentacceleration} it follows that the EoS for the effective DDE is
\begin{equation}\label{Eq:SemiAnalytical.EffEoS}
\weff(z)=\frac{p_{\Lambda}}{\rDE(H)}\simeq -1+\frac{3\nueff}{8\pi G \rL}\,H^2(z)=-1+\frac{\nueff}{\Omega_\CC}\,\frac{H^2(z)}{H_0^2}\,,
\end{equation}
where use has been made of \eqref{Eq:SemiAnalytical.rLeff}. It follows that the BD-RVM, in contrast to the original RVM, does not describe a  DE of pure vacuum form ($p_{\Lambda}=-\rL$) but a DE whose EoS departs slightly from the pure vacuum. In fact, for $\epsilon>0\ (\epsilon<0) $  we have $\nu_{\rm eff}>0\  (\nu_{\rm eff}<0)$ and the effective DDE behaves quintessence (phantom)-like.  For $\epsilon\to 0$ (hence  $\nu_{\rm eff}\to 0$) we have  $\weff\to -1$ ($\CC$CDM).   As could be expected, Eq.\,\eqref{Eq:BD.effEoS} is the BD-RVM version of the effective EoS that we obtained in \hyperref[Sect:EffectiveEoS]{Sect.\,\ref{Sect:EffectiveEoS}} -- see Eq.\,\eqref{Eq:BD.BDEoSz0}.  The two languages are consistent. Indeed, by comparison we see that $\nueff$ here plays the role of $\dvphi$ there. We know that $\dvphi=1-\varphi>0$, {\it i.e.}$\varphi<1$,  for $\eBD<0$, as we have shown previously, which is consistent with the fact that $\nueff\propto\epsilon\propto -\eBD>0$.  Finally, since $\weff$  approaches $-1$ from above (cf. \hyperref[Fig:BD.weffvarphi]{Fig.\ref{Fig:BD.weffvarphi}}) it corresponds to an effective quintessence behavior, which is more pronounced the more we explore the EoS into our past.

\newpage

\chapter{Fixed points in BD-$\CC$CDM cosmology of chapter 5}\label{Appendix:FixedPoints}

In order to study the fixed points of this system of differential equations we must define new variables such that the system becomes of first order in the derivatives when it is rewritten in terms of the new variables. It is useful though to firstly carry out the change $t\to N\equiv \ln(a)$. This preliminary step will help us to identify in an easier way how we must define the new variables. When written in terms of $N$ the system takes the following form\footnote{{Primes in this appendix stand for derivatives {\it w.r.t.}  to the variable $N=\ln a$, {\it i.e.}  $()^\prime \equiv d()/d N$. We consider, as in \hyperref[Appendix:Semi-Analytical]{Appendix\,\ref{Appendix:Semi-Analytical}}, three massless neutrinos.}}:
\begin{equation}\label{Eq:FixedPoints.System1}
\frac{\psi^{\pp}}{\psi}+\frac{H^\p}{H}\frac{\psi^\p}{\psi}+3\frac{\psi^\p}{\psi}=\frac{8\pi}{3+2\w}\left(\frac{\rho_{\rm m}+4\rho_\Lambda}{\psi H^2}\right)\,,
\end{equation}
\begin{equation}\label{Eq:FixedPoints.System2}
3+3\frac{\psi^\p}{\psi}-\frac{\w}{2}\left(\frac{\psi^\p}{\psi}\right)^2=\frac{8\pi}{\psi H^2}(\rho_{\rm r}+\rho_{\rm m}+\rho_\Lambda)\,,
\end{equation}
\begin{equation}\label{Eq:FixedPoints.System3}
3+\frac{H^\p}{H}\left(2+\frac{\psi^\p}{\psi}\right)+\frac{\psi^{\pp}}{\psi}+2\frac{\psi^\p}{\psi}+\frac{\w}{2}\left(\frac{\psi^\p}{\psi}\right)^2=\frac{8\pi}{\psi H^2}\left(\rho_\Lambda-\frac{\rho_{\rm r}}{3}\right)\,,
\end{equation}
\begin{equation}\label{Eq:FixedPoints.System4}
\rho^\p_{\rm r}+4\rho_{\rm r}=0\,,
\end{equation}
\begin{equation}\label{Eq:FixedPoints.System5}
\rho^\p_{\rm m}+3\rho_{\rm m}=0\,.
\end{equation}

\noindent
Now one can define the following quantities:
\begin{equation}\label{Eq:FixedPoints.xdef}
\xp\equiv \frac{\psi^\p}{\psi}\quad ; \quad x_i^2\equiv\frac{8\pi \rho_i}{H^2\psi}\,,
\end{equation}
\noindent
where $i=r,m,\Lambda$. In terms of these variables the system of equations can be easily written as follows:
\begin{equation}\label{Eq:FixedPoints.eq1}
\xp^\p+\xp^2+\frac{H^\p}{H}\xp+3\xp=\frac{\xm^2+4\xl^2}{3+2\w}\,,
\end{equation}
\begin{equation}\label{Eq:FixedPoints.eq2}
3+3\xp-\frac{\w}{2}\xp^2=\xr^2+\xm^2+\xl^2\,,
\end{equation}
\begin{equation}\label{Eq:FixedPoints.eq3}
3+\frac{H^\p}{H}\left(2+\xp\right)+\xp^\p+\xp^2+2\xp+\frac{\w}{2}\xp^2=\xl^2-\frac{\xr^2}{3}\,,
\end{equation}
\begin{equation}\label{Eq:FixedPoints.eq4}
\xr^\p=-\xr\left(\frac{H^\p}{H}+2+\frac{\xp}{2}\right)\,,
\end{equation}
\begin{equation}\label{Eq:FixedPoints.eq5}
\xm^\p=-\xm\left(\frac{H^\p}{H}+\frac{3}{2}+\frac{\xp}{2}\right)\,.
\end{equation}
\noindent
This system is of first order, as wanted. We have five equations and five unknowns, namely: $\xr,\xm,\xp,\xl,H^\p/H$. We can reduce significantly the complexity of the system if we just isolate $H^\p/H$ from \eqref{Eq:FixedPoints.eq1} and $\xl$ from \eqref{Eq:FixedPoints.eq2},
\begin{equation}\label{Eq:FixedPoints.Hder}
\frac{H^\p}{H}=\frac{1}{\xp}\left[\frac{\xm^2+4\xl^2}{3+2\w}-\xp^\p-3\xp-\xp^2\right]\,,
\end{equation}
\begin{equation}\label{Eq:FixedPoints.xL}
\xl^2=3+3\xp-\frac{\w}{2}\xp^2-\xr^2-\xm^2\,,
\end{equation}
\noindent
and substitute the resulting expressions in the other equations, {\it i.e.} in \eqref{Eq:FixedPoints.eq3}, \eqref{Eq:FixedPoints.eq4}, and \eqref{Eq:FixedPoints.eq5}. Doing this, and after a little bit of algebra, one finally obtains three equations written only in terms of $\xr,\xm,\xp$:
\begin{equation}\label{Eq:FixedPoints.DynSysPsi}
\xp^\p=-\xp\left[3+3\xp-\frac{1}{2}\w\xp^2-\frac{2}{3}\xr^2-\frac{\xm^2}{2}-\left(\frac{1}{\xp}+\frac{1}{2}\right)\left(\frac{12+12\xp-2\w\xp^2-4\xr^2-3\xm^2}{3+2\w}\right)\right]\,,
\end{equation}
\begin{equation}\label{Eq:FixedPoints.DynSysR}
\xr^\p=-\xr\left[2+\frac{5}{2}\xp-\frac{\w}{2}\xp^2-\frac{2}{3}\xr^2-\frac{\xm^2}{2}-\frac{1}{2}\left(\frac{12+12\xp-2\w\xp^2-4\xr^2-3\xm^2}{3+2\w}\right)\right]\,,
\end{equation}
\begin{equation}\label{Eq:FixedPoints.DynSysM}
\xm^\p=-\xm\left[\frac{3}{2}+\frac{5}{2}\xp-\frac{\w}{2}\xp^2-\frac{2}{3}\xr^2-\frac{\xm^2}{2}-\frac{1}{2}\left(\frac{12+12\xp-2\w\xp^2-4\xr^2-3\xm^2}{3+2\w}\right)\right]\,.
\end{equation}
\noindent
They allow us to search for the fixed points of the system. There is an important restriction produced by \eqref{Eq:FixedPoints.eq4} and \eqref{Eq:FixedPoints.eq5}. Supposing that $x_{\rm r} \neq 0$ and $x_{\rm m}\neq 0$, we see that the mentioned equations impose that
\begin{equation}\label{Eq:FixedPoints.xr}
\frac{x_{\rm r}^\prime}{x_{\rm r}}=-\frac{x_{\rm m}^\prime}{x_{\rm m}}+\frac{1}{2}.
\end{equation}
This equation is not compatible with the conditions of fixed point, so that we should assume that $x_{\rm r}=0$, $x_{\rm m}=0$ or both conditions at the same time. The fixed points are:
\newline
\newline
\textbf{RDE}
\begin{equation}\label{Eq:FixedPoints.xRDE}
(\xr,\xm,\xl,\xp)_{\rm RD} = \left(\sqrt{3},0,0,0\right)\,,
\end{equation}
the Jacobian of the nonlinear system \eqref{Eq:FixedPoints.DynSysPsi}, \eqref{Eq:FixedPoints.DynSysR}, \eqref{Eq:FixedPoints.DynSysM} has eigenvalues $\lambda^{RD}_1=-1$, $\lambda^{RD}_2 =1/2$, $\lambda^{RD}_3=4$ so that is an unstable point.
\newline
\newline
\textbf{MDE}
\begin{equation}\label{Eq:FixedPoints.xMDE}
(\xr,\xm,\xl,\xp)_{\rm MD} = \left(0,\frac{\sqrt{12+17\w+6\w^2}}{\sqrt{2}|1+\w|},0,\frac{1}{1+\w}\right)\,,
\end{equation}
the Jacobian of the nonlinear system \eqref{Eq:FixedPoints.DynSysPsi}, \eqref{Eq:FixedPoints.DynSysR}, \eqref{Eq:FixedPoints.DynSysM} has eigenvalues $\lambda^{MD}_1=-1/2,$ $\lambda^{MD}_2\approx-3/2,$ $\lambda^{MD}_3\approx 3$ so that is also an unstable point.
\newline
\newline
\textbf{$\Lambda$-dominated or VDE}
\begin{equation}\label{Eq:FixedPoints.xVDE}
(\xr,\xm,\xl,\xp)_{\rm \Lambda D} = \left(0,0,\frac{\sqrt{15+28\w+12\w^2}}{|1+2\w|},\frac{4}{1+2\w}\right)\,
\end{equation}
the Jacobian of the nonlinear system \eqref{Eq:FixedPoints.DynSysPsi}, \eqref{Eq:FixedPoints.DynSysR}, \eqref{Eq:FixedPoints.DynSysM} has eigenvalues $\lambda^{\Lambda}_1=-2,$ $\lambda^{\Lambda}_2 =-3/2,$ $\lambda^{\Lambda}_3 \approx -3$ so that is a stable point.

The first and second fixed points are unstable, whereas the latter is stable. We have assumed that $|\w|\gg1$ for approximating the eigenvalues.

The first one is very well-known, since regardless of the initial conditions for the BD scalar field we already know that the velocity of the scalar field decays during the RDE, and the solution tends to the attractor with $\psi^\p=0$ ($\xp=0$) and full domination of radiation, {\it i.e.} $\xr=\sqrt{3}$. The Hubble function and BD scalar field are:
\begin{equation}\label{Eq:FixedPoints.psiRD}
\psi_{RD}(a)=\psi_{\rm r}\,,
\end{equation}
\begin{equation}\label{Eq:FixedPoints.h2}
H^2_{RD}(a)=\frac{8\pi}{3\psi_{\rm r}}\rho^{0}_{\rm r}a^{-4}\,,
\end{equation}
where $\psi_{\rm r}$ is an arbitrary constant and $\rho^{0}_{\rm r}$ is the value of the radiation energy density at present.
This fixed point is unstable because at some moment nonrelativistic matter starts to dominate the expansion. When this happens the solution starts to look for the new attractor, the one of the MDE. We stress  that this is an exact solution. The  BD scalar field and the Hubble function during the MDE take the following form:
\begin{equation}\label{Eq:FixedPoints.psiMDE}
\psi_{MD}(a)=Ca^{\frac{1}{1+\w}}\,,
\end{equation}
\begin{equation}\label{Eq:FixedPoints.HsqMDE}
H^2_{MD}(a)=\frac{16\pi \rho_{\rm m}^{0}(1+\w)^2a^{-\left(\frac{4+3\w}{1+\w}\right)}}{C(12+17\w+6\w^2)}\,,
\end{equation}
\noindent
where $C$ is an arbitrary constant and $\rho_{\rm m}^{0}$ is the value of the matter energy density at present. The MD fixed point is, again, unstable, because the MDE finishes and the VDE starts. The solution searches now for the last fixed point, which is stable. In this last case, we have obtained the following Hubble function and BD-field,
\begin{equation}\label{Eq:FixedPoints.psiL}
\psi_\Lambda (a)= D a^{4/(1+2\w)}\,,
\end{equation}
\begin{equation}\label{Eq:FixedPoints.H2L}
H^2_\Lambda (a)=\frac{8\pi \rho_\Lambda }{D}\frac{(1+2\w)^2}{15+28\w +12 \w^2}a^{-4/(1+2\w)}\,,
\end{equation}
where $D$ is an arbitrary constant and $\rho_\Lambda$ is the constant value of vacuum energy.

One can easily check that the solutions computed in \hyperref[Appendix:Semi-Analytical]{Appendix\,\ref{Appendix:Semi-Analytical}}, of first order in $\epsilon_{\rm BD}=1/\w$, coincide (once the decaying modes become irrelevant) with the ones presented here when the latter are Taylor-expanded up to first order in this parameter as well. The results we have found here are consistent with previous studies on fixed points in cosmological dynamical systems, see e.g. \cite{Bahamonde:2017ize} and references therein.
\blankpage

\chapter{Cosmological perturbations in the synchronous gauge for Brans-Dicke models of chapter 5} \label{Appendix:CosmoPerturbationsSynchronous}

In this appendix we explicitly derive the set of perturbed equations for the Brans-Dicke model in the synchronous gauge, and up to linear order in the perturbed quantities. The perturbed FLRW metric written in conformal time $\eta$ reads
\begin{equation}\label{Eq:Synch.LineElementSyn}
ds^2=a^2(\eta)[-d\eta^2+(\delta_{ij}+h_{ij})dx^idx^j]\,,
\end{equation}
where {$\eta$ is the conformal time, $\vec{x}$ the spatial comoving coordinates, and} $h_{ij}$ is the metric perturbation. We apply here the usual metric formalism and therefore we assume that the Christoffel symbols are unequivocally determined by the metric through the Levi-Civita connection. Thus, the perturbed Chrystoffel symbols together with the perturbed Riemann and Ricci tensors, and the perturbed Ricci scalar, can be easily written in terms of $h_{ij}$, its trace $h\equiv \delta^{ij}h_{ij}$, and their spacetime derivatives. Although the corresponding expressions can be already found in the literature, we have opted to provide them together with the perturbed energy-momentum tensor in \hyperref[Sect:PertGeoEM]{Sect.\,\ref{Sect:PertGeoEM}} for completeness. In \hyperref[Sect:PertEqPos]{Sect.\,\ref{Sect:PertEqPos}} and \hyperref[Sect:PertEqMom]{Sect.\,\ref{Sect:PertEqMom}} we write the main perturbed equations in position and momentum space, respectively, and in \hyperref[Sect:PertEqLowScales]{Sect.\,\ref{Sect:PertEqLowScales}} we derive and discuss the equation that rules the growth of matter perturbations in the matter and $\Lambda$-dominated Universe at deep subhorizon scales. {In \hyperref[Sect:IC]{Sect.\,\ref{Sect:IC}} we provide a brief note on the initial conditions for all the perturbed quantities. Finally, in \hyperref[Sect:GWs]{Sect.\,\ref{Sect:GWs}} we discuss tensor perturbations in the BD-$\Lambda$CDM cosmology.}


\section{Perturbed geometric quantities, energy-momentum tensor, and other relevant terms appearing in the field equations}\label{Sect:PertGeoEM}

The elements of the metric tensor and its inverse can be straightfordwardly obtained from \eqref{Eq:Synch.LineElementSyn}. They read as follows,
\begin{equation}\label{Eq:Synch.g00}
\begin{split}
&g_{00}=-a^2\qquad g_{ij}=a^2(\delta_{ij}+h_{ij})\quad ; \quad g^{00}=-1/a^2\qquad g^{ij}=\frac{1}{a^2}(\delta_{ij}-h_{ij}).
\end{split}
\end{equation}
Plugging them into the formula of the Levi-Civita connection one gets the following expressions for the Christoffel symbols:
\begin{equation}\label{Eq:Synch.Gamma}
\begin{split}
&\Gamma^{0}_{00}=\mathcal{H}\,,\qquad \Gamma^{0}_{i0}=\Gamma^{i}_{00}=0\,,\qquad \Gamma^{0}_{ij}=\mathcal{H}(\delta_{ij}+h_{ij})+\frac{h^\prime_{ij}}{2}\,,\\
&\Gamma^{i}_{j0}=\mathcal{H}\delta_{ij}+\frac{h^\prime_{ij}}{2}\,,\qquad \Gamma^{i}_{jl}=\frac{1}{2}\left(h_{ij,l}+h_{il,j}-h_{jl,i}\right)\,,\\
\end{split}
\end{equation}
where each prime denotes a derivative with respect to the conformal time, {\it i.e.} \,$d/d\eta$,  the lower commas partial derivatives with respect to spatial (comoving) coordinates, and $\mathcal{H}\equiv a^\prime/a$. The contributions of all the second and higher order terms in perturbation theory have been neglected since we are not interested here in analyzing nonlinear structure formation processes, {in which also the details of the screening mechanism acting in the nonlinear regime could become important}. The Ricci tensor components can be computed making use of the above formulas,
\begin{equation}\label{Eq:Synch.Rmunu}
\begin{split}
R_{00}=&-3\mathcal{H}^\prime-\frac{h^{\prime \prime}}{2}-\frac{\mathcal{H}}{2}h^\prime,\qquad R_{0i}=\frac{1}{2}\left(\partial_jh^\prime_{ij}-\partial_{i}h^\prime\right)\,,\\
R_{ij}=&(\delta_{ij}+h_{ij})(\mathcal{H}^\prime+2\mathcal{H}^2)+\frac{h^{\prime\prime}_{ij}}{2}+\frac{\mathcal{H}}{2}h^\prime\delta_{ij}+\mathcal{H}h^\prime_{ij}+\frac{1}{2}(h_{li,jl}+h_{lj,il}-h_{ij,ll}-h_{,ij})\,.
\end{split}
\end{equation}
Notice that we have applied Einstein's summation convention. Contracting the Ricci tensor with the metric we finally obtain the scalar curvature,
\begin{equation}\label{Eq:Synch.Ricci}
\begin{split}
a^2R=6(\mathcal{H}^\prime+\mathcal{H}^2)+h^{\prime\prime}+3\mathcal{H}h^\prime+h_{li,li}-h_{,ll}\,.
\end{split}
\end{equation}
Equipped with these tools we can proceed to compute the components of the Einstein tensor, which read
\begin{equation}\label{Eq:Synch.PerturbEinstTensor}
\begin{split}
G_{00}=&3\mathcal{H}^2+\mathcal{H}h^\prime+\frac{1}{2}\left(h_{li,li}-h_{,ll}\right)\,,\\
G_{i0}=&\frac{1}{2}\left(\partial_jh^\prime_{ij}-\partial_{i}h^\prime\right)\,,\\
G_{ij}=&-(\delta_{ij}+h_{ij})(2\mathcal{H}^\prime+\mathcal{H}^2)+\frac{h^{\prime\prime}_{ij}}{2}-\frac{h^{\prime\prime}}{2}\delta_{ij}-\mathcal{H}h^\prime\delta_{ij}+\mathcal{H}h^\prime_{ij}\\
&+\frac{1}{2}(h_{li,jl}+h_{lj,il}-h_{ij,ll}-h_{,ij}-h_{lt,lt}\delta_{ij}+h_{,ll}\delta_{ij})\,.
\end{split}
\end{equation}
It is also convenient to obtain the trace of $G_{ij}$, since it will be employed in subsequent calculations,
\begin{equation}\label{Eq:Synch.Gmunu}
G_{ii}=-(3+h)(2\mathcal{H}^\prime+\mathcal{H}^2)-h^{\prime\prime}-2\mathcal{H}h^\prime+\frac{1}{2}(h_{,ll}-h_{li,li})\,.
\end{equation}
As in Ref.\,\cite{Ma:1995ey}, we can express $h_{ij}$ as follows,
\begin{equation}\label{Eq:Synch.hFourier}
h_{ij}(\eta,\vec{x})=\int d^3k\, e^{-i\vec{k}\cdot\vec{x}}\left[\hat{k}_i\hat{k}_j h(\eta,\vec{k})+\left(\hat{k}_i\hat{k}_j-\frac{\delta_{ij}}{3}\right)6\xi(\eta,\vec{k})\right]\,,
\end{equation}
where $\hat{k}_i=k_i/k$ with $k=|\vec{k}|$, and $h(\eta,\vec{k})$ the Fourier transform of the trace of $h_{ij}(\eta,\vec{x})$. When we work in Fourier space we will denote it $h$, like in position space, without specifying its dependence on the wave number $\vec{k}$ explicitly. Plugging \eqref{Eq:Synch.hFourier} into the perturbed part of \eqref{Eq:Synch.PerturbEinstTensor} we obtain the elements of the perturbed Einstein tensor in Fourier space. We will employ them later on. They read,
\begin{equation}\label{Eq:Synch.PerturbEinstTensor2}
\begin{split}
\delta G_{00}=&\mathcal{H}h^\prime-2\xi k^2\,,\\
\delta G_{i0}=&-ik_i2\xi^\prime\,,\\
\delta G_{ii}=&-h(2\mathcal{H}^\prime+\mathcal{H}^2)-h^{\prime\prime}-2\mathcal{H}h^\prime+2\xi k^2\,.
\end{split}
\end{equation}
We do not write here the Fourier transform of $\delta G_{ij}$ because we will not use it later.

In order to compute the perturbed energy-momentum tensor of the perfect fluids that fill the Universe with Eq. \eqref{Eq:BD.EMT} we must know which is the form of their perturbed 4-velocities. It is easy to show that they are just given by
\begin{equation}\label{Eq:Synch.umu}
u^\mu=\frac{1}{a}(1,v^i)\qquad ;\qquad u_\mu=a (-1,v^i)\,,
\end{equation}
with $v^i=\frac{dx^i}{d\eta}$. Using this in Eq. \eqref{Eq:BD.EMT} and splitting the total energy density and pressure in a background and a perturbed parts, {\it i.e.} considering $\rho(\eta,\vec{x})=\bar{\rho}(\eta)+\delta\rho(\eta,\vec{x})$ and $p(\eta,\vec{x})=\bar{p}(\eta)+\delta p(\eta,\vec{x})$, we obtain the following elements of the perturbed energy-momentum tensor,
\begin{equation}\label{Eq:Synch.Tmunu}
\begin{split}
T_{00} =&\, a^2(\bar{\rho}+\delta\rho)\,,\\
T_{ij} =&\, a^2\bar{p}(\delta_{ij}+h_{ij})+a^2\delta_{ij}\delta p\,,\\
T_{0i} =&\, -a^2(\bar{p}+\bar{\rho})v^i\,,\\
T \equiv &\, g^{\mu\nu}T_{\mu\nu} = 3(\bar{p}+\delta p)-\bar{\rho}-\delta\rho\,,
\end{split}
\end{equation}
where a sum over all the species in the Universe is taken for granted. The following quantities will also be useful in subsequent calculations.
\begin{equation}\label{Eq:Synch.PertQuant}
\begin{split}
\partial_\alpha\varphi \partial^\alpha\varphi=&\,-\frac{1}{a^2}\left[(\bar{\varphi}^\prime)^2+2\bar{\varphi}^\prime\delta\varphi^\prime\right] \,,\\
a^2\Box\varphi =&\, -\bar{\varphi}^{\prime\prime}-2\mathcal{H}\bar{\varphi}^\prime-\delta\varphi^{\prime\prime}+\nabla^2\delta\varphi-2\mathcal{H}\delta\varphi^\prime-\frac{h^\prime}{2}\bar{\varphi}^\prime\,,\\
\nabla_0\nabla_0\varphi=&\,\bar{\varphi}^{\prime\prime}+\delta\varphi^{\prime\prime}-\mathcal{H}\bar{\varphi}^\prime-\mathcal{H}\delta\varphi^\prime\,,\\
\nabla_i\nabla_0\varphi =&\,\partial_i\delta\varphi^\prime-\mathcal{H}\partial_i\delta\varphi\,,\\
\nabla_i\nabla_j =& \partial_i\partial_j\delta\varphi-\bar{\varphi}^\prime\left[\mathcal{H}(\delta_{ij}+h_{ij})+\frac{h_{ij}^\prime}{2}\right]-\delta_{ij}\mathcal{H}\delta\varphi^\prime\,.
\end{split}
\end{equation}
Here we have split the BD-field as the sum of the mean (background) field $\bar{\varphi}$ and its corresponding perturbation $\delta\varphi$, {\it i.e.} $\varphi(\eta,\vec{x})=\bar{\varphi}(\eta)+\delta\varphi(\eta,\vec{x})$.


\section{Perturbation equations in position space}\label{Sect:PertEqPos}

We apply now the machinery derived in the previous subsection to perturb the modified Einstein's Eqs. \eqref{Eq:BD.BDFieldEquation1}, the covariant conservation of the energy-momentum tensor and the Klein-Gordon equation \eqref{Eq:BD.BDFieldEquation2}. The last one reads,
\begin{equation}\label{Eq:Synch.pertKG}
-\delta \varphi^{\prime \prime}-2\mathcal{H}\delta \varphi^\prime+\nabla^2 \delta \varphi-\frac{h^\prime}{2}\bar{\varphi}^\prime =\frac{8\pi G_N}{3+2\omega_{\rm BD}}a^2(3\delta p-\delta\rho)\, ,
\end{equation}
where $\nabla^2 \equiv \sum\limits_{i=1}^{3}\partial_i^2$. The perturbed $00$, $0i$, and $ij$ components of Einstein's equation lead, respectively, to
\begin{equation}\label{Eq:Synch.pertT00}
\begin{split}
\bar{\varphi}\left(\mathcal{H}h^\prime+\frac{h_{li,li}-h_{,ll}}{2}\right)+& 3\mathcal{H}^2\delta\varphi-\nabla^2\delta\varphi+3\mathcal{H}\delta\varphi^\prime+\frac{h^\prime}{2}\bar{\varphi}^\prime\\
&+\frac{\omega_{\rm BD}}{2\bar{\varphi}}\left[\frac{\delta\varphi}{\bar{\varphi}}(\bar{\varphi}^\prime)^2-2\bar{\varphi}^\prime\delta\varphi^\prime\right] = 8\pi G_N a^2\delta \rho\,,
\end{split}
\end{equation}
\begin{equation}\label{Eq:Synch.pertT0i}
\begin{split}
\bar{\varphi}\left(\frac{\partial_{j}h^\prime_{ij}-\partial_ih^\prime}{2}\right)-\partial_i\delta\varphi^\prime+\mathcal{H}\partial_i\delta\varphi-\frac{\omega_{\rm BD}}{\bar{\varphi}}\bar{\varphi}^\prime\partial_i\delta\varphi=-8\pi G_N a^2(\bar{\rho}+\bar{p})v^i\,,
\end{split}
\end{equation}
\begin{equation}\label{Eq:Synch.pertTij}
\begin{split}
\delta_{ij} & \left[  -\delta\varphi(2\mathcal{H}^\prime+\mathcal{H}^2)-\bar{\varphi}\left(\frac{h^{\prime\prime}}{2}+\mathcal{H}h^\prime\right)-\mathcal{H}\delta\varphi^\prime-\delta\varphi^{\prime\prime}+\nabla^2\delta\varphi-\frac{h^\prime}{2}\bar{\varphi}^\prime+\bar{\varphi}\left(\frac{h_{,ll}-h_{lt,lt}}{2}\right)\right.\\
&\left. +\frac{\omega_{\rm BD}}{2\bar{\varphi}}\left(\frac{\delta\varphi}{\bar{\varphi}}(\bar{\varphi}^\prime)^2-2\bar{\varphi}^\prime\delta\varphi^\prime\right)\right]+h_{ij}^\prime\left(\mathcal{H}\bar{\varphi}+\frac{\bar{\varphi}^\prime}{2}\right)-\partial_i\partial_j\delta\varphi+\frac{h_{ij}^{\prime\prime}}{2}\bar{\varphi}\\
&-h_{ij}\left[\bar{\varphi}^{\prime\prime}+\bar{\varphi}(2\mathcal{H}^\prime+\mathcal{H}^2) +\mathcal{H}\bar{\varphi}^\prime+\frac{\omega_{\rm BD}}{2\bar{\varphi}}(\bar{\varphi}^\prime)^2\right]+\frac{\bar{\varphi}}{2}\left(h_{li,jl}+h_{lj,il}-h_{ij,ll}-h_{,ij}\right)=8\pi G_N \delta T_{ij}\,.
\end{split}
\end{equation}
Finally, the covariant conservation of the energy-momentum tensor leads to the following pair of extra equations, {\it i.e.} $\nabla^\mu T_{\mu \nu}=0$, for $\nu=0$ and $\nu=i$, respectively,
\begin{equation}\label{Eq:Synch.covariant0}
\sum_{j} \bar{\rho}_{(j)}\left[\delta^\prime_{(j)}+3\mathcal{H}\left(\frac{\delta p_{(j)}}{\delta \rho_{(j)}}-{w}_{(j)}\right)\delta_{(j)}+(1+{w}_{(j)})\left(\theta_{(j)}+\frac{h^\prime}{2}\right)\right]=0\,,
\end{equation}
\begin{equation}\label{Eq:Synch.covarianti}
\sum_{j}\bar{\rho}_{(j)}(1+{w}_{(j)})\left[\theta^\prime_{(j)}+\left( \mathcal{H}(1-3{w}_{(j)})+\frac{{w}_{(j)}^\prime}{1+{w}_{(j)}} \right)\theta_{(j)}+\frac{\nabla^2\delta p_{(j)}}{(1+{w}_{(j)})\bar{\rho}_{(j)}}+\frac{{w}_{(j)}}{1+{w}_{(j)}}\partial_i\partial_l h_{il}\right]=0\,,
\end{equation}
where $\theta_{(j)}\equiv \partial_i v^i_{(j)}$, $\delta_{(j)}\equiv \delta \rho_{(j)}/\bar{\rho}_{(j)}$, ${w}_{(j)}\equiv \bar{p}_{(j)}/\bar{\rho}_{(j)}$, and the sums run over all the species $j$ that fill the Universe.


\section{Perturbation equations in momentum space}\label{Sect:PertEqMom}

As it is well-known, working in momentum space simplifies a lot the treatment of the cosmological perturbations, basically because at linear order in perturbation theory the different modes of the perturbed quantities do not couple to each other, {\it i.e.} there is no mixture of wave numbers and we can safely omit the subscript $k$ for the modes. Here we limit ourselves to just write the expressions provided in the previous subsection, but in momentum space. The calculations are straightforward and no further details are thus needed.
\newline
\newline
\noindent Equation \eqref{Eq:Synch.pertKG}
\begin{equation}\label{Eq:Synch.pertKGk}
\delta \varphi^{\prime \prime}+2\mathcal{H}\delta \varphi^\prime+k^2\delta \varphi+\frac{h^\prime}{2}\bar{\varphi}^\prime =\frac{8 \pi G_N}{3+2\omega_{\rm BD}}a^2(\delta \rho-3\delta p)\,.
\end{equation}
Equation \eqref{Eq:Synch.pertT00}
\begin{equation}\label{Eq:Synch.pertT00k}
\begin{split}
\bar{\varphi}(\mathcal{H}h^\prime-2\xi k^2)+(3\mathcal{H}^2+&k^2)\delta\varphi+3\mathcal{H}\delta\bar{\varphi}^\prime+\frac{h^\prime}{2}\bar{\varphi}^\prime\\
&+\frac{\omega_{\rm BD}}{2\bar{\varphi}}\left[\frac{\delta\varphi}{\bar{\varphi}}(\bar{\varphi}^\prime)^2-2\bar{\varphi}^\prime\delta\varphi^\prime\right]=8\pi G_N a^2 \delta \rho\,.
\end{split}
\end{equation}
Equation \eqref{Eq:Synch.pertT0i}
\begin{equation}
-2\bar{\varphi}k^2\xi^\prime+k^2\delta\varphi^\prime-\mathcal{H}k^2\delta \varphi+\omega_{\rm BD}k^2\delta \varphi\left(\frac{\bar{\varphi}^\prime}{\bar{\varphi}}\right)=-8\pi G_N a^2 (\bar{\rho}+\bar{p})\theta\,.
\end{equation}
Trace of equation \eqref{Eq:Synch.pertTij}, and after making use of the pressure equation \eqref{Eq:BD.pressureequation}
\begin{equation}
\begin{split}
-\delta \varphi(6\mathcal{H}^\prime+3\mathcal{H}^2+2k^2)-3\delta \varphi^{\prime \prime}-3\mathcal{H}\delta \varphi^\prime&-\bar{\varphi}^\prime h^\prime+\frac{3\omega_{\rm BD}\bar{\varphi}^\prime}{2\bar{\varphi}}\left(\frac{\bar{\varphi}^\prime}{\bar{\varphi}}\delta\varphi-2\delta \varphi^\prime\right) \\
&+\bar{\varphi}\left( -h^{\prime \prime}-2h^\prime \mathcal{H}+2k^2 \xi \right)=24\pi G_N a^2 \delta p\,.
\end{split}
\end{equation}
Equation \eqref{Eq:Synch.covariant0}
\begin{equation}\label{Eq:Synch.covariant0k}
\sum_{j} \bar{\rho}_{(j)}\left[\delta^\prime_{(j)}+3\mathcal{H}\left(\frac{\delta p_{(j)}}{\delta \rho_{(j)}}-{w}_{(j)}\right)\delta_{(j)}+(1+{w}_{(j)})\left(\theta_{(j)}+\frac{h^\prime}{2}\right)\right]=0\,.
\end{equation}
Equation \eqref{Eq:Synch.covarianti}
\begin{equation}\label{Eq:Synch.covariantiK}
\sum_{j}\bar{\rho}_{(j)}(1+{w}_{(j)})\left[\theta^\prime_{(j)}+\left( \mathcal{H}(1-3{w}_{(j)})+\frac{{w}_{(j)}^\prime}{1+{w}_{(j)}} \right)\theta_{(j)}-\frac{k^2\delta_{(j)}}{(1+{w}_{(j)})}\frac{\delta p_{(j)}}{\delta\rho_{(j)}}-\frac{k^2{w}_{(j)}}{1+{w}_{(j)}}(h+4\xi)\right]=0\,.
\end{equation}
Another useful and compact relation can be obtained from the $\hat{k}_i\hat{k}_j$ part of Eq.\,\eqref{Eq:Synch.pertTij} in momentum space. The result, after using again the pressure equation\,\eqref{Eq:BD.pressureequation}, reads
\begin{equation}\label{Eq:Synch.kikjk}
\begin{split}
&h^{\prime \prime}+6\xi^{\prime \prime}+(h^\prime +6\xi^\prime)\left(2\mathcal{H}+\frac{\bar{\varphi}^\prime}{\bar{\varphi}}\right)+2k^2\left(\frac{\delta\varphi}{\varphi}-\xi\right)=0\,.
\end{split}
\end{equation}
%

\section{Matter density contrast equation at deep subhorizon scales}\label{Sect:PertEqLowScales}

Let us restrict us now to the matter and $\Lambda$-dominated epochs and see what is the evolution of matter perturbations in the late stages of the Universe's expansion and deeply inside the horizon. Using the fact that vacuum does not cluster when it is described by a cosmological constant and matter is covariantly conserved, and also taking into account that radiation has only very mild impact on the Large-Scale Structure formation processes we want to study, we can obtain the following relation from \eqref{Eq:Synch.covariantiK},
\begin{equation}\label{Eq:Synch.thetam}
\theta^\prime_{\rm m}=-\mathcal{H}\theta_{\rm m}\,.
\end{equation}
This leads to a decaying solution for the velocity potential gradient, $\theta_{\rm m}=\theta_{\rm m}^{0}/a$, and in practice we can take $\theta_{\rm m}\sim 0$. By doing this in \eqref{Eq:Synch.covariant0k} we find
\begin{equation}\label{Eq:Synch.simpli0}
\delta_{\rm m}^\prime=-\frac{h^\prime}{2}\,.
\end{equation}
At low scales, Eqs. \eqref{Eq:Synch.pertKGk}, \eqref{Eq:Synch.pertT00k} and \eqref{Eq:Synch.kikjk} simplify, after neglecting some terms which are clearly subdominant at large $k$'s, giving rise to
\begin{equation}\label{Eq:Synch.simpli1}
k^2\delta \varphi+\frac{h^\prime}{2}\bar{\varphi}^\prime =\frac{8 \pi G_N}{3+2\omega_{\rm BD}}a^2 \bar{\rho}_{\rm m}\delta_{\rm m}\,,
\end{equation}
\begin{equation}\label{Eq:Synch.simpli2}
\bar{\varphi}(\mathcal{H}h^\prime-2\xi k^2)+k^2\delta\varphi+\frac{h^\prime}{2}\bar{\varphi}^\prime=8\pi G_N a^2 \bar{\rho}_{\rm m}\delta_{\rm m}\,,
\end{equation}
\begin{equation}\label{Eq:Synch.simpli3}
2k^2\delta \varphi+\bar{\varphi}^\prime h^\prime+\bar{\varphi}\left(h^{\prime \prime}+2h^\prime \mathcal{H}-2k^2 \xi\right)=0\,.
\end{equation}
Using \eqref{Eq:Synch.simpli0} in \eqref{Eq:Synch.simpli1} one can isolate $\delta\varphi=\delta\varphi(\delta_{\rm m},\delta_{\rm m}^\prime)$, and doing the same in \eqref{Eq:Synch.simpli2} one gets $\xi=\xi(\delta_{\rm m},\delta_{\rm m}^\prime)$. Introducing these expressions in \eqref{Eq:Synch.simpli3}, and after making use again of \eqref{Eq:Synch.simpli0}, one finally obtains the equation for the matter density contrast at deep subhorizon scales,
\begin{equation}\label{Eq:Synch.deltam}
\delta_{\rm m}^{\prime\prime}+\mathcal{H}\delta_{\rm m}^\prime-\frac{4\pi G_N a^2}{\bar{\varphi}}\bar{\rho}_{\rm m}\delta_{\rm m}\left(\frac{4+2\omega_{\rm BD}}{3+2\omega_{\rm BD}}\right)=0\,.
\end{equation}
 The expression in terms of the scale factor is given in the main text, Eq.\,\eqref{Eq:BD.ExactPerturScaleFactor}, where we recall that primes there mean $d/da$ whereas here $d/d\eta$. A quick comparison of the last term of this equation with the one that is obtained in the GR-$\Lambda$CDM allows us to note that the effective value of the gravitational constant that is controlling the formation of linear structures at subhorizon scales is
\begin{equation}\label{Eq:Synch.Geffective}
G_{{\rm eff}}(\bar{\varphi})=\frac{G_N}{\bar{\varphi}}\left(\frac{4+2\omega_{\rm BD}}{3+2\omega_{\rm BD}}\right)\,.
\end{equation}
{More details are provided in the main body of the thesis, see \hyperref[Sect:StructureFormationBD]{Sect.\,\ref{Sect:StructureFormationBD}}.}


\section{Brief note on the initial conditions}\label{Sect:IC}

We consider adiabatic initial conditions for the various species filling the Universe. For the DM velocity divergence, we use the usual synchronous condition $\theta_{cdm,ini}=0$.  We would like to point out here that the initial perturbation of the BD-field and its time derivative can also be set to zero. This is because the modes of interest were superhorizon modes during the radiation-dominated epoch, and in that period of the Universe's expansion Eq.\,\eqref{Eq:Synch.pertKG} reduces to
\begin{equation}\label{Eq:Synch.deltavarphi}
\delta\varphi^{\pp}+\frac{2}{\eta}\delta\varphi^\p+k^2\delta\varphi=0\,,
\end{equation}
where we have used $\mathcal{H}=\eta^{-1}$. The solution of this equation reads,
\begin{equation}\label{Eq:Synch.deltavarphi2}
\delta\varphi(k,\eta)=\frac{A(k)}{k\eta}\cos(k\eta+\beta(k))\,,
\end{equation}
with $A(k)$ and $\beta(k)$ being an amplitude and a phase, respectively. This solution corresponds to a damped oscillation, which is decaying fastly and can be naturally set to zero \cite{Chen:1999qh}. The initial conditions are thus equal to the ones in the GR-$\Lambda$CDM scenario \cite{Ma:1995ey}, but substituting $G_N$ by $G(\bar{\varphi}_{\rm ini})$.


\section{Gravitational waves in BD-\texorpdfstring{$\Lambda$CDM}{LCDM} cosmology}\label{Sect:GWs}

Gravitational waves (GWs) are given by the traceless and transverse part of the metric fluctuations, $h^T_{ij}$, which contains two degrees of freedom (corresponding to the two polarization states, usually denoted as $\times$ and $+$). Hence, they satisfy $h^T=0$ and $\partial_i h^T_{ij}=0$. As scalar, vector and tensor cosmological perturbations decouple from each other at linear order, we can consider the line element
\begin{equation}\label{Eq:Synch.ds2}
ds^2=a^2(\eta)[-d\eta^2+(\delta_{ij}+h^T_{ij})dx^i dx^j]
\end{equation}
and study the ij component of Einstein's equations. In order to do so we can directly take the traceless and transverse part of equation \eqref{Eq:Synch.pertTij}. We obtain,
\begin{equation}\label{Eq:Synch.Gij}
-h_{ij}\left[\bar{\varphi}(2\mathcal{H}^\prime+\mathcal{H}^2)+\bar{\varphi}^\pp+\mathcal{H}\bar{\varphi}^\p +\frac{\omega_{\rm BD}}{2\bar{\varphi}}(\bar{\varphi}^\p)^2\right]+h_{ij}^\p\left(\mathcal{H}\bar{\varphi}+\frac{\bar{\varphi}^\p}{2}\right)+\frac{\bar{\varphi}}{2}h^\pp_{ij}+\frac{\bar{\varphi}}{2} k^2 h_{ij}=8\pi G_N a^2\bar{p}\,h_{ij}\,.
\end{equation}
Notice that we have omitted the superscript $T$ for simplicity, doing $h^T_{ij}\rightarrow h_{ij}$. This equation can be reduced by using the background pressure equation \eqref{Eq:BD.pressureequation}, yielding
\begin{equation}\label{Eq:Synch.hdiffeq}
h^\pp_{ij}+h^\p_{ij}\left(2\mathcal{H}+\frac{\bar{\varphi}^\p}{\bar{\varphi}}\right)+k^2 h_{ij}=0\,.
\end{equation}
For a general scalar-tensor theory of gravity one has
\begin{equation}\label{Eq:Synch.hdiffeq2}
h^\pp_{ij}+\mathcal{H}h^\p_{ij}\left(2+\alpha_{\rm M}\right) + k^2(1+\alpha_T)h_{ij} = 0\,,
\end{equation}
where $\alpha_{\rm M}$ and $\alpha_T$  are functions that parametrize the deviations from standard GR. The former modifies the friction term, and is basically the running of the effective Planck mass, whereas the latter is directly related with the speed of propagation of the GWs,  $c_{gw}$, since $\alpha_T=c^2_{gw}-1$. In BD, $\alpha_{\rm M}=d\ln(\bar{\varphi})/d\ln(a)$ (e.g. at leading order in $\epsilon_{\rm BD}$, we have $\alpha_{\rm M}=\epsilon_{\rm BD}$ in the pure MDE and $\alpha_{\rm M}=2\epsilon_{\rm BD}$ in the VDE, cf. \hyperref[Appendix:FixedPoints]{Appendix\,\ref{Appendix:FixedPoints}}) and $\alpha_T=0$, so $c_{gw}=1$. This function, $\alpha_T$, has been recently constrained to be $|\alpha_T(z\approx 0)|\lesssim 10^{-15}$ using the measurement of the gravitational wave event GW170817 and the accompanying electromagnetic counterpart GRB170817A \cite{TheLIGOScientific:2017qsa}, located both at a distance of $40^{+8}_{-14}$ Mpc from us. BD theory automatically satisfies this constraint \cite{Creminelli:2017sry,Ezquiaga:2017ekz}, since GWs propagate exactly at the speed of light.


\blankpage

\chapter{Pertubation theory in Newtonian gauge for Brans-Dicke Models of chapter 5}\label{Appendix:PerturbationTheoryNewtonian}
In the main text, we have been working with the synchronous gauge in BD linear perturbation theory. For completeness we are going to provide the equations in the conformal Newtonian (or longitudinal) gauge. {This appendix has the same structure as the previous one (we only skip the recomputation of tensor perturbations in this gauge, and the discussion of the initial conditions)}. In particular, we will show that the same density contrast differential equation for subhorizon scales that we found in \hyperref[Sect:PertEqLowScales]{Sect.\,\ref{Sect:PertEqLowScales}} arises also for the Newtonian gauge, as expected. In the last section we provide the transformation equations of the perturbed quantities from one gauge to another in analogy to \cite{Ma:1995ey}.


\section{Perturbed geometric quantities, energy-momentum tensor, and other relevant terms appearing in the field equations}\label{PertGeometricNewton}
The square of the line element in the perturbed flat FLRW Universe in the Newtonian Gauge reads as follows,
\begin{equation}\label{Eq:NewtGauge.ds2}
ds^2=a^2[-(1+2\Phi)d\eta^2+(1+2\Psi)\delta_{ij}dx^idx^j]\,,
\end{equation}
where the pair $\Phi$ and $\Psi$ the so-called Bardeen potentials, which are functions of conformal time and space. In the following we provide the perturbed expressions (at linear order) for the various geometrical quantities that will be used later. As before, the primes denote a derivative with respect to the conformal time, and $\mathcal{H}\equiv a^\prime/a$.
Thus, the metric elements are
\begin{equation}\label{Eq:NewtGauge.g00}
\begin{split}
g_{00}=-a^2 (1+2\Phi),\qquad g_{ij}=a^2 (1+2\Psi)\delta_{ij} \quad; \quad g^{00}=-\frac{1}{a^2}(1-2\Phi), \qquad g^{ij}=\frac{1}{a^2}(1-2\Psi)\delta_{ij}.
\end{split}
\end{equation}
We can compute the Christoffel symbols associated to the metric in the usual way:
\begin{equation}\label{Eq:NewtGauge.Gamma}
\Gamma^0_{00}=\mathcal{H}+\Phi^\prime, \qquad \Gamma^0_{0i}=\Gamma^i_{00}=\partial_i\Phi, \qquad \Gamma^0_{ij}=\delta_{ij}[\mathcal{H}(1+2\Psi-2\Phi)+\Psi^\prime],
\end{equation}
$$\qquad \Gamma^i_{j0}=\delta^i_j (\mathcal{H}+\Psi^\prime) ,\qquad \Gamma^i_{jl}=\delta^i_j\partial_l\Psi+\delta^i_l\partial_j\Psi-\delta_{jl}\partial_i\Psi .$$
The components of the Ricci tensor are
\begin{equation}\label{Eq:NewtGauge.RiemannTens}
R_{00}=-3\mathcal{H}^\prime+\nabla^2\Phi-3\Psi^{\prime\prime}+3\mathcal{H}(\Phi^\prime-\Psi^\prime), \qquad R_{0i}=-2\partial_i\Psi^\prime+2\mathcal{H}\partial_i\Phi,
\end{equation}
$$R_{ij}=-\partial_i\partial_j(\Psi+\Phi)+\delta_{ij}\left[(2\mathcal{H}^2+\mathcal{H}^\prime)(1+2\Psi-2\Phi)-\nabla^2\Psi+\Psi^{\prime\prime}+5\mathcal{H}\Psi^\prime-\mathcal{H}\Phi^\prime\right].$$
Contracting the indices of the previous tensor we are able to compute the Ricci scalar
\begin{equation}\label{Eq:NewtGauge.Ricci}
Ra^2=6(\mathcal{H}^2+\mathcal{H}^\prime)(1-2\Phi)-2\nabla^2(\Phi+2\Psi)+6\Psi^{\prime\prime}-6\mathcal{H}\Phi^\prime+18\mathcal{H}\Psi^\prime.
\end{equation}
The components of the Einstein tensor entering Einstein's equations are
\begin{equation}\label{Eq:NewtGauge.EinsteinTensor}
G_{00}=3\mathcal{H}^2+6\mathcal{H}\Psi^\prime-2\nabla^2\Psi,
\end{equation}
$$G_{ij}=-\partial_i\partial_j(\Psi+\Phi)+\delta_{ij}\left[-(\mathcal{H}^2+2\mathcal{H}^\prime)(1+2\Psi-2\Phi)+2\mathcal{H}(\Phi^\prime-2\Psi^\prime)+\nabla^2(\Psi+\Phi)-2\Psi^{\prime\prime}\right],$$
$$G_{0i}=-2\partial_i\Psi^\prime+2\mathcal{H}\partial_i\Phi.$$
As we have done for the synchronous gauge, we consider the energy-momentum tensor of a perfect fluid, Eq. \eqref{Eq:BD.EMT}, and split the energy densities and pressures as before. The perturbed 4-velocity $u^\mu$ and its covariant form $u_\mu$ read now, respectively,
\begin{equation}\label{Eq:NewtGauge.umu}
u^\mu=\frac{1}{a}(1-\Phi,v^i)\qquad u_\mu=a(-[1+\Phi],v^i)\,,
\end{equation}
where $\vec{v}$ is the physical 3-velocity of the fluid, whose modulus is much lower than 1, so we can treat it as a linear perturbation. Taking all this into account one can compute the perturbed elements of $T_{\mu\nu}$ and its trace:
\begin{equation}\label{Eq:NewtGauge.Tij}
\begin{split}
T_{00}&=a^2[(1+2\Phi)\bar{\rho}+\delta\rho],\\
T_{ij}&=a^2\delta_{ij}[\bar{p}(1+2\Psi)+\delta p],\\
T_{0i}&=-a^2 v^i(\bar{\rho}+\bar{p}),\\
T&=3(\bar{p}+\delta p)-\bar{\rho}-\delta\rho .
\end{split}
\end{equation}
Now, we provide the formulas of some other perturbed expressions depending on $\varphi$ that will be also useful in subsequent computations:
\begin{equation}\label{Eq:NewtGauge.BoxVarphi}
a^2\Box\varphi = -\bar{\varphi}^{\prime\prime}-2\mathcal{H}\bar{\varphi}^\prime-\delta\varphi^{\prime\prime}+2\bar{\varphi}^{\prime\prime}\Phi+\nabla^2\delta\varphi-2\mathcal{H}\delta\varphi^\prime+\bar{\varphi}^\prime(\Phi^\prime-3\Psi^\prime)+4\mathcal{H}\Phi\bar{\varphi}^\prime ,
\end{equation}
\begin{equation}\label{Eq:NewtGauge.partialpartial}
\partial_\alpha\varphi\partial^\alpha\varphi=-\frac{(\bar{\varphi}^{\prime})^2}{a^2}(1-2\Phi)-\frac{2}{a^2}\bar{\varphi}^\prime\delta\varphi^\prime ,
\end{equation}
\begin{equation}\label{Eq:NewtGauge.nabla0nabla0}
\nabla_0\nabla_0\varphi = \bar{\varphi}^{\prime\prime}+\delta\varphi^{\prime\prime}-\bar{\varphi}^{\prime}(\mathcal{H}+\Phi^\prime)-\mathcal{H}\delta\varphi^\prime ,
\end{equation}
\begin{equation}\label{Eq:NewtGauge.nablainablaj}
\nabla_i\nabla_j\varphi=\partial_i\partial_j\delta\varphi-\bar{\varphi}^\prime\delta_{ij}[\mathcal{H}(1+2\Psi-2\Phi)+\Psi^\prime]-\mathcal{H}\delta\varphi^\prime\delta_{ij} ,
\end{equation}
\begin{equation}\label{Eq:NewtGauge.nablainabla0}
\nabla_i\nabla_0\varphi=\partial_i(\delta\varphi^\prime-\bar{\varphi}^\prime\Phi-\mathcal{H}\delta\varphi) .
\end{equation}


\section{Perturbation equations in position space}\label{PertEqPosNewton}

The perturbed Einstein equations read as follows,
\newline
\newline
\noindent
For $\mu=i,\nu=j$, $i\neq j$:
\begin{equation}\label{Eq:NewtGauge.gFieldij1}
\Psi+\Phi=-\frac{\delta\varphi}{\bar{\varphi}}\,.
\end{equation}
As we can see, the presence of the perturbation of the BD-field, $\delta\varphi\neq0$ induces anisotropic stress since it  violates the usual $\Phi=-\Psi$ setting of  GR-$\CC$CDM, which holds good  only in the absence of anisotropic stress (induced e.g. by massive neutrinos).

\noindent For $i=j$:
\begin{equation}\label{Eq:NewtGauge.gFieldiequalj1}
\begin{split}
&(\Phi^\prime-2\Psi^\prime)(2\mathcal{H}\bar{\varphi}+\bar{\varphi}^\prime)+(\Phi-\Psi)\left[2\bar{\varphi}(\mathcal{H}^2+2\mathcal{H}^\prime)+2\bar{\varphi}^{\prime\prime}+2\mathcal{H}\bar{\varphi}^\prime+\frac{\omega_{\rm BD}}{\bar{\varphi}}(\bar{\varphi}^\prime)^2\right]\\
  {}&-2\Psi^{\prime\prime}\bar{\varphi}-\delta\varphi^{\prime\prime}-\mathcal{H}\delta\varphi^\prime-\delta\varphi(\mathcal{H}^2+2\mathcal{H}^\prime)+\frac{\omega_{\rm BD}}{2\bar{\varphi}}\left[\frac{\delta\varphi}{\bar{\varphi}}(\bar{\varphi}^\prime)^2-2\bar{\varphi}^\prime\delta\varphi^\prime\right]= 8\pi G_N a^2(2\bar{p}\Psi+\delta p)\,.
\end{split}
\end{equation}
For $\mu=0,\nu=0$:
\begin{equation}\label{Eq:NewtGauge.gField00}
\begin{split}
\bar{\varphi}(6\mathcal{H}\Psi^\prime-2\nabla^2\Psi)+3\mathcal{H}^2\delta\varphi-&\nabla^2\delta\varphi+3\mathcal{H}\delta\varphi^\prime+3\Psi^\prime\bar{\varphi}^\prime\\
  {}&+\frac{\omega_{\rm BD}}{2\bar{\varphi}}\left[\frac{\delta\varphi}{\bar{\varphi}}(\bar{\varphi}^\prime)^2-2\bar{\varphi}^\prime\delta\varphi^\prime\right]=8\pi G_N a^2(2\Phi\bar{\rho}+\delta\rho)\,.
\end{split}
\end{equation}
The perturbed covariant conservation equations leads to:
\begin{equation}\label{Eq:NewtGauge.consEq0}
\sum_j \bar{\rho}_{(j)}\left[\delta^\prime_{(j)}+\delta_{(j)}\frac{\rho_{(j)}^\prime}{\rho_{(j)}}+3\mathcal{H}\delta_{(j)}\left(1+\frac{\delta p_{(j)}}{\delta \rho_{(j)}} \right)+(1+{w}_{(j)})(\theta_{(j)}+3\Psi^\prime) \right]=0\,,
\end{equation}
\begin{equation}\label{Eq:NewtGauge.consEq1}
\sum_j \left[4\mathcal{H}\theta_{(j)}\bar{\rho}_{(j)}(1+{w}_{(j)})+\frac{d}{d\eta}(\theta_{(j)}\bar{\rho}_{(j)}(1+{w}_{(j)})) +\bar{\rho}_{(j)}(1+{w}_{(j)})\nabla^2\Phi+\nabla^2 \delta p_{(j)}\right]=0,
\end{equation}
where again, as in \hyperref[Appendix:CosmoPerturbationsSynchronous]{Appendix\,\ref{Appendix:CosmoPerturbationsSynchronous}} $\theta_{(j)}\equiv \partial_i v^i_{(j)}$, $\delta_{(j)}\equiv \delta \rho_{(j)}/\bar{\rho}_{(j)}$, ${w}_{(j)}\equiv \bar{p}_{(j)}/\bar{\rho}_{(j)}$, and the sums run over all the species $j$ that fill the Universe.

On the other hand, the perturbed part of the Klein-Gordon equation can be written as
\begin{equation}\label{Eq:NewtGauge.BDfield2}
-\delta\varphi^{\prime\prime}+2\bar{\varphi}^{\prime\prime}\Phi+\nabla^2\delta\varphi-2\mathcal{H}\delta\varphi^\prime+\bar{\varphi}^\prime(\Phi^\prime-3\Psi^\prime)+4\mathcal{H}\Phi\bar{\varphi}^\prime=\frac{8\pi G_N}{3+2\omega_{\rm BD}}a^2(3\delta p-\delta\rho).
\end{equation}

\section{Perturbation equations in momentum space}\label{PertEqMomentumNewton}
From now on we will work in momentum space. Let us show first the Einstein equations.
For $\mu=i,\nu=j$, $i\neq j$:
\begin{equation}\label{Eq:NewtGauge.gFieldij1Mom}
\Psi+\Phi=-\frac{\delta\varphi}{\bar{\varphi}}\,.
\end{equation}
For $i=j$:
\begin{equation}\label{Eq:NewtGauge.gFieldij2Mom}
\begin{split}
&(\Phi^\prime-2\Psi^\prime)(2\mathcal{H}\bar{\varphi}+\bar{\varphi}^\prime)+(\Phi-\Psi)\left[2\bar{\varphi}(\mathcal{H}^2+2\mathcal{H}^\prime)+2\bar{\varphi}^{\prime\prime}+2\mathcal{H}\bar{\varphi}^\prime+\frac{\omega_{\rm BD}}{\bar{\varphi}}(\bar{\varphi}^\prime)^2\right] \\
  {}&-2\Psi^{\prime\prime}\bar{\varphi}-\delta\varphi^{\prime\prime}-\mathcal{H}\delta\varphi^\prime-\delta\varphi(\mathcal{H}^2+2\mathcal{H}^\prime)+\frac{\omega_{\rm BD}}{2\bar{\varphi}}\left[\frac{\delta\varphi}{\bar{\varphi}}(\bar{\varphi}^\prime)^2-2\bar{\varphi}^\prime\delta\varphi^\prime\right]= 8\pi G_N a^2(2\bar{p}\Psi+\delta p)\,. 
\end{split}
\end{equation}
Finally for $\mu=0,\nu=0$:
\begin{equation}\label{Eq:NewtGauge.gField00Mom}
\begin{split}
\bar{\varphi}(6\mathcal{H}\Psi^\prime+2k^2\Psi)+3\mathcal{H}^2\delta\varphi+&k^2\delta\varphi+3\mathcal{H}\delta\varphi^\prime+3\Psi^\prime\bar{\varphi}^\prime\\
  {}&+\frac{\omega_{\rm BD}}{2\bar{\varphi}}\left[\frac{\delta\varphi}{\bar{\varphi}}(\bar{\varphi}^\prime)^2-2\bar{\varphi}^\prime\delta\varphi^\prime\right]=8\pi G_N a^2(2\Phi\bar{\rho}+\delta\rho)\,.
\end{split}
\end{equation}
Notice that the previous equation yields the usual  perturbed Poisson equation for $\delta\varphi=0$  --  which implies $\Phi=-\Psi$, according to \eqref{Eq:NewtGauge.gFieldij1Mom}. Doing also $\bar{\varphi}=1$, at deep subhorizon scales it boils down to the expected simpler form   $k^2\Psi=-k^2\Phi=-4\pi G_N a^2\delta\rho$,  since $k^2\gg a^2 G_N\rho\sim  a^2 H^2= \mathcal{H}^2$.

The perturbation of the covariant conservation equation, with $\nu=0$, gives
\begin{equation}\label{Eq:NewtGauge.consEq0Mom}
\sum_j \bar{\rho}_{(j)}\left[\delta^\prime_{(j)}+3\mathcal{H}\delta_{(j)}\left(\frac{\delta p_{(j)}}{\delta \rho_{(j)}}-{w}_{(j)}\right)+(1+{w}_{(j)})(\theta_{(j)}+3\Psi^\prime) \right]=0.
\end{equation}
And for $\nu=i$, we obtain:
\begin{equation}\label{Eq:NewtGauge.consEqiMom}
\sum_j \left[4\mathcal{H}\theta_{(j)}\bar{\rho}_{(j)}(1+{w}_{(j)})+\frac{d}{d\eta}(\theta_{(j)}\bar{\rho}_{(j)}(1+{w}_{(j)})) -\bar{\rho}_{(j)}(1+{w}_{(j)})k^2\Phi-k^2 \delta p_{(j)}\right]=0.
\end{equation}
So far, these conservation equations take the same form as in the GR-$\Lambda$CDM. On the other hand, the perturbed Klein-Gordon equation reads
\begin{equation}\label{Eq:NewtGauge.BDfield2Mom}
-\delta\varphi^{\prime\prime}+2\bar{\varphi}^{\prime\prime}\Phi-k^2\delta\varphi-2\mathcal{H}\delta\varphi^\prime+\bar{\varphi}^\prime(\Phi^\prime-3\Psi^\prime)+4\mathcal{H}\Phi\bar{\varphi}^\prime=\frac{8\pi G_N}{3+2\omega_{\rm BD}}a^2(3\delta p-\delta\rho).
\end{equation}
%

\section{Matter density contrast equation at deep subhorizon scales}\label{Sect:PertEqLowScalesNewtonian}

As done in Appendix C for the synchronous gauge, we study now the evolution of matter perturbations at deep subhorizon scales, {\it i.e.} at those scales at which $k^2\gg \mathcal{H}^2$ (deep subhorizon scales). In this limit, Eq. \eqref{Eq:NewtGauge.consEq0Mom} boils down to
\begin{equation}\label{Eq:NewtGauge.deltam}
\delta_{\rm m}^\prime+\theta_{\rm m}+3\Psi^\prime=0\,,
\end{equation}
and \eqref{Eq:NewtGauge.consEqiMom} can be written as
\begin{equation}\label{Eq:NewtGauge.thetam}
\theta^\prime_{\rm m} +\mathcal{H} \theta_{\rm m}-k^2\Phi=0.
\end{equation}
These equations can be easily combined to make disappear the dependence on $\theta_{\rm m}$. If we do that, we obtain the following approximate second order differential equation for the matter density contrast,
\begin{equation}\label{Eq:NewtGauge.perturbk2Phi}
\delta_{\rm m}^{\prime \prime}+\mathcal{H}\delta_{\rm m}^\prime+k^2 \Phi=0.
\end{equation}
The problem is now reduced to find an expression for $k^2\Phi$ in terms of background quantities and $\delta_{\rm m}$. Collecting \eqref{Eq:NewtGauge.gFieldij1Mom}, \eqref{Eq:NewtGauge.gField00Mom} and \eqref{Eq:NewtGauge.BDfield2Mom},
\begin{equation}\label{Eq:NewtGauge.PhiplusPsi}
-\frac{\delta \varphi}{\bar{\varphi}}=\Psi+\Phi,
\end{equation}
\begin{equation}\label{Eq:NewtGauge.Poisson1}
2k^2 \Psi+k^2 \frac{\delta \varphi}{\bar{\varphi}}=\frac{8\pi G_N a^2}{\bar{\varphi}} \bar{\rho}_{\rm m} \delta_{\rm m},
\end{equation}
\begin{equation}\label{Eq:NewtGauge.Poisson2}
k^2 \delta \varphi=\frac{8\pi G_N a^2}{3+2\omega_{\rm BD}}\bar{\rho}_{\rm m} \delta_{\rm m},
\end{equation}
respectively. We can see, as expected, that for $\wBD\to\infty$ \textit{and} $\bar{\varphi}=1$ the first equation above gives $\Phi=-\Psi$ (no anisotropy stress) and the third one renders a trivial equality ($0=0$), whereas the second equation yields Poisson equation $k^2\Psi=-k^2\Phi=4\pi G_N a^2\delta\rho$.
In the general case, by combining the above equations one finds:
\begin{equation}\label{Eq:NewtGauge.Poisson3}
k^2\Phi=-\frac{4\pi G_N a^2 \bar{\rho}_{\rm m} \delta_{\rm m}}{\bar{\varphi}}\left( \frac{4+2\omega_{\rm BD}}{3+2\omega_{\rm BD}} \right)\,.
\end{equation}
So, finally, inserting the previous relation in \eqref{Eq:NewtGauge.perturbk2Phi} we are led to the desired equation for the density contrast at deep subhorizon scales:
\begin{equation}\label{Eq:NewtGauge.DiffEqdeltam}
\delta_{\rm m}^{\prime\prime}+\mathcal{H}\delta_{\rm m}^\prime-\frac{4\pi G_N}{\bar{\varphi}} a^2\bar{\rho}_{\rm m}\delta_{\rm m} \left(\frac{4+2\omega_{\rm BD}}{3+2\omega_{\rm BD}}\right)=0\,,
\end{equation}
or, alternatively, in terms of the cosmic time $t$,
\begin{equation}\label{Eq:NewtGauge.dcEq}
\ddot{\delta}_{\rm m}+2H\dot{\delta}_{\rm m}-\frac{4\pi G_N}{\bar{\varphi}}\bar{\rho}_{\rm m}\delta_{\rm m}\left(\frac{4+2\omega_{\rm BD}}{3+2\omega_{\rm BD}}\right)=0\,,
\end{equation}
with the dots denoting derivatives with respect to $t$.  As expected, these equations coincide with the density contrast equation at deep subhorizon scales for the synchronous gauge and we can recover the standard $\Lambda$CDM result for  $\omega_{\rm BD}\to \infty$ and $\bar{\varphi}\to 1$.

As already mentioned above, because of \eqref{Eq:NewtGauge.PhiplusPsi} non-null scalar field perturbations induce a deviation of the anisotropic stress, $-\Psi/\Phi$, from $1$, {\it i.e.} the GR-$\Lambda$CDM value. At scales well below the horizon,
\begin{equation}\label{Eq:NewtGauge.PsiPhi}
-\frac{\Psi}{\Phi} = \frac{1+\oD}{2+\oD}=1-\epsilon_{\rm BD}+\mathcal{O}(\epsilon^2_{\rm BD})\,,
\end{equation}
so constraints on the anisotropic stress directly translate into constraints on $\epsilon_{\rm BD}$, and a deviation of this quantity from $1$ at the linear regime would be a clear signature of non-standard gravitational physics. A model-independent reconstruction of the anisotropic stress from observations has been recently done in \cite{Pinho:2018unz}. Unfortunately, the error bars are still of order $\mathcal{O}(1)$, so these model-independent results cannot put tight constraints on $\epsilon_{\rm BD}$ (yet).

\section{Transformations between Gauges}\label{Sect:TransformationGauges}
It is possible to establish a set of equations relating the different perturbation quantities in both gauges at the same coordinates in momentum space. For the potentials, we have the well known relations of \cite{Ma:1995ey},
\begin{equation}\label{Eq:NewtGauge.Phi(k)}
\Phi ( \vec{k} , \eta)=\frac{1}{2k^2}\left\{ h^{\prime \prime}( \vec{k},\eta)+6\xi^{\prime \prime}(\vec{k},\eta)+\mathcal{H}\left[ h^{\prime }(\vec{k},\eta)+6\xi^{\prime }(\vec{k},\eta)\right] \right\}\,,
\end{equation}
\begin{equation}\label{Eq:NewtGauge.Phi(k)2}
\Psi (\vec{k},\eta)= -\xi(\vec{k},\eta)+\frac{1}{2k^3}\mathcal{H}\left[h^\prime (\vec{k},\eta)+6\xi^\prime(\vec{k},\eta)\right]\,,
\end{equation}
and for the other perturbed quantities we find,
\begin{equation}\label{Eq:NewtGauge.NewtonToSynch}
\begin{split}
\delta_S &= \delta_N-\alpha \frac{\bar{\rho}^\prime}{\bar{\rho}}\,,\\
\theta_S &= \theta_N- \alpha k^2\,,\\
\delta p_S &= \delta p_N-\alpha \bar{p}^\prime\,,\\
\delta \varphi_S &=\delta \varphi_N-\alpha \bar{\varphi}^\prime\,.
\end{split}
\end{equation}
Here, the subscripts $N$ and $S$ mean {\it newtonian} and {\it synchronous}, respectively, and
\begin{equation}\label{Eq:NewtGauge.DefAlpha}
\alpha \equiv \frac{1}{2k^2}\left[h^\prime+6\xi^\prime \right]\,.
\end{equation}

\blankpage
\chapter{Bayesian statistics and cosmology}\label{Appendix:Bayesian}

Cosmology is a science wrapped with an increasing amount of data and complexity, which needs of the most advanced techniques in order to analyse observations, give meaningful predictions of parameters and test cosmological models. It is not the goal to do a comprehensive nor exhaustive overview, but in this Appendix we will review some of the fundamental concepts in Bayesian statistics and related topics\,\cite{Trotta:2008qt,Amendola:2015ksp}. In this context, the probability is a measure of the degree of belief about a proposition.

The fundamental block of Bayesian statistics is the Bayes' theorem. One illustrative way to state this result is considering that we have a dataset consisting of several surveys or measures, $\mathcal{D} \equiv \left( D_1, D_2, \dots D_{\rm m}\right)$. On the other hand we want to test a cosmological model, $M$. The model is described by some parameters $\vec{\theta}=\left( \theta_1 , \theta_2 , \dots \theta_{\rm n} \right)$. Some of these parameters are physically relevant, in the sense that they have a role in the description of phenomena. On the other hand, we call a {\it nuisance parameter} those with influence over data and its uncertainties, playing an intermediate role in the analysis, but do not have interest for being determined at the end. The Bayes theorem can be written in the following way\footnote{A potential source of confusion in our notation is the use of $P$ to represent both probabilities and probability density functions. Specifically, if $X$ is a continuous random variable, $\int_{R} P(X) dx$ represents the probability of $X$ being within the region $R$, while if $X$ is a discrete variable, $P(X=x)$ refers to the probability of $X$ taking the value $x$. Typically, parameters and data are continuous variables. However, $P(M)$ represents the prior probability of a model (assuming a finite set of competing models).} 
\begin{equation}\label{Eq:Bayes.BayesTheorem}
 P (\vec{\theta} \,|\,\mathcal{D} , M  )= \frac{P(\mathcal{D} \,|\, \vec{\theta}, M ) P(\vec{\theta} \,|\,M )}{P (\mathcal{D} \,|\,M )} \,.
 \end{equation}
From the previous formula,

\begin{itemize}

\item[$\bullet$]$ P (\vec{\theta}  \,|\, \mathcal{D} , M )$ is the {\it posterior probability}, after considering the data.

\item[$\bullet$]$P(\vec{\theta} \,|\, M )$ is the \emph{prior probability}, which gives the degree of believe before incorporating any data. For instance, when constructing a prior probability, one may take into account the outcomes of a previous experiment or consider physical arguments of the model.

\item[$\bullet$]$P(\mathcal{D} \,|\,M )$ is the {\it Bayesian evidence} or marginal likelihood, which constitutes a normalization constant. The law of total probability let us to write it as a sum over the model space (i.e. over the region of possible values of the parameters),
\begin{equation}\label{Eq:Bayes.BayesianEvidence}
P(\mathcal{D} \,|\, M)= \int P(\mathcal{D} \,|\, \vec{\theta}, M )P(\vec{\theta} \,|\, M)d\vec{\theta}\,.
\end{equation}
\item[$\bullet$] $P(\mathcal{D} \,|\, \theta, M )$ is the sampling distribution of the data, provided that $M$ is true. For a fixed data $\mathcal{D}$ it is just a function of $\vec{\theta}$, called the {\it likelihood function}, $\mathcal{L}(\vec{\theta})$.

\end{itemize}

Our ultimate goal is to use statistical methods to estimate the parameters that characterize each cosmological model and to conduct a fair comparison between the candidates in order to determine which is most suitable for the data. 

\section{Bayesian parameter inference}\label{Sect:ParameterInference}

In chapters \hyperref[Chap:PhenomenologyofBD]{\ref{Chap:PhenomenologyofBD}} and \hyperref[Chap:PhenomenologyofRVM]{\ref{Chap:PhenomenologyofRVM}} we studied different extensions of the $\Lambda$CDM (or GR-$\Lambda$CDM, if we are more precise). Let us take the example of the BD-$\Lambda$CDM. It supposes an extension of the concordance cosmological model by the inclusion of a scalar field, $\phi$ taking the role of evolving gravitational constant. In particular, as a nested model, the BD-$\Lambda$CDM contains two more parameters with respect the standard GR-$\Lambda$CDM which describe new physics: $\epsilon_{\rm BD}$ and $\phi_{\rm ini}$. As an extension, one can recover the $\Lambda$CDM from the BD-$\Lambda$CDM for specific values of the extra parameters. That is, their GR-$\Lambda$CDM values, $\epsilon_{\rm BD}=0$ and $\varphi_{\rm ini}=1$. Although the final problem is actually a model comparison against the concordance model, the first step is to look for the constrains that the data imposes on the different parameters. In particular, those describing new physics or those related with the cosmological tensions such as $H_0$ can be of special interest. This is possible with an extensive and precise dataset, which makes possible to accurately fit cosmological models to observational data and estimate their values. 

The first step in the parameter inference process is to determine the likelihood function for the measurement, $\mathcal{L}(M)$, which describes the plausability of the data for a given model $M=(\vec{\theta},\vec{\phi})$, where $\vec{\theta}$ represents the vector of interest parameters and $\vec{\phi}$ is the vector of nuisance parameters. Furthermore, a prior for the parameters $\vec{\theta}$ and $\vec{\phi}$ is required, which can be obtained from a previous experiment, physical arguments, a flat prior if we admit any possible value of the parameters. The Bayesian evidence can be constructed from \eqref{Eq:Bayes.BayesianEvidence} as a normalization factor and it is not necessary to be given in this context. The posterior distribution is then just proportional to the product of the likelihood and the prior,
\begin{equation}
P(\vec{\theta},\vec{\phi}|\mathcal{D},M )\propto \mathcal{L}(\vec{\theta},\vec{\phi}) P(\vec{\theta},\vec{\phi} |M).
\end{equation}
It is usual to focus in the interest parameters and simplify the model by {\it marginalizing} the nuisance parameters at some point, by integration of the joint posterior pdf:
\begin{equation}\label{Eq:Bayes.Marginalization}
 P(\vec{\theta} | \mathcal{D},M )=\int P(\vec{\theta},\vec{\phi}| \mathcal{D},M )d \vec{\phi}\propto \int \mathcal{L}(\vec{\theta},\vec{\phi}) P(\vec{\theta},\vec{\phi}|M) d\vec{\phi}\,.
 \end{equation}
The final step of inference from the posterior pdf is to present information about the parameters, either by providing its mean, median, standard deviation, etc. or by presenting a marginalized one or two dimensional plot after the posterior probability density function with respect to the other parameters.

The process of finding the posterior distribution is not always straightforward, particularly when analytical methods are not possible. In such cases, sophisticated techniques and approximations may be utilized. For instance, Gaussian functions can be used to approximate the likelihood and priors. Alternatively, numerical tools such as Markov Chain Monte Carlo (MCMC) techniques can be used to evaluate the likelihood and sample from the posterior probability density function (pdf). MCMC is a widely used and well-known method for handling complex scenarios.

\subsection{The Fisher matrix}

The maximum likelihood (ML) estimators are values of the parameters, although depending on the data, that maximize the likelihood. In the simplest scenario, they can be computed through standard mutivariable calculus, by solving the set of equations
\begin{equation}
\frac{\partial \mathcal{L}(\vec{ \theta})}{\partial \theta_i}=0,\qquad \textrm{for } i=1,2,\dots,n\,.
\end{equation}%
The Fisher Matrix formalism holds for an unimodal likelihood which can be well approximated by a multidimensional Gaussian function near the maximum. They key idea  is to approximate the full likelihood as a multigaussian distribution
\begin{equation}\label{Eq:Bayes.LikelihoodApproxGaussian}
\mathcal{L} (\vec{\theta }) \approx \mathcal{L}_* \exp\left\{ -\frac{1}{2} \left[ \vec{\theta }- \vec{\theta }_*\right]^T F \left[\vec{\theta }-\vec{\theta }_* \right] \right\}\,,
\end{equation}
where $\vec{\theta}_*\left(\mathcal{D}\right) $ is the vector containing the ML estimators and $\mathcal{L}_*\equiv \mathcal{L}(\vec{\theta}_*)$, both functions of the data. The matrix $F$ is called the {\it Fisher Matrix}, and is well approximated by the inverse of the covariance matrix of the parameters, $F=C^{-1}$. The covariance matrix is defined as
\begin{equation}\label{Eq:Bayes.CovarianceMatrixFormula}
C\equiv \begin{pmatrix}
\sigma_1 & 0 & \cdots & 0 \\
0 & \sigma_2 & \cdots & 0 \\
\vdots & \vdots & \ddots &\vdots\\
0 & 0 & \cdots & \sigma_{\rm n}
\end{pmatrix}
\begin{pmatrix}
1 & \rho_{1,2} & \cdots & \rho_{\rm 1, n} \\
\rho_{\rm 2,1} & 1 & \cdots & \rho_{\rm 2,n} \\
\vdots & \vdots & \ddots &\vdots\\
\rho_{\rm n,1} & \rho_{\rm n,2} & \cdots & 1
\end{pmatrix}
\begin{pmatrix}
\sigma_1 & 0 & \cdots & 0 \\
0 & \sigma_2 & \cdots & 0 \\
\vdots & \vdots & \ddots &\vdots\\
0 & 0 & \cdots & \sigma_{\rm n}
\end{pmatrix}\,.
\end{equation}
where ${\rm diag}(\sigma_1,\dots,\sigma_n)$ is the matrix of the 1D-marginal variances and $\rho_{\rm i,k}$ is the correlation between the parameters $\theta_{\rm i}$ and $\theta_{\rm k}$. In general, the Fisher Matrix, can be written as
\begin{equation} \label{Eq:Bayes.FisherMatrix}
{F}_{ij}=- \frac{\partial^2 \ln \mathcal{L}\left(\vec{\theta}\right)}{\partial \theta_i \partial \theta_j} \,.
\end{equation}
Within the context of this approximation it is easy to perform parameter estimation and to compute their uncertainties, mean values, etc. This is a first method that may be useful for parameter estimation, albeit non-gaussianities can be a serious issue that may prevent this formalism to apply in a satisfactory way. In such a case it would be necessary to require another method such as the mentioned MCMC, in order to estimate the posterior pdf.

\subsection{The Joint likelihood Function \texorpdfstring{$\chi^2$}{x2}}

When dealing with a dataset $\mathcal{D}$ conformed by independent experiments $D_1,D_2, \dots D_{\rm m}$ the joint likelihood can be factorized as
\begin{equation}\label{Eq:Bayes.ProdLikelihood}
\mathcal{L}_{\rm tot}=\mathcal{L}_{1}\times \mathcal{L}_{\rm 2}\times \cdots\times \mathcal{L}_{\rm m} \,.
\end{equation}
It is useful to define the $\chi^2_{\rm i}$ function for each experiment $D_{\rm i}$ as $\chi_{\rm i}^2\equiv -2\ln \mathcal{L}_{\rm i}$.  With this definition, a maximization of the likelihood is equivalent to a minimization of the $\chi^2$. From \eqref{Eq:Bayes.ProdLikelihood}, it follows that the total $\chi^2$ can be expressed in an additive way,
\begin{equation}\label{Eq:Bayes.TotalChiSquare}
  \chi^2_{\rm tot}=\chi^2_{\rm 1}+\chi^2_{\rm 2}+\dots+\chi^2_{\rm m}\,.
\end{equation}
or, within the gaussian approximation,
\begin{equation}\label{Eq:Bayes.ChiSquareDatasets}
\chi^2_{\rm tot}(\vec{\theta})=\chi^2_*+\sum_{\rm i=1}^{\rm m}\left(g_{\rm i}(\vec{\theta})-\vec{d}_{\rm i}\right)^{\rm T} F_{\rm i} \left(g_{\rm i}(\vec{\theta})-\vec{d}_{\rm i}\right)\,,
\end{equation}
where the sum runs over the different datasets involved in the analysis (for instance $\textrm{i}=\textrm{CMB, BAO, LSS},\dots$), $g_{\rm i}(\vec{\theta})$ contains the theoretical predictions as a function of the free parameters and $\vec{d}_{\rm i}$ is the data vector containing the information of the data set $\mathcal{D}_{\rm i}$. The first term $\chi^2_*$ is the logarithm of the normalization factor of the likelihood \eqref{Eq:Bayes.LikelihoodApproxGaussian} and is a constant value. 

There is a clear advantage in using a comprehensive data set conformed of different kinds of observations. Combining independent and and compatible sources can aid in breaking degeneracies in the parameters of the fitted model. After obtaining the likelihood for the different observations and computing the total $\chi^2$, one can proceed to compute the 1D-marginalized posterior pdf of the interest parameters or the 2D posteriors contours of the joint distribution of two parameters.

\subsection{Markov chains and Monte Carlo estimates}
 
A classical method for cosmological parameter estimation is to use grid in order to search the minimum of the $\chi^2$ function. This is a brute force procedure which either covers the entire parameter space or covers a localized region containing the maximum, meaning that we need some previous information about its approximate location. This task usually demands a lot of computational resources once the number of dimensions of the parameter space grows. ofently making the process non-viable if we need a precise and fast evaluation of the $\chi^2$. An alternative procedure based on a Bayesian approach is Markov Chains and Monte Carlo (MCMC) methods. There are different codes, such as \verb!MontePython!\,\cite{Audren:2012wb} and \verb!CosmoMC!\,\cite{Lewis:2002ah}, that may implement this method. Working together with a Boltzmann code they may calculate the model parameters and matter power spectrum. In particular we used the pair \verb!MontePython!+\verb!CLASS!\,\cite{Blas:2011rf} to perform our analysis in chapters \hyperref[Chap:PhenomenologyofRVM]{\ref{Chap:PhenomenologyofRVM}} and \hyperref[Chap:PhenomenologyofBD]{\ref{Chap:PhenomenologyofBD}}. \verb!MontePython! runs a Metropolis-Hastings algorithm for sampling.

Basically, without going into the details, an MCMC code explores the parameter space through a random process, constructing a sequence of points referred to as {\it chains}. The distribution of these chains approaches the posterior probability density function for the parameters of the model with every step. After enough iterations, we can use the chains as samples of the posterior pdf, allowing us to perform parameter inference and computing different statistics as the mean value of a parameter,
\begin{equation}\label{Eq:Bayes.MCMC}
  \left\langle \theta \right\rangle \approx \int \theta\,p\left(\theta|d,M \right) d\theta \approx \frac{1}{N}\sum_{t=0}^{N-1} \theta^{(t)}\,,
\end{equation}
expected values of any function, standard deviations, correlations between parameters or the 1D or 2D marginalized distributions.

\section{Bayesian model comparison}

Fitting the parameters of our cosmological model is not always sufficient. For instance, if our model includes extra physics, such as the BD-$\Lambda$CDM or Running Vacuum Models presented in this dissertation, we are not only interested in determining the values of its parameters, but also in comparing its performance with the standard cosmological model to see if it fits the data better. 

This contest between models should be conducted in a fair manner, taking into account their complexity.  One guiding principle when choosing between different models is Occam's razor. In plain words, when comparing two models with the same predictivity and quality of fit, one should choose the simpler model. Of course, a model with a large number of parameters and/or wider ranges of values may fit the data just as well as a simpler model, but at the cost of increased complexity. This complexity should be avoided if a simpler model provides a satisfactory description of the phenomena that is consistent with observations. Bayesian model comparison provides the tools to objectively determine whether a more complex model is truly necessary based on the data.

New observations may update our degree of believe in a particular theory. However, one can not simple refuse it because of the implausibility of the data given the model, unless it is compared with an alternative one which gives better agreements. We have to start with a model $M$ characterized by a set of parameters $\vec{\theta}$ and its prior probability, $p(\vec{\theta}|M)$.

As a difference from the parameter inference, the key ingredient in model comparison is the so-called {\it Bayesian evidence} in Bayes' Theorem, 
\begin{equation}\label{Eq:Bayes.BayesianEvidenceII}
P(\mathcal{D}|M )= \int \mathcal{L}( \vec{\theta} ) P( \vec{\theta} | M ) d\vec{\theta}\,,
\end{equation}
with $\mathcal{L} \left( \theta \right)\equiv P \left(\mathcal{D} | \theta, M \right).$

Let's consider two different models $M_{\rm A}$ and $M_{\rm B}$ and the ratio of their posterior probabilities, also called {\it posterior odds},
\begin{equation}\label{posteriorodds}
  \frac{P(M_{\rm A}|\mathcal{D})}{P(M_{\rm B}|\mathcal{D})}=\frac{\frac{P(M_{\rm A}) P(\mathcal{D}|M_{\rm A})}{P(\mathcal{D})}}{\frac{P(M_ {\rm B}) P(\mathcal{D}|M_{\rm B})}{P(\mathcal{D})}}=\mathcal{B}_{\rm A,B}\frac{P(M_{\rm A})}{P(M_{\rm B})}\,,
\end{equation}
where $\mathcal{B}_{\rm A,B}$ is called {\it Bayes Factor} and is defined as
\begin{equation}\label{BayesFactor}
 \mathcal{B}_{\rm A,B}\equiv \frac{P(\mathcal{D}|M_{\rm A})}{P(\mathcal{D}|M_{\rm B})}\,.
\end{equation}
or, equivalently,
\begin{equation}\label{Eq:BayesFactorII}
  \mathcal{B}_{\rm A,B}=\frac{\frac{P(M_{\rm A}|\mathcal{D})}{P(M_{\rm B}|\mathcal{D})}}{\frac{P(M_{\rm A})}{P(M_{\rm B})}}\,,
\end{equation}
written in this way, we can understand the Bayes Factor as the factor that accounts for the change in the ratio odds between the two models, before and after the data has been taking into account. Thus, $\mathcal{B}_{\rm A,B}$ reflects an update of the relative degree of belief of the models. If there are no reasonable clues prior the data, it is natural to assign equal degree of plausability to both models, $P(M_{\rm A})=P(M_{\rm B})$. In that case, the Bayes factor coincides with ratio of posteriors odds. When $\mathcal{B}_{\rm A,B} \ll 1$, it is clear that the Bayesian evidence for model $M_{\rm A}$ is much greater than the Bayesian evidence for model $M_{\rm B}$. Equivalently, we may interpret this fact as an enhancement of the odds ratio due to the new information provided by the data, as shown in equation \eqref{Eq:BayesFactorII}. In such a case, we should favor model $A$ over model $B$.

Once again, it must be emphasized that calculating Bayesian evidence is not straightforward, as it requires integration over the entire parameter space, a task that can demand a lot of computational resources. Some approximations exist to mitigate the problem, but we will not focus on these in what follows. Instead, we will describe a simple procedure for Bayesian model selection that takes into account the complexity of the model.

\subsection{Bayesian complexity and information criteria}

We can naively associate the complexity of a model with the number of free parameters, but this may not be accurate. Some parameters may be poorly constrained by the data and effectively have no role. To take this into consideration when defining a measure for complexity, we need to account for the number of parameters that the data can support. We start by defining the Kullback-Leibler (KL) divergence,
\begin{equation}\label{Eq:Bayes.DivergenceKL}
D_{\rm KL} \equiv \int P(\theta|\mathcal{D}, M) \ln \frac{P(\theta | \mathcal{D},M)}{p(\theta| M)} d\theta = \left\langle \ln \frac{P\left(\theta| \mathcal{D},M \right)}{P\left(\theta,M\right)}\right\rangle \,,
\end{equation}
which is the expected value of the information gain (under the posterior distribution) from the prior to the posterior pdf. Remember the definition we did of $\chi^2 (\theta) \equiv -2\ln \mathcal{L}(\theta)$. So that, an equivalent form for the KL term coming from the Bayes theorem is
\begin{equation}\label{Eq:Bayes.DivergenceKLII}
D_{\rm KL}=-\ln P(\mathcal{D}|M)+\int p(\theta |\mathcal{D},M)\ln \mathcal{L}(\theta)d\theta=-\ln P \left(\mathcal{D}|M \right)-\frac{1 }{2}\left \langle \chi^2 \left(\theta\right) \right\rangle\,,
\end{equation}
where, the $\left \langle \chi^2 \right\rangle$ is the mean $\chi^2$ function over the posterior distribution, which can be obtained by MCMC methods. On the other hand, we can define $\widehat{D_{\rm KL}}$ as,
\begin{equation}\label{Eq:Bayes.DivergenceKLmin}
\widehat{D_{\rm KL}}=-\ln P(\mathcal{D},M) -\frac{1}{2}\chi^2 (\hat{\theta}) \,,
\end{equation}
where $\hat{\theta}$ is the value of the information gain at an estimated value of the parameters (the posterior mean, the ML estimator,etc.). The {\it Bayesian complexity} will be defined as
\begin{equation}\label{Eq:Bayes.BayesianComplexity}
C_{b}=-2\left(D_{\rm KL}-\widehat{D_{\rm KL}}\right)=\left\langle \chi^2 \left(\theta\right)\right\rangle-\chi^2 (\hat{\theta})\,,
\end{equation}
and gives the number of {\it effective parameters} that the data can constrain. Bayesian complexity provides an additional step when comparing several models with similar evidence. In such a case, we may choose the one with lesser Bayesian complexity (i.e., with a lesser number of effective parameters and greater simplicity) in order to respect the philosophy of Occam's razor.

Although we introduced the concepts of Bayesian evidence and complexity for completeness, when comparing the models in chapters \hyperref[Chap:PhenomenologyofRVM]{\ref{Chap:PhenomenologyofRVM}} and \hyperref[Chap:PhenomenologyofBD]{\ref{Chap:PhenomenologyofBD}}, we adopted a simpler approach using information criteria. These provide a method for model selection that penalizes models with more parameters, on the assumption that such models are more complex and therefore have a greater potential to overfit the data. Unlike Bayesian evidence, information criteria typically penalize models with a higher number of free parameters, whether or not they are well-constrained by the data. Consequently, model selection based on information criteria is oftenly more reliable when the data constrain all the free parameters satisfactorily, which is approximately the case for our models.

Different information criteria can be used in Cosmology, let us summarize the main three:

\begin{itemize}

\item[a)] The {\it Akaike Information Criterion} (AIC)\,\cite{akaike1974new} is a widely used criterion that adds a penalization of the double of independently adjusted parameters, $k$. To calculate the AIC for a particular model, one starts by fitting the model to the data and obtaining the maximum likelihood estimate of the model parameters. The likelihood function is then evaluated at this estimate, and the AIC is calculated using
\begin{equation}\label{Eq:Bayes.AIC}
  {\rm AIC} \equiv \chi^2_* + 2k + \frac{2k(k+1)}{N-k-1}\approx\chi^2_*+2k\,.
\end{equation}
Here $\chi^2_*\equiv -2 \ln \mathcal{L}_{*}$, where $\mathcal{L}_{*}\equiv p \left(d|\theta_{*}, M\right)$ is the value of the likelihood maximized by the parameter $\theta_{*}$,$\chi^2_*\equiv \chi^2 ( \theta_* )$ and $N$ is the number of data point. The approximation in the last equality is valid for $N\gg k$, which is usually the case.  When comparing different models, the model with the lowest AIC is usually considered the best fit, balancing goodness of fit and model complexity. AIC has the advantage of being relatively easy to calculate and widely applicable across a range of statistical models.

\item[b)] The {\it Bayesian Information Criterion} (BIC)\,\cite{Schwarz1978}, is also based on the maximum likelihood estimate (MLE) of the parameters of the model, and penalizes models with more parameters by adding a term proportional to the logarithm of the number of parameters, multiplied by a factor that depends on the sample size,
\begin{equation}\label{Eq:Bayes.BIC}
  {\rm BIC} \equiv -2 \ln \mathcal{L}_{*}+k\ln N\,,
\end{equation}
where $\mathcal{L}_{*}$ is the maximum likelihood value of the model, $k$ is the number of parameters in the model, and $N$ is the number of data points. This criterion is especially useful when $N\gg k$. In this regime, the BIC tends to favor simpler models that are more likely to generalize well to new data.  Again, the model with the lowest BIC is typically considered the best among the models being compared.

\item[c)] The {\it Deviance information Criterion} (DIC)\,\cite{Spiegelhalter:2002yvw},
\begin{equation}\label{Eq:Bayes.DIC}
  {\rm DIC} \equiv -2 \widehat{D_{\rm KL}}+2 C_{\rm b}=-2\left(\ln \mathcal{L}_{*}-\ln P \left(\mathcal{D}|M \right)\right)+2C_{\rm b}
\end{equation}
is a generalization of AIC, replacing the number of parameters, $k$, by the number of effective parameters, $C_{\rm b}$. In the limit of well-informative data $\ln P \left(\mathcal{D}|M \right)\approx 0$ and $C_{\rm b}\approx k$, so that recovering \eqref{Eq:Bayes.AIC}. Compared to the AIC, the DIC is more accurate as it can take into account unconstrained parameters, which the AIC cannot. However, it is also more complex than the AIC, and can be computationally intensive to calculate.
\end{itemize}
\newpage

\blankpage

\chapter{Description of the data}\label{Appendix:Description}

As a final step to rigorously evaluate the quality of a cosmological model it is vital to confront it against the existing empirical evidence. Our phenomenological works presented in chapters  \ref{Chap:PhenomenologyofBD} and \ref{Chap:PhenomenologyofRVM} represent our efforts to judge our theoretical investigations on the dynamical behaviour of the vacuum energy against the available cosmological data. In there, we try to give our best to perform an exhaustive and complete statistical analysis presenting not only the fit of the models, but also their fair model comparision under different datasets in order to decide which one is the most favored by the data. The variety of sources that we can enjoy in this era of precision cosmology is vast and, albeit it is impossible to consider all of them in our analysis, we decided to collect the most consistent and reliable sets under our own criteria. Needless to say that this is a delicate and nonstop process in which experience plays a crucial role as one has to carefully scrutinize multitudinous results in the literature along the years. This intrincate task implies to be aware of correlations and double counting between the outcome of different surveys, which are usually not completely independent and may be fundamented in overlapping samples.

In this appendix we summarize the different sources of cosmological data that have been used in our particular selection for the works presented in the main text.

\section*{CMB}

The discovery of the cosmic microwave background (CMB) in the 1960s was a turning point in the history of cosmology. It provided a definitive clue to the validity of the Big Bang framework and a priceless tool for obtaining information from the early stages of the Universe. The CMB is a remnant of the hot, dense state of the Universe shortly after the Big Bang. At the recombination epoch, (redshift of $z\approx 1100$, when the temperature was below 1 eV), the Universe had cooled enough for electrons and protons to combine into neutral hydrogen atoms. This caused a sudden drop in the opacity of the Universe to radiation, allowing it to freely stream across space. This radiation has been traveling through the Universe ever since, and has been stretched to longer wavelengths by the expansion of the Universe, resulting in its detection as microwaves today.

Prior to this time, the photon-to-baryon ratio was high enough to maintain photons coupled to electrons, even if it was not favored by the energetic distribution. This caused an ionization of any neutral hydrogen atom that could be bound. The effective decoupling occurred when the rate of photon-electron scattering became less frequent due to the expansion of the Universe. As a result, the mean free path of photons became much higher, producing the almost homogeneous and isotropic radiation that we can detect today, coming from the so-called surface of last scattering.

CMB observations are fundamental to constraint the free parameters of our cosmological models. In particular, CMB temperature and polarization anisotropies are a rich source of information, which are mainly originated by scalar perturbations of Einstein's equations\,\cite{Dodelson:2003ft,Amendola:2015ksp}.

Due to the spherical simmetry, the temperature anistropies can be decomposed in spherical harmonics,
\begin{equation}\label{Eq:Description.Harmonics}
\Theta\left(\vec{x}, z, \hat{n}\right)\equiv \frac{\delta T(\vec{x},z,\hat{n})}{T(z)}=\sum\limits_{\ell=1}^\infty \sum\limits_{m=-\ell}^\ell a_{\ell m }\left(\vec{x},z \right) Y_{\ell m}\left(\hat{n}\right)\,,
\end{equation}
where $\vec{n}$ indicates the direction of the incoming radiation and $a_{\ell m}\left(\vec{x},z \right)$ are spacetime dependent coefficients. The index $\ell$ is usually called multipole moment and is related to the angular size as $\theta=\pi/\ell$. For that reason, low (respt. high) multipoles are associated to large (respt. small scales).

The harmonics are normalized as
\begin{equation}\label{Eq:Description.HarmonicsNorm}
\int  Y_{\ell m}\left(\hat{n}\right) Y_{\ell^\prime m^\prime}^*\left(\hat{n}\right) d\hat{n}=\delta_{\ell \ell^\prime}\delta_{m m^\prime}\,.
\end{equation}
The latter expression let us to isolate $a_{\ell m}$ as
\begin{equation}\label{Eq:Description.alm}
a_{\ell m}=\int  \Theta\left(\vec{x}, z, \hat{n}\right) Y_{\ell m}^*\left(\hat{n}\right) d\hat{n}\,.
\end{equation}
The coefficients are understood to have a zero mean, $\langle a_{\rm \ell m}\rangle =0$, because our initial perturbations are likely to be either positive or negative. However, the variance is non-zero:
\begin{equation}\label{Eq:Description.Cl}
C_\ell \equiv \langle |a_{\rm \ell m} |^2 \rangle = \frac{1}{2\ell +1 }\sum\limits_{m=-\ell}^\ell |a_{\ell m}|^2\,.
\end{equation}
The CMB power spectrum is described through the $C_\ell$'s. In particular, we compare our theoretical predictions for $\ell(\ell+1 )C_\ell/(2\pi)$ as a function of the multipole (or the angular scale) with observations. Unfortunately, the exact shape of $C_\ell$ is not accessible through analytical computations. The photons and other matter components are coupled, and this situation is described by a system of Boltzmann differential equations that cannot be solved analytically without approximations. Nevertheless, numerical tools can do the job. In our case, we used \verb!CLASS! \cite{Blas:2011rf}, as described in chapters \hyperref[Chap:PhenomenologyofRVM]{\ref{Chap:PhenomenologyofRVM}} and \hyperref[Chap:PhenomenologyofBD]{\ref{Chap:PhenomenologyofBD}}. 

In both chapters we have worked with the results from Planck 2018\,\cite{aghanim2020planck}, considering different datasets: the full Planck 2018 TT + lowE likelihood (in the Baseline scenario), the full Planck 2018 + TTTEEE+ lowE likelihood to study the effect of the higher multipoles and full Planck 2018TTTEEE+lowE+lensing in order to study the lensing data. In our baseline dataset we consider the full Planck 2018 TT+lowE likelihood \cite{aghanim2020planck}. CMB part of the Baseline dataset is common for both chapters. In order to study the influence of the CMB high-$\ell$ polarizations and lensing we consider in \hyperref[Chap:PhenomenologyofBD]{Chap.\,\ref{Chap:PhenomenologyofBD}} two alternative (non-baseline) datasets, in which we substitute the TT+lowE likelihood by: (i) the TTTEEE+lowE likelihood, which incorporates the information of high multipole polarizations; (ii) the full TTTEEE+lowE+lensing likelihood, in which we also incorporate the Planck 2018 lensing data. In \hyperref[Table:BD.TableFitAlternativeDataset]{\ref{Table:BD.TableFitAlternativeDataset}} and \hyperref[Table:BD.TableFitGR]{\ref{Table:BD.TableFitGR}} these scenarios are denoted as B+$H_0$+pol and B+$H_0$+pol+lens, respectively.

\section*{Supernovas}

The late-time acceleration was confirmed by the observation of distant supernovae, as explained in the introduction. The independent reports of two different groups, the High-z Supernova Search Team (HSST) \cite{riess1998observational} and the Supernova Cosmology Project (SCP) \cite{perlmutter1999measurements}, marked a milestone in modern cosmology, confirming the existence of so-called dark energy as the agent responsible for this acceleration. By 1998, the HSST had discovered a set of 16 high-redshift supernovae and a set of 34 nearby supernovae, while the SCP had studied 42 supernovae in the redshift range of 0.18-0.83. The main subject of the study of supernovae was to set constraints on the cosmological parameters.

Supernovae are superluminous events that can be classified into different kinds. Among them, Type Ia supernovae (SnIa) share a nice feature in that their absolute luminosity is approximately constant at the peak of brightness. This makes it possible to relate their apparent luminosity to their distance; they are considered "standard candles" and can be used to construct Hubble diagrams. Type Ia supernovae are characterized by a spectral line of hydrogen and an absorption line of ionized silicon. They are superluminous events that occur due to explosions in white dwarfs composed of carbon and oxygen in binary systems. When the white dwarf absorbs gas from the other star, its mass eventually exceeds the Chandrasekhar limit (approximately 1.44 solar masses), and it explodes in an extremely bright release of energy. The peak absolute magnitude of a SnIa is around M=-19. After correcting the observed magnitudes, we may relate the apparent magnitude ($m$) to the absolute one ($M$) as:
\begin{equation}\label{Eq:Description.Magnitudes}
m-M=5\log_{10} \left( \frac{D_{\rm L}}{1 \textrm{Mpc}}\right)+25\,,
\end{equation}
here $D_{\rm L}$ is the luminosity distance measured in Mpc. In cosmologies based on the FLRW metric, $D_{\rm L}$ is a function of the cosmological parameters, making it of special interest. For instance, for null spatial curvature as assumed along this thesis, the luminosity distance up to redshift $z$ takes the form:
\begin{equation}\label{Eq:Description.LuminosityDistanceLCDM}
D_{\rm L}(z;\vec{\theta})= (1+z)\int_0^z \frac{d\tilde{z}}{H(\tilde{z})}\,.
\end{equation}
Here, $\vec{\theta}$ generically denotes the parameters of the model. Since the peak absolute magnitude is well-known, information regarding the luminosity distance can be inferred from Eq.\,\eqref{Eq:Description.Magnitudes}. In particular, what is fitted against data is the distance moduli,
\begin{equation}\label{Eq:Description.DistanceModuli}
\mu_{\rm p} (z;\vec{\theta}) \equiv 5\log_{10} D_{\rm L}(z;\vec{\theta})+\alpha\,,
\end{equation}
where $\alpha$ can be treated as a nuisance parameter and be marginalized in the corresponding $\chi^2$.

In this chapter we use the full Pantheon likelihood, which incorporates the information from 1048 SNIa \cite{Pan-STARRS1:2017jku}. In addition, we also include the 207 SNIa from the DES survey \cite{DES:2018paw}.

\section*{Large scale structures}

Understanding how large-scale structures evolve in an expanding Universe may lead us to a powerful tool for cosmological model testing.  In this context, an intereseting possibility is to study the quantity conformed by the product $f\sigma_8 (z)\equiv f(z)\sigma_8(z)$, known as {\it weighted linear growth}, where $f$ is the {\it growth rate}, defined as
\begin{equation}\label{Eq:Description.GrowthRate}
f(a,\vec{k})=\frac{d \ln D(a,\vec{k})}{d\ln a},
\end{equation}
being $D$ the {\it growth function} and $a$ the scale factor or, in terms of the density contrast for matter $\delta_{\rm m}\equiv \delta \rho_{\rm m}/\rho_{\rm m}$,
\begin{equation}\label{Eq:Description.GrowthRateDensitycont}
f(a,\vec{k})=\frac{d\ln \delta_{\rm m}(a,\vec{k})}{d \ln a}\,.
\end{equation}
If one assumes General Relativity, the growth rate can be expressed in a simple way as\,\cite{Linder:2007hg}
\begin{equation}\label{Eq:Description.GrowthRateParametrization}
f(a)\approx \Omega_{\rm m}(a)^\gamma\,,
\end{equation}
being $\Omega_{\rm m}(z)$ the energy density fraction of matter and $\gamma\approx 0.55$. This parametrization may not hold in an arbitrary cosmological model with modifications of gravity, as the BD model presented in \hyperref[Chap:PhenomenologyofBD]{Chap.\,\ref{Chap:PhenomenologyofBD}} or type II RRVM of \hyperref[Chap:PhenomenologyofRVM]{Chap.\,\ref{Chap:PhenomenologyofRVM}}. In fact, we described a method for obtaining $f(a)$ at subhorizon scales ($a H\ll k$) directly from \verb!CLASS!, without assuming any parametrization like \eqref{Eq:Description.GrowthRate} at the end of \hyperref[Sect:CosmPerturbations]{Sect.\,\ref{Sect:CosmPerturbations}}. In this case, we observed that $f$ is independent of $k$. The density contrast is seen to satisfy, in general, a second order differential equation,
\begin{equation}\label{Eq:Description.DensityContrastDiff}
\frac{d^2 \delta_{\rm m}}{d a^2}+F(a,\vec{\theta})\frac{d \delta_{\rm m}}{d a}+G(a,\vec{\theta})\delta_{\rm m}=0\,,
\end{equation}
where $F$ and $G$ are functions depending on the free parameters of the model, $\vec{\theta}$, and therefore change from one model to another.

On the other hand, $\sigma_8$ is the root mean square of the mass fluctuation amplitude in spheres of $8 h^{-1}$ Mpc radii at a particular redshift. It is related to the normalization of matter power spectrum. Its expression is
\begin{equation}\label{Eq:Description.Sigma8}
\sigma_8^2 (z)=\frac{1}{2\pi^2}\int_0^\infty dk k^2 P_{\rm m}(z,\vec{k})W^2 (|\vec{k}| R_8)\,,
\end{equation}
where $R_8\equiv 8h^{-1}{\rm Mpc}$ and $W$ is defined as
\begin{equation}\label{Eq:Description.W(x)}
W^2 (x)\equiv \frac{3(\sin(x)-x\cos (x))}{x^3}\,.
\end{equation}
The combination $f\sigma_8$ is measurable from galaxy surveys and its the observable that enters in the analysis. For instance, by measuring the two-point correlation function  of the mass density field or studying redshift space distorsions (RSD) of peculiar velocities as a probe of the gravitational interaction of galaxies and the density constrast. Our dataset for \hyperref[Chap:PhenomenologyofBD]{Chap.\,\ref{Chap:PhenomenologyofBD}} and \hyperref[Chap:PhenomenologyofRVM]{Chap.\,\ref{Chap:PhenomenologyofRVM}} is composed by a set of measures of $f\sigma_8(z_i)$ for different effective redshifts $z_i$. The datapoints entering in our analysis of \hyperref[Chap:PhenomenologyofBD]{Chap.\,\ref{Chap:PhenomenologyofBD}} are explicitly written in \hyperref[Table:BD.fs8]{Table\,\ref{Table:BD.fs8}} in \hyperref[Sect:MethodData]{Sect.\,\ref{Sect:MethodData}}, together with corresponding references and more details. For chapter \hyperref[Chap:PhenomenologyofRVM]{Chap.\,\ref{Chap:PhenomenologyofRVM}}, we use a similar dataset, with some differences remarked at the caption of \hyperref[Table:PhenomenologyRVM.Fit1]{Table\,\ref{Table:PhenomenologyRVM.Fit1}}.

\section*{Gravitational lensing}

General Relativity states that the presence of mass can bend the path of a nearby light ray. For instance, the image of a distant galaxy behind a massive structure in the same line of sight can be distorted. When this effect is accentuated it can produce noticiable distorsions such as Einstein rings or multiple images of the same source and we generically call this phenomena {\it Strong Lensing}. On the other hand, when the magnitude of these effects is not as significant the distorsions can only be tracked by statistical analysis to measure the {\it cosmic shear} around a particular region produced by this {\it Weak Lensing}. All together both Strong and Weak gravitational lensing can constitute an alternative tool for parameter estimation.

\subsection*{Weak gravitational lensing}

The Kilo-Degree Survey (KiDS) has measured the Weak-Lensing statistical distortion of angles and shapes of galaxy images caused by the presence of inhomogeneities in the low-redshift Universe \cite{Hildebrandt:2016iqg,Joudaki:2017zdt,Kohlinger:2018sxx,Wright:2020ppw}. The two-point correlation functions of these angle distortions are related to the power spectrum of matter density fluctuations, and from it it is possible to obtain constraints on the parameter combination $S_8=\sigma_8\sqrt{\Omega_{\rm m}/0.3}$. It seems not to perform great in general for non-$\Lambda$CDM scenarios, as nonlinear effects for small angular scales are calculated with the Halofit module \cite{Takahashi:2012em}, which only works fine for the GR-$\Lambda$CDM and minimal extensions of it, as the aforementioned XCDM \cite{Turner:1998ex} and also for the CPL parametrization of the DE EoS parameter\, \cite{Chevallier:2000qy,Linder:2002et}. Thus, it is not able to model accurately the potential low-scale particularities of the BD-$\Lambda$CDM model in our analysis in  \hyperref[]{chapter\,\ref{Chap:PhenomenologyofBD}}. 

\subsection*{Strong gravitational lensing}

As part of one of our non-baseline datasets in \hyperref[Chap:PhenomenologyofBD]{Chap.\,\ref{Chap:PhenomenologyofBD}} we use the data extracted from the six gravitational lensed quasars of variable luminosity reported by the H0LICOW team. They measure the time delay produced by the deflection of the light rays due to the presence of an intervening lensing mass. After modeling the gravitational lens it is possible to compute the so-called time delay distance $D_{\Delta{t}}$ (cf. \cite{Wong:2019kwg} and references therein). The fact of being absolute distances and not relative ones (as for the SNIa and BAO datasets) {allows them to directly constrain the Hubble parameter in the context of the GR-$\Lambda$CDM as follows:  $H_0=73.3^{+1.7}_{-1.8}$ km/s/Mpc} . It turns out that for the three sources B1608+656, RX51131-1231 and HE0435-1223, the posterior distribution of $D_{\Delta t}$ can be well approximated by the analytical expression of the skewed log-normal distribution,
\begin{equation}\label{Eq:BD.skewed}
\mathcal{L}(D_{\Delta t}) = \frac{1}{\sqrt{2\pi}(D_{\Delta t} - \lambda_D)\sigma_D}\,{\rm exp}\left[-\frac{(\ln(D_{\Delta t} -\lambda_D) - \mu_D)^2}{2\sigma^2_D}\right]\,,
\end{equation}
where the corresponding values for the fitted parameters $\mu_D$, $\sigma_D$ and $\lambda_D$ are reported in Table 3 of \cite{Wong:2019kwg}. On the other hand, the former procedure cannot be applied to the three remaining lenses, {\it i.e.} SDSS 1206+4332, WFI 2033-4723 and PG 1115+080. From the corresponding Markov chains provided by H0LICOW\footnote{\url{http://shsuyu.github.io/H0LiCOW/site/}} we have instead constructed the associated analytical posterior distributions of the time delay angular diameter distances for each of them. Taking advantage of the fact that the number of points in each bin is proportional to $\mathcal{L}(D_{\Delta t})$ evaluated at the average $D_{\Delta t}$ for each bin, we can get the values for $-\ln(\mathcal{L}/\mathcal{L}_{\rm max})$ and fit the output to obtain its analytical expression as a function of $D_{\Delta t}$. The fitting function can be as accurate as wanted, e.g. using a polynomial of order as high as needed. The outcome of this procedure is used instead of \eqref{Eq:BD.skewed} for the three aforesaid lenses.

\section*{BAO}
Probes of Primordial baryon acoustic oscillations (BAO) are also a fundamental tool in modern cosmology for imposing constraints to the parameters of the $\Lambda$CDM and its extensions. The strongly coupled Photon-Plasma fluid that almost homogeneously filled the early Universe was a medium for the propagation of acoustic oscillations resulting from the interplay between gravity and radiation pressure. After the Universe cooled down, photons decoupled from the fluid and acquired a huge mean free path bigger than the Hubble distance, in what is know as {\it recombination} ($z_{\rm dec}\approx 1100$). After this scenario, photon pressure cannot prevent gravitational instability and overdensities are formed, in which baryonic (and also dark matter due to gravitational interaction) were distributed around the Universe. At the drag epoch, when baryons effectively decouple from photons after recombination ($z_{\rm d}\lesssim z_{\rm dec})$, these overdensities had a comoving characteristic scale
\begin{equation}\label{Eq:Description.SoundHorizon}
r_{\rm s}(z_{\rm d})=\int_{z_{\rm d}}^\infty \frac{c_{s}(z)}{H(z)}dz\,,
\end{equation}
where $c_{\rm s}(z)=(3(1+3\rho_{\rm b}(z)/4\rho_\gamma(z)))^{-1/2}$ is the speed sound in the fluid, depending on the energy densities of photons and baryons. Its value is $r_{\rm s}\approx 150$ Mpc, and it is called {\it sound horizon} at the drag epoch. As a consequence, after the oscillations became frozen, there is an imprint in the CMB in the form of acoustic peaks in power spectrum. Additionally, it can be trace in the matter distribution, from the oscillations in the matter power spectrum (in Fourier space) or by the peak in the 2-point correlation functions around the BAO scale (in real space). 

More in detail, the matter power spectrum $P_{\rm m}(\vec{k},z)$ can be obtaned from galaxy surveys, which measure the angular (perpendicular to the line of sight) and redshift (along the line of sight) distributions of galaxies. Translating this information to distances depends on the cosmological model, which is usually the concordance model. For this study the comoving vector $\vec{k}$ is decomposed in a parallel to the line of sight component, $\vec{k}_{\parallel}$ and a transverse one, $\vec{k}_{\bot}$. The BAO standard ruler from the analysis of the 2D power spectrum serves as a measurement of the angular diamater distance as a function of redshift,
\begin{equation}\label{Eq:Description.AngularSeparation}
\Delta \theta_{\rm s} (z)=\frac{r_{\rm s}(z_{\rm d})}{(1+z) D_{\rm A}(z)}\,
\end{equation}
where $D_{A }$ is the proper diameter angular distance,
\begin{equation}
D_{\rm A}(z)\equiv \frac{1}{1+z}\int_0^z \frac{dz^\prime}{H(z^\prime )}\,.
\end{equation}
The BAO standard ruler can also be applied along the line of sight to obtain redshift separations:
\begin{equation}\label{Eq:Description.RedshiftSeparation}
\Delta z_{\rm s} (z)=r_{\rm s} (z_{\rm d})H (z)\,.
\end{equation}
Galaxy surveys impose constraints on the former quantities and to extract anisotropic BAO information. To be precise, a comparison with the fiducial cosmology the survey has used to convert angles and redshift to distance (we denote it by "fid.") is performed. We then define the perpendicular and parallel dilation scale factors:
\begin{equation}\label{Eq:Description.DilatationFactors}
\begin{split}
&\alpha_{\bot}\equiv  \frac{\Delta \theta_{\rm s}^{\rm fid.} (z)}{\Delta \theta_{\rm s} (z)}=\frac{r_{\rm s}^{\rm fid.} (z_{\rm d})/D_{\rm A}^{\rm fid.}(z)}{r_{\rm s} (z_{\rm d})/D_{\rm A}(z)}\,,\\
&\alpha_{\parallel} \equiv \frac{\Delta z_{\rm s}^{\rm fid.}(z)}{\Delta z_{\rm s}(z)}=\frac{r_{\rm s}^{\rm fid.} (z_{\rm d}) H^{\rm fid.} (z)}{r_{\rm s}(z_{\rm d}) H(z)}\,,
\end{split}
\end{equation}
and, hence, we can extract information of $H(z)$ and $D_{\rm A}(z)$. But BAO DATA is not only formulated in terms of these observables. Sometimes, when the data do no let us disentangle the perpendicular and parallel directions, it is convenient to define the volume-averaged spectrum in a particular volume $V$. The associated observable is in this case $r_{\rm s} (z_{\rm d})/ D_{\rm V}(z)$, where $D_{\rm V}(z)$ is the so--called dilation scale\,\cite{SDSS:2005xqv}, defined as
\begin{equation}\label{Eq:Description.Dilationscale}
D_{\rm V} (z)\equiv \left[ z D_{\rm M}^2 (z) D_{\rm H} (z) \right]^{1/3}\,.
\end{equation}
Here $D_{\rm M} (z)\equiv (1+z)D_{\rm A} (z)$ is the comoving angular diameter distance and $D_{\rm H}$ is the Huble radius, $D_H (z)\equiv 1/H(z)$. Similarly, we define a isotropic dilation scale factor as\,\cite{Beutler:2011hx}
\begin{equation}\label{Eq:Description.Istropicfactor}
\alpha_{\rm V}\equiv \frac{D_{\rm V}^{\rm fid.}(z)}{D_{\rm V}^{\rm fid.} (z)}\,,
\end{equation}
or, in other cases\,\cite{Kazin:2014qga,Ross:2014qpa}, as
\begin{equation}\label{Eq:Description.Isotropicfactor2}
\alpha_{\rm V} \equiv \frac{r_{\rm s}^{\rm fid.} (z_{\rm d})/D_{\rm V}^{\rm fid.}(z)}{r_{\rm s}(z_{\rm d})/D_{\rm V} (z)}\,.
\end{equation}
As a difference from the anisotropic dilation factors, which provide us information of $D_{\rm A} (z)$ and $H(z)$, for the estimators $\alpha_{\rm V}$ we cannot disentangle this information and as a consequence is not as valuable at the time of imposing constraints to the cosmological parameters. 

All in all we constructed a complete dataset conformed by anistropic and isotropic data from different surveys and tracers for BAO data. The particular data points together with the corresponding observable used for the analysis can be found in \hyperref[Table:BD.BAO]{Table\,\ref{Table:BD.BAO}}. As a final comment, galaxy surveys provide measurements of the BAO standard ruler distance measurements, but sometimes are combined with measurements of the growth of structure in the Universe obtained from the Redshift Space Distorsion (RSD) signature\,\cite{Gil-Marin:2016wya}. This implies underlying correlations that are need to be carefully accounted for when analyzing the data and interpreting the cosmological constraints.

 \section*{Cosmic chronometers}

The {\it Cosmic chronometers} (or cosmic clocks) method, allows us to obtain direct measurements of the Hubble parameter at different redshifts, $H(z_i)$, rather than inferring its value indirectly from observables such as the luminosity distance \cite{Jimenez:2001gg}. The method relies on the estimation of the variation of redshift with cosmic time, $dz/dt$. To do that, we make use of the aforementioned chronometers: spectroscopic datings of galaxy ages. Two nearby galaxies separated by $\Delta z$ in redshift space, with a relative age difference of $\Delta t$, are used to estimate the derivative from the ratio $\Delta z /\Delta t$. The galaxies selected for this method should possess similar metallicity and have a poor star formation rate with an old stellar population so that they are not fast-evolving. By studying the relative ages, we can find the Hubble rate,
\begin{equation}\label{Eq:Description.HubbleRateFromRatio}
H(z)\approx -\frac{1}{1+z}\frac{\Delta z}{\Delta t}\,,
\end{equation}
in a cosmological model-independent way. The observables we compare with the theory are then the $H(z_i)$ values. It is worth mentioning that, although relying on the theory of spectral evolution of galaxies, they are uncorrelated with the BAO data points. In our analysis in chapters \hyperref[Chap:PhenomenologyofBD]{\ref{Chap:PhenomenologyofBD}} and \hyperref[Chap:PhenomenologyofRVM]{\ref{Chap:PhenomenologyofRVM}}, we use a selection of 31 data points extracted from Table 2 of \cite{Gomez-Valent:2018gvm}.

\section*{$H_0$ prior from SH0ES collaboration}

Keeping in mind the tension surrounding the $H_0$ parameter, we explored an alternative dataset beyond our Baseline scenarios by examining the effect of incorporating a prior on the value of the $H_0$ parameter reported by the SH0ES collaboration, which is $73.5\pm 1.4 {\rm km/s/Mpc}$ in \cite{Reid:2019tiq}. This value was obtained through the estimation of the angular-diameter distance to the galaxy NGC 4258, by studying the dynamics of water masers near the galaxy nucleus. In conjunction with other geometric calibrators, such as Milky Way parallaxes and detached eclipsing binaries in the Large Magellanic Cloud, the distance ladder of Cepheids/SnIa can be calibrated to obtain a determination of $H_0$. The distance ladder is a concatenation of methods for measuring distance to farther objects in which each step serves as a calibrator for the next.

Our interest in this prior is to observe how the models respond when combined with the CMB dataset considered in our Baseline analysis. The results reported in\,\cite{aghanim2020planck} under TT,TE,EE+lowE +lensing data yield the value $H_0=67.36\pm 0.54 {\rm km/s/Mpc}$, with a disagreement of $4.1\sigma$ with SH0ES' value. This significant discrepancy between the two values has sparked considerable interest in the scientific community and has motivated the search for a new cosmological framework that can reconcile the discrepancy within an analysis including CMB data. Therefore, a model that can deal with this discrepancy is highly promising and could provide new insights into the nature of our Universe.

\end{appendices}
\pagebreak
\addcontentsline{toc}{chapter}{Bibliography}
\bibliography{Bibliography} 
\bibliographystyle{ieeetr}

\end{document}